\documentclass[12pt]{thesis}
\usepackage{epsfig,amssymb,amsmath,rotating,graphics}
\usepackage{indentfirst}
\usepackage{graphicx}
\usepackage{natbib}

\usepackage[colorlinks=true,linkcolor=black,citecolor=blue,urlcolor=blue]{hyperref}
\def\lsim{\lower.5ex\hbox{$\; \buildrel < \over \sim \;$}}
\def\gsim{\lower.5ex\hbox{$\; \buildrel > \over \sim \;$}}

\usepackage{makeidx}
\makeindex
\usepackage[explicit]{titlesec}
\usepackage{lipsum}
\usepackage{type1cm}
\usepackage{xcolor}
\usepackage{pdflscape}
\usepackage{marvosym}
\usepackage[T1]{fontenc}
\usepackage{wrapfig}

\titleformat{\chapter}[display]
  {\normalfont\Large\rmfamily}
  {\sffamily\flushright\fontsize{60}{0}\textbf{\textcolor{blue!70}{{\Huge\chaptername}~\thechapter\vskip0pt\rule{\textwidth}{2pt}}}}{0pt}
  {\flushleft\fontsize{30}{0}{\Huge\textcolor{blue!80}{#1}}\vskip15pt}{\huge}
\titlespacing*{\chapter}
  {0pt}{-40pt}{0pt}

\begin{document}
\setcounter{section}{0}
\setcounter{figure}{0}
\setcounter{table}{0}
\vskip 3cm
\thispagestyle{empty}

\centerline{\Huge\bf Extraction Of Physical Properties}
\vskip 0.5cm
\centerline{\Huge\bf Of Interstellar Medium}
\vskip 0.5cm
\centerline{\Huge\bf From The Observed Line Profiles}

\vskip 3.5cm
\begin{center}
{\large \bf Thesis submitted for the degree of\\
        {\Large Doctor of Philosophy} (Science)\\
            in Physics (Theoretical) \\}
\end{center}

\vskip 2.5cm

\centerline{\large by}
\vskip 0.5cm
\centerline{\Large\bf Ms. Bratati Bhat}

\vskip 5.0cm
\centerline{\large \bf Department of Physics}
\vskip 0.5cm

\centerline{\Large \bf University of Calcutta}

\vskip 0.5cm
\centerline{\large\bf 2023}

\newpage

\thispagestyle{empty}

\vspace*{8cm}

\begin{center}
{\Large \it Dedicated to my parents.}
\end{center}

\newpage
\pagestyle{newheadings}
\pagenumbering{roman}
\setcounter{page}{1}

\chapter*{ABSTRACT}

Since molecules are ubiquitous in space, the study of the
`Molecular Universe' could unfold the mystery of the existing Interstellar medium. Star formation is linked to the chemical evolution processes. Thus, an analysis of the formation of stars coupled with the chemical evolution would give a clear insight into the entire process. For example, various evolutionary
stages of star formation could be probed by observing various
molecules. Chemical diagnostics of these regions could be used to extract the physical properties (e.g., density, temperature, ionization degree, etc.) of these regions. Radiative transfer calculations are worthwhile in estimating physical parameters of the region where molecules are detected. However, the radiative transfer calculations are limited due to insufficient molecular data, such as spectroscopic information or collisional excitation probabilities of many interstellar species. Complex organic molecules are detected in various environments ranging from the cold gas in prestellar cores to the warm gas on solar system scales close to individual protostars. A comparative study of the relative abundances of molecules could provide insights into the beginning of chemical complexity and the link to our solar system.
In my thesis, I would mainly investigate the physical properties
and kinematics of different star-forming regions using radiative
transfer modeling. The observed spatial differentiation between various key molecules is used to explain their physical structure or evolution and various microphysical effects. In addition, some key molecules are used to study the various evolutionary phases. This simulated data is useful for interpreting the observed data of
different telescopes like IRAM 30m, GBT, ALMA, Herschel, SOFIA, etc.

Chapter 1 presents a brief history of the interdisciplinary subject “Astrochemistry” a new branch of astrophysics. It also discusses various stages of star formation and different astrochemical aspects to study the conditions of the ISM. Observational technique and detailed modeling technique are discussed to understand the chemical compositions of various star-forming regions. We briefly explain the role of radiative transfer to reveal the mystery of our “molecular Universe”.

Chapter 2 discusses the detailed method of Markov Chain Monte Carlo (MCMC) simulation to obtain the physical characteristics from the observed line profiles of different COMs. We implemented MCMC method to fit different molecular lines observed towards different hot molecular cores (G31.41, G10.47 etc.) to infer the physical characteristics and the excitation conditions (temperature, FWHM) of different sources.

Chapter 3 presents examples of observations of COMs in the hot molecular core (HMC, G31.41+0.31). Along with the chemical composition, the physical conditions
associated with this source are obtained using detailed radiative transfer modeling.

Phosphorus (P) related species are not known to be as omnipresent in space as hydrogen, carbon, nitrogen, oxygen, and sulfur-bearing species. Astronomers spotted very few P-bearing molecules in the interstellar medium and circumstellar envelopes. In Chapter 4, we showed a radiative transfer model to investigate the transitions of some of the P-bearing species in the diffuse cloud and hot core regions and estimate the line profiles. 

To investigate the evolution of a Sun-like protostar, understanding the COMs’ physical and chemical origins would be helpful by observing a variety of star-forming regions. In Chapter 5, data from the Large Programme ’Astrochemical Surveys At IRAM’ (ASAI), using the IRAM 30 m telescope, is analyzed, and several COMs were identified. It was a millimeter line survey that included the protoplanetary disc phase, outflow region, prestellar core, and protostars. We chose four sample sources, namely L1544 (prestellar core), B1b (first hydrostatic core), IRAS4A (class 0), SVS13A (class I), to understand the typical evolution of COMs as it is not possible to examine all the stages of a single star.

\chapter*{ACKNOWLEDGMENTS}

{\it Ph. D. Is not about just getting a degree; it is a roller coaster journey to learn about your passion for research. This journey could not be possible without the love and constant support of many people. I want to acknowledge all of them from the bottom of my heart.

First and foremost, I want to acknowledge my supervisor Dr. Ankan Das and my joint supervisor Prof. Sandip K. Chakrabarti. Words would fail to say how I am grateful to him for his constant guidance, encouragement, and support and for introducing me to the beautiful world of Astrochemistry. He always motivated and encouraged me throughout my Ph.D. period and guided me to handle different new problems related to Astrochemistry. His dedication to research, responsibility toward students, and friendly behavior always makes me surprised. I learned a lot of things from him, which I will cherish throughout my life. He always cared as an elder brother and helped me through many difficult times during my Ph. D. I am also thankful to my joint supervisor Prof. Sandip K. Chakrabarti. His dedication and knowledge of science always motivated me. I am thankful to him for many important scientific suggestions throughout my Ph. D.

I want to thank my Astrochemistry department at ICSP for creating such a wonderful homely atmosphere for research. I always feel it is a family and a home-like atmosphere to share each and every hard and good time together. I want to thank my labmates, Suman Mondal and Rana Ghosh. We had many wonderful scientific discussions together during this period which always helped me to learn. I want to thank my former senior lab mates, Dr. Prasanta Gorai and Dr. Milan Sil. Your hard work and dedication toward research have always motivated me in a different way. I always get help and support from my seniors as an elder brother whenever I ask for it.

This research was performed in collaboration with other scientists all over the world. My thanks go to Professor Paola Caselli, Dr. Sergio Ioppolo, Dr. Takashi Shimonishi, Dr. Naoki Nakatani, Dr. Kenji Furuya, Dr. Bhalamurugan Sivaraman, and Dr. Amit Pathak. They were always so helpful and provided me with their support and collaboration throughout my dissertation. A special thanks to Dr. Takashi Shimonishi, sir, for his wonderful hospitality during our stay in Japan (JSPS collaboration). The wonderful scientific discussions during the visit enriched my knowledge of this subject a lot. I also want to acknowledge the COSPAR scientific assembly for the financial support to attend the 44th COSPAR in person, held in Athens, Greece. It was my first international travel as well as the first international conference I attended in person.

I also want to thank Dr. Dipak Debnath, Dr. Sudipta Sasmal, and Dr. Ritabrata Sarkar for all their support and cooperation. I also learned many things from them. At ICSP, we spent very beautiful times together and have many wonderful memories to carry. I want to thank Dr. Kaushik Chatterjee, Abhijeet Roy, Sujoy Nath, Riya Bhowmick, Subrata Kundu, and Swati Chowdhury Chatterjee. We had many beautiful times together and enjoyed ourselves a lot. A special thanks to Kaushik, who helped me in many difficult times and supported me. I also want to thank my senior scholars of ICSP, Arka da, Dusmanta da, Suman Chakraborty da, Argha da, Debjit da, Soujan da, and some juniors Abhrajit, Binayak, Rupnath, Sagar, Ashim, Shyam, and Pabitra. I also want to thank some of my friends from school, college, and other places – Aritra Bandyopadhyay, Sanchari Bhattacharya, Abhisikta Barman, Ronson Dey, Biswajit Ghosh, Ananya Das, Jagannath Das, Panchami Roy and many more. I want to thank all the office members of ICSP, Rajkumar Maiti, Ram Chandra Das, Jyotisman Moitra, and Uttam Sardar.

A special thanks to my Ph. D. time roommate Pallabi Dutta Choudhury and her mother, Maya Dutta Choudhury. I will always cherish the memory of our stay in Kolkata. I will always be grateful to you for how you helped me and understood me to heal my wounds. Aunty always makes me feel the room like my home away from home, and I will always be grateful for her love and care.

I want to dedicate this thesis to my parents. I am always indebted to my mother, Mrs. Santi Bhat, who always supported me in higher studies. In our society, being a woman, it is always extra hard to possess research as a career and to continue your passion. My mother always supported me in every possible way to complete my Ph. D. and follow my dreams. Nothing would be possible without your prayers, sacrifice, love, and care. I want to thank my father Late Sankar Lal Bhat, for his blessings. I always followed you to prepare myself as a human being. I want to thank my elder sister Mrs. Aditi Bhat for her love and support, and she always plays a vital role in motivating me to complete my Ph.D and pursue research as a career. I want to thank my brother Mr. Sourav Bhat for his love and support in completing my Ph.D. and for always being there in my bad times. I also want to thank Mr. Abhishek pal and  Mr. Angshuman Mukherjee and many other family members who supported me during my Ph.D.

Last but not least, I am thankful to the Department Of Science and Technology (DST) inspire fellowship for the financial support. I would also like to acknowledge the Indian Centre for Space Physics for allowing me to work here. This journey could not be so beautiful without this support.

\begin{flushright}
Bratati Bhat \\
July 2023, Kolkata.
\end{flushright}
}

\chapter*{Scientific contributions}

\section*{Publications}

\subsection*{\bf List of Publications in Peer Reviewed Journals}
\begin{enumerate}
\item \href{https://doi.org/10.1016/j.asr.2021.07.011}{\bf Radiative transfer modeling of the observedline profiles in G31.41+0.31}, \underline{\bf Bratati Bhat}, Prasanta Gorai, Suman K. Mondal, Sandip, K, Chakrabarti, \& Ankan Das, 2021, \emph{Advances in Space Research}. \textbf{(Journal Impact Factor 2022: 2.152)} \\

\item \href{https://doi.org/10.3847/1538-4357/aa984d}{\bf Chemical Modeling for Predicting the Abundances of Certain Aldimines and Amines in Hot Cores}, Milan Sil, Prasanta Gorai, Ankan Das, \underline{\bf Bratati Bhat}, \& Sandip, K, Chakrabarti, 2018, \emph{The Astrophysical Journal}, 853, 2. \textbf{(Journal Impact Factor 2022: 5.874)} \\

\item \href{https://doi.org/10.3847/1538-4357/ab8871}{\bf Identification of Pre-biotic Molecules Containing Peptide-like Bond in a Hot Molecular Core, G10.47+0.03}, Prasanta Gorai, \underline{\bf Bratati Bhat}, Milan Sil, Suman K. Mondal, Rana Ghosh, Sandip K. Chakrabarti, \& Ankan Das, 2020, \emph{The Astrophysical Journal}, 895, 86. \textbf{(Journal Impact Factor 2022: 5.874)} \\

\item \href{https://doi.org/10.3847/1538-4357/abc9c4}{\bf Identification of Methyl Isocyanate and Other Complex Organic Molecules in a Hot Molecular Core, G31.41+0.31}, Prasanta Gorai, Ankan Das, Takashi Shimonishi, Dipen Sahu, Suman K. Mondal, \underline{\bf Bratati Bhat}, \& Sandip K. Chakrabarti, 2021, \emph{The Astrophysical Journal}, 907, 108. \textbf{(Journal Impact Factor 2022: 5.874)} \\

\item \href{https://doi.org/10.3847/1538-4357/abb5fe}{\bf Exploring the Possibility of Identifying Hydride and Hydroxyl Cations of Noble Gas Species in the Crab Nebula Filament}, Ankan Das, Milan Sil, \underline{\bf Bratati Bhat}, Prasanta Gorai, Sandip K. Chakrabarti, \& Paola Caselli, 2020, \emph{The Astrophysical Journal}, 902, 131. \textbf{(Journal Impact Factor 2022: 5.874)} \\

\item \href{https://doi.org/10.3847/1538-3881/ac09f9}{\bf Chemical complexity of phosphorous bearing species in various regions of the Interstellar medium}, Milan Sil, Satyam Srivastav, \underline{\bf Bratati Bhat}, Suman Kumar Mondal, Prasanta Gorai, Rana Ghosh, Takashi Shimonishi, Sandip K. Chakrabarti, Bhalamurugan Sivaraman, Amit Pathak, Naoki Nakatani, Kenji Furuya, Ankan Das, 2021, \emph{The Astronomical Journal}, 162, 119. \textbf{(Journal Impact Factor 2022: 6.263)} \\

\item \href{https://www.aanda.org/articles/aa/full_html/2023/01/aa43802-22/aa43802-22.html}{\bf Investigating the hot molecular core, G10.47+0.03, a pit of nitrogen-bearing complex organic molecules}, Suman Kumar Mondal, Wasim Iqbal, Prasanta Gorai, \underline{\bf Bratati Bhat}, Valentine Wakelam, Ankan Das, 2022, \emph{Astronomy \& Astrophysics}. \textbf{(Journal Impact Factor 2022: 5.802)} \\

\item \href{https://iopscience.iop.org/article/10.3847/1538-4357/acfc4d}{\bf Chemical Evolution of Some Selected Complex Organic Molecules in Low-mass Star-forming Regions}, \underline{\bf Bratati Bhat}, Rumela Kar, Suman Kumar Mondal, Rana Ghosh, Prasanta Gorai, Takasi Shimonishi, Kei Tanaka, Kenji Furuya, Ankan Das, 2023, \emph{The Astrophysical Journal}, 958, 111. \textbf{(Journal Impact Factor 2022: 5.874)} \\

\end{enumerate}

\subsection*{\bf Publication in Proceedings}
\begin{enumerate}
 \item \href{https://link.springer.com/chapter/10.1007\%2F978-3-319-94607-8_39}{\bf  Radiative Transfer Modeling of Some Relevant Interstellar Species}, 
\underline{\bf Bratati Bhat}, 2018, Exploring the Universe: From Near Space to Extra-Galactic, 503-510, Astrophysics and Space Science Proceedings, vol 53. Springer, Cham., Online ISBN:978-3-319-94607-8.
\end{enumerate}

\clearpage

\section*{Oral presentations}

\begin{enumerate}

\item
Astrochemistry in the THz domain, October 2017,
Chennai, India

Title: ``Radiative transfer modeling to extract the physical parameters from the observed line profiles''

\item
\href{https://www.bose.res.in/Conferences/EXPUNIV2018/}{Exploring the Universe: Near Earth Space Science to
Extra-Galactic Astronomy}, November 2018, Kolkata, India

 Title: ``Radiative transfer modeling of some observable Interstellar species''
 
\item
43rd COSPAR Scientific Assembly (COSPAR-2021-Hybrid), 28 January - 4 February 2021, Sydney Australia.

 Title: ``\href{https://ui.adsabs.harvard.edu/abs/2021cosp...43E1921B}{Radiative transfer modeling to explain the observed Inverse P-Cygni profile in a high mass star-forming region}''
 
 \item
 Atomic Molecular and Optical Physics Division Seminar, 29 July 2021 (virtually).

Title: ``Radiative transfer modeling to explain the observed line profiles of a hot molecular core.''

\item
44th COSPAR Scientific Assembly (COSPAR-2022), 16 - 24 July 2022, Athens, Greece (Attended in person).

 Title: ``\href{https://ui.adsabs.harvard.edu/abs/2022cosp...44.2750B/abstract}{Radiative transfer model to explain the observed line profiles of a hot molecular core, G31.41+0.31}''

 \end{enumerate}

\section*{Poster presentations}

\begin{enumerate}
\item
\href{https://sites.google.com/site/irastrondust2019/home}{International Conference on Infrared Astronomy and Astrophysical Dust}, October 2019, IUCAA Pune, India

 Title: ``Radiative transfer modeling to extract the physical parameters from the observed line profile''
 
\item
42nd COSPAR Scientific Assembly (COSPAR-2018), 14 July - 22 July 2018, Pasadena, California.

 Title: ``\href{https://ui.adsabs.harvard.edu/abs/2018cosp...42E.322B}{Radiative transfer modeling of some potentially observable Interstellar species}''

 \end{enumerate}

\newpage
\tableofcontents
\listoffigures
\listoftables

\newpage
\pagestyle{myheadings}
\pagenumbering{arabic}

\mainmatter
 \chapter{Introduction} \label{chap:intro}
\section{What is Astrochemistry?}
Astrochemistry is the study of chemical elements and molecules throughout the vast Universe. It studies the abundance and reactions of chemical species in the Universe and their interaction with radiations (\url{https://en.wikipedia.org/wiki/Astrochemistry}), which includes the study of the formation, destruction, and excitation of molecules in astronomical environments and their influence on the structure, dynamics, and evolution of astronomical objects. Astrochemistry may also be termed Molecular Astrophysics \citep{dalg08}. It is a highly interdisciplinary subject involving scientists from different research areas, from laboratory-based experiments and telescopic observations at different wavelengths to theoretical modeling to understand the physical and chemical processes.
\section{Historical overview of Astrochemistry}
After the Big Bang, the chemical processes of the universe were initiated. Matter, which consists of electrons, protons, neutrons, and photons, is created within the first four seconds of the universe through pair creation. For example, deuterium was created in the first two minutes of the universe's existence. As a result, the average photon energy at that moment equals the deuterium binding energy, 1.8$\times10^{14}$ Joule. Thus it is destroyed more quickly than it was created. Therefore, deuterium breaks into its parts when photons strike it. After the first three minutes of the Big Bang, when the temperature fell much below 10$^{10}$ Kelvin, hydrogen, helium, and a small number of lithium nuclei were formed. The recombination era started at redshift, Z=1100.

Matter and radiation were separated beyond the recombination era. After the Big Bang, electrons were only confined in orbits around nuclei for about 3,80,000 years before the first atom was formed. Then, for the first time, any molecular bond formed during the period when the universe's average temperature fell to around $\sim$ 4000 degrees Kelvin. The first molecule to form in the universe was the helium hydride ion (HeH$^+$), followed by H$_2$, the first neutral molecule.

\begin{figure}[htbp]
\begin{center}
\includegraphics[width=\textwidth]{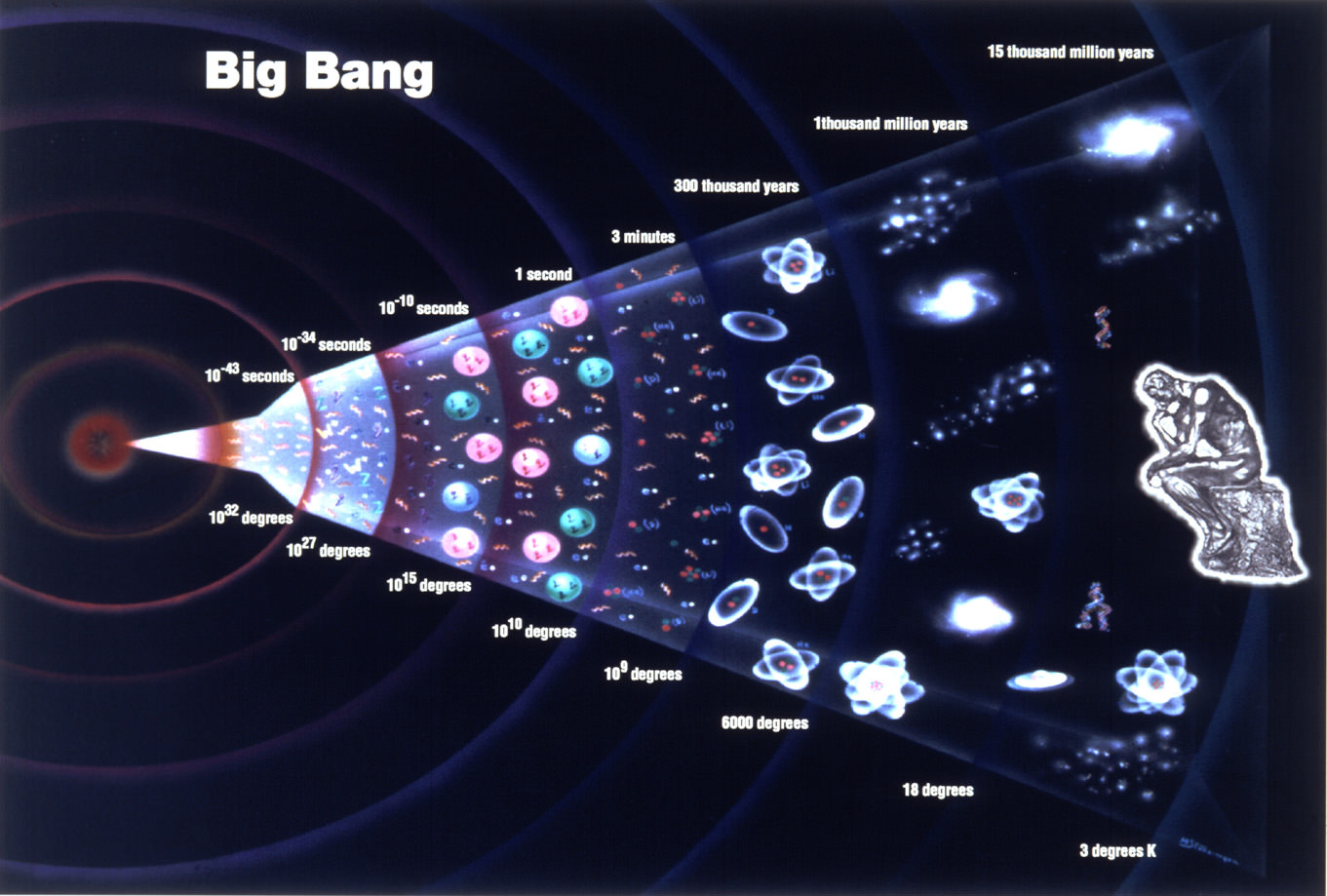}
\caption{History of Universe from Big Bang to life formation. 
\citep[Courtesy:][\url{https://www.universetoday.com/54756/what-is-the-big-bang-theory/}]{}
}
\label{fig:bigbang}
\end{center}
\end{figure}

Recently,  \cite{gust19} has reported the first detection of 
HeH$^{+}$  in the young and dense planetary nebula NGC 7027 situated at the constellation Cygnus. All other molecules start to evolve after the formation of H$_2$. It is the most abundant interstellar molecule in the Universe. CO is the second most abundant species after H$_2$ and is also ubiquitous. It is regarded as a tracer of interstellar molecular gas. Figure \ref{fig:bigbang} depicts the development of our Universe from the time of the Big Bang to the present.

Thomas Alva Edison first attempted to receive radio waves from any part of the sky in the early 1890s. Between 1930 and 1932, Karl Jansky, a telephone engineer working for Bell Laboratories in America, was interested in finding the radio wave interference at 20.5 Mz. Jansky constructed an antenna and found three sources of radio static : (1) nearby thunderstorms, (2) distant thunderstorms, and (3) a faint, steady hiss of unknown origin. Jansky spent over a year investigating the third type of static and figured out that radiation was coming from the Milky Way. Radio astronomy began just before the Second World War and matured in the 1950s, mainly through the pioneering efforts of scientists with backgrounds in radio science, electrical engineering, or wartime radar. Radio observation sheds light on the physical condition of the star-forming regions. After discovering HI 21 cm line (1420.4 MHz) in space, the millimeter and sub-millimeter observation focus on the possible identification of more and more species through radio telescopes. This kind of study got its required momentum when OH radical was confirmed using the 84-ft. parabolic antenna of the Millstone Hill Observatory of Lincoln Laboratory \cite{wein63}. In the 1970s, radio astronomers started building new facilities to explore this exciting area. They developed new techniques for Very long baseline interferometry, millimeter-wavelength spectroscopy, fast data acquisition, and signal processing. The first carbon-containing simple and the second
most abundant molecule in space, CO, was identified in the Orion nebula by \cite{wils70} using National Radio Astronomy Observatory (NRAO) 36-feet telescope. Among the 270 molecules identified to date, $70-80\%$ were observed in the mm and sub-mm regions. Experiments with long baseline interferometry led to construction of the transcontinental Very Long Baseline Array. At the same time, astronomy in the millimeter-wave has opened up new windows to understanding the evolution of stars, the Universe itself, and galaxies in a better way. The composition of the interstellar medium, the earliest stages of star formation, and the internal kinematics of luminous galaxies are uniquely revealed at millimeter wavelengths. The present array-type radio telescopes of the millimeter and sub-millimeter wavelengths significantly improve the sensitivity and resolution. Millimeter wavelength observations of the gaseous envelopes around ancient stars give insight into their morphology, dynamics, and abundance. The improved sensitivity and resolution at millimeter and sub-millimeter wavelengths also led to a much better understanding of star-forming regions' structure, dynamics, and chemistry.
During the 1990s, the massive star-forming regions such as Orion-KL (d$\sim$400 pc) and Sgr B2 (d$\sim$8kpc) had been prime targets for line surveys \cite{turn91,hwan08,luca90}. Among the modern facility, Green Bank Telescope (GBT), IRAM 30m, APEX, MOPRA telescope, etc., are the single-dish telescope. In contrast, the major interferometric facilities are Atacama Large Millimeter Array (ALMA) and PDBI (presently known as NOEMA). 

The origin of life is a long-standing mystery. Scientists are curious about how life emerged on Earth and how the biomolecules are synthesized in space and delivered to the early Earth. It is speculated that bio-molecules such as amino acids, simple sugar molecules, etc., could be synthesized in space \citep{chak00a,chak00b}, and they were transferred to
the solar nebula and later on to the early Earth during meteorites shower. Comets are also the reservoirs of various COMs and pre-biotic molecules. \cite{chak00a,chak15} studied the chemical evolution of numerous species, including some biomolecules such as glycine, alanine, adenine, etc., during molecular cloud collapse. \cite{maju15} also studied the synthesis of pyrimidine bases in the interstellar regions.

\section{Star formation and its evolutionary stages}
 
\subsection{Low mass star}
\begin{figure}
\label{fig:lowmass}
\centering
\includegraphics[height=15cm,width=10cm]{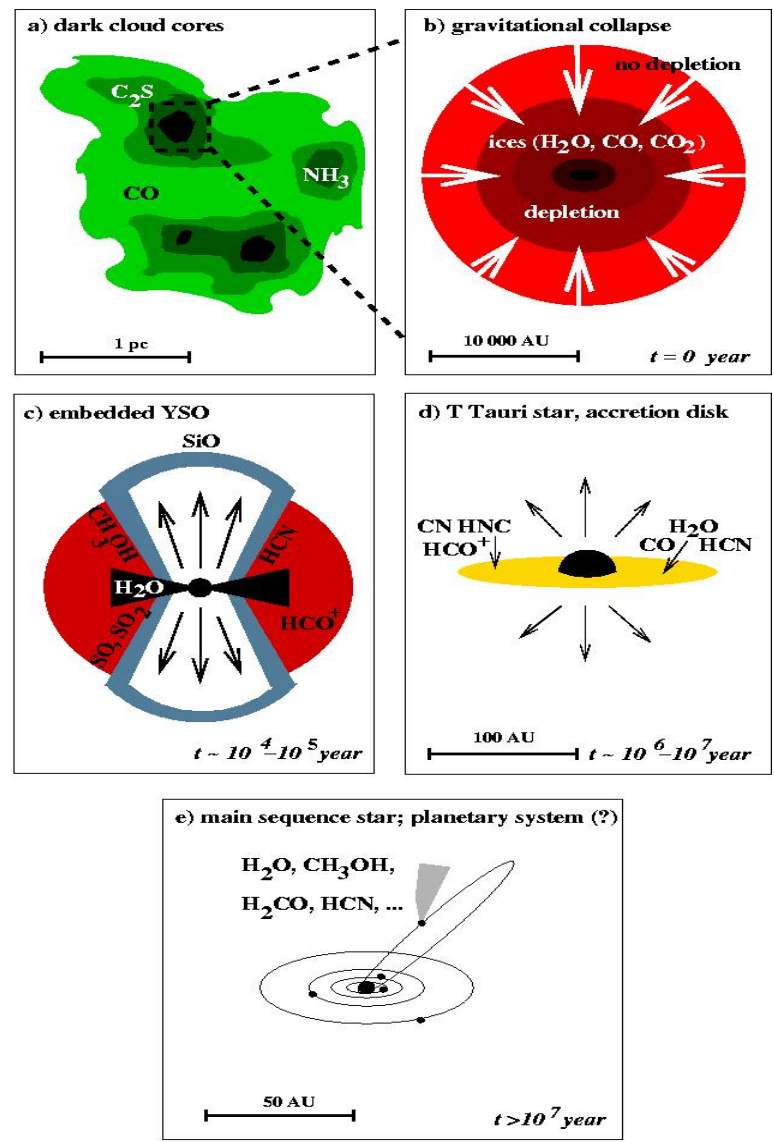}
\caption{A cartoon depiction of the stages of low-mass star formation. a) An
inhomogeneous molecular cloud with several over-dense regions, called cloud
cores. b) A cloud core is triggered into gravitational collapse when it reaches
its Jeans mass. The material moves radially toward the center of the core under gravity. c) Due to turbulence and shear motions of the molecular
cloud, the core has a net angular momentum which makes the material spin up as
the core contracts. The result is a rotating disk of material onto which material is accreted from the still-collapsing molecular envelope. d) A T Tauri star emerges
when the whole envelope has either been accreted or blown away by the bipolar
outflow that characterizes these objects. The remaining material is located in a
protoplanetary disk. \citep[Courtesy:][]{vand98}}
\end{figure}
The low-mass star formation process has been extensively studied in the last 50 years. According to \cite{shuf77}, gravitational collapse starts in a centrally concentrated configuration and spreads outward at the speed of sound ($a$); after a certain amount of time $t$, matter inside the infall radius (r$_{inf}$ = $at$) starts to fall inward in free-fall motion. The density distribution follows the r$^{-2}$ law holds in the static or nearly static outer envelope and r$^{-3/2}$ law holds for the freely falling inner envelope. This model's collapse is self-similar, hence the structure can always be determined from a single solution. This situation is known as inside-out collapse as the front of accretion expands radially outward in time. Since the inflow region expand outward, it was termed the inside-out collapse solution \cite{shuf77}. Among some of the earlier studies, \cite{shuf77} studied star formation by gravitational collapse with an inside-out collapse process \citep{shuf77}. As gravitational collapse starts, matter falls toward the center and eventually reaches a free-fall speed. The region covering the in-fall of matter extends outwardly with a speed of sound. But \cite{shuf77} model was unable to explain the binary star, disk-like structure, and stellar rotation as it was a spherically symmetric model, and therefore unable to consider angular momentum rather than rotation. 
The observational evidence suggests that due to the conservation of angular momentum, the rotational motion breaks the spherical symmetry and the matter forms a disk-like structure in the center.
Later series of papers of Shu included angular momentum in the model \citep{shu87,shu88a,shu88b,shu94,shu97}. 

As the star formation takes place in a long time scale ($>10^6$ yrs), it is impossible to observe all the stages of star formation in a single star-forming region. An alternate way is to observ different evolutionary stages of the similar-type star-forming regions and sequentially linking them up carefully to develop a meaningful scenario. The Mass of a stellar object can-not be an accurate tracer of the evolutionary stage always because a star ends up with a wide range of mass. A more massive star can be in the same evolutionary stage as a less massive star, if the initial mass was different.

Prevailing chemistry can be used to estimate the low-mass star formation time scale. For example, \cite{char92} suggested using the abundance ratio of some species, \cite{Jorg04} suggested using the abundance profiles of some species as a 'chemical clock'.

The present understanding of the different stages of low-mass star formation is described below (see Fig. \ref{fig:lowmass}):\\
(a)  Star formation is initiated with condensation in molecular clouds. Turbulent motions and magnetic fields resist the interstellar clouds from collapsing. Only charged particles are coupled to the magnetic field lines. Due to ambipolar diffusion, the materials concentrate in the center, forming a cloud core on a time scale of $10^6$ years. Generally, the density in the cloud core is predicted by a singular isothermal sphere $\rho\propto r^{-2}$ \citep{shuf77}. Some observations reveal that the core is slightly elongated in dark clouds. Cores without an embedded star are flatter than the clouds with an embedded star in the core though almost the same density distribution holds for both of them.\\
(b) At a certain point when magnetic energy equals the gravitational energy, called critical magnetic mass, and when it meets the supercritical magnetic condition, the collapse starts. When the initial central gas pressure is not enough to hold the gravitational pressure, the system starts to collapse, and the collapse information goes outward at a local speed of sound. As the radiative cooling by spectral line emission and the dust emission is efficient in this phase, the core's temperature remains constant ($\sim10$k). This is known as the isothermal phase of the collapse. The duration of this phase is approximately equal to the free fall time.\\

(c) The density of $\rm{H_2}$ increases to a certain value $\sim 10^{11}$ cm$^{-3}$, and the dust emission becomes opaque, and the emerged photons are again absorbed within the cloud; and thus the cooling efficiency diminishes. The temperature of the central part starts to increase; this high-temperature core is known as the first core. The temperature of the core increases until it reaches $\sim 1500$k. It takes approximately $100$ years. Subsequently, due to high-temperature thermal dissociation of $\rm{H_2}$ molecules starts, followed by the ionization of the hydrogen and helium atoms. This process takes a lot of energy, and the collapse process accelerates. This phase is known as the second collapse, and the second core is formed.\\
(d) The protostar phase starts. A thin disk-like structure forms in the central region, and gas starts to accrete into the central protostar via an envelope. This is the main accretion phase. As the accreting material has certain angular momentum, this breaks the spherical symmetry of the in-falling envelope and flattens the disk along the equatorial region. At the same time, an outflow occurs along the rotation axis, which is perpendicular to the protostellar disk. The cause of this outflow is considered an interplay between magnetic fields and gas dynamics, which also plays a role in extracting the angular momentum from the accreting material. Outflow usually consists of an inner ionized jet and outer molecular outflow. The outflow velocity can range from $10$ to $100$ km/s.\\
(e) With increasing time, the accretion of gas and dust onto the central disk almost ends up with the consumption of the whole gas present in the envelope. The next phase is the T-Tauri phase. It is a quasi-static contraction phase. During this time, the central star is heated up due to contraction, and heat is transferred to the outer region by convection. This phase is known as the Hayashi phase, and the young stellar object is known as a T-Tauri star. A main sequence star forms after $10^7-10^8$ years from this phase. As the parent cloud mostly dissipated by the outflow and by other activities, only a protostellar disk is formed, which evolved into the protoplanetary disk in a later stage which, causes the birth of a planetary system.\\

The results of core accretion models are also known to be significantly impacted by disc metallicity \citep{mord12,hase14}. The global disc dust-to-gas ratio is impacted by disc metallicity, which also has an impact on the solid accretion time scale during planet formation. Setting the timeline for the initial phases of core accretion before gas accretion is important. The sample of G-type stars containing observed planets has stellar metallicities ranging from -0.6 $\leq$ [Fe/H] $\leq$ 0.6 \citep{han14}. The metallicities of the gaseous discs are a reflection of the metallicities of the parent stars that gave birth to them. The observed planet-metallicity relationship demonstrates that the gas giants frequency scales with the metallicity of the host star \citep{vale08,wang15}, emphasising the significance of disc metallicity in the context of planet formation.

\subsection{High mass star}
High-mass stars are the essential building blocks of our Galaxy and the universe, though they only make up a small fraction of all stars \citep{zinn07}. These high-mass stars hold a big energy budget and are very luminous compared to the other stars. They generate a large amount of energy during the formation via energetic jets, during their lifetime via intense UV radiation or supernovae explosion. The high mass star forming regions are the source of the heavy elements, and no formation of life would be possible without these heavy elements. The theory behind the formation of high-mass stars ($>8_{\odot}$) is an active area of research. It has a shorter formation time scale compared to the low-mass stars. The high-mass stars start hydrogen burning before the accretion of gas has stopped. So the emitted radiation pressure is enough to resist the further gas accretion by radiation pressure. This is why the high-mass star formation can not be considered a scaled-up version of a low-mass star formation. 
Three patterns of high-mass star formation are considered in the literature. In the first type, the gravitational collapse of a turbulent, isolated massive core takes place. This is also known as monolithic collapse. Initially, a high-mass starless core is formed without turbulence; otherwise, it would have collapsed before gathering enough mass from its envelope. The high magnitude of turbulence causes a high accretion rate of gas onto the protostar and fuels the rapid growth of the protostar. For instance, a turbulent core may be formed during the collision between two clouds. Though the radiation pressure suppresses accretion, it continues to some extent through a disk structure formed around the protostar. \\

\begin{figure}[htbp]
\begin{center}
\includegraphics[width=\textwidth]{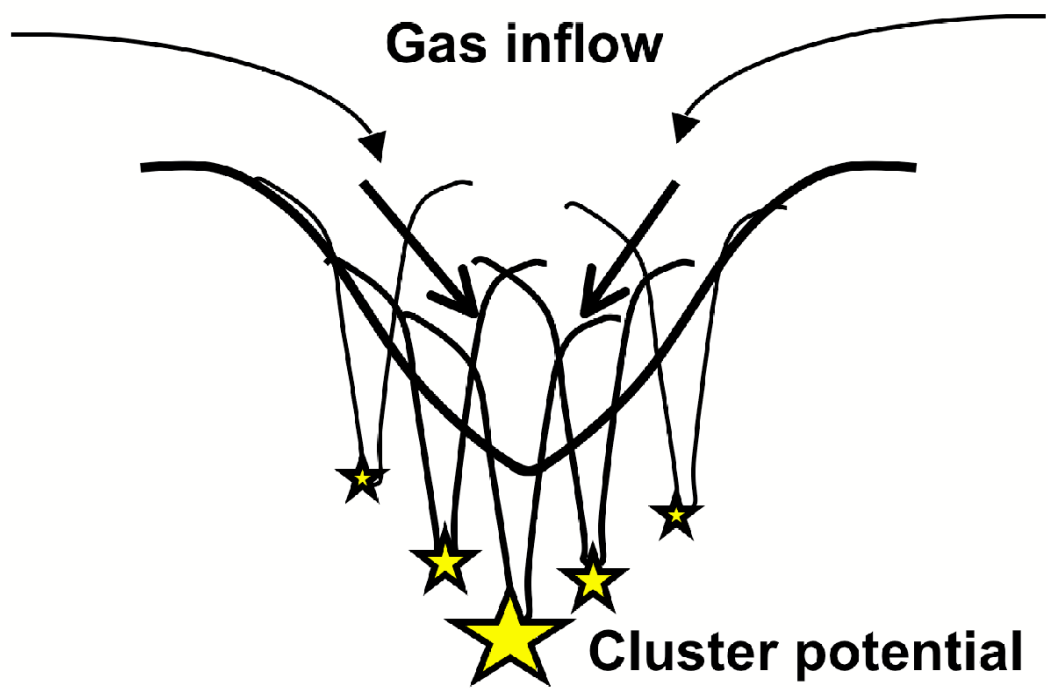}
\caption{The gravitational potentials of the individual stars combine to create a larger-scale potential that directs gas down to the cluster centre, this is a schematic representation of the mechanics of accretion in a stellar cluster. The stars in the centre are consequently able to accrete more gas and develop into stars with higher masses. .\citep[Courtesy:][]{bonn08}.}
\label{fig:hm_potential}
\end{center}
\end{figure}

The second type is when 'competitive accretion' takes place. In this case, a high-mass star is born as a member of a low-mass protostar cluster in a protocluster cloud. Small-N clusters are formed when individual stars collide and their mutual potential pulls gas towards the centre. The few stars found in the cluster centres have substantially faster accretion rates due to the higher gas density there and the fact that this gas is constantly replenished. Suppose the low-mass protostar is in the central region of the cloud. In that case, its parent core can gather more gas from the cloud than the other cores because of its low gravitational potential \citep{bonn08}. Hence accumulating more and more mass, it becomes massive with time. In figure \ref{fig:hm_potential} the scenario is described.\\
The third one is growth by the merger of low-mass protostars. Many observational and theoretical studies have been conducted, but no unique theory has been identified. The initial condition for the high-mass star formation needs to be better understood than the low-mass star formation.
\section{Chemistry in the Interstellar medium}
\begin{figure}[htbp]
\begin{center}
\includegraphics[width=\textwidth]{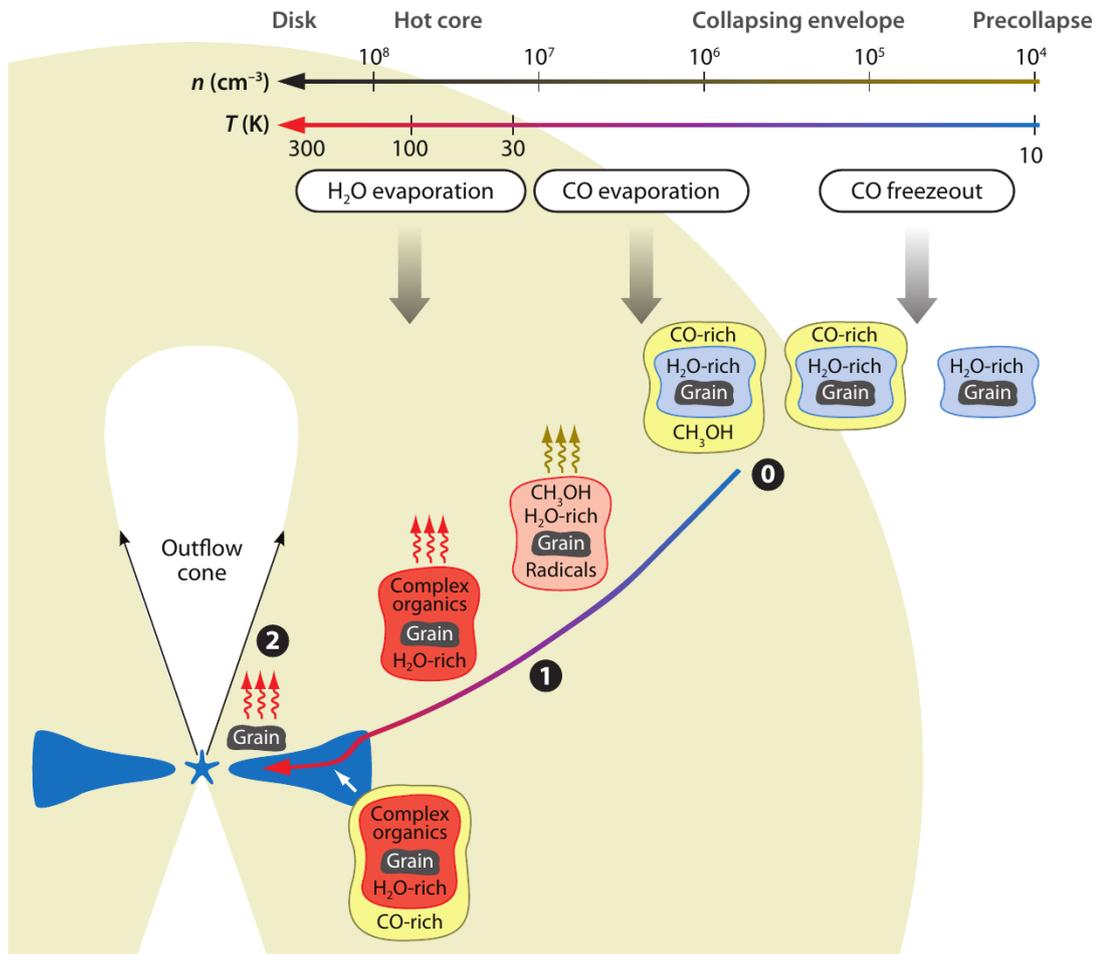}
\caption{Diagram illustrating the composition, chemical evolution, sublimation into the gas phase, and collapse into a $0.05$ pc envelope for the dust, ice, and mantle. According to \cite{char92}, the formation of 0th, 1st, and 2nd generation molecules are denoted by 0, 1, and 2 accordingly in the diagram. A location where the dust ice mantle is entirely sublimated is typically characterized as 100 K by the hot corino at (2). The molecules will once more freeze out onto the dust grains in the PP-disk's dense and cold midplane, enabling more grain surface chemistry to take place. The top panel displays the envelope's temperature and density.\citep[Courtesy:][]{herb09}.}
\label{fig:cartoon_evolutioncoms}
\end{center}
\end{figure}
Following the discovery of ammonia (NH$_3$) in the ISM in 1968 \citep{cheu68}, 272 more gas phase molecules were found using rotational emission spectroscopy. Known interstellar detected molecules are listed in Table \ref{tab:known_molecules}. These discoveries marked the beginning of modern astrochemistry (see \cite{herb09}). Studies of infrared absorption and emission have shown the complicated chemistry of the ISM. It includes discrete infrared emission spikes whose carriers are still unknown. Although no significant detections have been obtained, polycyclic aromatic hydrocarbons (PAHs) are suggested as the potential carriers of these features. For instance, the high dipole moment of the molecules such as CS could indicate high-density regions. High local densities are required to maintain the molecular energy levels by collisional excitation because high dipole moments cause high rates of spontaneous emission. The ions, such as HCO$^+$, naturally examine ionized areas in protostellar cores (e.g., \cite{stah04}). Organic chemistry, which relies on molecules made of carbon, hydrogen, oxygen, nitrogen, and other elements to a lesser extent, is the foundation of terrestrial life (see discussion of \cite{cecc17}, and references therein). We can calculate the likelihood that life, as we know it, will eventually exist on planets created in these star-forming areas by searching the ISM and star-forming regions for primordial chemicals like glycolaldehyde and assessing their abundances.
To accomplish this, we must follow the chemistry of the formation of a planetary system from a prestellar core in a molecular cloud from beginning to end. Cosmic rays and the UV radiation from the surrounding stars, which generate ions, impact the chemistry of the cold outer envelope of a protostellar core. These ions interact with the neutral atoms and molecules in the gas phase. The low temperature (10-20 K) causes molecules to fall onto the dust surface when the dust density is high enough. Some of these molecules interact and stick together, forming an ice mantle on the grain. Only a few simple molecules exist in the gas because of this freezing out of molecules on the grains of dust; only a few simple molecules exist in the gas phase of the cold envelope (e.g., $\rm{N_2H^+, HCO^+, H_2CO}$, see \cite{vand06}, and references therein).
The tightly bonded molecules, like H$_2$O, could sublimate at temperatures exceeding 90-100 K. This temperature is feasible in the hot inner region of the envelope closest to the protostar \citep{sand93}. Many molecules are released into the gas phase during the sublimation of ice. As a result, complex organic molecules (COMs) are seen in the gas phase of this hot corino in a low-mass protostellar core (referred to as a hot core in the case of high-mass star formation) (see Fig. \ref{fig:cartoon_evolutioncoms}). COMs are molecules with six or more atoms, containing at least one carbon \citep{herb09}.

\begin{landscape}
\begin{table}
\scriptsize
\caption{Known interstellar molecules ($\sim 272$ molecules; last updated: June 2022).}
\label{tab:known_molecules}
\vskip 0.2 cm
\hskip -2.5cm
\begin{tabular}{|ll|ll|ll|ll|l|l|l|ll|}
\hline
\multicolumn{2}{|c|}{\bf 2 atoms (45)} & \multicolumn{2}{|c|}{\bf 3 atoms (45)} & \multicolumn{2}{|c|}{\bf 4 atoms (34)} & \multicolumn{2}{|c|}{\bf 5 atoms (33)} & {\bf 6 atoms (27)} & {\bf 7 atoms (18)} & {\bf 8 atoms (19)} & \multicolumn{2}{c|}{\bf 9 atoms (16)} \\
\hline
$\rm{H_2}$ & SiS & $\rm{C_3}$ & $\rm{NH_2}$ & $\rm{c-C_3H}\ ^a$ & $\rm{c-SiC_3}\ ^a$ & $\rm{C_5}$ & $\rm{C_4H^-}$ & $\rm{C_5H}$ & $\rm{C_6H}$ & $\rm{CH_3C_3N}$ & $\rm{CH_3C_4H}$ & $\rm{C_3H_6}$ \\
AlF & CS & $\rm{C_2H}$ & $\rm{H_3^+}$ & $\rm{l-C_3H}\ ^b$ & $\rm{CH_3}$ & $\rm{C_4H}$ & HC(O)CN & $\rm{l-H_2C_4}\ ^b$ & $\rm{CH_2CHCN}$ & $\rm{HC(O)OCH_3}$ & $\rm{CH_3CH_2CN}$ & $\rm{CH_3CH_2SH}$ \\
AlCl & HF & $\rm{C_2O}$ & SiCN & $\rm{C_3N}$ & $\rm{C_3N^-}$ & $\rm{C_4Si}$ & HNCNH & $\rm{C_2H_4}$ & $\rm{CH_3C_2H}$ & $\rm{CH_3COOH}$ & $\rm{(CH_3)_2O}$ & $\rm{CH_3NHCHO}$ \\
$\rm{C_2}$ & HD & $\rm{C_2S}$ & AlNC & $\rm{C_3O}$ & $\rm{PH_3}$ & $\rm{l-C_3H_2}\ ^b$ & $\rm{CH_3O}$ & $\rm{CH_3CN}$ & $\rm{HC_5N}$ & $\rm{C_7H}$ & $\rm{CH_3CH_2OH}$ & $\rm{HC_7O}$ \\
CH & FeO & $\rm{CH_2}$ & SiNC & $\rm{C_3S}$& HCNO & $\rm{c-C_3H_2}\ ^a$ & $\rm{NH_4^+}$ & $\rm{CH_3NC}$ & $\rm{CH_3CHO}$ & $\rm{C_6H_2}$ & $\rm{HC_7N}$ & $\rm{H_2C_3HCCH}$ \\
$\rm{CH^+}$ & $\rm{O_2}$ & HCN & HCP & $\rm{C_2H_2}$ & HOCN & $\rm{H_2CCN}$ & $\rm{H_2NCO^+}$ & $\rm{CH_3OH}$ & $\rm{CH_3NH_2}$ & $\rm{CH_2OHCHO}$ & $\rm{C_8H}$ & $\rm{HC_3HCHCN}$ \\
CN & $\rm{CF^+}$ & HCO & CCP & $\rm{NH_3}$ & HSCN & $\rm{CH_4}$ & $\rm{NCCNH^+}$ & $\rm{CH_3SH}$ & $\rm{c-C_2H_4O}\ ^a$ & $\rm{l-HC_6H}\ ^b$ &$\rm{CH_3C(O)NH_2}$ &  $\rm{H_2CCHC_3N}$ \\
CO & SiH & $\rm{HCO^+}$ & AlOH & HCCN & $\rm{H_2O_2}$ & $\rm{HC_3N}$ & $\rm{CH_3Cl}$ & $\rm{HC_3NH^+}$ & $\rm{H_2CCHOH}$ & $\rm{CH_2CHCHO}$ & $\rm{C_8H^-}$ & $\rm{HOCHCHCHO}$ \\
\cline{12-13}
$\rm{CO^+}$& PO & $\rm{HCS^+}$ & $\rm{H_2O^+}$ & $\rm{HCNH^+}$ & $\rm{C_3H^+}$ & $\rm{HC_2NC}$ & $\rm{MgC_3N}$ & $\rm{HC_2CHO}$ & $\rm{C_6H^-}$ & $\rm{CH_2CCHCN}$ & {\bf 10 atoms (11)} & {\bf 11 atoms (7)} \\
\cline{12-13}
CP & AlO & $\rm{HOC^+}$ & $\rm{H_2Cl^+}$ & HNCO & HMgNC & HCOOH & $\rm{NH_2OH}$ & $\rm{NH_2CHO}$ & $\rm{CH_3NCO}$ & $\rm{H_2NCH_2CN}$ & $\rm{CH_3COCH_3}$ & $\rm{HC_9N}$ \\
SiC & $\rm{OH^+}$ & $\rm{H_2O}$ & KCN & HNCS & HCCO & $\rm{H_2CNH}$ & $\rm{HC_3O^+}$  & $\rm{C_5N}$ & $\rm{HC_5O}$ & $\rm{CH_3CHNH}$ & $\rm{HOCH_2CH_2OH}$ & $\rm{CH_3C_6H}$ \\
HCl & $\rm{CN^-}$ & $\rm{H_2S}$ & FeCN & $\rm{HOCO^+}$ & CNCN & $\rm{H_2C_2O}$ & $\rm{HC_3S^+}$ & $\rm{l-HC_4H}\ ^b$ & $\rm{HOCH_2CN}$ & $\rm{CH_3SiH_3}$ & $\rm{CH_3CH_2CHO}$ & $\rm{C_2H_5OCHO}$ \\
KCl & $\rm{SH^+}$ & HNC & $\rm{HO_2}$ & $\rm{H_2CO}$ & HONO & $\rm{H_2NCN}$ & $\rm{H_2CCS}$ & $\rm{l-HC_4N}\ ^b$ &$\rm{HC_4NC}$ & $\rm{(NH_2)_2CO}$ & $\rm{CH_3C_5N}$ & $\rm{CH_3OC(O)CH_3}$ \\
NH & SH & HNO & $\rm{TiO_2}$ & $\rm{H_2CN}$ & $\rm{MgC_2H}$ & $\rm{HNC_3}$ & $\rm{C_4S}$ & $\rm{c-H_2C_3O}\ ^a$ & $\rm{HC_3HNH}$ & $\rm{HCCCH_2CN}$ & $\rm{CH_3CHCH_2O}$ & $\rm{CH_3COCH_2OH}$ \\
NO & $\rm{HCl^+}$ & MgCN & $\rm{C_2N}$& $\rm{H_2CS}$& HCCS & $\rm{SiH_4}$ & HC(O)SH  & $\rm{H_2CCNH}$ & $\rm{c-C_3HCCH}\ ^a$ & $\rm{HC_5NH^+}$ & $\rm{CH_3OCH_2OH}$ & $\rm{c-C_5H_6}\ ^a$ \\
NS & TiO & MgNC & $\rm{Si_2C}$ & $\rm{H_3O^+}$ & HNCN & $\rm{H_2COH^+}$ & HC(S)CN  & $\rm{C_5N^-}$ & $\rm{l-H_2C_5}\ ^b$ & $\rm{CH_2CHCCH}$ & $\rm{c-C_6H_4}\ ^a$& $\rm{NH_2CH_2CH_2OH}$ \\
NaCl & $\rm{ArH^+}$ & $\rm{N_2H^+}$ & $\rm{HS_2}$ &&$\rm{H_2NC}$&  &HCCCO  & HNCHCN & $\rm{MgC_5N}$  &$\rm{MgC_6H}$& $\rm{H_2CCCHC_3N}$ &  \\
OH & $\rm{N_2}$ & $\rm{N_2O}$ & NCO &&$\rm{HCCS^+}$&&& $\rm{SiH_3CN}$ &$\rm{CH_2C_3N}$& $\rm{C_2H_3NH_2}$&$\rm{C_2H_5NCO}$  &  \\
PN & $\rm{NO^+}$ & NaCN & HSC &&&&& $\rm{C_5S}$ && $\rm(CHOH)_2$& $\rm{C_2H_5NH_2}$ &   \\
SO & $\rm{NS^+}$ & OCS & HCS &&&&& $\rm{MgC_4H}$ && & $\rm{HC_7NH^+}$ &   \\
\cline{12-13}
$\rm{SO^+}$& $\rm{HeH^+}$ & $\rm{SO_2}$ & CaNC &&&&& $\rm{CH_3CO^+}$ && & {\bf 12 atoms (9)} & {\bf $>$12 atoms (10)}  \\
\cline{12-13}
SiN & $\rm{PO^+}$& $\rm{c-SiC_2}\ ^a$ & NCS &&&&& $\rm{C_3H_3}$ && & $\rm{c-C_6H_6}\ ^a$ & $\rm{HC_{11}N}$ \\
SiO & & $\rm{CO_2}$ &&&&&& $\rm{H_2C_3S}$  && & $\rm{C_2H_5OCH_3}$ & $\rm{c-C_6H_5CN}\ ^a$  \\
&&&&&&&& $\rm{HCCCHS}$ &&& $\rm{n-C_3H_7CN}$ & $\rm{1-C_{10}H_7CN}$  \\
&&&&&&&& C$_5$O&&& $\rm{i-C_3H_7CN}$ & $\rm{2-C_{10}H_7CN}$  \\
&&&&&&&& C$_5$H$^+$&&& $\rm{1-c-C_5H_5CN}\ ^a$ & $\rm{c-C_9H_8}\ ^a$ \\
&&&&&&&& HCCNCH$^+$&&&$\rm{2-c-C_5H_5CN}\ ^a$ & $\rm{C_{60}}$ \\
&&&&&&&& &&& $\rm{CH_3C_7N}$&$\rm{C_{60}^+}$ \\
&&&&&&&& &&& $\rm{n-C_3H_7OH}$&$\rm{C_{70}}$ \\
&&&&&&&& &&& $\rm{i-C_3H_7OH}$& $\rm{1-c-C_5H_5CCH}\ ^a$\\
&&&&&&&& &&& & $\rm{2-c-C_5H_5CCH}\ ^a$\\
\hline
\hline
\end{tabular} \\
\vskip 0.2cm
{\bf Notes:}
Total number of molecules for each category is provided in parentheses. These statistics are obtained by avoiding the isotopomers. Deuterium isotopic species are given separately only if their method of detection is intrinsically different from that of pure hydrogen ones. \\
$^a$ The `c' refers to cyclic form. \\
$^b$ The `l' refers to cyclic form.
\end{table}
\end{landscape}

\section{Different approach to study the physical condition of ISM}
\subsection{Laboratory experiment}
Recreating the interstellar conditions (temperature, density, pressure, grain surface property, different time scales, etc.) is challenging. There are uncertainties in mimicking an interstellar condition. For example, there are significant uncertainties regarding the grain surface properties, time scales, pressures, temperatures, etc. The typical experimental timescale in the laboratory is one to a few days, whereas the astrophysical phenomenon happens in a time scale of $10^5$ - $10^7$ years. A temperature of 5 K can be obtained in the laboratory and regulated up to 200k to mimic the interstellar condition. Pressure as low as the diffuse interstellar medium cannot be obtained, but ultra-high vacuum can generate an intimate nature to the dense interstellar medium. 
Therefore, we must infer the experimental output carefully to resemble the interstellar scenario.
Among the most pioneering works, \cite{mill59}  carried out an experiment using H$_2$O, CH$_4$, NH$_3$, and H$_2$ (assuming these are the major constituents of the Earth's atmosphere) at the University of Chicago and recreating the primitive Earth-like condition. At the end of his experiment, he noticed that 20 amino acids related to the basic building blocks of life were produced.
There are modern experimental facilities to obtain the spectroscopic details of different molecules for astrochemical purpose. Across the world laboratory facilities include large synchrotron and powerful computers. Two significant developments of the past few decades are the so-called Selected Ion Flow Tube (SIFT; \cite{mart08}) and Reaction kinetics in uniform supersonic flow (CRESU; \cite{sims93,chas01}), or crossed molecular beams \citep{mora11} method which enable the measurements of gas-phase reaction rates. Ice-phase reactions are characterized using surface-science apparatuses designed to mimic the high vacuum and range of temperatures (down to 4 K) characteristic of interstellar environments.

\subsection{Theoretical study}
\subsubsection{\bf Chemical modeling}
Simulations are an essential means of providing the chemical abundances of molecules as a function of the physical conditions of the cloud. From the numerical point of view, chemical models solve a system of differential equations of the type: 
\begin{equation}
 \frac{dn_i}{dt} = \sum production-\sum destruction, 
\end{equation}
where $\rm{n_i}$ is the number density of the species $i$ ($n$ cm$^{-3}$). Here, 'production' and 'destruction' refer to all physical and chemical processes that produce and destroy the species. Many parameters are used as inputs of a chemical model. Among them, initial elemental abundances, geometry, external cosmic-ray ionization rate ($\zeta$), radiation field strength ($\chi$), the number density of the gas, temperature of the gas and the dust, properties of the dust grains, freeze out of complex molecules, desorption processes, and databases of the reactions are some of the basic needs to construct such models. Chemical models provide the gas phase and ice phase fractional abundances of all the molecules included in the chemical network. Over the years, various types of chemical models were built, and some of the most important references of these are \cite{kram46,bate51,hase92,chak00a,garr06b,chak06a,chak06b,das08a,das11,dasa15,gor17b,das19,sil21,bhat22}.

\subsubsection{\bf Quantum chemical modeling}
Scientists start from the basic principles of quantum mechanics using ab initio quantum chemical calculations. This theory explains how subatomic particles behave to decide molecular properties based on the motions of the protons, neutrons, and electrons in the atoms that form the molecule. Numerous simulations are carried out by considering multiple molecular structures and determining the ideal geometry of the molecule. The spectral information of the species falling in various electromagnetic wavelengths could also be predicted by such calculations with quantum chemical software like Gaussian \citep{fris13}. However, such calculations are not very economical in time and thus have some limitations on the size of the molecule.
Many other software is available for quantum chemical calculations. GAMESS \citep{barc20}, has most of the functionality of Gaussian software (ab initio quantum chemistry, density functional theory, CI, MP calculations, transition state calculations, solvent effects and IR and NMR calculations). NWChem \citep{apra20} can calculate a smaller set of properties, but it can handle mixed QM/MM calculations and periodic systems like solids. ACES is another program which is specialized in high-level quantum chemistry calculation. Quantum-mechanical computations of molecular features, such as structures, transition frequencies, etc., must support astrochemical investigations in order to guide and interpret observations, line assignments, and data processing in the most challenging and unusual environments. Spectroscopic methods, which are frequently utilized to infer knowledge about molecule structure and dynamics and are thus playing a vital role in the exploration of atmospheric chemistry and the interstellar medium over the last decades, serve as a typical example.
\section{Observations}
\subsection{Molecular spectroscopy and energy levels}
\begin{figure}[htbp]
\begin{center}
\includegraphics[width=\textwidth]{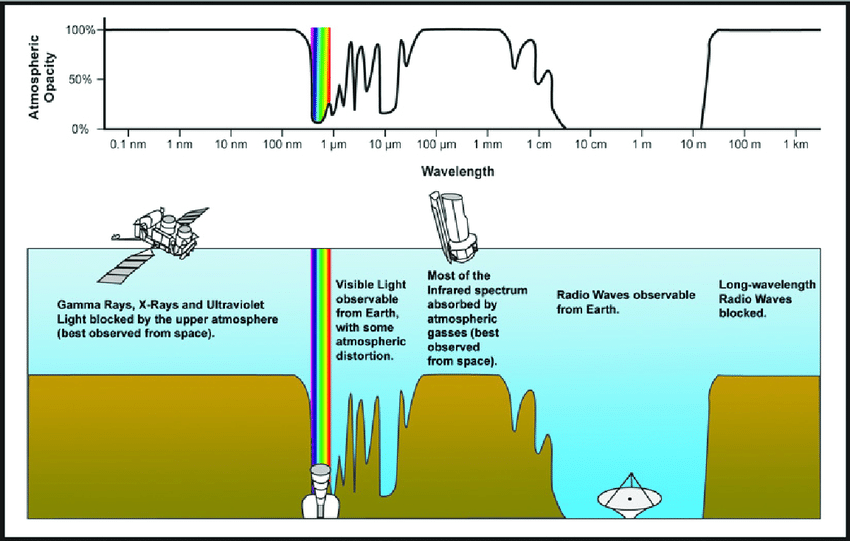}
\caption{Atmospheric absorption through the electromagnetic spectral range \citep[Courtesy:][NASA, public domain]{}.}
\label{fig:atm_abs}
\end{center}
\end{figure}

The light emitted by the various objects in different wavelengths is the sole information an astronomer receives from space. According to its wavelength, this electromagnetic wave travels through space with the energy E (in units of Joule) as follows:
\begin{equation}
\label{eqn:ene}
    E=\frac{hc}{\lambda}=h\nu,
\end{equation}
where $\lambda$ is the wavelength (in unit of meter), and $\nu$ is the frequency in units of Hz, h is the Planck constant ($6.6260 \times 10^{-34} m^2 .kg .s^{-1}$), and the speed of light, c, which is 299792458 m s$^{-1}$. 
Figure \ref{fig:atm_abs}) represents the electromagnetic spectrum for reference. This diagram indicates that the visible light spectrum only makes up a small portion. The energy of a photon E (in units of Joule) can also be described in terms of temperature as follows:
\begin{equation}
\label{eqn:kt}
    E = k_B T, 
\end{equation}
where $T$ is the absolute temperature (in Kelvin, K), and k$_B$ is the Boltzmann constant, $1.3806\times 10^{-23}$ m$^2$ kg s$^{-2}$ K$^{-1}$. A photon with a high energy is termed 'hot', whereas one with a low energy is termed 'cool'. Molecules interact with photons arriving from all directions in the interstellar medium only when a photon's energy (or wavelength or frequency) matches the energy of a molecule. 

\begin{figure}[htbp]
\begin{center}
\includegraphics[width=\textwidth]{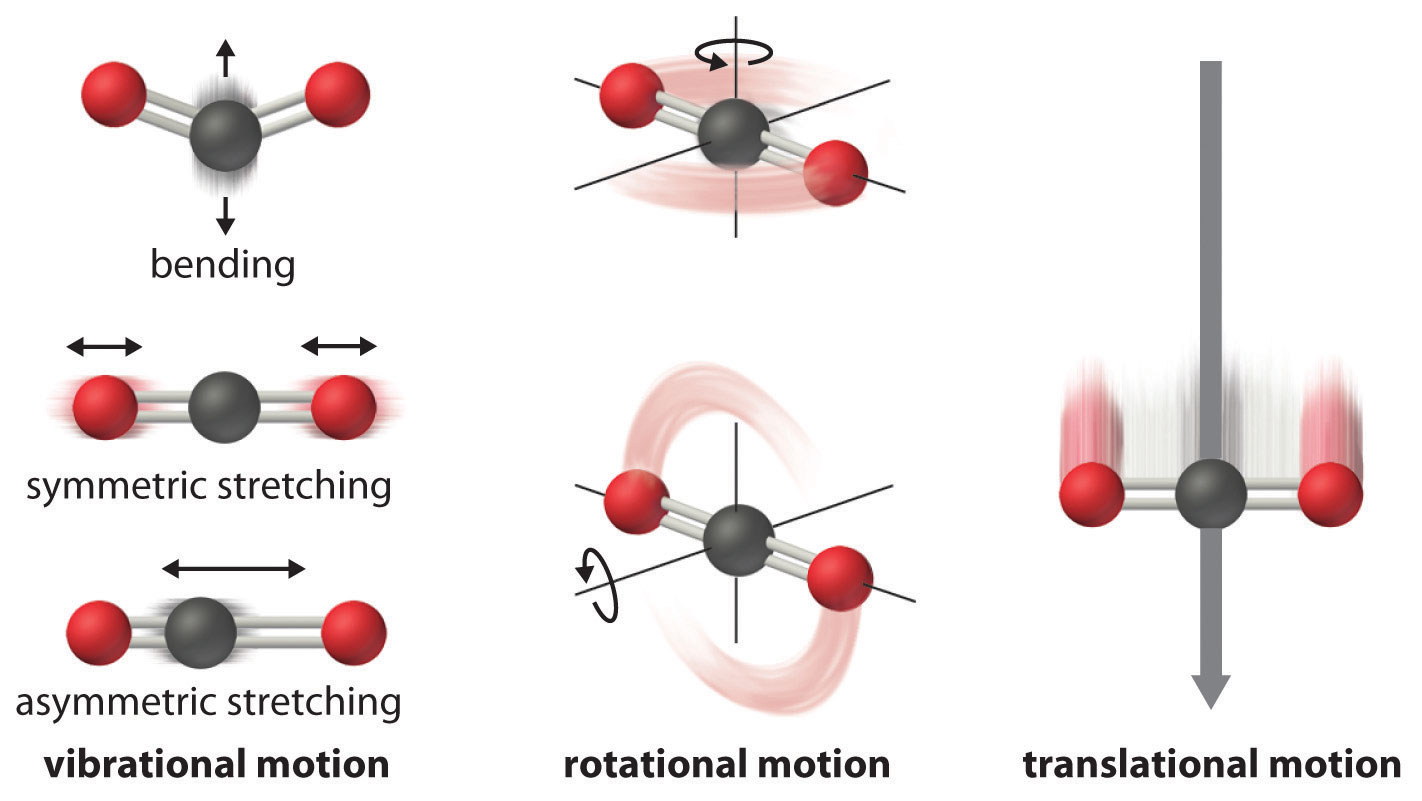}
\caption{Representation of the different movements a molecule can adopt. \citep[Courtesy:][Principles of general chemistry \(Martin Silberberg\)]{}.}
\label{fig:mol_motion}
\end{center}
\end{figure}

\begin{figure}[htbp]
\begin{center}
\includegraphics[width=\textwidth]{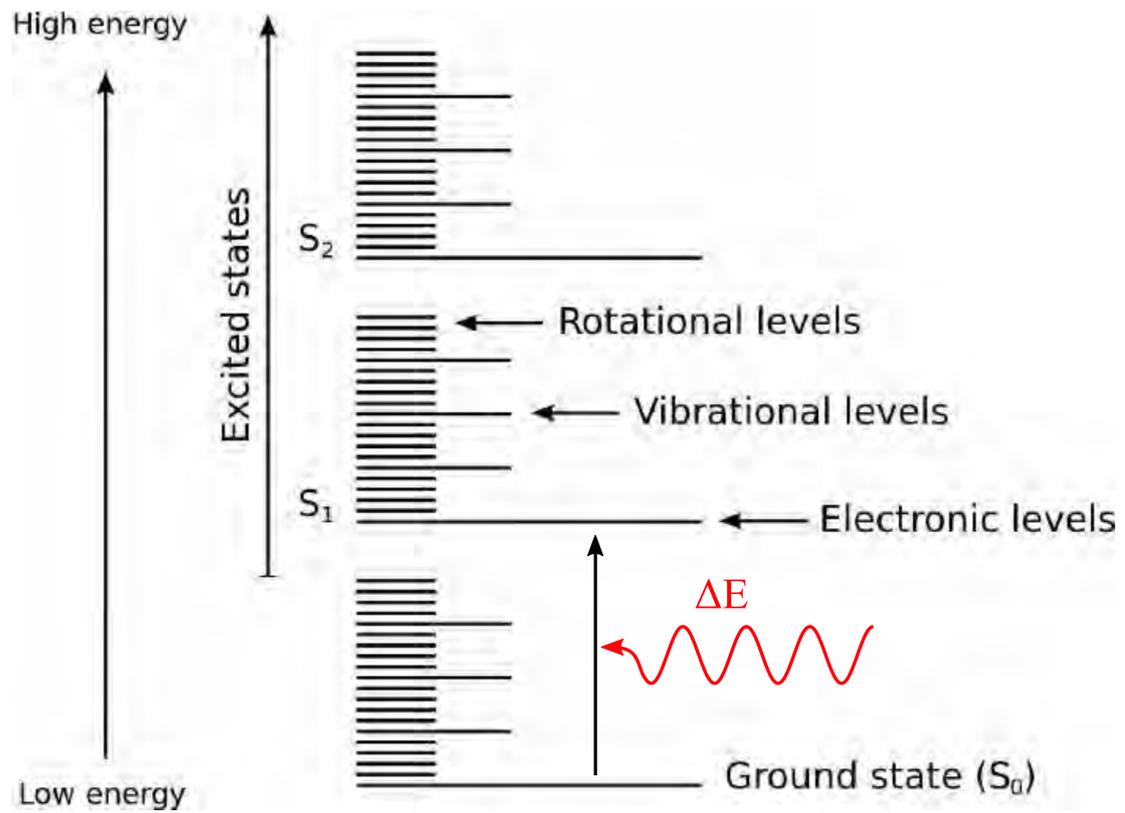}
\caption{Representation of the many transitional forms and how they are nested. This molecule absorbs a photon (red) with energy E.}
\label{fig:ener_lev}
\end{center}
\end{figure}

Molecules in space may involve translations, stretching, bending, scissoring, rocking, wagging, or rotations around various axes (see Fig. \ref{fig:mol_motion}). The quantified energies that correspond to these various processes (electronic excitation, vibrations, rotations, and translations) define the different energy levels of a molecule (see Fig. \ref{fig:ener_lev}). Each electronic level has several vibrational levels, each of which has several rotational levels.

The energy of a molecule in its ground state, level S$_0$ in Figure \ref{fig:ener_lev}, is denoted as E$_0$. If a photon with the energy $\Delta E = E_1 - E_0$ is absorbed by this molecule, it can, for example, reach the excited electronic state S$_1$ (see Fig. \ref{fig:ener_lev}). Additionally, $\Delta$E represents the energy of the electronic transition between levels S$_0$ and S$_1$. Transitions involving vibration, rotation, and translation are all defined similarly. Electronic transitions are more energetic than vibrational transitions ($\Delta E_{el} \gg \Delta E_{vib}$) because electronic energy levels are more widely spread. Similarly, the vibrational transitions are more energetic than rotational ones ($\Delta E_{vib} \gg \Delta E_{rot}$).
At the low temperature (less than 10 K to a few 100 K),  
rotational transitions can happen in the interstellar medium because the energy needed to move from one level to another is insufficiently low. Because of Eqs. \ref{eqn:ene} and \ref{eqn:kt}, molecules can only emit radio frequencies with wavelengths in the millimeter/submillimeter range:
\begin{equation}
    \lambda[mm]=\frac{14.388}{T},
\end{equation}
where $\lambda$ is the photon's wavelength, and T is the absolute temperature. Also, where the temperature is comparatively higher, molecules may emit in the infrared range. 

The quantum number J acts as an identifier for rotational transitions. The ground state is at the J = 0 level, and rotational transitions are only allowed if $\Delta J = \pm$ 1. As a result, a molecule will move from rotational level J (of energy E$_J$) to level J + 1 (of energy E$_{J+1}$) when it absorbs the appropriate photon. In contrast, a molecule emitting a photon will move from the rotational level J (of energy E$_J$) to the level J - 1 (of energy E$_{J-1}$).

These energy levels can be estimated using quantum mechanics and solving the Schrodinger equation. The Cologne Database for Molecular Spectroscopy (CDMS; \cite{mull05, mull01}), the Jet Propulsion Laboratory (JPL; \cite{pick98}), or the National Institute of Standards and Technology \url{https://www.nist.gov/pml/atomic-spectra-database} are spectroscopic databases that gather information about the properties of molecules (NIST).

Due to the challenges of investigating interstellar ices in the mid-IR, only a few species can be clearly observed there. Using ground-based observation, it is challenging to find interstellar atomic or molecular spectral signatures in the IR regions. The Earth's atmosphere weakens the majority of the signals as they travel through it. Absorption requires a background illumination source, such as a protostar or field star. As of now, IR measurements lead to the conclusion that the few molecules compose the majority of the ice mantles in molecular clouds \citep{herb09}. However, it is also anticipated that more sophisticated species, like COMs, will be frozen on ice grains in dense cores. Weak features resulting from solid COMs that are hidden by more common ice species can be caused by the low sensitivity or low resolution of available observations combined with spectral uncertainty in the IR region. In our attempt to understand the origins of molecules in space, the future NASA JWST \url{https://www.stsci.edu/jwst/} space project, which will investigate the molecular nature of the Universe and the habitability of planetary systems, promises to be a huge step forward. 

\begin{figure}[htbp]
\begin{center}
\includegraphics[width=\textwidth]{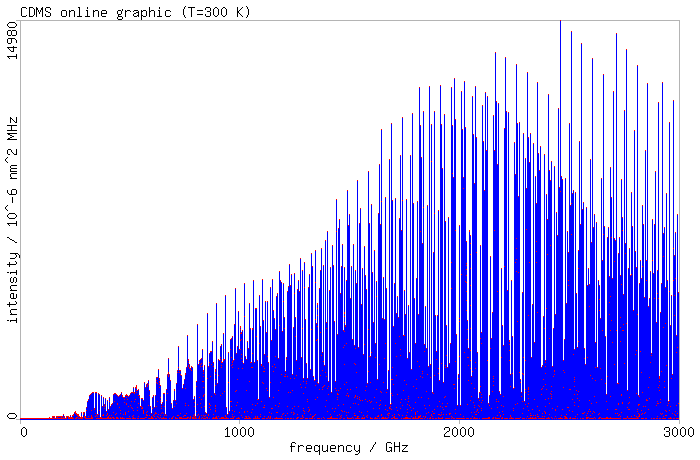}
\caption{Spectrum of first 40 transitions of CH$_3$OH. The intensities are dependent on temperature, taken to be 1000K in this example. \citep[Courtesy:][Taken from CDMS website)]{}.}
\label{fig:met_rot}
\end{center}
\end{figure}
In Figure \ref{fig:met_rot}, an example of rotational transitions of CH$_3$OH is shown as a function of frequency. 

\subsection{Radio interferometry}
\begin{figure}[htbp]
\begin{minipage}{0.45\textwidth}
\includegraphics[width=\textwidth]{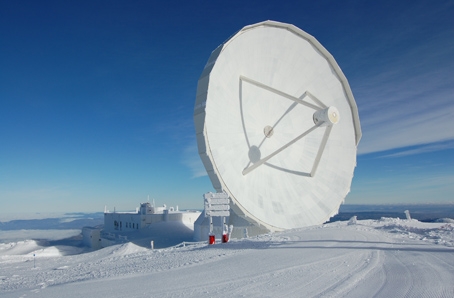}
\end{minipage}
\hskip 0.01 cm
\begin{minipage}{0.45\textwidth}
\includegraphics[width=\textwidth]{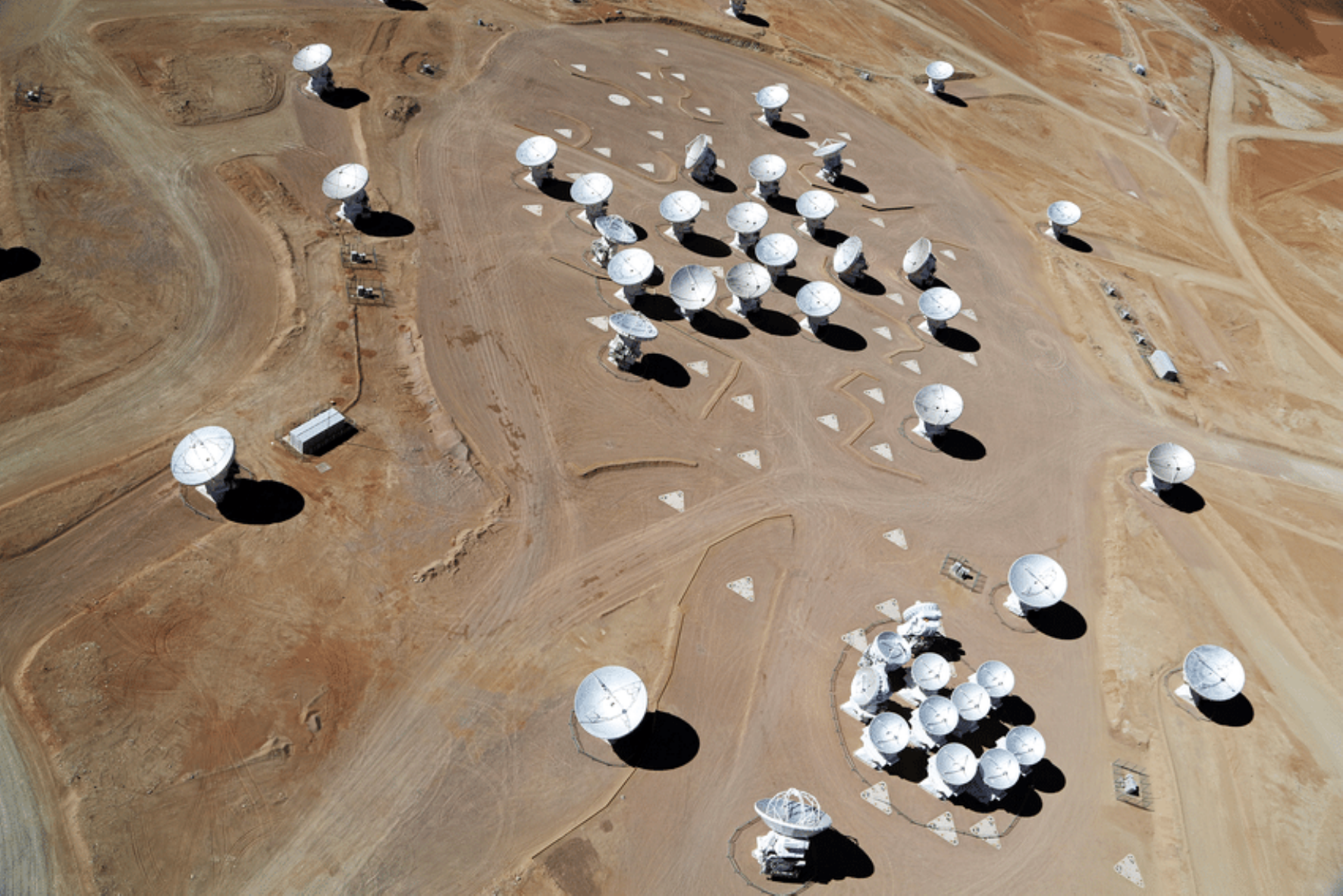}
\end{minipage}
\caption{Single dish radio telescope IRAM 30M (left) and radio interferometer telescope (ALMA). \citep[Courtesy:][Internet]{}.}
\label{fig:telescope}
\end{figure}
Single-dish and interferometric observations are carried out in the millimeter and sub-millimeter radio bands to unfold the astronomical mystery.
\begin{itemize}
    \item Single-dish telescopes, having a single antenna (e.g., IRAM-30m, Effelsberg 100-m Radio Telescope, Green Bank Telescope (GBT), APEX, etc.)
    
    \item Interferometers are combined with many antennas. (e.g., ALMA, PdBI, SMA, etc.)
\end{itemize}
 
An interferometric telescope array is a set of separate telescopes, mirror segments, or radio telescope antennas that work together as a single unit to provide higher angular resolution to images of astronomical objects. The location of the telescope, observable frequencies available from the instruments (both spectral bandwidth and resolution to detect the signature of a targeted molecule), and the antenna size are the primary factors in choosing the basic needs. The next factor is to look at the telescope's lowest resolvable spatial scale or angular resolution. In general, the angular resolution B (in rad) for a single-dish telescope with an antenna diameter D is given by: 
\begin{equation}
 \theta_B=1.22 \frac{\lambda}{D},
\end{equation}
where $\lambda$ is the wavelength. The observed object is poorly resolved, and the angular resolution decreases with increasing wavelength. A larger antenna enhances the angular resolution. The baseline distance B$_{max}$ between the two farthest antennas in an interferometer is what determines the angular resolution $\theta_I$:
\begin{equation}
 \theta_I= \frac{\lambda}{B_{max}}.
\end{equation}

Even for an interferometer, it is ideal for observing with larger antennas since more photons are gathered, enhancing the sensitivity of the signal detected. Single-dish telescopes typically have an angular resolution between 5 and 5000 arcseconds; however, interferometers can have an angular resolution as low as a few milliarcseconds, depending on the frequency. Single-dish telescopes can collect prolonged emissions due to their weak spatial resolution, whereas interferometers are superior at examining the small-scale structures of objects. Interferometers, however, are unable to collect extended emissions more than
\begin{equation}
 \theta_{LAS} \approx 0.6 \frac{\lambda}{B_{min}},
\end{equation}
where B$_{min}$ is the baseline distance between the two closest antennas and $\theta_{LAS}$ is the largest angular structure's angular resolution.

\subsection{Data analysis with CASSIS}\label{sec:cassis}
In my thesis, I analyze the observed spectra using the CASSIS software \citep{vast15}. I have primarily used CASSIS as the line identification tool. It has an excellent graphical interface to choose a molecule from the spectroscopic database (implemented in CASSIS and built on JPL, CDMS, etc.). The spectral analyzer tool displays the frequency of every possible transition for that molecule. A molecule is regarded to have been identified if its multiple transitions are detected. The line bending can also be determined
by simultaneously displaying transitions of numerous molecules. The spectroscopic properties of a specific molecular transition, including its frequency ($\nu$), quantum numbers (J, K$_a$, K$_b$, F, etc.), Einstein coefficient (A$_{ij}$), and upper energy level (E$_{up}$ ), are displayed by CASSIS from the spectroscopic database. 
With the CASSIS, it is possible to fit the line profile with a Gaussian distribution by adjusting the parameters of the following formula:
\begin{equation}
 T(v)=I_0 exp\left( -\frac{(v-v_0)}{2\sigma^2}\right),
\end{equation}
where $\nu_0$ (km/s) is the frequency at the peak position, I$_0$ (K) is the peak intensity of the Gaussian, and $\sigma$ is a parameter related to the FWHM (Full Width Half Maxima) by the following relation:  
\begin{equation}
\label{eqn:sigma}
 \sigma=\frac{FWHM}{2\sqrt{2 ln2}}.
\end{equation}
The velocity of the source with respect to the Local Standard of Rest, or V$_{LSR}$, is given as v$_0$. This velocity can be determined from the central frequency, of the detected transition, by the following relation, 
\begin{equation}
 V_{LSR} = c\frac{\Delta \nu}{\nu_0} = c \frac{\nu_0 - \nu}{\nu_0}.
\end{equation}
where $c$ is the velocity of light (c=$3\times10^8$ m/s), and $\nu_0$ is the frequency of the emitted photon in the rest frame of the source.
The total flux (integrated intensity) of the line transition is calculated from Gaussian fitting in CASSIS using the following relation,
\begin{equation}
 \int T(v) dv = I_0 \times \sqrt{2 \pi \sigma^2}.
\end{equation}
Combining with the Eqn. \ref{eqn:sigma} it becomes,
\begin{equation}
 \int T(v) dv = I_0 \times FWHM \times \sqrt {\frac{\pi}{4 ln2}}\approx 1.065 \times I_0 \times FWHM.
\end{equation}
It is also possible to extract the RMS amplitude (root mean square) of the spectra using baseline fitting. According to the following equation, this is determined in CASSIS:
\begin{equation}
 rms= \sqrt{\frac{1}{N}\sum_{i=1}^{N}(T_i - \langle T_i \rangle)^2},
\end{equation}
Here, N denotes the number of spectra data points, and T$_i$ denotes their intensity. As a function of the system temperature T$_{sys}$, which is the observed antenna temperature modified for the opacity of the environment, the RMS can also be written for a single-dish telescope as below, where the spectral resolution and the on-source integration time of the observations are determined by $\Delta\nu$ and t$_{int}$,
\begin{equation}
 rms\approx \frac{T_{sys}}{\sqrt{2\Delta \nu t_{int}}}.
\end{equation}
In the case of an interferometer with N antennas, the equation becomes
\begin{equation}
 rms\approx \frac{T_{sys}}{\sqrt{N (N-1)\Delta \nu t_{int}}}.
\end{equation}
The noise is only reduced by $\sqrt{2}$ when the on-source time is multiplied by a factor of 2. Because the noise and spectral resolution of the spectra are anti-correlated, flattening the spectra also lowers the noise level. If a line's total flux exceeds at least three times the uncertainty ($\sigma$) of its integrated intensity, we consider the line to have been observed in the spectra, with x being the calibration uncertainty and $\Delta$v being the spectral resolution in the velocity frame (measured in units of m/s),
\begin{equation}
 \sigma=\sqrt{((1+x)rms \sqrt{2 \Delta v FWHM}})+\left(x \int T(v) dv\right)^2.
\end{equation}

The assumption of Local Thermodynamical Equilibrium (LTE, see Sect. \ref{sec:LTE}) or Large Velocity Gradient (LVG, see Sect. \ref{sec:nonlte}) can be used with CASSIS to perform line modeling. While the LVG approach is carried out via the RADEX \citep{vand07} code, the LTE method is implemented directly in CASSIS. The required parameter is the column density of molecules (N$_{mol}$), excitation temperature (T$_{ex}$) for LTE modeling or kinetic temperature (T$_{kin}$) for LVG modeling, FWHM, and the source size ($\theta_S$).
To determine the best-fit LTE model, use a regular grid of models or the Monte-Carlo Markov Chain (MCMC) method.

\subsection{Monte-Carlo Markov Chain method (MCMC)}\label{sec:MCMC}
For the species for which several transitions are detected, the rotational diagram is employed; however, the excitation temperature and column density are constrained for the comparison using a different approach, Markov chain Monte Carlo (MCMC). The probabilistic behavior of a group of atomic particles was the initial focus of the development of the MCMC algorithm. Analytically, it was challenging to do this. As a result, an iterative technique was used to simulate a solution. The probabilities of each feasible occurrence in this stochastic model only depend on the state obtained in the preceding. The MCMC method is a dynamic procedure that uses a random walk to iteratively examine all line parameters, such as molecular column density, excitation temperature, source size, line width, and heads into the solution's space.
Considering the N number of spectra, we applied the $\chi^2$ minimization approach to identify the best-fitted set that can suit the observational findings. This python script calculates the difference in squares ($\chi^2$) between the observed and simulated data and identifies the minimal value of $\chi^2$ regarding the following relation,
\begin{equation}
 \chi_i^2 = \sum_{j=1}^{N_i} \frac{(I_{obs,ij} - I_{model,ij})^2}{rms_i^2+(cal_i \times I_{obs,ij})^2}.
\end{equation}
where I$_{obs,ij}$ and I$_{model,ij}$ are the observed and modeled intensity in the channel j of transition i, respectively. Cal$_i$ is the calibration error, and rms$_i$ is the RMS of the spectrum. Reduced $\chi^2$ is calculated using the below formula,
\begin{equation}
 {\rm \chi^2_{red}=\frac{1}{ \sum_{i}^{N_{spec}} N_i} \sum_{i=1}^{N_{spec}} \chi^2_i}.
\end{equation}
The initial physical values used in the MCMC computation are selected at random from a range between the minimum and maximum values (X$_{max}$ and X$_{min}$) of the user set. The iteration number l, along with other parameters ($\alpha$ and $\nu$), determine the MCMC computation step ($\theta_l$), where $\theta_{l+1} = \theta_l + \alpha (\nu - 0.05)$. ($\nu$ is a random number between 0 and 1). Here $\alpha$ is,
$$
{\rm \alpha=\frac{k(X_{max}-X_{min})}{k^\prime}},
$$
and k is defined as,
 $$
k=r_c \hskip 3.5cm \rm{when \ l>c},
$$
$$
k=\frac{(r_c-1)}{c}l+1  \hskip 2cm  \rm{when \ l<c},
$$ 
The user sets the cutoff parameter (c) and the ratio at cutoff (r$_c$), both variables during modeling. A reduced physical parameter called k$^{\prime}$ is described as being given a value during calculation. $\alpha$ defines the steps' amplitude, starting with a larger step at the beginning of the computation to locate a good $\chi^2$ and smaller steps at the conclusion to extract the value of the probable best $\chi^2$. I have rigorously used this method in my thesis to obtain physical parameters from line fitting of observed lines.

\subsection{Inverse P-Cygni: A special type of spectral line profile}

\begin{figure}
\centering
\includegraphics[height=8cm,width=11cm]{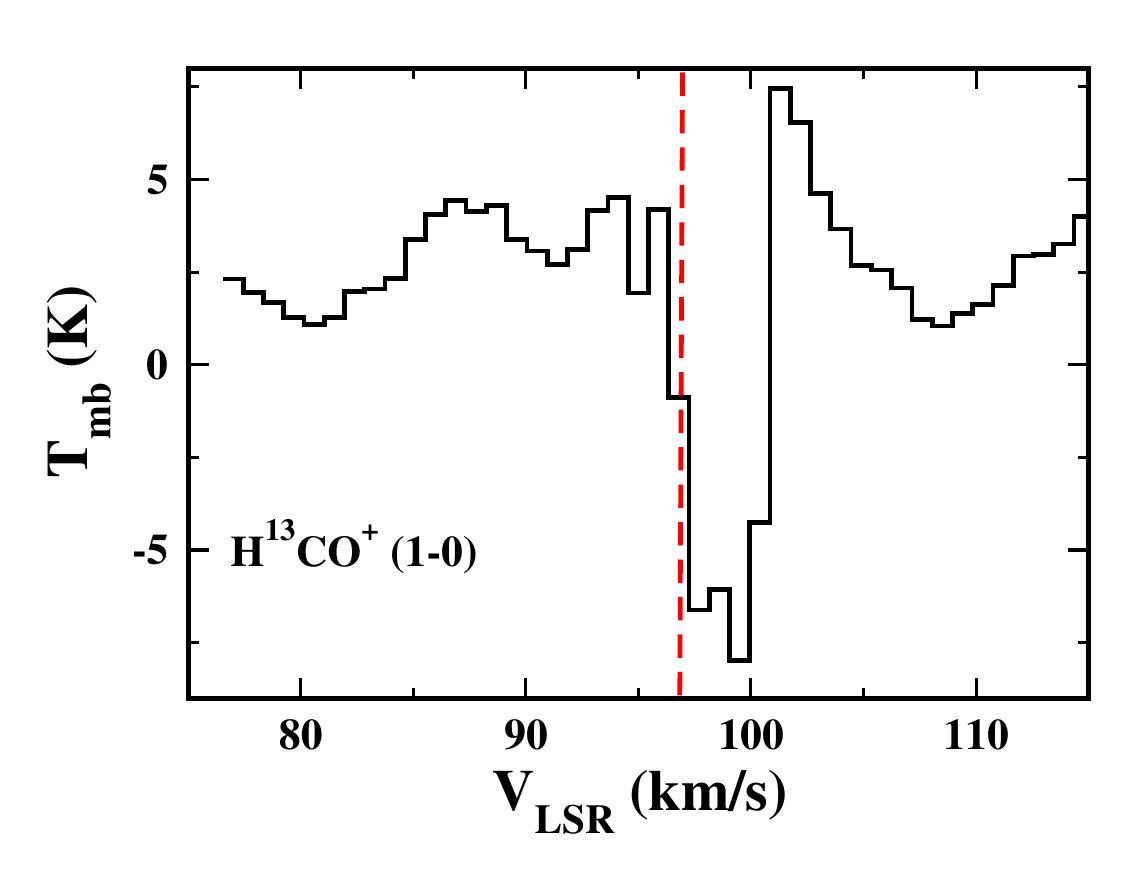}
\caption{Observed spectra of $\rm{H^{13}CO^+}$ (inverse P-Cygni) towards G31 hot molecular core. \citep[Courtesy:][]{gora21}.}\label{fig:h13co_intt}
\end{figure}

P-Cygni is a variable star situated in the constellation of Cygnus. It is located about $5000-6000$ ly away from the Earth. It is the hyper-giant luminous blue variable star of spectral type B1la+
that is one of the brightest stars in the Milky Way. Johann Bayer, a German lawyer, assigned the name "P". Spectral analysis of electromagnetic radiations from distant objects carries lots of information about the physical properties. From the Doppler shift, it is possible to measure the local standard of rest velocity or the mean motion of any celestial object in the Milky way in the neighborhood of the sun.

The name 'P-Cygni profile' refers to the fact that this type of spectra was the first observed towards the P-Cygni star. A P-Cygni profile is a combination of emission and
absorption spectra where emission is red-shifted and absorption is blue-shifted. This type of spectral (P-Cygni) profile indicates the existence of a gaseous envelope expanding away from the central star.

The inverse P-Cygni profile is the opposite of the P-Cygni profile, where both the emission and absorption lobes are present and the emission lobe is blue-shifted, whereas the absorption lobe is red-shifted. This type of profile mainly signifies the in-falling nature of astronomical sources. An example of an inverse P-Cygni profile is shown in Fig. \ref{fig:h13co_intt}. The detailed kinematics behind this type of spectral profile is discussed later.

\begin{figure}
\centering
\includegraphics[height=8cm,width=10cm]{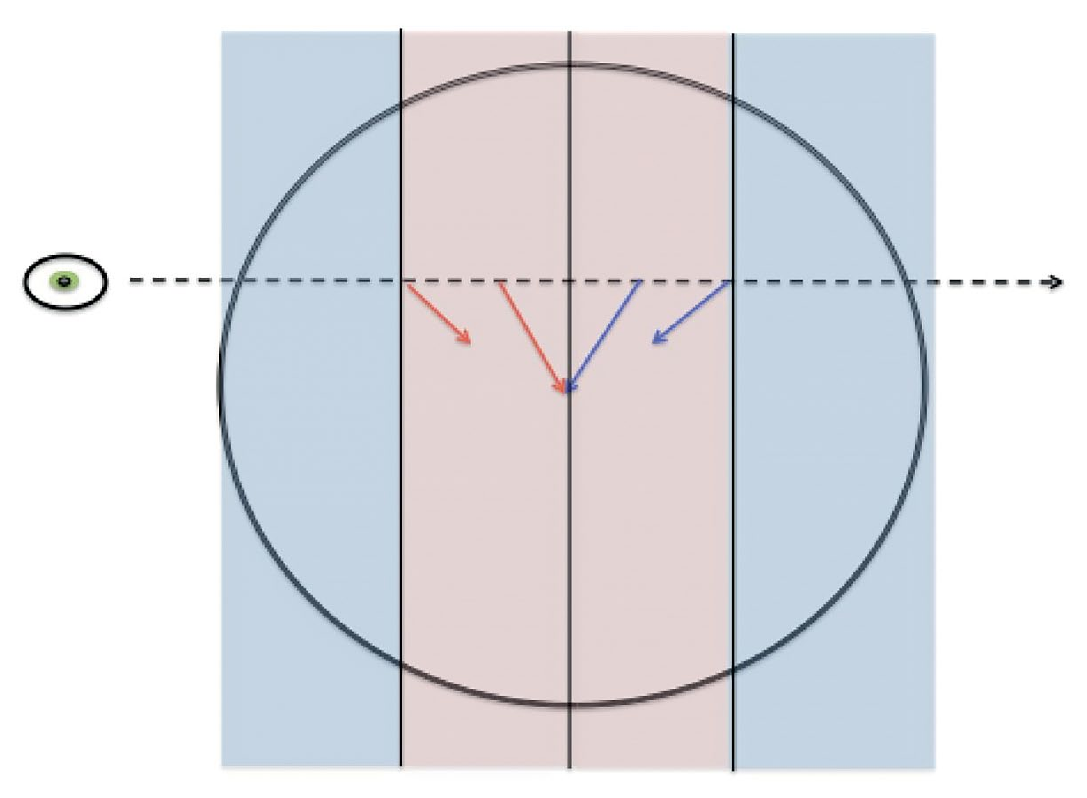}
\caption{The in-falling source is divided into four sub-regions. Here, blue represents hot, while red represents cold. The position of the observer is shown on the left eye. \citep[Courtesy:][Calahan, J., 2017. Inverse P Cygni profiles and their use in astronomical observations. MERICI, p.9.]{}}\label{fig:slab}
\end{figure}
There are mainly two types of temperature which play the role to generate P-Cygni spectra. One is the kinetic temperature of the gas, which is related to the motion of the gas molecules; another one is the excitation temperature which is the temperature difference between a molecule at an upper energy level with a molecule at a lower energy level. In the case of LTE (local thermodynamic equilibrium), these two temperatures can be assumed to be almost equal, whereas when the density is low, mostly in the case of the interstellar medium, this assumption fails. This condition is known as the non-LTE (nonlocal thermodynamic equilibrium) condition.\\
To get an inverse P-Cygni profile, the interior part of the source should be warmer than the exterior part. We consider the source divided into four regions, as shown in Fig. \ref{fig:slab}. In the red-shifted region, there is a cool sheet in front of a hot sheet, and then a hot sheet in front of a cool sheet in the blue-shifted region. According to thermodynamics, when there is a cool sheet in front of a hot sheet, the cool sheet will
absorb a lot of energy from the hot sheet but in the red-shifted region. But in the blue-shifted region, as it is hot in front of a cool sheet, such absorption did not occur. This phenomenon generates the inverse P-Cygni profile having an absorption lobe on the red-shifted side and emission on the blue-shifted side. The kinetic temperature throughout the cloud is considered constant. The reason behind this is that in the case of young objects, there is no central source of heat in the beginning when in-fall starts, and the outer part of the source mainly gets heated by the neighboring starlights. In the central portion when the density becomes high adiabatic process starts and gets heated up. That is why the kinetic temperature of the system is constant and nearly about $10$k. This suggests that the cause behind the generation of the inverse P-Cygni profile is the excitation temperature. In the inner portion, as the density is high, it holds the LTE condition, and the excitation temperature is high, whereas in the outer portion, the density is less, and the LTE condition is no more valid, which suggests low excitation temperature.\\
To observe this phenomenon, certain molecules can be used as a tracer which must possess certain properties. It must be relatively abundant and can have certain energy levels so it can be excited in a cold environment. It must be somewhere in between optically thick and optically thin. Optically thick means it has a chance of less than $50 \%$ to escape the cloud, and optically thin is just the opposite. If a molecule is too optically thick, the asymmetry of inverse P-Cygni cannot be observed; if too optically thin, it would be difficult to detect from the cloud at all. One can think that $\rm{H_2}$ is a very good tracer as it is the most abundant molecule in ISM and quite sure will be abundant in clouds also but $\rm{H_2}$ have no active transitions in this temperature regime. Some other molecules may have different problems, also. We need certain molecules which are abundant as well as have active transitions at that temperature and also must be in between optically thick and thin. A few examples of such types of molecules are $\rm{HCN, CS, N_2H^+, HCO^+}$, etc. By observing such types of profiles towards astronomical sources, we can detect the in-falling nature of the source, which is in a very early stage of star formation. 

\section{Radiative transfer model}
Electromagnetic radiation carries information about astronomical objects. The interstellar dust obscures the electromagnetic radiations at the optical and near-IR wavelengths, but millimeter wavelengths pass freely through the same regions. Dust radiates widely as blackbody radiation across the IR and sub-mm. A thorough understanding of the dust continuum and molecular line emission from star-forming regions is necessary to properly understand the source. 
In the early stages of YSO, the emission from the embedded core protostar is absorbed by dust and re-emitted at longer wavelengths. In-depth knowledge of radiative excitation, spontaneous emission, and other mechanisms affecting the molecular energy level populations is also necessary to fully understand the innermost areas of protostellar. Since each molecular line transition depends on the gas density and temperature of the surrounding environment and has unique properties, they can be used as efficient probes of the protostellar environment.

\subsection{Thermal emission}
\begin{figure}[htbp]
\begin{center}
\includegraphics[width=0.6\textwidth]{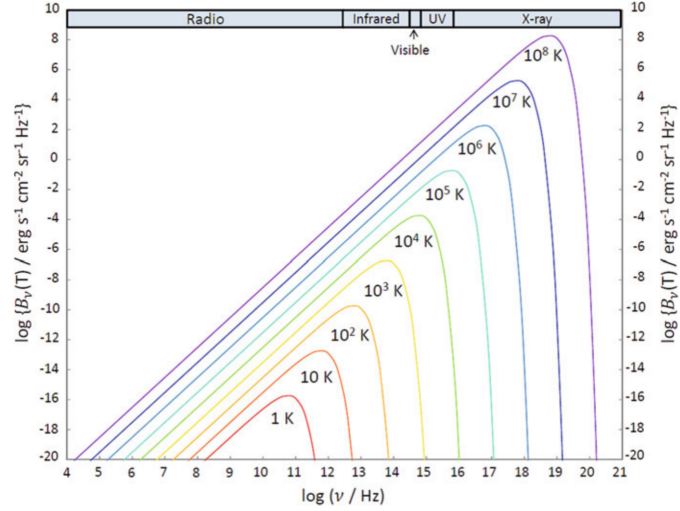}
\caption{Spectral signature of blackbody radiation at various temperatures \citep[Courtesy:][]{}.}
\label{fig:weins}
\end{center}
\end{figure}

The radiation emitted by a matter is related to the temperature of the matter. A matter is considered dark by human if it does not emit radiation in the visible range. Instead, matter at different temperatures emits electromagnetic radiation of different wavelengths. Low temperature objects emit radiation of higher wavelengths. For example, the interstellar matter with low temperature ($\sim$ 10 K) emits radio waves instead of infrared. 
Radiation emitted from any matter at non-zero temperature is known as thermal emission. In an ideal case, an entirely opaque matter would absorb all radiation at all wavelengths and emits nothing. This type of matter is called a blackbody. However, in reality, no perfect blackbody exists. The emission from a blackbody only depends on its temperature, not its composition. Gustav Kirchhoff first established this important characteristic in 1860. Depending on this, later Max Planck established the theory of blackbody radiation and introduced the Planck constant in 1901. This theory was a milestone in the foundation of quantum theory. 
Energy emitted per area ds, frequency d$\nu$, time dt, solid angle d$\Omega$ is described as,
\begin{equation}
\label{eqn:energy}
    dE=B_{\nu}(T)dS dt d\nu d\Omega,
\end{equation}
where B$_{\nu}$(T) is the intensity of the radiation emitted from the blackbody at temperature T. The intensity described by the Planck function is,
\begin{equation}
\label{eqn:planck}
    B_{\nu}(T)=\frac{2h\nu^3}{c^2} \frac{1}{e^{h\nu/k_BT}-1}
\end{equation}
where $\nu$ is the frequency of radiation, k$_B$ is the Boltzmann constant, c is the speed of light, and h is the Planck constant. The units of B$_\nu$ is erg $s^{-1} cm^{-2} Hz^{-1} sr^{-1}$. This function only depends on the blackbody's temperature and not on the blackbody's structure or composition. Figure \ref{fig:weins} shows a plot of the function B$_{\nu}$ with frequency ($\nu$) at different temperatures. The frequency corresponding to the peak intensity increases with increasing temperature. It is known as Wein's displacement law.  

If $h\nu$/k$_BT \ll$ 1, expanding the exponential part of Eq. \ref{eqn:planck} we obtain,
\begin{equation}
    B_{\nu}(T) \cong \frac{2\nu^2}{c^2}k_BT,
\end{equation}
which is known as Rayleigh-Jeans law. In this case, the intensity of radiation emitted by a body is directly proportional to the temperature of the body. This equation is independent of the Planck constant (h) and is known as the classical limit of blackbody radiation. When $h\nu$/k$_BT \gg$ 1, the Eq. \ref{eqn:planck} becomes,
\begin{equation}
     B_{\nu}(T)\cong \frac{2h\nu^3}{c^2} exp \left(-\frac{h\nu}{k_BT}\right). 
\end{equation}
In this case, with increasing frequency, the intensity decreases exponentially.

As previously mentioned, blackbody radiation solely depends on the temperature of the emitting source. However, thermal radiation also provides rich details about the composition of matter (e.g., atoms and molecules). For instance, the strength of their spectral lines can be used to estimate the number of atoms and molecules in interstellar clouds. In detail discussions on the thermal radiation of atoms and molecules are discussed later in this thesis.

\subsection{Radiative transfer equation}
\begin{figure}[htbp]
\begin{center}
\includegraphics[width=0.6\textwidth]{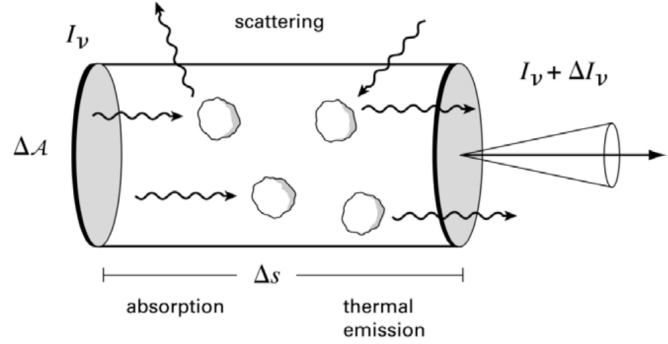}
\caption{Atmospheric absorption through the electromagnetic spectral range \citep[Courtesy:][]{stah04}.}
\label{fig:radtrans}
\end{center}
\end{figure}

Let us consider an electromagnetic wave with frequency $\nu$ assuming it is traveling in a straight line. This assumption is justified when the beam diameter is much larger than the wavelength of radiation ($\lambda$=c/$\nu$). Similar to the Eqn. \ref{eqn:energy}, the intensity I$_{\nu}$ of the wave is defined as,
\begin{equation}
    dE=I_{\nu} dS dt d\nu d\Omega.
\end{equation}
The unit of intensity I$_\nu$ is s$^{-1}$cm$^{-2}$sr$^{-1}$Hz$^{-1}$. The radiation intensity varies due to the emission or absorption by the atoms and molecules in the intervening medium. Here, we consider the direction of propagation of the wave along the x-axis. For a small propagation dx, change in the intensity dI$_\nu$ is described as,
\begin{equation}
\label{eqn:di}
    dI_\nu=-\alpha_\nu I_\nu dx+j_\nu dx,
\end{equation}
where $\alpha_\nu$ is the absorption coefficient and j$_\nu$ is the emission coefficient. So the equation \ref{eqn:di} becomes,
\begin{equation}
    \frac{dI_\nu}{dx}=-\alpha_\nu I_\nu+j_\nu.
\end{equation}
$\alpha_\nu$ and $j_\nu$ are functions of $x$. Dividing both sides with $\alpha_\nu$ and defining \begin{equation}
    d\tau_\nu=\alpha_\nu dx,
\end{equation}
the simplified equation becomes
\begin{equation}
    \frac{dI_\nu}{d\tau}=-I_\nu+S_\nu.
\end{equation}
Here, we define a new function called the source function, defined as the ratio of emission coefficient to absorption coefficient,
\begin{equation}
    S_\nu=j_\nu/\alpha_\nu.
\end{equation}
By solving this radiative transfer equation, we obtain the intensity at any position if the source function S$_\nu$ is supplied. The intensity at x=0 ($\tau_\nu$=0) is set to B$_\nu$(T$_b$), indicating that the source radiation is blackbody radiation at T$_b$. The cosmic microwave background radiation ($T_b=2.73$ K) should be the background radiation if the cloud is only present when x$\geq$0. We assume that the cloud is physically and chemically homogeneous for simplicity. S$_\nu$ therefore remains constant at all points where x$\geq$0, and the following equation gives us the intensity at x=L,
\begin{equation}
\label{eqn:inu}
    I_\nu=S_\nu + exp(-\tau_\nu)[B_\nu(T_b)-S_\nu],
\end{equation}
where 
\begin{equation}
  \tau_\nu=\alpha_\nu L.  
\end{equation}
The optical depth or optical thickness is denoted as $\tau_\nu$ in this case. The above relation relates this quantity to the actual cloud thickness, L. If the absorption coefficient is low, the optical depth can be extremely low, even for substantial physical thicknesses. 

Assume that a blackbody at a temperature T fully encloses a cloud. The cloud will reach thermal equilibrium at this temperature after sufficient time. Because the system (the cloud enclosed by the blackbody) continues to be a blackbody, the intensity we detect when we create an infinitesimally small hole to view within the cloud is B$_\nu$(T). Additionally, we can set T$_b$=T. So, Eqn. \ref{eqn:inu} becomes,
\begin{equation}
    B_\nu(T)=S_\nu+exp(-\tau_\nu)[B_\nu(T)-S_\nu].
\end{equation}
Blackbody radiation is independent of the internal structure of the emitting body, so this equation holds for any value of $\tau_\nu$. So, the condition will be satisfied only when the source function is the Planck function. We obtain,
\begin{equation}
    S_\nu=B_\nu (T).
\end{equation}
This relation suggests that the absorption coefficient and emissivity are interdependent. The ratio of absorption coefficient to emissivity (source function, S$_\nu$) depends on the temperature through the Planck function. This relation is known as Kirchhoff's law of thermal radiation, which is based on the emission and the absorption in thermal equilibrium.
Substituting S$_\nu$ by B$_\nu(T)$, from equation \ref{eqn:inu} we have,
\begin{equation}
    I_\nu=B_\nu(T)+exp(-\tau_\nu)[B_\nu(T_b)-B_\nu(T)].
\end{equation}
In actual observations, the gain fluctuation of the telescope changes with the atmospheric conditions. This is minimized by alternately seeing the target and nearby positions free of cloud emission and absorption. Their difference is then calculated. The intensity toward the off-source position is expressed as,
\begin{equation}
    I_\nu=B_\nu(T_b),
\end{equation}
The difference represents the actual intensity observed by the radio telescopes, calculated using the following equation,
\begin{equation}
   \bigtriangleup I_\nu=[B_\nu(T)-B_\nu(T_b)]{1-exp(-\tau_\nu)}.
\end{equation}
In radio astronomy, a temperature scale is often used to represent intensity. The intensity of a blackbody monotonically grows with increasing temperature at a fixed frequency, as shown by differentiating the Planck function by T. Therefore, the temperature at which a blackbody emission of the same intensity can be used to indicate intensity. Specifically, the Rayleigh-Jeans law (h$\nu$/k$_B$T$\ll$1) states that intensity is inversely related to temperature. Therefore, regardless of its applicability, the Rayleigh-Jeans law is used to establish the temperature scale of the intensity, T. Next, we have,
\begin{equation}
\label{eqn:inten}
    \bigtriangleup T=\frac{h\nu}{k_B}\Bigl\{\frac{1}{exp(h\nu/k_BT)-1}-\frac{1}{exp(h\nu/k_BT_b)-1} \Bigr\}\big\{1-exp(-\tau_\nu)\big\}.
\end{equation}
In the limit h$\nu$/k$_B$T$\ll$1, the equation simplifies to,
\begin{equation}
\label{eqn:t}
    \bigtriangleup T= (T-T_b)\big\{ 1- exp(-\tau_\nu)\big\}.
\end{equation}

We may avoid using the rather complicated quantities of erg s$^{-1}$ cm$^{-2}$ Hz$^{-1}$ sr$^{-1}$ in equation \ref{eqn:inten} and can easily imagine a physical temperature of an emitter using equation \ref{eqn:t}.

\subsection{Absorption Coefficient}
\begin{figure}[htbp]
\begin{center}
\includegraphics[width=0.6\textwidth]{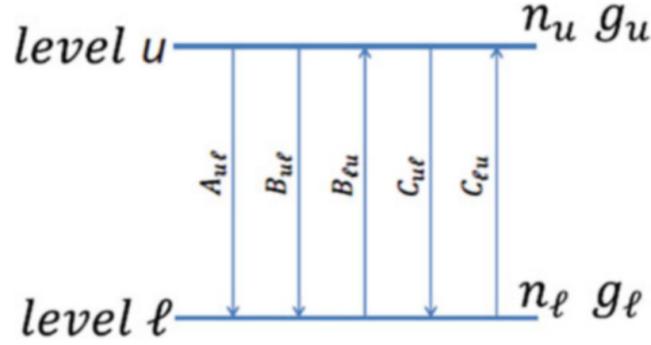}
\caption{Two-level system: Populations for the upper and lower levels are denoted by n$_u$ and n$_l$, respectively; degeneracies by g$_u$ and g$_l$, respectively.}
\label{fig:abs_coef}
\end{center}
\end{figure}

The absorption coefficient (which reflects the electromagnetic characteristics of an emitter) and the optical depth are connected. Here, we express the absorption coefficient in terms of atom or molecule characteristics. For the sake of simplicity, we first consider a two-level system with a lower level (level l) and an upper level (level u), which are determined by the energy levels of a molecule or an atom (Fig. \ref{fig:abs_coef}). Then the absorption coefficient for the transition from level l to level u. 
B$_{lu}$J$_\nu$ represents the absorption probability per unit time. Here, B$_{lu}$ is the Einstein B coefficient for a transition from level l to level u, and J$_\nu$ is the mean intensity of the molecule represented by
\begin{equation}
    J_\nu=\frac{1}{4\pi} \int I_\nu d\Omega.
\end{equation}
A similar relation holds in the case of stimulated emission from level u to level l. The energy emitted from an electromagnetic wave in a solid angle d$\Omega$, in a frequency range d$\nu$, time dt, and volume dV is given by,
\begin{equation}
\label{eqn:phi}
    -dI_\nu d\nu d\Omega dt dS=\frac{h\nu}{4\pi}\phi(\nu)(n_lB_{lu}-n_uB_{ul})I_\nu d\nu d\Omega dt dV,
\end{equation}
where n$_j$ is the number of molecules populating level j, and $\phi(\nu)$ is the line profile function that follows the normalization relation,
\begin{equation*}
    \int \phi(\nu) d\nu= 1.
\end{equation*}
As dV=dS dx, and from the definition,
\begin{equation}
    dI_\nu= -\alpha_\nu I_\nu dx,
\end{equation}
we can compare with Eqn. \ref{eqn:phi} and have the expression for absorption coefficient,
\begin{equation}
    \alpha_\nu = \frac{h\nu}{4\pi} \phi(\nu)(n_lB_{lu}-n_uB_{ul}).
\end{equation}
The Einstein B coefficients B$_{lu}$ and B$_{ul}$ are related by the following relation and proportional to the Einstein A coefficients,
\begin{equation}
    g_l B_{lu}=g_u B_{ul},
\end{equation}
and,
\begin{equation}
    B_{ul}=\frac{c^2}{2h\nu_{ul}^3} A_{ul}.
\end{equation}
$\nu_{ul}$ is the resonance frequency for the two-level system, and g$_l$ is the degeneracy of the level l. Using this relation for a two-level system, the absorption coefficient becomes,
\begin{equation}
\label{eqn:alpha}
    \alpha_{ul}=\frac{c^2 n_u}{8\pi \nu_{ul}^2 \bigtriangleup\nu}\Big\{exp\left(\frac{h\nu_{ul}}{k_BT}\right)-1\Big\}A_{ul},
\end{equation}
where it follows the Boltzmann distribution at a temperature T, the populations at level l and level u are described by,
\begin{equation}
\label{eqn:nunl}
    \frac{n_u}{n_l}=\frac{g_u}{g_l}exp\left(-\frac{h\nu_{ul}}{k_BT}\right).
\end{equation}

\begin{figure}[htbp]
\begin{center}
\includegraphics[width=0.7\textwidth]{chap1fig/line_profilefun.pdf}
\caption{Line profile function. $\nu_{ul}$ is the line center frequency, and $\bigtriangleup\nu$ is line width \citep[Courtesy:][]{yama17}.}
\label{fig:lineprofile}
\end{center}
\end{figure}

When $\bigtriangleup\nu$ is much smaller than $\nu_{ul}$, the line profile function $\phi(\nu)$ (see Figure. \ref{fig:lineprofile}) is described as,  
\begin{equation}
    \phi(\nu) = \frac{1}{\bigtriangleup\nu}.
\end{equation}
The Einstein A coefficient, A$_{ul}$ is described by,
\begin{equation}
\label{eqn:eina}
    A_{ul}=\frac{64\pi^4\nu_{ul}^{3}S_{ul}\mu^2}{3hc^3g_u},
\end{equation}
where S$_{ul}$ is the line strength, $\mu$ is the dipole moment responsible for the transition. In case of linear or diatomic molecules, S$_{ul}$ equals J+1 for the transition J+1$\longleftrightarrow$J. For many molecules, these values are tabulated in spectral line databases. 

assuming the level population has a Boltzmann distribution at temperature T, the number of molecules in level u per unit volume can be expressed in terms of the total number of molecules, n.
\begin{equation}
\label{eqn:part}
    n_u=\frac{g_u}{U(T)}n . exp\left(-\frac{E_u}{k_BT}\right),
\end{equation}
here E$_{u}$ is the upper state energy at level u and U(T) is the partition function at temperature T. From Equation \ref{eqn:alpha}, \ref{eqn:eina}, \ref{eqn:part} we have,
\begin{equation}
    \alpha_{ul}=\frac{8\pi^3\nu_{ul}S_{ul}\mu^2}{3hc\bigtriangleup\nu U(T)} n \biggl\{exp \left(\frac{h\nu_{ul}}{k_BT}\right)-1\biggr\}exp\left(-\frac{E_u}{k_BT}\right).
\end{equation}
Since the Doppler effect plays a significant role in determining line width, it is more common to express line width in terms of velocity than frequency. The velocity width is expressed as,
\begin{equation}
    \bigtriangleup v=\frac{\bigtriangleup\nu}{\nu_{ul}} c.
\end{equation}
Finally, we obtain the expression for optical depth in terms of column density N, where N=nL,
\begin{equation}
\label{eqn:tauul}
    \tau_{ul}=\frac{8\pi^3S_{ul}\mu^2}{3h\bigtriangleup vU(T)}\biggl\{exp \left(\frac{h\nu_{ul}}{k_BT}\right)-1\biggr\}exp\left(-\frac{E_u}{k_BT}\right)N.
\end{equation}
The above equation shows that the column density, line strength, and square of the dipole moment are all related to the optical depth. Therefore, the column density can be calculated from observations when the cloud's depth (L) along the line of sight is known.

\subsection{LTE modeling}\label{sec:LTE}
To link the population of the upper level to the overall number of molecules, Eqn. \ref{eqn:tauul} is derived under the assumption that the Boltzmann distribution at a temperature T represents the level population. The system is said to be in local thermodynamic equilibrium (LTE). This temperature is the rotation temperature for a rotational energy-level system.
From the following relation the excitation temperature T$_{ex}$ for the rotational transition from level u $\rightarrow$ l can be estimated,
\begin{equation}
    \frac{n_ug_l}{n_lg_u}=exp \left(\frac{h\nu}{k_BT_{ex}}\right).
\end{equation}
The excitation temperature and rotation temperature are the same under LTE conditions. When molecules collide with H$_2$ in interstellar clouds, they are excited and de-excited, while collisions with H atoms and electrons can also be significant in some circumstances. The LTE requirement is met if the rotation temperature is close to the kinetic temperature of H$_2$ and the collisions are sufficiently frequent compared to radiation probability. Radiative cooling becomes more significant as collision frequency decreases. When this happens, the level population significantly differs from LTE and is no longer modeled by a single Boltzmann distribution. Excitation temperatures in this environment vary from transition to transition and are often lower than the gas kinetic temperatures.

\subsection{Optically thin case: Rotational diagram analysis}
\begin{figure}[htbp]
\begin{center}
\includegraphics[width=0.7\textwidth]{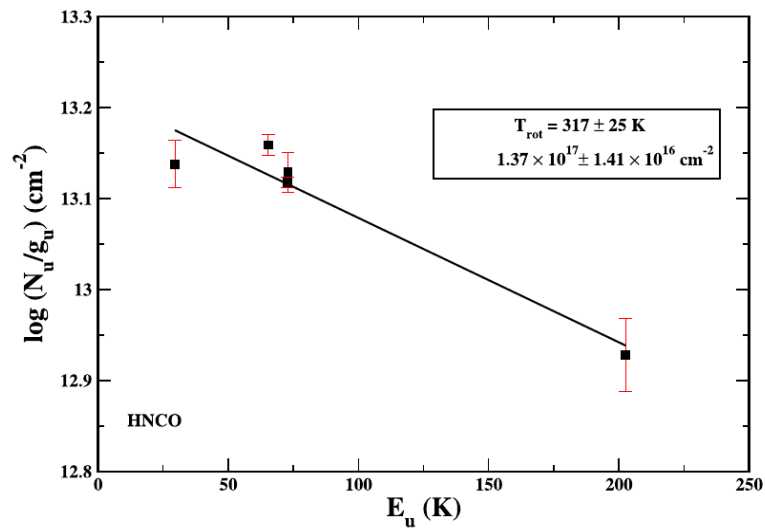}
\caption{ Rotational diagrams of HNCO, Black filled squares are the data points, and red lines represent the error bars. Best-fit rotational temperature and column density are mentioned inside the small
box in the right corner of the figure.\citep[Courtesy:][]{}.}
\label{fig:rotationdia}
\end{center}
\end{figure}

We refer to a spectral line as optically thin when its optical depth is much below unity. Optically thin typically refers to having a low column density or line strength because the optical depth is related to both. From Equation. \ref{eqn:inten} expanding the exponential part and approximating we have,
\begin{equation}
    \bigtriangleup T=\frac{h\nu_{ul}}{k_B}\Bigl\{\frac{1}{exp(h\nu_{ul}/k_BT)-1}-\frac{1}{exp(h\nu_{ul}/k_BT_b)-1} \Bigr\} \tau_{ul}.
\end{equation}
From the above equation, we can see that the intensity is proportional to the optical depth and eventually to the column density. The photons emitted in the cloud escaped from the cloud without getting absorbed in an optically thin case.
In the limit when, h$\nu$/k$_B$T $\ll$ 1 and T \\gg T$_b$, in LTE condition we have,
\begin{equation}
    \bigtriangleup T \bigtriangleup v=\frac{8\pi^3\nu_{ul}S_{ul}\mu^2}{3k_BU(T)}exp\left(-\frac{E_u}{k_BT}\right)N.
\end{equation}
The above relation can be written as,
\begin{equation}
\label{eqn:rot}
    ln\frac{3k_BW}{8\pi^3\nu_{ul}S_{ul}\mu^2}=ln\frac{N}{U(T)}-\frac{E_u}{k_BT},
\end{equation}
 \noindent where W represents the integrated intensity. W=$\bigtriangleup$ T$\bigtriangleup$ v applies to the idealized box-shaped line profile function depicted in Fig \ref{fig:lineprofile}. The observed values for the left side of the Equation \ref{eqn:rot} can be plotted against the upper-state energy E$_u$ if several transitions for the same species are observed using a telescope. If the LTE condition is met, then the relationship should be linear. In this case, the intercept may be used to generate N/U(T), whereas the slope can derive 1/k$_B$T. The partition function is then calculated at the determined temperature to estimate the column density. In warm, thick clouds, such as star-forming cores, this technique, known as the rotation diagram approach, can be used to determine molecules' column density and rotation temperature. However, it only works if the assumptions necessary for this technique are reasonably met. An example of this method is shown in Figure \ref{fig:rotationdia} toward a high-mass star-forming region G10.47+0.03. It represents a rotational diagram for HNCO, which contains peptide-like bonds and plays a vital role in prebiotic chemistry.
The fit often fails because the observed cloud consists of several components with distinct physical characteristics. 

\subsection{Optically thick case}
The optical depth for abundant species like CO, HCN, and HCO$^+$ can sometimes be much higher than unity. The line is said to be optically thick in this situation. As the exponential part of Equation \ref{eqn:t} can be neglected in the limit $\tau_\nu$ $\gg$ 1, h$\nu$/k$_B$T $\ll$ 1, and T $\gg$ T$_b$, the intensity is, 
\begin{equation}
    \bigtriangleup T=T,
\end{equation}
i.e., observed temperature equals the excitation temperature. In an optically thick regime, the photons emitted in the cloud get reabsorbed within the cloud and can not escape from it, and the surface temperature determines the intensity. In other words, in the line profile the cloud behaves like a blackbody at the molecule's excitation temperature, and the cloud surface can be thought of as a "photosphere." As a result, the optically thick line cannot be used to determine the column density. In this case, the isotopes are used to determine the column density if the molecules of rarer atoms are optically thin.

\subsection{Non-LTE moddeling}\label{sec:nonlte}
Although it is the simplest, the LTE method must adequately explain most physical cases. The level population calculation can be separated from the radiative transfer equation using the LVG approach (Large Velocity Gradient), also known as the escape probability. \cite{sobo60} was the first to present this technique for an expanding sphere. This technique functions as follows: The statistical equilibrium equation for population densities is modified by adding a factor, $\beta(\tau)$. The probability that a photon generated locally at the optical depth $\tau$ can escape the system under study is represented by this factor. Only when the properties of the gas remain constant along the Sobolev length L$_S$, which is defined as,
\begin{equation}
    L_S=\frac{v_{th}}{dv/dr},
\end{equation}
where v$_{th}$ is the thermal broadening of the spectral line, and dv/dr is the velocity gradient. When the LVG cell size L is less than the Sobolev length, the Sobolev approximation is applicable,
\begin{equation}
    L < L_S.
\end{equation}
This assumption is true if the turbulence dominates the line's natural width. According to the Sobolev approximation, photons released into the cloud can only interact with nearby molecules. This approximation turns the photon transport problem from a global problem to a local one. In this approximation, it is necessary to determine the escape probability of a photon. One fraction of emitted photons remains in the medium, while the other fraction will leave. The current expression for the local mean radiation is,
\begin{equation}
    J^{loc}_\nu=(1-\beta)S_\nu+\beta J_\nu(T_{bg}).
\end{equation}
If $\beta$ = 1, J$^{loc}_\nu$ = J$_\nu$(T$_{bg}$) corresponding to the optically thin case, and if $\beta$ = 0, J$^{loc}_\nu$ = S$_\nu$ (LTE case). As a result, the equation for the equilibrium at the local level is net absorption = emitted photons that do not escape, applying this we have,
\begin{equation}
    (n_lB_{lu} - n_uB_{ul})U_\nu=n_u A_{ul}(1-\beta(\tau)).
\end{equation}
From this, when two levels are in equilibrium, we know,
\begin{equation}
\label{eqn:nlc}
    n_lC_{lu} - n_uC_{ul} - \beta(\tau)n_uA_{ul} = 0.
\end{equation}
This new form makes the population density calculation simpler by eliminating the effect of the energy density U$_\nu$. This $\beta(\tau)$ parameter depends on the geometry of the source:
\begin{itemize}
    \item For an expanding sphere:
    \begin{equation}
        \beta(\tau) = \frac{1-e^{-\tau}}{\tau}
    \end{equation}
    \item For a homogeneous slab:
    \begin{equation}
        \beta(\tau) = \frac{1-e^{-3\tau}}{3\tau}
    \end{equation}
    \item For a uniform sphere:
     \begin{equation}
        \beta(\tau) = \frac{1.5}{\tau}\Bigl[1-\frac{2}{\tau^2}+\left(\frac{2}{\tau}+\frac{2}{\tau^2}\right)e^{-\tau}\Bigr]
    \end{equation}
\end{itemize}
Due to the "trapping" of emitted photons, the probability that a photon would escape the medium decreases noticeably as the gas gets optically thick ($\tau \gg$ 1). The number of photons leaving the system in this instance must be subtracted from the effective rate of spontaneous emission:
\begin{equation}
    A_{ul}^{eff}=A_{ul}\beta(\tau).
\end{equation}
So, the critical density is expressed as,
\begin{equation}
\label{eqn:ncr}
    n_{cr}^{eff}=\frac{A_{ul}\beta(\tau)}{\gamma_{ul}}
\end{equation}
Because the critical density for thermalization is lower in this condition, molecular levels can thermalize more readily ($\beta(\tau$) < 1). The resolution of the LVG technique is relatively similar to the LTE approach discussed in the preceding section (\ref{sec:LTE}), with the exception that the terms A$_{ul}$ and U$_\nu$ have been substituted out for A$_{ul}\beta(\tau)$ and (1-$\beta$)S$_\nu$, respectively.
From Equation \ref{eqn:nlc}, we can write,
\begin{equation}
    n_ln\gamma_{lu}=n_u(n\gamma_{ul+\beta A_{ul}}).
\end{equation}
Considering this equation with \ref{eqn:nunl},
\begin{equation}
    n_ln\gamma_{ul}\frac{g_u}{g_l}e^{-\frac{h\nu_0}{k_BT_{kin}}}=n_u(n\gamma_{ul}+\beta A_{ul})
\end{equation}
further manipulating we have,
\begin{equation}
     \frac{n_u}{n_l}=\frac{g_u}{g_l}e^{-\frac{h\nu_0}{k_BT_{kin}}}\Bigl[\frac{\beta A_{ul}}{n\gamma_{ul}}+1\Bigr]^{-1}.
\end{equation}
Considering Equation \ref{eqn:ncr}
\begin{equation}
    \frac{n_u}{n_l}=\frac{g_u}{g_l}e^{-\frac{h\nu_0}{k_BT_{kin}}}\Bigl[\frac{n_{cr}^{eff}}{n}+1\Bigr]^{-1}.
\end{equation}

\begin{equation}
    \frac{g_u}{g_l}e^{-\frac{h\nu_0}{k_BT_{ex}}}=\frac{g_u}{g_l}e^{-\frac{h\nu_0}{k_BT_{kin}}}\Bigl[\frac{n_{cr}^{eff}}{n}+1\Bigr]^{-1}
\end{equation}

\begin{equation}
    \frac{h\nu_0}{k_BT_{ex}}=\frac{h\nu_0}{k_BT_{kin}}ln\left(\frac{n_{cr}^{eff}}{n}+1\right).
\end{equation}
Finally, we obtain,
\begin{equation}
    T_{ex}=\frac{T_{kin}}{1+\frac{k_BT_{kin}}{h\nu_0}ln\left(\frac{n_{cr}^{eff}}{n}+1\right)}.
\end{equation}
In the regime, n$\gg$n$_{cr}^{eff}$, T$_{ex}$=T$_{kin}$ and the lines follows the LTE case.

\subsection{Dust thermal equilibrium}
From equation \ref{eqn:planck}, we can rewrite the Planck function in terms of wavelengths,
\begin{equation}
  B_{\lambda}(T)=\frac{2hc^2}{\lambda^5} \frac{1}{e^{hc/\lambda k_BT}-1} .
\end{equation}

\begin{figure}[htbp]
\begin{center}
\label{fig:dust}
\includegraphics[width=0.7\textwidth]{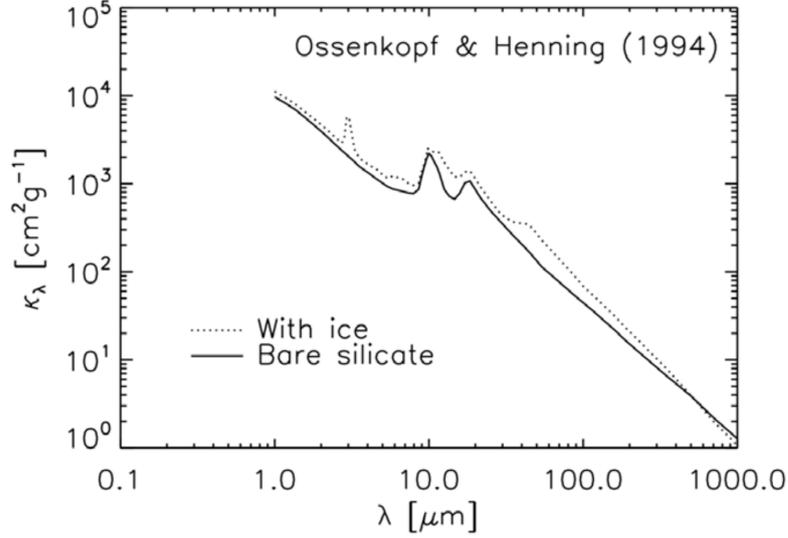}
\caption{ Dust absorption opacity with an ice-mantle and bare silicate grain. \citep[Courtesy:][]{osse94}.}
\end{center}
\end{figure}

The thermal emissivity of dust using the blackbody approximation can be defined as:
\begin{equation}
\label{eqn:kapp}
 j_{\nu,therm}=\rho \kappa_{\nu,abs} B_\nu(T).
\end{equation}
The volumetric cooling rate due to emission from the dust grains, $\Lambda_d$, determines the additional emission from a dusty medium to a beam of radiation traveling across it. By substituting $\rho \kappa_{\nu,abs}$ with $n_d \sigma_d Q_\nu,abs$ in equation \ref{eqn:kapp}'s expression for the emissivity of matter and integrating over all solid angles,
\begin{equation}
 \Lambda_d=4\pi n_d \sigma_d  \int_{0}^{\infty}Q_{\nu,abs} B_\nu(T_d) d\nu,
\end{equation}
where $\sigma_d$ represents the dust cross-section, n$_d$ is the dust number density, and Q$_{\nu,abs}$ is the frequency dependent absorption efficiency. The dust grain cross-section is denoted by,
\begin{equation}
 \sigma_d=\sigma_abs + \sigma_scat.
\end{equation}
It describes the total cross-section as a sum of a cross-section of absorption and scattering. Q$_\nu,abs$ is related to the $\sigma_abs$, the absorption cross-section, by,
\begin{equation}
 \sigma_abs=Q_\nu,abs \sigma_{geo}.
\end{equation}
The geometric cross section is defined by $\sigma_{geo}$=$\pi r^2$. The dust absorption opacity $\kappa_{\nu,abs}$ is related to the absorption cross-section as, 
\begin{equation}
 \kappa_{\nu,abs}=\sigma_{abs}/m,
\end{equation}
where m defines the dust particle mass. The size, chemical content, and morphology of the dust particles can affect the absorption property. A spherical dust particle made up of an ice mantle, and a refractory core is adopted for simplicity. The typical assumption is that the frozen mantle is made up of a mixture of H$_2$O, CO, and other molecules, while the core is thought to be made up of silicates that can replicate the distinctive 10 $\mu$m feature seen in SEDs of YSOs (Fig. \ref{fig:dust}).

Figure \ref{fig:dust} and Eqn. \ref{eqn:kapp} shows that the dust emissivity is highly dependent on wavelength and temperature.

Based on the received light, it is required to determine the dust temperature to define the distribution of dust density in the dense envelope surrounding a protostar. If the other energy sources, such as friction in viscous accretion, are neglected, the radiation from the protostars heats the dust in protoplanetary environments. The dust grains are highly efficient at absorbing optical and IR photons, which heats the grains (Fig. \ref{fig:dust}). A single particle of dust heats up depending on its protostellar radiation F$_\nu$ and opacity,
\begin{equation}
 \Gamma_d=\int_{0}^{\infty} \kappa_{\nu,abs} F_\nu d\nu.
\end{equation}
The cooling of a dust grain is given by
\begin{equation}
 \Lambda_d=4\pi \int_{0}^{\infty} \kappa_\nu B_\nu (T_d) d\nu.
\end{equation}
The dust must be in thermal equilibrium in order that these two terms be equal. Computer code like RADMC-3D can perform Monte Carlo simulations of radiation from a protostar propagating through a dusty medium to solve the thermal dust equilibrium. Using such numerical codes, astronomers can mimic the physical conditions of protoplanetary disks and protostellar cores around young protostars. To constrain the environment of protostellar cores and protoplanetary disks, synthetic observations of these modeled environments can be created via ray tracing and compared with actual observations.

\section{Radiative transfer codes}
The RADEX \citep{vand07} code applies the escape probability algorithm (see Sect. \ref{sec:cassis}). This algorithm requires as inputs the kinetic temperature T$_{kin}$, the collision partner density n, and the molecular total column density N$_{tot}$. The outputs include the integrated flux of the line $\int T_b dv$, the excitation temperature T$_{ex}$, and the opacity at the line's center. The code makes use of molecular data files from the LAMDA database, which provide collisional rates ($\gamma_{ul}$) for one or more collision partners as well as spectroscopic properties of the molecule ($\nu_{ul}, A_{ul}, g_u, g_l, E_l$, etc.). These programs are constrained by extremely high opacities and improperly handle maser effects. Furthermore, as was previously said, it is essential to choose a geometry to compute $\beta$ in advance because the excitation depends on it. The size and shape of this geometry can alter the required total input column density, N$_{tot}$. The limitation of RADEX is its inability to directly and effectively handle the comparison of models and observational data. The program's inability to manage more than one molecule simultaneously prevents it from accounting for the effects of line overlap. Such overlaps can happen at radio and infrared wavelengths (\cite{fonf06}, for example). In certain circumstances, the overlap between lines of the same molecule, such as the hyperfine components of HCN or N$_2$H$^+$, may affect their excitation \citep{dani06}. In my thesis, I used this code to check the physical environment causing the molecular excitation.

\begin{figure}[htbp]
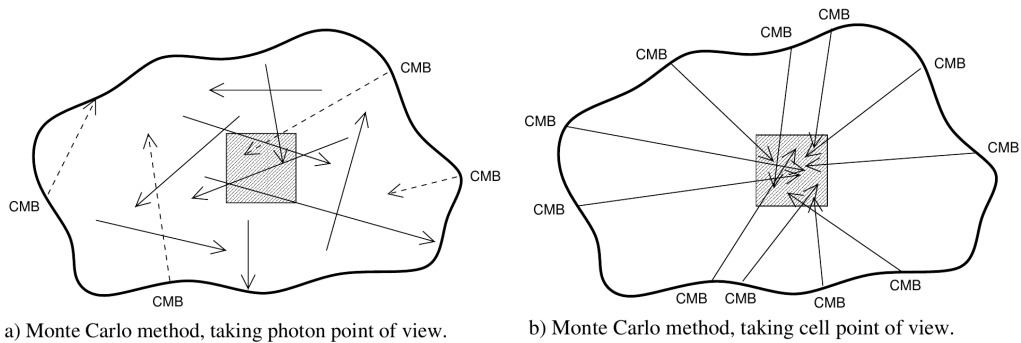

\includegraphics[width=0.45\textwidth]{chap1fig/ratran_mc1.pdf}
\includegraphics[width=0.45\textwidth]{chap1fig/ratran_mc2.pdf}
\caption{ Views of the Monte-Carlo method from the perspective of the photons (top panel) and the cells (bottom panel). \citep[Courtesy:][]{hoge00}.}
\label{fig:monte}
\end{figure}

I have also used the RATRAN code for radiative transfer modeling. The Monte-Carlo method is the backbone of the non-LTE radiative transfer code known as RATRAN \citep{hoge00}. It is a 1-dimensional (1D) algorithm because it only considers spherically symmetric systems. Instead of using the point of view of photons, this code uses the point of view of cells. It aids in separating the radiation field's local and external contributions. The benefit of this code is that it considers the source structure for the various physical input parameters. Since numerous gradients of density and temperature are present depending on the radius (such as in collision partner density and kinetic temperature), it is crucial to specify a precise physical structure of the source. To create a spherical shell from this 1D structure, it is essential to provide some critical parameters for each layer. The density of the collision partner, the temperature of the gas and the dust, the radial velocity, the doppler parameter, and the density of the considered molecule (related to its abundance [X] = n$_X$/n(H$_2$)) are these critical parameters.
The dust continuum and line profile are calculated by this code using the input data on dust opacities as a function of wavelength. For example, these dust opacities can be considered from \citep{osse94}. This algorithm can handle up to 10$^3$ - 10$^4$ optical opacities.
Combining radiative transfer and statistical equilibrium, we have \citep{hoge00},
\begin{equation}
 J_\nu=\Lambda [S_{ul}(J_\nu)],
\end{equation}
where the $\Lambda$ operator operates on the source function S$_{ul}$ which is a function of J$_\nu$ which depends on level population. S$_{ul}$ is defined as,
\begin{equation}
 S_{ul}\equiv \frac{j_{\nu0}(dust) + \int j_\nu^{ul}(gas) d\nu}{\alpha_{\nu0}(dust) + \int \alpha_\nu^{ul}(gas) d\nu}.
\end{equation}
The following J$_\nu$ value is generated from the earlier population densities by iteratively solving this equation:
\begin{equation}
 J_\nu=\Lambda [S_{ul}^\dagger (J_\nu)],
\end{equation}
where $\dagger$ signifies the adjoint matrix. Iterative methods for non-LTE radiative transfer codes are usually referred to as $Lambda$-iteration even though no $Lambda$-operator is actually formed. This is because of the notation used in the above equation. J$_\nu$ is approximated in a cell from the many contributions coming from all the different angles from different directions. Contributions of various cells to the radiation field, J$_\nu$ are added together along the propagation path of photon's. The Monte-Carlo method often permits the model to select the random direction of these photons (see Fig. \ref{fig:monte}), but this method is not followed in RATRAN. In RATRAN the radiation field is calculated using cell point of view, i.e.,    for each cell, only the incoming rays are traced back to their origin at the edge of the cloud where the boundary condition is satisfied (considered to be the CMB, see Fig. \ref{fig:monte}).

The origin of this photon was randomly selected from the edge of the model by the Monte-Carlo algorithm employed in RATRAN. The Accelerated Lambda Iteration (ALI, \cite{rybi91}) is combined with this procedure and is defined as follows:
\begin{equation}
 J_\nu=(\Lambda - \Lambda^*) [S_{ul}^\dagger (J_\nu)]+\Lambda^* [S_{ul} (J_\nu)],
\end{equation}
where $\Lambda^*$ is an approximated operator. The calculation time is decreased, especially at large opacities, by effectively separating the local radiative contribution from the global radiative transfer. \cite{hoge00} provided additional information on the ALI approach.

\newpage
\begin{wrapfigure}[]{l}{0.58\textwidth}
\begin{center}
\includegraphics[width=8cm,height=12cm]{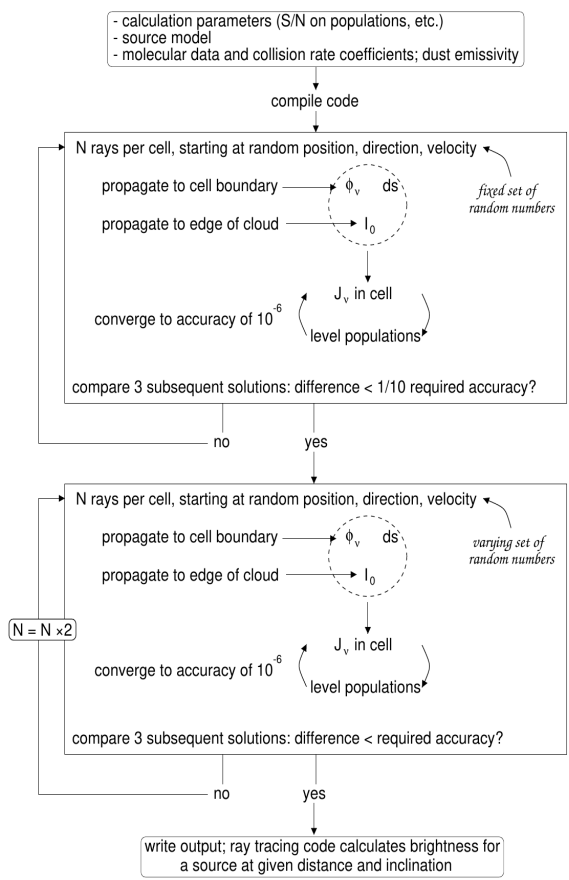}
\end{center}
\caption{Schematic diagram of the steps involved in Monte Carlo calculations. \citep[Courtesy:][]{hoge00}.}
\label{fig:rat_schem}
\end{wrapfigure}

The "amc" and "sky" are two consecutive stages of RATRAN's calculation. The population level densities are calculated in the first stage ("amc"). Then, using the ray-tracing technique, the second stage ("sky") determines the emission distribution for user-defined transitions. This results in a hyper-spectral cube with three dimensions, two of which are spatial and one of which is the velocity (giving the line profiles). This line profile (T$_{mb}$ as a function of velocity), can be compared with the observations by convolving the calculated line profile with the telescope beam captured at the transitional frequency. In Fig. \ref{fig:rat_schem}, the step-by-step calculation of RATRAN is summarised.

The resolution of the fine-scale structures of a source increases as the number of interferometric measurements increases. But a simple spherical symmetric model could not explain these results. The public version of the RATRAN code can be used by considering the rotational motion with some limitations.  

A 3D non-LTE radiative transfer code called LIME (Line Modelling Engine, \cite{brin11}) is based on the Monte-Carlo method combined with an ALI treatment. Although completely redesigned and modified to handle a 3-dimensional problem, it is mostly based on the RATRAN framework. Therefore, most of the code and the approach to solving the problem are shared with RATRAN.
Nonetheless, there are differences between LIME and RATRAN. The photon propagation method, which significantly cuts the calculation time, makes it possible to solve 3D models in a reasonable amount of time. It also provides a graphical window with customizable model inputs. LIME is very useful for examining various astronomical objects, including protoplanetary discs, envelopes, outflows, clusters of proto-stars, etc.

 \chapter{MCMC Fitting: Extracting Physical Parameters of Astrophysical Sources} \label{chap:mcmc_all}

\section*{Overview}
According to LTE modeling, the gas is assumed to be in local thermodynamic equilibrium, which means that the density is high enough to favor collisions over other forms of excitation. N$_t$, T$_{ex}$, V$_{LSR}$, $\Delta$V$_{FWHM}$, and $\Omega$ are the five input variables for the LTE. N$_t$ represents total column density, T$_{ex}$ represents temperature and $\Omega$ represents the size of the emitting region (which couples to the beam sizes to consider beam dilution effects). It should be noted that the gas kinetic temperature or rotation temperature, which describes the populations of all rotational levels, and the excitation temperature, which determines the relative populations of the upper and lower levels of a spectral line, are all equal by definition under LTE conditions. Each combination of these variables results in a Gaussian model spectrum for each transition of the chosen species. In contrast to the Gaussian fitting procedure, which fits the V$_{LSR}$ and $\Delta$V$_{FWHM}$ to each line separately, we obtain a single average value of V$_{LSR}$ and $\Delta$V$_{FWHM}$ for all transitions of a given species when using LTE modeling.

\section{MCMC fitting}
We used a Markov Chain Monte Carlo (MCMC) method implemented in CASSIS to identify the set of parameters that results in synthetic spectra that best match the observed spectral line profiles (e.g., \cite{guan07}). The MCMC method selects a seed randomly from the X$_0$ state, a five-dimensional parameter space. Then, according to a variable step size determined for each iteration, it randomly selects one of the closest neighbors (known as the X$_1$ state). It calculate the  $\chi^2$ for the new state. If p = $\chi ^2$ (X$_0$)/ $\chi^2$ ($X_1$) > 1, the new state is accepted. This new state might still be accepted with a particular acceptance probability if p = $\chi^2 (X_0)/\chi^2 (X_1) < 1$. The X$_0$ state will remain if the new state is rejected, and a different nearby state will be chosen randomly to become the X$_1$ state. Having a finite probability of accepting a new position even if the $\chi^2$ is worse ensures that we do not directly converge to a local minimum but instead force better sampling of the entire parameter space. The code starts with several initial random states. It is typically assumed to have reached the right answer when the variance between different clusters of states is less than the variance of each cluster \citep{hast70}. When the code is getting close to convergence, it calculates many models and $\chi^2$ values in a small cluster close to the "best" answer. This enables us to determine the statistical standard deviation and the median value for each fitted parameter. To manage the probabilistic behaviour of a group of atomic particles, the MCMC method was created. The probability of each event in this stochastic model depends on the state obtained in the preceding event, \citep{gagn17}. The MCMC approach is an interactive procedure that uses a random walk to iterate over all line parameters (for example, excitation temperature, source size, and line width) before moving into the solutions space. $\chi^2$ minimization then yields the ideal solution. Here, the MCMC fitting is mostly used for species with more than two confirmed transitions. The MCMC method benefits species with several transitions across a wide range of energy, just like the rotation diagram method \citep{vast18}. The MCMC method is an iterative approach that uses the $\chi^2$ minimization procedure over parameter space to move toward the solution. Statistically estimating the physical parameters from fitting numerous transitions will be more accurate.

\section{Explanation of the line profiles obtained towards G10.47+0.03}
\subsection{Peptide-like bond related molecules}
Nitrogen stands among the most chemically active components in the interstellar medium after hydrogen, oxygen, and carbon (ISM). Due to their active participation in synthesizing biomolecules, nitrogen molecules are extremely significant. Isocyanic acid (HNCO), formamide ($\rm{NH_2CHO}$), and methyl isocyanate ($\rm{CH_3NCO}$) containing -NH-C(=O)- bond were identified in hot molecular core, G10.47+0.03 (hereafter, G10). To determine the physical characteristics of G10, the MCMC method is employed. Additionally, we calculated the hydrogen column density and many other physical parameters from the observed line profiles. \cite{gora20} found that HNCO, $\rm{NH_2CHO}$, and $\rm{CH_3NCO}$ are chemically linked with each other.

\begin{figure}
\begin{minipage}{0.32\textwidth}
\includegraphics[width=\textwidth]{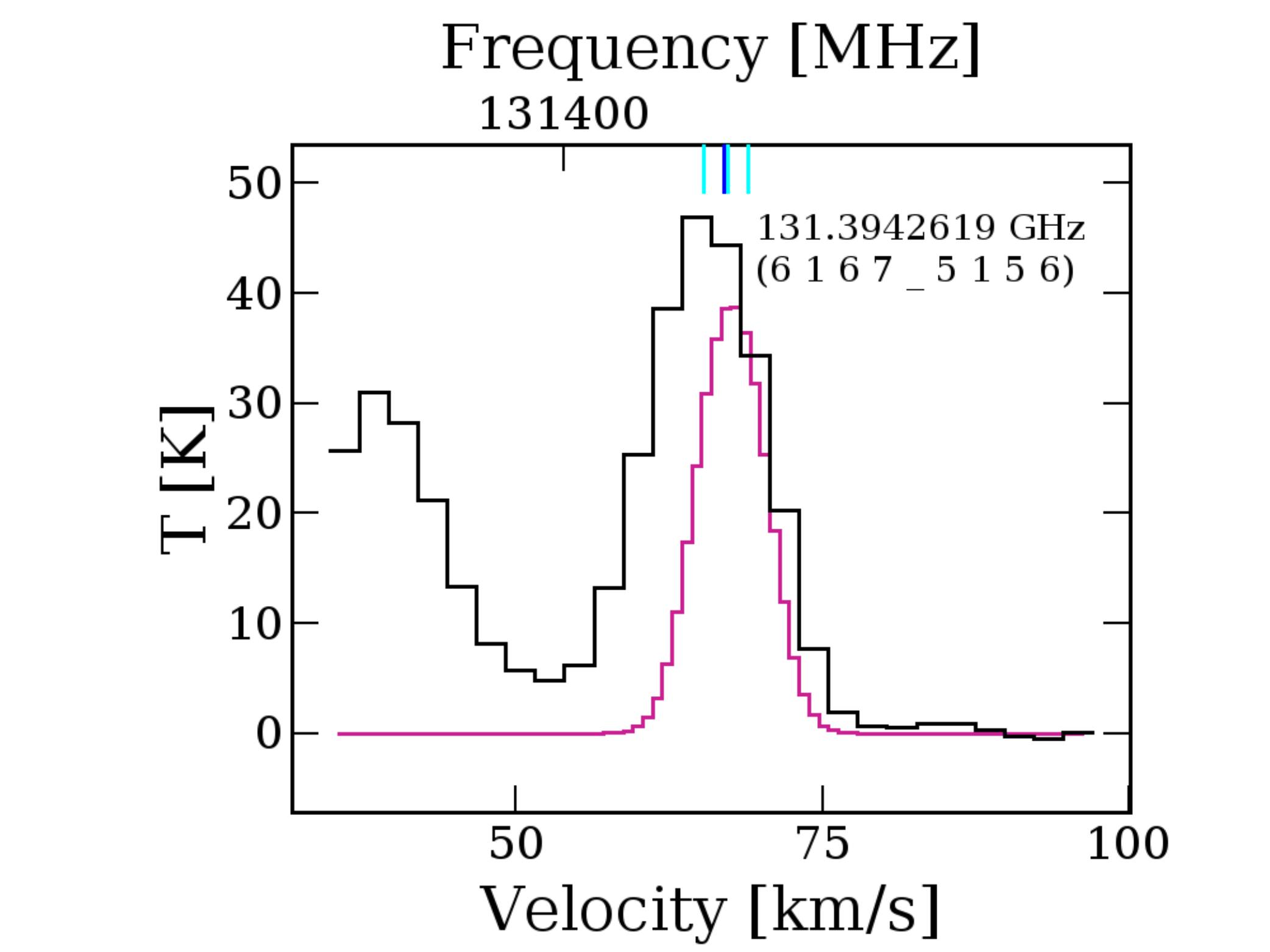}
\end{minipage}
\begin{minipage}{0.32\textwidth}
\includegraphics[width=\textwidth]{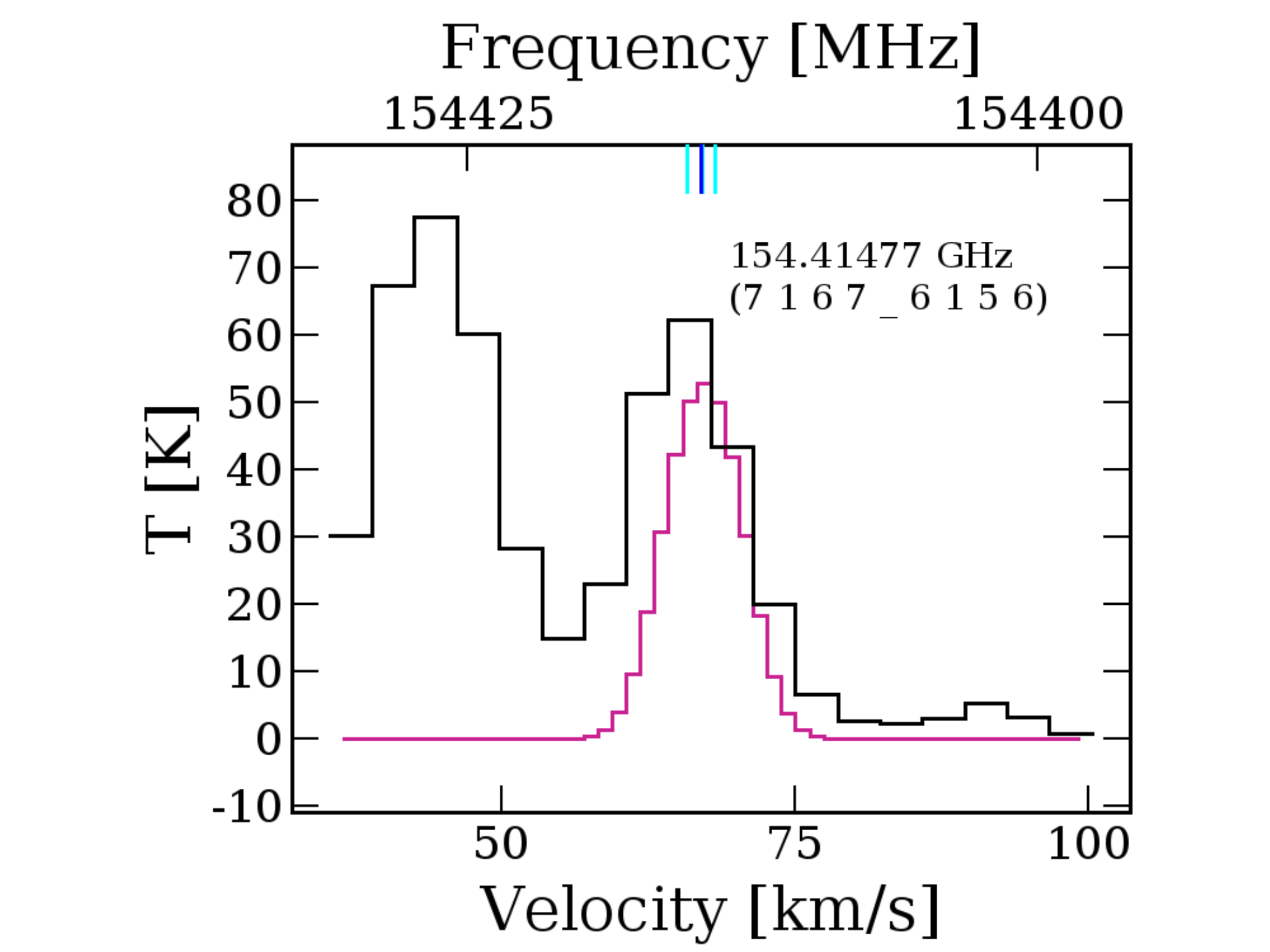}
\end{minipage}
 \begin{minipage}{0.32\textwidth}
 \includegraphics[width=\textwidth]{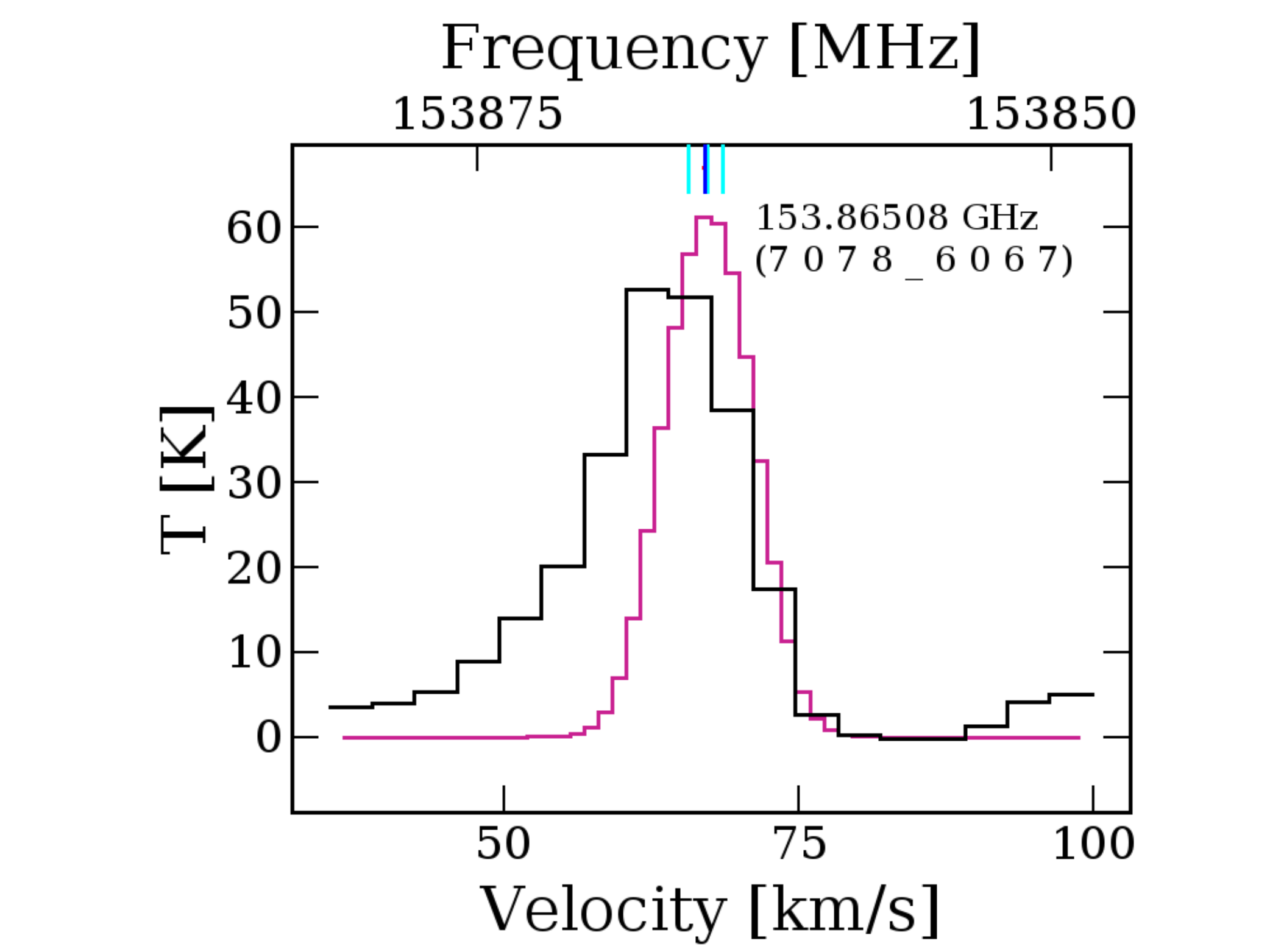}
 \end{minipage}
\begin{minipage}{0.32\textwidth}
\includegraphics[width=\textwidth]{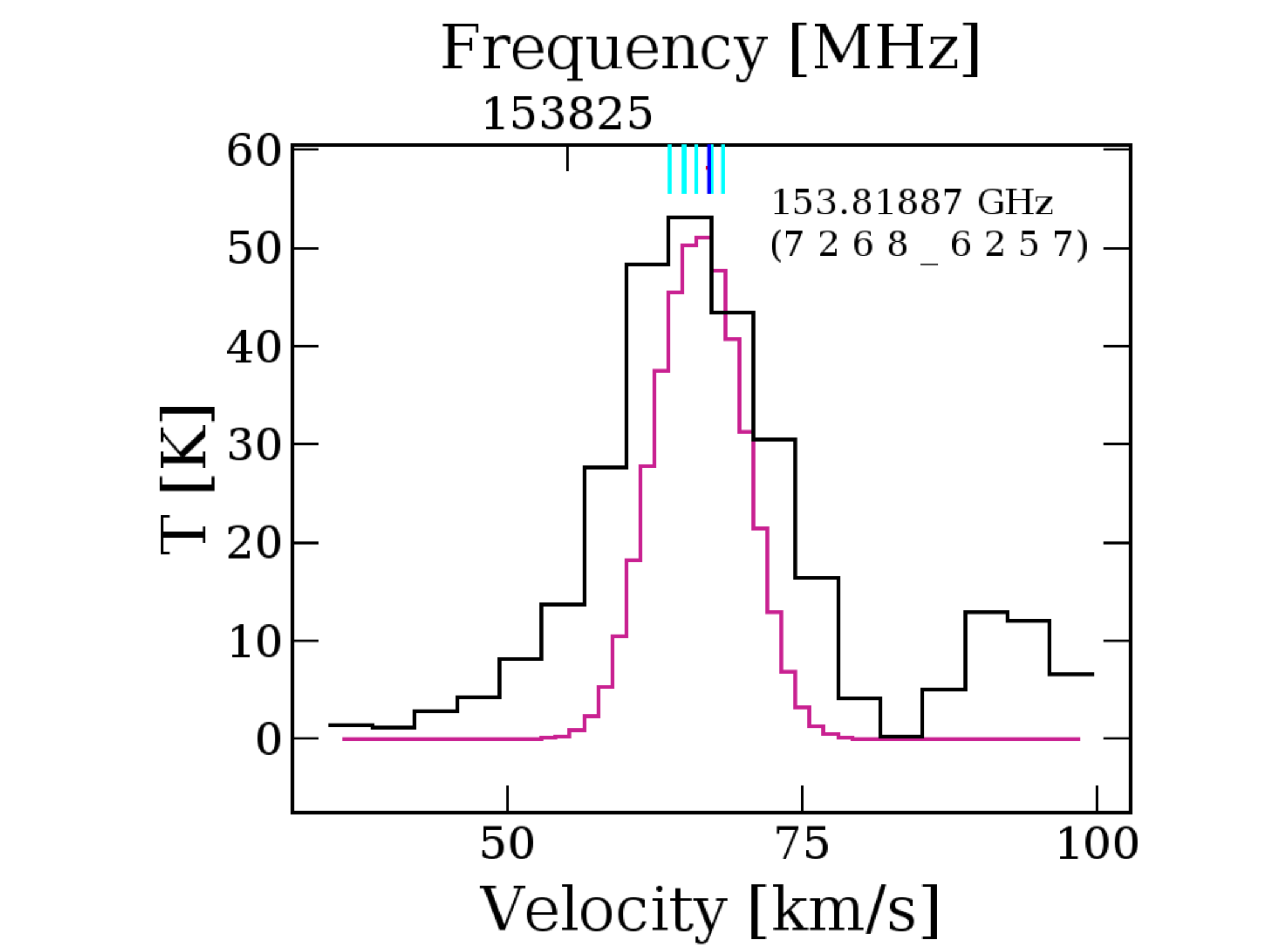}
\end{minipage}
\begin{minipage}{0.32\textwidth}
\includegraphics[width=\textwidth]{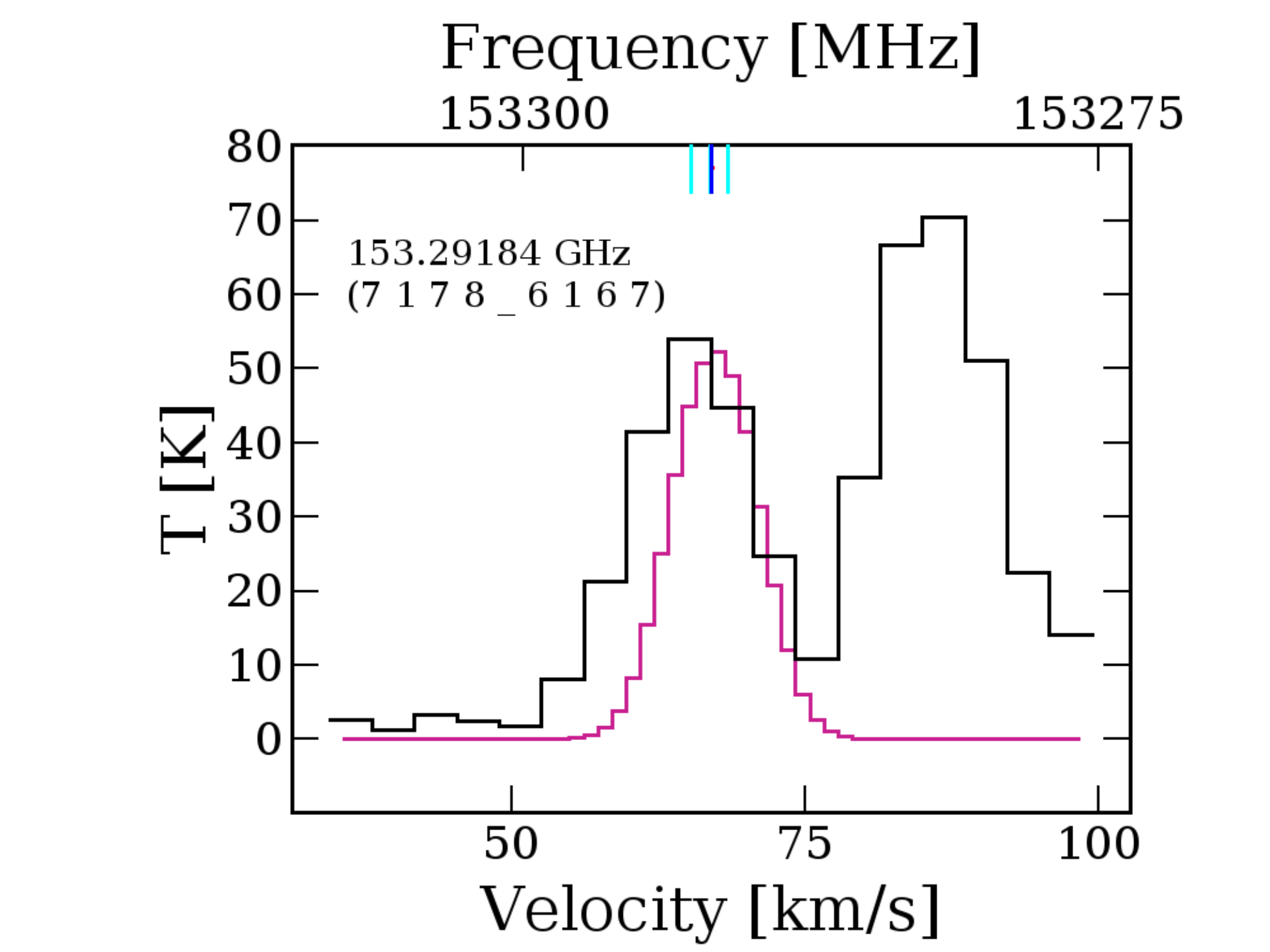}
\end{minipage}
\caption{LTE fitting of observed lines of HNCO towards G10.  Black line represents the observed spectra and pink line 
is the fitted profile. \citep[Courtesy:][]{gora20}}
\label{fig:HNCO}
\end{figure}
To match the observed line profiles of $\rm{HNCO}$, $\rm{NH_2CHO}$, and $\rm{CH_3NCO}$ towards the hot core G10, we employed the Markov Chain Monte Carlo (MCMC) approach. We considered that the source is in LTE. Column density, excitation temperature, FWHM, optical depth, and source velocity are the physical characteristics that the fitting procedure returned as the best-fitted parameters of the model. To determine these parameters, we employed the Python scripting interface offered by CASSIS. Considering the N number of spectra, we applied the $\chi^2$ minimization technique to identify the best-fitted set that suit the observational findings. The $\chi^2$ between the observed and simulated data is calculated with the Python script, and the minimum $\chi^2$is considered from the following equations:
\begin{equation}
 {\rm {\chi_i}^2=\sum_{j=1}^{N_i} \frac{(I_{obs,ij}-I_{model,ij})^2}{rms^2_i+(cal_i\times I_{obs,ij})^2}},
\end{equation}
where rms$_i$ is the rms of the spectrum i, cal$_i$ is the calibration error, and ${\rm I_{obs,ij}}$ and ${\rm I_{model,ij}}$ are, respectively, the observed and modeled intensity in the channel j of the transition i. This relation is used to compute the reduced $\chi^2$:
\begin{equation}
 {\rm \chi^2_{red}=\frac{1}{ \sum_{i}^{N_{spec}} N_i} \sum_{i=1}^{N_{spec}} \chi^2_i}.
\end{equation}
The initial physical values used in the MCMC computation are selected at random from a range between the user's specified minimum and maximum (${\rm X_{min}}$) and (${\rm X_{max}}$). The iteration number ($l$) and other parameters ($\alpha$ and $v$) affect the MCMC computation step ($\theta_l$), where,
$\theta_{l+1}=\theta_{l}+\alpha (v-0.05)$
(v is a random number between 0 to 1). Here $\alpha$ is defined as,
\begin{equation}
{\rm \alpha=\frac{k(X_{max}-X_{min})}{k^\prime}},
\end{equation}
\noindent
where, k is defined as 
$$
k=r_c \hskip 3.5cm \rm{when \ l>c},
$$
$$
k=\frac{(r_c-1)}{c}l+1  \hskip 2cm  \rm{when \ l<c},
$$ 
where c and r$_c$ are the user-set parameters for cutoff and ratio at cutoff, respectively. A reduced physical parameter, or k$^{\prime}$, is assigned to a value during computation. With a larger step at the beginning of the computation to identify a good $\chi^2$ and smaller steps at the end of the computation to extract the value of the probable best $\chi^2$, is used, where $\alpha$ sets the amplitude of the steps.

\begin{figure}
\begin{minipage}{0.32\textwidth}
\includegraphics[width=\textwidth]{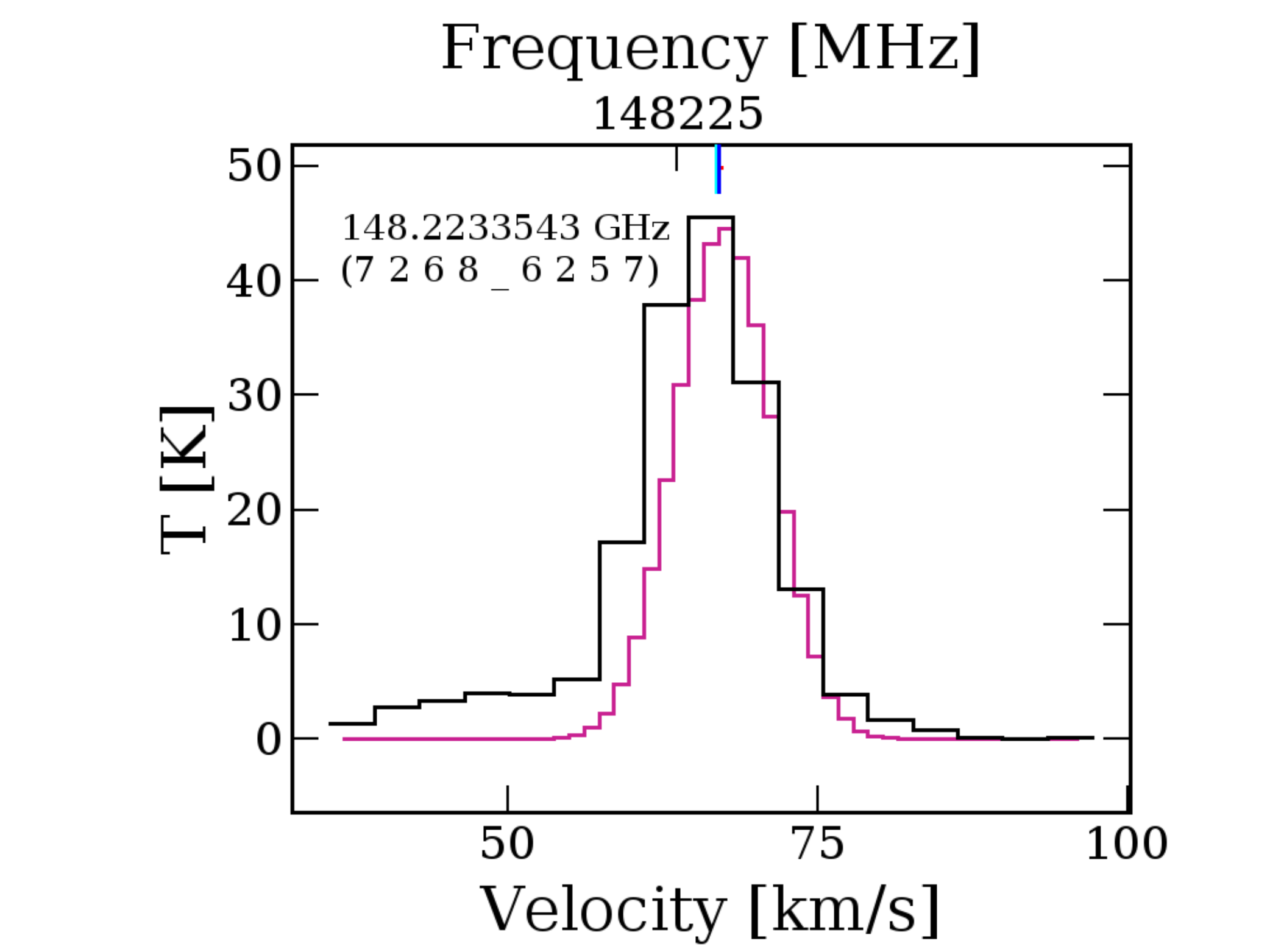}
\end{minipage}
\begin{minipage}{0.32\textwidth}
\includegraphics[width=\textwidth]{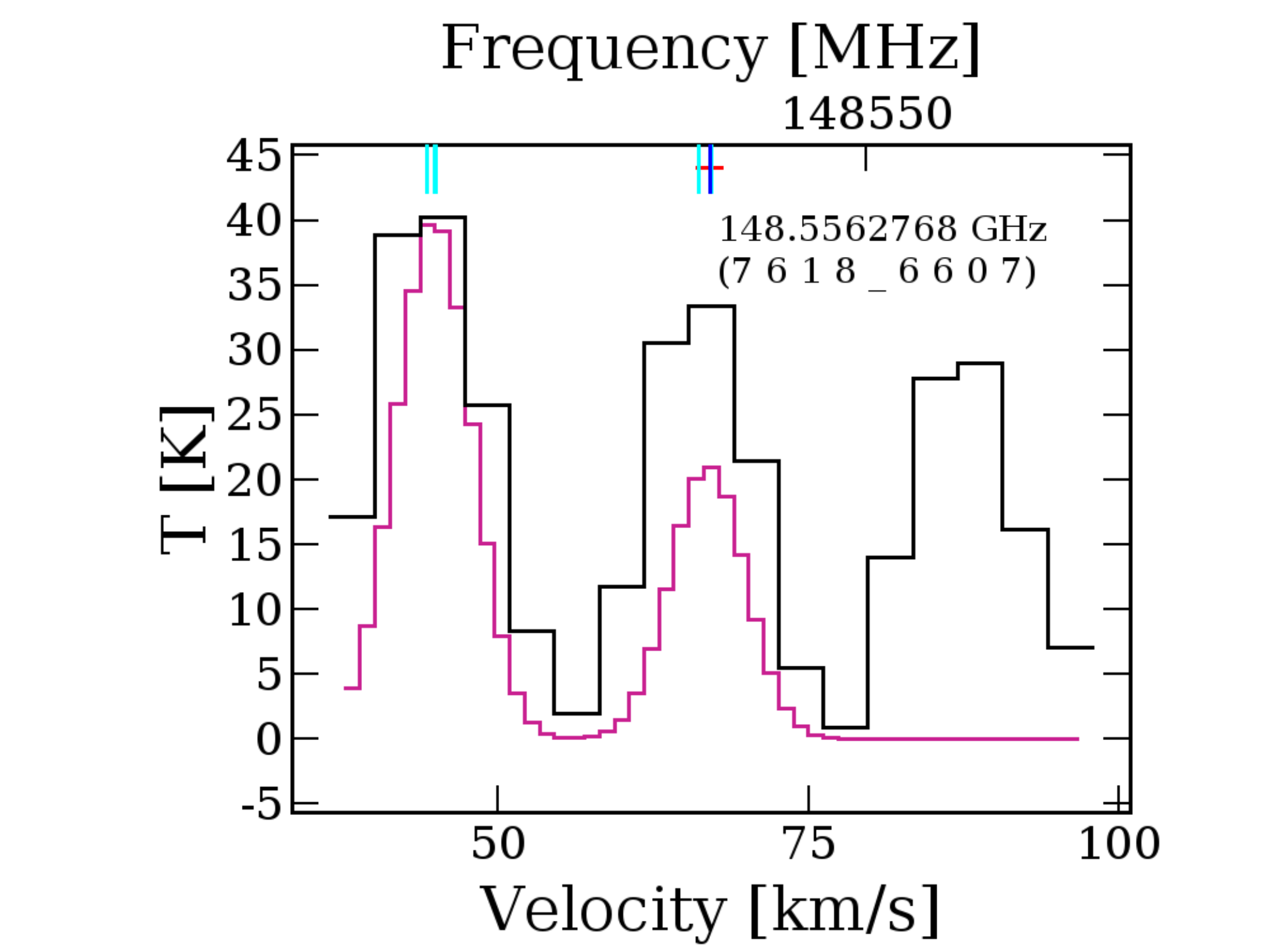}
\end{minipage}
 \begin{minipage}{0.32\textwidth}
 \includegraphics[width=\textwidth]{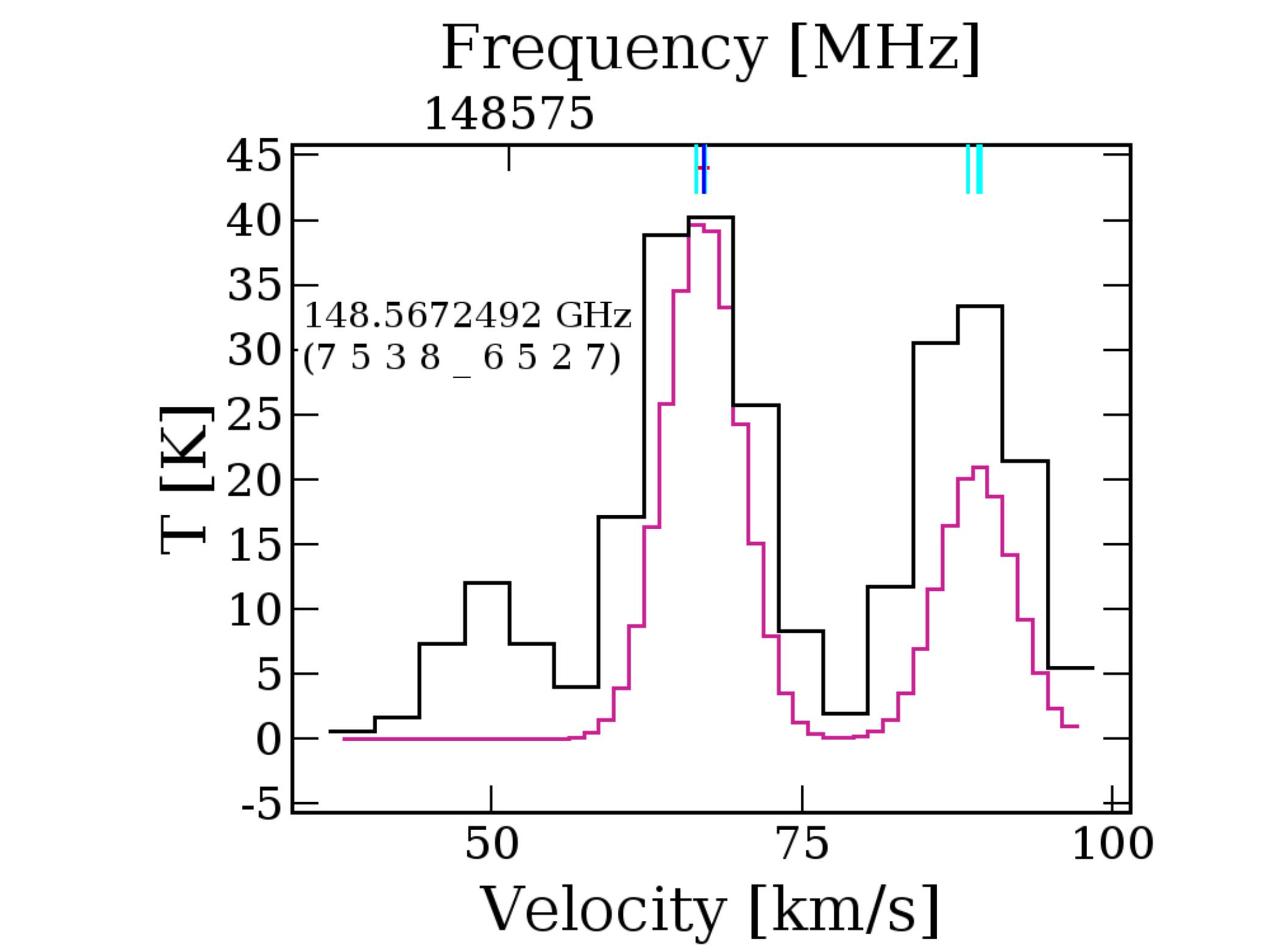}
 \end{minipage}
\begin{minipage}{0.32\textwidth}
\includegraphics[width=\textwidth]{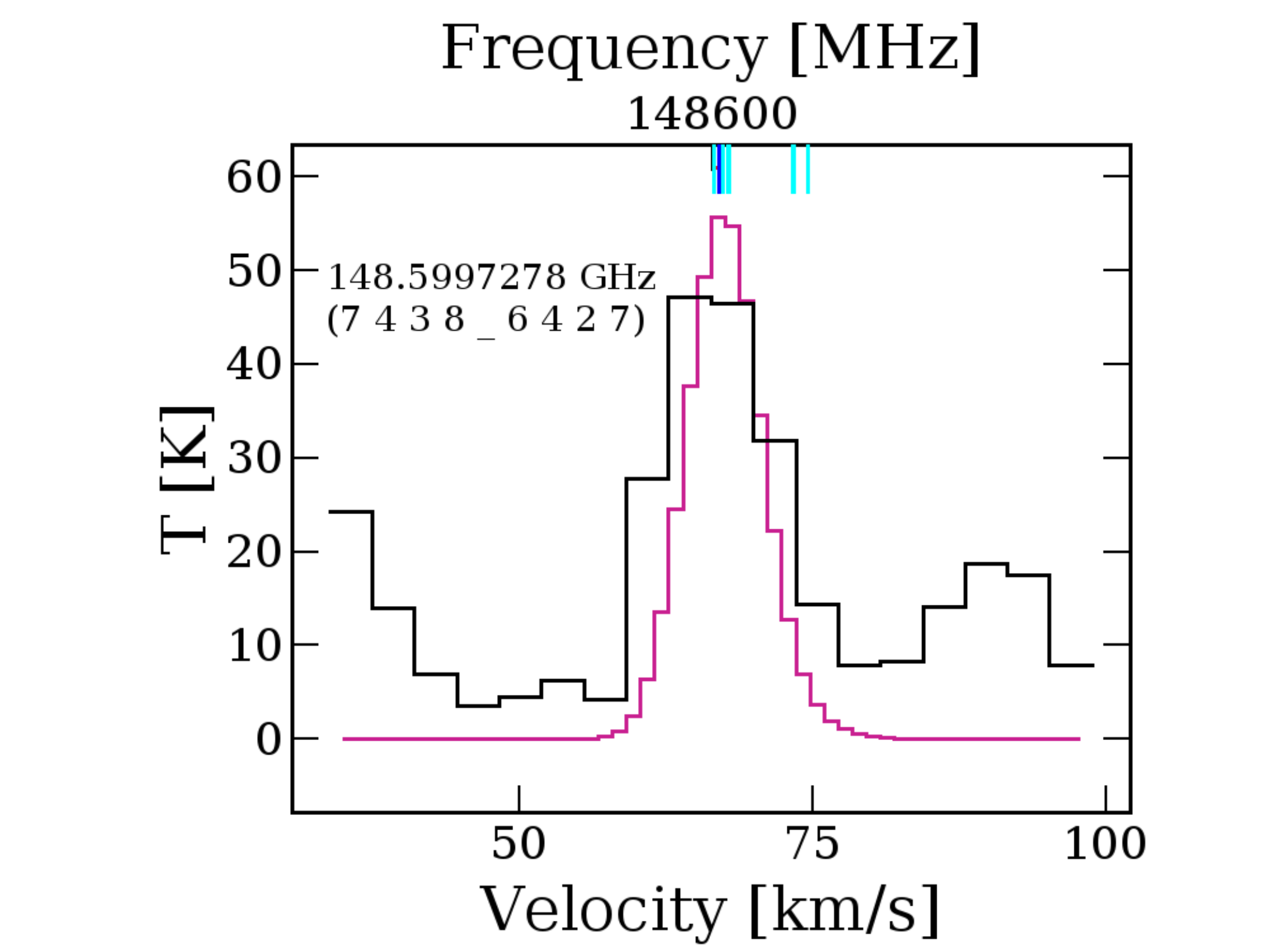}
\end{minipage}
\begin{minipage}{0.32\textwidth}
\includegraphics[width=\textwidth]{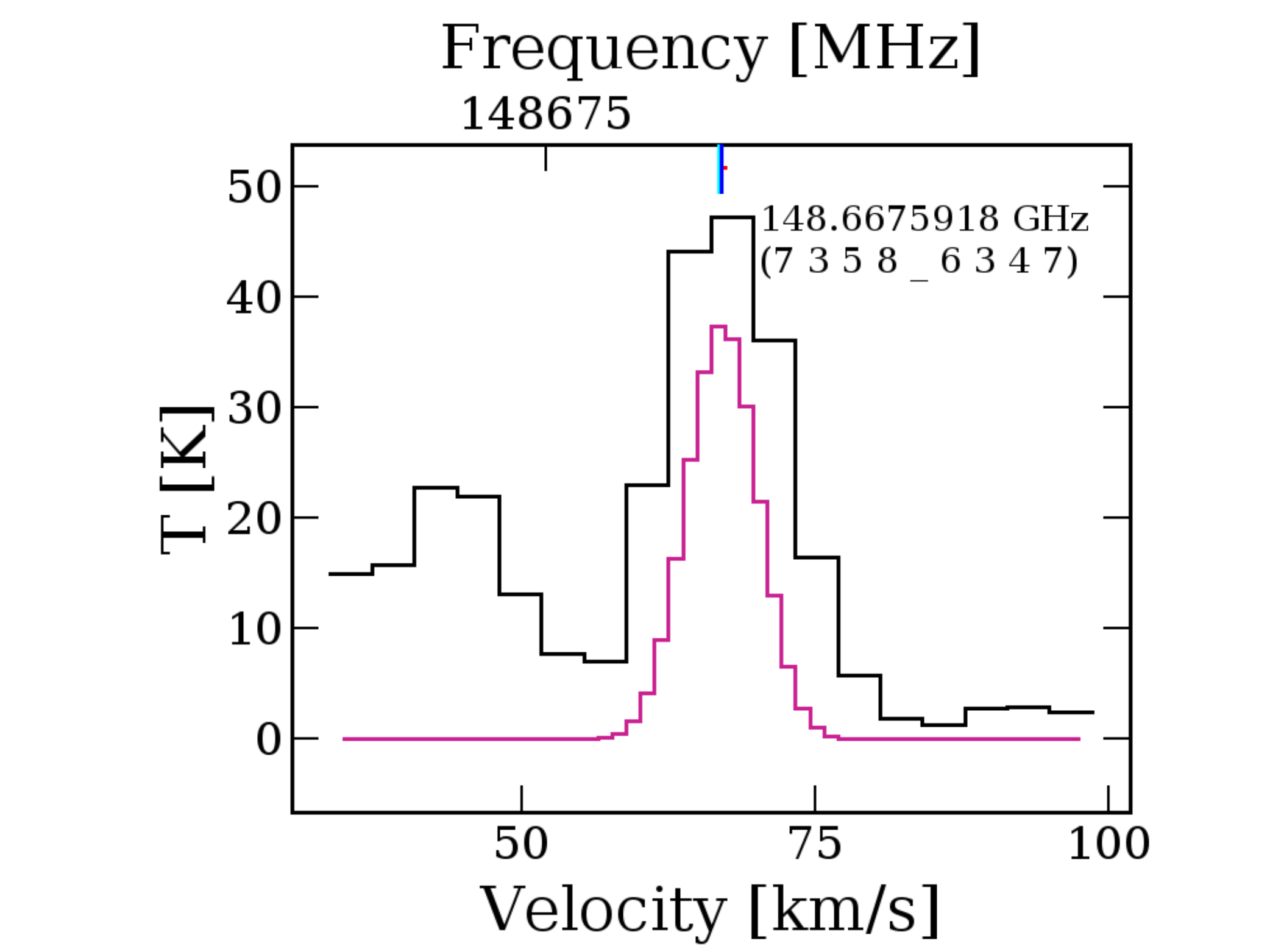}
\end{minipage}
\begin{minipage}{0.32\textwidth}
\includegraphics[width=\textwidth]{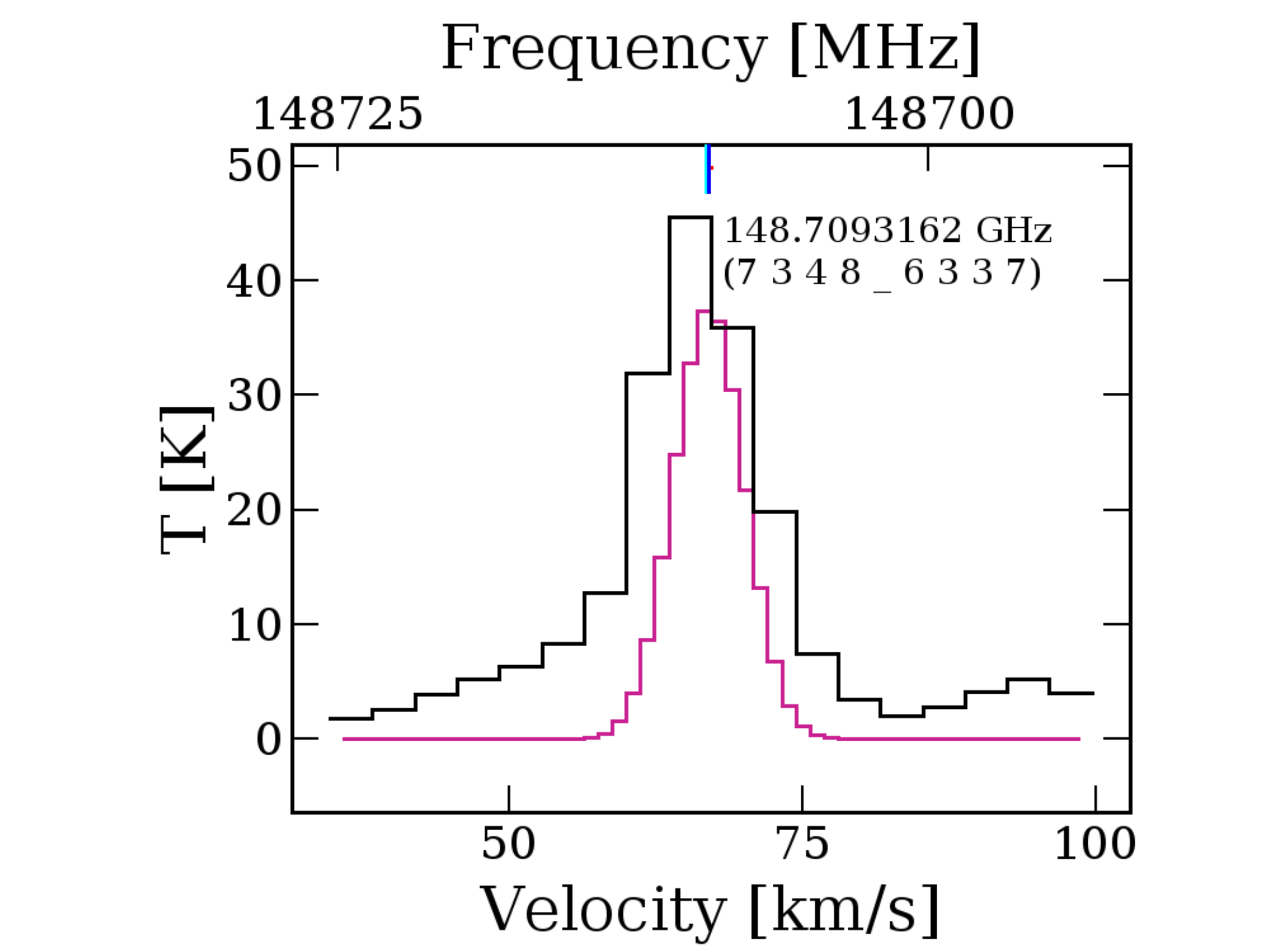}
\end{minipage}
\begin{minipage}{0.32\textwidth}
\includegraphics[width=\textwidth]{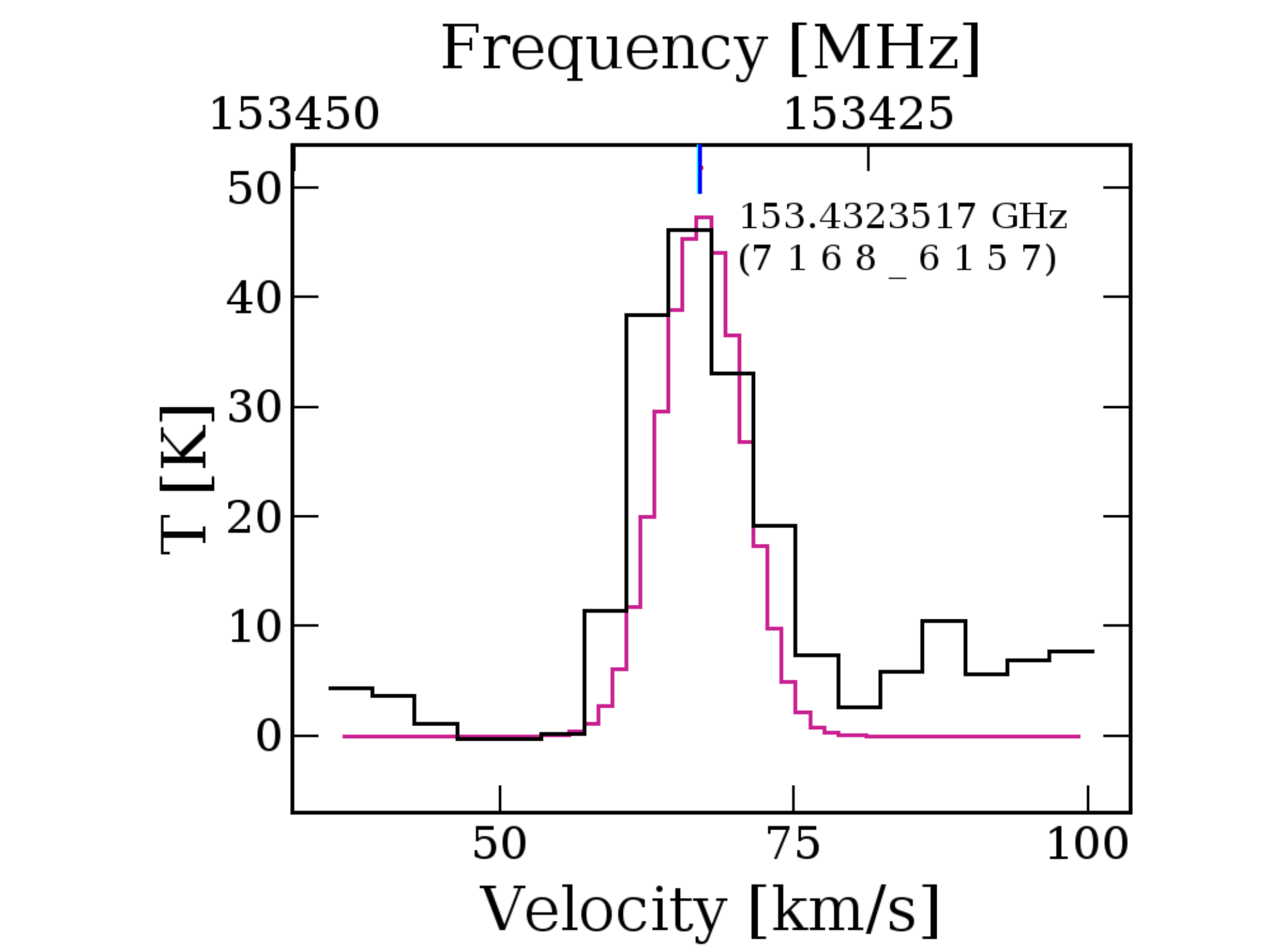}
\end{minipage}
\caption{LTE fitting of observed lines of $\rm{NH_2CHO}$ towards G10. Black line represents the observed spectra and pink line is the fitted profile.\citep[Courtesy:][]{gora20}}
\label{fig:NH2CHO}
\end{figure}

\begin{figure}[t]
\begin{minipage}{0.24\textwidth}
\includegraphics[width=\textwidth]{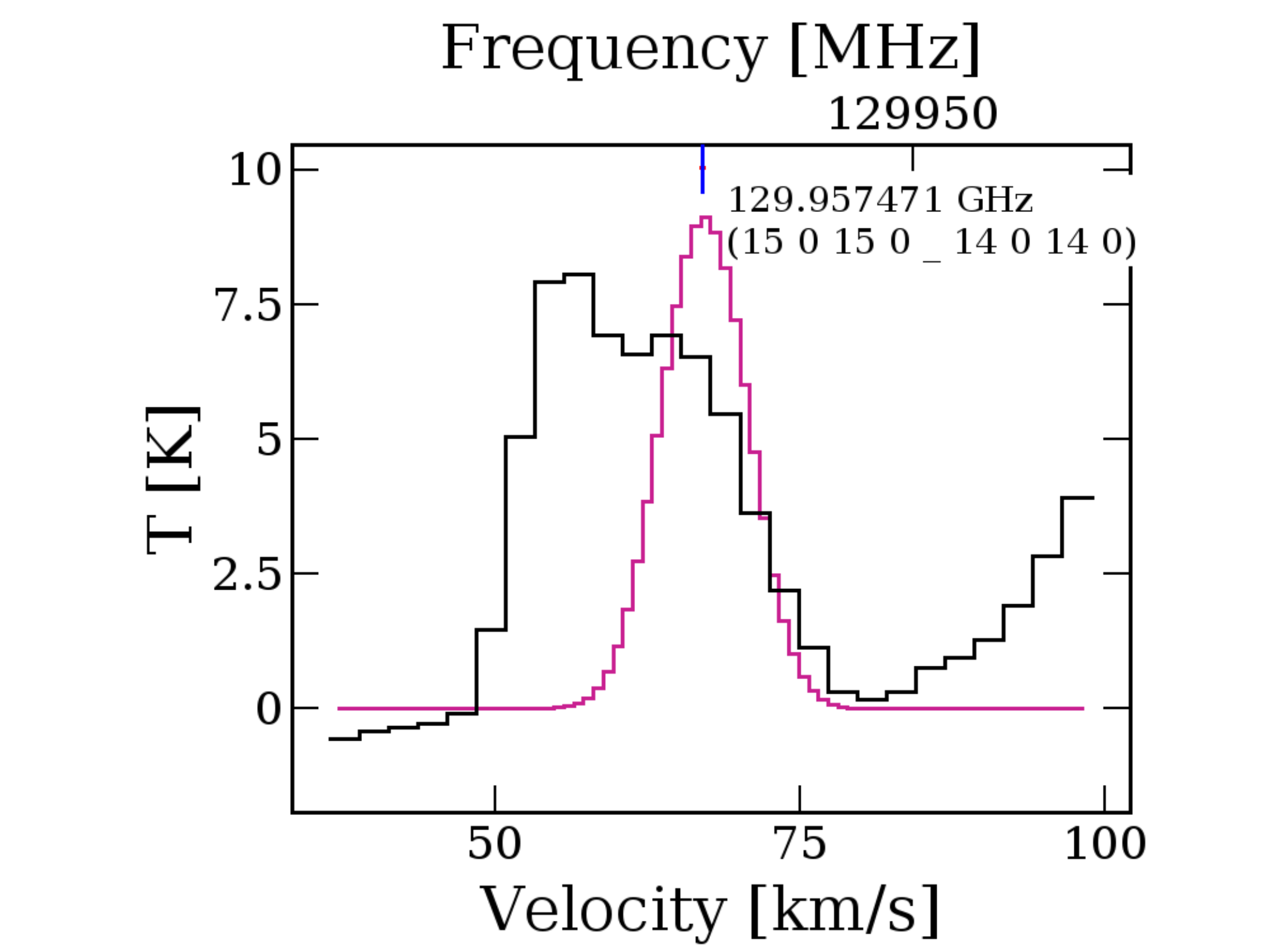}
\end{minipage}
\begin{minipage}{0.24\textwidth}
\includegraphics[width=\textwidth]{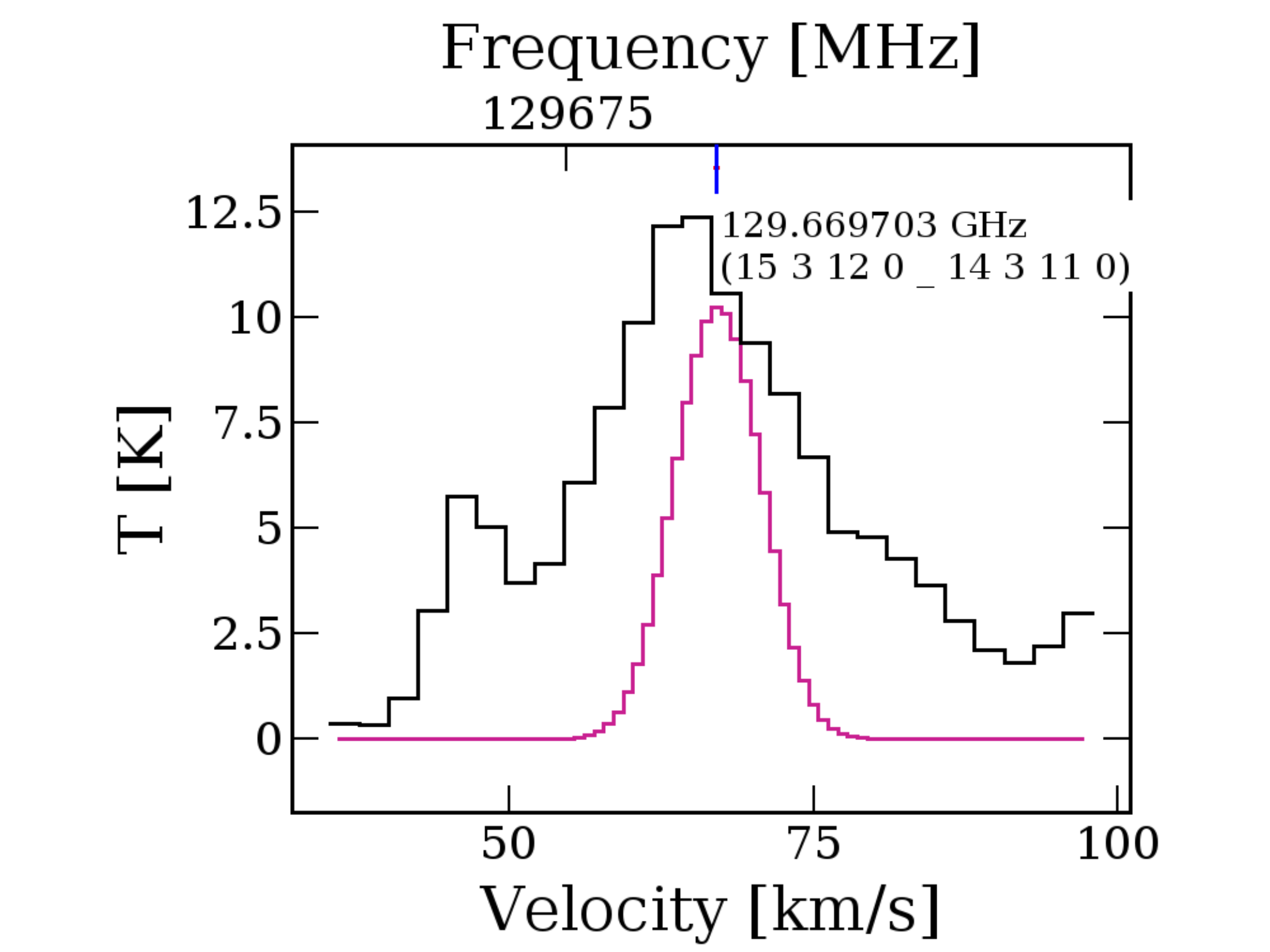}
\end{minipage}
 \begin{minipage}{0.24\textwidth}
 \includegraphics[width=\textwidth]{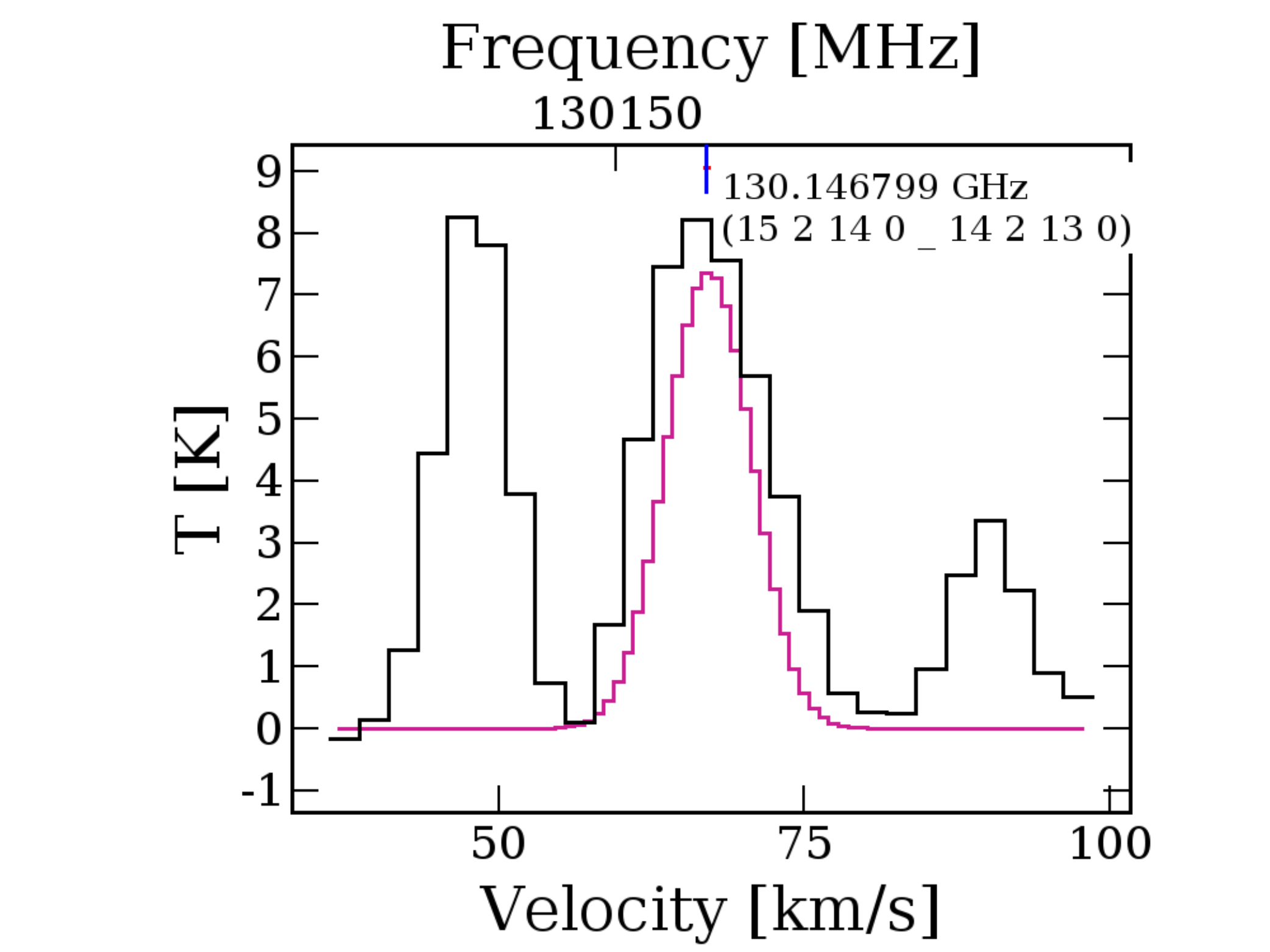}
 \end{minipage}
\begin{minipage}{0.24\textwidth}
\includegraphics[width=\textwidth]{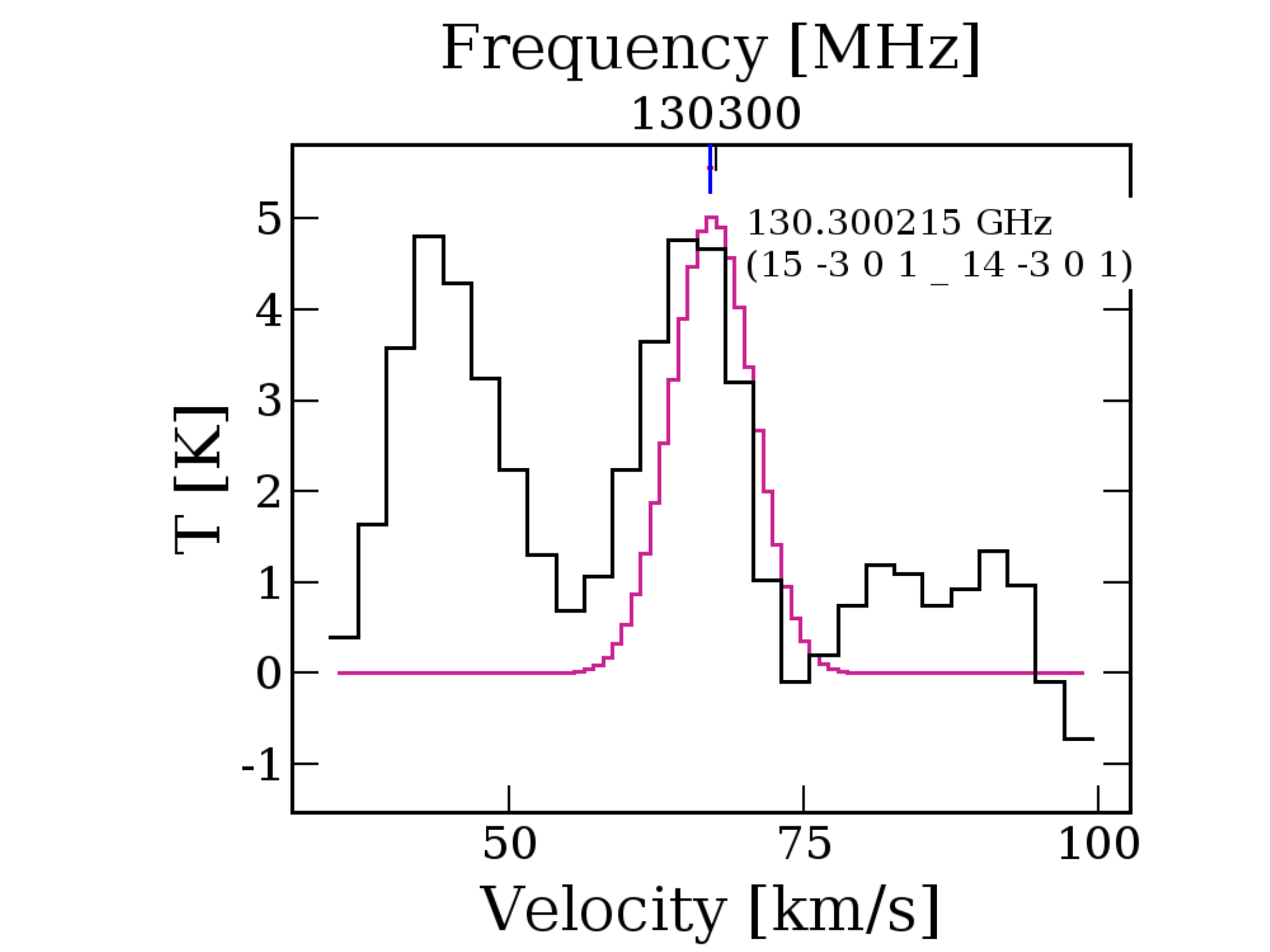}
\end{minipage}
\begin{minipage}{0.24\textwidth}
\includegraphics[width=\textwidth]{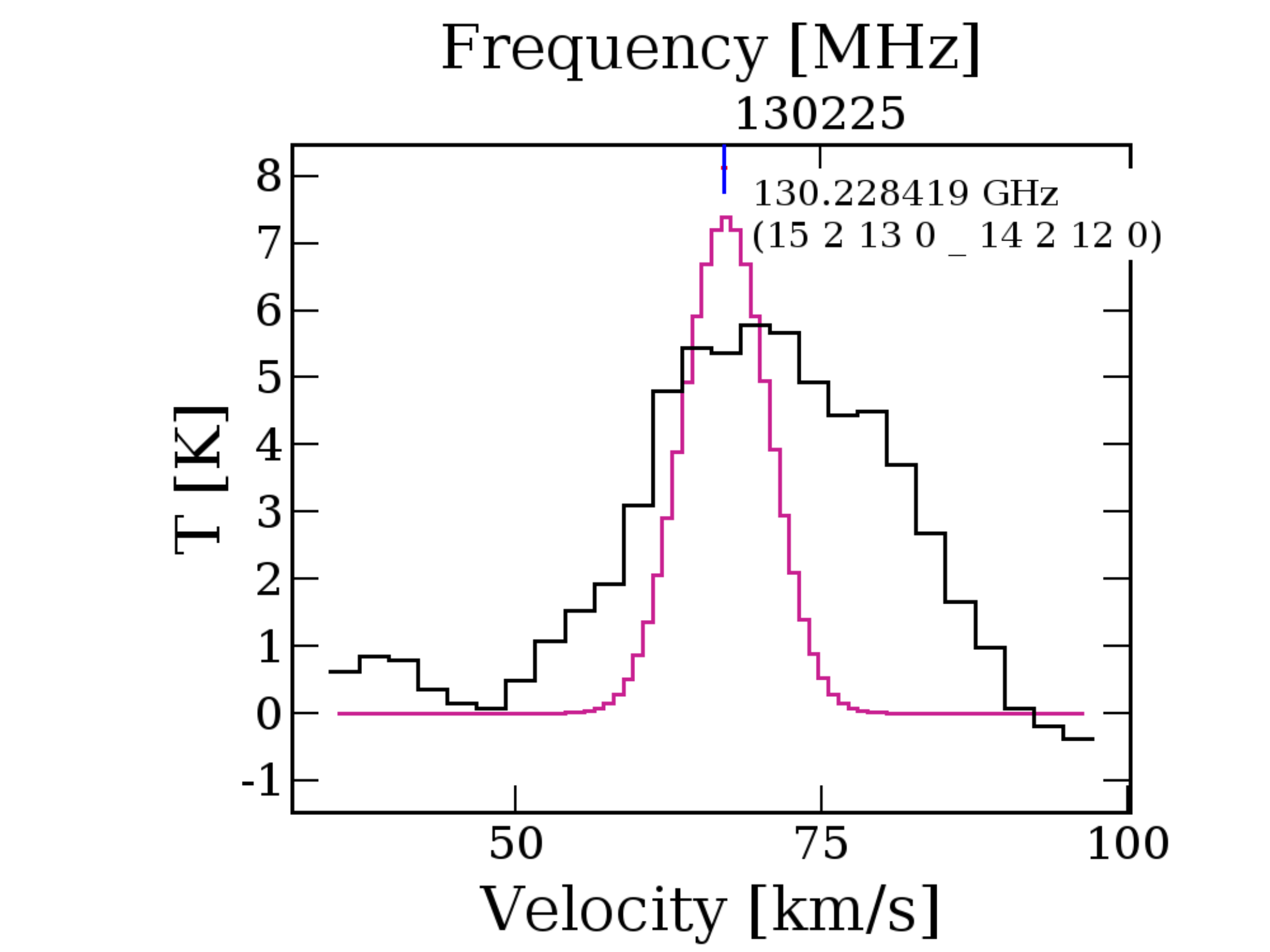}
\end{minipage}
\begin{minipage}{0.24\textwidth}
\includegraphics[width=\textwidth]{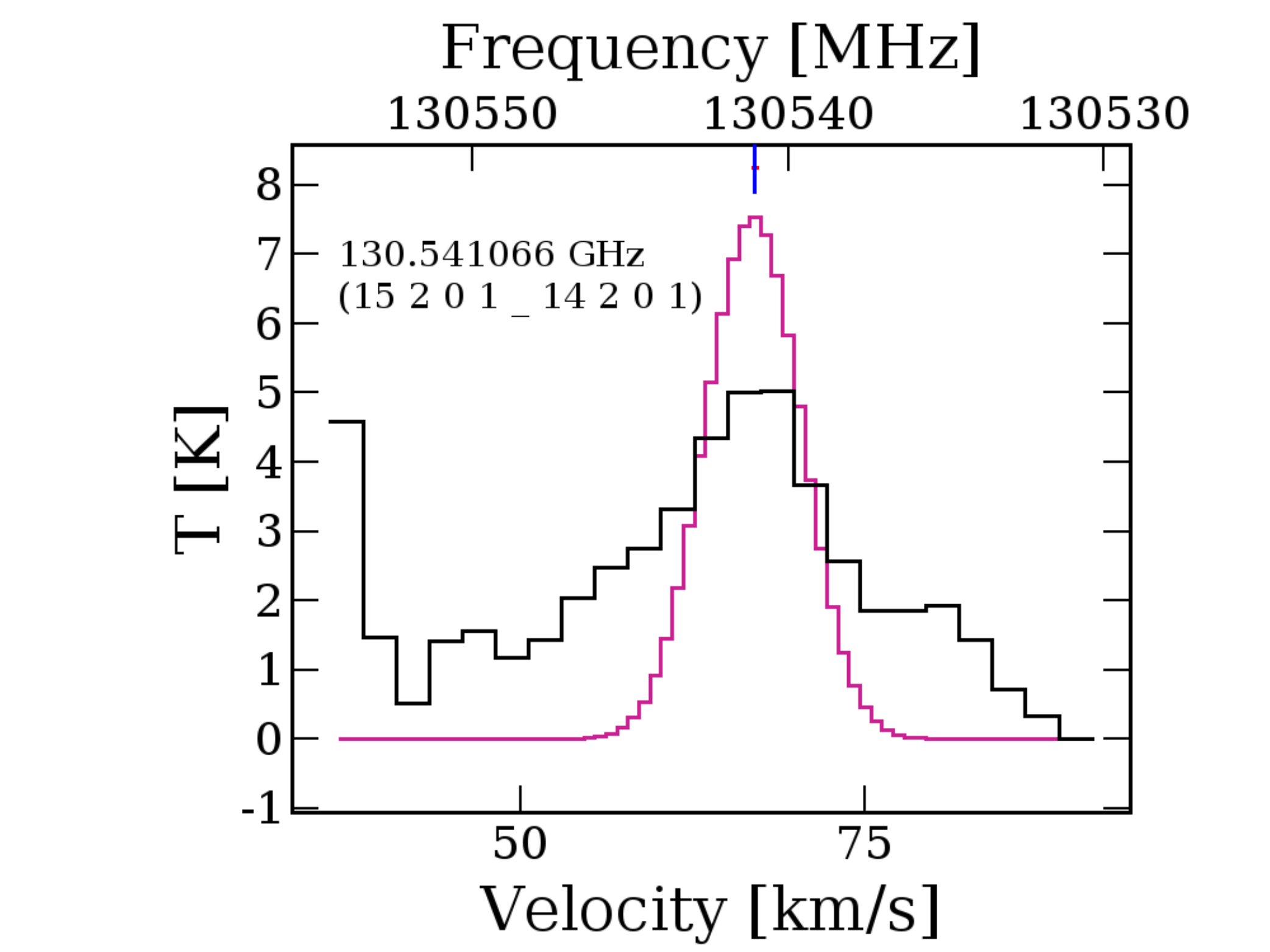}
\end{minipage}
\begin{minipage}{0.24\textwidth}
\includegraphics[width=\textwidth]{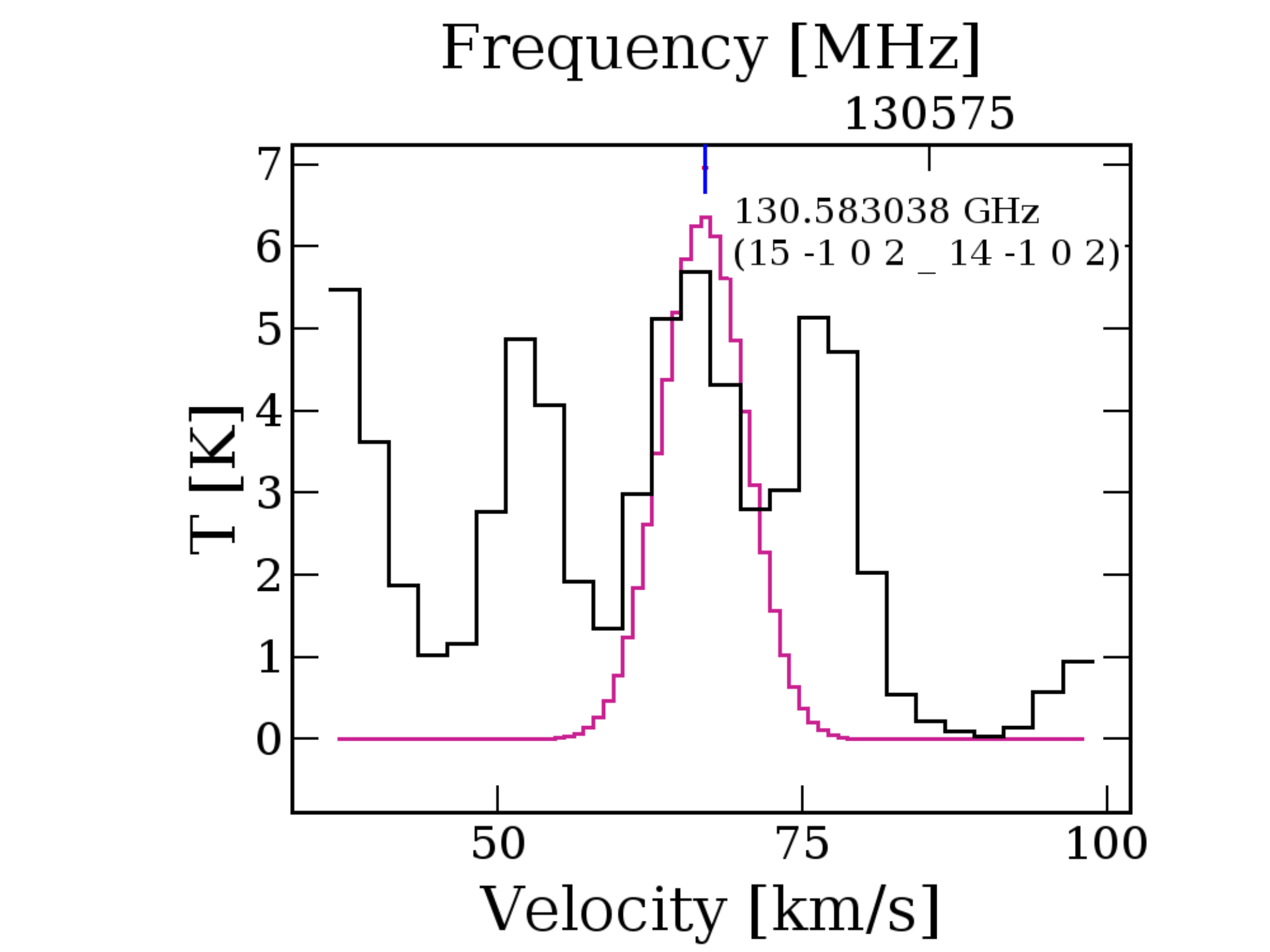}
\end{minipage}
\begin{minipage}{0.24\textwidth}
\includegraphics[width=\textwidth]{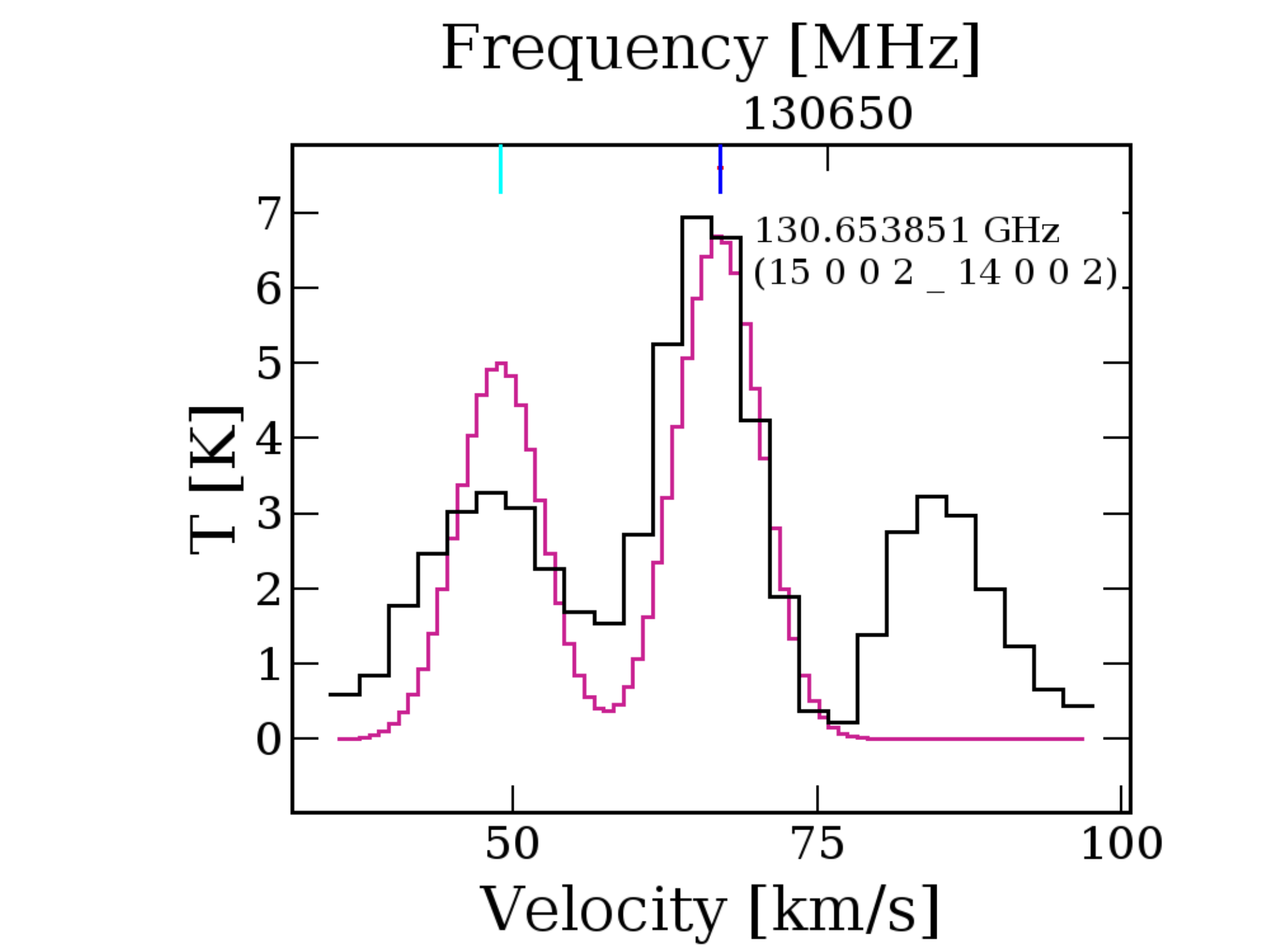}
\end{minipage}
\begin{minipage}{0.24\textwidth}
\includegraphics[width=\textwidth]{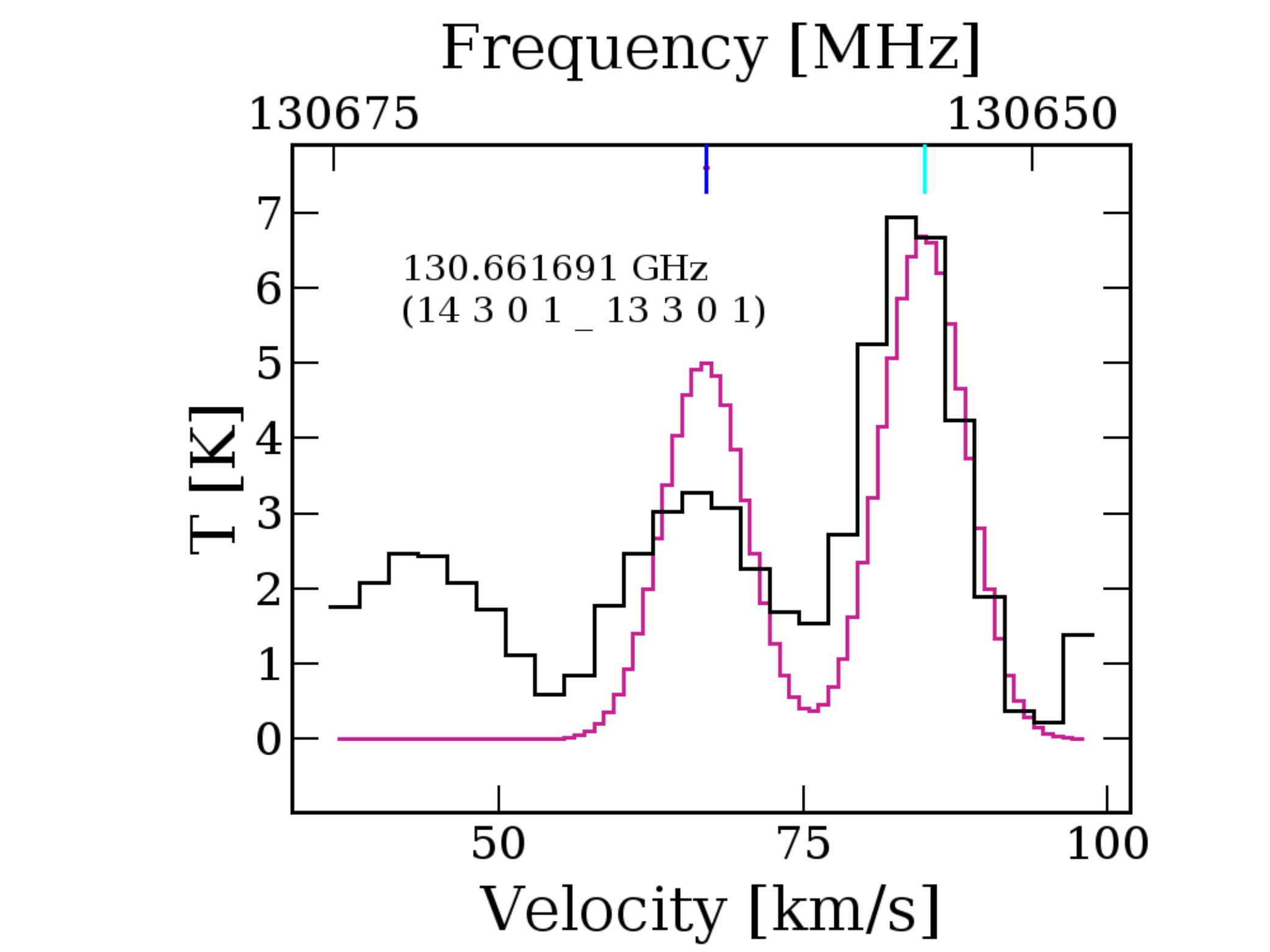}
\end{minipage}
\begin{minipage}{0.24\textwidth}
\includegraphics[width=\textwidth]{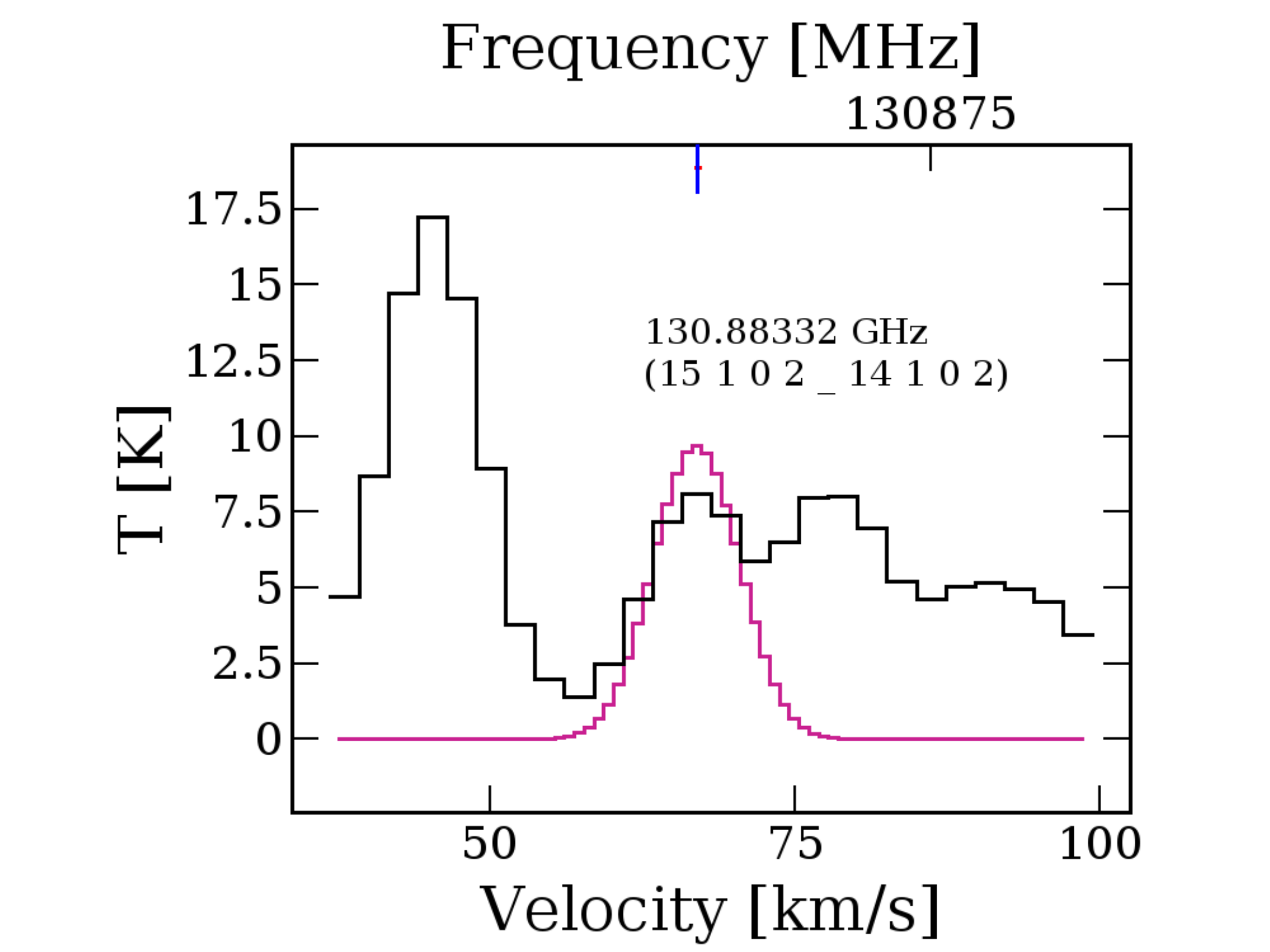}
\end{minipage}
\begin{minipage}{0.24\textwidth}
\includegraphics[width=\textwidth]{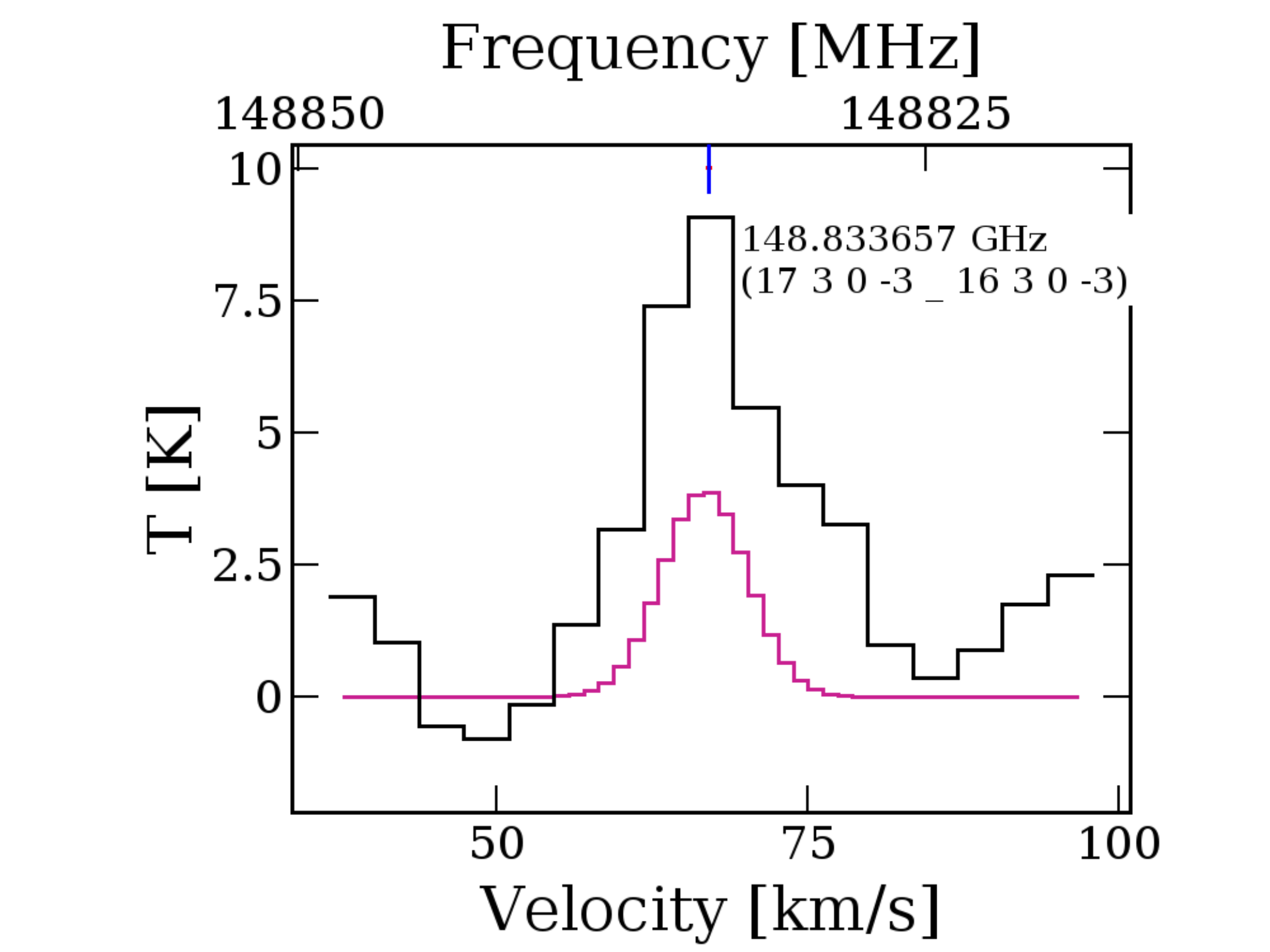}
\end{minipage}
\begin{minipage}{0.24\textwidth}
\includegraphics[width=\textwidth]{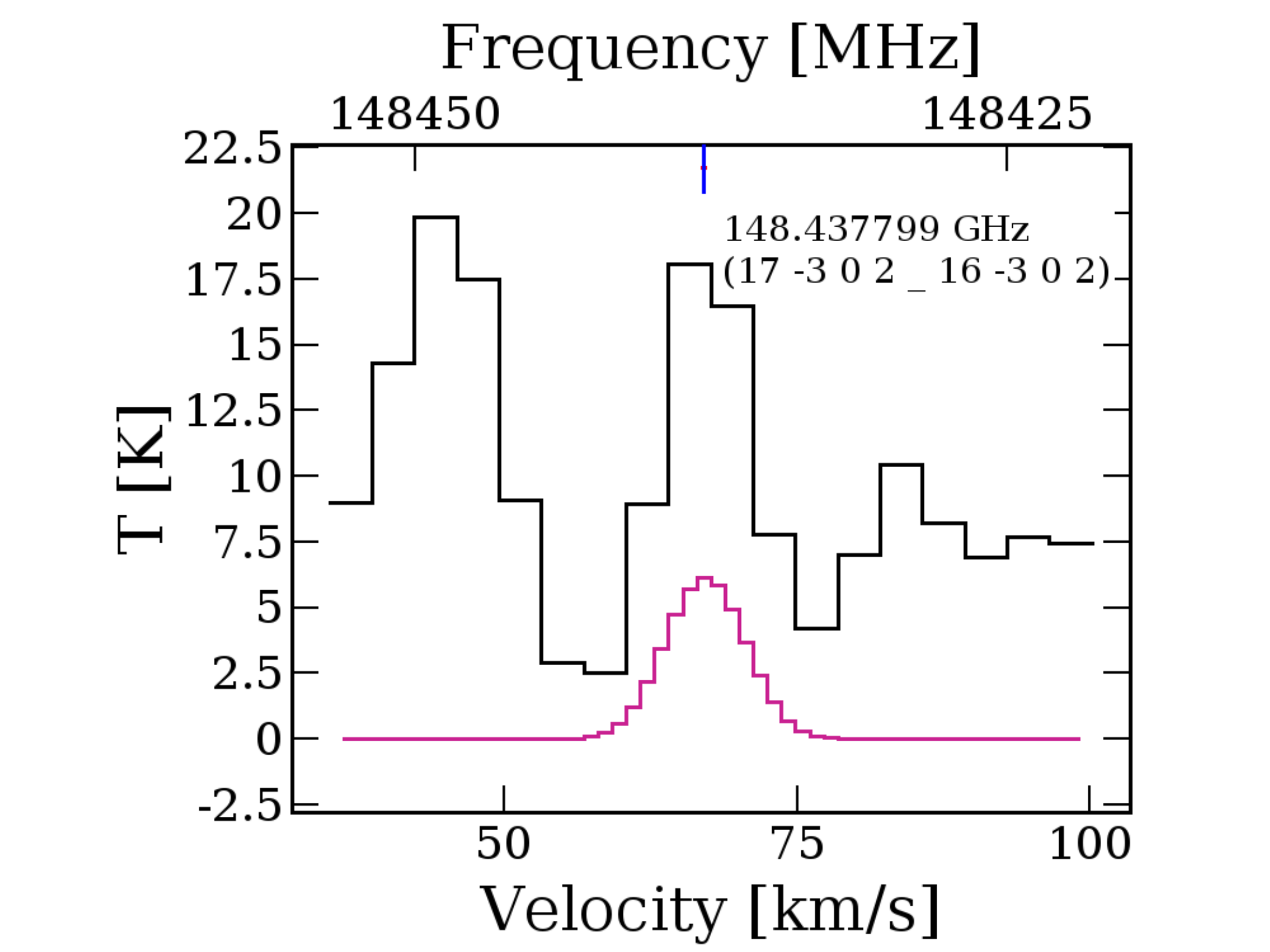}
\end{minipage}
\begin{minipage}{0.24\textwidth}
\includegraphics[width=\textwidth]{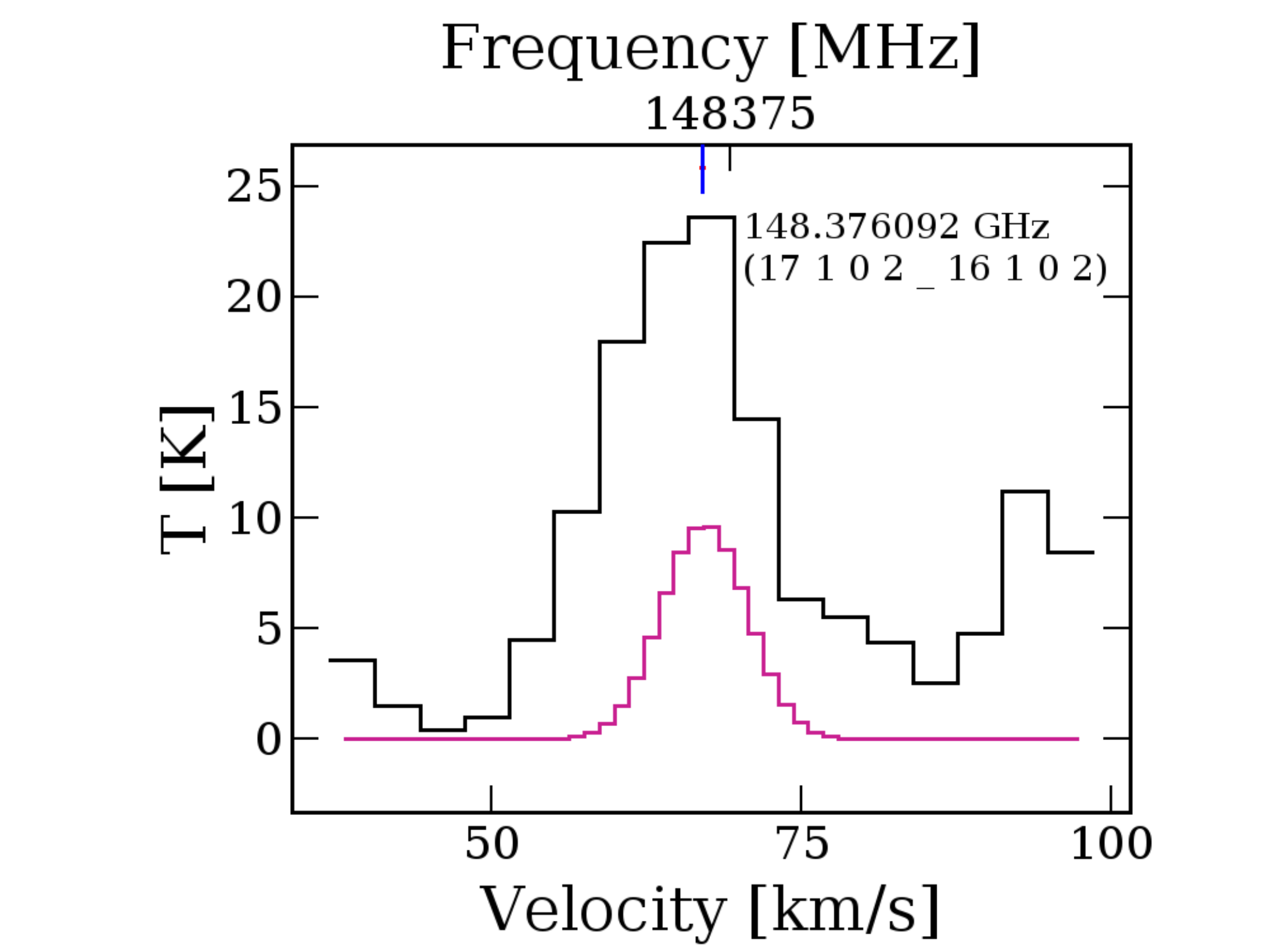}
\end{minipage}
\begin{minipage}{0.24\textwidth}
\includegraphics[width=\textwidth]{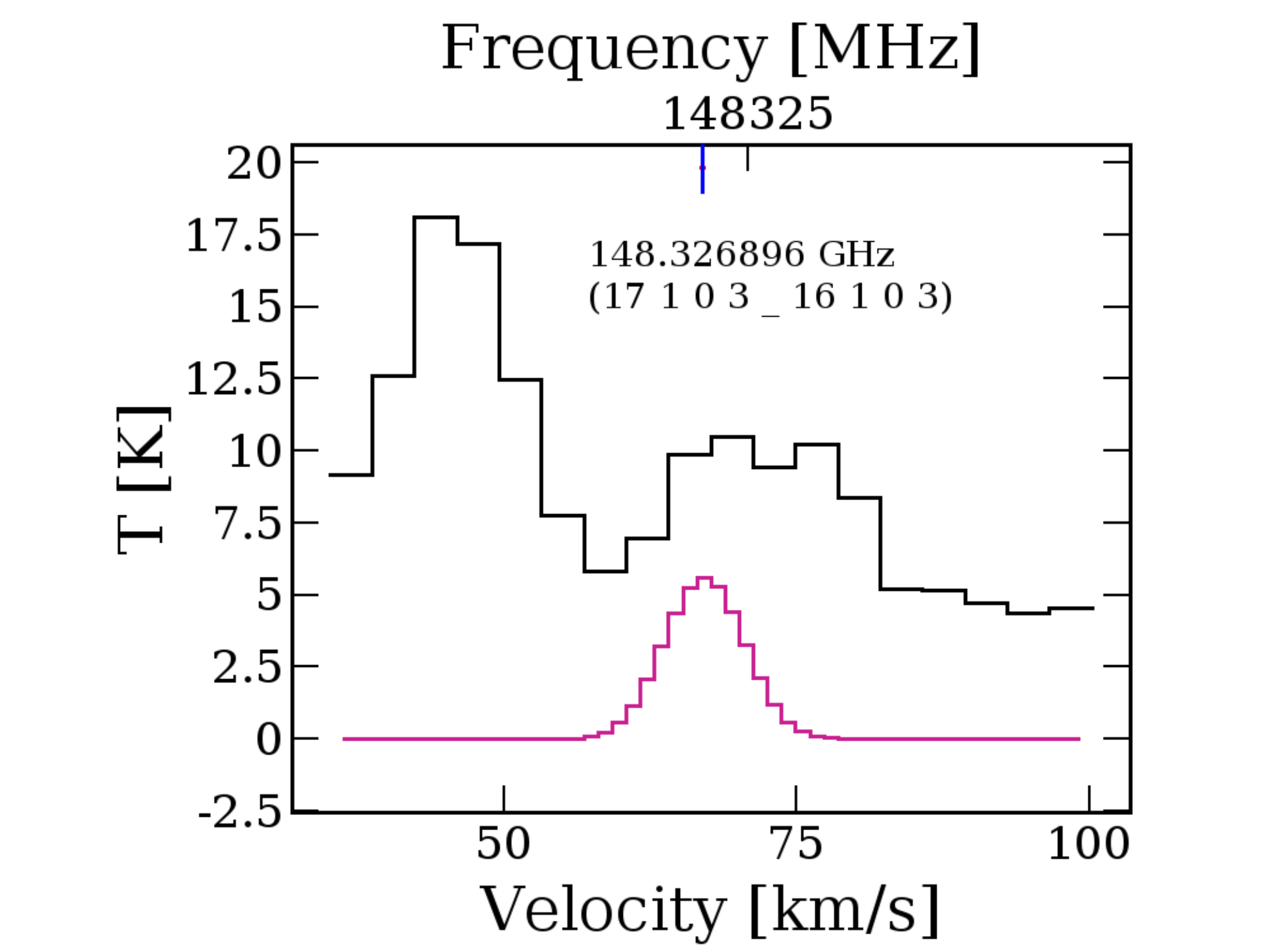}
\end{minipage}
\begin{minipage}{0.24\textwidth}
\includegraphics[width=\textwidth]{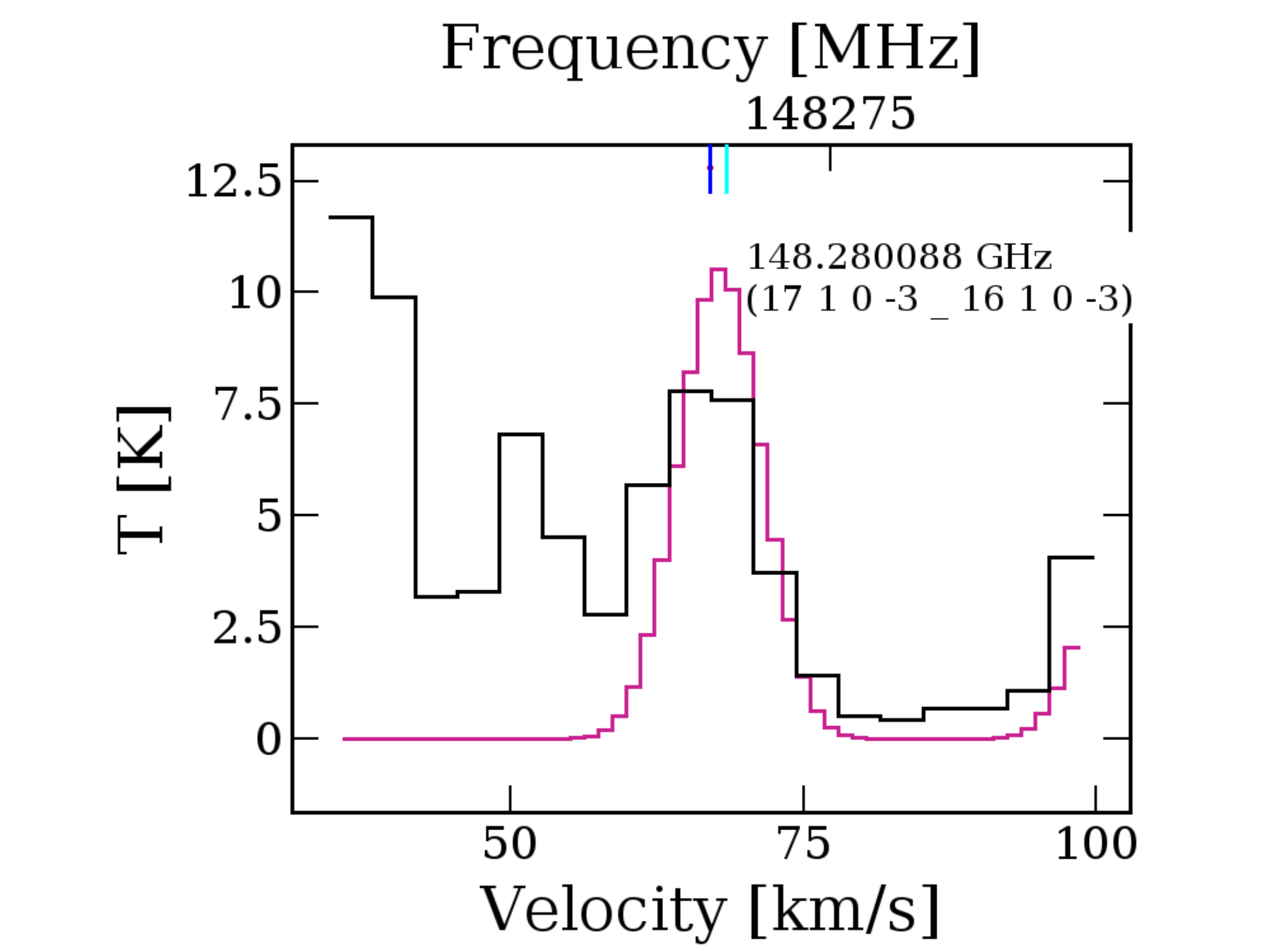}
\end{minipage}
\begin{minipage}{0.24\textwidth}
\includegraphics[width=\textwidth]{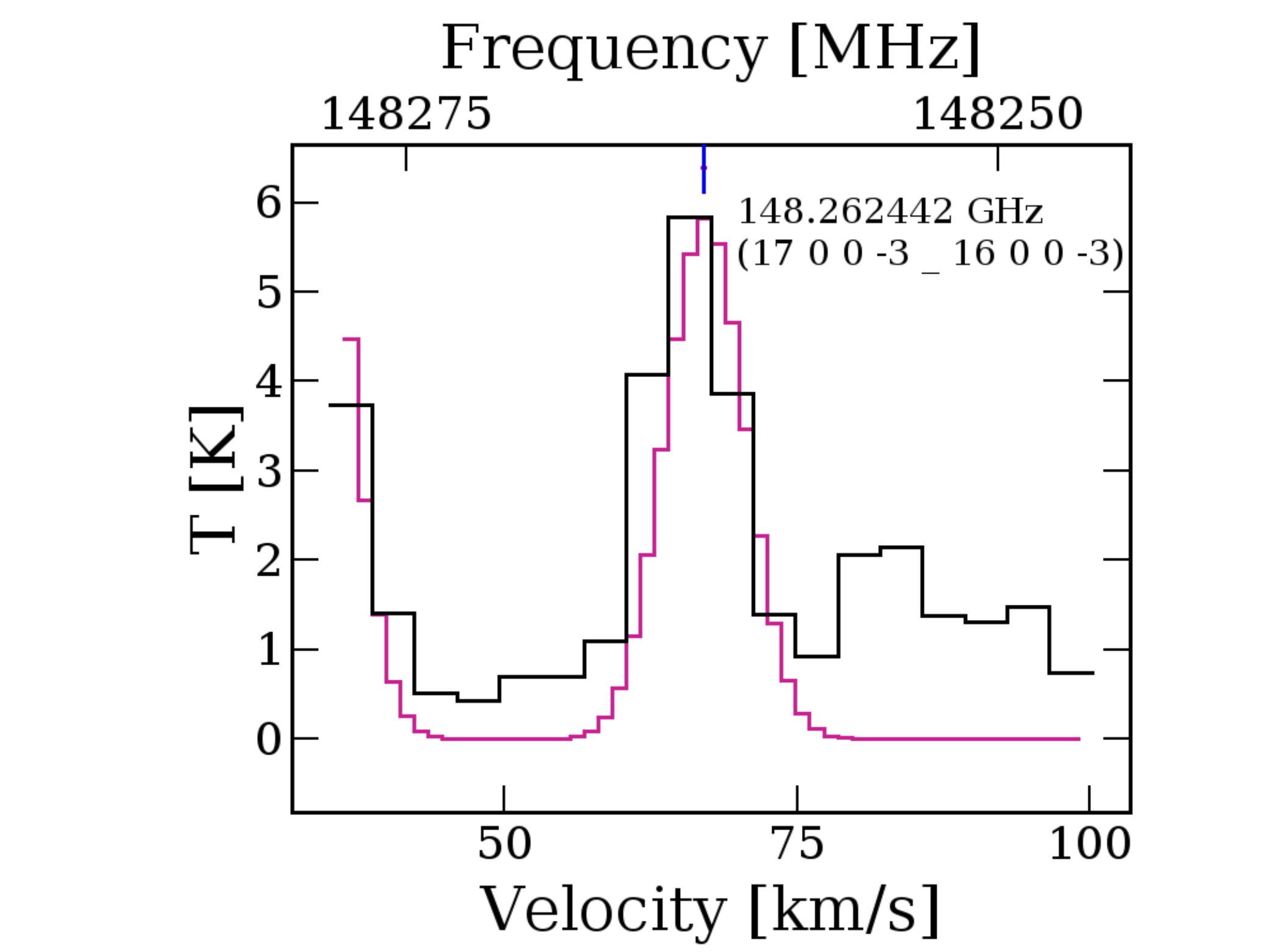}
\end{minipage}
\begin{minipage}{0.24\textwidth}
\includegraphics[width=\textwidth]{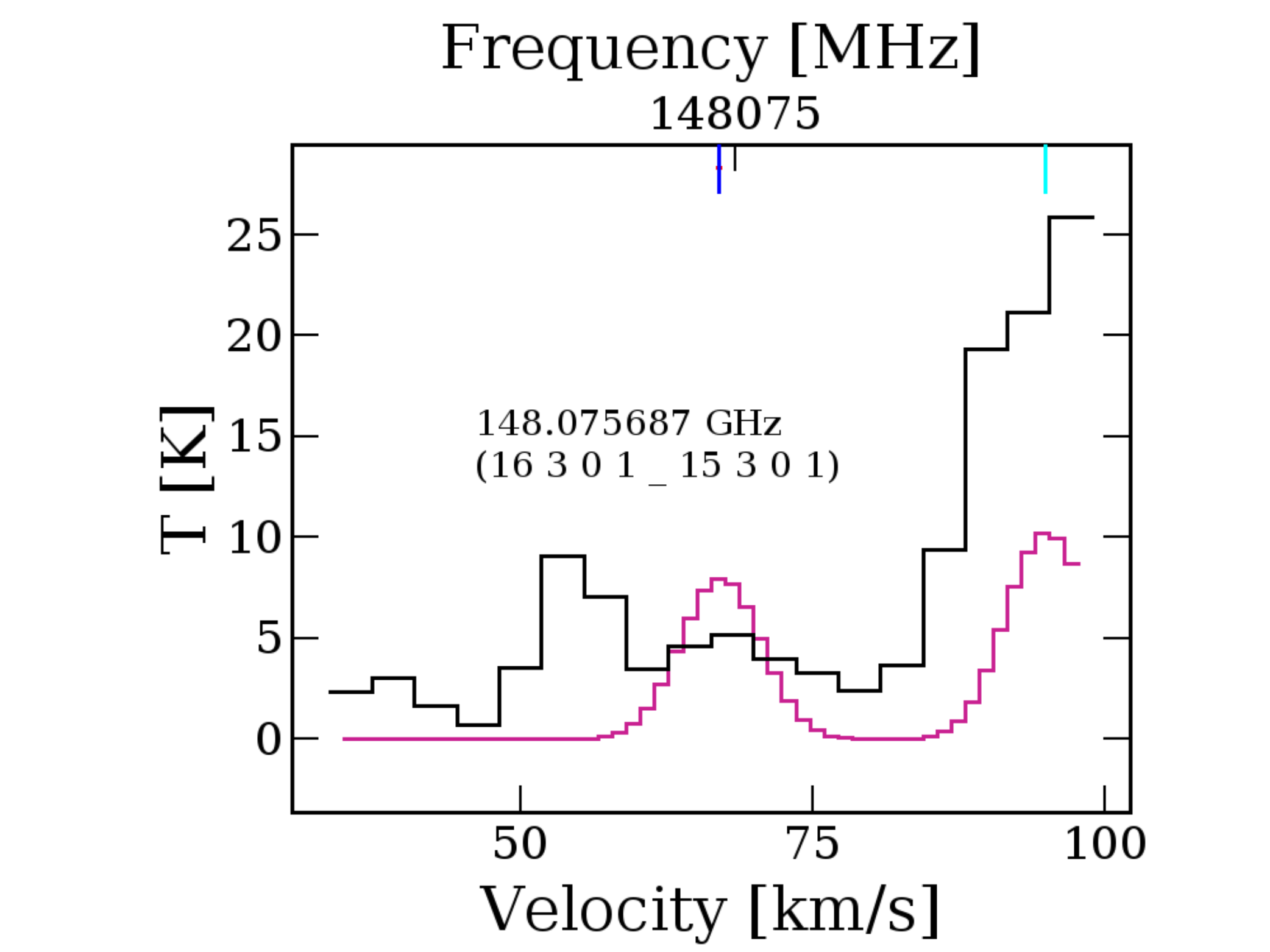}
\end{minipage}
\begin{minipage}{0.24\textwidth}
\includegraphics[width=\textwidth]{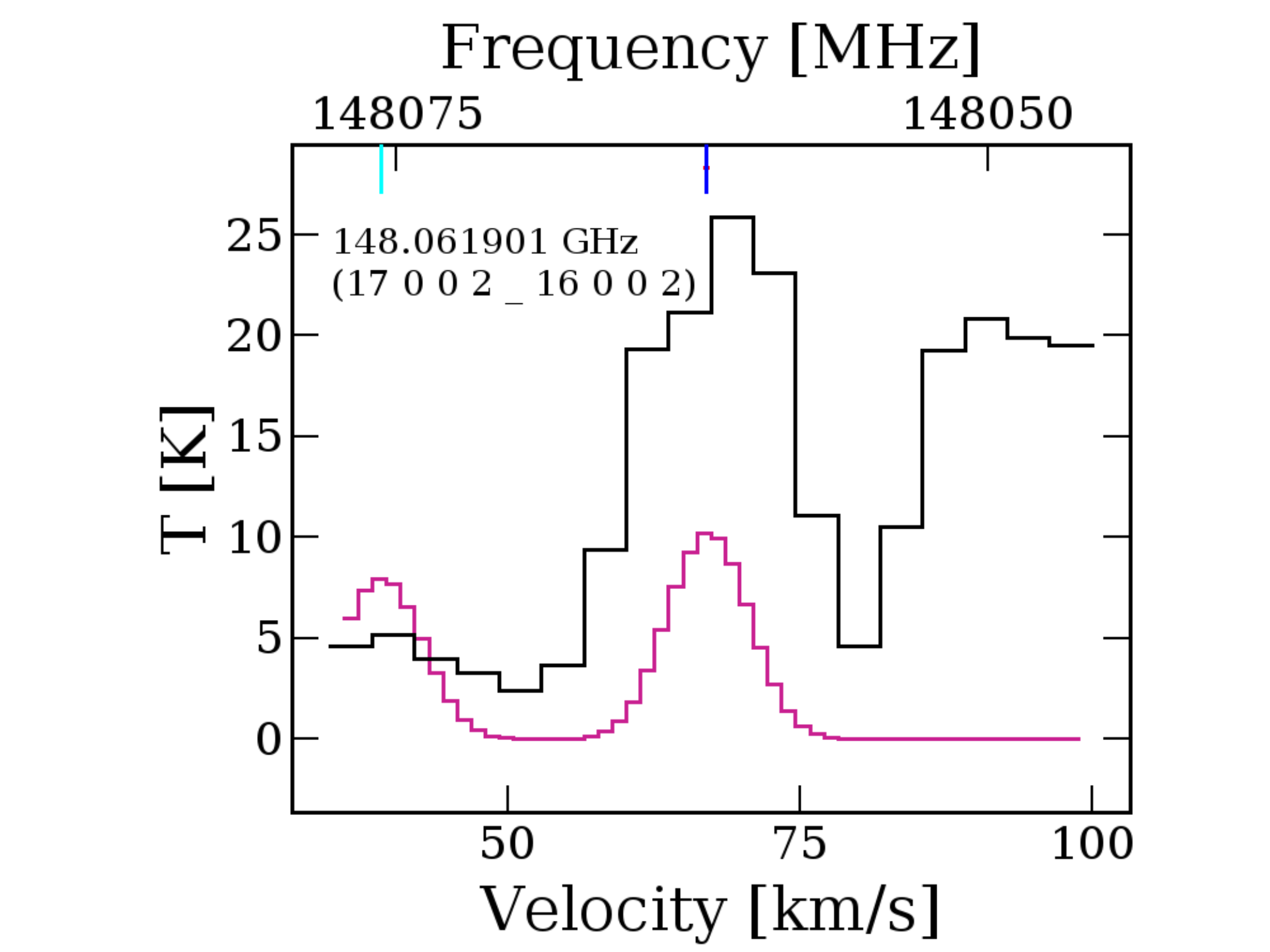}
\end{minipage}
\begin{minipage}{0.24\textwidth}
\includegraphics[width=\textwidth]{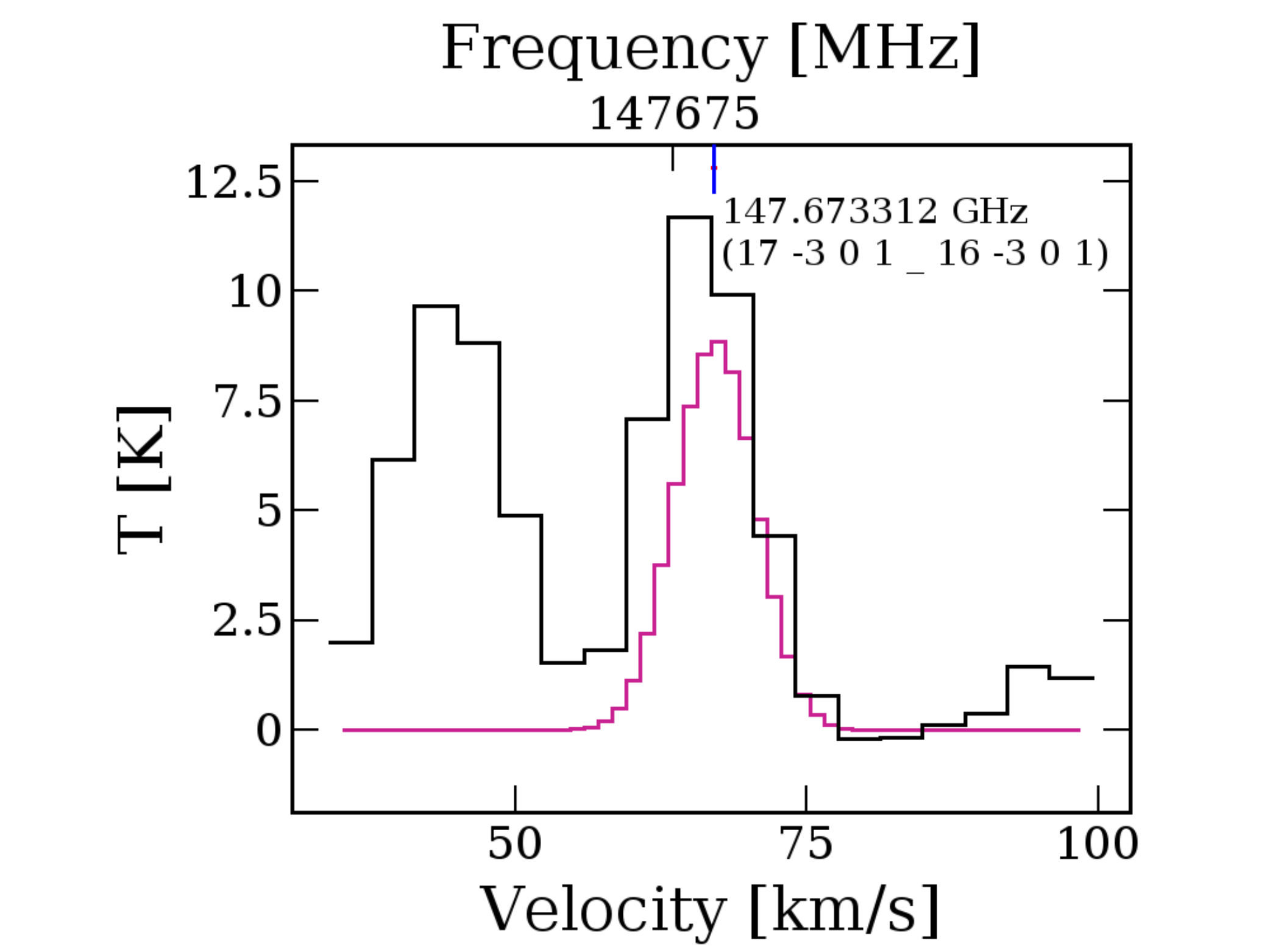}
\end{minipage}
\begin{minipage}{0.24\textwidth}
\includegraphics[width=\textwidth]{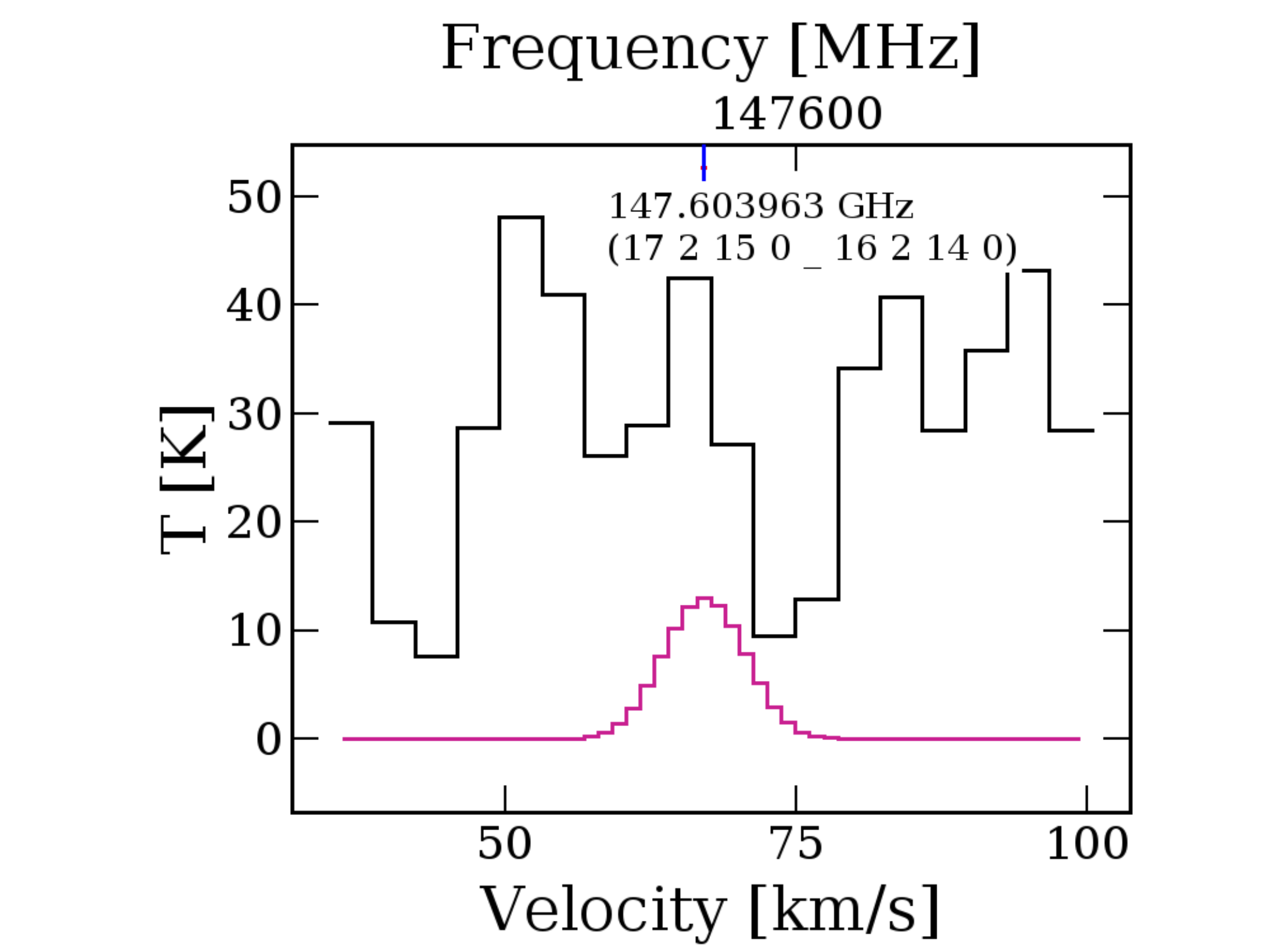}
\end{minipage}
\begin{minipage}{0.24\textwidth}
\includegraphics[width=\textwidth]{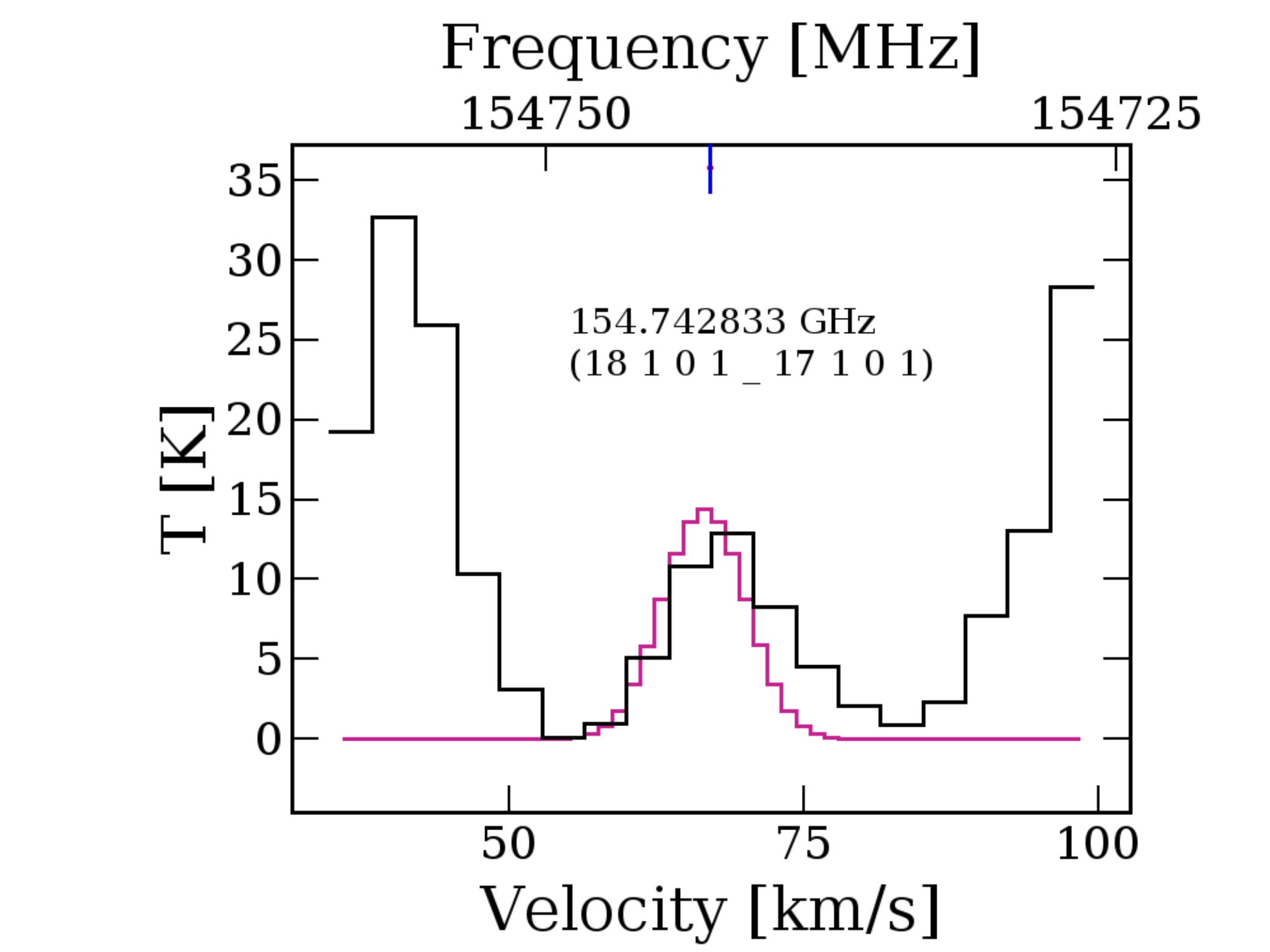}
\end{minipage}
\begin{minipage}{0.24\textwidth}
\includegraphics[width=\textwidth]{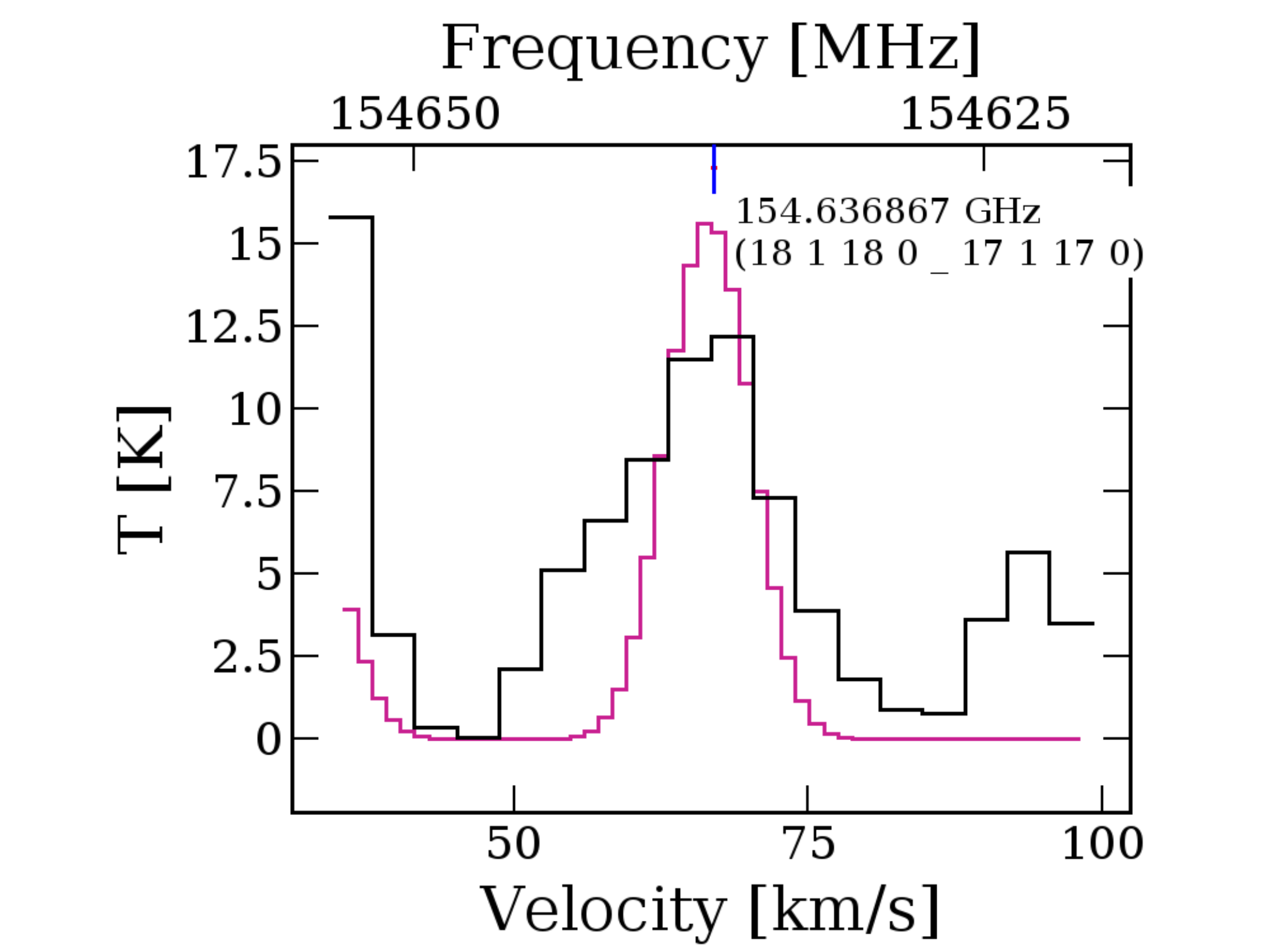}
\end{minipage}
\caption{LTE fitting of observed transitions of $\rm{CH_3NCO}$ towards G10. Black line represents the observed spectra and pink line is the fitted profile. \citep[Courtesy:][]{gora20}}
\label{fig:CH3NCO}
\end{figure}

All the LTE model-fitted line parameters are shown in Table \ref{table:fitted}. The observed and fitted spectra for $\rm{HNCO}$, $\rm{NH_2CHO}$, and $\rm{CH_3NCO}$, respectively, are illustrated in Figs. \ref{fig:HNCO}, \ref{fig:NH2CHO}, and \ref{fig:CH3NCO}. Because some of the $\rm{CH_3NCO}$ transitions are blended, we did not find good fits for those transitions. Multiple hyperfine transitions can be found in some of the spectra depicted in these pictures. We only considered the transition with the highest Einstein coefficient value for the LTE fitting. The results of LTE fitting show a small offset from some observable spectra because some of the transitions with the highest Einstein coefficients are somewhat off from the peak position. Table \ref{table:fitted} extracted physical parameters demonstrate that the optical depths ($\tau$) of all the lines are less than $1$, and are optically thin. The column densities of these three species we determined to be the best fits are displayed in Table \ref{table:fitted}. We used the various source sizes for various species as determined by their two-dimensional Gaussian fitting for our MCMC fitting. We were able to achieve higher excitation temperatures (Table \ref{table:fitted}), which are in line with the high rotational temperatures of these molecules obtained by the rotational diagram analysis discussed in \cite{gora20}.

\begin{landscape}
\begin{table}
\centering
\tiny
\caption{Summary of the best-fitted line parameters observed towards G10. \label{table:fitted} \citep[Courtesy:][]{gora20}}
\begin{tabular}{|c|c|c|p{1.3cm}|p{1.0cm}|c|c|c|c|c|c|c|}
\hline
\hline
Species&Frequency&{Range used}
&Range used &Best fit FWHM &Best fit column&Optical depth&Range used &Best fitted
&Source size&Range used &Best fitted \\
&(GHz)&{Frequency (GHz)}&FWHM (Km s$^{-1}$) &(Km s$^{-1}$)&density (cm$^{-2}$)& ($\tau$) &T$_{ex}$ (K)&T$_{ex}$ (K)&($^{''}$)&V$_{lsr}$ (Km s$^{-1}$)&V$_{lsr}$ (Km s$^{-1}$)\\
\hline\hline
&131.394262&130.53092 - 131.46660&3-6&5.97&1.6$\times$10$^{17}$&2.85$\times$10$^{-1}$&200-350&201.19&1.12&67.2-67.9&67.59\\
HNCO&154.414770&154.03118 - 154.96636&5-7&6.97&1.6$\times$10$^{17}$&2.64$\times$10$^{-1}$&200-350&205.65&1.16&66.5-67.5&67.20\\
&153.865080&153.03117 - 153.96523&5-8&7.96&1.5$\times$10$^{17}$&2.94$\times$10$^{-1}$&200-330&211.98&1.35&66.5-67.5&67.32\\
&153.818870&153.03117 - 153.96523&5-8&7.96&1.5$\times$10$^{17}$&1.19$\times$10$^{-1}$&200-330&211.98&1.35&66.5-67.5&67.32\\
&153.291840&153.03117 - 153.96523&5-8&7.96&1.5$\times$10$^{17}$&2.34$\times$10$^{-1}$&200-330&211.98&1.35&66.5-67.5&67.32\\
\hline
&148.223354&148.46523 - 147.53117&3-9&8.98&1.3$\times$10$^{17}$&7.73$\times$10$^{-2}$&400-600&472.07&1.33&66.5-67.5&67.43\\
&148.556276&148.53115 - 149.46522&3-7&7.00&9.5$\times$10$^{16}$&1.78$\times$10$^{-2}$&400-550&450.09&1.37&66.5-67.5&67.09\\
&148.567249&148.53115 - 149.46522&3-7&7.00&9.5$\times$10$^{16}$&3.52$\times$10$^{-2}$&400-550&450.09&1.37&66.5-67.5&67.09\\
NH$_2$CHO&148.599727&148.53115 - 149.46522&3-7&7.00&9.5$\times$10$^{16}$&5.15$\times$10$^{-2}$&400-550&450.09&1.37&66.5-67.5&67.09\\
&148.667591&148.53115 - 149.46522&3-7&7.00&9.5$\times$10$^{16}$&6.53$\times$10$^{-2}$&400-550&450.09&1.37&66.5-67.5&67.09\\
&148.709316&148.53115 - 149.46522&3-7&7.00&9.5$\times$10$^{16}$&6.54$\times$10$^{-2}$&400-550&450.09&1.37&66.5-67.5&67.09\\
&153.432351&153.96523 - 153.03118&3-8&7.99&1.2$\times$10$^{17}$&6.70$\times$10$^{-2}$&400-600&474.33&1.18&66.5-67.5&67.43\\
\hline

&129.957471&129.53092 - 130.46674&4-8&7.99&6.9$\times$10$^{16}$&3.08$\times$10$^{-1}$&100-300&104.04&1.05&66.5-67.5&67.14\\
&129.669703&129.53092 - 130.46674&4-8&7.99&6.9$\times$10$^{16}$&1.77$\times$10$^{-1}$&100-300&104.04&1.05&66.5-67.5&67.14\\
&130.146799&129.53092 - 130.46674&4-8&7.99&6.9$\times$10$^{16}$&2.41$\times$10$^{-1}$&100-300&104.04&1.05&66.5-67.5&67.14\\
&130.300215&129.53092 - 130.46674&4-8&7.99&6.9$\times$10$^{16}$&1.57$\times$10$^{-1}$&100-300&104.04&1.05&66.5-67.5&67.14\\
&130.228419&129.53092 - 130.46674&4-8&7.99&6.9$\times$10$^{16}$&2.41$\times$10$^{-1}$&100-300&104.04&1.05&66.5-67.5&67.14\\
&130.541066&130.53092 - 131.46660&4-8&5.99&7.8$\times$10$^{16}$&1.91$\times$10$^{-1}$&100-300&115.48&1.15&66.5-67.5&66.89\\
&130.583038&130.53092 - 131.46660&4-8&7.99&7.8$\times$10$^{16}$&1.59$\times$10$^{-1}$&100-300&115.48&1.15&66.5-67.5&66.89\\
&130.653851&130.53092 - 131.46660&4-8&7.99&7.8$\times$10$^{16}$&1.68$\times$10$^{-1}$&100-300&115.48&1.15&66.5-67.5&66.89\\
&130.661691&130.53092 - 131.46660&4-8&7.99&7.8$\times$10$^{16}$&1.23$\times$10$^{-1}$&100-300&115.48&1.15&66.5-67.5&66.89\\
&130.88332&130.53092 - 131.46660&4-8&7.99&7.8$\times$10$^{16}$&1.58$\times$10$^{-1}$&100-300&115.48&1.15&66.5-67.5&66.89\\
CH$_3$NCO&148.833657&148.53115  - 149.46522&4-8&7.99&8.8$\times$10$^{16}$&7.25$\times$10$^{-2}$&100-300&122.21&1.23&66.5-67.5&66.68\\
&148.437799&148.46523  - 147.53117&4-8&7.92&7.7$\times$10$^{16}$&1.41$\times$10$^{-1}$&100-300&101.24&1.23&66.5-67.5&67.13\\
&148.376092&148.46523  - 147.53117&4-8&7.92&7.7$\times$10$^{16}$&2.32$\times$10$^{-1}$&100-300&101.24&1.23&66.5-67.5&67.13\\
&148.326896&148.46523  - 147.53117&4-8&7.92&7.7$\times$10$^{16}$&1.28$\times$10$^{-1}$&100-300&101.24&1.23&66.5-67.5&67.13\\
&148.280088&148.46523  - 147.53117&4-8&7.92&7.7$\times$10$^{16}$&1.27$\times$10$^{-1}$&100-300&101.24&1.23&66.5-67.5&67.13\\
&148.262442&148.46523  - 147.53117&4-8&7.92&7.7$\times$10$^{16}$&1.33$\times$10$^{-1}$&100-300&101.24&1.23&66.5-67.5&67.13\\
&148.075687&148.46523  - 147.53117&4-8&7.92&7.7$\times$10$^{16}$&1.88$\times$10$^{-1}$&100-300&101.24&1.23&66.5-67.5&67.13\\
&148.061901&148.46523  - 147.53117&4-8&7.92&7.7$\times$10$^{16}$&2.48$\times$10$^{-1}$&100-300&101.24&1.23&66.5-67.5&67.13\\
&147.673312&148.46523  - 147.53117&4-8&7.92&7.7$\times$10$^{16}$&2.12$\times$10$^{-1}$&100-300&101.24&1.23&66.5-67.5&67.13\\
&147.603962&148.46523  - 147.53117&4-8&7.92&7.7$\times$10$^{16}$&3.27$\times$10$^{-1}$&100-300&101.24&1.23&66.5-67.5&67.13\\
&154.742833&154.03118  - 154.96636&4-8&7.97&7.5$\times$10$^{16}$&3.53$\times$10$^{-1}$&100-300&101.70&1.21&66.5-67.5&66.54\\
&154.636867&154.03118  - 154.96636&4-8&7.97&7.5$\times$10$^{16}$&3.92$\times$10$^{-1}$&100-200&101.70&1.21&66.5-67.5&66.54\\
\hline
\end{tabular}
\end{table}
\end{landscape}

\subsection{Nitrogen bearing molecules}
\cite{mond23} analyzed the ALMA archival data of the hot molecular core, G10.47+0.03. Local thermodynamic equilibrium was used (LTE) to examine the extracted spectra. For molecules for which many transitions have been detected, robust techniques like Markov chain Monte Carlo (MCMC) and rotational diagram approaches are employed to constrain the temperature and column density. Various transitions of nitrogen-containing species (NH$_2$CN, HC$_3$N, HC$_5$N, C$_2$H$_3$CN, C$_2$H$_5$CN, and H$_2$NCH$_2$CN), as well as some of their isotopologues and isomers, are reported by \cite{mond23}. The identification of CH$_3$CCH and one of its isotopologues was also reported by \cite{mond23}. Different transitions of nitrogen-bearing species are identified and explained in this study. Astrochemical modelling may be used to explain the compact emissions from cyanamide, cyanoacetylene, vinyl cyanide, and ethyl cyanide.

\begin{table}
\tiny{
\caption{Summary of the best-fit line parameters obtained using MCMC method considering a source size = 2$^{''}$ and a V$_{lsr}$ = 68 km/s.\label{table:mcmc_lte}} 
\begin{center}
\addtolength{\leftskip} {-2cm}
\addtolength{\rightskip}{-2cm}
\begin{tabular}{|c|c|c|c|c|c|c|c|}
\hline
Species&Quantum numbers&Frequency&E$_u$&FWHM&Best fitted column&Best fitted&Optical depth\\
&&(GHz)&(K)&(Km s$^{-1}$)&density (cm$^{-2}$) & T$_{ex}$ (K)&($\tau$)\\
\hline
\multicolumn{8}{|c|}{}\\
\multicolumn{8}{|c|}{\it Vinyl cyanide and its isotopologues}\\
\hline
$\rm{C_2H_3CN}$&$6_{2,4}\rightarrow6_{1,5}$,v=0&130.763576&18.2&&&&0.02\\
&$29_{2,27}\rightarrow29_{1,28}$,v=0&131.168737&209.5&&&&0.05\\
&$12_{2,11}\rightarrow12_{1,12}$,v=0&153.42175&44.1&8.2&$(3.4\pm0.4 )\times10^{17}$&209.8$\pm$18.02&0.03\\
&$19_{0,19}\rightarrow18_{1,18}$,v=0&158.773785&85.5&&&&0.06\\
&&&&&&&\\
$^{13}$CH$_2$CHCN&$16_{7,9}\rightarrow15_{7,8}$&147.927607&164.8&&&&0.03\\
&$16_{3,14}\rightarrow15_{3,13}$&147.986686&79.5&9.0&(5.3$\pm0.4)\times10^{15}$&$125.8\pm$13.67&0.07\\
&$16_{2,14}\rightarrow15_{2,13}$&149.423715&69.2&&&&0.07\\
&$17_{1,16}\rightarrow16_{1,15}$&159.954635&71.2&&&&0.08\\
\hline
\multicolumn{8}{|c|}{}\\
\multicolumn{8}{|c|}{\it Ethyl cyanide with its isotopologues and isomer}\\
\hline
C$_2$H$_5$CN&$14_{1,14}\rightarrow13_{0,13}$,v=0&129.768140&44.8&&&&0.20\\
&$21_{3,19}\rightarrow21_{2,20}$,v=0&130.693882&109.4&&&&0.12\\
&$24_{4,20}\rightarrow24_{3,21}$,v=0&148.293988&147.1&&&&0.12\\
&$10_{2,9}\rightarrow9_{1,8}$,v=0&148.36276&28.1&7.8&$(1.7\pm0.10)\times10^{17}$&104.1$\pm$9.9&0.10\\
&$28_{3,26}\rightarrow28_{2,27}$,v=0&153.589524&184.6&&&&0.08\\
&$21_{4,17}\rightarrow21_{3,18}$,v=0&154.47598&117.2&&&&0.14\\
&$26_{2,25}\rightarrow26_{1,26}$,v=0&154.55736&154.0&&&&0.07\\
&$20_{8,13}\rightarrow21_{7,14}$,v=0&158.742366&161.4&&&&0.02\\
&$17_{4,13}\rightarrow17_{3,14}$,v=0&159.3308&83.6&&&&0.15\\
&&&&&&&\\
$^{13}$CH$_3$CH$_2$CN&$15_{5,10}\rightarrow14_{5,9}$,v=0&131.155447&77.8&&&&0.05\\
&$15_{11,4}\rightarrow14_{11,3}$,v=0&131.161428&183.4&&&&0.01\\
&$15_{4,12}\rightarrow14_{4,11}$,v=0&131.233679&67.9&8.4&$(7.8\pm1.1)\times10^{15}$&130.8$\pm$23.2&0.06\\
&$15_{4,11}\rightarrow14_{4,10}$,v=0&131.249984&67.9&&&&0.06\\
&$18_{1,17}\rightarrow17_{1,16}$,v=0&159.10002&74.2&&&&0.08\\
\hline
\multicolumn{8}{|c|}{}\\
\multicolumn{8}{|c|}{\it Methylacetylene and its isotopologue}\\
\hline
CH$_3$CCH&9$_6\rightarrow8_6$&153.71152&296.1&&&&0.015\\
&9$_3\rightarrow8_3$&153.790769&101.7&10.0&$(8.7\pm0.7)\times10^{16}$&219.1$\pm$14.8&0.06\\
&9$_2\rightarrow8_2$&153.805457&65.7&&&&0.04\\
\hline
\end{tabular}
\end{center}}
\end{table}

For the species for which several transitions are detected, the rotational diagram is employed by \cite{mond23}; however, the excitation temperature and column density are constrained for the comparison using a different approach, namely, the Markov chain Monte Carlo (MCMC). The probabilistic behavior of a group of atomic particles was the initial focus of the development of the MCMC algorithm. Analytically, it was challenging to do this. As a result, an iterative technique was used to simulate a solution. The probabilities of each conceivable occurrence in this stochastic model only depend on the state obtained in the preceding condition. The MCMC method is a dynamic procedure that uses a random walk to iteratively examine all line parameters, such as molecular column density, excitation temperature, source size, and line width, and gives the solution using the $\chi^2$ minimization process.

For MCMC fitting, the source size is held constant at 2$^{''}$, v$_{LSR}$ is held steady at $68$ km/s, and the FWHM of all transitions within a species is held constant at their average measured value. The outcomes for these species are then compared with those of the rotation diagram approach obtained by \cite{mond23}. Table 1 of \cite{mond23} shows the results obtained using these two techniques.

\begin{figure*}[t]
\hskip -1.5cm
\includegraphics[width=10cm,height=17cm,angle =270]{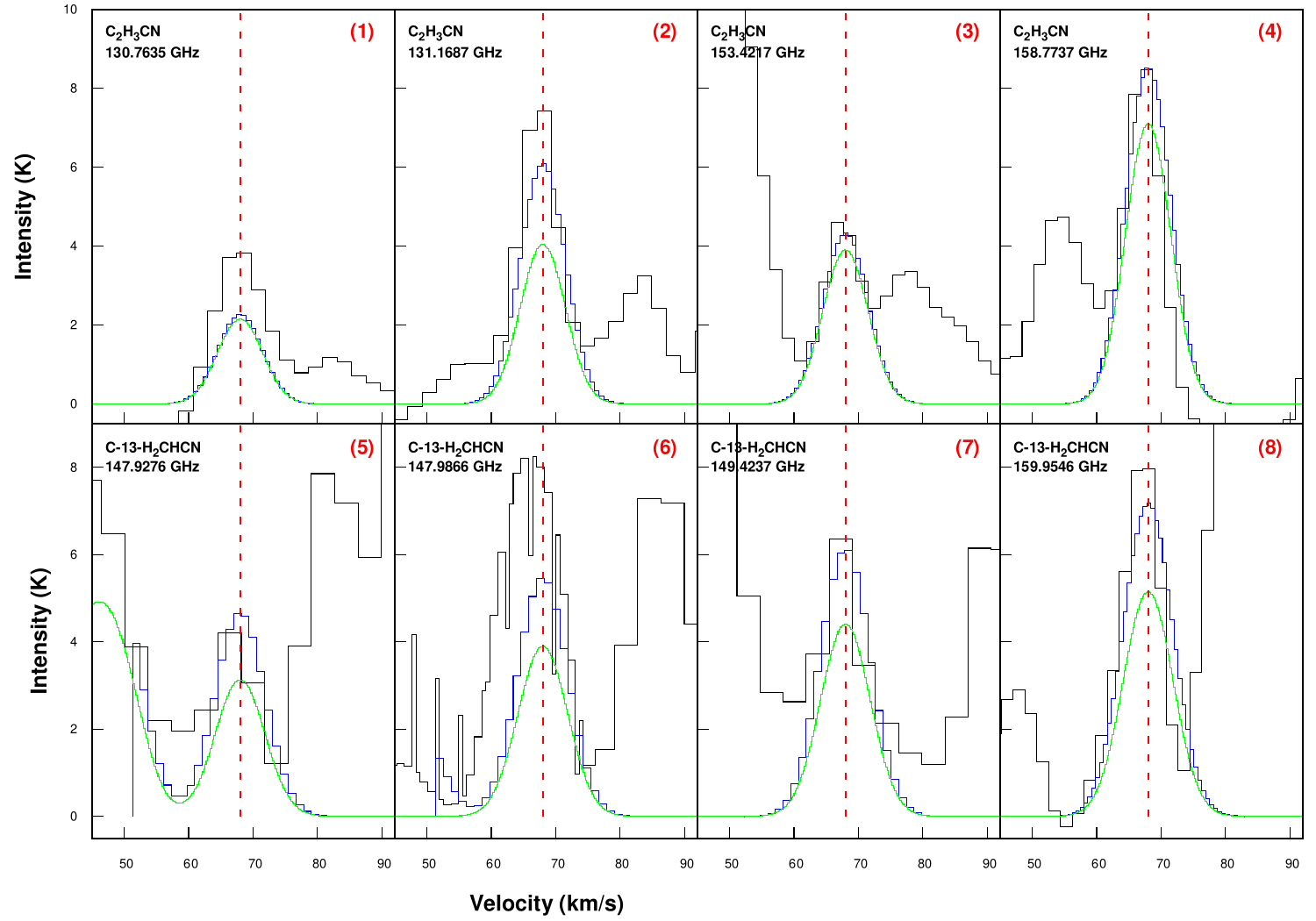}
\caption{MCMC fit of observed, unblended, optically thin transitions of vinyl cyanide and its isotopologues (solid green lines). The blue lines represent the modeled spectra, whereas the observed spectra are shown in black. The vertical red dashed line shows the position of V$_{lsr} = 68$ km/s. The green solid lines represent the LTE spectra of some species by considering the rotational temperature and column density obtained from the rotational diagram analysis (Details are in \cite{mond23}).} 
\label{fig:c2h3cn-mcmc}
\end{figure*}

\begin{figure*}[t]
\hskip -1.5cm
\includegraphics[width=13cm,height=18cm,angle =270]{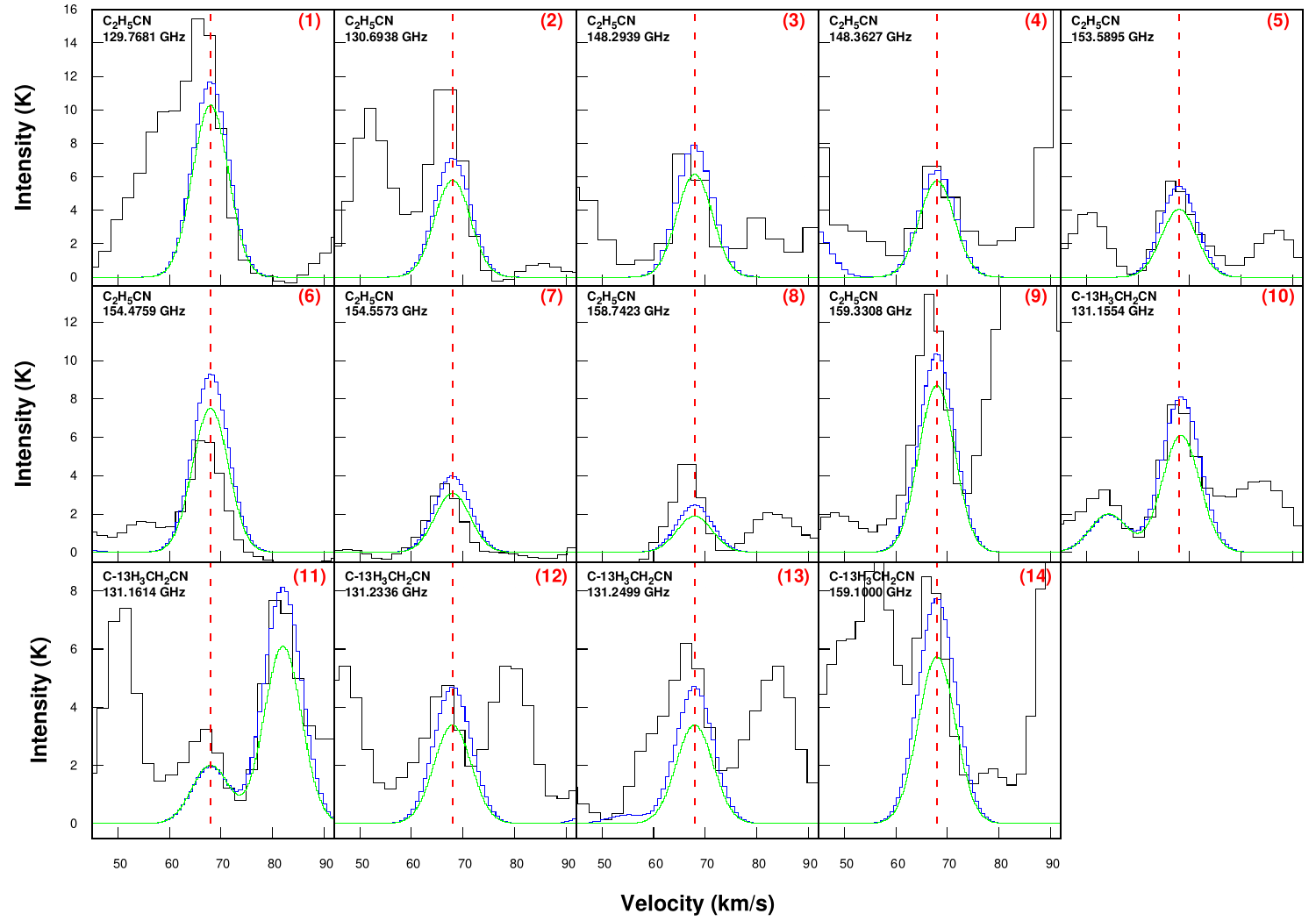}
\caption{MCMC-fit spectra (same as Figure \ref{fig:c2h3cn-mcmc}) of the observed, unblended, optically thin transitions of ethyl cyanide, its one isotopologue.}
\label{fig:c2h5cn-mcmc}
\end{figure*}

\begin{figure*}
\includegraphics[width=9cm,height=9cm]{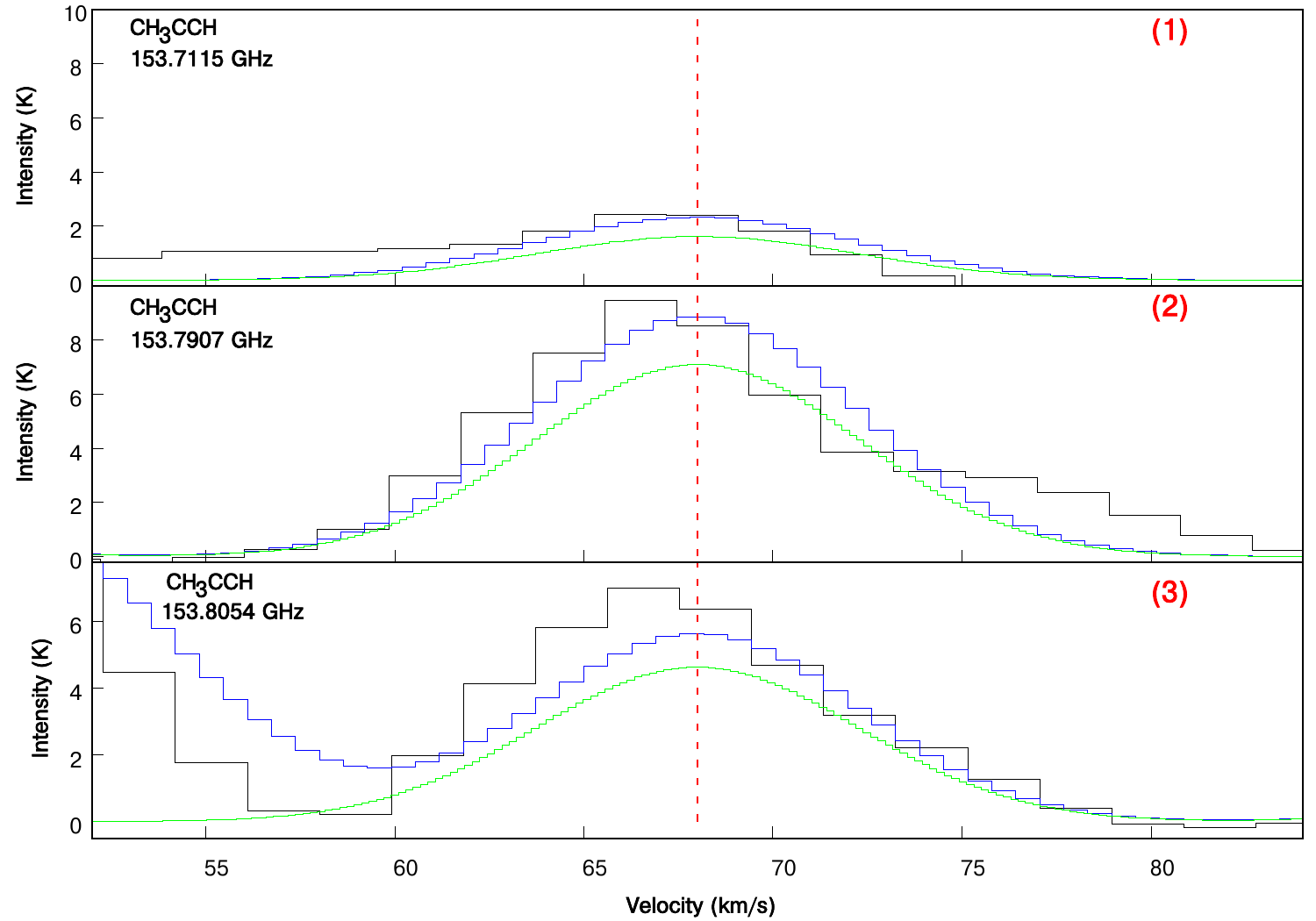}
\caption{MCMC-fit spectra of CH$_3$CCH transitions.}
\label{fig:ch3cch-mcmc}
\end{figure*}

The obtained line parameters of the MCMC fitting of five species are noted in Table \ref{table:mcmc_lte}. MCMC fitted spectra are shown in Figures \ref{fig:c2h3cn-mcmc}, \ref{fig:c2h5cn-mcmc}, and \ref{fig:ch3cch-mcmc}. Table 1 of \cite{mond23} provides an overview of the estimated column densities and temperatures using the rotational diagram, MCMC, and LTE methods. The estimated parameters all agree fairly well with one another. However, they both produce more realistic values because the MCMC method is an alternative to the rotational diagram. So we used the MCMC method to compare the estimated parameters and to constrain them better.

\section{MCMC fitting to the observed lines towards G31.41+0.31}
\label{sec:MCMC_G31}

\begin{figure}
\centering
\includegraphics[width=14cm, height=16cm, angle=270]{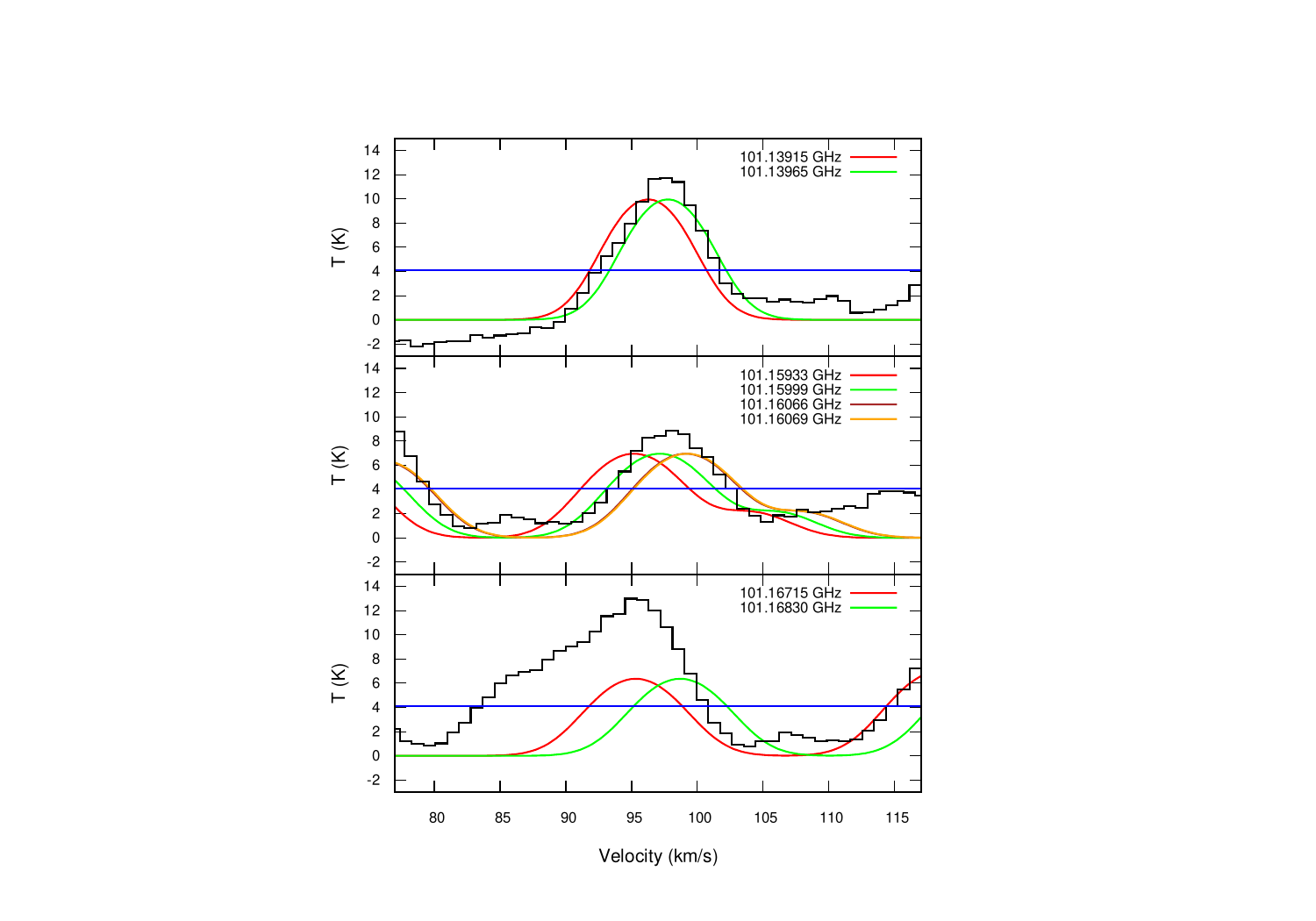}
\caption{The MCMC method is used to fit the observed transitions of $\rm{CH_3SH}$ in G31. The
observed spectral profile is shown with black lines, whereas the modeled profile is shown with the other colours. The horizontal blue line represents the 3$\sigma$($4.1$ K) RMS noise level.\citep[Courtesy:][]{bhat22}} 
\label{fig:ch3sh-mcmc}
\end{figure}

\begin{figure}
\hskip -1cm
\includegraphics[width=14cm,height=16cm,angle =270]{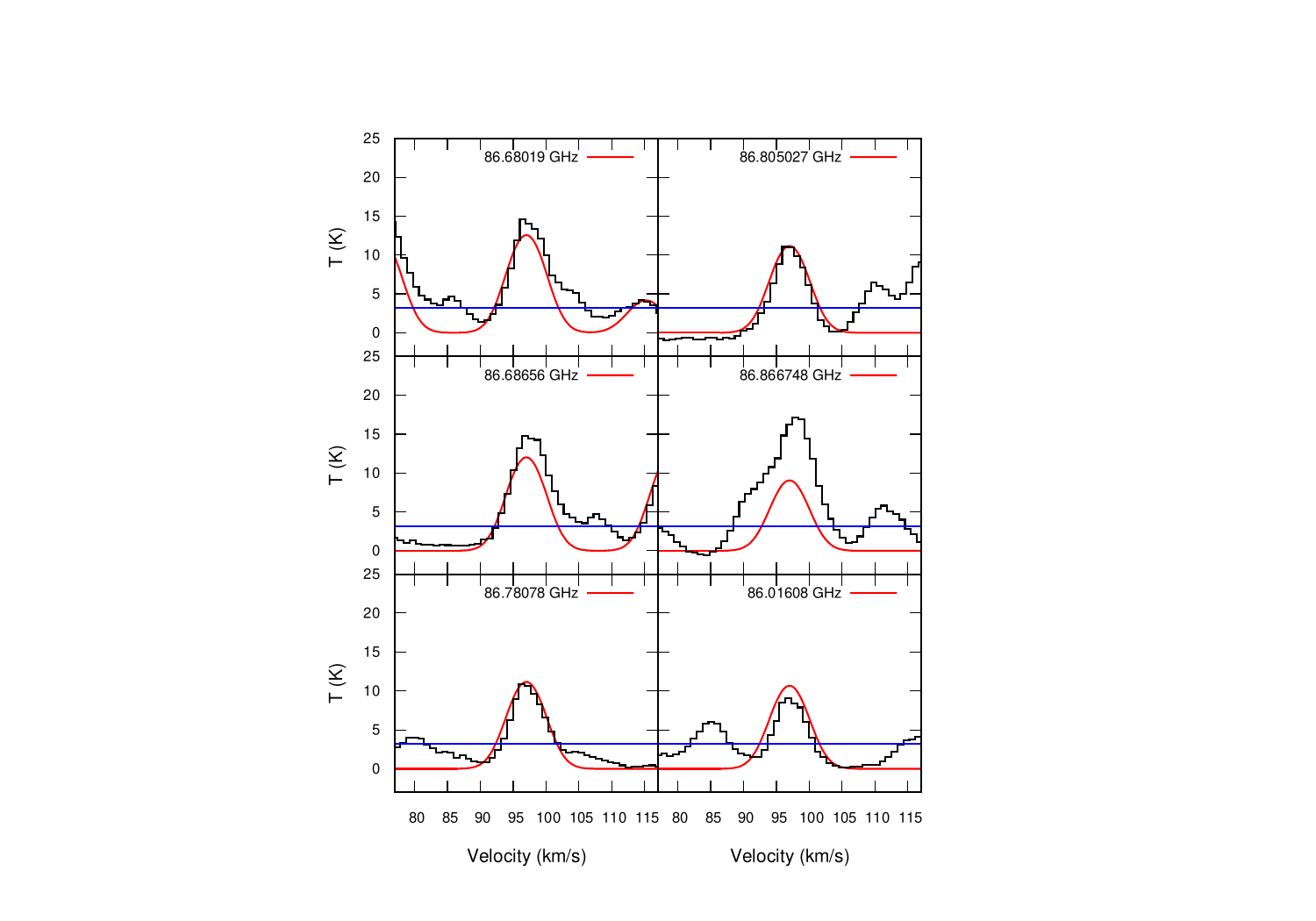}
\caption{The MCMC method used to fit the observed transitions of $\rm{CH_3NCO}$ in G31. The red lines represent the modeled spectral profile, whereas the observed spectra are shown in black. The horizontal blue line represents the 3$\sigma$(3.2K) RMS noise level. \citep[Courtesy:][]{bhat22}}
\label{fig:ch3nco-mcmc}
\end{figure}

\begin{figure}
\hskip -1cm
\includegraphics[width=14cm,height=18cm,angle=270]{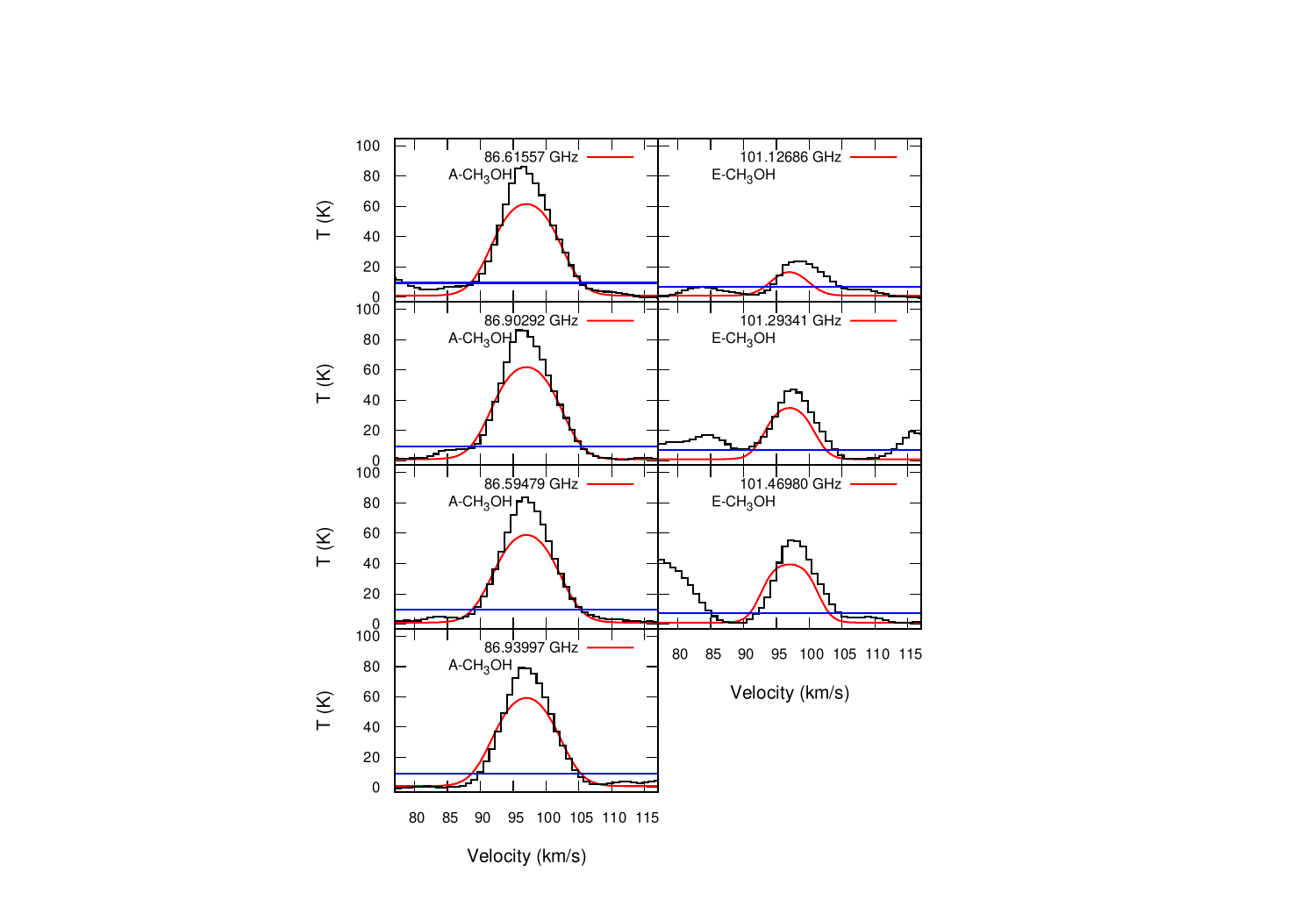}
\caption{MCMC method used to fit the observed transitions of methanol ($\rm{CH_3OH}$) towards G31. The red lines represent the modeled spectral profile to the observed spectra in black. The horizontal blue line represents the 3$\sigma$(9.4K for A-CH$_3$OH and 7K for E-CH$_3$OH) RMS noise level. \citep[Courtesy:][]{bhat22}}
\label{fig:ch3oh-mcmc}
\end{figure}

Methanethiol ($\rm{CH_3SH}$) and methyl isocyanate ($\rm{CH_3NCO}$) are two complex sulfur-bearing species that have recently been observed in G31 \citep{gora21}. Additionally, they detected several transitions of methanol ($\rm{CH_3OH}$) in G31. The observations' details are depicted in chapter \ref{chap:g31_ratran}. They used a rotation diagram and LTE model analysis to derive the excitation temperature, column density, FWHM, etc. The MCMC method estimates those parameters and compares them to the results obtained by \cite{gora21}.

Table \ref{table:mcmc_lte_G31} lists the initial constraints used for this fitting technique, the best-fitted values of the physical parameters, and the uncertainties determined from the $\chi^2$ reduction. Figures \ref{fig:ch3sh-mcmc}, \ref{fig:ch3nco-mcmc}, and \ref{fig:ch3oh-mcmc}, respectively, display all of the fitted spectra of $\rm{CH_3SH}$, $\rm{CH_3NCO}$, and $\rm{CH_3OH}$. It is observed that all transitions of CH$_3$SH and CH$_3$NCO are optically thin, with the best-fitted line parameters as presented in Table \ref{table:mcmc_lte_G31}. The transitions of A-CH$_3$OH and two transitions of E-CH$_3$OH, are all optically thick. For CH$_3$SH and CH$_3$NCO, column densities of $1.4 \times 10^{17}$ cm$^{-2}$ and $1.2 \times 10^{18}$ cm$^{-2}$ are achieved, respectively. Compared to CH$_3$NCO, the transitions of CH$_3$SH are well fitted with an excitation temperature of $\sim 77$ K. It is impossible to fit all the transitions simultaneously for methanol. We isolate the transitions resulting from A/E methanol as \cite{gora21}. A-CH$_3$OH and E-CH$_3$OH both had column densities of $2.1 \times 10^{19}$ cm$^{-2}$ and $8.2 \times 10^{19}$ cm$^{-2}$, respectively. Comparatively, A-CH$_3$OH is related with a greater temperature ($\sim 173$ K) than E-CH$_3$OH ($\sim 82$ K). Deep inside the envelope ($5840$ AU), as indicated by the peak abundance profile of CH$_3$OH (Figure \ref{fig:abundance} and Table \ref{table:abundances}), the abundance of methanol steadily declines. Thus, getting a somewhat greater abundance with the cold component of methanol (i.e., transitions of E-CH$_3$OH) is valid. These column densities, calculated using the rotational diagram approach and LTE analysis, are also stated in the footnote of the Table.

\begin{table*}
\tiny{
\caption{Summary of the best fitted line parameters obtained by using MCMC method.\label{table:mcmc_lte_G31} \citep[Courtesy:][]{bhat22}} 
\begin{center}
\addtolength{\leftskip} {-2cm}
\addtolength{\rightskip}{-2cm}
\begin{tabular}{|c|c|c|c|c|p{1.4cm}|c|p{0.9cm}|c|}
\hline
\hline
Species&Quantum numbers&Frequency&E$_u$&Best fit FWHM&Best fitted column&Best fitted
&Best fitted&Optical depth \\
&&(GHz)&(K)&(Km s$^{-1}$)&density$^a$ (cm$^{-2}$) & T$_{ex}$ (K)& source size ($^{''}$)& ($\tau$)\\
\hline\hline
&4(0,4)-3(0,3), m=0&101.13915&12.14&&&&&0.526\\
CH$_3$SH&4(0,4)-3(0,3), m=1&101.13965&13.56&&&&&0.517\\
&4(-2,3)-3(-2,2), m=0&101.15933&31.26&&&&&0.308\\
&4(-3,2)-3(-3,1), m=1&101.15999&52.39&&&&&0.137\\
&4(3,1)-3(3,0), m=0&101.16066&52.55&6.5&(1.4$\pm$1.0)$\times$10$^{17}$&77.20$\pm$30.08&0.97$\pm$0.11&0.136\\
&4(-3,2)-3(-3,1), m=0&101.16069&52.55&&&&&0.136\\
&4(-2,3)-3(-2,2), m=1&101.16715&29.62&&&&&0.315\\
&4(2,2)-3(2,1), m=1&101.16830&30.27&&&&&0.312\\
\hline
&10(0,10)-9(0,9), m=0(v$_t$=0)&86.68019&22.88&&&&&0.568\\
&10(0,0)-9(0,0), m=1(v$_t$=0)&86.68656&34.97&&&&&0.536\\
$\rm{CH_3NCO}$&10(2,9)-9(2,8), m=0(v$_t$=0)&86.78078&46.73&6.5&(1.2$\pm$ 0.4)$\times$10$^{18}$&202.3$\pm$16.05&0.90$\pm$0.11&0.485\\
&10(2,8)-9(2,7), m=0(v$_t$=0)&86.80503&46.73&&&&&0.485\\
&10(-3,0)-9(-3,0), m=1(v$_t$=0)&86.86675&88.58&&&&&0.373\\
&10(2,0)-9(2,0), m=1(v$_t$=0)&87.01608&58.81&&&&&0.455\\
\hline
&$7(2)^{-}-6(3)^{-},V_{t}=0$&86.61557&102.70&&&&&1.576\\
A-$\rm{CH_3OH}$&$7(2)^{+}-6(3)^{+},V_{t}=0$&86.90292&102.72&8.8&$(2.1 \pm 0.82) \times 10^{19}$&$172.51 \pm 15.46$&$1.98 \pm 0.25$&1.582\\
&$15(3)^{+}-14(4)^{+},V_{t}=0$&88.59479&328.26&&&&&1.350\\
&$15(3)^{-}-14(4)^{-},V_{t}=0$&88.93997&328.28&&&&&1.356\\
\hline
&$5(-2)-5(1),V_{t}=0$&101.12686&60.73&&&&&0.478\\
E-$\rm{CH_3OH}$&$7(-2)-7(1),V_{t}=0$&101.29341&90.91&6.0&$(8.2 \pm 1.47) \times 10^{19}$&81.74 $\pm$ 7.74&$1.97 \pm 0.14$ &1.75\\
&$8(-2)-8(1),V_{t}=0$&101.46980&109.49&&&&&2.754\\
\hline
\hline
\end{tabular}
\end{center}}
\hskip 2cm {$^a$ With the LTE analysis, \cite{gora21} obtained the column density of $4.13 \times 10^{17}$ cm$^{-2}$, $7.22 \times 10^{17}$ cm$^{-2}$, and $1.84 \times 10^{19}$ cm$^{-2}$ respectively for  CH$_3$SH, CH$_3$NCO, and CH$_3$OH.} \\
{\noindent With the rotational diagram analysis, \cite{gora21} obtained it $2.85 \times 10^{16}$ cm$^{-2}$, $1.58 \times 10^{16}$ cm$^{-2}$ and $2.94 \times 10^{19}$ cm$^{-2}$ respectively.}\\
\end{table*}

\section{Summary}
Monte Carlo Markov Chain (MCMC) is a very useful method to extract the physical condition from observed line transitions. In this chapter we extracted the physical conditions of two hot cores, G10 and G31 from the observed line profiles.

 \chapter{Radiative transfer modeling of observed line profiles: A Young Hot Molecular Core G31.41+0.31} \label{chap:g31_ratran}

\section*{Overview}
Recently, the 1 $\rightarrow$ 0 transition of H$^{13}$CO$^+$ was discovered in G31.41+0.31. It suggested the presence of an infalling gas envelope. An outflow tracer, SiO, was also observed in this source. 
Many hot core tracers, like, CH$_3$CN, CH$_3$OH, CH$_3$SH, and CH$_3$NCO, were also obtained in this source. It makes this a desirable source for observation. We used these observed line profiles and compared them with our synthetic spectra to infer the excitation conditions of the environment where they were observed. We implemented non-LTE calculations for this purpose. For the non-LTE analysis, the RATRAN radiative transfer code and the CASSIS LTE computation are considered. The best-fitted line parameters, which reflect the current physical state of the gas envelope, are derived. Our findings imply that, except SiO, all the reported line profiles might be explained by an infalling gas \citep{bhat22}. Therefore, one additional outflow component is included to model the SiO line profile.

\clearpage
\section{General discussion}
High-mass ($>$8 $M_\odot$) star-forming regions are frequently detected based on the presence of the ultra-compact HII (UC HII) regions (HMSFRs). However, the evolutionary paths of the high-mass star-forming region are not well constrained. In particular, the consequences of fragmentation and creation of different systems in HMSFRs and mass accumulation processes toward the cores are not yet well-established \citep{bosc19,pala14, beut18}. 

High gas temperature ($\geq 100$ K), high density ($\geq 10^6$ cm$^{-3}$), and small size ($\leq 0.1$ pc) are the main characteristics of Hot Molecular Cores (HMCs), which exhibit a variety of simple and complex organic molecular emissions \citep{kurt00}. The formation of a hot molecular core is the basic fundamental process of high-mass star formation. A fascinating HMC, G31.41+0.31 (from now on G31), has various ongoing kinematics and chemical compositions. The Main core and NE core of G31 as its two components \citep{belt18} are identified from the dust continuum emission. It appears uniform and featureless, with no indication that the main core is fragmented \citep{belt18}. However, the high opacity of the dust emission restrict a precise observation of the central core. 
The presence of the red-shifted absorption, rotational spin-up, and two embedded free-free emitting sources in the centre suggests that this core may have fragmentation with infall and rotation present.

The main core of G31 is located $\sim 5^{''}$ away from the UC HII zone \citep{cesa10}. This source is located at a distance of 7.9 kpc from the Sun. This source has a luminosity of $3\times10^5$ L$_\odot$ \citep{chur90}. According to the recent parallax measurements, it is situated at 3.7 kpc distant and would have a luminosity of $4\times 10^4$ L$_\odot$ \citep{reid19,belt19}. According to \cite{cesa10} and \cite{rivi17}, its systematic velocity (V$_{LSR}$) is 97 km s$^{-1}$. \cite{gira09} explain their observed inverse P-Cygni profile of the low excitation line of C$^{34}$S(J=7-6) in G31 by considering the infall. \cite{cesa11} observed C$^{17}$O and CH$_3$CN and suggested infall in this source. Later, using the Atacama Large Millimeter/submillimeter Array (ALMA), \cite{belt18} observed this source with excellent angular and spatial resolution and discussed the related kinematics. They recognized the J=5–4 transition of SiO and discussed the existence of molecular outflow and its possible orientations in this source. Additionally, they noticed the inverse P-Cygni profile in H$_2$CO and CH$_3$CN lines with varied upper state energies (E$_{up}$) and discovered the rotation and accelerating infall present in the HMC. Different transitions of SiO, HCN, and H$^{13}$CO$^+$ in G31, which trace outflow and infall, have recently been observed by \cite{gora21}.
The time variability of the infall is not easily explained. The infall might happen outside-in \citep{fost93} or inside-out \citep{shuf77}. \cite{osor09} modeled the observed spectral energy distribution (SED) in detail for G31. They determined the essential physical parameters responsible for the resulting spectral profile and fitted the SED using several models. The density and temperature profile obtained from the singular logatropic sphere solution were used to model the source. They calculated the core mass to be $\sim 25M_\odot$ and the outer radius of the envelope to be $\sim30000$ AU by taking into account a distance of $7.9$ kpc (\cite{vand13} utilized an outside radius of $\sim 119000$ AU for continuum modeling). \cite{maye14} detected infall motion in G31. The spectral signatures of the red and blue asymmetries present in blue-shifted and red-shifted portions relative to the rest frame velocity, are used to identify this infall. \cite{maye14} also reported the inversion transitions of ammonia ((2,2), (3,3), (4,4), (5,5), and (6,6) in addition to the infall motion present in the source. They added a new signature to first-order moment maps called "central blue spot," which establishes the infalling motion. Recently, \cite{esta19} estimated the range of the central mass of G31 $\sim 70-120M_\odot$, using this central blue spot signature.
The observed line profiles are commonly explained by using the radiative transfer code. For instance, \cite{wyro16} used the RATRAN radiative transfer code to simulate the reported $\rm{HCO^+}$ line profile. With the Stratospheric Observatory for Infrared Astronomy (SOFIA), they focused on nine massive molecular clumps (regions of a molecular cloud where density is high, there are many dust and gas cores present) and detected numerous $\rm{NH_3}$ transitions as well as some $\rm{HCO^+}$, $\rm{HCN}$, and $\rm{HNC}$ transitions. From red-shifted absorption, they obtained the infall signature and other physical properties. Ammonia ($\rm{NH_3}$) absorption lines were found in all nine sources, indicating the presence of infall with an infall rate between $3-10\times10^{-3} M_\odot$/yr. The large molecular outflow in G331.512-0.103 (a compact radio source in the center of an energetic molecular outflow and an active and extreme high-mass star-forming environment in our Galaxy) was explained by \cite{herv19} using the one-dimensional radiative transfer code, MOLLIE \citep{keto10}. SiO was reported in this source using band 7 data from the ALMA by \cite{herv19}. They obtained the outflow characteristics of the source and replicated the observed SiO spectral signature using the radiative transfer model.
Many observations were made earlier to investigate the chemical composition of G31. This source has been found to contain a large number of COMs, making it an exciting target for astronomers. \cite{belt09} found methyl formate (HCOOCH$_3$), glycolaldehyde (CH$_2$OHCHO), and methanol using the IRAM facility. Dimethyl ether ($\rm{CH_3OCH_3}$), glycolaldehyde, methyl formate, ethylene glycol ((CH$_2$OH)$_2$), and ethanol ($\rm{C_2H_5OH}$) were recently discovered by \cite{rivi17} in the main molecular core of G31 using Green Bank Telescope (GBT), IRAM $30$ m telescope, and Submillimeter Array (SMA) interferometric observations. Then, for the first time in G31, \cite{gora21} detected complex sulfur-bearing species (methanethiol, CH$_3$SH) and a molecule that contains peptide bonds (methyl isocyanate, CH$_3$NCO).
This chapter describes a modeling effort that uses chemical and radiative transfer codes to explain the line profiles seen in G31. To explain the observed abundances and spectral fingerprints, a chemical model and radiative transfer codes are being used. By looking at the observed line profiles of H$^{13}$CO$^+$, HCN, SiO, and NH$_3$, it is not always possible to determine the physical characteristics of G31 (infall, outflow, temperature, etc.). Some line profiles of $\rm{CH_3SH}$, $\rm{CH_3NCO}$, and $\rm {CH_3OH}$, are also modelled. The same spectral data from observations shown in \cite{gora21} is used. The spectral signature of NH$_3$ is used from \cite{osor09}.

\section{Observations}
\begin{figure}
\centering
\includegraphics[height=11cm,width=12cm]{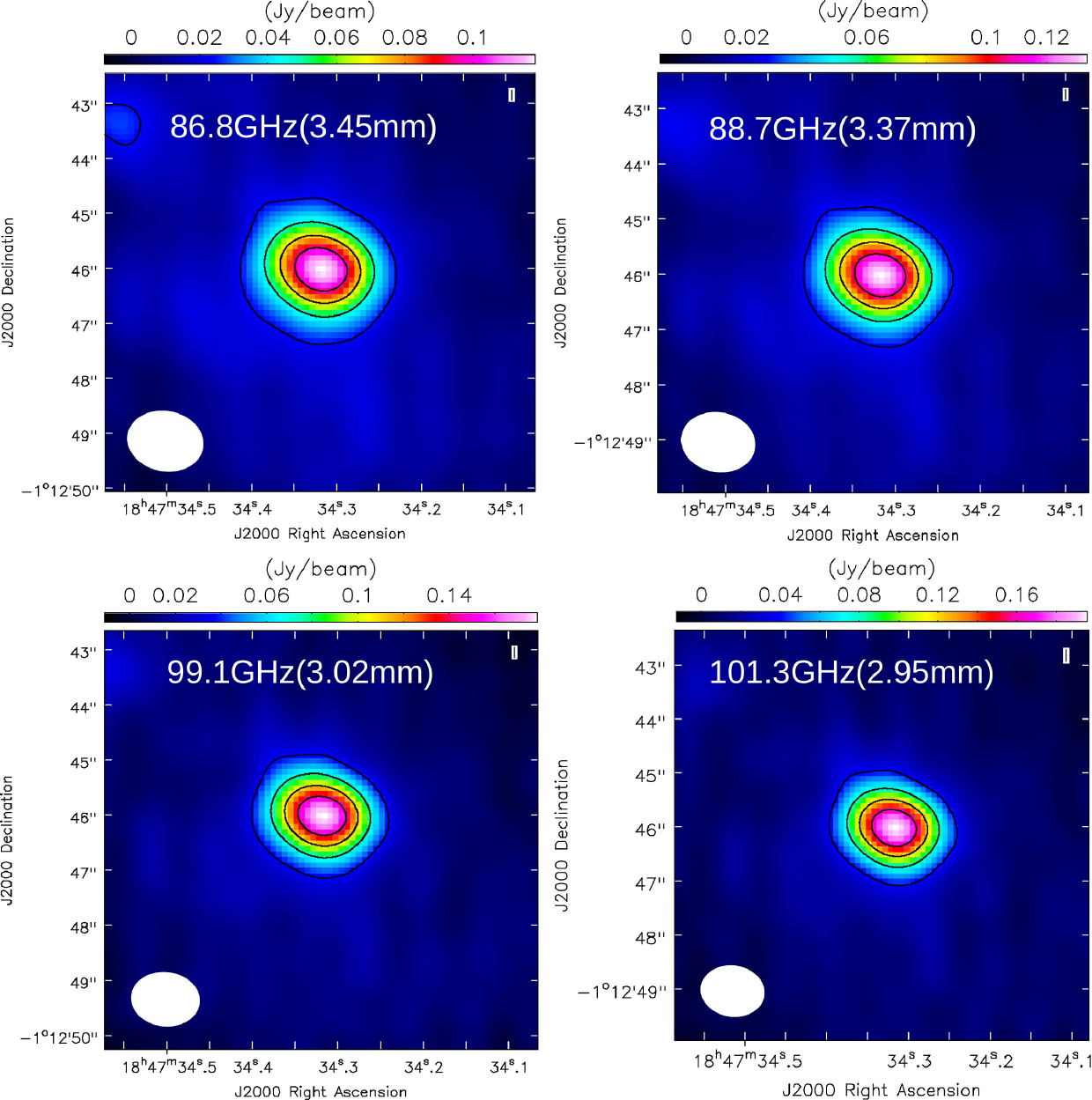}
\caption {Continuum emission observed towards G31 with ALMA band 3 at (i) 86.8 GHz (peak intensity = 118.34 mJy/beam), (ii) 88.7 GHz (peak intensity = 133.51 mJy/beam),  (iii) 99.1 GHz (peak intensity 179.2 mJy/beam), and (iv) 101.3 GHz (peak intensity = 196.2 mJy/beam). Contour levels are at 20\%, 40\%, 60\%, and 80\% of the peak flux. The synthesized beam is shown in the lower left-hand corner of each figure. \citep[Courtesy:][]{bhat22}} 
\label{fig:cont_four}
\end{figure}

The Atacama Large Millimeter/Submillimeter Array (ALMA) ($\#$2015.1.01193.S) cycle 3 archived data is examined in this chapter. The dataset has four spectral windows (86.559-87.028 GHz, 88.479-88.947 GHz, 98.977-99.211 GHz, and 101.076-101.545 GHz). The continuum emission images at these four spectral windows are displayed in Figure \ref{fig:cont_four}. With a synthesized beam of $\sim 1.19^{''} \times 0.98^{''}$ ($9402 \times 7743$ AU considering 7.9 kpc distance and $4403 \times 3636$ AU considering 3.7 kpc distance) and a position angle of 76$^\circ$, the data cube has a spectral resolution of 244 kHz ($\sim 0.84$ km s$^{-1}$). Unless otherwise specified, the calculations are based on a distance of 3.7 kpc. The systematic velocity of this source is 97 km s$^{-1}$. According to \cite{gora21}, the dust continuum's integrated flux and peak intensity were both $158.4$ mJy/beam and $242.4$ mJy/beam, and the RMS noise of the continuum map  $\sim 5$ mJy/beam. They obtained the molecular hydrogen column density of $\sim 1.53 \times 10^{25}$ cm$^{-2}$.

The observed spectral energy distribution and ammonia emission from the G31 hot core were modeled by \cite{osor09}. They considered that the hot core was in a rapid accretion phase with an infalling envelope. By fitting the observed spectral energy distribution, they determined the physical characteristics of the envelope and stellar component. The G31 HMC has been observed extensively. However, most of the observations cannot distinguish the G31 HMC and from the UC HII area, whose peaks are only 5$^{''}$ apart. The overall spectral resolution of the VLA data in \cite{cesa94}, \cite{cesa98} was $\sim 0^{''}.63$. \ used this high spatial resolution data \cite{osor09} to analyze spectral signatures. To test the predictions of their model, \cite{osor09} additionally employed low angular resolution data (40$^{''}$) from \cite{chur90}. They found the best fit when they employed a central mass of 25 M$_\odot$, a distance of 7.9 kpc, a mass accretion rate of $\sim 3 \times 10^{-3} M _{\odot}$, and an index of the power-law that characterizes the dust absorption coefficient ($\beta$) of 1.

\section{Physical conditions}
\label{sec:phys}
\begin{figure}
\centering
\includegraphics[height=6cm]{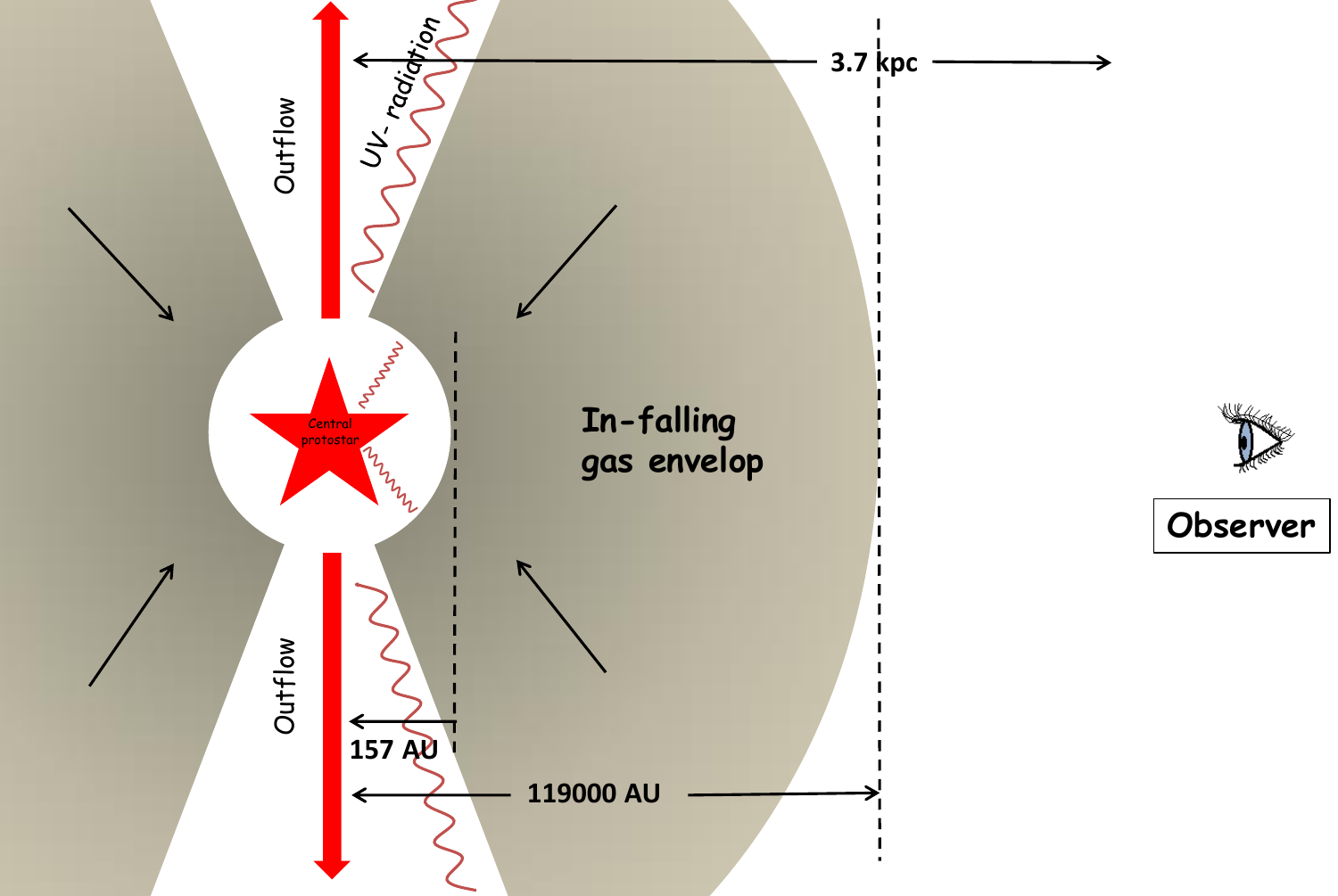}
  \caption{Cartoon diagram to represent the modeled region of the cloud. \citep[Courtesy:][]{bhat22}}
  \label{fig:cartoon}
\end{figure}

\begin{figure}
\centering
\includegraphics[height=8cm]{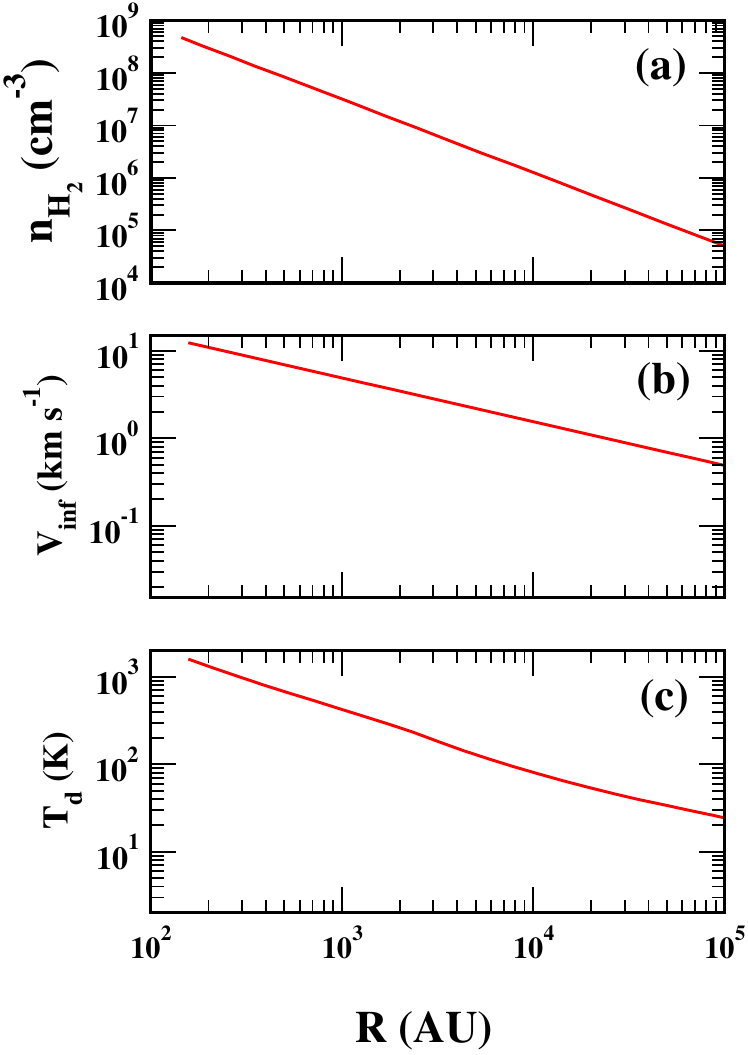}
  \caption{Spatial distribution of the physical parameters (a. H$_2$ density, b. infall velocity, and c. dust temperature) considered here are shown. The density and temperature variation are taken from \cite{vand13} and infall velocity variation is derived from the Equation \ref{eqn:infall} by using $v_{1000}= 4.9$ km s$^{-1}$. \citep[Courtesy:][]{bhat22}}
  \label{fig:physical_profiles}
\end{figure}

The cloud envelope is divided into $23$ grids. \cite{vand13} used the envelope beyond $156$ AU. This selection was made since \cite{osor09} found that the dust temperature was $\sim 1200$ K in this area. It represents the dust sublimation temperature. Therefore, the infalling envelope region between 156 AU and 119000 AU ($7.56 \times 10^{-4} - 5.769 \times 10^{-1}$ pc) has been taken into account in our model. A cartoon diagram of the modeled region is displayed in Figure \ref{fig:cartoon}. It shows a gas envelope infalling toward the main protostar. An outflow component is also present in the cloud. \cite{mars10} reported the presence of a cold foreground cloud in the line of sight of G31. However, the foreground clouds are moving at completely different speeds. Therefore, any foreground cloud component is not considered to be combined with this source. The grids considered to model the total cloud are logarithmically spaced. A density exponent ($p$) $\sim 1.40$ (obtained from the dust continuum model of \cite{vand13}) is used to define the density distribution of the envelope. Likewise, the temperature distribution with spatial variation is obtained from \cite{vand13}. Assuming gas and dust are closely connected, the kinetic gas and dust temperatures are considered the same. Figure \ref{fig:physical_profiles} displays the spatial distribution of the physical parameters.

\cite{belt18} used the following free-fall condition:
\begin{equation}
\label{eqn:infall}
v_{inf}(r)=v_{1000}\left(\frac{r}{1000 \ AU}\right)^{-0.5}.
\end{equation}
\noindent where v$_{1000}$ is the infall velocity at 1000 AU and r is the radius. The observed spectral energy distribution was fitted to this hot core by \cite{osor09}. Since most of the observations had a resolution of $>5"$ and the nearby HII region and central emitting region of G31 are separated by $\sim 5^{''}$, the emission from the HII region had contaminated the emission from the majority of the observations. The upper limit of several physical parameters was thus estimated from the SED fitting by \cite{osor09} by considering numerous data points using archival data. Except where otherwise noted, v$_{1000}$ $\sim 4.9$ km s$^{-1}$ is taken from \cite{osor09}. The rotation of the infalling gas envelope is not considered in our model.

 \section{Chemical modelling}
\label{sec:chemical}
\begin{figure}
\includegraphics[height=10.0cm]{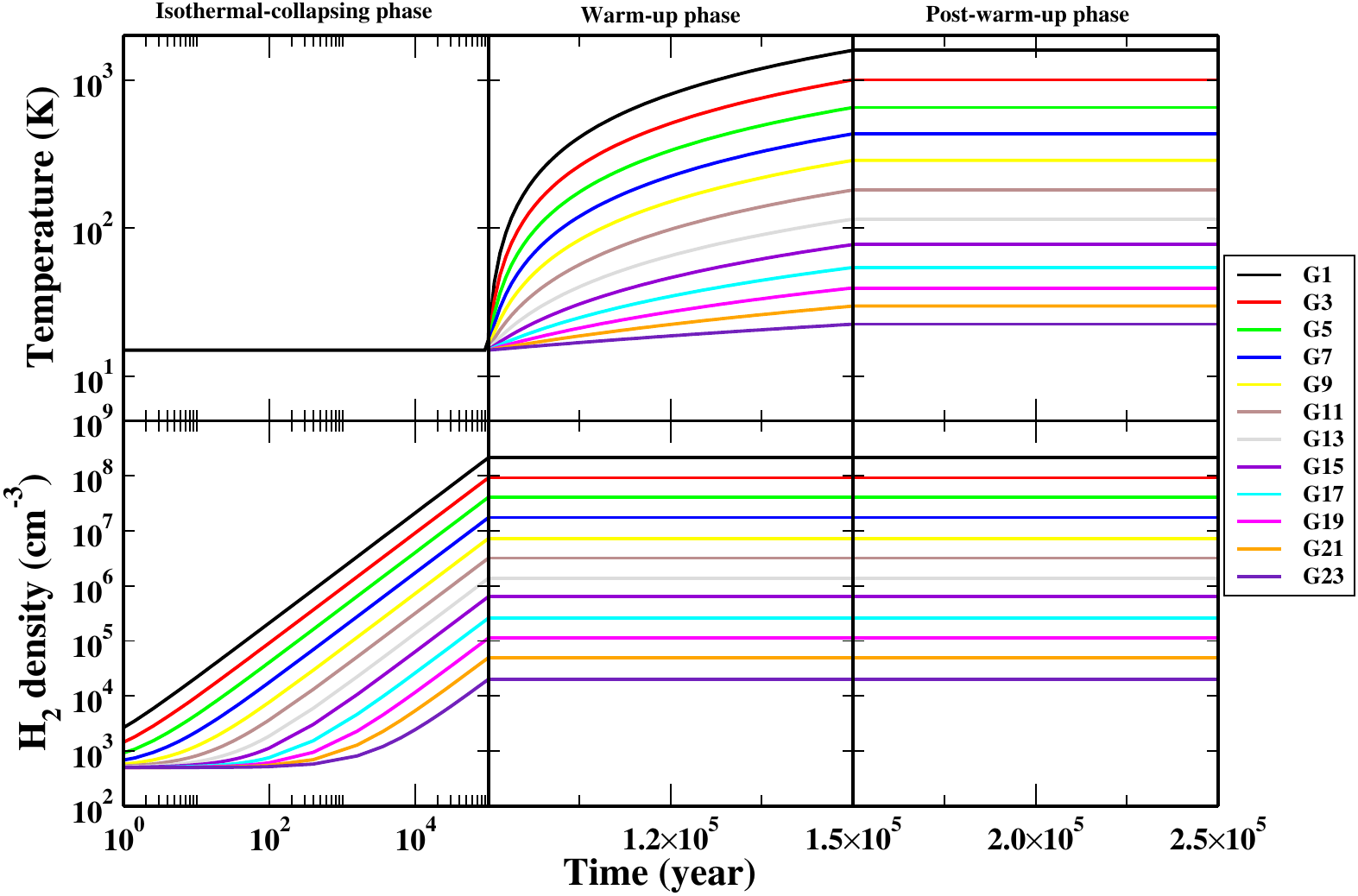}
\caption{Physical evolution considered in this simulation is shown. The cloud envelope is divided the cloud into $23$ spherical shells. The density and temperature evolution in the alternative shells (G1-G23) are shown for the representation. The entire simulation time scale is divided into three parts: The isothermal (gas and grain at $15$ K) and the collapsing phase, which extends up to $10^5$ years. The second phase corresponds to the warm-up time scale, whose span is for $5 \times 10^4$ years, and finally, the third phase corresponds to the post warmup phase, whose span is for $10^5$ years.\citep[Courtesy:][]{bhat22}}
\label{fig:physical_model}
\end{figure}

\begin{figure}
\includegraphics[height=16cm,width=10cm,angle=270]{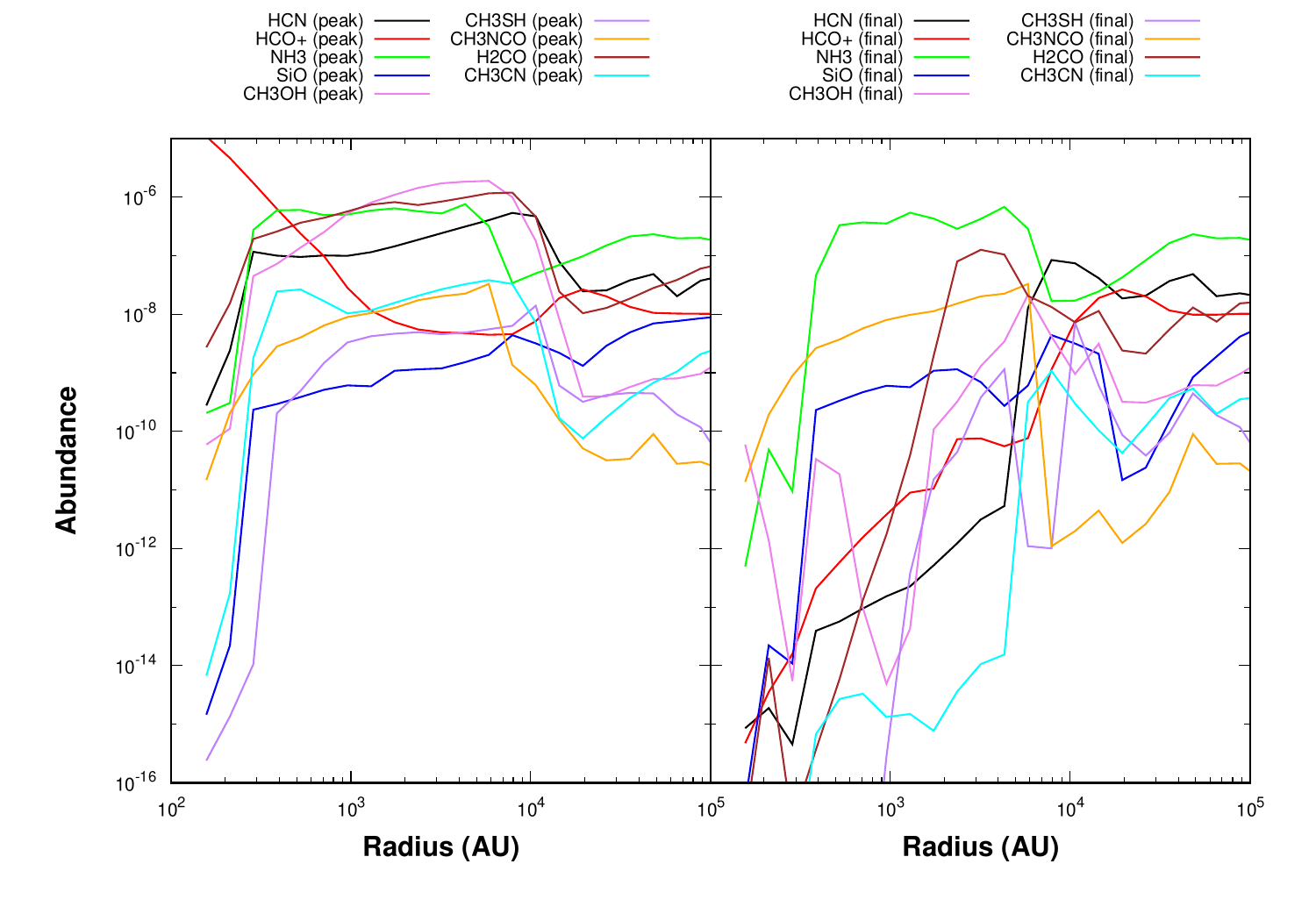}
\caption{The spatial distribution of the peak abundances (left panel) and final abundances (right panel) of some key interstellar species are shown. The peak abundances are taken beyond $10^5$ years, whereas the final abundances are taken at the end of the simulation time scale $\sim 2.5 \times 10^5$ years. \citep[Courtesy:][]{bhat22}}
\label{fig:abundance}
\end{figure}

The abundances of the species observed in G31 have been studied using complex chemical modeling.
This modeling uses our Chemical Model for Molecular Cloud (CMMC) code \citep{dasa15,das15a,das16,das21,gor17a,gor17b,gora20,sil18,sil21}. Most of the gas-phase pathways are taken from the UMIST database \citep{mcel13}. In contrast, the ice phase pathways and binding energies (BEs) of the surface species are taken from the KIDA database \citep{ruau16} and \citep{das18,sil17}.
Three key phases- the isothermal collapse phase, the warm-up phase, and the post-warm-up phase are considered while analyzing the time evolution of the physical parameters (H$_2$ density and temperature). The chemical evolution of hot cores \citep{gora20,sil21} is best studied using this kind of straightforward model. The gas cloud envelope is subdivided into 23 spherical shells, as described in Section \ref{sec:phys}. The shells in the first phase, often known as the isothermal phase, are maintained at a constant temperature of $\sim 15$ K. The $\rm{H_2}$ number density of the cloud is permitted to grow from $5 \times 10^2$ cm$^{-3}$ to a final value within this time period ($\sim 10^5$ years). The shorter collapse time scale represents the high-mass star formation scenario. The final densities of each shell are calculated based on the data shown in Figure \ref{fig:physical_profiles}. For instance, in this phase, the density of the outermost shell is permitted to evolve up to $4 \times 10^4$ cm$^{-3}$, whereas the density of the innermost shell is allowed to evolve up to $4 \times 10^8$ cm$^{-3}$. In the warm-up stage, all shells are permitted to evolve from their starting temperatures to their final temperatures in $5 \times 10^4$ years. The findings shown in Figure \ref{fig:physical_profiles} serve as a basis for determining the final temperature of each shell. For instance, it is assumed that the temperature will increase to $23$ K in the outermost shell and up to $1593$ K in the innermost shell.
The post-warmup phase begins after the warm-up phase is over. The shells are permitted to maintain the same temperature and density as in their previous stages. The post-warmup phase is estimated to last $10^5$ years. Our complete simulation time scale, which includes the period spent for isothermal phase, warm-up, and post-warmup is $2.5 \times 10^5$ years. In Figure \ref{fig:physical_model}, we can see how our three-phase model's density and temperature change. Figure \ref{fig:abundance} displays the estimated abundances of various chemical species with respect to H$_2$ molecules. The peak abundance achieved beyond the collapse time scale is shown in the left panel of Figure \ref{fig:abundance}, while the abundances obtained at the completion of the simulation time scale ($\sim 2.5 \times 10^5$ years) are shown in the right panel.
The final and peak abundances of these species are reported at various radii in Table \ref{table:abundances} for easier understanding. Notably, abundances of complex organic molecules, particularly CH$_3$SH and CH$_3$NCO, reach their maximum abundance at the end of the simulation around the center of the cloud, where the temperature is higher. During the intermediate stages, peak abundance is taken for the remaining molecules (NH$_3$, HCO$^+$, HCN, and SiO). 

\begin{landscape}
\begin{table}
\tiny \setlength{\tabcolsep}{2pt}
\caption{Final and peak abundances at different radius obtained from our chemical model. The peak values are considered beyond the collapsing time scale. So our time uncertainty in the peak abundances is $1.5 \times 10^5$ years. \citep[Courtesy:][]{bhat22}} \label{table:abundances}
\hskip -4.5cm
\begin{tabular}{|c|c|c|c|c|c|c|c|c|c|c|c|c|c|c|c|c|c|c|}
\hline
\hline
\multicolumn{3}{|c|}{HCO$^+$}&\multicolumn{2}{|c|}{HCN}&\multicolumn{2}{|c|}{SiO}&\multicolumn{2}{|c|}{NH$_3$}&\multicolumn{2}{|c|}{CH$_3$OH}&\multicolumn{2}{|c|}{CH$_3$SH}&\multicolumn{2}{|c|}{CH$_3$NCO}&\multicolumn{2}{|c|}{CH$_3$CN}&\multicolumn{2}{|c|}{H$_2$CO}\\
\hline
\hline
Radius (AU)&Peak value &Final value&Peak value&Final value&Peak value&Final value&Peak value&Final value&Peak value&Final value&Peak value&Final value&Peak value&Final value&Peak value&Final value&Peak value&Final value\\
\hline
\hline
1.56$\times10^{2}$&1.08$\times10^{-5}$ &4.76$\times10^{-16}$ &2.76$\times10^{-10}$&8.58$\times10^{-16}$&1.46$\times10^{-15}$ &4.66$\times10^{-17}$ &2.05$\times10^{-10}$ &4.91$\times10^{-13}$ &5.93$\times10^{-11}$&5.93 $\times10^{-11}$&2.41$\times10^{-16}$&8.45$\times10^{-23}$ &1.47$\times10^{-11}$ &1.37$\times10^{-11}$ &6.72$\times10^{-15}$&5.77$\times10^{-18}$&2.71$\times10^{-9}$&1.88$\times10^{-17}$\\
2.12$\times10^{2}$&4.62$\times10^{-6}$&3.56$\times10^{-15}$&2.39$\times10^{-9}$&1.88$\times10^{-15}$&2.21$\times10^{-14}$ &2.21$\times10^{-14}$ &3.04$\times10^{-10}$ &4.83$\times10^{-11}$ &1.10$\times10^{-10}$&1.38$\times10^{-12}$ &1.35$\times10^{-15}$&3.10$\times10^{-25}$ &2.06$\times10^{-10}$ &1.93$\times10^{-10}$ &1.73$\times10^{-13}$&2.29$\times10^{-17}$&1.52$\times10^{-8}$&1.36$\times10^{-14}$\\
2.87$\times10^{2}$&1.74$\times10^{-6}$&1.58$\times10^{-14}$&1.15$\times10^{-7}$&4.54$\times10^{-16}$&2.33$\times10^{-10}$ &1.08$\times10^{-14}$ &2.73$\times10^{-7}$ &9.50$\times10^{-12}$ &4.44$\times10^{-8}$& 5.47$\times10^{-15}$&1.06$\times10^{-14}$&4.14$\times10^{-39}$ &9.35$\times10^{-10}$ &8.83$\times10^{-10}$ &1.79$\times10^{-9}$&7.76$\times10^{-19}$&1.91$\times10^{-7}$&1.98$\times10^{-17}$\\
3.88$\times10^{2}$&6.24$\times10^{-7}$&2.07$\times10^{-13}$&9.96$\times10^{-8}$ & 3.94$\times10^{-14}$&2.92$\times10^{-10}$ &2.30$\times10^{-10}$ &5.89$\times10^{-7}$ &4.57 $\times10^{-8}$&7.18$\times10^{-8}$&3.37$\times10^{-11}$ &2.03$\times10^{-10}$&9.59$\times10^{-34}$ &2.81$\times10^{-9}$ &2.62$\times10^{-9}$ &2.44$\times10^{-8}$&6.73$\times10^{-16}$&2.57$\times10^{-7}$&3.67$\times10^{-16}$\\
5.24$\times10^{2}$&2.37$\times10^{-7}$&5.83$\times10^{-13}$&9.47$\times10^{-8}$ &5.67$\times10^{-14}$&3.84$\times10^{-10}$ &3.31$\times10^{-10}$ &5.98$\times10^{-7}$ &3.30$\times10^{-7}$ &1.37$\times10^{-7}$&1.85$\times10^{-11}$ &4.89$\times10^{-10}$&1.18$\times10^{-31}$ &3.99$\times10^{-9}$ &3.69$\times10^{-9}$ &2.64$\times10^{-8}$&2.70$\times10^{-15}$&3.62$\times10^{-7}$&5.88$\times10^{-15}$\\
7.09$\times10^{2}$&9.71$\times10^{-8}$&1.55$\times10^{-12}$&1.00$\times10^{-7}$ &9.45$\times10^{-14}$&5.10$\times10^{-10}$ &4.67$\times10^{-10}$ &4.90$\times10^{-7}$ &3.68$\times10^{-7}$ &2.52$\times10^{-7}$&9.69$\times10^{-14}$ &1.47$\times10^{-9}$&3.58 $\times10^{-20}$&6.41$\times10^{-9}$ &5.71$\times10^{-9}$ &1.67$\times10^{-8}$&3.34$\times10^{-15}$&4.39$\times10^{-7}$&1.31$\times10^{-13}$\\
9.58$\times10^{2}$&2.81$\times10^{-8}$&3.79$\times10^{-12}$&9.85$\times10^{-8}$ &1.52$\times10^{-13}$ &6.07$\times10^{-10}$ &5.97$\times10^{-10}$ &5.02$\times10^{-7}$ &3.51$\times10^{-7}$ &5.21$\times10^{-7}$&4.90$\times10^{-15}$ &3.28$\times10^{-9}$&2.82$\times10^{-16}$ &8.92$\times10^{-9}$ &7.96$\times10^{-9}$ &1.03$\times10^{-8}$&1.33$\times10^{-15}$&5.67$\times10^{-7}$&1.75$\times10^{-12}$\\
1.29$\times10^{3}$&1.15$\times10^{-8}$&8.98$\times10^{-12}$&1.14$\times10^{-7}$ &2.26$\times10^{-13}$ &5.84$\times10^{-10}$ &5.65$\times10^{-10}$ &5.86$\times10^{-7}$ &5.37$\times10^{-7}$ &7.96$\times10^{-7}$&4.25$\times10^{-14}$ &4.19$\times10^{-9}$&3.73$\times10^{-13}$ &1.05$\times10^{-8}$ &9.72$\times10^{-9}$ &1.16$\times10^{-8}$&1.50$\times10^{-15}$&7.36$\times10^{-7}$&3.87$\times10^{-11}$\\
1.75$\times10^{3}$&7.29$\times10^{-9}$&1.05$\times10^{-11}$&1.44$\times10^{-7}$ &5.14$\times10^{-13}$ &1.08$\times10^{-9}$ &1.08$\times10^{-9}$ &6.41$\times10^{-7}$ &4.26$\times10^{-7}$ &1.09$\times10^{-6}$&1.07$\times10^{-10}$ &4.65$\times10^{-9}$&1.51$\times10^{-11}$ &1.28$\times10^{-8}$ &1.12$\times10^{-8}$ &1.56$\times10^{-8}$&7.76$\times10^{-16}$&8.16$\times10^{-7}$&1.93$\times10^{-9}$\\
2.37$\times10^{3}$&5.45$\times10^{-9}$& 7.33$\times10^{-11}$&1.85$\times10^{-7}$ &1.23$\times10^{-12}$ &1.14$\times10^{-9}$ &1.14$\times10^{-9}$ &5.68$\times10^{-7}$ &2.85$\times10^{-7}$ &1.42$\times10^{-6}$&3.20$\times10^{-10}$ &4.90$\times10^{-9}$&4.42$\times10^{-11}$ &1.72$\times10^{-8}$ &1.51$\times10^{-8}$ &2.07$\times10^{-8}$&3.64$\times10^{-15}$&7.26$\times10^{-7}$&7.92$\times10^{-8}$\\
3.20$\times10^{3}$&4.88$\times10^{-9}$&7.55$\times10^{-11}$&2.41$\times10^{-7}$ &3.11$\times10^{-12}$ &1.19$\times10^{-9}$ &6.92$\times10^{-10}$ &5.21$\times10^{-7}$ &4.22$\times10^{-7}$ &1.71$\times10^{-6}$&1.29 $\times10^{-9}$&4.54$\times10^{-9}$&3.78$\times10^{-10}$ &2.02$\times10^{-8}$ &2.00$\times10^{-8}$ &2.65$\times10^{-8}$&1.06$\times10^{-14}$&8.33$\times10^{-7}$&1.25$\times10^{-7}$\\
4.32$\times10^{3}$&4.75$\times10^{-9}$& 5.58$\times10^{-11}$&3.11$\times10^{-7}$&5.32$\times10^{-12}$ &1.51$\times10^{-9}$ &2.71$\times10^{-10}$ &7.53$\times10^{-7}$ &6.77$\times10^{-7}$ &1.82$\times10^{-6}$&3.39$\times10^{-9}$ &4.84$\times10^{-9}$&1.14$\times10^{-9}$ &2.24$\times10^{-8}$ &2.24$\times10^{-8}$ &3.25$\times10^{-8}$&1.55$\times10^{-14}$&9.73$\times10^{-7}$&1.04$\times10^{-7}$\\
5.84$\times10^{3}$&4.43$\times10^{-9}$& 7.64$\times10^{-11}$&4.02$\times10^{-7}$ &1.27$\times10^{-8}$& 2.01$\times10^{-9}$&6.05$\times10^{-10}$ &3.18$\times10^{-7}$ &2.87$\times10^{-7}$ &1.88$\times10^{-6}$&2.13 $\times10^{-8}$&5.49$\times10^{-9}$&1.09$\times10^{-12}$ &3.27$\times10^{-8}$ &3.27$\times10^{-8}$ &3.81$\times10^{-8}$&3.16$\times10^{-10}$&1.15$\times10^{-6}$&2.05$\times10^{-8}$\\
7.90$\times10^{3}$&4.53$\times10^{-9}$& 1.14$\times10^{-9}$& 5.35$\times10^{-7}$&8.35$\times10^{-8}$&4.39$\times10^{-9}$ &4.39$\times10^{-9}$ &3.40$\times10^{-8}$ &1.67$\times10^{-8}$ &9.92$\times10^{-7}$&4.16$\times10^{-9}$ &6.32$\times10^{-9}$&1.00$\times10^{-12}$ &1.36$\times10^{-9}$ &1.10$\times10^{-12}$ &3.27$\times10^{-8}$&1.07$\times10^{-9}$&1.18$\times10^{-6}$&1.32$\times10^{-8}$\\
1.07$\times10^{4}$& 7.56$\times10^{-9}$& 7.50$\times10^{-9}$& 4.66$\times10^{-7}$&7.36$\times10^{-8}$& 3.17$\times10^{-9}$&3.14$\times10^{-9}$ &4.96$\times10^{-8}$ &1.70$\times10^{-8}$ &1.78$\times10^{-7}$&9.47$\times10^{-10}$ &1.39$\times10^{-8}$&7.15$\times10^{-9}$ &6.16$\times10^{-10}$ &1.99$\times10^{-12}$ &7.13$\times10^{-9}$&2.96$\times10^{-10}$&4.61$\times10^{-7}$&7.21$\times10^{-9}$\\
1.44$\times10^{4}$& 1.87$\times10^{-8}$& 1.87$\times10^{-8}$&7.82$\times10^{-8}$ &4.11$\times10^{-8}$ &2.18$\times10^{-9}$ &2.10$\times10^{-9}$ &6.89$\times10^{-8}$ &2.44$\times10^{-8}$ &8.21$\times10^{-9}$&3.13 $\times10^{-9}$&6.02$\times10^{-10}$&6.02$\times10^{-10}$ &1.57$\times10^{-10}$ &4.44$\times10^{-12}$ &1.67$\times10^{-10}$&1.04$\times10^{-10}$&2.41$\times10^{-8}$&1.13$\times10^{-8}$\\
1.95$\times10^{4}$&2.63$\times10^{-8}$&2.63$\times10^{-8}$& 2.45$\times10^{-8}$&1.85$\times10^{-8}$& 1.31$\times10^{-9}$&1.47$\times10^{-11}$ &9.73$\times10^{-8}$ &4.24$\times10^{-8}$ &3.91$\times10^{-10}$&3.19 $\times10^{-10}$&3.18$\times10^{-10}$&8.67$\times10^{-11}$ &5.09$\times10^{-11}$ &1.25$\times10^{-12}$ &7.55$\times10^{-11}$&4.26$\times10^{-11}$&1.04$\times10^{-8}$&2.39$\times10^{-9}$\\
2.64$\times10^{4}$&2.01$\times10^{-8}$&2.01$\times10^{-8}$&2.55$\times10^{-8}$ &2.07$\times10^{-8}$&2.88$\times10^{-9}$ &2.40$\times10^{-11}$ &1.48$\times10^{-7}$ &8.30$\times10^{-8}$ &3.94$\times10^{-10}$&3.08$\times10^{-10}$ &4.15$\times10^{-10}$&3.85 $\times10^{-11}$& 3.19$\times10^{-11}$&2.63 $\times10^{-12}$&1.72$\times10^{-10}$&1.23$\times10^{-10}$&1.28$\times10^{-8}$&2.12$\times10^{-9}$\\
3.56$\times10^{4}$&1.33$\times10^{-8}$& 1.15$\times10^{-8}$&3.78$\times10^{-8}$ &3.68$\times10^{-8}$ &4.86$\times10^{-9}$ &1.48$\times10^{-10}$ &2.12$\times10^{-7}$ &1.63$\times10^{-7}$ &5.71$\times10^{-10}$&4.12 $\times10^{-10}$&4.53$\times10^{-10}$& 9.38$\times10^{-11}$& 3.40$\times10^{-11}$&9.18$\times10^{-12}$ &3.67$\times10^{-10}$&3.67$\times10^{-10}$&1.85$\times10^{-8}$&5.46$\times10^{-9}$\\
4.82$\times10^{4}$&1.05$\times10^{-8}$&9.78$\times10^{-9}$&4.83$\times10^{-8}$ &4.83$\times10^{-8}$ &6.94$\times10^{-9}$ &8.37$\times10^{-10}$ &2.30$\times10^{-7}$ &2.30$\times10^{-7}$ &7.83$\times10^{-10}$&6.10$\times10^{-10}$ &4.43$\times10^{-10}$& 4.43$\times10^{-10}$& 9.02$\times10^{-11}$&9.02$\times10^{-11}$ &6.77$\times10^{-10}$&5.36$\times10^{-10}$&2.79$\times10^{-8}$&1.29$\times10^{-8}$\\
6.51$\times10^{4}$&1.02$\times10^{-8}$&9.81$\times10^{-9}$&2.02$\times10^{-8}$ &2.02$\times10^{-8}$ &7.62$\times10^{-9}$ &1.88$\times10^{-9}$ &1.97$\times10^{-7}$ &1.97$\times10^{-7}$ &7.97$\times10^{-10}$&6.02$\times10^{-10}$ &1.94$\times10^{-10}$&1.88$\times10^{-10}$ &2.79$\times10^{-11}$ &2.79$\times10^{-11}$ &1.05$\times10^{-9}$&1.99$\times10^{-10}$&3.84$\times10^{-8}$&7.42$\times10^{-9}$\\
8.80$\times10^{4}$& 1.02$\times10^{-8}$& 1.02$\times10^{-8}$&3.72$\times10^{-8}$ &2.26$\times10^{-8}$ &8.49$\times10^{-9}$ &4.12$\times10^{-9}$ &1.99$\times10^{-7}$ &1.99$\times10^{-7}$ &9.53$\times10^{-10}$&9.53$\times10^{-10}$ &1.17$\times10^{-10}$&1.17$\times10^{-10}$ &3.02$\times10^{-11}$ &2.84$\times10^{-11}$ &2.07$\times10^{-9}$&3.51$\times10^{-10}$&5.99$\times10^{-8}$&1.52$\times10^{-8}$\\
1.19$\times10^{5}$&1.01$\times10^{-8}$&1.01$\times10^{-8}$&4.57$\times10^{-8}$ &1.93$\times10^{-8}$ &9.22$\times10^{-9}$ &6.32$\times10^{-9}$ &1.69$\times10^{-7}$ &1.69$\times10^{-7}$ &1.73$\times10^{-9}$&1.67$\times10^{-9}$ &2.83$\times10^{-11}$&2.83$\times10^{-11}$ &2.19$\times10^{-11}$ &1.38$\times10^{-11}$ &2.83$\times10^{-9}$&3.97$\times10^{-10}$&7.35$\times10^{-8}$&1.63$\times10^{-8}$\\
\hline
\hline
\end{tabular}
\end{table}
\end{landscape}

\section{Radiative transfer model}
The identification of multiple COMs (CH$_3$SH, CH$_3$NCO, and CH$_3$OH) in G31 was published by \cite{gora21}. They also demonstrated that H$^{13}$CO$^+$, HCN, and SiO have particular unique spectral characteristics. Here, the 1D RATRAN programme \citep{hoge00} and CASSIS (created by IRAP-UPS/CNRS, \cite{vast15}, \url{http://cassis.irap.omp.eu}) and the spectroscopic database 'Cologne Database for Molecular Spectroscopy' \citep[CDMS,][]{mull01,mull05,endr16} \url{(https://cdms.astro.uni-koeln.de)} are used. Also the Jet Propulsion Laboratory (JPL \citep{pick98} \url{(http://spec.jpl.nasa.gov/)} database are used to simulate the observed line profiles.
Markov Chain Monte Carlo Method is not employed since H$^{13}$CO$^+$, HCN, and SiO exhibit some distinctive line profiles and very few transitions are found. The peculiar properties of H$^{13}$CO$^+$ (inverse P-Cygni profile, a representation of the infalling envelope) are explained by the two slab model of \cite{myer96} and further improved by \cite{difr01}. The infalling gas envelope is taken into account in this model in two different, spherically symmetric areas with different excitation temperatures. It sets two parallel slabs with varying excitation temperatures along the line of sight. The excitation temperature is $T_f$ in the front layer and $T_r$ in the rear layer, respectively. It takes into account the region's maximum optical depth, $\tau_0$. The infalling envelope takes into account an infall velocity, $V _{in}$.
The continuum temperature ($T_c$), which corresponds to the peak flux density of the continuum image. The symbol $\Phi$ stands for this beam filling factor. The core is regarded as a black body, while the rear layer is irradiated by background radiation with temperature ($T_b$ = $2.7$ K). The radiative transfer solution was defined in terms of $T_f$, $T_r$, $T_b$, and T$_B$ in \cite{myer96} (brightness temperature). Later, \cite{difr01} added the continuum source into consideration.
The two slab model has two main drawbacks: it is very simple and has a lot of free variables. However, this model works well for reproducing the line profile representing infalling envelope. Based on these two slab model, the CASSIS software is used to fit the observed spectral signature.
All of the observed line profiles are explained using the 1D RATRAN model. Table \ref{table:RATRAN-input-parameters} is a list of the many input parameters used for RATRAN modelling, including the inner and outer radius of the envelope, dust opacity, distance from the Sun, dust to gas mass ratio, background radiation, and bolometric temperature. Assuming a power-law emissivity model, we investigate the dust emissivity ($\kappa$). The dust emissivity power law index $\beta$ can range from $1$ to $1.4$.

\begin{table}
{\scriptsize
 \caption{Key parameters used for our 1D-RATRAN modeling. \citep[Courtesy:][]{bhat22} \label{table:RATRAN-input-parameters}}
 \hskip -2.0cm
\begin{tabular}{|l|l|}
  \hline
  \hline
  Input parameters & Used \\
  \hline
  \hline
 Inner radius of the envelope& $156$ AU\\
Outer radius of envelope & $119000$ AU\\
T$_{cmb}$ & $2.73$ K\\
Gas to dust mass ratio& $100$\\
 $\kappa$ (dust emissivity) & $\kappa=\kappa_0(\frac{\nu}{\nu_0})^\beta$, where $\kappa_0= 19.8$ cm$^2$/gm \citep{osse94},
 $\nu_0= 6 \times 10^{11}$ Hz,
 and $\beta=1 - 1.4$ \\
Distance & $3700$ pc\\
Bolometric temperature & $55$ K \citep{muel02}\\
  \hline
  \hline
 \end{tabular}
}
 \end{table}

It has been observed that the line profiles can be significantly affected by the velocity components present in a collapsing gas envelope. In RATRAN, the thermal broadening of lines is taken into account automatically. Infall motion, expansion, and non-thermal turbulent motion are additional elements that impact the line profile. Our model include these contributions into the calculation. To evaluate how turbulence present in the envelope affects the line profile, here we consider the Doppler '$b$' parameter (characterizes the spectral line width) is taken to be a constant across the envelope. This parameter relates to line broadening and is directly related to the FWHM by, 
$$
\frac{b}{FWHM}=\frac{1}{2\times\sqrt{ln(2)}}=0.60.
$$
In our RATRAN model, the entire cloud region is considered with a constant linewidth. However, this parameter would radially vary in reality. Therefore, the line intensity would be reduced by increasing this value while the other way around for the reverse case.
\section{Results and discussions}\label{sec:result-discussions}
\subsection{Two slab model}\label{sec:twoslab}
A red-shifted absorption lobe and a blue-shifted emission lobe can be seen in an inverse P-Cygni profile. The observation of $\rm{H^{13}CO^+}$ at G31 indicates to an infalling gas envelope toward the source center \citep{gora21}. Here, the two-slab model, proposed by \cite{myer96} and then further modified by \cite{difr01} is used to explain the emission and absorption nature of the inverse P-Cygni profile.
The best-fitted parameters are presented in Table \ref{table:H13CO+_fit} along with the input variables used to fit the inverse P-Cygni profile. The value of $\chi^2$ is minimized to obtain these parameters. A popular and effective tracer of the envelope region is $\rm{HCO^+}$. An inverse P-Cygni profile for the $1-0$ transition of $\rm{H^{13}CO^+}$ in G31 was recently published by \cite{gora21}. They had concluded from their study of this profile that there was substantial evidence of an infalling gas envelope toward the center of G31.
The dynamic behavior of the infalling envelope (infall velocity, mass infall rate, velocity dispersion, etc.) is easily extracted using inverse P-Cygni line profiles. In this case, a two-slab model explains the observed inverse P-Cygni profile of H$^{13}$CO$^+$. There are a total of eight parameters in the solution. There may be many best-fit solutions as a result of this freedom. Some parameters are fixed to a realistic estimation to prevent any misleading results. The best-fitted parameters are listed in Table \ref{table:H13CO+_fit}.

\begin{table}
{\scriptsize
 \caption{Best fitted parameters obtained with the two slab model for $\rm{H^{13}CO^+}$. \citep[Courtesy:][]{bhat22}\label{table:H13CO+_fit}}
 \hskip 2.0cm
   \begin{tabular}{|l|l|l|}
  \hline
  \hline
  Input parameters & Range of values used as input & Best fitted parameters \\
  \hline
  \hline
 $T_r$ (K)& 10.0 - 200.0& 29.0\\
$\tau_0$ & 0.1 - 11.0& 1.19\\
$V_{in}$ (km s$^{-1}$) & 1.0 - 4.0 & 2.50 \\
$\sigma_v$ (km s$^{-1}$)& 0.1 - 3.0 & 0.97\\
$v_{lsr}$ (km s$^{-1}$) & 96.0 - 97.5 & 96.5 \\
$T_f$ (K) & 10.0 - 10.0 & 10.0\\
$T_c$ (K)& 36.0 & 36.0\\
$\Phi$ & 0.5 - 0.9 & 0.62\\
  \hline
  \hline
 \end{tabular}}
 \end{table}

 \begin{figure}
 \centering
\includegraphics[height=7cm,width=8cm]{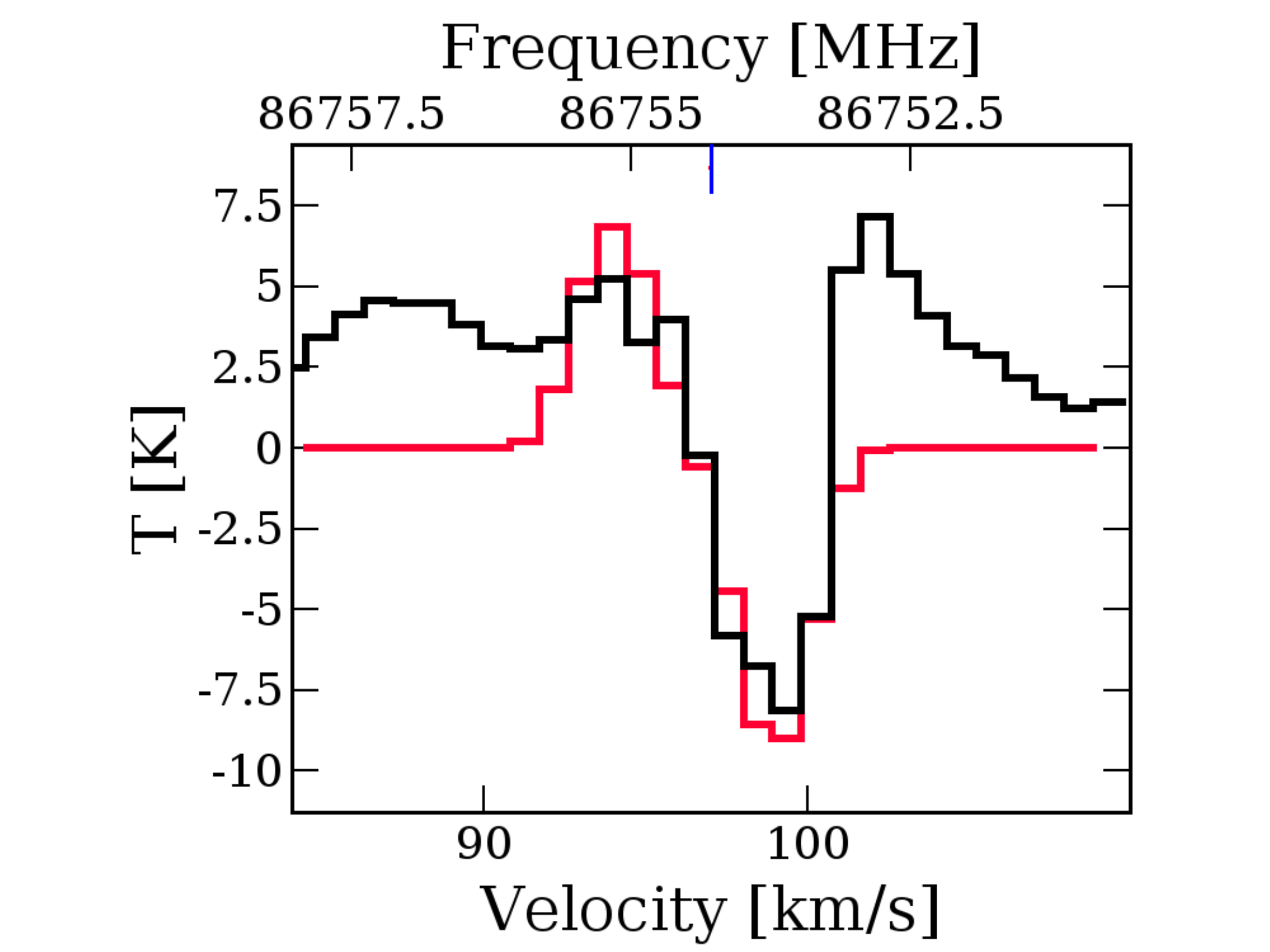}
\caption{Fitted inverse P-Cygni profile using two slab model is shown. Black line is the observed spectrum, whereas red line is the fitted spectrum. \citep[Courtesy:][]{bhat22}}
\label{fig:inP-cygni}
\end{figure}
 
From the best-fitted infall velocity, the mass infall rate is calculated using a straightforward approach given in \cite{klaa07}. Assuming the source is collapsing as a spherically symmetric collapse, \cite{klaa07} proposed the following relation:
 \begin{equation}
\dot{M}=\frac{dM}{dt}\simeq\frac{M}{t}=\frac{\rho V v_{in}}{r_{gm}}=\frac{4}{3}\pi n_{H_2}\mu m_H r^{2}_{gm} v_{in},
\end{equation}
  where $\rm{v_{in}}$ denotes the infall velocity, $\rm{m_H}$ denotes the mass of a hydrogen atom, $n_{H_2}$ denotes the ambient source density ($5.4 \times 10^6$ cm$^{-3}$), $\mu$ denotes the mean molecular weight of the gas ($2.35$), and $r_{gm}$ denotes the geometric mean radius of the emitting region ($\rm{H^{13}CO^+}$) derived from the equation $r_{gm}=\sqrt{f} r_{beam}$. The beam filling factor in this case is $f$ (0.62 from Table \ref{table:H13CO+_fit}) and the beam radius is r$_{beam}$ (1.1$^{''}$ from \cite{gora21}). A best-fit infall velocity of $\rm{H^{13}CO^+}$ $\sim 2.5$ km s$^{-1}$ is obtained in this case. This value is utilized in the equation above to calculate the mass infall rate, which is $\sim 1.3 \times 10^{-3}$ M$_\odot$ yr$^{-1}$ for the 3.7 kpc distance and $\sim 5.9 \times 10^{-3}$ M$_\odot$ yr$^{-1}$ for the 7.9 kpc distance.
 These results are consistent with the earlier prediction made by \cite{osor09} for their best-fitted model with a core mass of 25M$_\odot$, which was $\sim 3 \times 10^{-3}$ M$_\odot$ yr$^{-1}$. According to the infall rate, the source is in a high accretion phase. The reported inverse P-Cygni profile of \cite{gora21} and the best-fitted spectra are displayed in Figure \ref{fig:inP-cygni}.
 
 \subsection{1D spherically symmetric RATRAN model}
 \label{sec:RATRAN}
 The physical characteristics of the envelope and surrounding medium are further modeled using the RATRAN code using the reported inverse P-Cygni profile ($\rm{H^{13}CO^+}$) and the feature of SiO, HCN, and NH$_3$. Additionally, this code also models the transitions of COMs.
 
 \subsubsection{\rm{\bf{H$^{13}$CO$^+$}} \label{sec:hco+}}
\begin{figure}
\centering
\includegraphics[height=6cm]{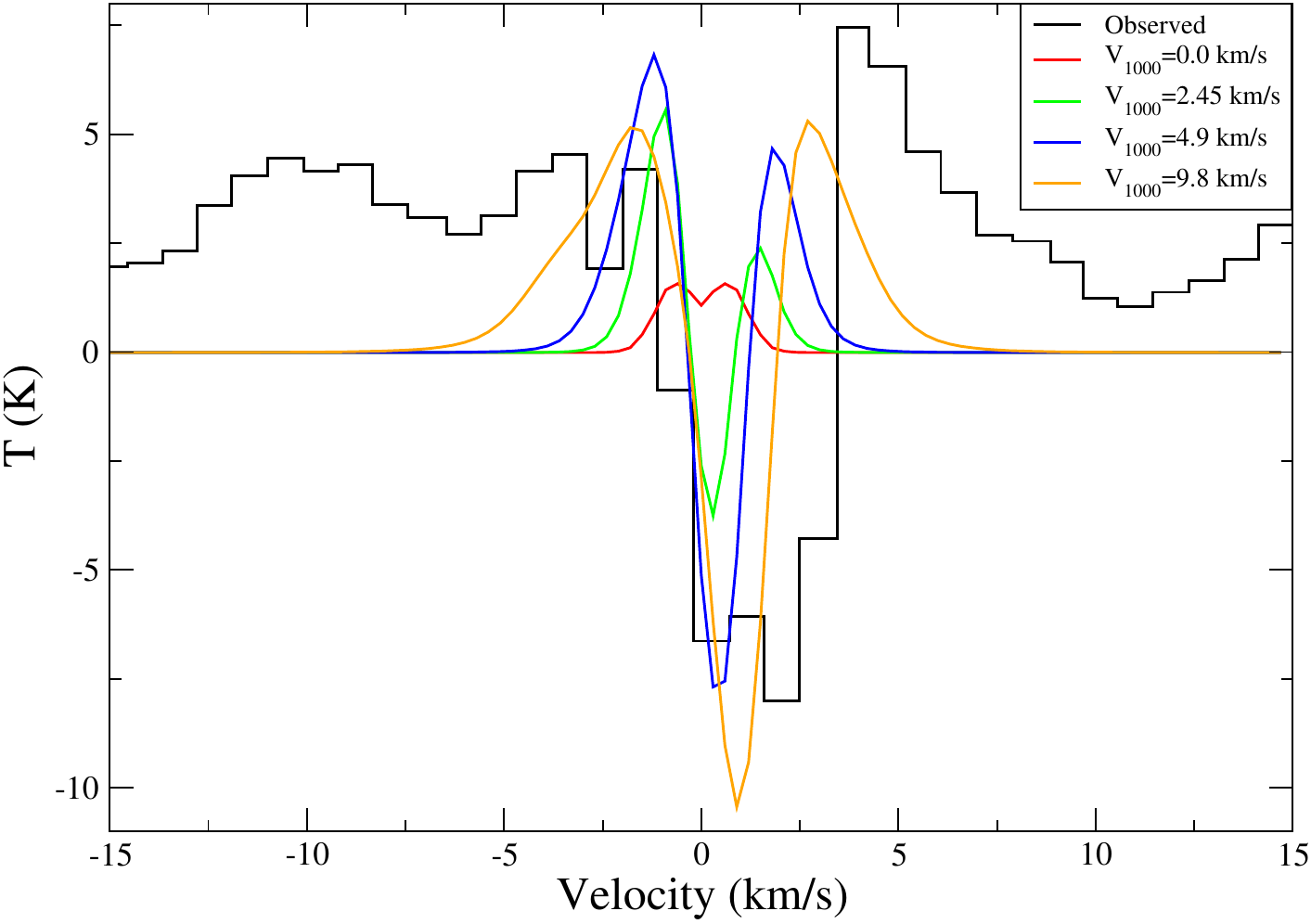}
\caption{A comparison between the observed line profile (black line) and modeled line profile with a constant abundance of H$^{13}$CO$^+$ ($7.07 \times 10^{-11}$) is shown. The best fit is obtained when $\beta=1$ is used. Different cases with the infall velocity are shown. It is evident that with the static envelope (red line), there is no asymmetry, and asymmetry increases with the increase in velocity. The best fit is obtained when an infall velocity of $4.9$ km/s at 1000AU is used. \citep[Courtesy:][]{bhat22}}
\label{fig:h13co+_best}
\end{figure}

\begin{figure}
\centering
\includegraphics[height=6cm]{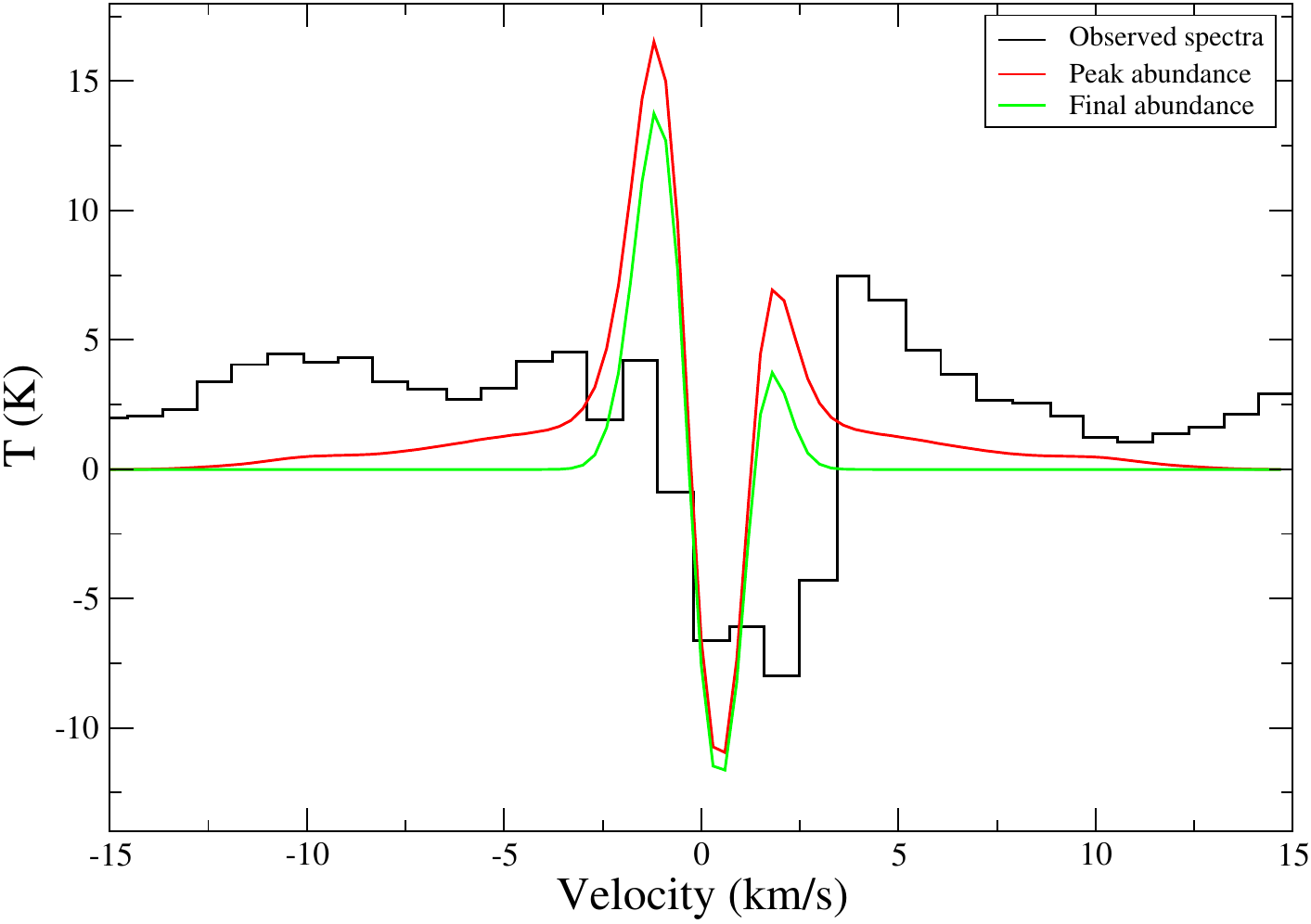}
\caption{ A comparison between the observed line and modeled line with the abundances obtained from our chemical modeling is shown. The abundance profile of HCO$^+$ obtained from our chemical modeling is used. Here, this abundance profile is further reduced by a factor of 65 to have the peak and final abundance profile of H$^{13}$CO$^+$. The red line shows the modeled line profile with the peak value, whereas the green line shows the modeled line profile with the final value obtained from our chemical model. \citep[Courtesy:][]{bhat22}}
\label{fig:h13co+_chemmodel}
\end{figure}

\begin{figure}
  \centering
\begin{minipage}{0.42\textwidth}
    \includegraphics[height=7cm,width=9cm]{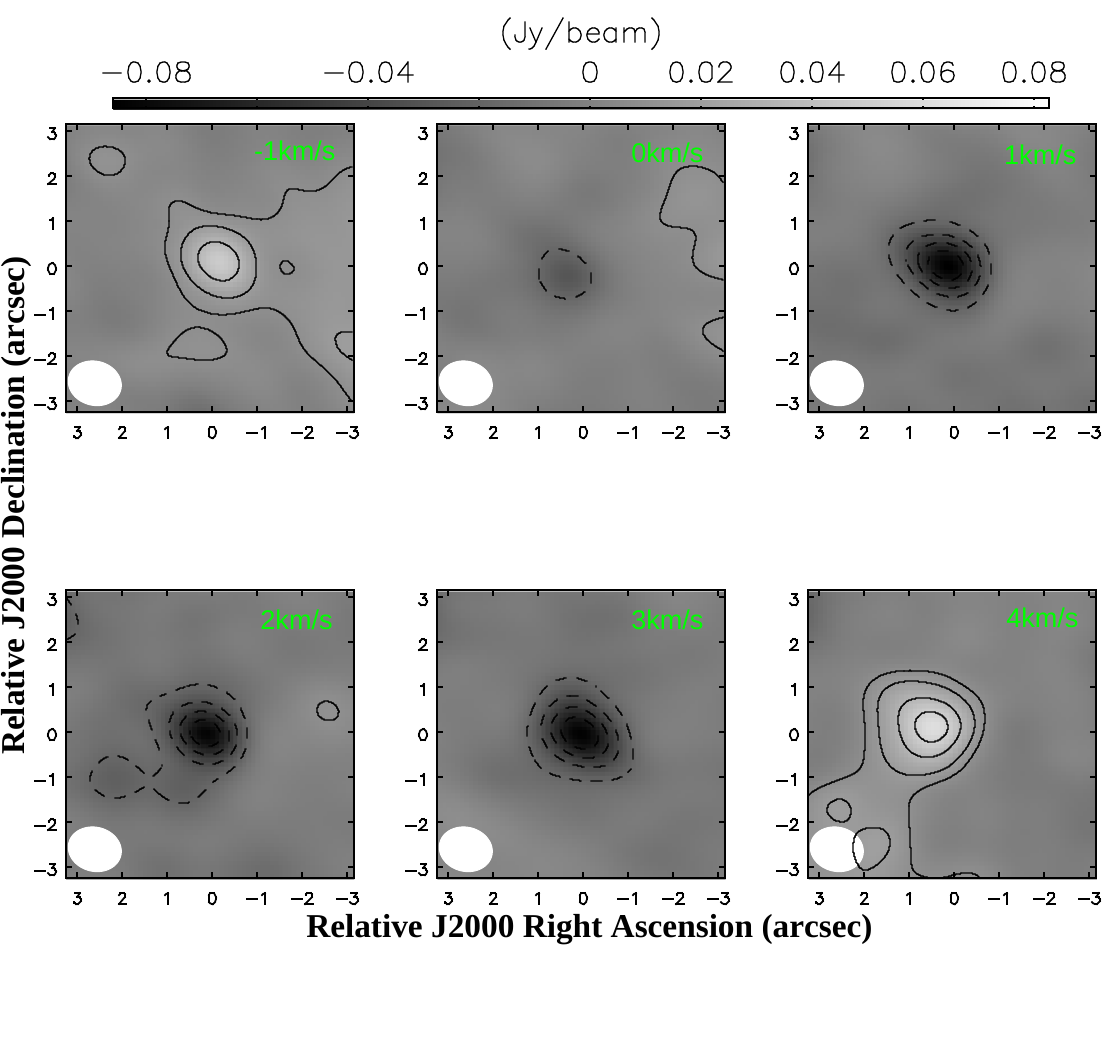}
  \end{minipage}
\hskip 2.5cm
\begin{minipage}{0.42\textwidth}
    \includegraphics[height=7cm,width=9cm]{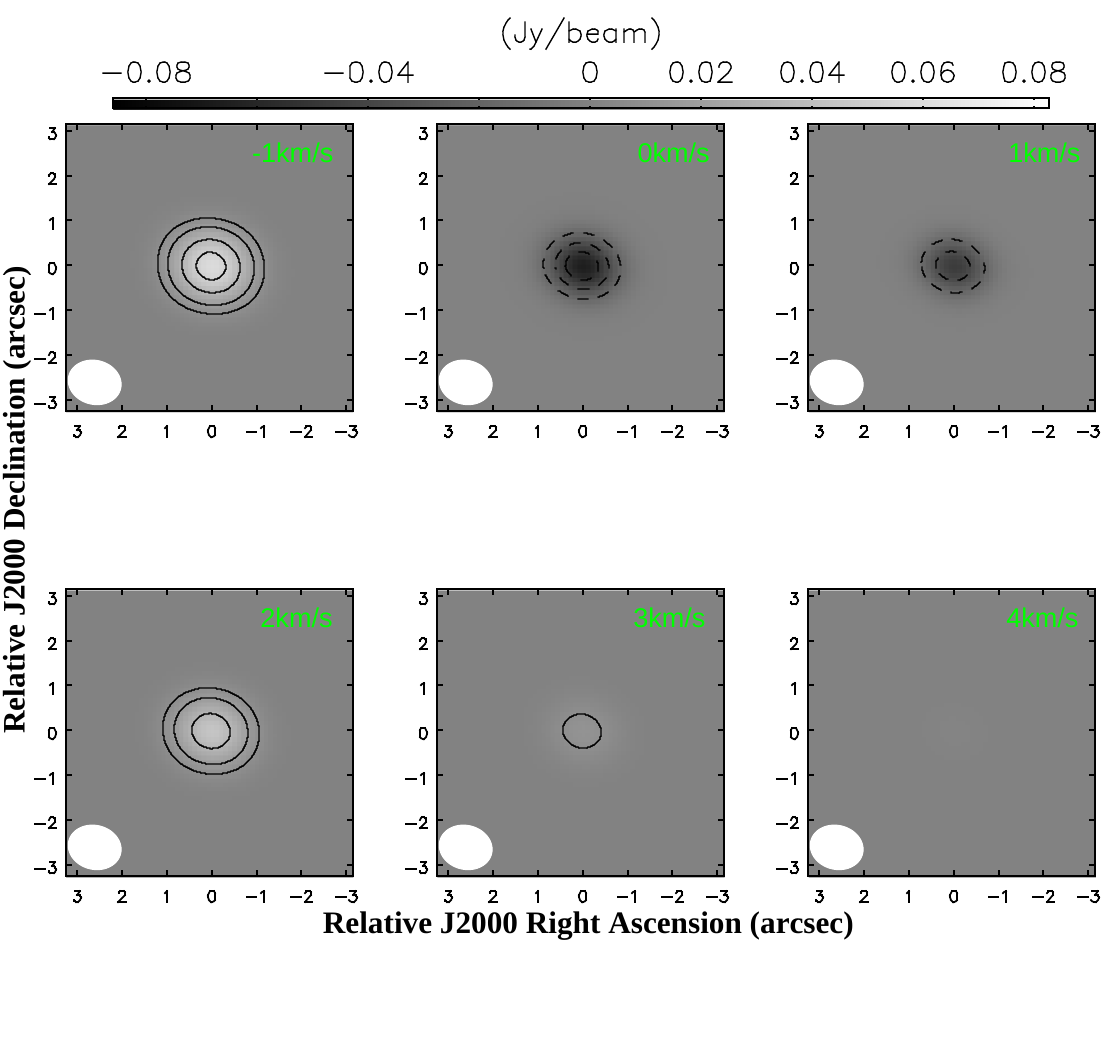}
  \end{minipage}
\caption{ A comparison between the observed channel map emission of  H$^{13}$CO$^+$ 
(top) and modeled channel map emission (bottom) by considering a constant abundance ($\sim 7.07 \times 10^{-11}$) profile is shown. The contour levels are drawn at -10\%, -20\%, -40\%, -60\%, 80\%, 10\%, 20\%, 40\%, 60\%, and 80\% of the  peck observed intensity ($0.086$ mJy/beam). The solid and dashed contours represent the emission and absorption, respectively. \citep[Courtesy:][]{bhat22}} 
\label{fig:channel-h13co+}
\end{figure}

 The detected inverse P-Cygni profile of $\rm{H^{13}CO^+}$ signifies a star-forming region's infalling envelope. The 1D RATRAN code is used to model the synthetic spectrum of the $1-0$ transition ($86.753970$ GHz) of $\rm{H^{13}CO^+}$. From the LAMDA database (\url{https://home.strw.leidenuniv.nl/~moldata/}), the collisional rates of $\rm{H^{13}CO^+}$ with H$_2$ are taken. Section \ref{sec:phys} describes the model's physical parameters. At first, it is assumed that there is a constant abundance of H$^{13}$CO$^+$ in the region. The comparison of the observed (black) and modeled (orange) line profiles are shown in Figure \ref{fig:h13co+_best}. To determine the best-fit line profile, the constant abundances of H$^{13}$CO$^+$, FWHM, and $\beta$ are varied. When an FWHM of $0.97$ km/s is employed, the two-slab model provides the best fit (see Table \ref{table:H13CO+_fit}).
 When using the 1D RATRAN modeling, an FWHM of $\sim 1.42$ km s$^{-1}$ yields the best fit. It is determined that $\beta=1$ and a constant abundance of $\sim 7.07 \times 10^{-11}$ constitute the best-fit line profile. The fractionation ratio used is $\rm{H^{13}CO^+}$:HCO$^+$=$1:65$ \citep{hoge00}, which results in the abundance of HCO$^+$ that fits the data the best, which is $4.6 \times 10^{-9}$. The best-fit HCO$^+$ abundance is substantially within the range of our modeled abundance, which is listed in Table \ref{table:abundances} and depicted in Figure \ref{fig:abundance}.
 When the envelope is static (i.e., the infall velocity is 0, red curve shown in Figure \ref{fig:h13co+_best}), it produces a symmetric line profile. But increasing the infall velocity the asymmetry increases. The rate of infall accelerates the red-shifted absorption and the blue-shifted emission nature. Figure \ref{fig:h13co+_best} shows that the asymmetry is gradually increased by the predicted line profile when $v_{1000} = 2.45$ (green line), $4.9$ (blue line), and $9.8$ (orange line) km/s are taken into account. The infall velocity of $\sim 4.9$ km/s (blue line) at 1000 AU is consistent with the observation. We found that $\beta$ significantly impacts the spectral profile. As \cite{osor09} obtained $\beta=1$ from their analysis, here we use $\beta=1$. Furthermore, we use the abundance distribution of HCO$^+$  determined from our chemical model (see section \ref{sec:chemical}). Figure \ref{fig:abundance}
 depicts that beyond $10,000$ AU, the peak abundance of HCO$^+$ varies between ($\sim 7.6 \times 10^{-9}-10^{-8}$). The peak abundance of HCO$^+$ significantly increased in the innermost grid. Beyond $10,000$ AU, the final abundance and peak abundance of HCO$^+$ essentially coincide. The final abundance of HCO$^+$ significantly decreased inside the $10,000$ AU. Because carbon fractionation is not taken into account in our chemical model, no abundance of H$^{13}$CO$^+$ is predicted by our model. However, a guess is made, and the spatial distribution of H$^{13}$CO$^+$ is produced using an atomic fractionation ratio of $^{13}$C and $^{12}$C $\sim 1:65$. As a result, the abundance of HCO$^+$ is diminished by a factor of $65$ to have an estimation of H$^{13}$CO$^+$ abundance.
 The difference between our observed and modeled line profiles using the abundances from our chemical model is shown in Figure \ref{fig:h13co+_chemmodel}. When the peak abundance (measured from the warmup to post-warmup time scale) from our chemical model is employed, the red curve shows the modeled line profile. When considering the final abundance from our chemical model, the green curve represents the modeled line profiles. The abundance profile obtained from our chemical model accurately reproduces the observed absorption profile, as shown in Figure \ref{fig:h13co+_chemmodel}. The deuterium fractionation of the molecule makes it clear that the molecules \citep{case02,das13a,das15a,das16,maju14a} do not necessarily reflect the initial atomic D/H ratio.
 As a result, there might be some differences when H$^{13}$CO$^+$ abundances are considered. Additionally, Figures \ref{fig:h13co+_best} and \ref{fig:h13co+_chemmodel} show the absorption nature obtained from model is narrower than the observed. Our radiative transfer model did not consider the rotational motion. Our modeled H$^{13}$CO$^+$ "fits" file is further processed to simulate the interferometric observations with the Common Astronomy Software Applications package \citep[hereafter, CASA,][]{mcmu07}, to show the difference between our computed line profile and the actual observation. In this case, instead of convolving the model image with a Gaussian beam, it is simulated what ALMA would observe with the same array configuration as for the observation. Here, the data in UV plane are generated using CASA's 'simobserve'  job. First, a model image is created to illustrate the sky brightness distribution (for simplicity, only the line profile generated with the constant abundance $\sim 7.07 \times 10^{-9}$ is presented). The CASA task called "tclean" is then used for the convolution. The velocity channel maps obtained with original observation and using our model are shown in Figure \ref{fig:channel-h13co+}. The observed spatial distribution with a few velocity channels, including both red-shifted (channels with a velocity greater than the systematic velocity) and blue-shifted (channels with a velocity less than the systematic velocity) velocity channels together, are shown in the left panel of Figure \ref{fig:channel-h13co+} to understand the distribution of molecules in G31. The channel map of the observed line replicates the absorption profile of the H$^{13}$CO$^+$ by having a negative contour (i.e., $0$, $1$, $2$, and $3$, km s$^{-1}$). The emission is observed in all other channel maps. The right panel of Figure \ref{fig:channel-h13co+} shows our simulated velocity channel maps, which are well-matched with the actual observation. In agreement with the observed channel maps, the modeled channel maps at $0$ km s$-1$ and $1$ km s$^{-1}$ show the absorption. However, the absorption of the observed line is broader (our simulated image lacks the absorption at $2$ and $3$ km s$^{-1}$). Discrepancies could be explained by a more accurate physical model that includes infall, outflow, and rotating motion.
 
\begin{figure}
 \centering
   \includegraphics[height=5.5cm,width=8cm]{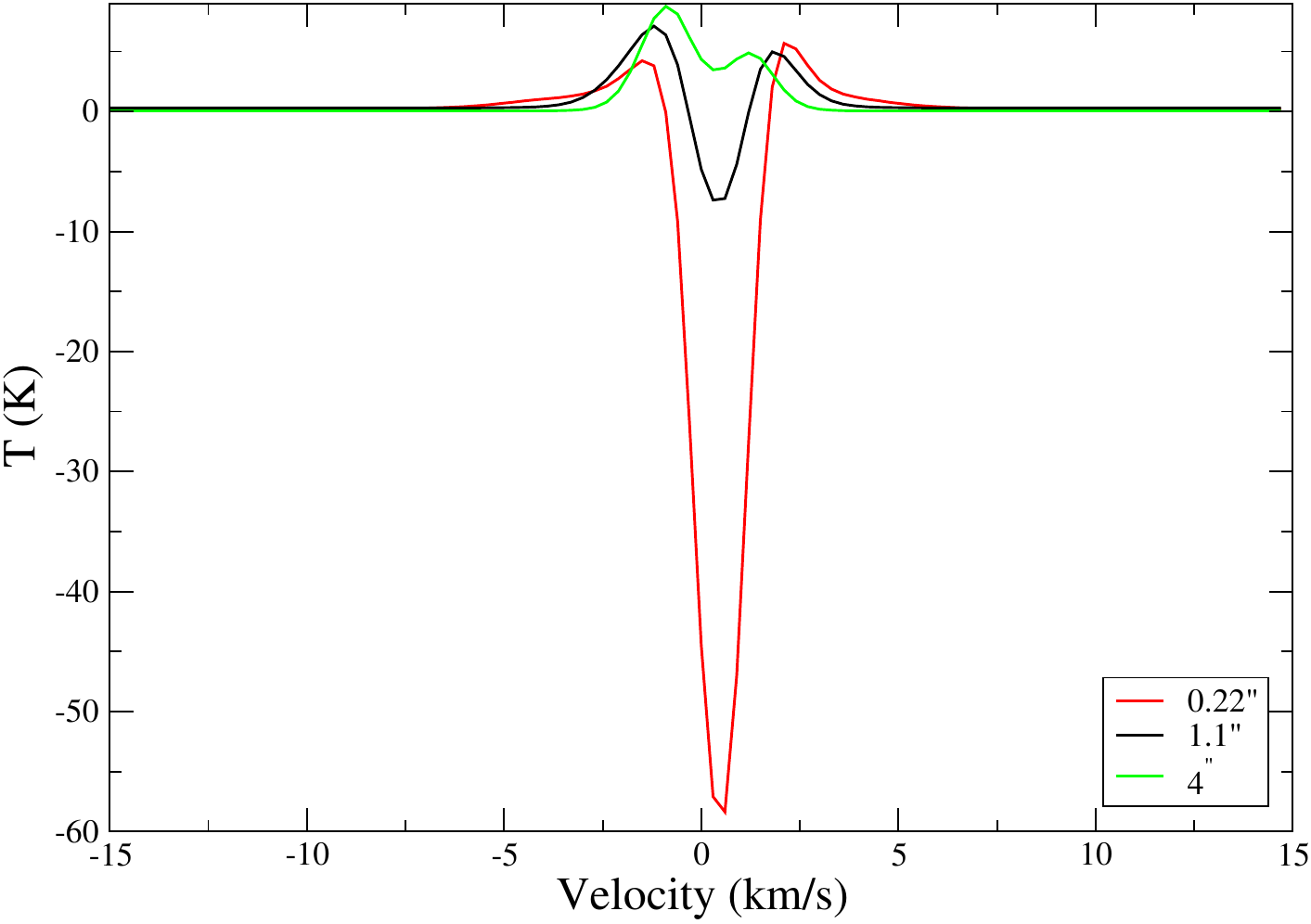}
 \caption{The modeled line profiles of H$^{13}$CO$^+$ with the $0.22^{''}$ (red line), $1.1^{''}$ (black line), and $4^{''}$ (green line) resolution are shown. It depicts that the inverse P-Cygni profile is only visible when $1.1^{''}$ resolution is used. \citep[Courtesy:][]{bhat22}}
\label{fig:resolution}
\end{figure}

 H$^{13}$CO$^+$ basically traces the envelope region, which is best suited to our present resolution. With the further better resolution ($0.22^{''}$), \cite{belt18} observed more inner region, and hence no such profile of H$^{13}$CO$^+$ was expected. The extended emission characteristics are filtered out by interferometric observations with excessively high angular resolution. On the other side, beam dilution will occur from a shallow resolution. It isn't easy to investigate the infall and outflow properties using interferometric data simultaneously. Therefore, molecules with extended emission are sensitive to the possibility that interferometric data filtered out the emission. Using an interferometer: the VLA and a single dish telescope, the IRAM-30M, \cite{cesa11} imaged the same transition to recover the filtered VLA data. However, the angular resolution of our observation, \citep{gora21}, is only about 1.1$^{''}$, which is only marginally or not at all resolving the source. When analyzing the line profiles of HCN, HCO$^+$, and N$_2$H$+$ using interferometric (SMA-1, SMA-2) data with $\sim 2.5^{''}$ resolution, \cite{zhul11}, for instance, reported the results on infall and outflow in the star-forming region W3-SE (it is unrelated to G31, but the linear resolution of these two sources is comparable). A comparison between the low resolution ($4^{''}$, green line, and $1.1^{''}$, black line) and high resolution ($0.22^{''}$, red line) modelled spectra of H$^{13}$CO$^+$ is illustrated in Figure \ref{fig:resolution}. With an improvement in resolution, it is seen that the emission peak steadily decreases. Only when $1.1^{''}$ resolution is used the inverse P-Cygni nature can be seen. It supports the necessity of taking into account the extended H$^{13}$CO$^+$ emission profile in our relatively low-resolution ALMA data.

\begin{figure}
 \centering
\includegraphics[height=7cm,width=9cm]{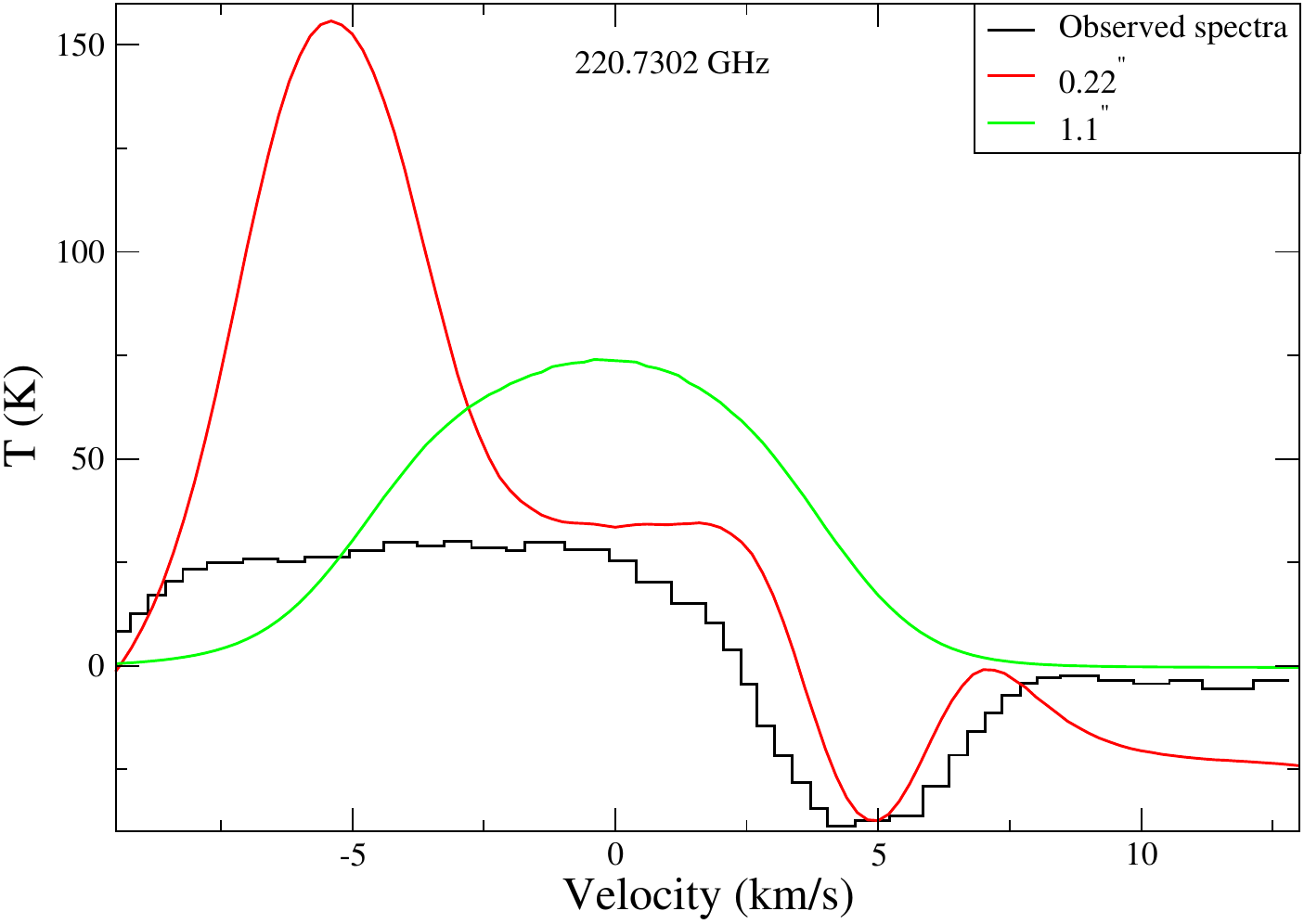}
\caption{A comparison between the observed (0.22$^"$, black line) (data taken from \cite{belt18}) and modeled (0.22$^"$ in red and 1.1$^"$ in green) line profiles of CH$_3$CN is shown. It depicts that the inverse P-Cygni nature is not visible with our low-resolution data. For the best-fitted case, $\beta=1$, FWHM = $2.5$ km/s, V$_{1000}$=6.0 km/s and a constant abundance of $6 \times 10^{-8}$ are used. \citep[Courtesy:][]{bhat22}}
\label{fig:ch3cn}
\end{figure}

 \cite{belt18} was able to observe numerous transitions of $\rm{CH_3CN}$ and its isotopologues having distinct upper state energies due to the better resolution (0.22$^{''}$). CH$_3$CN mostly forms on the grain surface and fills the gas phase when it is warmer. Therefore, it mainly traces the hot inner zone \citep{hung19}. The deviation of the red-shifted absorption peak from the systematic velocity represents the infall velocity. This shift is enhanced by an increase in up-state energy, as noted by \cite{belt18} (K $<$ 10). It suggests that this source has an accelerated infall present. Although interferometric filtering severely affects molecules with extended emission (such H$^{13}$CO$^+$), in our case, the angular resolution is considerably lower (1.1$^{''}$) than the resolution of \cite{belt18} (0.22$^{''}$). Prior research has identified $\rm{CH_3CN}$ transitions toward the main core of G31 using interferometry (SMA) and single dish (IRAM-30m). Here, the observed line profile of CH$_3$CN in the J=12 - 11 (K=2) transition (220.7302 GHz) by \cite{belt18} is reproduced using the 1D RATRAN radiative transfer model. It is crucial to remember that our observational resolution differs significantly from that of \cite{belt18}. A comparison of the predicted high-resolution and low-resolution spectra with the observed profile of CH$_3$CN is shown in Figure \ref{fig:ch3cn}. The abundance of CH$_3$CN obtained by \cite{belt18} was $1.0 \times 10^{-8}$. With a constant abundance of $\sim 6 \times 10^{-8}$, FWHM of $2.5 $km/s, and $\beta \sim$ 1, the best fit with the observed spectra is found. According to our chemical modeling, the highest abundance of CH$_3$CN can reach up to $\sim 3.8 \times 10^{-8}$ (Figure \ref{fig:abundance} and Table \ref{table:abundances}). Figure \ref{fig:ch3cn} makes it clear that the inverse P-Cygni profile in high resolution ($0.22^{''}$) modeled data matches the observed profile of CH$_3$CN. In contrast, a strong emission pattern is obtained in the blue-shifted area of our model. On the other hand, modeled data with low resolution ($1.1^{''}$) has a much lower intensity than data with high resolution. Because CH$_3$CN traces the infall materials into the source's inner area, it does not exhibit any inverse P-Cygni profile.
  
\subsubsection{\rm{\bf{HCN}}} 
 
 \begin{figure}
 \centering
 \includegraphics[height=7cm,width=9cm]{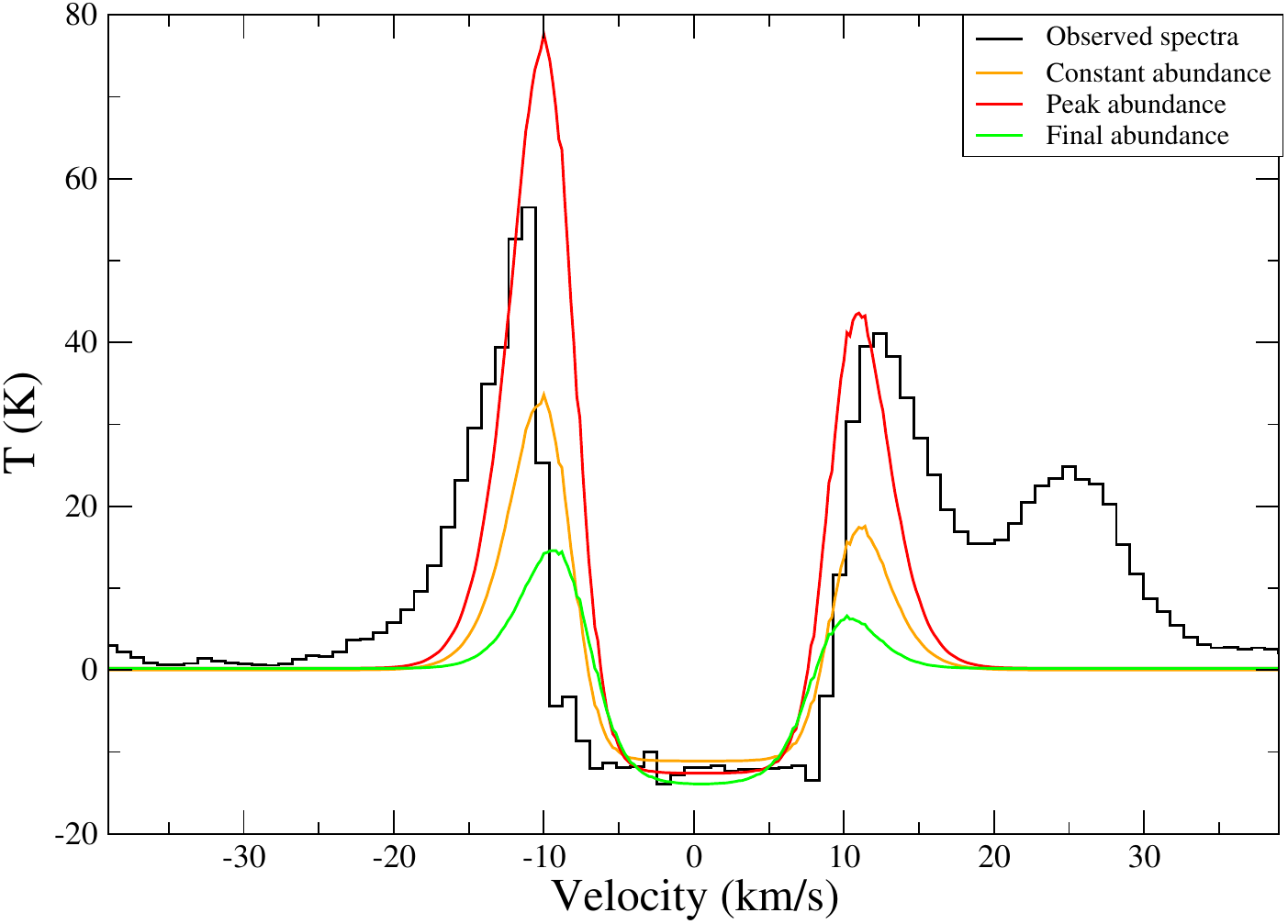}
 \caption{ A comparison between the synthetic spectrum of HCN generated using 1D RATRAN radiative transfer code and line profile observed towards G31 is shown. The solid black line represents the observed spectrum, whereas the orange line is the line profile of the transition of HCN obtained by using constant abundance ($7.6 \times 10^{-8}$) and red by using the peak abundance profile obtained from the chemical model, green by using the final abundance profile obtained from chemical model. We have obtained the best fit with $\beta=1.4$ and an FWHM of $10$ km/s.\citep[Courtesy:][]{bhat22}}
\label{fig:cassis_hcn}
\end{figure} 

\begin{figure}
  \centering
  \begin{minipage}{0.42\textwidth}
    \includegraphics[height=7cm,width=9cm]{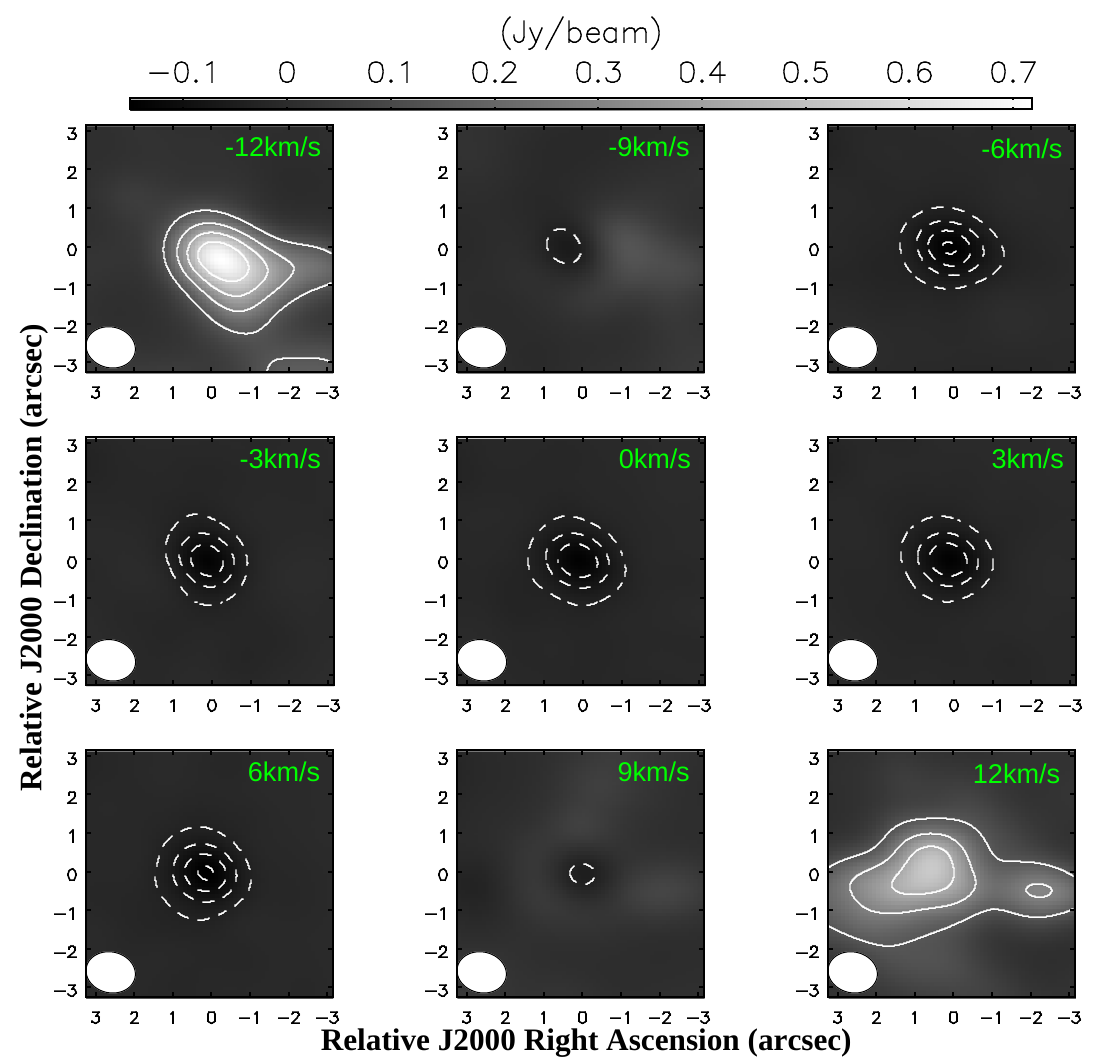}
  \end{minipage}
\hskip 2.5cm
 \begin{minipage}{0.42\textwidth}
    \includegraphics[height=7cm,width=9cm]{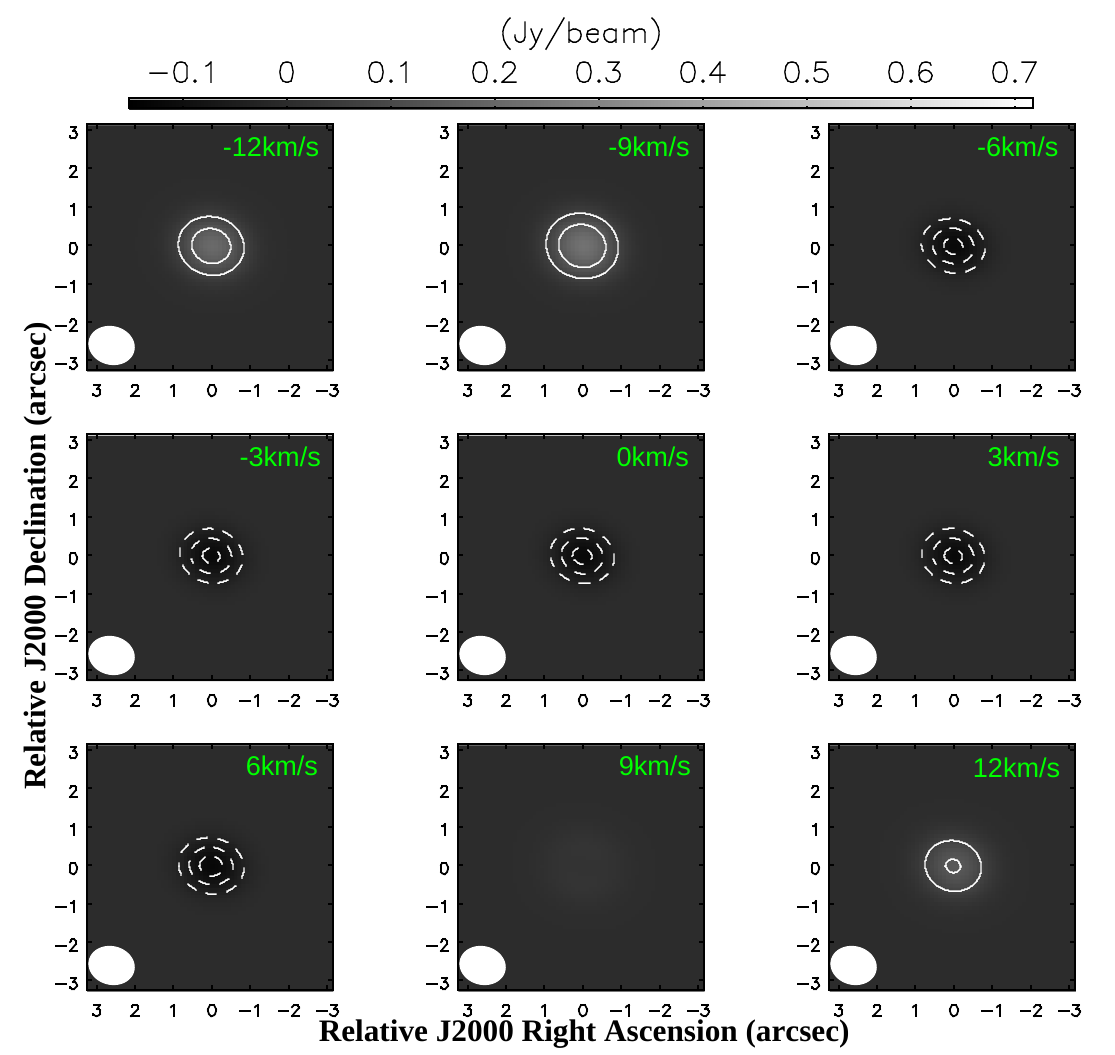}
  \end{minipage}
\caption{ Comparison of channel map emission of HCN between observation in top and model (constant abundance $\sim 7.6 \times 10^{-8}$) in bottom. Contour levels are at -5\%, -10\%, -15\%, -20\%, 20\%, 40\%, 60\%, and 80\% of the  peak observed intensity (0.719 mJy/beam). Solid and dashed contours represent the emission and absorption, respectively. \citep[Courtesy:][]{bhat22}}
\label{fig:hcn-channel}
\end{figure}
 
\cite{gora21} observed three HCN hyperfine transitions (F = 1$\rightarrow$1, 2$\rightarrow$1, and 1$\rightarrow$0) in absorption. These three transitions were blended. The physical input parameters mentioned in Section \ref{sec:RATRAN} are considered for the 1D RATRAN model. When an FWHM of $\sim 10$ km s$^{-1}$, constant abundance of $\sim 7.6 \times 10^{-8}$, and $\beta=1.4$ are employed, a good match between the modeled profile (orange line) and the observed profile (black line) is obtained. For the hyperfine transitions of HCN, the collisional rate between HCN and H$_2$ is taken from the LAMDA database. Among the three hyperfine transitions of HCN, it is noted that the F = 2$\rightarrow$1 (88.63184 GHz) transition is more robust. Our chemical model predicted abundance distribution of HCN is used for further analysis. A small increase in the abundance of HCN is depicted in Figure \ref{fig:abundance}. Its peak abundance ranges from $2.80 \times 10^{-10}$ to $5.40 \times 10^{-7}$. Our best-fit constant abundance easily meets this limit, $\sim 7.6 \times 10^{-8}$. The final abundance of HCN within the cloud exhibits a consistent declining trend. The line profile of HCN's F = 2$\rightarrow$1 transition is modeled using the peak and final abundances obtained through chemical modeling. These are shown by the red and green curves, respectively, in Figure \ref{fig:cassis_hcn}. The red curve perfectly matches the HCN line profile that has been observed. The comparison of channel map emission between the observation and modeling (by considering best-fitted constant abundance) is shown in Figure \ref{fig:hcn-channel}. This simulated emission is produced using a similar technique to that described in section \ref{sec:hco+}. It shows that the extended nature of our modeled channel map emissions is similar to the observed HCN channel map emissions.
 
 \subsubsection{\rm{\bf {SiO}} \label{sec:sio}}
 \begin{figure}
 \centering
\includegraphics[height=6cm]{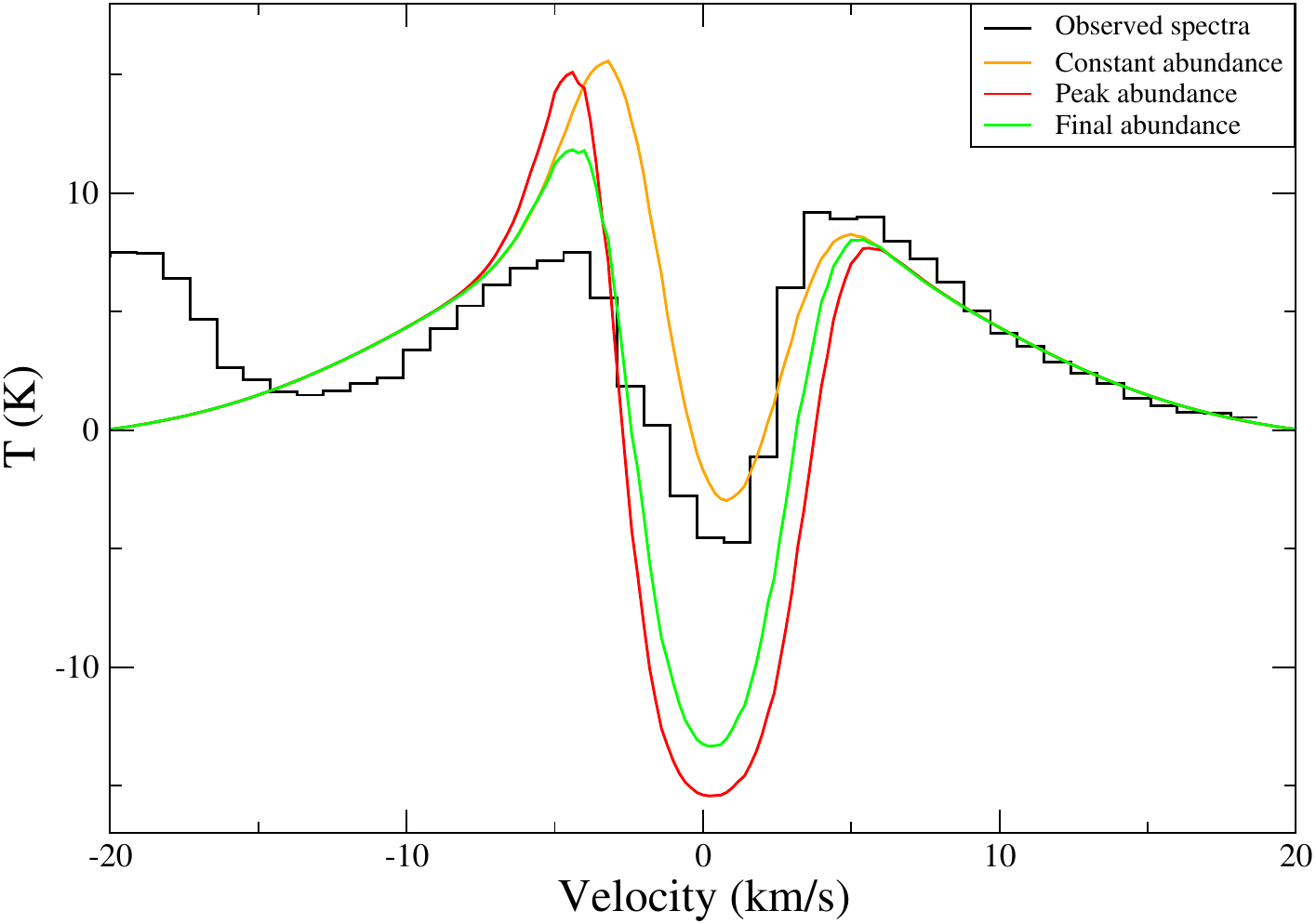}
\caption{A comparison between the observed and modeled SiO spectral profile in G31 is shown. The modeled line profile with the constant abundance ($9.5 \times 10^{-10}$) is shown in orange. The modeled line profiles with the peak and final abundance profiles are indicated with the red and green lines. The best fit was obtained when $\beta=1$ and an FWHM of $\sim 4.67$ km/s is used.\citep[Courtesy:][]{bhat22}}
\label{fig:sio-best}
\end{figure}
 
In G31, the 5-4 transition \citep{belt18} and 2-1 \citep{maxi01,gora21} transition of SiO were observed. SiO is a great indicator of outflow that is present in the source. It is now known that the G31 core has at least three outflows connected: E-W outflow to the south of the main cores's dust continuum emission peak (not connected to the two free-free embedded sources), the N-S outflow (may or may not be connected to the main core), and the NE-SW outflow (in the south of the main core). From the velocity obtained from the red-shifted and blue-shifted lobes of the $\rm{SiO}$ spectrum, \cite{gora21} computed the dynamical timescale. By considering the distances of 7.9 and 3.7 kpc, respectively, they calculated the average dynamic timescale from the SiO observation as $\sim 4.1 \times 10^{-3}$ years and $1.92 \times 10^{-3}$ years, respectively. Previously, $\rm{CO}$ molecule was used to study the molecular outflow present in the hot molecular core G31 \citep{olmi96}. Numerous outflow directions are suggested by other studies in the literature, such as \citep{aray08,belt18}. Here, the 1D RATRAN radiative transfer model simulates the reported 2-1 transition of $\rm{SiO}$. From the LAMDA database, the collisional rates of $\rm{SiO}$ and $\rm{H_2}$ are employed. In Section \ref{sec:phys}, the physical structure taken into account for the modeling is described. In the simulation, an additional outflow component is also taken into account because the outflow significantly impacts SiO spectra. It should be highlighted that our physical model does not consider outflows. Instead, using the ray-tracing method, a wide Gaussian component is added as the ray moves from the back to the front half. Next,  it affects the intensity in each velocity channel \citep{mott13}. Both the radial abundance profile produced from our chemical model and the constant abundance of $\rm{SiO}$ are used in the model. The best fit using the constant abundance is achieved with $\beta=1$, an FWHM of $4.67$ km/s, and a constant abundance of $9.5 \times 10^{-10}$. Here, we consider the FWHM of $\sim$ 20 km s$^{-1}$ for the gaussian component used for outflow and intensity of $\sim$ 10 K. The observed line profile of the SiO (2-1) transition is depicted in black in Figure \ref{fig:sio-best}, while the modeled (constant abundance) line profile is depicted in orange. It shows how well the modeled spectra may mimic the observed absorption. The emission is a little stronger than what was observed, though. The radial distribution of the SiO abundance is depicted in the left panel of Figure \ref{fig:abundance}. It demonstrates how quickly $\rm{SiO}$'s peak abundance rises from 156 AU to 287 AU and how, after this, it continued to rise at a prolonged rate ($2.2 \times 10^{-10}-9.2 \times 10^{-9}$). The best fitted constant abundance of SiO, $9.5 \times 10^{-10}$, lies between the predicted peak abundance range. The final abundance deviates greatly from its peak abundance and exhibits ups and downs in the abundance profile. Modeled spectra with the peak abundance profile (red line) and final abundance profile (green line) are displayed in Figure \ref{fig:sio-best}. The shape of the observed line profile can be replicated using the modeled peak abundance profile, as shown in Figure \ref{fig:sio-best}. Shock is not considered in our chemical model even though it might be essential for the chemistry of SiO.
 
\subsubsection{$\rm{\bf{NH_3}}$}
\label{sec:NH3}

 \begin{figure*}
  \centering
\begin{minipage}{0.43\textwidth}
    \includegraphics[width=\textwidth]{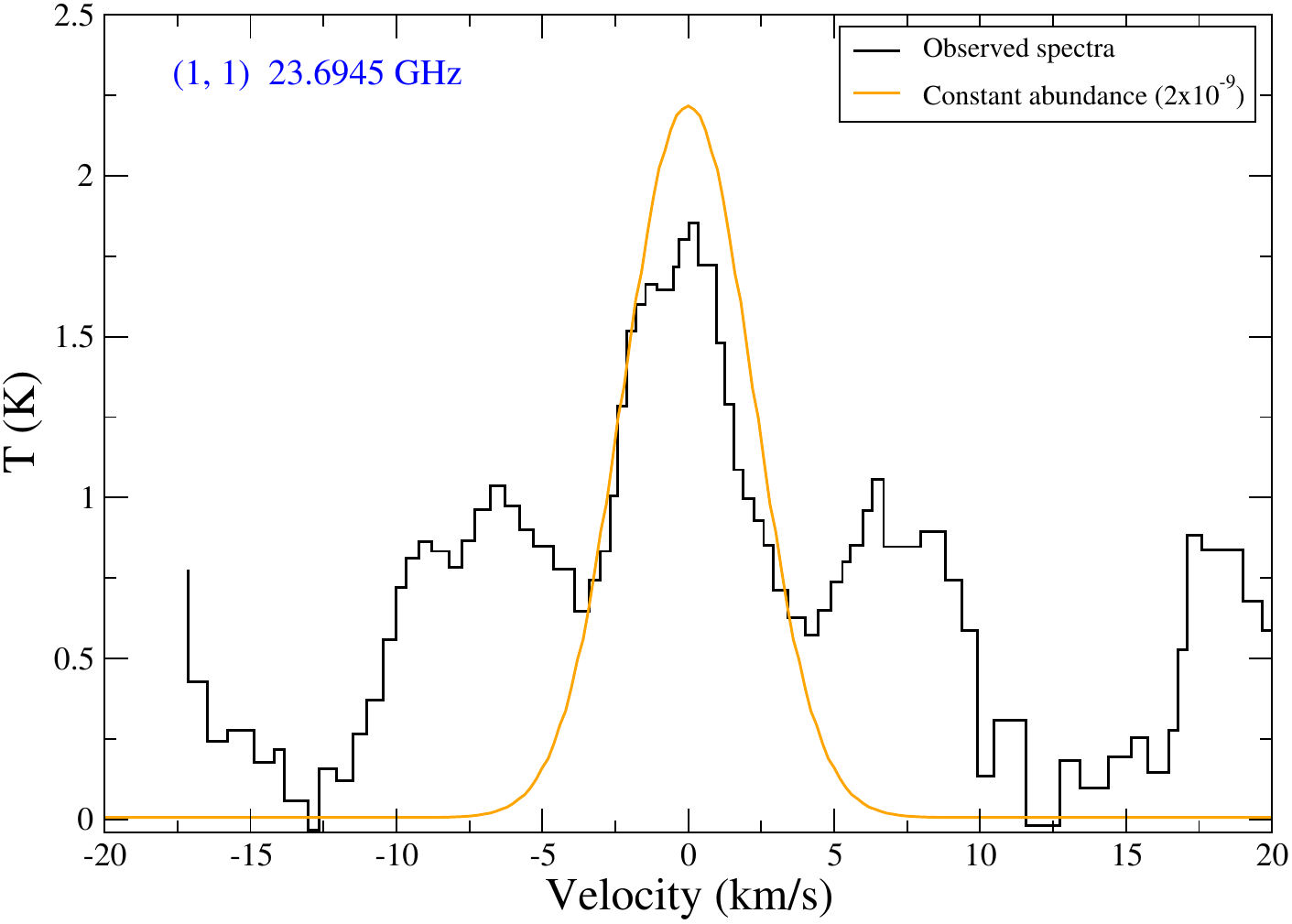}
  \end{minipage}
\hskip 0.5cm
  \begin{minipage}{0.43\textwidth}
    \includegraphics[width=\textwidth]{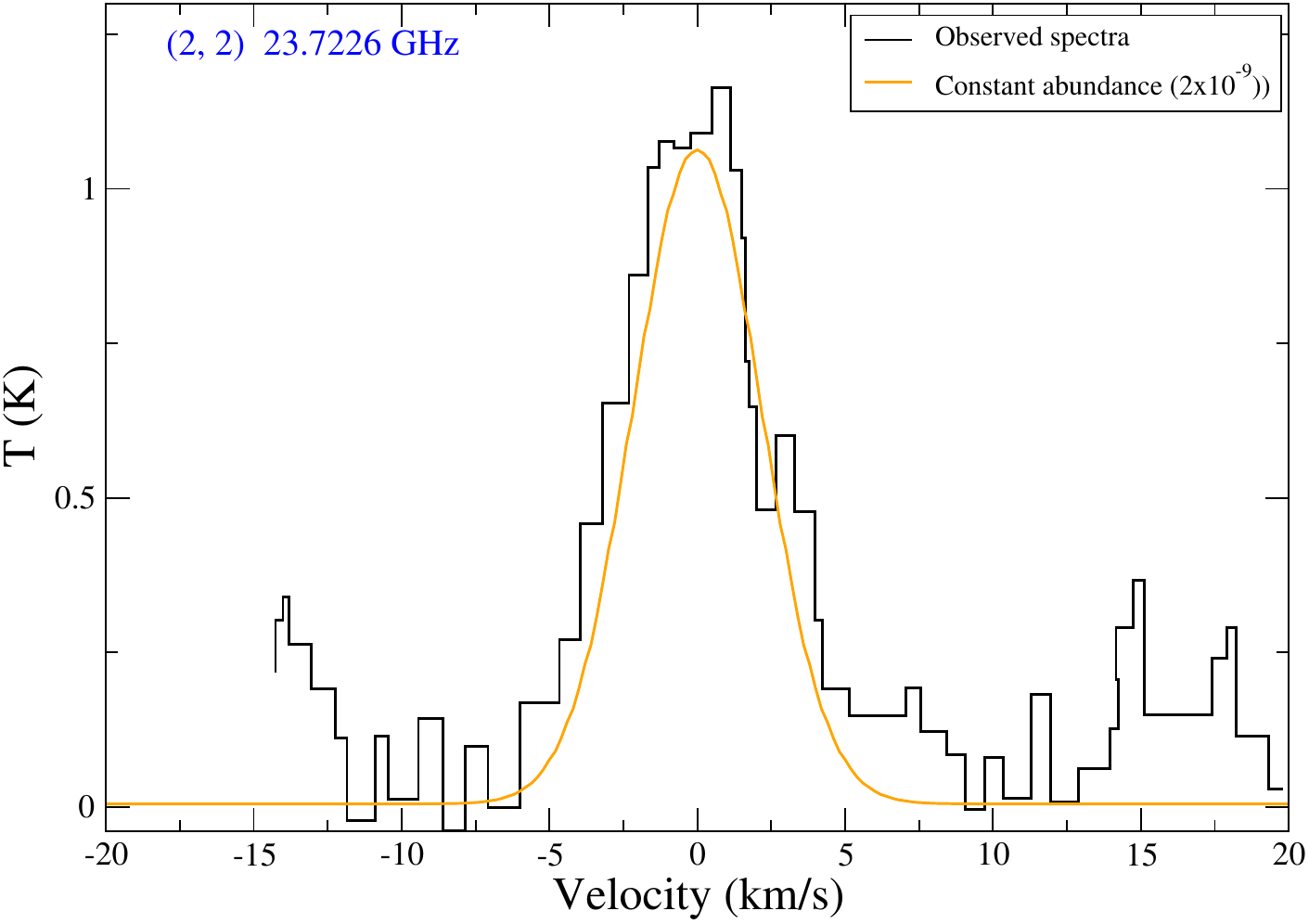}
  \end{minipage}
  \begin{minipage}{0.43\textwidth}
    \includegraphics[width=\textwidth]{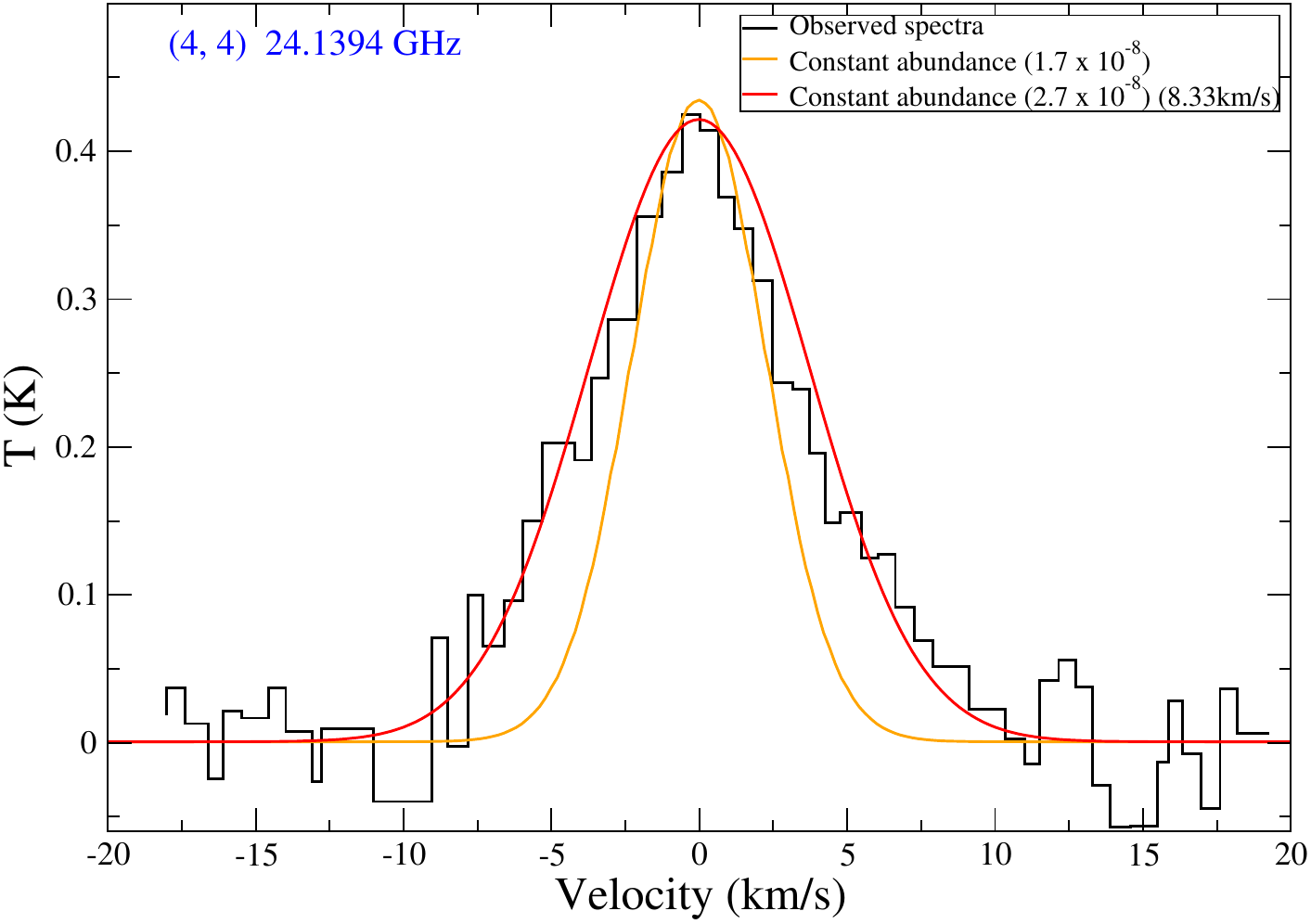}
  \end{minipage}
\hskip 0.6cm
  \begin{minipage}{0.43\textwidth}
    \includegraphics[width=\textwidth]{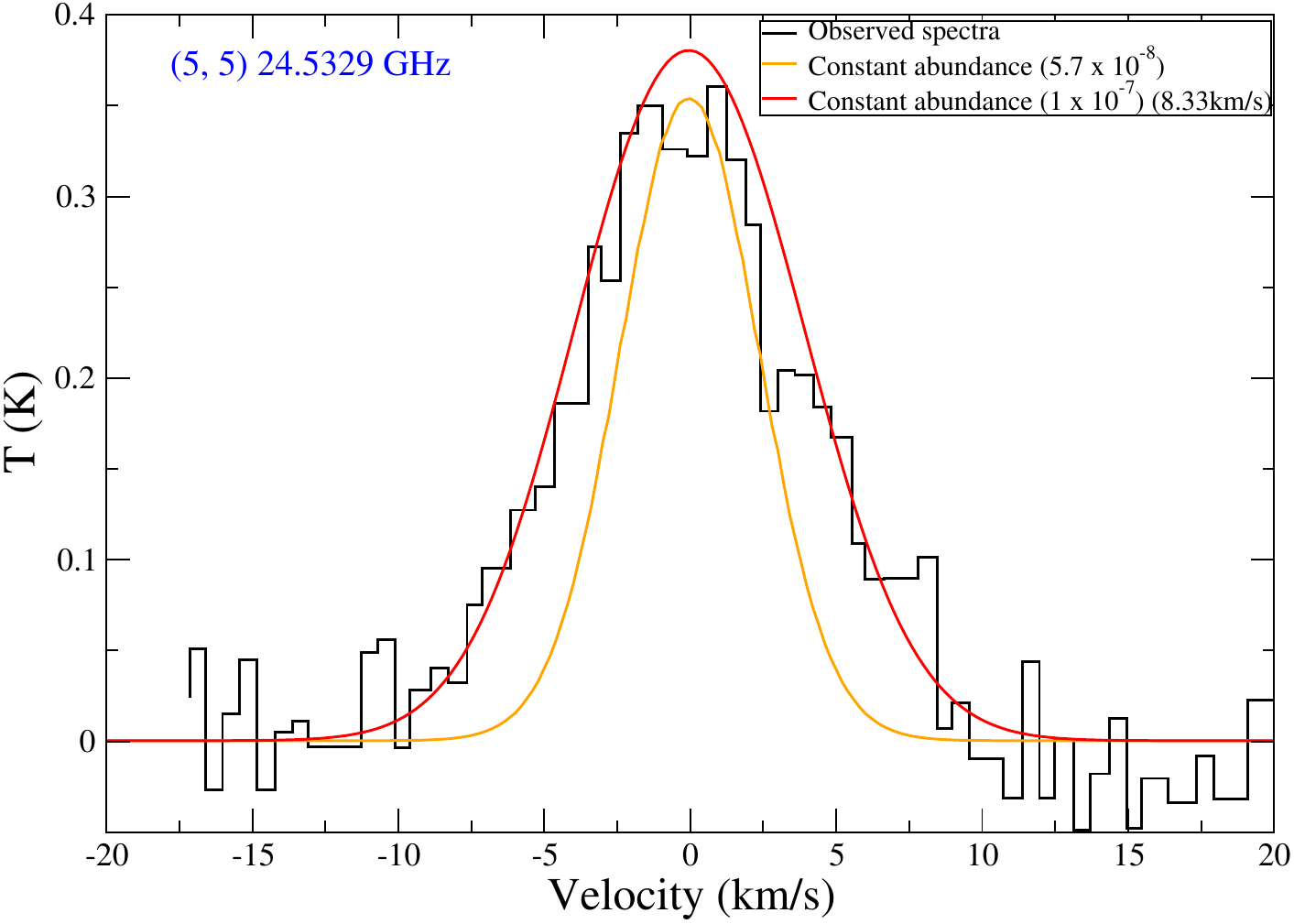}
  \end{minipage}
 \caption{ A comparison between the observed line profile (black) and synthetic line profiles (orange) of
 (a) 1,1 , (b) 2,2 , (c) 4,4 , (d) 5,5 transitions of $\rm{NH_3}$ are shown. These line profiles are generated with an FWHM of $\sim 4.9$ km/s, $\beta=1.4$ and a constant abundance of $\sim 2 \times 10^{-9}$ for 1,1 and 2,2 transitions, a constant abundance $\sim 1.7 \times10^{-8}$ for 4,4 transition and $ \sim 5.7\times10^{-8}$ for 5,5 transition. For 4,4 and 5,5, the best line profiles (red curves) with little higher FWHM ($8.33$ km/s) appeared with the constant abundance $2.7 \times 10^{-8}$ and $1.0 \times 10^{-7}$, respectively.\citep[Courtesy:][]{bhat22}}
 \label{fig:nh3_best}
\end{figure*}

\begin{figure}
  \centering
    \includegraphics[width=9cm]{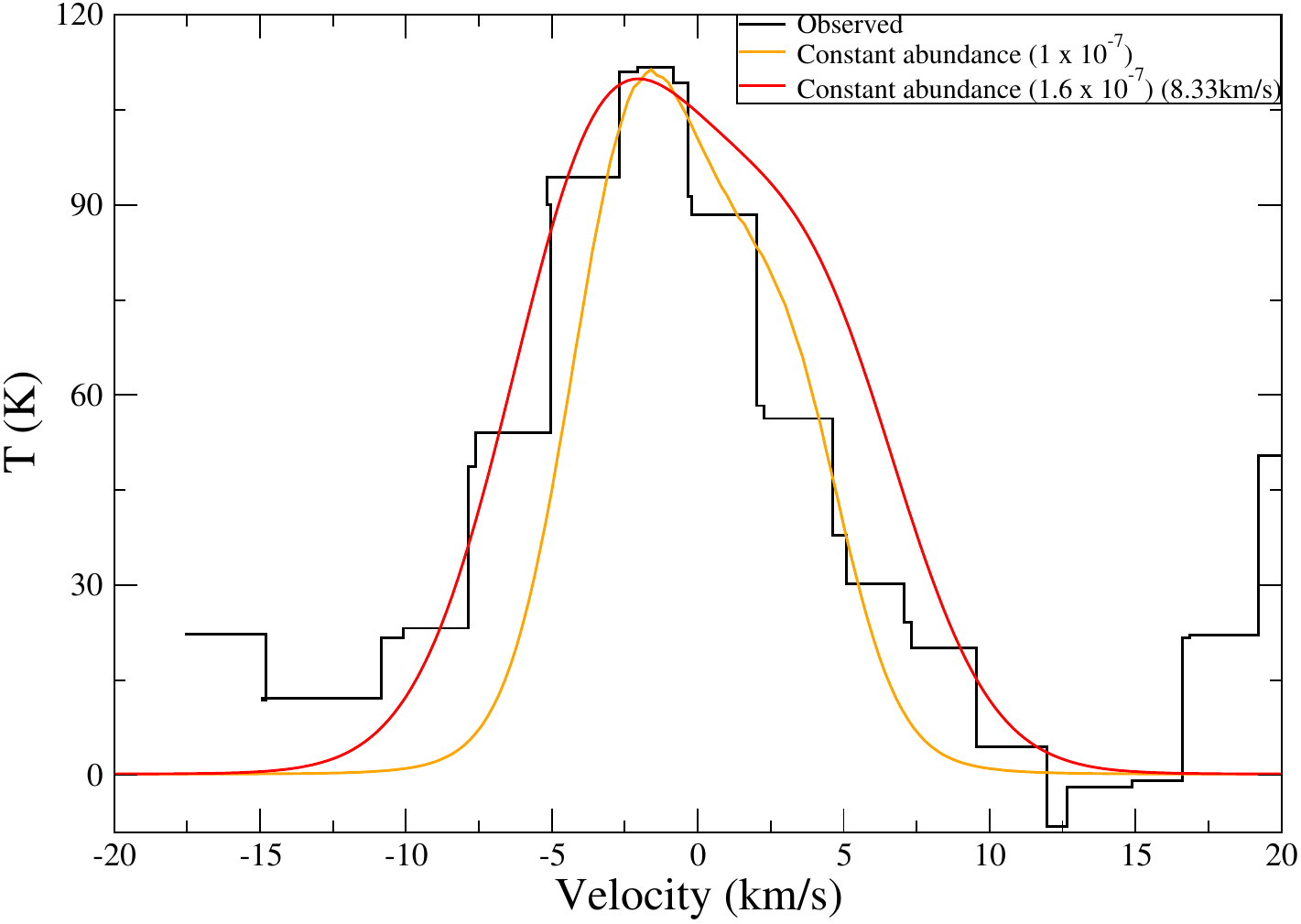}
 \caption{ A comparison between the high spatial resolution (0.63$^{''}$) VLA observation of the (4,4) transition (black) of NH$_3$ and modeled line profile
 (orange) of $\rm{NH_3}$ is shown. The synthetic line profile is generated with an FWHM of $\sim 4.9$ km/s, $\beta=1.4$, and a constant abundance $\sim 1 \times 10^{-7}$. With a little higher FWHM (8.33 km/s, red curve), the best fitted abundance is $1.6 \times 10^{-7}$.\citep[Courtesy:][]{bhat22}}
 \label{fig:nh3_best44}
 \end{figure}
 
 The G31 main core is near to the UC HII region, which is $5^{''}$ away from it from an angular perspective. Due to this close distance, the HII area frequently affects the observations with an angular resolution of $>5^{''}$. \cite{cesa92} studied the G31 HMC region using a single dish $100$ m telescope (GBT) with a beam size of $\sim 40^{''}$. \cite{cesa92} and \cite{chur90} detected many ammonia inversion transitions. All of these inversion transitions had been modeled by \cite{osor09} using the radiative transfer model. These transitions were originally modeled starting with a constant abundance of $\rm{NH_3}$. However, they also parameterized the abundance profile based on the temperature at which NH$_3$ and water condense and sublimate. The abundance profile of $\rm{NH_3}$ was created using the relation in \cite{osor09},
 \begin{equation}
 X_{mol}=\frac{X_{max}-X_{min}}{1+\eta}+X_{min}.
\end{equation}
 The ratio of the species in the solid phase to those in the gas phase is shown here by the symbol $\eta$. It was believed that the overall number of molecular species would remain constant. They considered that the variation of density, condensation, sublimation temperatures, and $\eta$ depend on each other. They took into account the minimum gas-phase abundance ($X_{min}$) and the maximum gas-phase abundance ($X_{max}$) using $\eta \gg 1$ and $\eta=0$, respectively. It was also considered how NH$_3$ would sublimate. When they used $X_{max}=3 \times 10^{-6}$ and $X_{min}=2 \times 10^{-8}$, \cite{osor09} found a good correlation with the observational results. A step profile for the abundance of NH$_3$ is shown in Figure \ref{fig:abundance}. Our model predicts a peak abundance variation of $3.2 \times 10^{-8}$ to $6.4 \times 10^{-7}$. Deep inside the cloud, it is more abundant. A minimum constant abundance (X$_{min}$) $\sim 2 \times 10^{-8}$ was considered in \cite{osor09} when the temperature is roughly below, $90$ K. They assumed a constant maximum abundance ($X_{max}$) of $\sim 3 \times 10^{-6}$ for temperatures larger than $100$ K in the inner envelope, where the temperature rises steadily. According to our chemical model, peak NH$_3$ abundance ranges in the outer section of the envelope from $\sim 5.0 \times 10^{-8}$ to $\sim 2.3 \times 10^{-7}$ (see the left panel of Figure \ref{fig:abundance}). This abundance is for the region beyond $\sim 8000$ AU, which corresponds to the area where the temperature is less than $93$ K. Inside, there is an abrupt change in the abundance profile that ranges from $\sim 3.4 \times 10^{-8}$ to $\sim 6.0 \times 10^{-7}$. The peak abundance of NH$_3$ significantly reduced inside $300$ AU. Peak and final abundances of NH$_3$ exhibit more or less comparable patterns. It is fascinating to notice that when the gas and grain temperature at the starting stage of our model is preserved at $20$ K instead of the $15$ K described here, a striking match between our obtained peak abundance profile and that employed in \cite{osor09} is obtained. The peak abundance is $\sim 1.7 \times 10^{-6}$, which is closer to the X$_{max}=3 \times 10^{-6}$ used in \cite{osor09}, when an initial temperature of $\sim 20$ K is employed. \cite{osor09} did not apply a chemical model to reach this abundance. Therefore, a perfect match with \cite{osor09} is not anticipated. For the identification of the $\rm{NH_3}$(1,1), $\rm{NH_3}$(2,2), $\rm{NH_3}$(4,4), and $\rm{NH_3}$(5,5) transitions, \cite{chur90,cesa92} used the data obtained from $100$ m telescope. Here, a collapsing envelope is used to model all of these transitions. \cite{cesa92,chur90} used $100$ m single dish telescope to study the transitions of NH$_3$. As a result, the obtained spectra were convolved using a 40 $^{''}$ beam size, and the results were compared to the observed one. The modeled (1,1), (2,2), (4,4), and (5,5) transitions of NH$_3$ are depicted in Figure \ref{fig:nh3_best} together with their observed (black) line profiles. From the LAMDA database, the collisional rate of $\rm{p-NH_3}$ is obtained. To directly compare the produced line profile with the observed spectra, it is convolved with the GBT $100$ m telescope's 40$^{''}$ beam size. Figure \ref{fig:nh3_best} displays the modeled spectra (orange for constant abundance) together with the observed one (black) is shown. It is noted that the four observed line profiles cannot all be explained by a single constant abundance value. \cite{cesa92} found that the FWHM for the $(4,4)$ transition was $\sim 4.9\pm 0.1$ km s$^{-1}$. An FWHM of $\sim 4.9$ km s$^{-1}$ is used to obtain a good fit in this case. A constant NH$_3$ abundance of $\sim 2 \times 10^{-9}$ yields the best fit to the observed spectrum for (1,1) and (2,2) transitions. When a constant abundance of $1.7 \times 10^{-8}$ and $5.7 \times 10^{-8}$ is applied, respectively, the good fit for the (4,4) and (5,5) transitions is attained. The (4,4) and (5,5) transitions have a better match with a higher FWHM ($\sim 8.33$ km/s) with $2.7 \times 10^{-8}$ and $1.0 \times 10^{-7}$, respectively. Similar to our situation, \cite{osor09} had trouble matching all of the peak intensities at once with a single abundance profile (see Figure 6 of \cite{osor09}). To explain the (4,4) and (5,5) transitions in this instance, a higher constant abundance is required. It is because the upper state energies of the (1,1) and (2,2) transitions ($1.14$ K and $42.32$ K, respectively) are lower than those of the (4,4) and (5,5) transitions ($178.39$ K and $273.24$ K). It is predicted that the (4,4) and (5,5) transitions would come from the warmer (i.e., inner) area of the envelope in comparison because of these variations in upper state energy. Figure \ref{fig:abundance} shows that the abundance of NH$_3$ is larger in the inner than the outer portion of the envelope. So it is justified to use the higher abundance for (4,4)and (5,5) transition. In addition to the constant abundance, peak abundance obtained from our model is also used. It is noted that our modeled line profiles cannot fit the observed transitions well in that case. With our modeled abundance profile, all transitions have intensities that are overproduced.

Our modeled (4,4) transition is compared to that obtained with the VLA observation (angular resolution 0.63$^{''}$) in Figure \ref{fig:nh3_best44}. A constant abundance of $\sim 1 \times 10^{-7}$, $\beta = 1.4$, and FWHM $4.9$ km/s used to obtain the best fit. When the FWHM is somewhat higher ($\sim 8.33$ km/s), the abundance used is $\sim 1.6 \times 10^{-7}$, which yields the best fit.
 
 \begin{figure}
 \centering
\includegraphics[height=7cm,width=9cm]{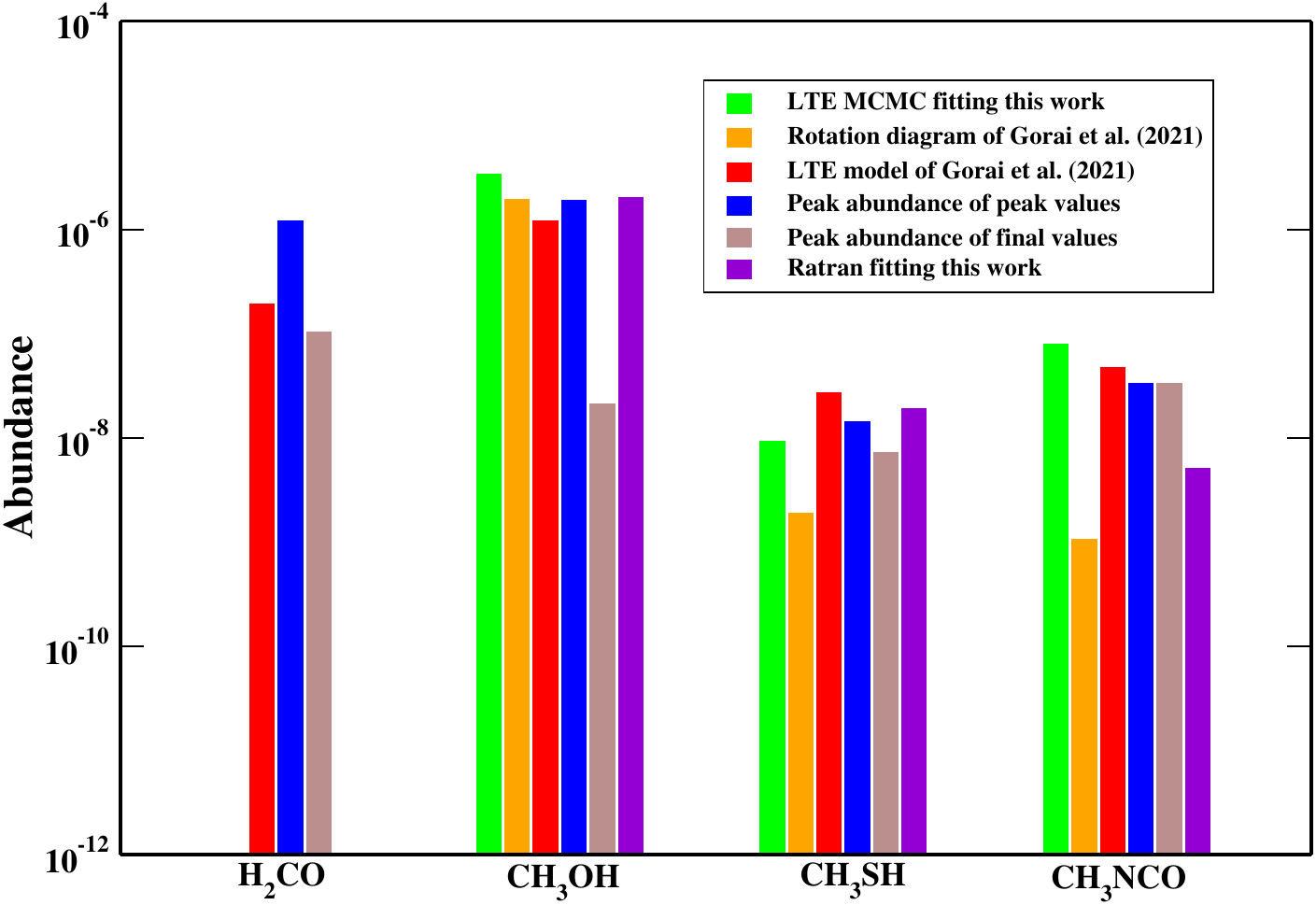}
\caption{Comparison between the observed (LTE and rotation diagram method of \cite{gora21} and MCMC fitting discussed in section \ref{sec:MCMC_G31} (for CH$_3$OH average of A-CH$_3$OH and E-CH$_3$OH is taken from Table \ref{table:mcmc_lte_G31}), RATRAN fitting) and simulated abundance is shown. Maximum values obtained from the peak abundance profile shown in Table \ref{table:abundances} is shown in blue, whereas the maximum abundance obtained from the final abundances is shown in brown. \citep[Courtesy:][]{bhat22}}
\label{fig:comp}
\end{figure}

\begin{figure*}
\begin{minipage}{0.32\textwidth}
\includegraphics[width=\textwidth]{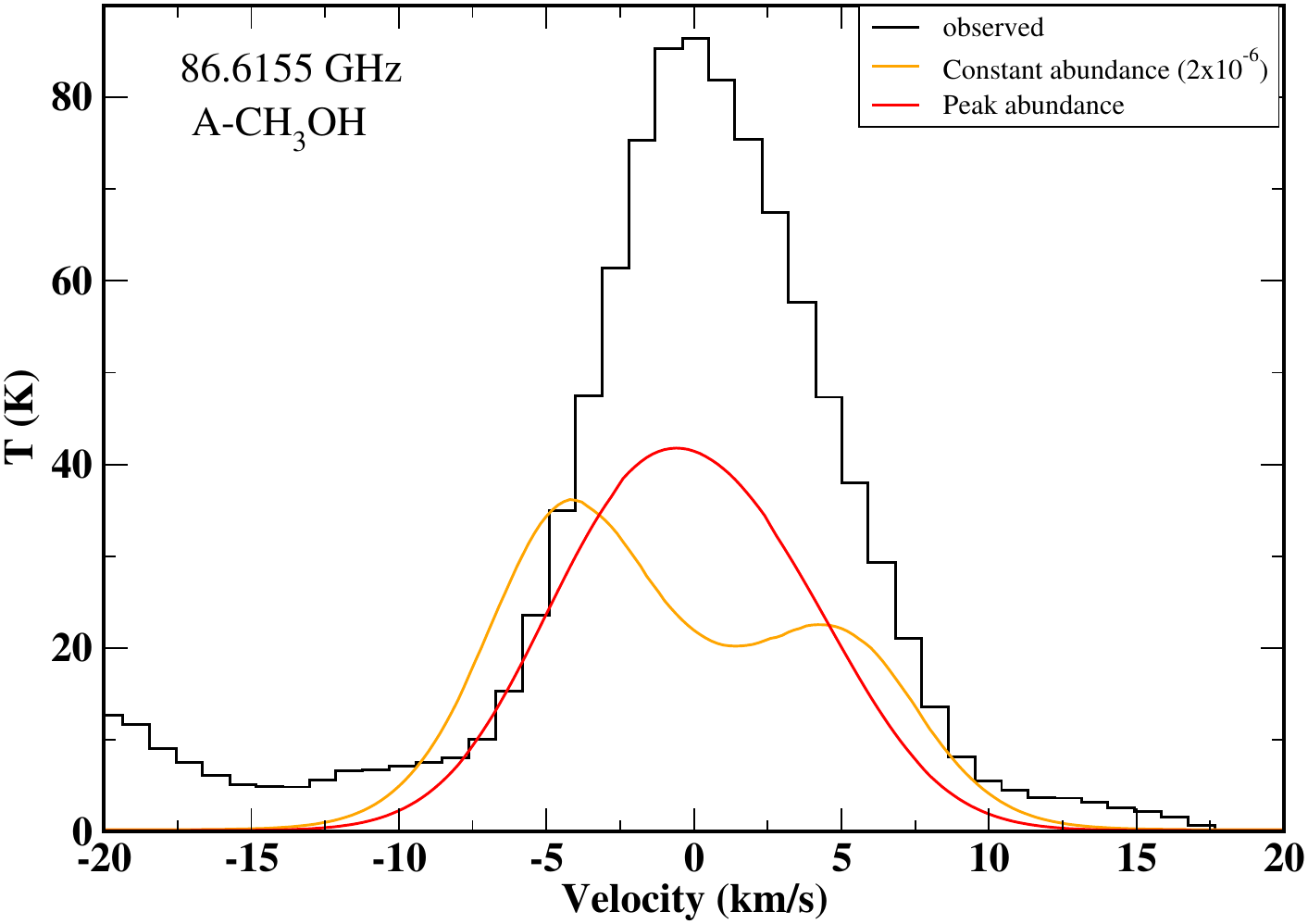}
\end{minipage}
\begin{minipage}{0.32\textwidth}
\includegraphics[width=\textwidth]{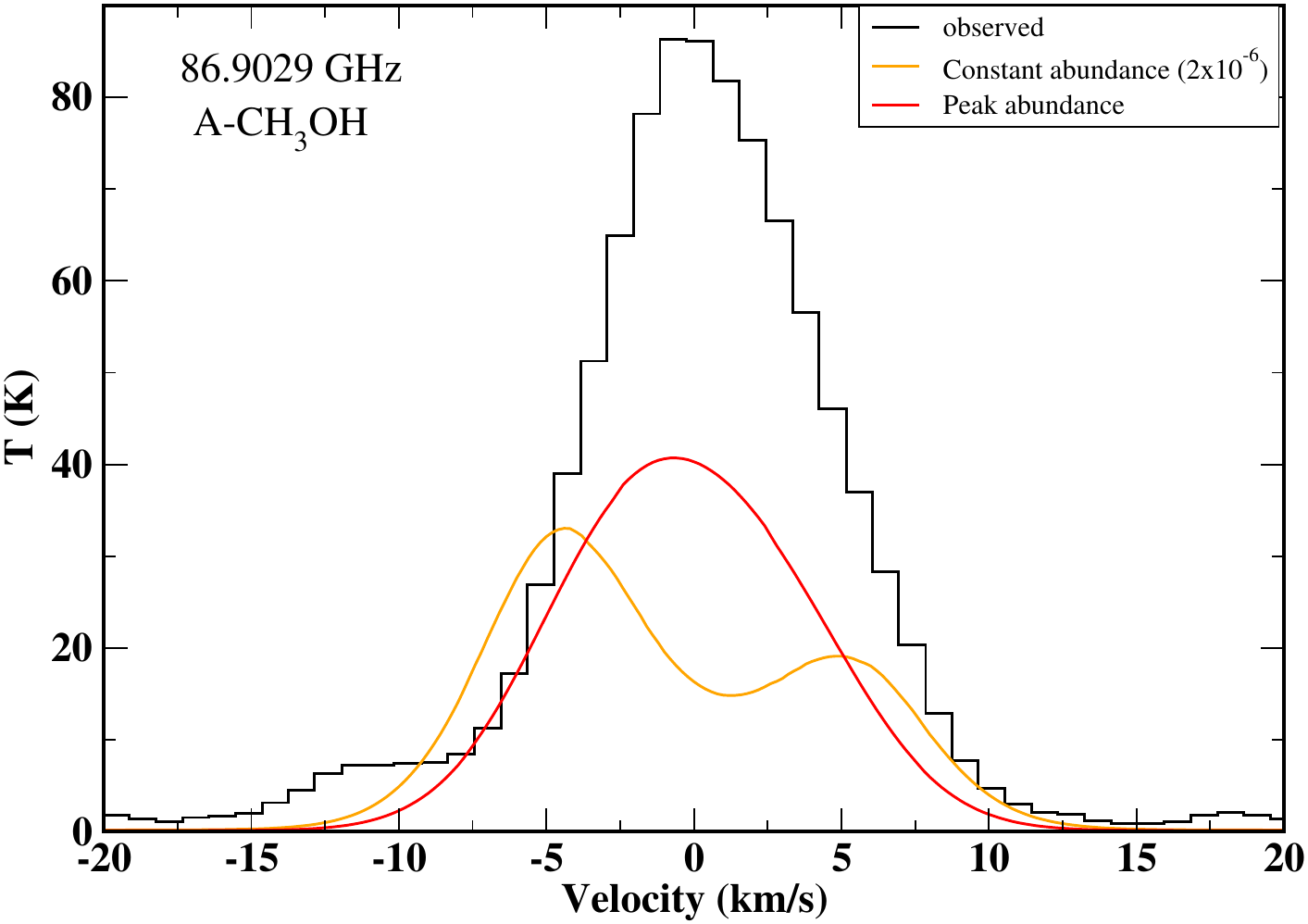}
\end{minipage}
 \begin{minipage}{0.32\textwidth}
 \includegraphics[width=\textwidth]{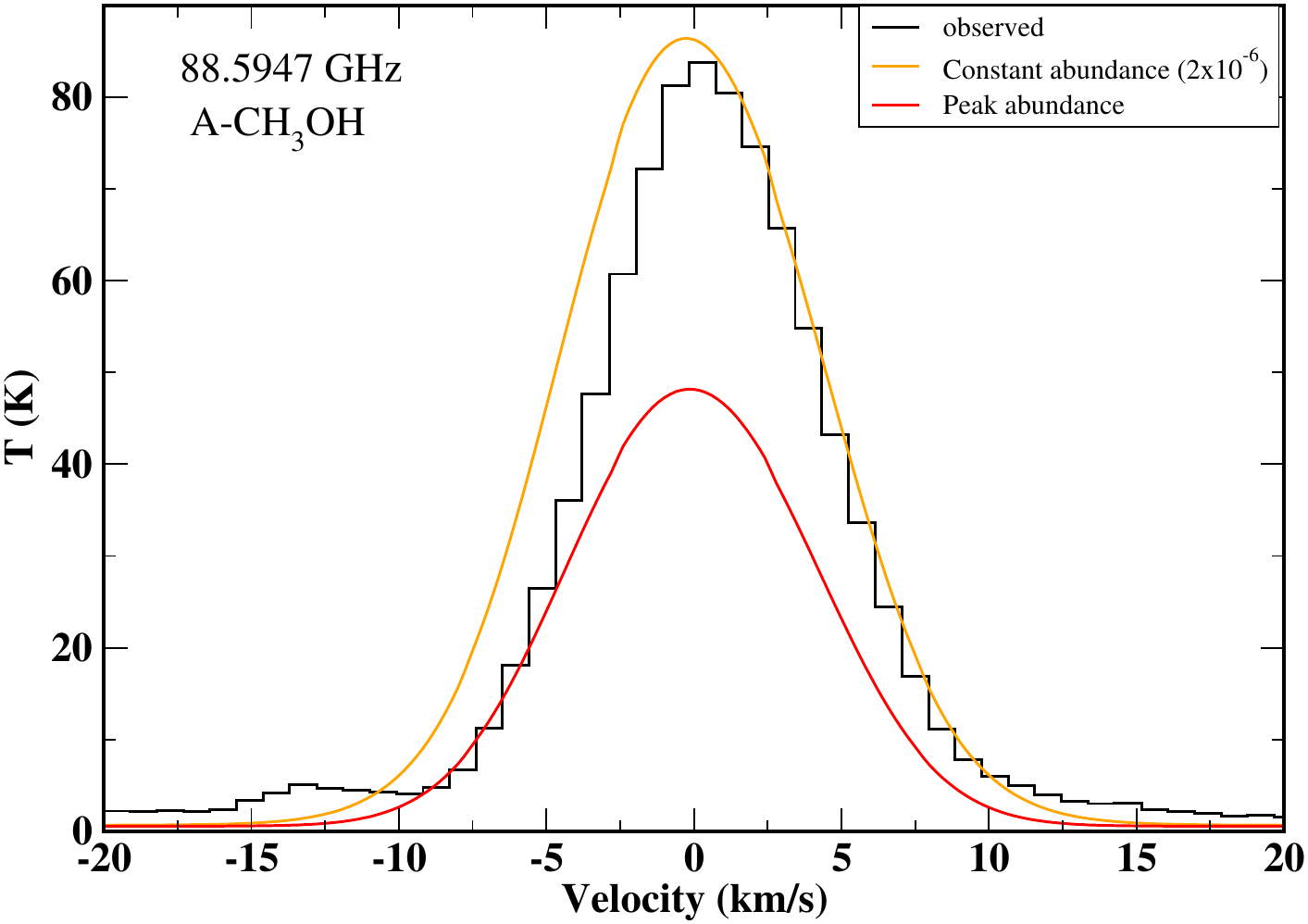}
 \end{minipage}
\hskip -0.9cm
\begin{minipage}{0.32\textwidth}
\includegraphics[width=\textwidth]{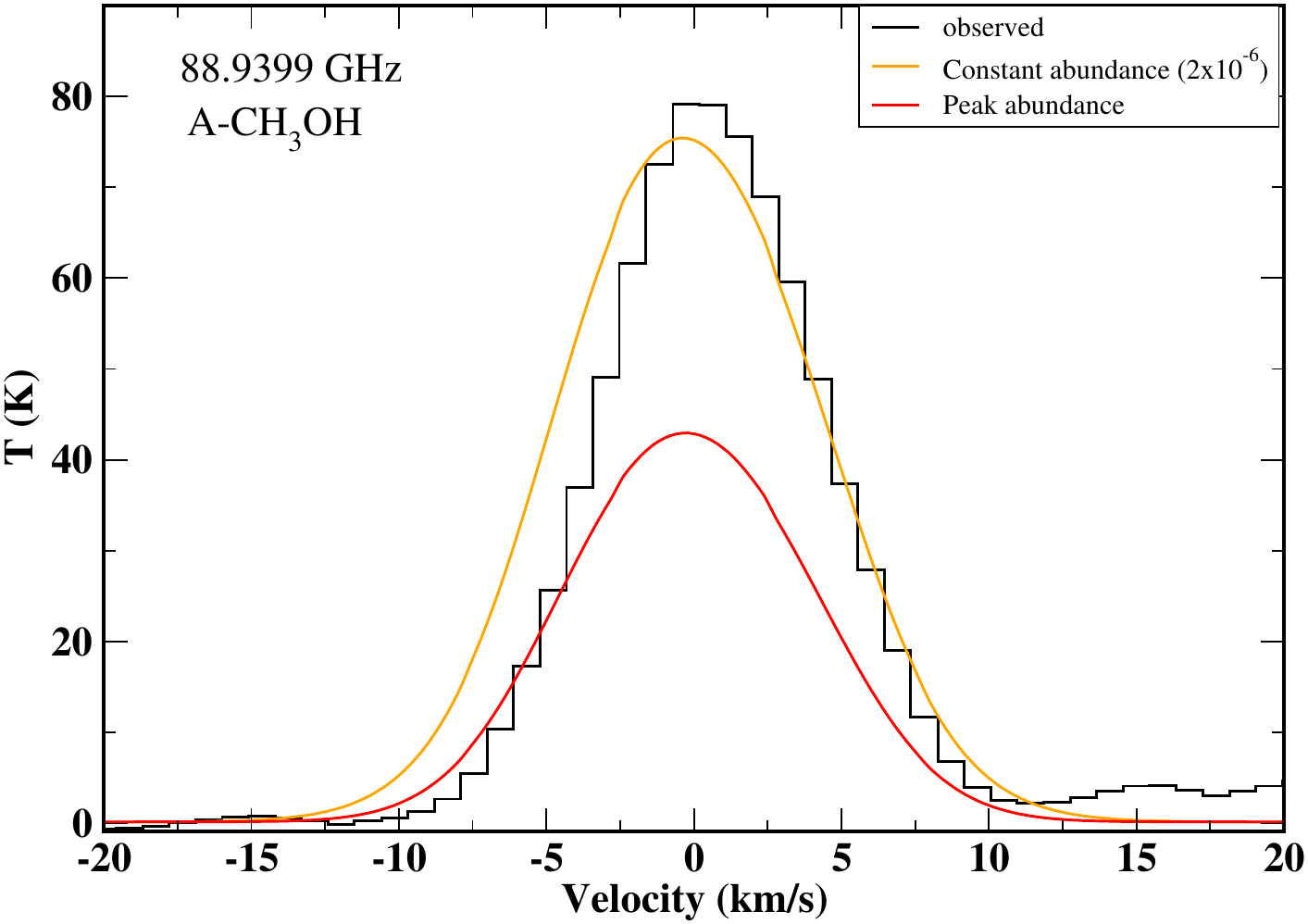}
\end{minipage}
\begin{minipage}{0.32\textwidth}
\includegraphics[width=\textwidth]{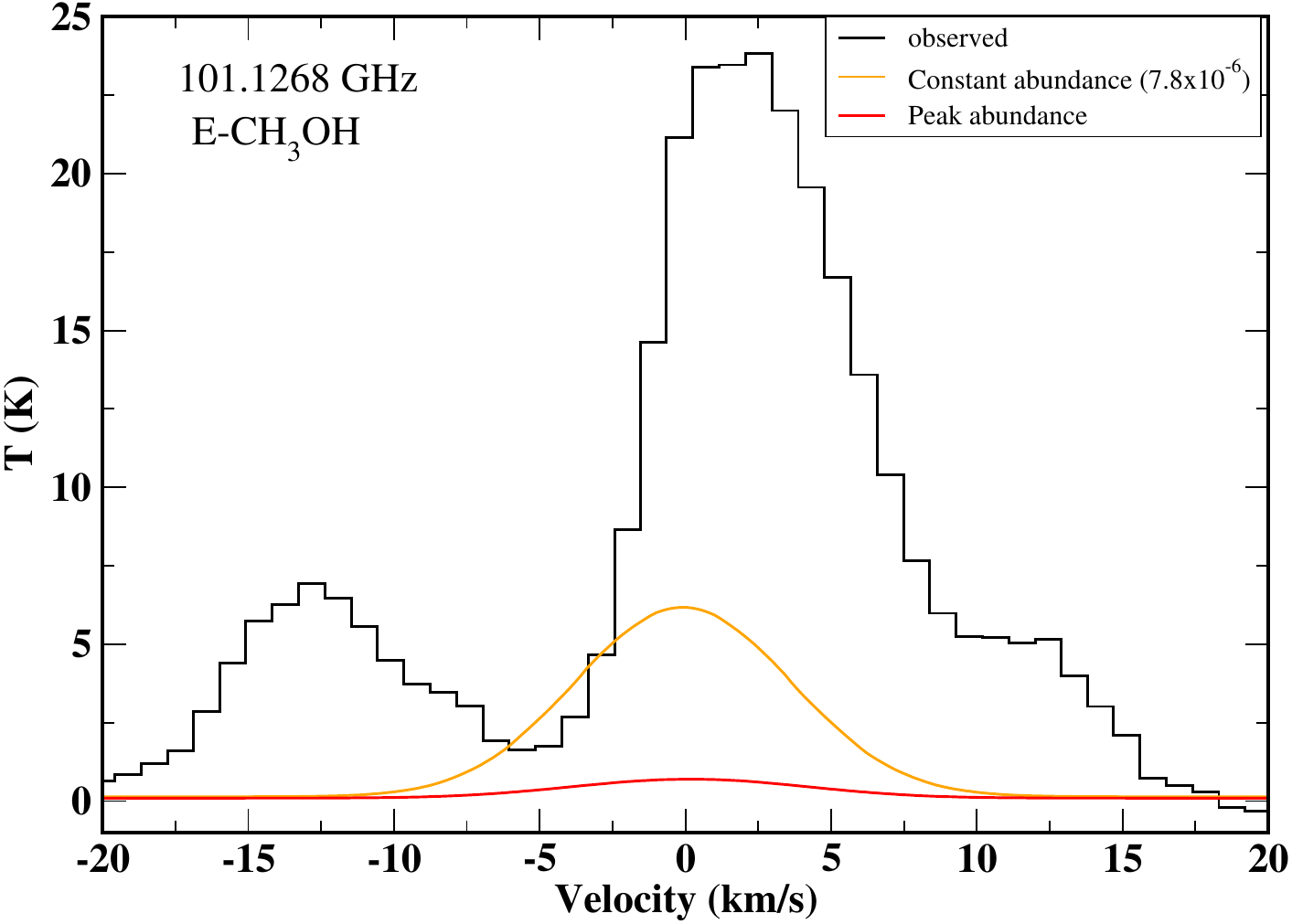}
\end{minipage}
\begin{minipage}{0.32\textwidth}
\includegraphics[width=\textwidth]{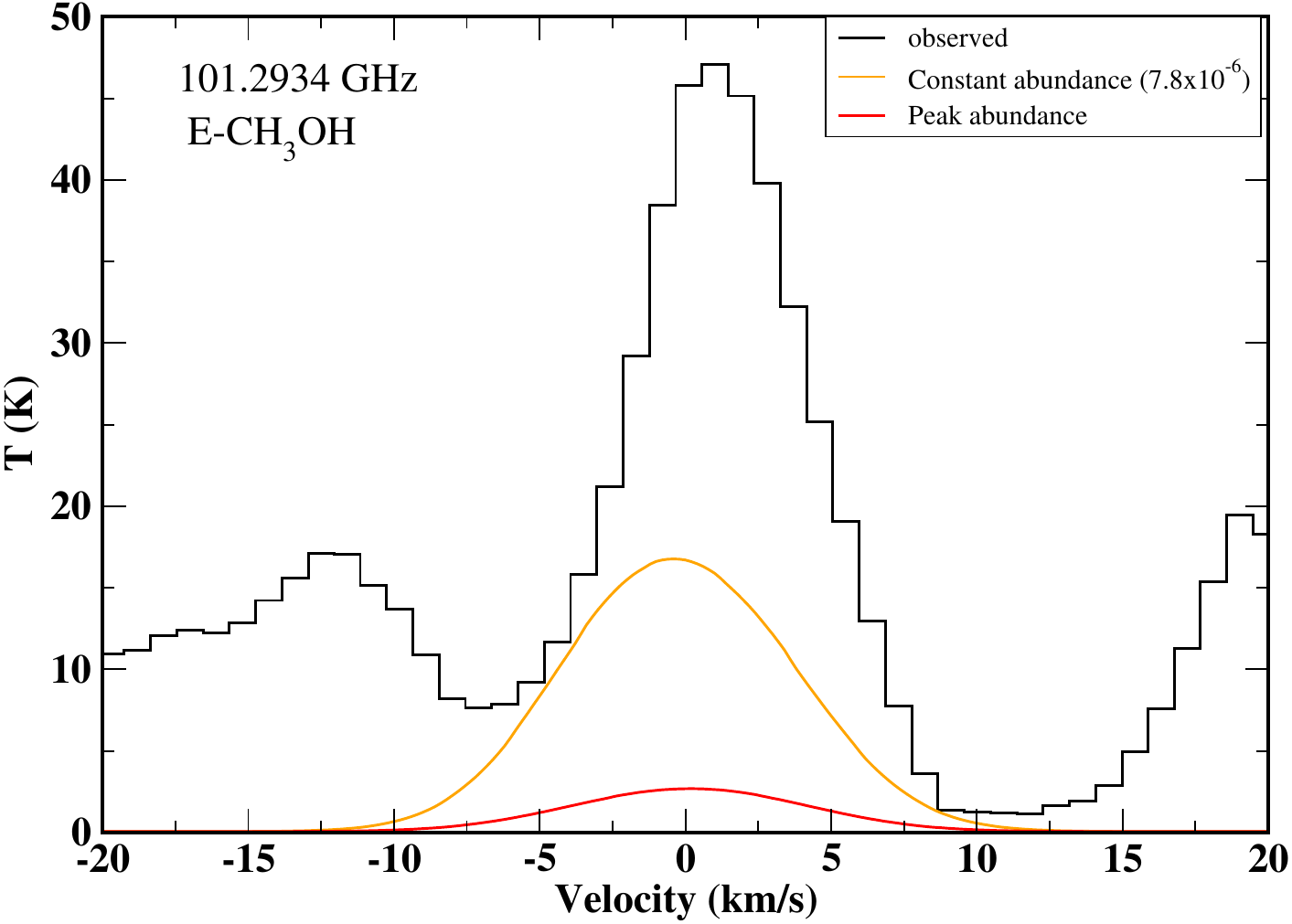}
\end{minipage}
\hskip -0.9 cm
\begin{minipage}{0.32\textwidth}
\includegraphics[width=\textwidth]{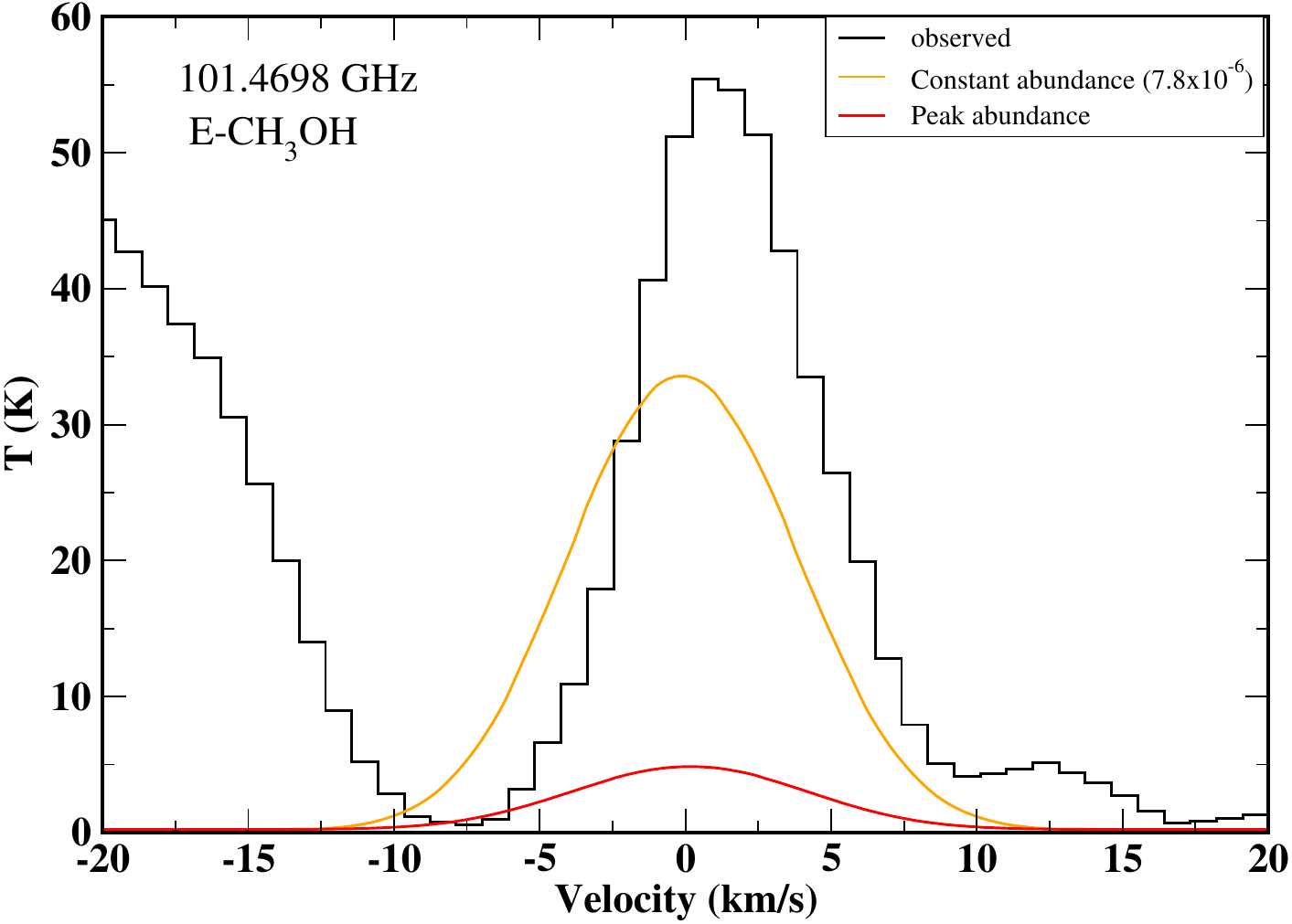}
\end{minipage}
\hskip 0.6cm
\caption{The modeled transitions of $\rm{A-CH_3OH}$ (first four transitions) and $\rm{E-CH_3OH}$ (last three transitions) along with their observed transitions (black lines) are shown. The line profile obtained with the constant abundance is displayed with the orange lines. The best fit for A-CH$_3$OH is obtained when a methanol abundance of $\sim 2 \times 10^{-6}$, an FWHM $\sim 8$ km/s, and $\beta=1.4$ are used. For E-CH$_3$OH, the FWHM and $\beta$ are kept the same as A-CH$_3$OH, but a constant abundance of $\sim 7.8 \times 10^{-6}$ is used. The red lines in the figure show the modeled line profiles when the peak spatial distribution of the methanol abundances from Fig. \ref{fig:abundance} is used. \citep[Courtesy:][]{bhat22}}
\label{fig:ch3oh-ratran}
\end{figure*}

\begin{figure*}
\begin{minipage}{0.32\textwidth}
\includegraphics[width=\textwidth]{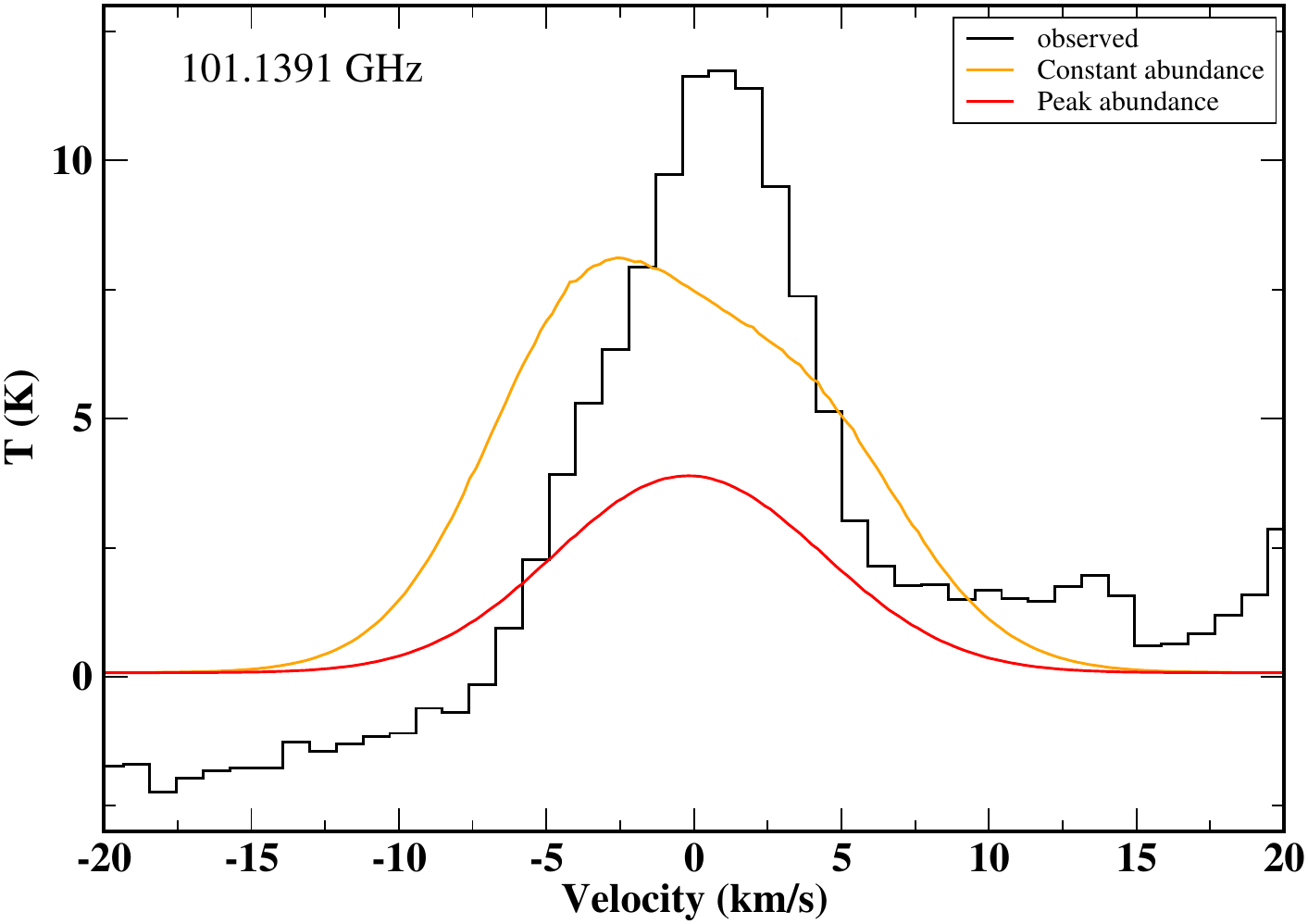}
\end{minipage}
\begin{minipage}{0.32\textwidth}
\includegraphics[width=\textwidth]{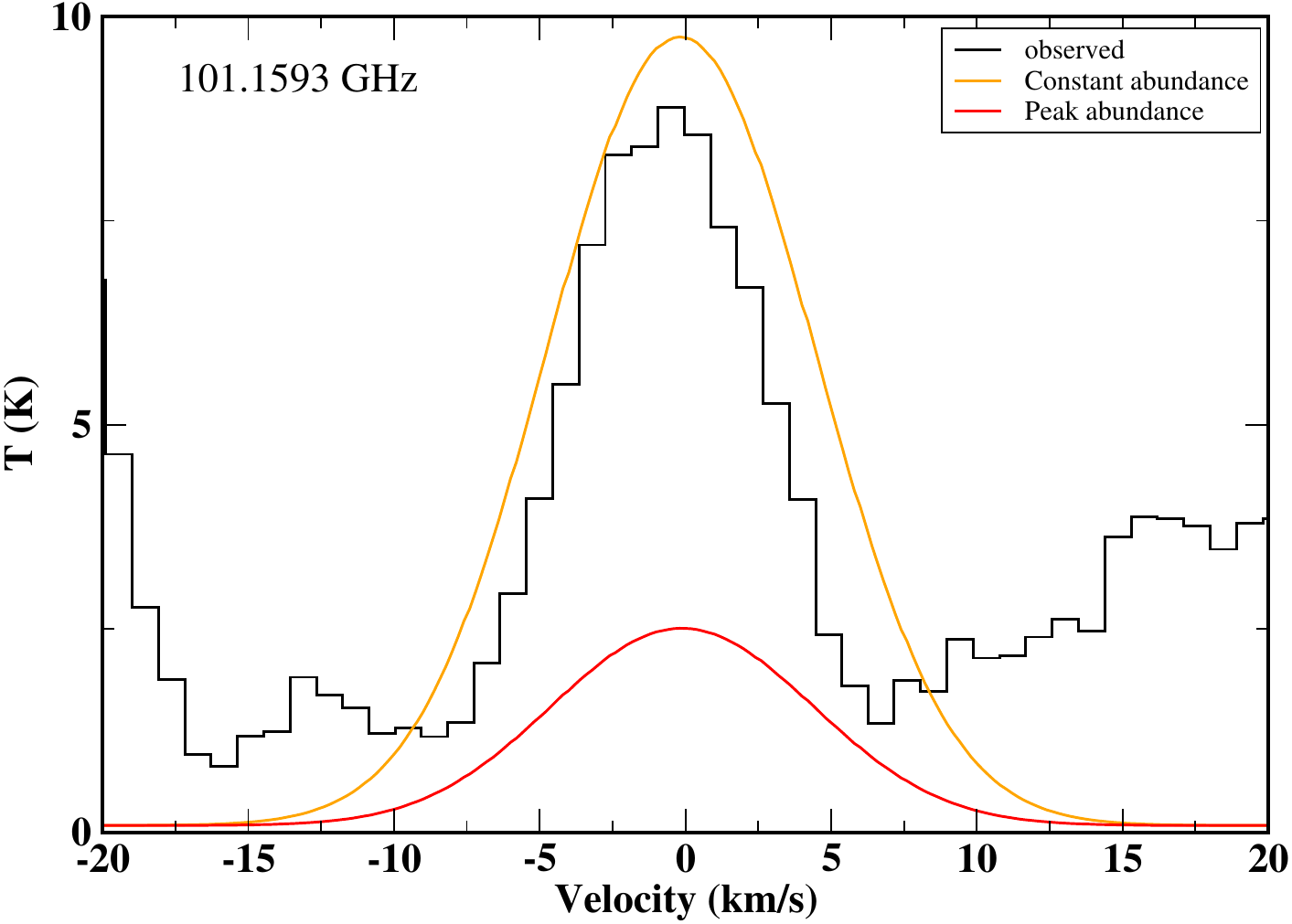}
\end{minipage}
 \begin{minipage}{0.32\textwidth}
 \includegraphics[width=\textwidth]{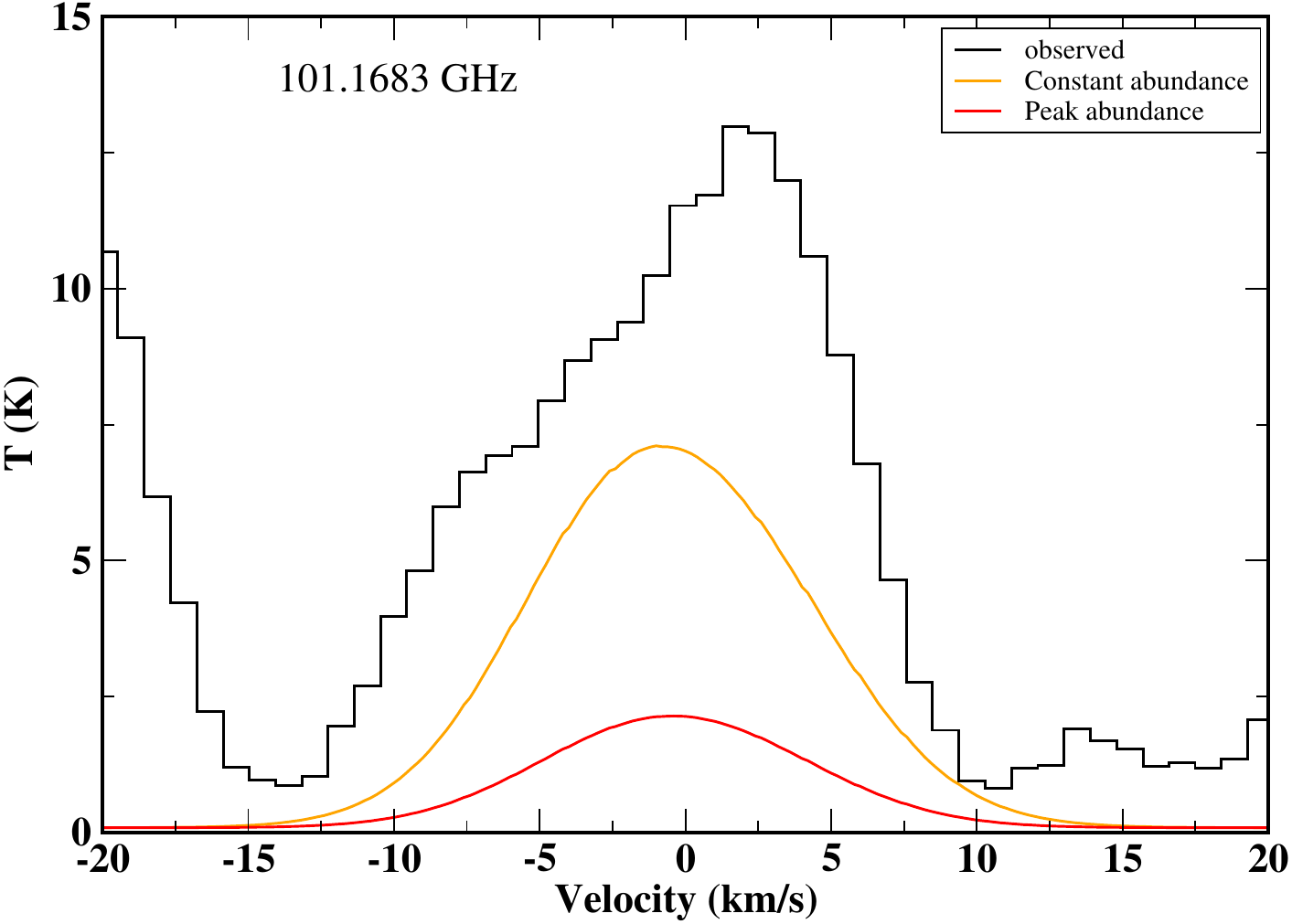}
 \end{minipage}
\hskip -0.9cm
\caption{The modeled line profiles of CH$_3$SH along with its observed line profiles are shown.
The best fit is obtained when a constant abundance $\sim 1.9 \times 10^{-8}$, an FWHM $\sim 9.45$ km/s, and $\beta=1.4$ are used. Observed line profiles are shown in black, whereas the modeled line profiles with the constant abundance are shown in orange. The line profiles obtained with the peak spatial distribution of the abundance profile are shown in red. \citep[Courtesy:][]{bhat22}}
\label{fig:ch3sh-ratran}
\end{figure*}

\begin{figure*}
\begin{minipage}{0.32\textwidth}
\includegraphics[width=\textwidth]{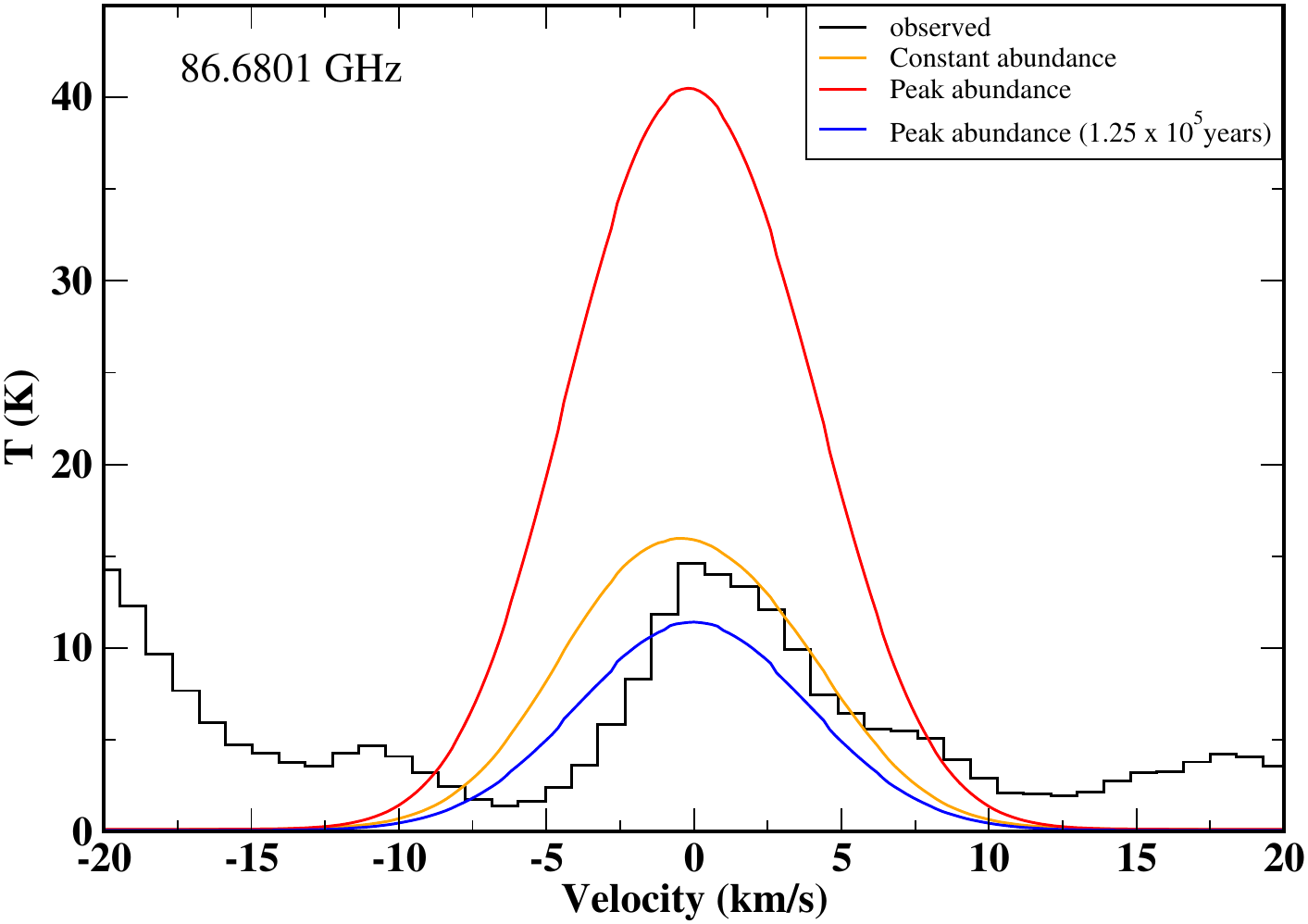}
\end{minipage}
\begin{minipage}{0.32\textwidth}
\includegraphics[width=\textwidth]{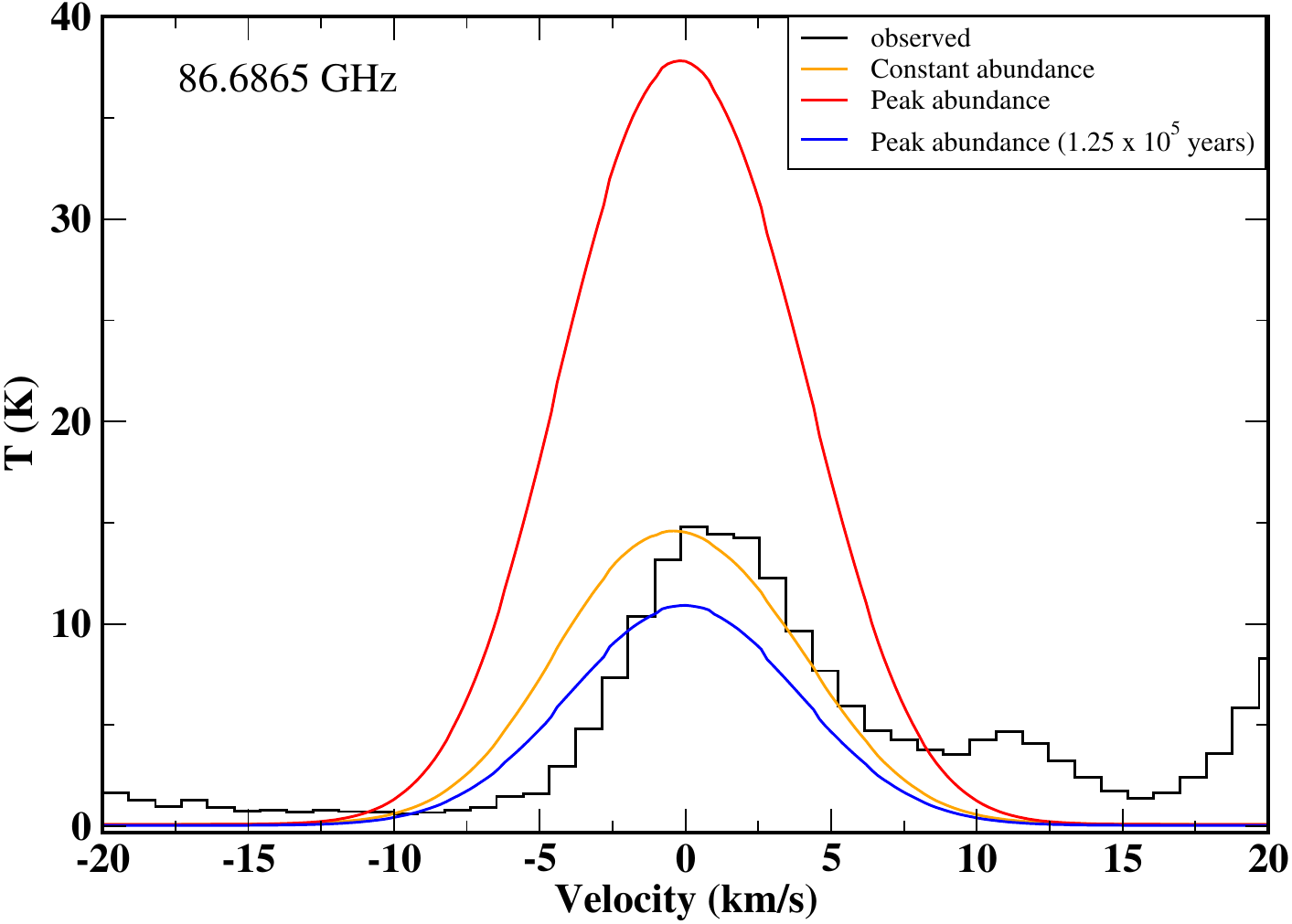}
\end{minipage}
 \begin{minipage}{0.32\textwidth}
 \includegraphics[width=\textwidth]{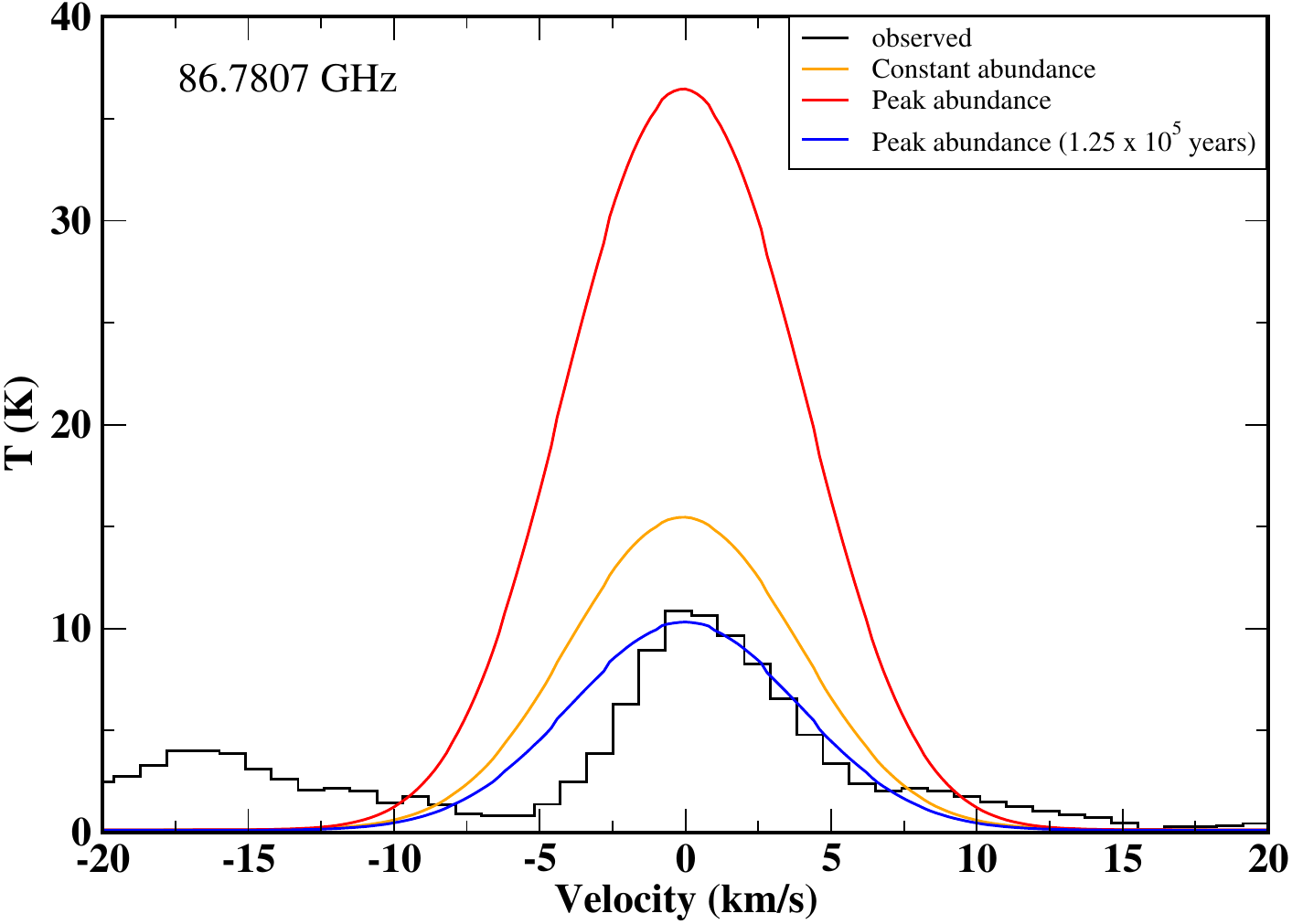}
 \end{minipage}
\hskip -0.9cm
\begin{minipage}{0.32\textwidth}
\includegraphics[width=\textwidth]{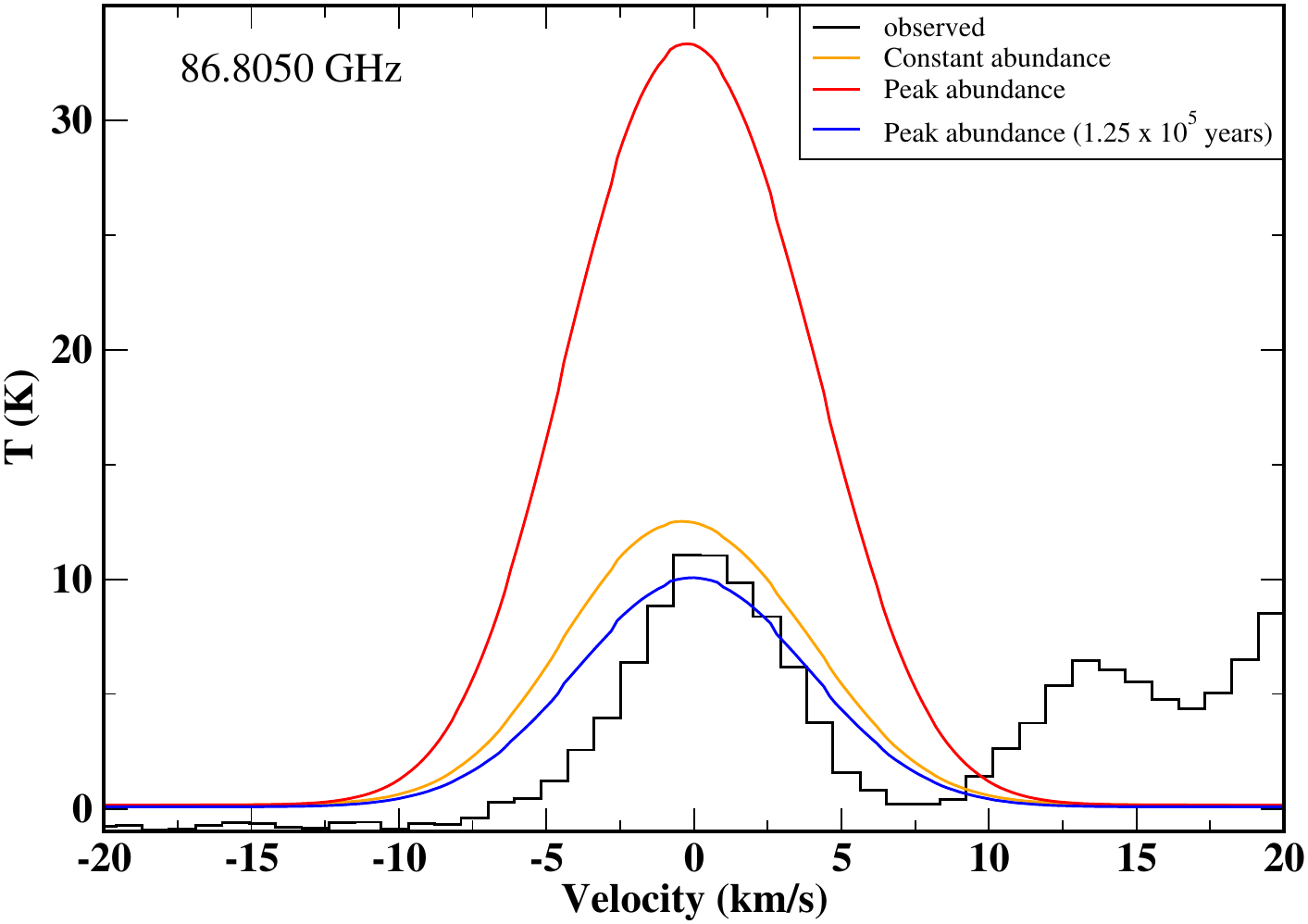}
\end{minipage}
\hskip 0.90cm
\begin{minipage}{0.32\textwidth}
\includegraphics[width=\textwidth]{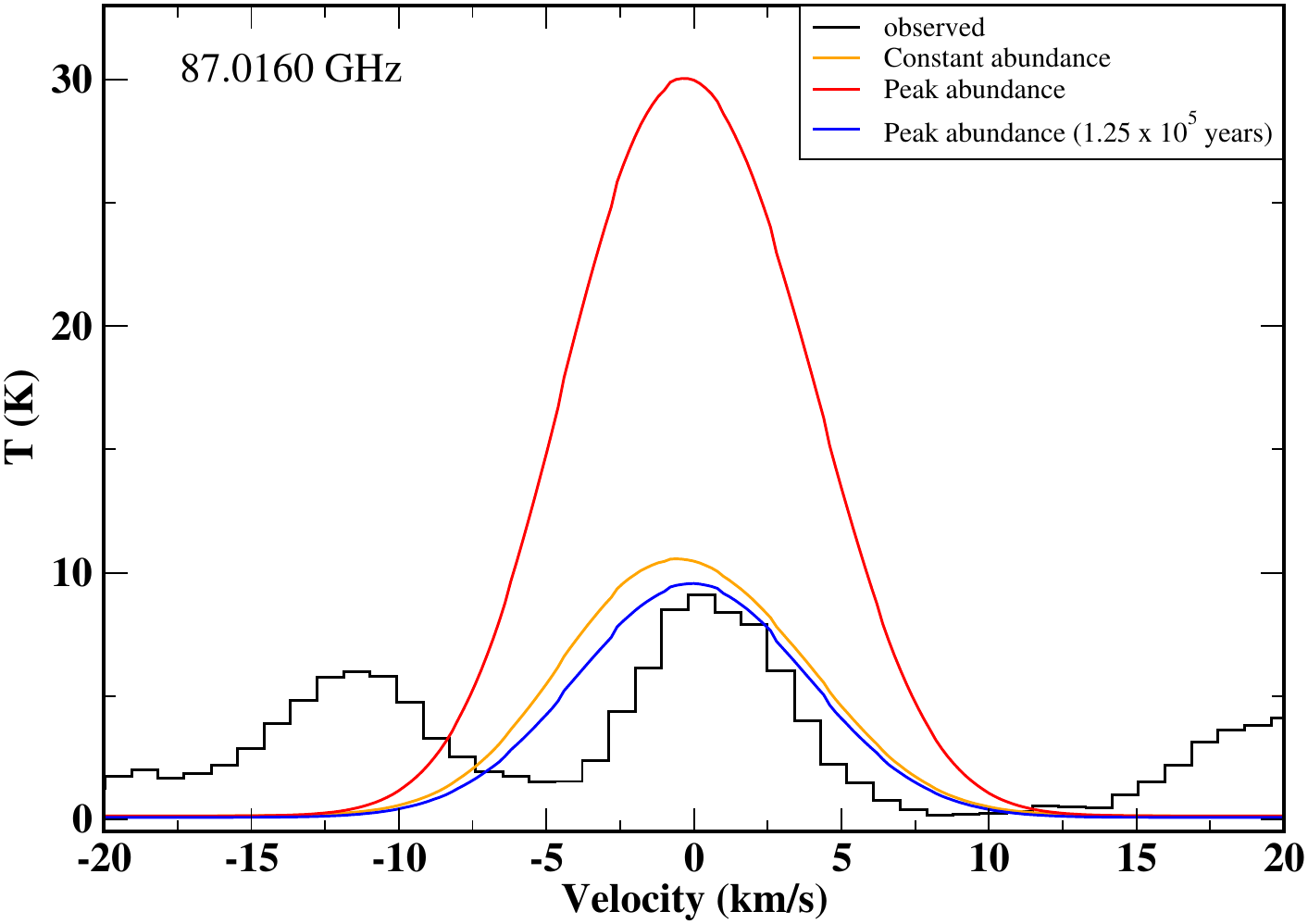}
\end{minipage}
\hskip -0.9 cm
\caption{A comparison between the observed (black line) CH$_3$NCO line profiles and the modeled line profiles is shown.  The modeled line profiles with the constant abundance are shown with the orange lines. The best fit is obtained when an FWHM of $\sim 7.5$ km/s, a constant abundance of $\sim 5 \times 10^{-9}$, and $\beta=1.4$ are used. The modeled line profiles with the peak values are shown with the red lines. It is noticed that our peak abundance profile over-predict the intensity of these transitions. An additional peak abundance profile (extracted at $1.25 \times 10^5$ years from our CMMC model) is used which shows a good fit (blue curve) with the observation. \citep[Courtesy:][]{bhat22}}
\label{fig:ch3nco-ratran}
\end{figure*}
 
 \subsubsection{\rm{\bf{Complex organic molecules}}}
 Recently, many transitions of various complex organic compounds in G31 were reported by \cite{gora21}. Here, modeled spectra of the CH$_3$OH, CH$_3$SH, and CH$_3$NCO emission features that have been observed are presented. To recognize the outflow present in this source, \cite{aray08} observed the transition of $\rm{CH_3OH}$ in G31. The peak abundance of CH$_3$OH has a jump at $\sim 10000$ AU and reaches a maximum value ($\sim 1.9 \times 10^{-6}$) (see Figure \ref{fig:abundance}). Methanol's peak abundance and final abundance obtained from our chemical modeling are very different. Deep inside the cloud, it significantly dropped from its peak value. A transition of H$_2$CO was detected, and its column density was calculated by \cite{gora21}. The abundance variation of H$_2$CO is displayed in Figure \ref{fig:abundance}. A jump profile similar to methanol is also shown and has a peak abundance $1.2 \times 10^{-6}$. To explain the observed abundance of CH$_3$SH and CH$_3$NCO in G31, this is the first exclusive modeling result presented. Here, the pathways indicated in \cite{gora20} and \cite{gor17a} explain the chemical evolution of CH$_3$NCO and CH$_3$SH. A similar pattern may be seen in the peak abundance profile of CH$_3$SH and CH$_3$OH. It exhibits a surge at $10000$ from its lowest value and reaches its highest abundance of $\sim 1.4 \times 10^{-8}$. The final abundance of CH$_3$SH in the inner region was significantly lower than that of CH$_3$OH. The formation of CH$_3$NCO in the late stages of the simulation via the reaction between HNCO and CH$_3$ in the gas phase (with a rate of $\sim 5 \times 10^{-11}$ cm$^3$ s$^{-1}$) results in a comparable value for the peak and final abundances. According to \cite{gora20}, the activation barrier for the reaction between H and NH$_2$CHO is applied here at its highest value ($3130$ K). At about $6000$ AU, CH$_3$NCO reaches a peak abundance of $\sim 3.3 \times 10^{-8}$. Figure \ref{fig:comp} compares the abundances obtained from our simulation and observation (obtained from \cite{gora21}) using LTE fitting, rotation diagrams, and MCMC(\ref{sec:MCMC_G31}) fitting for the COMs. It demonstrates that the observation and the abundance obtained from our model for CH$_3$OH, H$_2$CO, CH$_3$SH, and CH$_3$NCO are consistent. A molecular hydrogen column density of $\sim 1.53 \times 10^{25}$ cm$^{-2}$ is used to calculate the molecular abundance from the column density \citep{das08a,das10,das11,gor17a}. In the interstellar grain surface, where CH$_3$OH and CH$_3$SH are primarily generated. These species can move rapidly to the gas phase in the warmer area (temperature $>100$ K). After the release of grain phase HNCO, a sizable amount of gas-phase CH$_3$NCO can be produced \citep{gora20}. Therefore, the chemical process related to the grain significantly influences the abundance of these three species in the gas phase. We have divided the entire cloud into $23$ shells. For instance, the temperature of the outermost shell is permitted to change up to $23$ K. It is up to $1593$ K in the innermost shell. Therefore, it continues to be below $100$ K in the case of the outer grids (beyond grid number $13$). The peak value of these complex organic molecules for these shells appears at the end of the simulation. Inside the $13$th grid, the peak value slowly rises in a shorter amount of time. Briefly stated, our time uncertainty in projecting the peak value is $\sim 1.5 \times 10^5$ years because the peak values noted in Table \ref{table:abundances} and depicted in Figure \ref{fig:abundance} are outside the collapsing time scale. These time scales could be affected by various factors (physical and chemical characteristics). These COMs' observed line profiles are initially modeled using constant spatial abundances. As can be seen, all the transitions of these complex organic molecules can be found in emission (see Figures \ref{fig:ch3oh-ratran}, \ref{fig:ch3sh-ratran}, and \ref{fig:ch3nco-ratran}). The LAMDA database is used for the collisional data of E-CH$_3$OH and A-CH$_3$OH with H$_2$. The best fit is obtained for the methanol transitions depicted in Figure \ref{fig:ch3oh-ratran} when an A-CH$_3$OH abundance of $\sim 2 \times 10^{-6}$ and an FWHM of 8 km/s are utilized. It agrees with the abundance of methanol (A-CH$_3$OH) determined by the MCMC fit ($\sim 1.4 \times 10^{-6}$) of these transitions presented in Table \ref{table:mcmc_lte_G31}. An excellent fit can be seen between the observation and the modeled profile for the A-CH$_3$OH transitions at $88.594787$ and $88.939971$ GHz. However, the transitions of the first two panels in Figure \ref{fig:ch3oh-ratran} do not fit well. The intensities of each transition are underproduced with the peak abundance profile. According to table \ref{table:mcmc_lte_G31}, E-CH$_3$OH requires a $3.9$ times higher column density than A-CH$_3$OH. The abundance $\sim 7.8 \times 10^{-6}$ is used here for E-CH$_3$OH after that. For A-CH$_3$OH an FWHM of 8 km/s is used as well. However, we cannot match all three transitions of E-CH$_3$OH (the last three panels of Figure \ref{fig:ch3oh-ratran}) using the 1D RATRAN model. Similar to A-CH$_3$OH, using the peak abundance profile in this example also shows that the intensities of all three transitions are underproduced. These three transitions are a little off-center (101.126857 GHz, 101.293415 GHz, and 101.469805 GHz). Additionally, HOCH$_2$CN, CH$_3$OCN, and s-propanal may be detected at 101.1269 GHz, 101.2927 GHz, and 101.2935 GHz. Since \cite{gora21} did not find any other transitions of these species in the other part of their spectrum, they excluded these from their analysis. Only three of the seven CH$_3$SH observed transitions are depicted in Figure \ref{fig:ch3sh-ratran}, and only five of the six CH$_3$NCO observed transitions are depicted in Figure \ref{fig:ch3nco-ratran}. The non-LTE calculation requires the collisional data file. The collisional data files were available for the molecules H$^{13}$CO$^+$, HCN, NH$_3$, SiO, and E-CH$_3$OH, A-CH$_3$OH. Since no such collisional data were available for CH$_3$NCO and CH$_3$SH, the collisional data of HNCO and $\rm{A-CH_3OH}$ with H$_2$ are considered in their place. These files are prepared in the prescribed format of the RATRAN modeling. We must accept that these are extremely crude assumptions, but they are used to verify the reasonable guesses of the line profiles. Our file does not contain all of the transitions of the CH$_3$SH and CH$_3$NCO because collisional rates for only a few levels were provided.

A comparison of the CH$_3$SH line profiles from the observed and modeled data is shown in Figure \ref{fig:ch3sh-ratran}. A constant abundance of $\sim 1.9 \times 10^{-8}$ and an FWHM of $9.45$ km/s for CH$_3$SH are utilized to obtain the best-fitted model. Constant abundance does not suit well, and the peak abundance profile underproduces the intensity, as seen in Figure \ref{fig:ch3sh-ratran}. This is due to the fact that three transitions of CH$_3$SH are blended (see Figure \ref{fig:ch3sh-mcmc}). Similarly, Figure \ref{fig:ch3nco-ratran} compares the observed and predicted line profile. The constant abundance used for the best-fitted model is $\sim 5.0 \times 10^{-9}$, and the FWHM for CH$_3$NCO is $7.5$ km/s. Additionally, our radiative transfer model uses the radial distribution of the abundance profiles (from Figure \ref{fig:abundance}) to calculate the simulated line profiles of CH$_3$OH, CH$_3$SH, and CH$_3$NCO. Derived line profiles using peak values are displayed in red. The peak abundance profile overestimates the intensity of CH$_3$NCO transitions. At the final stages of our simulation the abundance of CH$_3$NCO peaks. By considering distinct abundance profiles extracted at various times, several studies using the RATRAN model are conducted. The peak abundance profile of CH$_3$NCO is extracted at $1.25 \times 10^5$ years, which yields the best fit (corresponding to a warmup time of $2.5 \times 10^4$ years after the isothermal collapse phase).
 
\section{Summary}
This chapter develops a thorough chemical and radiative transfer model to account for various observed line profiles in G31. Our CMMC model was combined with the physical characteristics found in this region to simulate a realistic scenario. A spatial distribution of the abundances of certain important interstellar species in G31 has been produced by this coupled model. The obtained quantities were adequately explained by our coupled model.
From our radiative transfer models, different physical properties like infall velocity, mass infall rate, FWHM, etc. are extracted. Table \ref{table:RATRAN-best-fit} provides a summary of the modeled and observed parameters. The following are the main highlights of this work:

$\bullet$
The abundance for CH$_3$NCO and CH$_3$SH using the LTE model \cite{gora21} was $4.72 \times 10^{-8}$ and $2.7 \times 10^{-8}$, respectively. They obtained $\sim 1.04 \times 10^{-9}$ and $1.86 \times 10^{-9}$ for CH$_3$NCO and CH$_3$SH, respectively, using the rotational diagram analysis.
For these two species, the MCMC model presented in section \ref{sec:MCMC_G31} obtained abundance values of $7.84 \times 10^{-8}$ and $9.15 \times 10^{-9}$, respectively. A maximum peak abundance of CH$_3$NCO and CH$_3$SH has been reported by our CMMC model to be $\sim 3.3 \times 10^{-8}$ and $\sim 1.4 \times 10^{-8}$, respectively. \\

$\bullet$
An infall velocity of $2.3$ km/s from the H$^{13}$CO$^+$ observation was reported by \cite{gora21}. An infall velocity of $2.5$ km/s is found here using the two slab model of H$^{13}$CO$^+$. Using this infall velocity and taking G31's distance of 3.7 kpc into account, a mass infall rate of $1.3 \times 10^{-3}$ M$_\odot$ yr$^{-1}$ is calculated. With the 1D RATRAN model, an infall velocity of $4.9$ km/s at 1000 AU yields the best match, which is consistent with the results from \cite{osor09}.

$\bullet$ It is observed that the dust emissivity significantly influences the line profile. The power-law emissivity relation of dust emission can be used to explain every line profile that has been seen. The observed line profiles of H$^{13}$CO$^+$, SiO, and CH$_3$CN can be reproduced using $\beta=1$, whereas the observed line profiles of other species may be generated using $\beta=1.4$. Best fitted modeled abundance, FWHM, and $\beta$ are summarized in Table \ref{table:RATRAN-best-fit}.\\

$\bullet$ The observed line profiles of H$^{13}$CO$^+$, HCN, SiO, NH$_3$, CH$_3$CN, CH$_3$OH, CH$_3$SH, and CH$_3$NCO are explained using the 1D RATRAN code. All of the line profiles can be well explained by an infalling envelope. Our model must include an additional outflow component only to explain the line profile of SiO. \\

$\bullet$ For various resolutions, the line profiles of H$^{13}$CO$^+$ and CH$_3$CN are modelled. It is noted that an inverse P-Cygni nature of CH$_3$CN might be anticipated with a higher angular resolution. In contrast, an inverse P-Cygni nature of the H$^{13}$CO$^+$ could be anticipated with lower angular resolution. These results are in line with \cite{gora21} and \cite{belt18}.

\begin{table}
{\scriptsize
 \caption{Abundance, linewidth, and $\beta$ obtained from our best-fitted RATRAN model are noted. For the comparison, we have reported the abundances obtained by the other methods (i.e., MCMC method discussed in section \ref{sec:MCMC}, and LTE and rotational method carried out by \cite{gora21}). Moreover, the abundances obtained from our chemical model are also noted (peak abundance noted in Table \ref{table:abundances}). \label{table:RATRAN-best-fit}. \citep[Courtesy:][]{bhat22}}
 \hskip -2.0cm
\begin{tabular}{|l|l|l|l|l|l|l|l|}
  \hline
  \hline
  Species & \multicolumn{5}{|c|}{Abundance}& Line width &$\beta$\\
  &RATRAN&LTE$^g$&Rotational diagram$^g$&MCMC$^h$&Chemical model&(km.s$^{-1}$)&\\
  \hline
  \hline
 H$^{13}$CO$^+$& 7.08$\times$10$^{-11}$&-&-&-&1.16$\times$10$^{-10}$-1.66$\times$10$^{-7}$&1.42&1.0\\
 HCN& 7.6$\times$10$^{-8}$&-&-&-&2.76$\times$10$^{-10}$-5.35$\times$10$^{-7}$&10.00&1.4\\
 SiO&9.5$\times$10$^{-10}$ &9.54$\times$10$^{-11}$&-&-&1.46$\times$10$^{-15}$-9.22$\times$10$^{-9}$&4.67&1.0\\
 A-CH$_3$OH&2.0$\times$10$^{-6}$&1.2$\times$10$^{-6}$&1.92$\times$10$^{-6}$&1.37$\times$10$^{-6}$&5.93$\times$10$^{-11}$-1.88$\times$10$^{-6}$&8.00&1.4\\
 E-CH$_3$OH&7.8$\times$10$^{-6}$&1.2$\times$10$^{-6}$&1.92$\times$10$^{-6}$&5.36$\times$10$^{-6}$&5.93$\times$10$^{-11}$-1.88$\times$10$^{-6}$&7.67&1.4\\
 CH$_3$SH&1.9$\times$10$^{-8}$&2.7$\times$10$^{-8}$&1.86$\times$10$^{-9}$&9.15$\times$10$^{-9}$&2.41$\times$10$^{-16}$-1.39$\times$10$^{-8}$&9.45&1.4\\
 CH$_3$NCO&5.0$\times$10$^{-9}$&4.72$\times$10$^{-8}$&1.04$\times$10$^{-9}$&7.84$\times$10$^{-8}$&9.02$\times$10$^{-11}$-3.27$\times$10$^{-8}$&7.49&1.4\\
 NH$_3$ (100M)&2.0$\times$10$^{-9}$-1.0$\times10^{-7}$&-&-&-&3.04$\times$10$^{-10}$-7.53$\times$10$^{-7}$&4.90-8.33&1.4\\
 NH$_3$ (VLA)&1.0$\times$10$^{-7}$-1.6$\times$10$^{-7}$&-&-&-&3.04$\times$10$^{-10}$-7.53$\times$10$^{-7}$&4.90-8.33&1.4\\
 CH$_3$CN&6.0$\times$10$^{-8}$&&-&-&6.72$\times$10$^{-15}$-3.81$\times$10$^{-8}$&2.5&1.0\\
  \hline
  \hline
 \end{tabular}}\\
 {\noindent $^g$ \cite{gora21}\\$^h$ Using N$_{H2}$ = 1.53$\times$10$^{25}$ from \cite{gora21}}
 \end{table}

 \chapter{Radiative Transfer Model: Phosphorous Bearing Species in Various Regions of the Interstellar medium} \label{chap:RATRAN_phosphorus}

Compared to species containing hydrogen, carbon, nitrogen, oxygen, and sulfur, phosphorus (P)- containing species are not as widely distributed in space. Consequently, not many P-containing molecules are observed in the interstellar medium and circumstellar envelopes. Severe uncertainties are placed on simulating the P-chemistry due to the limited discovery of the P-bearing species. We use a radiative transfer model to examine the transitions of various P-bearing species in the diffuse cloud and hot core regions and to estimate the line profiles.

The chemical evolution of galaxies strongly relies on the element phosphorus and its compounds. Phosphorus is one of the primary biogenic elements and a necessary component of life. The Atacama Large Millimeter/submillimeter Array (ALMA) and European Space Agency Probe Rosetta suggested evidence that P-related species may have migrated to the Earth through comets \citep{altw16,rivi20}. P-bearing molecules perform a wide range of biochemical activities as the primary building blocks of every living system. Large biomolecules or living organisms are formed by P-bearing molecules, which act as precursors in the creation of RNA and DNA and store and transfer genetic information in nucleic acids and nucleotides \citep{maci97}. Additionally, these molecules are primary building blocks \citep[main characteristic features of cellular membranes,][]{maci05} of phospholipids.

In the interstellar medium (ISM), phosphorus is comparatively rare. But it is very common in many meteorites \citep{jaro90,pase19,lodd03}. P is the eleventh most common element in the Earth's crust and the thirteenth most abundant element in a typical meteoritic substance.

In hot regions ($\sim 1200$ K), P$^+$ was found to have a cosmic abundance of $\sim 2\times10^{-7}$. Around evolved stars \citep{guel90,tene07,agun07,agun08,tene08,half08,mila08,debe13,agun14,ziur18}, P-bearing molecules like PN, PO, HCP, CP, CCP, and PH$_3$ have been observed in circumstellar envelopes. PN was found in various star-forming areas, including \citep{turn87, ziur87, turn90, caux11, yama11, font16, mini18, font19}. For a long time, \citep{turn87, ziur87, turn90, font16}, it was the only P-containing species found in dense ISM. Using the IRAM 30m telescope, \cite{rivi16} reported the first observation of PO near two large star-forming areas, W51 1e/2e and W3(OH), as well as PN.

\cite{mill91, char94, aota12, lefl16, rivi16, jime18} have discussed the P-chemistry comprehensively. In the dense cloud region,  P-chemistry is not well-constrained. The depletion factor of the initial elemental abundance of P is the biggest uncertain factor. The precise depletion of various elements onto the grains and the degree of complexity of gas-phase abundance inside the gas are still unknown. \cite{nguy20} made the chemical desorption of phosphine a recent discovery. This kind of research is essential for limiting the modeling parameters. The phosphorus cousin of ammonia (NH$_3$), phosphophine (PH$_3$), is a relatively stable molecule that could contain a sizable portion of phosphorus in various astronomical conditions. Researchers regard PH$_3$ as a biosignature \citep{sous20}. PH$_3$ have been found in the planets in our solar system having a reducing atmosphere. The Voyager data confirmed PH$_3$ to be present in Jupiter and Saturn, with volume mixing ratios of 0.6 and 2 ppm, respectively. The P/H ratio is in agreement with the solar value. Using the 2.5 cm$^{-1}$ spectral resolution of the Cassini/CIRS data, \cite{flet09} calculated the global distribution of PH$_3$ on Jupiter and Saturn. High temperatures and pressures inside the reducing atmospheres of the large planets \citep{breg75,tarr92} might generate PH$_3$, which would then be dredged upward by convection \citep{noll97,viss06}. Very interestingly, Venus lacks a reducing atmosphere of this kind, though the deck of the atmosphere there was assumed to have $\sim 20$ ppb of PH$_3$.

The steady-state chemical models, including the photochemical routes, could not account for such a large amount of PH$_3$. They also looked for various abiotic explanations for this high abundance. However, none of them appear to be appropriate. They assumed it had a mysterious photo or geochemical origin. More broadly, it was impossible to rule out the possibility of its high abundance by some biological means. Based on their most recent knowledge, \cite{bain20} could not explain the existence of PH$_3$ in Venus's clouds by any abiotic mechanism. This finding prompted several discussions. The analysis and interpretation of the spectroscopic data used in \cite{grea20} were the subject of a very recent \cite{vill20} query. These authors asserted that Venus's atmosphere is devoid of PH$_3$. \cite{snell20} also found no statistical evidence for PH$_3$ in the atmosphere of Venus.

The simple P-bearing species PH$_3$ was tentatively found in the envelope of the carbon-rich stars IRC+10216 and CRL 2688 \citep{tene08,agun08} (J = 1 $\rightarrow$ 0, $266.9$ GHz). Later, \cite{agun14} verified that PH$_3$ was present. Using the HIFI instrument on board Herschel, they noticed the J = 2 $\rightarrow$ 1 rotational transition of PH$_3$ (at $534$ GHz) in IRC+10216. They projected that PH$_3$ would be extremely abundant in this area. If so, it should happen via a gas phase reaction with no barrier or one with a minimal barrier that takes place at low temperatures. Here, we use a variety of cutting-edge chemical models to understand how PH$_3$ forms under diverse interstellar circumstances, such as dilute clouds, interstellar photon-dominated regions or photodissociation regions (PDRs), and hot core regions.

\cite{chan20} recently attempted to use the IRAM 30m telescope to observe the small extragalactic quasar B0355+508 along the line of sight to HCP (2-1), CP (2-1), PN (2-1), and PO (2-1). Unfortunately, they failed to notice these changes in their line of sight. However, based on their observations, they projected 3$\sigma$ upper limits for these transitions. However, along the same line of sight, they were able to successfully identify HNC (1-0), CN (1-0), and C$^{34}$S (2-1) in absorption and $^{13}$CO (1-0) in emission.

\section{1D-RATRAN radiative transfer model \label{sec:RATRAN_P}}
In this study, we used the 1D radiative transfer model \citep{hoge00} to investigate the transitions of some relevant P-bearing species in the diffuse cloud region to  hot core/corino region.

\begin{figure} 
\centering
\includegraphics[width=8cm, height=6cm]{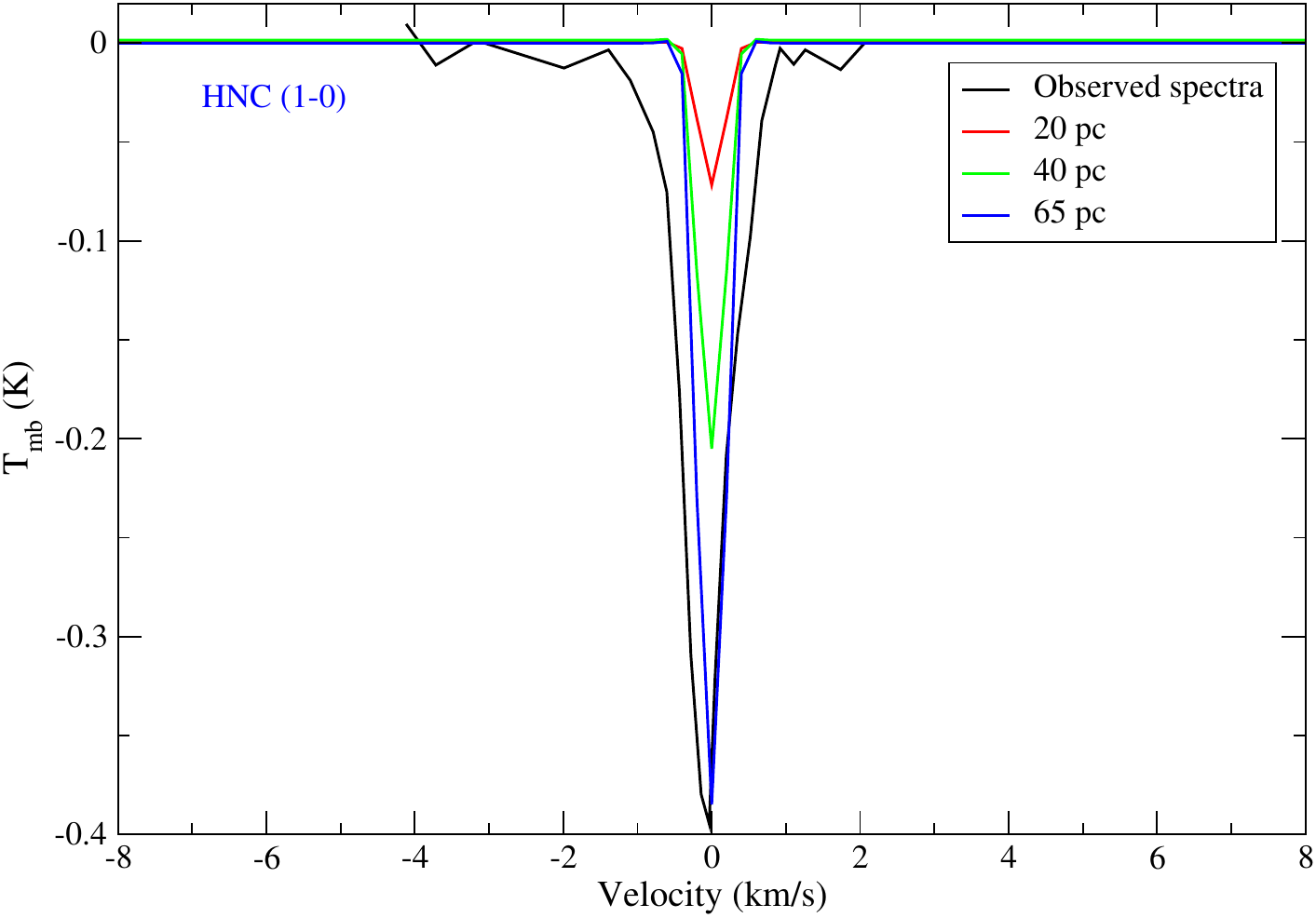}
\caption{Modeled $1 \rightarrow 0$ transitions of HNC by considering the different sizes of the diffuse cloud in the RATRAN model. It is interesting to note that the intensity of the absorption increases with the increase in the size of the cloud. Abundance of $2\times10^{-10}$ and Doppler parameter of 0.2 km/s are used. \citep[Courtesy:][]{sil21}}
\label{fig:cloud_size}
\end{figure}

\begin{figure*} 
\centering
\includegraphics[width=7cm, height=6cm]{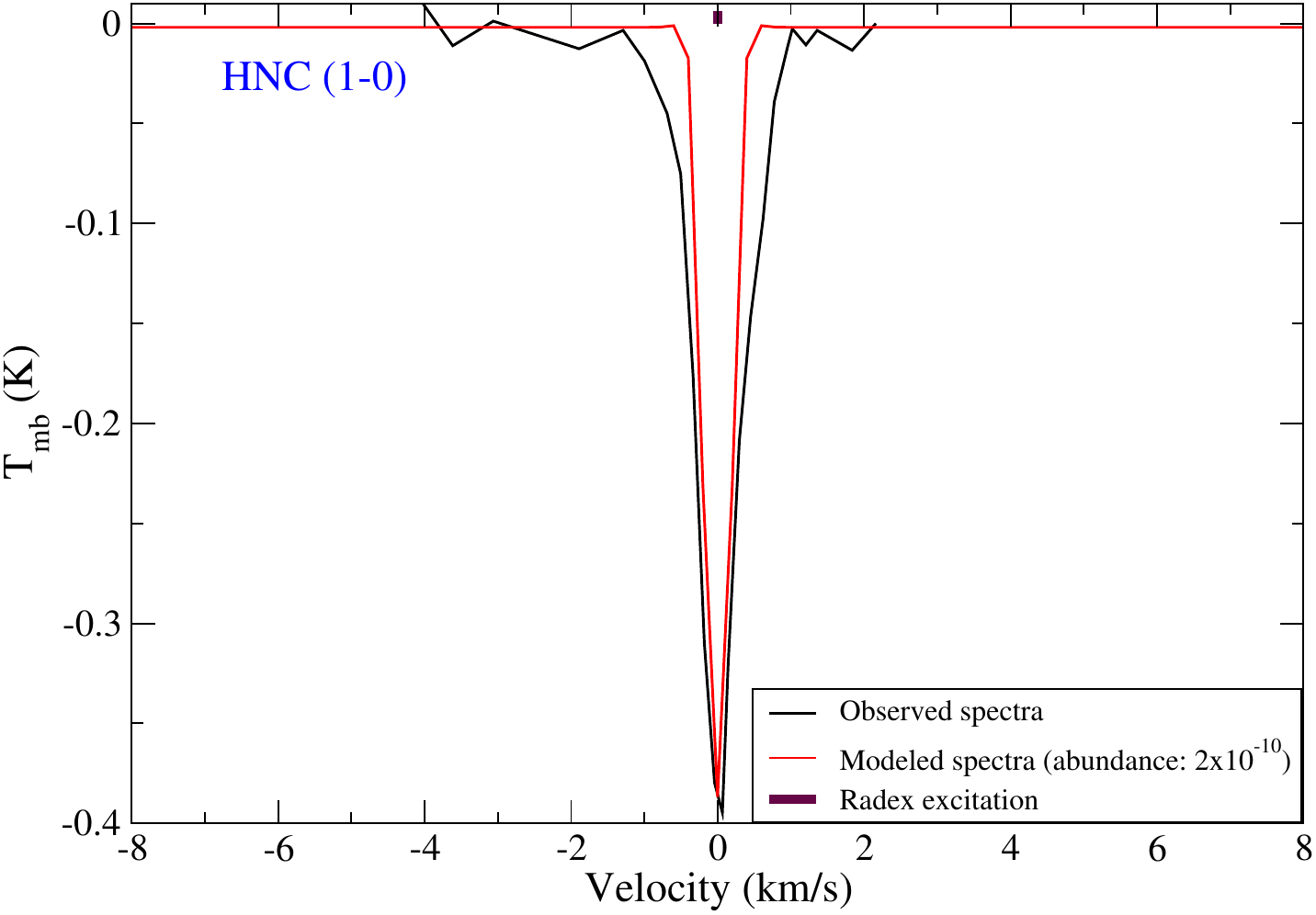}
\includegraphics[width=7cm, height=6cm]{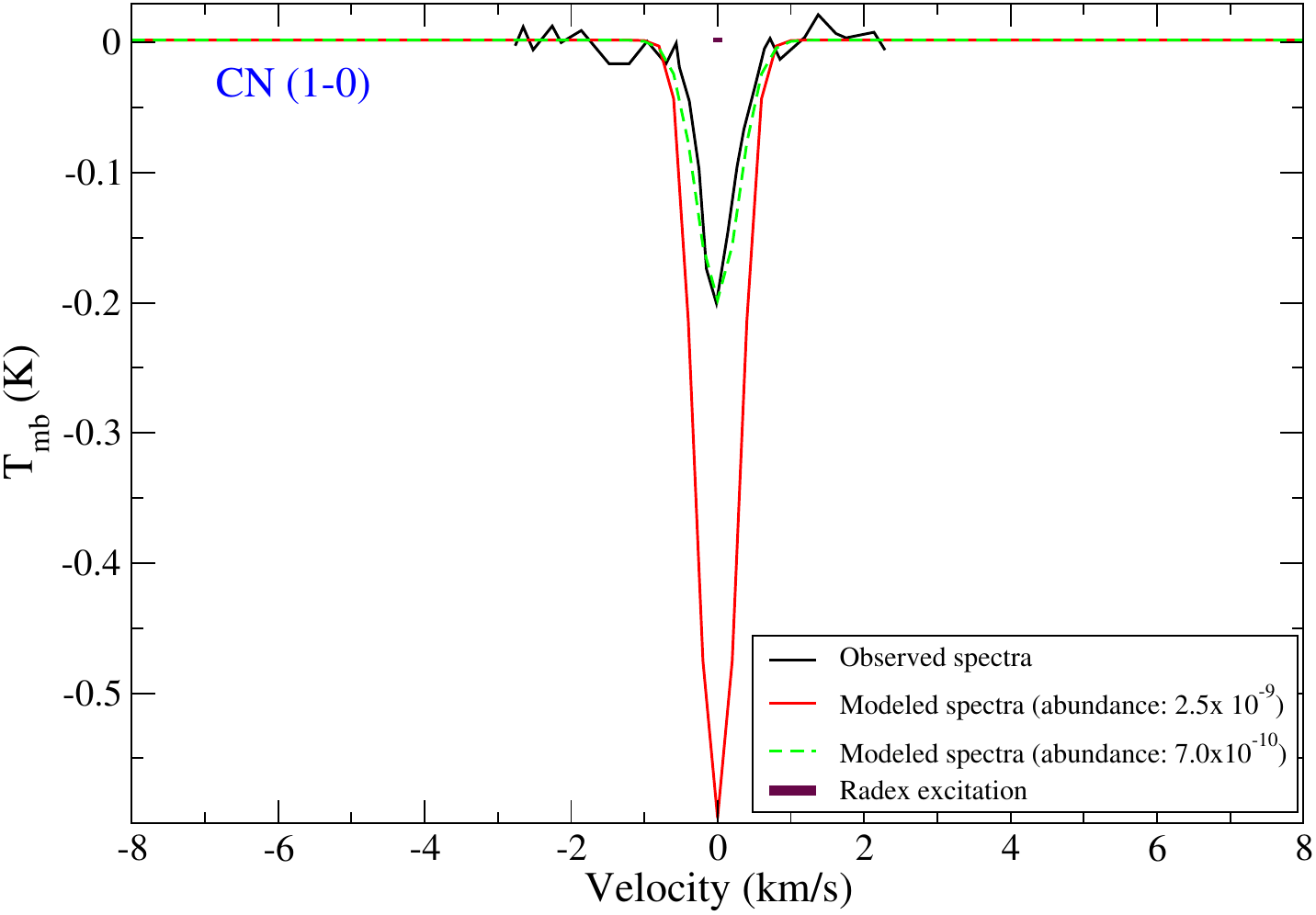}
\includegraphics[width=7cm, height=6cm]{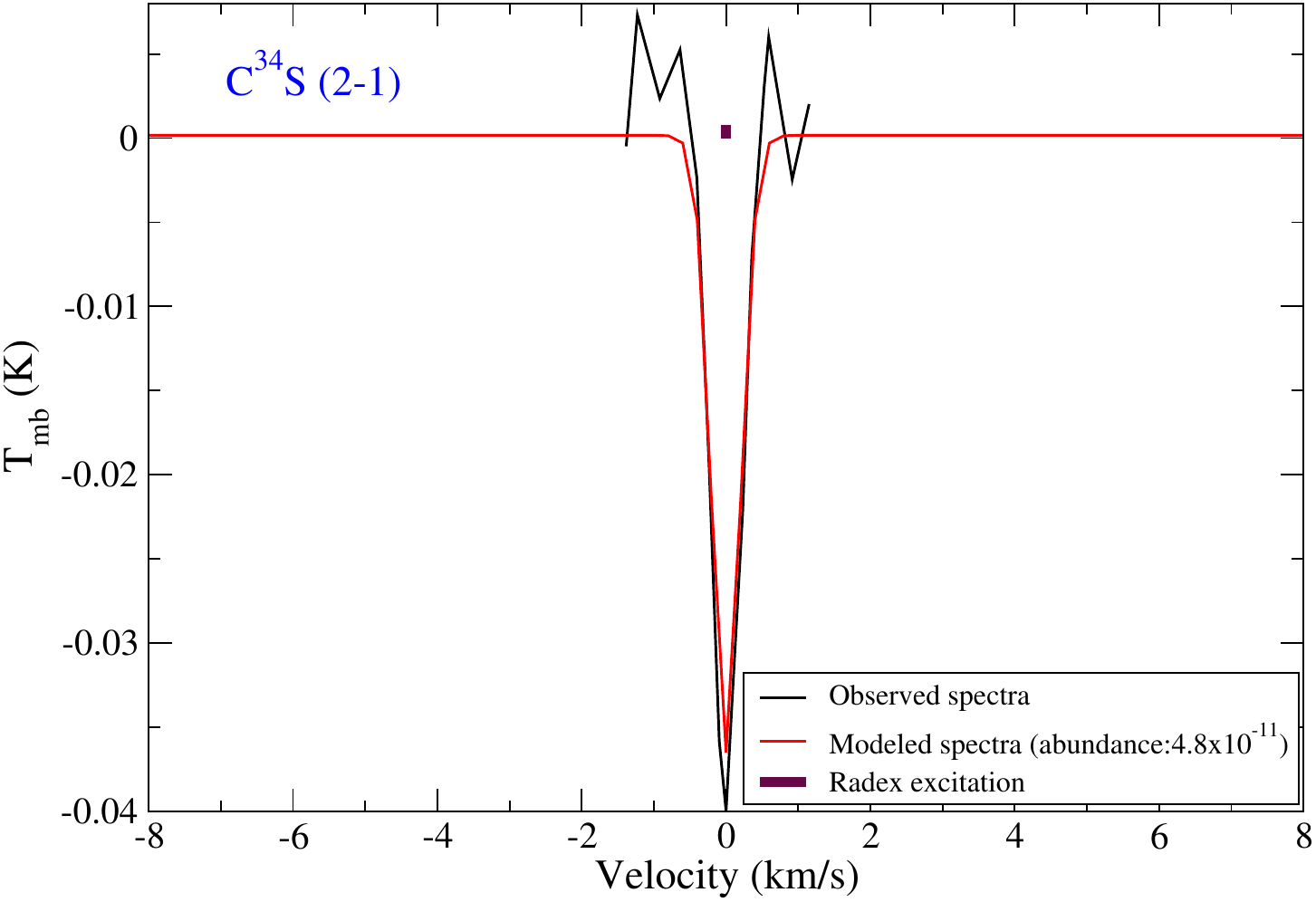}
\includegraphics[width=7cm, height=6cm]{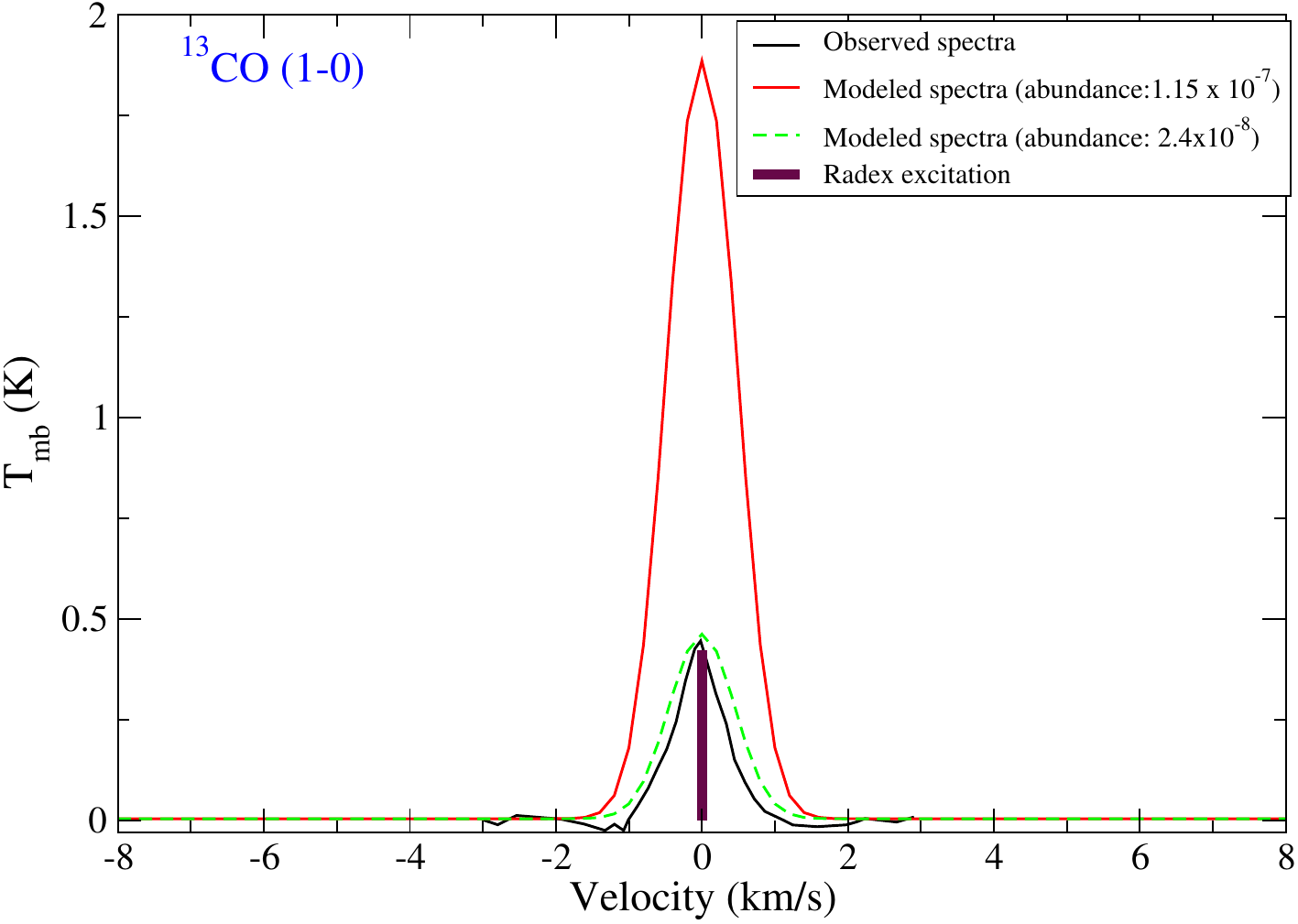}
\caption{Modeled line profile (with the RATRAN code) of HNC (1-0), CN (1-0), C$^{34}$S (2-1), and $^{13}$CO (1-0) for the diffuse cloud region. The black curve represents the observed line profile, whereas the red curve shows the modeled line profiles. Here, we consider H and H$_2$ as the collision partners having number density $100$ and $400$ cm$^{-3}$ for H and H$_2$ respectively. A temperature of $70$ K is considered for both gas and ice phases. The maroon vertical line shows the excitation with RADEX.
\label{fig:diffuse_observed} \citep[Courtesy:][]{sil21}}
\end{figure*}

\begin{figure*} 
\centering
\includegraphics[width=15cm, height=10cm]{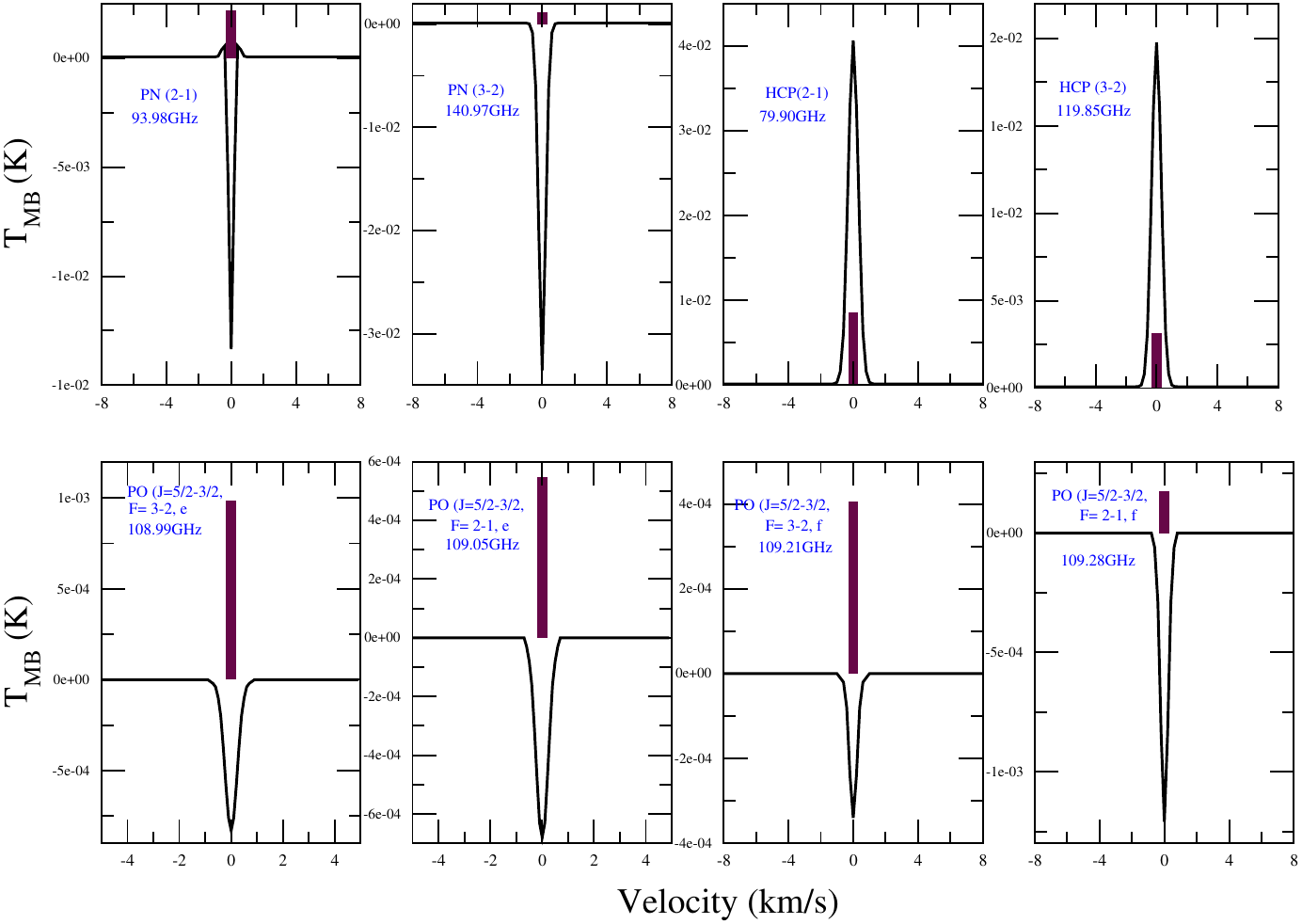}
\caption{Modeled line profile (with the RATRAN code) of PN, HCP, PO which can be observed in the diffuse cloud region. The black curve represent the modeled spectra whereas the maroon line represents the line excitation using RADEX. Here, we consider H and H$_2$ as the collision partners having number density of $100$ and $400$ cm$^{-3}$ for H and H$_2$ respectively. A temperature of $70$ K is considered for both gas and ice phases.
\label{fig:diffuse_proposed} \citep[Courtesy:][]{sil21}}
\end{figure*}

\subsection{Diffuse cloud model}
The galactic diffuse cloud region in the line of sight of the powerful continuum source B0355+508 \citep{chan20} is modeled here. We take into account a dust absorption coefficient with the bare grain (MRN model with no ice mantle) with a coagulation time of $10^8$ years \citep[see Table 1 of][]{osse94} because only a few mono-layers of ice could persist surrounding the diffuse cloud region. Finally, by taking into account a power-law emissivity model, we have used the dust emissivity ($\kappa$).
\begin{equation}
    \kappa=\kappa_0(\nu/\nu_0)^\beta,
\end{equation}
where $\kappa$ is equal to 1520 cm$^2$/gm and $\nu_0$ is equal to $1.62 \times 10^{13}$ Hertz (\citep{osse94}). It is assumed that the power-law index $\beta$ is $\sim 1.0$. The number density of the collision partners (atomic and molecular hydrogen), the kinetic temperature of the gas, the abundance profile, the Doppler broadening parameter, the size of the cloud, etc., have all varied in several ways. According to the diffuse cloud model, a large amount of hydrogen would transform into H$_2$ at the end of the simulation ($\sim 10^7$ years). In order to simulate this, we use $\rm{n_H=100}$ cm$^{-3}$ and $\rm{n_{H_2}=400}$ cm$^{-3}$.

The absorption profile of the $1 \rightarrow 0$ transition of HNC determined by the 1D RATRAN model is shown in Figure \ref{fig:cloud_size}. The absorption of HNC (1-0) noted by \cite{chan20} is to be represented by considering the various sizes of the cloud. We observe that the absorption becomes stronger as the cloud size grows. We achieve a good match with the observed intensity for the 65 pc size of the cloud. For HNC, we utilize an abundance of $2 \times 10^{-10}$ and a Doppler parameter of $\sim 0.2$ km/s. The extracted observed spectra (in black) of HNC(1-0), CN (1-0), C$^{34}$S (2-1), and $^{13}$CO (1-0) \citep{chan20} are shown in Figure \ref{fig:diffuse_observed} along with the modeled spectra in red. Interestingly, with the selected parameter, the observed absorption features of HNC, CN, and C$^{34}$S and the emission feature of $^{13}$CO are successfully explained.

A combination of increased gas temperature ($\sim 70$ K) and density is necessary to explain the observed line profiles. This is contrary to the modeling result of \cite{chan20} and the modeling results mentioned using spectral synthesis code CLOUDY \citep{ferl17} and CMMC model in \citep{sil21}. In our computation, we assume the gas and dust temperature to be the same. Here, we make use of the collisional data file for HNC, CN, and $^{13}$CO with H$_2$ from the LAMDA database (\url{https://home.strw.leidenuniv.nl/~moldata/}). There were no collisional rates available with H. Therefore, we consider it identical to the collisional rate with H$_2$ for simplicity.
The scaled rate for H \citep{scho05} was also investigated, but no noticeable variation in the synthetic spectra was found. In the absence of the datafile of C$^{34}$S, we consider it to be the same as C$^{32}$S from the LAMDA database.

For HNC, CN, C$^{34}$S, and $^{13}$CO, respectively, we find the best-fitted Doppler broadening parameter of 0.2, 0.4, 0.25, and 0.592 km/s, which is similar to the FWHM values reported by \cite{chan20}. Our findings may simultaneously account for the absorption characteristics of HNC, CN, C $ ^{34}$S, and the emission characteristics of $^{13}$CO.

The ideal beam size derived for IRAM-30m observation is convolved with the synthetic spectra in this case. We find a best-fitted abundance of $\sim 2\times 10^{-10}$ for HNC. \cite{chan20} obtained an abundance of HNC $\sim 2.1 \times 10^{-10}$. According to \cite{chan20}, with the $-17$ km/s cloud, estimated $^{32}$S/$^{34}$S isotopic ratio is $18.7$, while for $^{12}$CO/$^{13}$CO isotopic ratio it is $16.7$. Given this fractionation ratio, the abundance ratio for HNC:CN:C$^{34}$S:$^{13}$CO predicted by \cite{chan20} is $1:12.5:0.24:576.35$. Based on this, we have $2.5 \times 10^{-9}$, $4.8 \times 10^{-11}$, and $1.15 \times 10^{-7}$ for CN, C$^{34}$S, and $^{13}$CO, respectively.

The solid red lines in Figure \ref{fig:diffuse_observed} represent the modeled spectra with these abundances. Regarding C$^{34}$S, we perfectly match this abundance. But with these abundances, we have significant emission and absorption for CN and $^{13}$CO, respectively. An abundance of $7 \times 10^{-10}$ for CN and $2.4 \times 10^{-8}$ for $^{13}$CO, respectively, reveals a strong match (see the green dashed lines).

The line profiles of the P-bearing species, PN, HCP, and PO in the diffuse cloud region are displayed in Figure \ref{fig:diffuse_proposed}. We produce the synthetic spectra for these three P-bearing molecules by using the same diffuse cloud model and considering abundances in the CMMC model, $4.6 \times 10^{-11}$ for PN, $9.1 \times 10^{-12}$ for HCP, and $4.4 \times 10^{-12}$ for PO \citep{sil21}. 

To do this, we use the Doppler parameter to be $0.3$ km/s for PN, HCP, and PO, respectively, from \cite{chan20}, which is consistent with the low FWHM values shown for other molecules in the diffuse cloud. 
The collisional data for HCP and PN are found from the Basecol database \footnote{\url{(https://basecol.vamdc.eu/index.html)}}. The LAMDA \footnote{\url{(https://home.strw.leidenuniv.nl/~moldata/)}} database was used to determine the collisional rate of PO.

The collisional rate with the H atom and H$_2$ are assumed to be equivalent. Using the beam sizes specified in section \ref{sec:append_b}, the beam convolution for the IRAM-30m telescope is taken into account (see Tables \ref{table:PN_observation}, \ref{table:PO_observation}, \ref{table:HCP_observation}, and \ref{table:PH3_observation}). It is interesting to note that while we obtained the spectra for PN and PO in absorption, we did so for the HCP in emission. We use the RADEX code \citep{vand07} to examine further the excitation conditions for the molecules seen and the targeted phosphorus-bearing compounds. To calculate using RADEX, we take into account the column densities for HNC, CN, C$^{34}$S, $^{13}$CO as $0.69 \times 10^{12}$, $0.87 \times 10^{13}$, $1.64 \times 10^{11}$, and $3.98 \times 10^{14}$, respectively, from Table 4 of \cite{chan20} at an offset velocity $\simeq$ -17 km/s with respect to the source.

We consider the input kinetic temperature of 40 K, H number density of 300 cm$^{-3}$, and the extremely low H$_2$ density of 10 cm$^{-3}$. $2.73$ K is a typical background that is taken into account. The line excitation obtained with the RADEX is depicted with the maroon vertical line at the 0 km/s in Figures \ref{fig:diffuse_observed} and \ref{fig:diffuse_proposed}. Interestingly, the excitation for HNC, CN, and C$^{34}$S (for which absorption was seen) in Figure \ref{fig:diffuse_observed} is very weak for the previously observed molecules. In contrast, the excitation with the RADEX perfectly fits the measured spectra for $^{13}$CO, which were seen in emission. Under the optically thin approximation and considering a very crude assumption, for the two-level system, the critical density can be represented by the ratio between the Einstein A coefficient in s$^{-1}$ and collisional rates in cm$^3$ s$^{-1}$ \citep{shir15}. In Table \ref{tab:critical}, we have presented the Einstein A coefficient, collisional rate, and critical density of these transitions. With the RATRAN code, we have achieved the transitions of HCP and $^{13}$CO in emission (critical density low, see Figures \ref{fig:diffuse_observed} and \ref{fig:diffuse_proposed}), while the others are in absorption (critical density high).

\begin{table}
{\scriptsize
 \caption{Critical density of some transitions under the optically thin approximation. \label{tab:critical} \citep[Courtesy:][]{sil21}}
 \hskip -2.0cm
\begin{tabular}{l|l|l|l|l|l}
  \hline
  \hline
\bf{Species} &\bf{Transitions} &\bf{Frequency}&\bf{Einstein coefficient} &\bf{Collision rate}&\bf{Critical density}\\  
  &&\bf{(GHz)}&\bf{(s$^{-1})$}&\bf{ at $\sim 10$ K (cm$^{3} s^{-1}$)}&\bf{(cm$^{-3}$)}\\
  \hline
  \hline
CN&$N=1-0,\ J=1/2-1/2,\ F=3/2-1/2$&113.169&$1.182\times10^{-5}$&$9.10\times10^{-12}$&$1.299\times10^{6}$\\
HNC&$J=1-0$&90.664&$2.69\times10^{-5}$&$9.71\times10^{-11}$&$2.770\times10^{5}$\\
C$^{34}$S&$J = 2 - 1$ &96.413&$1.60\times10^{-5}$&$5.06\times10^{-11}$&$3.162\times10^{5}$\\
$^{13}$CO&$J = 1 - 0$ &110.201&$6.294\times10^{-8}$&$3.302\times10^{-11}$&$1.906\times10^{3}$\\
\hline
HCP&$J = 2 - 1$ &79.903&$3.61\times10^{-7}$&$6.884\times10^{-11}$&$5.244\times10^{3}$\\
HCP&$J = 3 - 2$ &119.854&$1.31\times10^{-6}$&$7.33\times10^{-11}$&$1.786\times10^{4}$\\
PN& $J = 2 - 1$ &93.979&$2.92\times10^{-5}$&$4.538\times10^{-11}$&$6.534\times10^{5}$\\
PN& $J = 3 - 2$ &140.968&$1.05\times10^{-4}$&$5.218\times10^{-11}$&$2.012\times10^{6}$\\
PO&$J = 5/2 - 3/2,\ \Omega = 1/2,\ F = 3 - 2,\ e$ &108.998&$2.132\times10^{-5}$&$5.10\times10^{-12}$&$4.179\times10^{6}$\\
PO&$J = 5/2 - 3/2,\ \Omega = 1/2,\ F = 2 - 1,\ e$ &109.045&$1.92\times10^{-5}$&$1.93\times10^{-11}$&$9.952\times10^{5}$\\
PO&$J = 5/2 - 3/2,\ \Omega = 1/2,\ F = 3 - 2,\ f$ &109.206&$2.143\times10^{-5}$&$1.161\times10^{-10}$&$1.846\times10^{5}$\\
PO&$J = 5/2 - 3/2,\ \Omega = 1/2,\ F = 2 - 1,\ f$ &109.281&$1.93\times10^{-5}$&$6.30\times10^{-12}$&$3.062\times10^{6}$\\
\hline
  \hline
 \end{tabular}}\\
 \end{table}

The line excitation of the P-bearing molecules is depicted in Figure \ref{fig:diffuse_proposed}. For HCP, which is in emission, the excitations obtained from the RADEX are well matched. We have observed emissions for PN. The findings produced with the RATRAN code demonstrate these in absorption for PO, but we have achieved a strong emission using the RADEX.

 \subsection{Hot core/corino model}
 To understand the expected line profiles of the P-bearing molecules in the hot core/corino, we ran the RATRAN radiative transfer model. \cite{rolf10} presented the best-fitted spatially varying density, temperature, and velocity from the profiles of HCN toward Sgr B2(M). They discovered that multiple outflows were expelling matter from the inner region while in-fall predominated in the outer parts. As one moves into the center of the hot core zone, temperature and density rise linearly. We produce the synthetic spectra for various P-bearing molecules (PN, PO, HCP, and PH$_3$) by this physical parameter as input to the RATRAN model. We also consider the same physical conditions for the hot corino model to keep things simple (line profiles that take into account more practical physical requirements for the hot corino region are also presented in section \ref{sec:append_c}). Using the IRAM-30m telescope, \cite{rivi16} discovered PO for the first time in the emission of the two star-forming areas, W51 and W3(OH). PO was previously found in the envelopes of evolved stars but not in the star-forming regions.
Additionally, \cite{rivi16} identified certain PN transitions in emission. They used a chemical model to explain the observed column densities of PN and PO. For the heated core/corino model, we use the abundances of PO, PN, HCP, and PH$_3$ from Table 6 of \cite{sil21}. Following \cite{rolf10}, here, we assume a bare dust grain \citep{osse94} due to the high temperature of the source.
 The large hot core Sgr B2(M) is located 7.8 kpc ($\simeq$) away from the star. FWHM for PO and PN were determined by \cite{rivi16} to be 7.0 and 8.2 km/s, respectively. Based on these choices, we employ the line-broadening parameter (also known as the Doppler parameter) for PO and PN as 4.2 and 4.9 km/s. In terms of HCP, we treat it similarly to PN (i.e., 4.9 km/s). For PH$_3$, we take 1.9 km/s into account. The collisional data file for PN and HCP with H$_2$ from the BASECOL database \citep{dube13} is considered. For PH$_3$, there were no collisional data files available. Therefore, we consider the collisional rate of NH$_3$ with H$_2$ instead of PH$_3$ while analyzing due to the structural similarity. For the hot corino, the same numerical quantities are also considered.
 
\begin{figure*}
\centering
\includegraphics[width=12cm, height=16cm, angle=270]{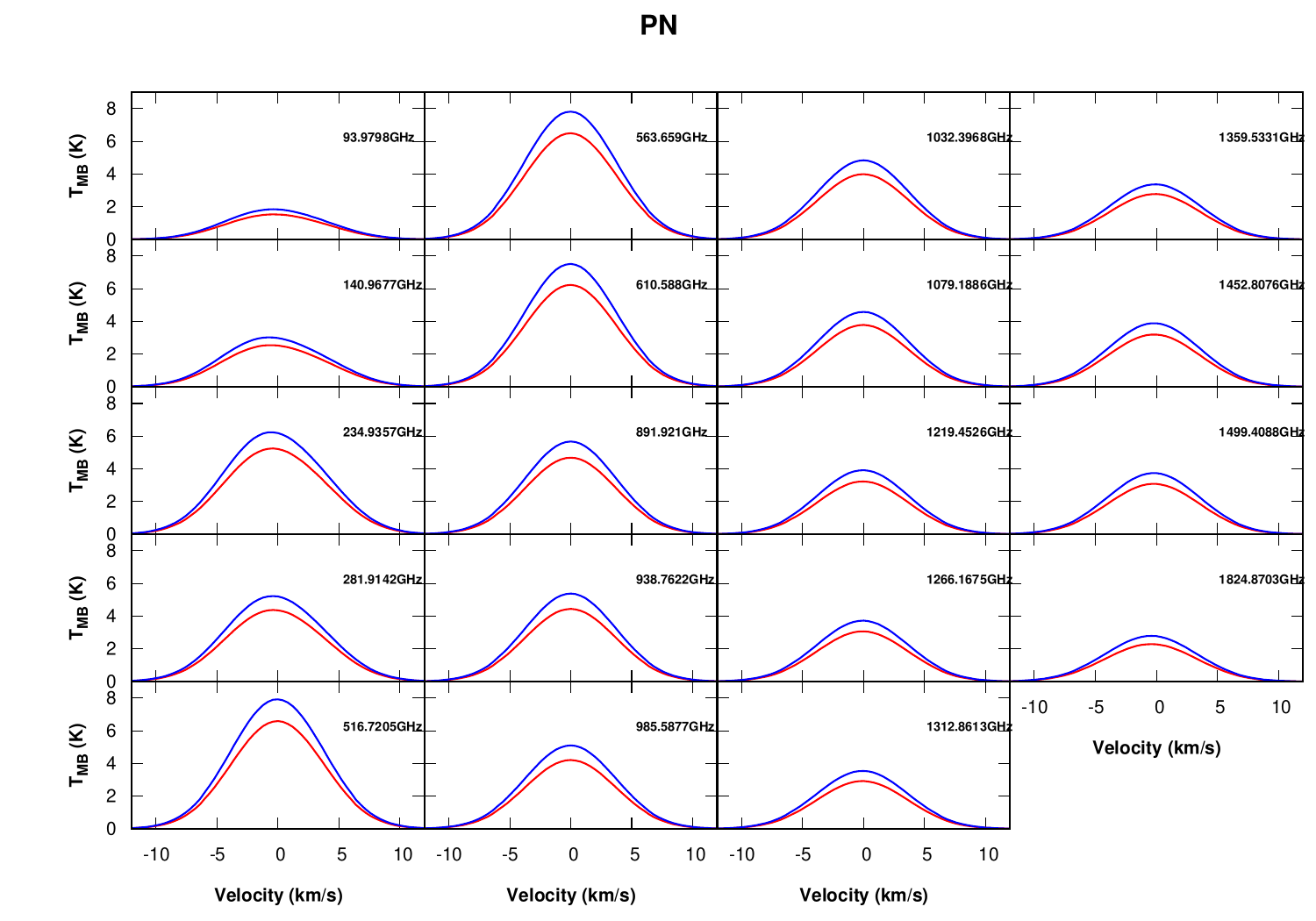}
\caption{Table \ref{table:PN_observation} shows possible transitions of PN, which could be observed with the IRAM-30m or SOFIA (GREAT) toward hot core/corino. The estimated line profile of these transitions obtained with the RATRAN model is shown here for hot core (blue) and hot corino (red) using the abundances from Table 6 of \cite{sil21}.}
\label{fig:PN}
\end{figure*}

\begin{figure*}
\centering
\includegraphics[width=12cm, height=16cm, angle=270]{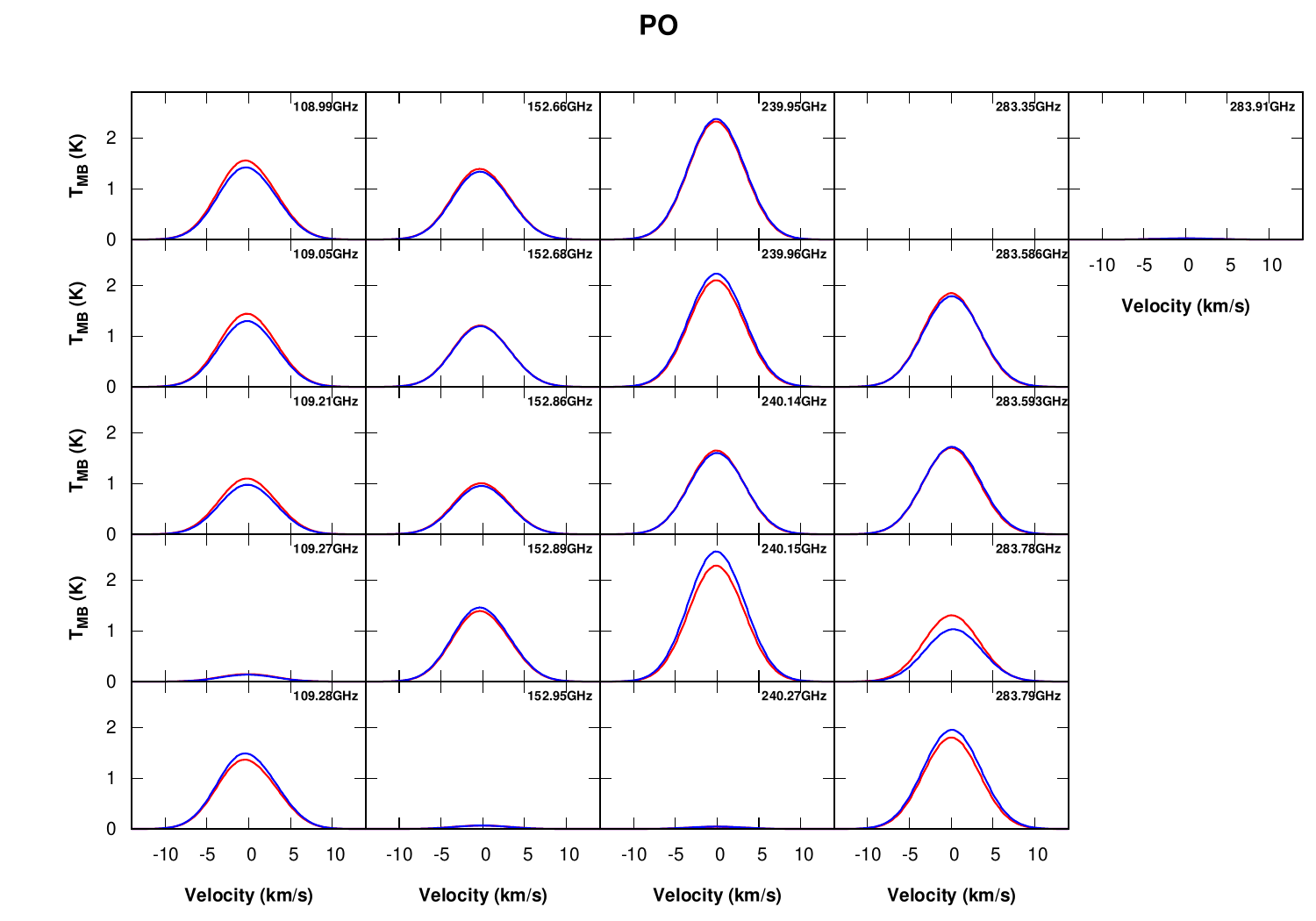}
\caption{Table \ref{table:PO_observation} shows the possible transitions of PO, which could be observed with the IRAM-30m toward Sgr-B2(M). The estimated line profile of these transitions with the RATRAN model is shown here for hot core (blue) and hot corino (red) using the abundances from Table 6 of \cite{sil21}.
\label{fig:PO}}
\end{figure*}

\begin{figure*}
\centering
\includegraphics[width=10cm, height=14cm, angle=270]{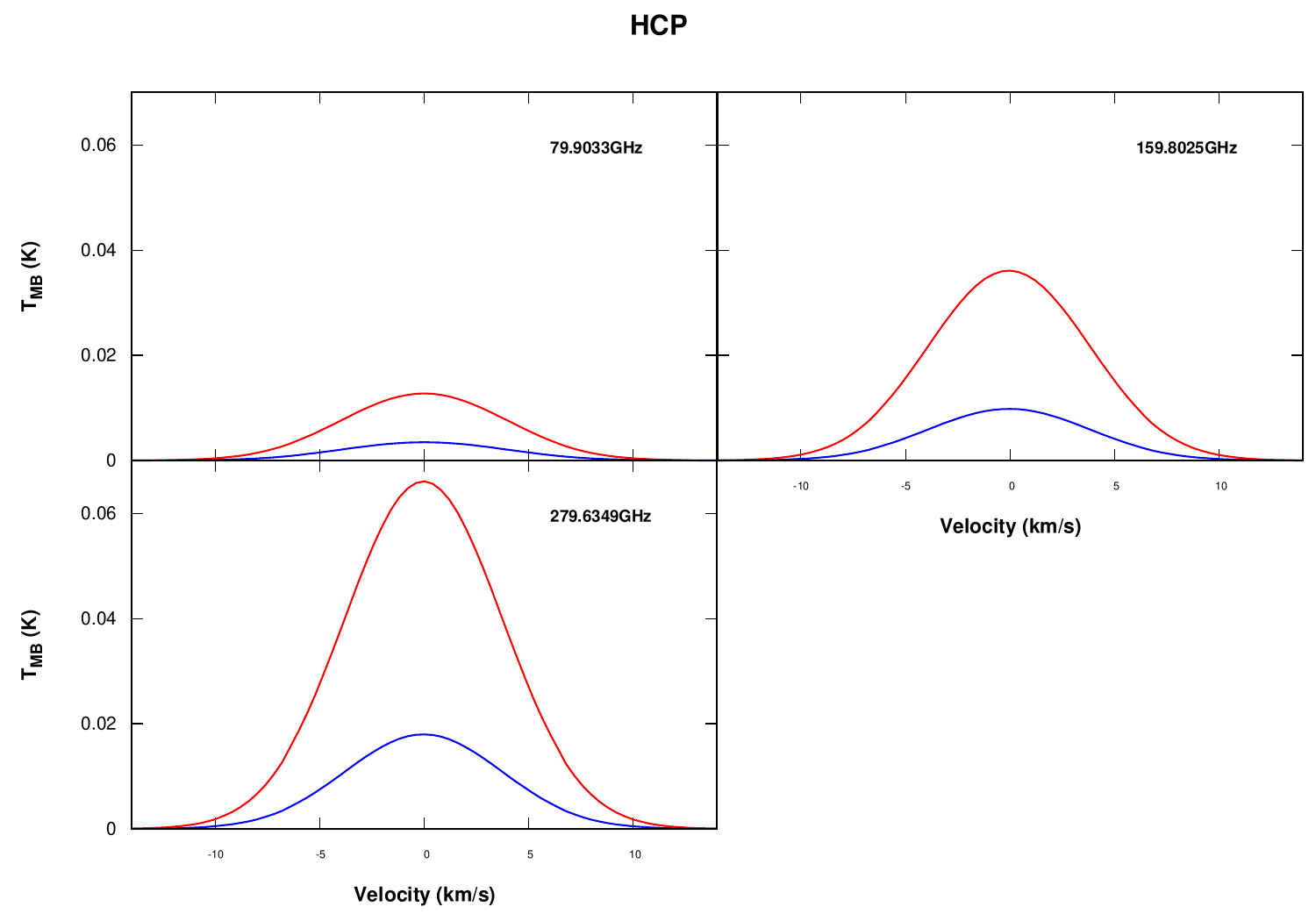}
\caption{Table \ref{table:HCP_observation} shows the possible transitions of HCP, which could be observed with the IRAM-30m toward Sgr-B2(M). The estimated line profile of these transitions is shown here for Hot core (blue) and Hot corino (red) using the abundances from Table 6 of \cite{sil21}.}
\label{fig:HCP}
\end{figure*}

\begin{figure*}
\centering
\includegraphics[width=10cm, height=16cm, angle=270]{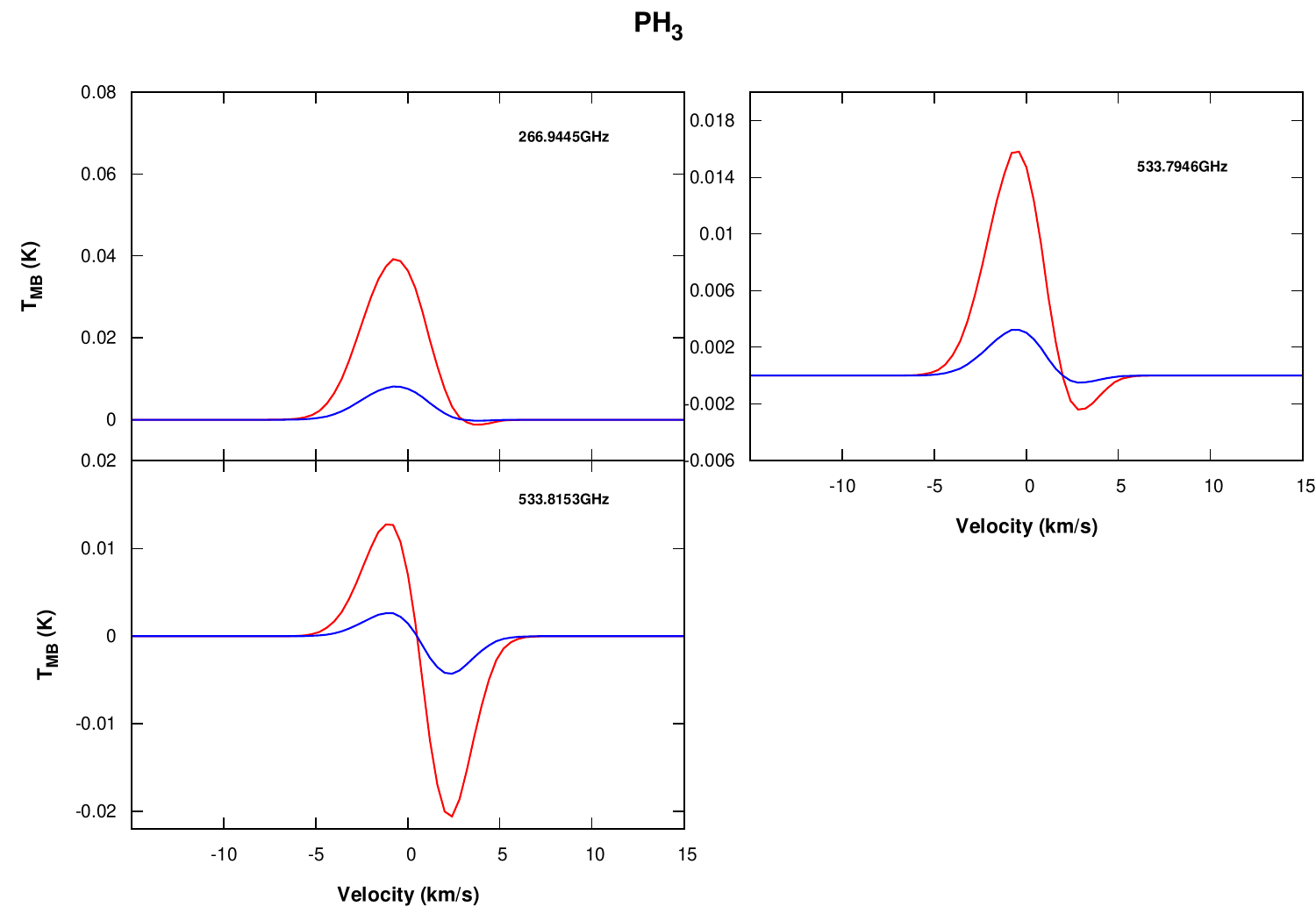}
\caption{Table \ref{table:PH3_observation} shows the possible transitions of PH$_3$, which could be observed with the IRAM-30m and SOFIA (GREAT) toward Sgr-B2(M). The estimated line profile of these transitions with the RATRAN model is shown here for hot core (blue) and hot corino (red) using the abundances from Table 6 of \cite{sil21}.}
\label{fig:PH3}
\end{figure*}

 We determine the beam size for SOFIA-GREAT\footnote{\url{https://www.sofia.usra.edu/science/proposing-and-observing/observers-handbook-cycle-7/7-great/72-planning-observations}} and IRAM 30m\footnote{\url{http://www.iram.es/IRAMES/mainWiki/Iram30mEfficiencies}}. Tables \ref{table:PN_observation}, \ref{table:PO_observation}, \ref{table:HCP_observation}, and \ref{table:PH3_observation} for PN, PO, HCP, and PH$_3$, respectively, list the estimated telescopic parameters (main beam temperature, beam size, integration time, and visibility) for the diffuse cloud and hot core/corino area. The tables only list the transitions that might be observable in the hot core/corino area. Figures \ref{fig:PN}, \ref{fig:PO}, \ref{fig:HCP}, and \ref{fig:PH3} of the Section \ref{sec:append_b} display the expected line profiles of these transitions by considering the effect of beam convolution. It is interesting to note that we were able to acquire "inverse P-Cygni type" spectral profiles for a few PH$_3$ transitions (see Figure \ref{fig:PH3}). This particular spectral profile shows that both in-fall and outflow is present in the source.
 
 Theoretically, \cite{rolf10} predicted the existence of accelerating in-fall having a density power-law index of $\sim$1.5 to support a spherically symmetric constant mass accretion rate in Sgr B2(M). But they also failed to gather observational proof for accelerating infall. We showed the line profiles obtained when the physical state associated with IRAS4A was considered as a representation of a hot corino case is shown in section \ref{sec:append_c}. We see that the resulting line profiles have a significant influence on the physical input parameters. While using the physical parameters of IRAS4A, we are unable to obtain the inverse P-Cygni profile. However, it is not within the scope of this work to elaborate on such issues at this time.
 
 \section{Estimated intensities from the radiative transfer model} \label{sec:append_b}
 In the diffuse and hot core regions, we measure the intensities of several P-bearing species. We further draw attention to the fact that similar transitions can be seen with IRAM 30m, Herschel (not in operation), and SOFIA. For both ground-based and space-based observations, we test the atmospheric transmission around these frequencies using the ATRAN software\footnote{\url{https://atran.arc.nasa.gov/cgi-bin/atran/atran.cgi}}.\\
 
\begin{table*}
\setlength{\tabcolsep}{2pt}
{\scriptsize
 \caption{The telescopic parameters (IRAM and SOFIA) of some transitions of PN obtained toward a hot core/corino  or diffuse cloud regions obtained with the RATRAN model are shown. The footnote shows the adopted parameters for these calculations. \label{table:PN_observation} \citep[Courtesy:][]{sil21}}
 \hskip -2.5cm
\begin{tabular}{l|l|l|l|l|l|l|l}
  \hline
  \hline
\bf{Frequency}&\bf{Telescope}&\bf{Beam size}&\bf{Atmospheric}&\bf{T$_{MB}$(K)}&\bf{Integration time}&\bf{T$_{MB}$(K)}&\bf{Integration time}\\
 \bf{(GHz)}&\bf{}&\bf{($^{''}$)}&\bf{transmission$^a$}&\bf{(Diffuse)}&\bf{(Diffuse$^b$)}&\bf{Hot core (Hot corino)}&\bf{Hot core$^c$ (Hot corino$^c$)}\\
  \hline
  \hline
 $^\star$93.9798 & IRAM-30m & 26.18 &0.7&$-0.0134$&4.7 hr&1.85 (1.54)&11.3 min (11.3 min)\\
 $^\star$140.9677&IRAM-30m & 17.45 &0.65&$-0.0337$&47.4 min&3.03 (2.55)&11.9 min (11.9 min)\\
 234.9357 & IRAM-30m & 10.47 &0.4&$-0.00216$&high&6.23 (5.24)&25.6 min (25.6 min)\\
 $^\star$281.9142 & IRAM-30m & 8.73 &0.35&0.0038&high&5.23 (4.38)&46.6 min (46.6 min)\\
 \hline
  516.7205&SOFIA-GREAT&18.2&0.97&0.0032&high&7.90 (6.58)&0.6sec (0.9 sec)\\
  563.6590&SOFIA-GREAT&18.2&0.72&0.0024&high&7.80 (6.48)&2.3 sec (3.3 sec)\\
  610.5880&SOFIA-GREAT&18.2&0.94&0.0019&high&7.51 (6.23)&0.8 sec (1.2 sec)\\
 891.9210&SOFIA-GREAT&18.2&0.99&0.0008&high&5.67 (4.68)&2.2 sec (3.2 sec)\\
 938.7622&SOFIA-GREAT&18.2&0.98&0.0008&high&5.37 (4.44)&2.4 sec (3.4 sec)\\
 985.5877&SOFIA-GREAT&18.2&0.17&0.0007&high&5.10 (4.20)&140.6 sec(207.4 sec)\\
 1032.3968&SOFIA-GREAT&18.2&0.7&0.0007&high&4.84 (4.00)&3.4 sec(5.0 sec)\\
 1079.1886&SOFIA-GREAT&18.2&0.81&0.0005&high&4.59 (3.78)&6.5 sec (9.6 sec)\\
 1219.4526&SOFIA-GREAT&18.2&0.71&0.0004&high&3.91 (3.23)&88.1 sec (129.1 sec)\\
 1266.1675&SOFIA-GREAT&18.2&0.95&0.0004&high&3.72 (3.07)&47.8 sec(70.2 sec)\\
 1312.8613&SOFIA-GREAT&18.2&0.88&0.0003&high&3.54 (2.92)&62.5 sec(91.8 sec)\\
 1359.5331&SOFIA-GREAT&18.2&0.96&0.0004&high&3.37 (2.78)&55.1 sec (81.0 sec)\\
 1452.8076&SOFIA-GREAT&14&0.96&0.0002&high&3.89 (3.20)&37.3 sec(55.2 sec)\\
 1499.4088&SOFIA-GREAT&14&0.98&0.0002&high&3.74 (3.08)&37.9 sec(55.9 sec)\\
 1824.8703&SOFIA-GREAT&14&0.93&0.0001&high&2.79 (2.29)&52.5 sec (77.9 sec)\\
 \hline
\hline
 \end{tabular}}
\hskip 2cm {$^a$ ATRAN (\url{https://atran.arc.nasa.gov/cgi-bin/atran/atran.cgi}) program is used for the calculation of the atmospheric transmission.\\ 
$^b$ For the diffuse cloud region, we use a frequency resolution of $1$ km/s and a signal to noise ratio $\geq$ 3 with IRAM-30m and a frequency resolution of 1 km/s and a signal to noise ratio $3$ with SOFIA.}\\
{\noindent $^c$ For the hot core/hot corino region, we use a frequency resolution of 1 km/s with a signal to noise ratio $10$ with SOFIA and a frequency resolution of 1 km/s and   a signal to noise ratio $\geq$ 3 with IRAM 30m.}\\
{\noindent $^\star$Observed by \cite{mini18}}\\
 \end{table*}

 \begin{table*}
\setlength{\tabcolsep}{2pt}
{\scriptsize
 \caption{The telescopic parameters (IRAM and SOFIA) of some transitions of PO toward hot core/corino or diffuse cloud regions obtained with the RATRAN model. The footnote shows the adopted parameters for these calculations. \label{table:PO_observation}\citep[Courtesy:][]{sil21}}
 \hskip -2.5cm
\begin{tabular}{l|l|l|l|l|l|l|l}
  \hline
  \hline
  \bf{Frequency}&\bf{Telescope}&\bf{Beam size}& \bf{Atmospheric} & \bf{T$_{MB}$} & \bf{Integration time$^b$} & \bf{T$_{MB}$(K)} & \bf{Integration time$^c$}\\
  \bf{(GHz)}&\bf{}&\bf{($^{''}$)}&\bf{transmission$^a$}&\bf{(Diffuse)}&\bf{(Diffuse)}&\bf{Hot core (Hot corino)}&\bf{Hot core (Hot corino)}\\
  \hline
  \hline
 108.7072&IRAM-30m&22.63&0.65&$-0.02793$&1.3 hr&- (-)& - (-)\\
 108.9984&IRAM-30m&22.57&0.65&$-8.32\times10^{-4}$&high&1.4164 (1.5525)& 12.6 min (12.6 min)\\
 109.0454&IRAM-30m&22.56&0.65&$-6.81\times10^{-4}$&high&1.295 (1.4385)& 12.6 min (12.6min)\\
 109.2062&IRAM-30m&22.53&0.64&$-3.41\times10^{-4}$&high&0.9740 (1.0954)& 12.8 min (12.8 min)\\
 109.2714&IRAM-30m&22.51&0.65&$-1.2\times 10^{-4}$&high&0.1358 (0.1490)&12.8 min (12.8 min)\\
 109.2812&IRAM-30m&22.51&0.64&$-1.21\times10^{-3}$&high&1.3576 (1.48)& 12.8 min (12.8 min)\\
 152.3891&IRAM-30m&16.14&0.6&$-0.0213$&2.3 hr&1.4532 (1.3870)& 12.5 min (12.5 min)\\
 $^\star$152.6570&IRAM-30m&16.11&0.6&$-4.11\times10^{-4}$&high&1.3359 (1.3907)&12.5 min (12.5 min)\\
 $^\star$152.6803&IRAM-30m&16.11&0.6&$-3.71\times10^{-4}$&high&1.1911 (1.2072)& 12.5 min (12.5 min)\\
 $^\star$152.8555&IRAM-30m&16.09&0.61&$-1.41\times10^{-4}$&high&0.9519 (1.0044)& 12.5 min (12.5 min)\\
 $^\star$152.8881&IRAM-30m&16.09&0.61&$-6.01\times10^{-4}$&high&1.4532 (1.3870)&12.5 min (12.5 min)\\
 152.9532&IRAM-30m&16.08&0.61&$-2.01\times10^{-5}$&high&0.0655 (0.0671)&12.5 min (12.5 min)\\
 239.7043&IRAM-30m&10.26&0.01&$-0.0392$&2.7 hr&- (-)&- (-)\\
 239.9490&IRAM-30m&10.25&0.02&$-1.3\times 10^{-4}$&high&2.3684 (2.3166)& 25.3min (25.3 min)\\
 239.9581&IRAM-30m&10.25&0.01&$-1.00\times10^{-4}$&high&2.2247 (2.0924)&25.3 min (25.3 min)\\
 240.1411&IRAM-30m&10.24&0.01&$-3.00\times 10^{-5}$&high&1.5976 (1.6473)&25.1 min (25.1 min)\\
 240.1525&IRAM-30m&10.24&0.01&$-1.3\times 10^{-4}$&high&2.5592 (2.2764)&25.1 min (25.1 min)\\
 240.2683&IRAM-30m&10.24&0.02&-&-&0.0450 (0.0418)&39.2 min (39.2 min)\\
 283.3487&IRAM-30m&8.68&0.33&$-0.0117$&high&- (-)&- (-)\\
 283.5868&IRAM-30m&8.67&0.32&$-2.01\times10^{-5}$&high&1.7843 (1.8441)&47.6 min (47.6 min)\\
 283.5932&IRAM-30m&8.67&0.32&$-3.00\times10^{-5}$&high&1.7176 (1.6969)& 47.6min (47.6min)\\
 283.7776&IRAM-30m&8.67&0.32&$-1.00\times10^{-5}$&high&1.0301 (1.3038)&47.7 min (47.7min)\\
 283.7854&IRAM-30m&8.67&0.33&$-4.01\times10^{-5}$&high&1.9453 (1.7935)&47.7 min (47.7min)\\
 283.9125&IRAM-30m&8.66&0.33&-&-&0.0245 (0.0233)&7.5 hr (7.5 hr)\\
 \hline
\hline
 \end{tabular}}
\hskip 2cm {$^a$ ATRAN (\url{https://atran.arc.nasa.gov/cgi-bin/atran/atran.cgi}) program is used for the calculation of the atmospheric transmission.\\ 
$^b$ For the diffuse cloud region, a frequency resolution of $1$ km/s and a signal to noise ratio$\geq$ 3 are used with IRAM-30m.}\\
{\noindent $^c$ For the hot core/hot corino region,  a frequency resolution of 1 km/s and a signal to noise ratio$\geq$ 3  are used  with IRAM 30m.}\\
{\noindent $^\star$Observed by \cite{font16}}\\
 \end{table*}

 \begin{table*}
\setlength{\tabcolsep}{2pt}
{\scriptsize
 \caption{The telescopic parameters (IRAM and SOFIA) of some transitions of HCP toward hot core/corino  or diffuse cloud regions obtained with the RATRAN model. The footnote shows the adopted parameters for these calculations. \label{table:HCP_observation} \citep[Courtesy:][]{sil21}}
 \hskip -2.5cm
\begin{tabular}{l|l|l|l|l|l|l|l}
  \hline
  \hline
 \bf{Frequency}&\bf{Telescope}&\bf{Beam size}&\bf{Atmospheric}&\bf{T$_{MB}$}&\bf{Integration time$^c$}&\bf{T$_{MB}$(K)}&\bf{Integration time$^c$}\\
 \bf{(GHz)}&\bf{}&\bf{($^{''}$)}&\bf{transmission$^a$}&\bf{(Diffuse)}&\bf{(Diffuse)$^b$}&\bf{Hot core(Hot corino)}&\bf{Hot core(Hot corino)}\\ 
  \hline
  \hline
 79.9033 & IRAM-30m &30.79 &0.63&0.0005&high&0.0035 (0.0127)&high (6.1 hr)\\
 159.8025 & IRAM-30m &15.39 &0.55&-$3.01\times10^{-5}$&high&0.0098 (0.0359)&10.3 hr (55.6 min)\\
 279.6349 & IRAM-30m &8.80 &0.33&-&-&0.0179 (0.0658)&12.4 hr (46.5 min)\\
 \hline
\hline
 \end{tabular}}
\hskip 2cm {$^a$ ATRAN (\url{https://atran.arc.nasa.gov/cgi-bin/atran/atran.cgi}) program is used for the calculation of the atmospheric transmission.\\ 
$^c$ In this case, we use a frequency resolution of 1 km/s and a signal to noise ratio $\geq$ 3 with IRAM 30m.}\\
 \end{table*}
 
  \begin{landscape}
 \begin{table*}
\setlength{\tabcolsep}{2pt}
{\scriptsize
 \caption{The telescopic parameters (IRAM and SOFIA) of some transitions of PH$_3$ toward hot core/corino  or diffuse cloud regions obtained with the RATRAN model. The footnote shows the adopted parameters for these calculations. \label{table:PH3_observation} \citep[Courtesy:][]{sil21}}
 \hskip -3.0cm
\begin{tabular}{l|l|l|l|l|l|l|l}
  \hline
  \hline
\bf{Frequency}&\bf{Telescope}&\bf{Beam size}&\bf{Atmospheric}&\bf{T$_{MB}$}&\bf{Integration time}&\bf{T$_{MB}$(K)}&\bf{Integration time}\\  
\bf{(GHz)}&\bf{}&\bf{($^{''}$)}&\bf{transmission$^a$}&\bf{(Diffuse)}&\bf{(Diffuse)}&\bf{Hot core/Hot core$^x$(Hot corino/Hot corino$^y$)}&\bf{Hot core/Hot core$^x$ (Hot corino/Hot corino$^y$)} \\ 
  \hline
  \hline
  266.9445&IRAM-30m&9.22&0.23&-&-&0.008/0.8634 (0.039/0.9395)&31.5 hr$^c$/32.0 min$^c$(127.8 min$^c$/ 32.0 min$^c$)\\
 \hline
 533.7946&SOFIA-GREAT&18.2&0.95&-&-&0.0032/0.3415 (0.0157/0.3702)&high$^l$/7.24 min$^m$(5.1 hr$^l$/6.16 min$^m$)\\
 533.8153&SOFIA-GREAT&18.2&0.95&-&-&0.003/0.2794 (0.0125/0.3029)&high$^l$/11.3 min$^m$(8.5 hr$^l$/9.6 min$^m$)\\
 \hline
\hline
 \end{tabular}}
\hskip 2cm {$^a$ ATRAN (\url{https://atran.arc.nasa.gov/cgi-bin/atran/atran.cgi}) program is used for the calculation of the atmospheric transmission.\\
 $^x$ second values are considering PH$_3$ abundance $n_m$ = $2.1 \times 10 ^{-10}$ for the case when initial P$^+$ abundance is $ 1.8 \times 10^{-9}$ mentioned in Table 6 of \cite{sil21} not considering the destruction reactions.}\\
{\noindent $^y$ second values are considering PH$_3$ abundance $n_m$ = $2.3 \times 10 ^{-10}$ for the case when initial P$^+$ abundance is $ 5.6 \times 10^{-9}$ mentioned in Table 6 of \cite{sil21} not considering the destruction reactions.}\\
{\noindent $^c$ In this case, we use frequency resolution 1 km/s and a signal to noise ratio $\geq$ 3 with IRAM 30m.}\\
{\noindent $^l$ In this case, we use a frequency resolution of 1 km/s with a signal to noise ratio $3$ with SOFIA.}\\
{\noindent $^m$ In this case, we use a frequency resolution of 1 km/s with a signal to noise ratio $10$ with SOFIA.}\\
 \end{table*}
\end{landscape} 

\section{Radiative transfer model for the hot corino} \label{sec:append_c}
We used the spatial distribution of the physical input parameters for RATRAN that are relevant to the hot core region in Section \ref{sec:RATRAN_P}. We used the same physical criteria there to examine the line profiles of the hot corino region. We have used a spatial variant of the physical input parameters used in \cite{mott13} for low mass protostar IRAS4A in order to eliminate any possibility of misunderstanding when examining the same physical parameters for the low mass analogue.

\begin{figure*}
\centering
\includegraphics[width=7cm, height=11cm]{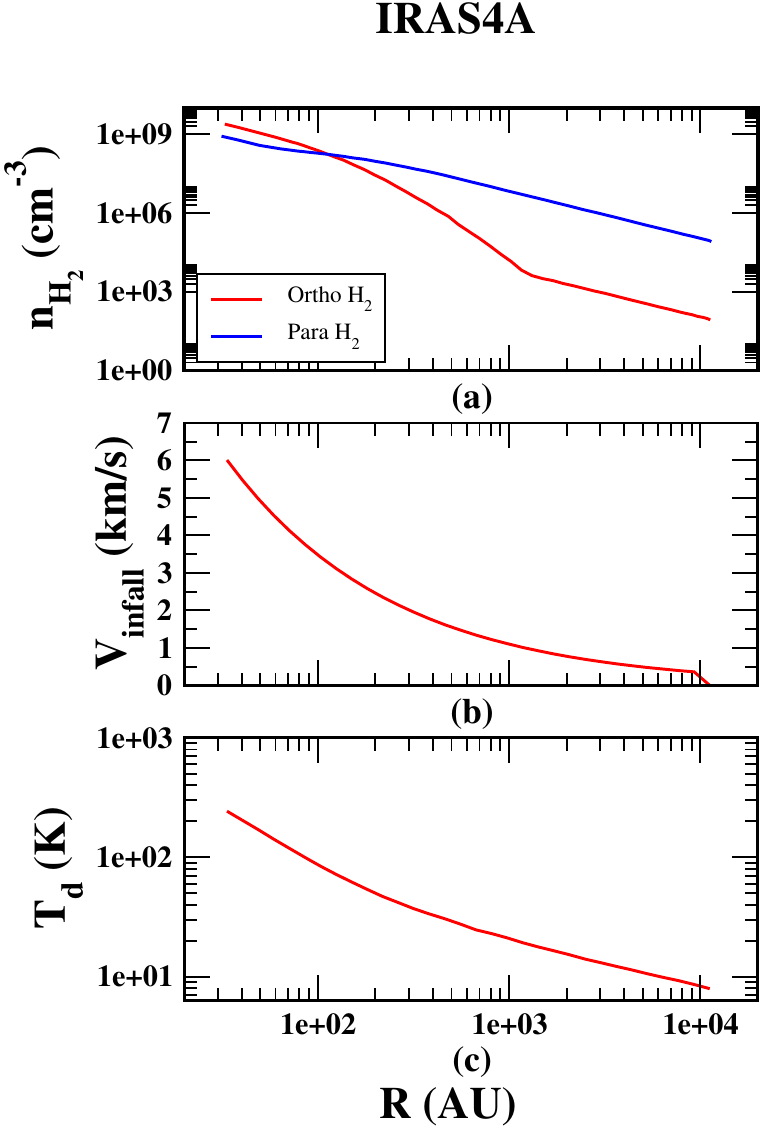}
\includegraphics[width=7cm, height=11cm]{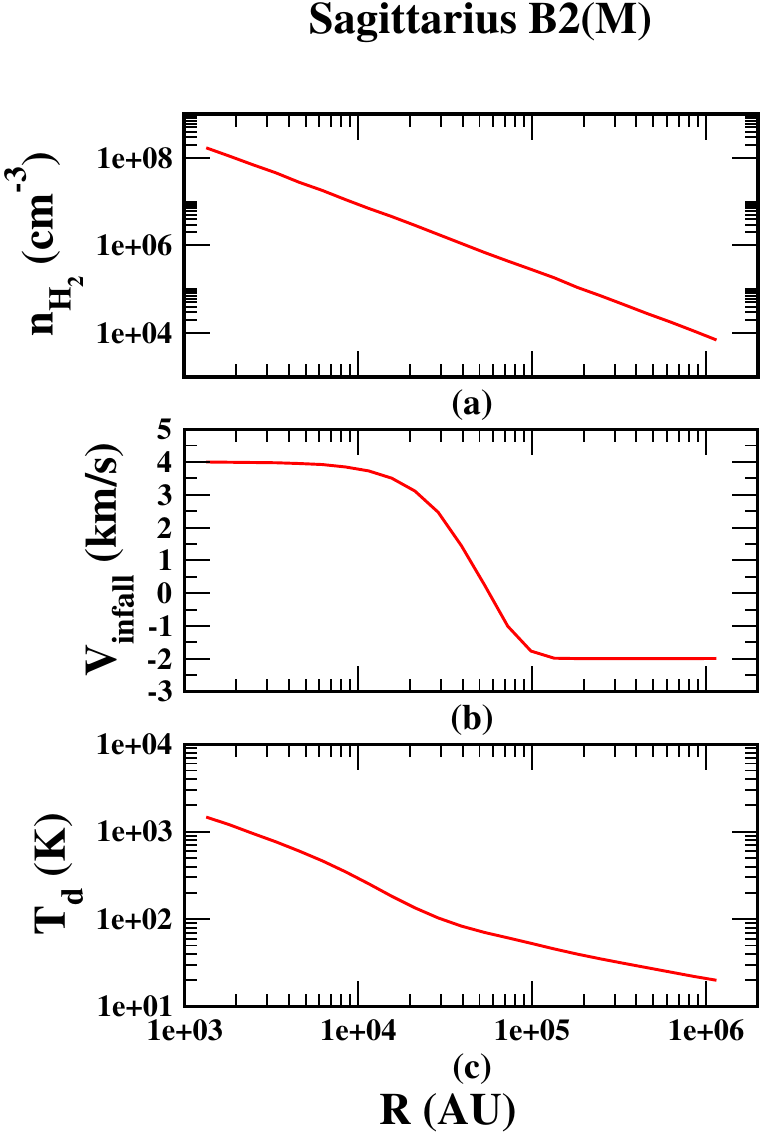}
\caption{Spatial distribution of physical parameters: (a) H$_2$ density, (b) In-fall velocity, and (c) Dust temperatures for hot corino (IRAS4A) and hot core (Sagittarius B2) are taken from \cite{mott13} and \cite{rolf10} respectively.
\label{fig:physical} \citep[Courtesy:][]{sil21}}
\end{figure*}

\begin{figure*}
\centering
\includegraphics[width=8cm, height=10cm, angle=270]{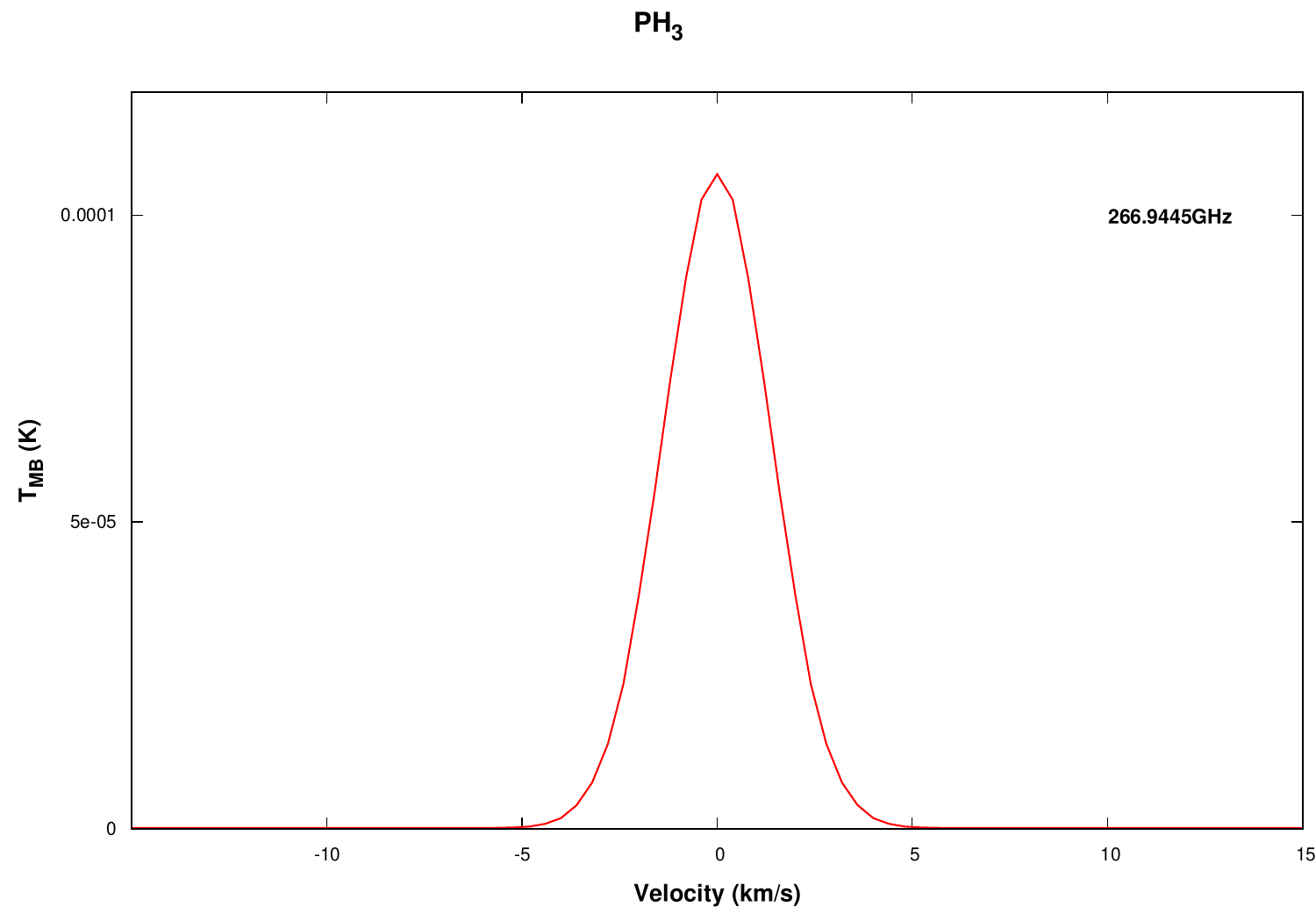}
\caption{The line profile of the transition of PH$_3$, which could be observed with the IRAM-30m toward IRAS4A is shown with the RATRAN model. Abundance of PH$_3$ is taken from the CMMC model of hot corino noted in Table 6 of \cite{sil21}.
\label{fig:irasph3} \citep[Courtesy:][]{sil21}}
\end{figure*}

The adopted spatial distribution of the physical input parameters for hot corino (IRAS4A, used here) and hot core (Sagittarius B2(M), used in section \ref{sec:RATRAN_P}) are both displayed in Figure \ref{fig:physical}. In addition to these physical input parameters, we also used our CMMC model's abundances (Table 6 of \cite{sil21}) to determine the main beam temperature of the P-bearing molecules (PN, PO, HCP, and PH$_3$) in the hot corino region.
The proper beam sizes for the SOFIA-GREAT instrument and IRAM-30m telescope are listed in Table \ref{tab:IRAS4Ap}. These beam sizes are used to convolve the intensities of other transitions. For the $266.9445$ GHz transition of the PH$_3$ molecule with this physical condition, no inverse P-Cygni type or red-shifted absorption spectral signature is obtained (see Figure \ref{fig:irasph3}).

 \begin{table*}
\setlength{\tabcolsep}{2pt}
{\scriptsize
 \caption{The telescopic parameters (IRAM-30m and SOFIA) of some transitions of PN, PO, HCP, PH$_3$ toward hot corino IRAS4A. \label{tab:IRAS4Ap} \citep[Courtesy:][]{sil21}}
\begin{tabular}{l|l|l|l|l|l|l}
  \hline
  \hline
  \bf{Molecule}&\bf{Frequency}&\bf{Telescope}&\bf{Beam size}&\bf{Atmospheric}&\bf{T$_{MB}$}&\bf{Integration time$^{b,c}$}\\
  \bf{ }&\bf{(GHz)}&\bf{ }&\bf{($^{''}$)}&\bf{transmission$^a$}&\bf{(K)}&\bf{}\\
  \hline
  \hline
&$^{\star \star}$93.9798 & IRAM-30m & 26.18 &0.7 &$2.83\times10^{-4}$&high\\
 &$^{\star \star}$140.9677 & IRAM-30m & 17.45 &0.65& $1.75\times10^{-3}$&high\\
 &234.9357 & IRAM-30m & 10.47 &0.4& $9.96\times10^{-3}$&19.0 hr\\
 &$^{\star \star}$281.9142 & IRAM-30m & 8.73 &0.35&$1.44\times10^{-2}$&15.3 hr\\
 &563.6590 &SOFIA-GREAT& 26.18 &0.72 &$3.22\times10^{-3}$&high\\
 &610.5880&SOFIA-GREAT&18.2&0.94& $3.06\times10^{-3}$&high\\
 &891.9210&SOFIA-GREAT&18.2&0.99& $2.88\times10^{-3}$&high\\
 &938.7622&SOFIA-GREAT&18.2& 0.98&$2.10\times10^{-3}$&high\\
 PN&985.5877&SOFIA-GREAT&18.2&0.17&$2.03\times10^{-3}$&high\\
 &1032.3968&SOFIA-GREAT&18.2&0.7& $1.93\times10^{-3}$&high\\
 &1079.1886&SOFIA-GREAT&18.2&0.81& $1.88\times10^{-3}$&high\\
 &1219.4526&SOFIA-GREAT&18.2&0.71& $1.8\times10^{-3}$&high\\
 &1266.1675&SOFIA-GREAT&18.2&0.95& $1.63\times10^{-3}$&high\\
 &1312.8613&SOFIA-GREAT&18.2&0.88& $1.56\times10^{-3}$&high\\
 &1359.5331&SOFIA-GREAT&18.2&0.96& $1.55\times10^{-3}$&high\\
 &1452.8076&SOFIA-GREAT&14&0.96&$2.27\times10^{-3}$&high\\
 &1499.4088&SOFIA-GREAT&14&0.98&2.25$\times10^{-3}$&high\\
 &1824.8703&SOFIA-GREAT&14&0.93&$1.84\times10^{-3}$&high\\
 \hline
 &108.7072&IRAM-30m&22.63&0.65& $1.13\times10^{-5}$&high\\
 &108.9984&IRAM-30m&22.57&0.65& $5.1\times10^{-4}$&high\\
 &109.0454&IRAM-30m&22.56&0.65& $4.59\times10^{-4}$&high\\
 &109.2062&IRAM-30m&22.53& 0.64&$2.24\times10^{-4}$&high\\
 &109.2714&IRAM-30m&22.51&0.65& $4.89\times10^{-5}$&high\\
 &109.2812&IRAM-30m&22.51&0.64& $6.12\times10^{-4}$&high\\
 &152.3891&IRAM-30m&16.14&0.6& $5.01\times10^{-3}$&high\\
 &$^\star$152.6570&IRAM-30m&16.11&0.6 &$5.64\times10^{-4}$&high\\
 &$^\star$152.6803&IRAM-30m&16.11&0.6& $5.15\times10^{-4}$&high\\
 &$^\star$152.8555&IRAM-30m&16.09&0.61& $2.36\times10^{-4}$&high\\
 &$^\star$152.8881&IRAM-30m&16.09&0.61& $6.73\times10^{-4}$&high\\
 &152.9532&IRAM-30m&16.08&0.61& $2.59\times10^{-5}$&high\\
 PO&239.7043&IRAM-30m&10.26&0.01& $1.34\times10^{-2}$&10.6 hr\\
 &239.9490&IRAM-30m&10.25&0.02& $6.19\times10^{-4}$&high\\
 &239.9581&IRAM-30m&10.25&0.01& $6.02\times10^{-4}$&high\\
 &240.1411&IRAM-30m&10.24&0.01& $1.59\times10^{-4}$&high\\
 &240.1525&IRAM-30m&10.24&0.01 &$4.62\times10^{-4}$&high\\
 &240.2683&IRAM-30m&10.24&0.02&$7.17\times10^{-6}$&high\\
 &283.3487&IRAM-30m&8.68&0.03 &$1.51\times10^{-2}$&19.8 hr\\
 &283.5868&IRAM-30m&8.67&0.32&$7.97\times10^{-4}$&high\\
 &283.5932&IRAM-30m&8.67&0.32 &$7.85\times10^{-4}$&high\\
 &283.7776&IRAM-30m&8.67&0.32& $6.68\times10^{-4}$&high\\
 &283.7854&IRAM-30m&8.67&0.33& $9.04\times10^{-4}$&high\\
 &283.9125&IRAM-30m&8.66&0.33& $1.04\times10^{-5}$&high\\
 \hline
 &79.9033 & IRAM-30m &30.79 &0.63& $4.04\times10^{-6}$&high\\
 HCP&159.8025 & IRAM-30m &15.39&0.55& $1.68\times10^{-5}$&high\\
 &279.6349 & IRAM-30m &8.80&0.33& $7.51\times10^{-6}$&high\\
 \hline
  &266.9445&IRAM-30m&9.22&0.23& $1.06\times10^{-4}$&high \\
 PH$_3$ &533.7946&SOFIA-GREAT&18.2&0.95& $5.53\times10^{-6}$&high\\ 
 &533.8153&SOFIA-GREAT&18.2&0.95& $4.90\times10^{-6}$&high\\
\hline
\hline
 \end{tabular}}
\hskip 2cm {$^a$ ATRAN (\url{https://atran.arc.nasa.gov/cgi-bin/atran/atran.cgi}) program is used for the calculation of the atmospheric transmission.\\ 
$^b$ We use a frequency resolution of $1$ km/s and a signal to noise ratio $\geq$ 3 with IRAM-30m.}\\
{\noindent $^c$ We use a frequency resolution of 1 km/s and a signal to noise ratio $3$ with SOFIA.}\\
{\noindent $^*$Observed by \cite{font16}.}\\
{\noindent $^{**}$ Observed by \cite{mini18}}\\
 \end{table*}

\section{Summary}
The observed line profiles of HNC, CN, C$^{34}$S in absorption, and one transition of $^{13}$CO in emission are satisfactorily explained by the radiative transfer model for the diffuse cloud environment. Some of the line profiles of P-bearing species are speculated and recommended for further study in the hot core region. The hot core region is predicted to have an inverse P-Cygni profile of PH$_3$.

 \chapter{Evolutionary Stages of Low-mass Star} \label{chap:asai_lowmass}

\section*{Overview}
The fate of Complex organic molecules (COMs) in star-forming regions is linked to different evolutionary stages. Therefore, understanding these COMs physical and chemical origins would be helpful from observing a variety of star-forming regions. Using data from the Large Programme 'Astrochemical Surveys At IRAM' (ASAI), using the IRAM 30 m telescope to conduct chemical surveys in Sun-like star-forming regions, several COMs were identified. It was a millimetre line survey that included the protoplanetary disc phase, outflow region, prestellar core, and protostar. Here, we reported some transitions with seven COMs, namely methanol (CH$_3$OH), acetaldehyde (CH$_3$CHO), methyl formate (CH$_3$OCHO), ethanol (C$_2$H$_5$OH), propynal (HCCCHO), dimethyl ether (CH$_3$OCH$_3$), and methyl cyanide (CH$_3$CN) in some sources L1544 (prestellar core), B1-b (first hydrostatic core), IRAS4A (class 0), and SVS13A (class I). Applying the rotational diagram approach and MCMC fit, we established a trend among these species based on the derived abundances. With the exception of HCCCHO, which showed a peak for the class I phase based on the observed upper limits, we found that the abundances of these COMs steadily increased from the prestellar core and peaked in the class 0 phase before decreasing at the stage of further evolution. It is observed that the luminosity of the sources and the abundance of these compounds are connected. The resulting trend can be seen in the earlier interferometric measurements as well.
\clearpage

\section{General discussions}
\begin{figure}
\hskip -1.0 cm
\includegraphics[width=17cm,height=22cm]{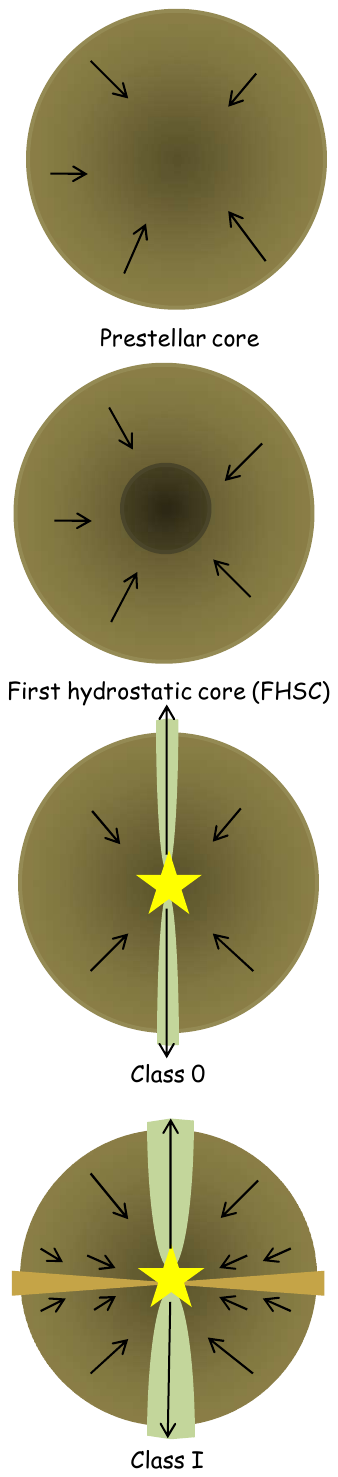}
\caption{Cartoon diagram of the evolutionary stages from prestellar core to class I.}
\label{fig:cartoon_asai}
\end{figure}

Understanding the chemical origins and evolution of matter during the formation of stars is one of the most fascinating topics in modern astrophysics \citep{case12}. The interstellar matter is made up of the molecules and dust created during a stellar cycle. These molecules are crucial to the gas cooling process and start the process of gravitational collapse that produces many stars. Thus, it is crucial to understand and describe the evolutionary processes of our solar system. The formation of low-mass stars through various evolutionary stages is of special interest to this subject. It starts with a molecular cloud breaking up into multiple gravitationally bound cores that are resisted by gravity and thermal, magnetic, and turbulence pressures. The temperature of the pre-stellar core is $\leqslant$30K \citep{garr06b,garr09}. It is now known that the active grain catalysis mechanism may have been responsible for the formation of many complex molecules or their precursors. Unfortunately, pre-stellar core has a low temperature and is covered in interstellar ice, making it difficult to see. However, numerous COMs were being discovered in pre-stellar cores L1544, L183, L1512, L1498 and others due to the improvement of current observational tools \citep{latt20,case17,vast18}. The gravitational energy freely radiates away as the pre-stellar cores become unstable and gravitational collapse begins, keeping the collapsing fragment isothermal. This isothermal collapse results in the first formation of a substantial central concentration of matter. As a result, an envelope with an opaque, hydrostatic protostellar dense core in the middle is still present. Because of the thickness of the envelope, at first, the main object is obscured. The spectral energy distribution (SED) is controlled by the envelope's cool outer sections. This phase is represented by the class 0 phase of star formation \citep{andr93}. Even if the majority of the envelope is frozen, there appears to be a depletion of the molecules that carry heavy elements and have been frozen into grain mantles, such as in the pre-stellar core. However, the existence of the inner core is fueled by gravitational energy, which warms the innermost part of the envelope. As a result, in pre-stellar cores, the grain mantles get evaporate, and the molecules locked in the ice are released into the gas phase, where they may experience additional reactions. Expanding over 100 AU, the "Hot Corino" would be formed. Hot Corinos and Hot Cores of high-mass protostars have similarities but are quite different objects. For instance, they differ slightly chemically according to \citep{bott07}. The most notable hot Corinos are the following: IRAS 16293-2422, IRAS4A and IRAS4B, HH212, L483, B335 \citep{sahu18,jorg04a,jabe14,sant15,jaco19}. After the accretion of more than half of its envelope onto the central core, a Class 0 core begins to transform into a Class I core. The strong stellar wind is created after a million years with the start of thermonuclear fusion in this core, which limits the in-fall of new masses. Evidence implies that the minor solar system constituents (comets, asteroids, etc.) derive a portion of their chemical makeup from the initial stages of solar-type protostar formation. Class I sources would serve as a link between class 0 and the visible disc that is evident in class II and class III phases of star formation. In comparison to the protoplanetary disc phase, more complex molecules (HCOOCH$_3$, $\rm{CH_3OCH_3}$, CH$_3$CHO, CH$_3$OH, HCOOH, $\rm{CH_3CH_2OH}$, NH$_2$CHO, CH$_3$CN, etc.) were seen in hot corino phase. Modern astrophysics greatly benefits from the use of far-infrared and radio observatories. The millimetre waveband radio-telescopes study the icy universe around, enabling in-depth views of systems at various phases of evolution, which ultimately sheds light on the most important chemical processes governing evolution. Systematic spectral line surveys, particularly in the millimetre range, are one of the most crucial investigation techniques to thoroughly examine the evolution of star-forming regions. A new age of molecular detection in star-forming regions has begun as a result of recent outstanding advancements in observational facilities in the radio and far-infrared regimes. Understanding the chemical evolution along the stages of star formation will be understood from focused line surveys of various astronomical sources at various evolutionary stages of star formation. To better understand the chemical evolution through the various stages of low-mass star-forming regions, we have examined the extensive survey of IRAM 30 data for L1544 (pre-stellar core), B1b (first hydrostatic core), IRAS4A (class 0), L1157-mm (class 0), and SVS13A (class I). In Figure \ref{fig:cartoon_asai}, these evolutionary stages are depicted in a cartoon diagram. It is impossible to observe every stage of a single star's evolution. However, we are able to observe and understand many stages of similar kinds of star-forming environments.

\section{Observations}\label{sec:observation_asai}
We used some of the archival data from (PI: Bertrand Lefloch and Rafael Bachiller) Large Programme of Astrochemical Surveys at IRAM (ASAI). To comprehend the chemical and dynamical evolution of solar-type protostars, a comprehensive line study was conducted. The observation was conducted between September 2012 - March 2015 using the IRAM 30 m's EMIR receivers. In this study, four sources - L1544, B1-b, IRAS4A, and SVS13A, each belonging to one of the four distinct stages—from the prestellar phase to class I are taken into account. Ten template sources at various stages of evolution were included in the survey. This line study covered the 3 mm (80–116 GHz), 2 mm (130–170 GHz), and 1.3 mm (200–276 GHz) bands. For B1b, IRAS4A, and SVS3A, all of these bands were covered; however, for L1544, only the 3 mm (80-116 GHz) range was covered. \cite{lefl18} already covered the observational information in extensive detail. The equation: HPBW($^{''}$) = 2460/frequency(GHz) would be used to calculate the half power beam width (HPBW) or beam size of the IRAM 30 m telescope. Here, T$_{MB}$ = T$_A^*$ /$\eta_{MB}$, where $\eta_{MB}$ is the antenna efficiency, was used to convert the antenna temperature (T$_A^*$) to the main beam temperature (T$_{MB}$). The main beam temperature is used to define all intensities in tables and graphs. The section below discusses the specifics of the sources that were taken into consideration. The required information about these sources and a summary of the targeted positions are in Table \ref{table:source}.

\begin{table}
\centering
\tiny
\caption{ \textbf {Targeted positions and relevant information of the sample sources. \label{table:source}}}
\begin{tabular}
{|c|c|c|c|c|c|}
\hline
Source name&Stage of the source&Coordinates&Distance&Luminosity&{$V_{LSR}$}\\
&&(J2000)&(pc)&(L$_\odot$)&km s$^{-1}$\\
\hline
L1544&Evolved prestellar core&$05^{h}04^{m}17.^{s}21$+$25^{\circ}10^{'}42^{''}.8$&140&1.0&7.2 \citep{jime16}\\
B1-b&First hydrostatic core (FHSC)&$03^{h}33^{m}20.^{s}80$+$31^{\circ}07^{'}34^{''}.0$&230&0.77&6.5 \citep{lope15}\\
IRAS4A&Class 0&$03^{h}29^{m}10.^{s}42$+$31^{\circ}13^{'}32^{''}.2$&260&9.1&7.2 \citep{geri09}\\
SVS13A&Class I&$03^{h}29^{m}03.^{s}73$+$31^{\circ}16^{'}03^{''}.8$&260&34.0&8.6 \citep{chen09}\\
\hline
\end{tabular}
\end{table}

\subsection{L1544 as pre-stellar core}
In Taurus molecular cloud, L1544 is a dense starless core that is in the early stages of collapse \citep{tafa98,ciol00}. The sun is 140 pc away from this proto-type evolved pre-stellar core \citep{cern87}. Matters accumulating in the core of the cloud during the pre-stellar core phase result in a decrease in temperature and an increase in density. As a result, the atoms and molecules in gas phase freeze on the dust, forming icy grain mantles. A very low temperature of $\sim$ 7 K is attained coupled with a central density of $\sim$ 10$^6$ cm$^{-3}$. A significant degree of deuterium fractionation of $\rm{N_2H^+}$ compared to HCO$^+$ is seen as a result of heavy depletion \citep{caseb02,rade19}. In L1544 \citep{vast14}, numerous sulfur-related compounds, as well as pre-biotic molecules \citep{vast18,lope15}, have been identified to date. The reported temperature decline inside the 2000 AU could not be replicated without taking into account a high dust opacity, according to detailed modeling of L1544 by \cite{keto10b}. \cite{case99}, suggested that it might be due to fluffy grains in the core's central region, where CO is deeply frozen and volume densities exceed 10$^6$ cm$^{-3}$. The physical composition of this source is well constrained by a number of observations. It is a particularly interesting target to investigate opacity variations due to its large volume densities and centrally concentrated shape.

\subsection{B1b as first hydrostatic core}
Barnard 1 (B1) is a member of the Perseus molecular cloud complex, which is located 235 pc away \citep{fuen16}. A number of active sources at different stages of star formation make it an intriguing source from a chemical and kinematic perspective. For instance, class 0 sources B1-a and B1-c are connected to outflows. The main core B1-b is made up of three components: B1-b-N and B1-b-S, two immature star objects, and B1-b-W, a more evolved object \citep{huan13,fuen16}. B1-b-N and B1-b-S, which are in the first hydrostatic core (FHSC) stage, are separated by around 18$^{''}$. The spectral energy distribution and outflow of the FHSC are used to categorize it \citep{geri15,pezz12}. The source has a total luminosity of 0.77L$_\odot$ \citep{lefl18}. B1-b has received the greatest attention because of its rich molecular spectrum. Previously, \cite{marc09} and \cite{cern12} conducted a study toward B1-b with the IRAM 30m radio telescope in the 3 millimeter band. Numerous species were discovered, including CH$_3$O \citep{cern12}, $\rm{NH_3D^+}$ \citep{cern13}, and HCNO \citep{marc09}. A high deuteration seen toward this source is confirmed by the identification of D$_2$CS, ND$_2$H, and ND$_3$. Many complex organic molecules, including CH$_3$OCOH, CH$_3$CHO, and many COMs ($\rm{NH_2CN, NH_2CHO, CH_3CH_2OH, CH_2OHCHO}$, and $\rm{CH_3CH_2OCOH}$) are tentatively found toward B1b \citep{marc18b}, according to recent observations made with the ALMA interferometer. Compared to other Barnard 1 components, B1b-S exhibits COMs rich spectra.

\subsection{NGC 1333-IRAS4A as class 0 object}
A proto-binary system called IRAS4A can be found in the NGC1333 reflection nebula situated in Perseus cloud. According to a recent GAIA result, it is located at a distance of $\sim$293$\pm$22 pc (\cite{orti18} and \cite{zuck18}). The gas envelop of IRAS4A has a total mass of 3.5M$_\odot$ and a total luminosity of 9.1L$_\odot$ \citep{lefl18}. IRAS4A \citep{marv08} is separated from another component, IRAS4B, by an angle of 31$^{''}$. Many COMs have been found to date that are aimed to observe in IRAS4A, including CH$_3$OH, $\rm{CH_3OCH_3, C_2H_5CN}$, and CH$_2$OHCHO. It is seen that IRAS4A is connected with a very high collimated outflow from CO, CS, and SiO emission \citep{blak95,lefl98}. By \cite{fran01} and \cite{choi99}, infall motion was discovered in this source, which has an estimated accretion rate of $1.1\times10^{-4}$ M$_\odot$.yr$^{-1}$, an inner mass of 0.7 M$_\odot$, and an age of $\sim$ 6500 yr (see also \cite{mare03}). Recent IRAM 30m observations categorise IRAS4A as a hot corino protostar. Recent high-resolution interferometric data shows that IRAS4A is made up of two parts, IRAS4A1 and IRAS4A2, which are spaced apart by an angle of 1.8$^{''}$ ($\sim$527 AU) \citep{des20a}. The chemical composition of IRAS4A1 and IRAS4A2 varies dramatically. From two of them \cite{sant15} and \cite{lope17} confirmed that IRAS4A2 is a hot corino protostar; however, further investigation is required to establish this for IRAS4A1. Recently, \cite{des20a} came to the conclusion that IRAS4A1 is a hot corino and the absence of iCOMS detected toward ths is caused by a significant increase in dust optical depth toward the centre. \cite{sahu19} illustrated two possible scenarios of IRAS4A1: a) In A1, the observed absorption features are likely the result of a hot-corino atmosphere colliding with a very small ($\leq$ 36 AU) disc. b) Different layers of a dense, temperature-stratified envelope in A1 may be responsible for the absorption.

\subsection{SVS13A as class I protostar}
 The comparatively advanced protostar SVS13A is in the class I phase. The source has a luminosity of 34L$_\odot$ and is situated $\sim$ 260pc \citep{schl14} away. It contains the NGC 1333-SVS13 system. It is made up of sources A, B, and C. SVS13A is separated from its two counterparts, SVS13B and SVS13C, at an angles 15 and 20 degrees, respectively \citep{bach98,loon07}. Extended outflows are present in SVS13A, which are connected to well-known Herbig-Haro chain (HH) objects 7–11 \citep{reip93}. According to the VLA observation, SVS13A is a closed binary system with VLA4A and VLA4B spaced apart by 0$^{''}.3$ ($\sim$70AU) \citep{angl00}. According to a recent observation by \cite{diaz22}, SVS13A is located at a distance of $\sim$ 300pc, and VLA4A and VLA4B are separated by $\sim$ 90AU. Numerous COMs were already found in this source \citep{bian19} using the ASAI data, including acetaldehyde (CH$_3$CHO), methyl formate (HCOOCH$_3$), dimethyl ether (CH$_3$OCH$_3$), ethanol (CH$_3$CH$_2$OH), and formamide (NH$_2$CHO).

\subsection{Line identification}\label{sec:data-analysis}
Software named CASSIS (footnote url: http://cassis.irap.omp.eu/?page=cassis) is used to identify the lines. For the spectroscopic details, we used the Jet Propulsion Laboratory (JPL) and the Cologne Database for Molecular Spectroscopy (CDMS; footnote url: https://cdms.astro.uni-koeln.de/; citealt: mull01, mull05). The transitions of several complex molecules that have been detected are listed in Table. \ref{tab:observation_1}, \ref{tab:observation_2}, \ref{tab:observation_3}, \ref{tab:observation_4} along with their quantum numbers, upper state energies (E$_{up}$), $V_{LSR}$, line parameters such as line width (FWHM), and integrated intensity ($\int$T$_{mb}$dV). A single Gaussian fit to the observed spectral profile of each unblended transition is used to determine the line parameters. The lines that have been identified are plotted in black color in Fig. \ref{fig:ch3oh_l1544mcmc}, \ref{fig:ch3oh_barnardmcmc}, \ref{fig:ch3oh_irasmcmc}, \ref{fig:ch3oh_svsmcmc}, \ref{fig:ch3cho_mcmc}, \ref{fig:ch3ocho_mcmc}, \ref{fig:c2h5oh_mcmc}, \ref{fig:hcccho_mcmc}, \ref{fig:ch3och3_mcmc}.

\subsection{H$_2$ column density} \label{sec:H2_col}
Due to the lack of the continuum observation data, we used the H$_{2}$ column density from the literature to derive the abundances of species. For the L1544, we used a H$_{2}$ column density of $8.9 \times 10^{22}$ cm$^{-2}$ (calculated by \citealt{hily22}) for a beam size of $26^{''}$ in L1544. \cite{dani13} obtained an average H$_2$ column density (1.2 mm observation with IRAM) of $7.6 \times 10^{22}$ cm$^{-2}$  within the 30$^{''}$ beam in B1. \cite{john10} estimated a H$_{2}$ column density $\sim 8.2 \times 10^{22}$ cm$^{-2}$ for the same beam. Following \cite{lope15}, we used an average H$_2$ column density $\sim 7.9 \times 10^{22}$ cm$^{-2}$ for B1-b in estimating the abundances. For IRAS4A, \cite{mare02} obtained a H$_2$ column density of $2.9 \times 10^{22}$ cm$^{-2}$ for a 30$^{''}$ beam, whereas with a 0.5 $^{''}$ beam it was $2.5 \times 10^{24}$ cm$^{-2}$. Since we analyzed the data obtained from the IRAM 30 m telescope, we use $2.9 \times 10^{22}$ cm$^{-2}$ for abundance estimation. For SVS13A, we used the H$_2$ column density of $10^{23}$ cm$^{-2}$ estimated by \cite{lefl98} for 20$^{''}$ beam. Due to this uncertainty in the H$_2$ column density, the derived abundances and the chemical trend have an uncertainty also. To minimize this effect, we used the values of H$_2$ column density in different sources calculated for $\sim$ 30$^{''}$ beam. The H$_2$ column density used in deriving the abundances is summarized in Table \ref{tab:rotdiag}.

\begin{landscape}
\begin{table}
\tiny{
\caption{Observed transitions toward some sources. \label{tab:observation_1}}
\begin{tabular}{|l|l|l|l|l|l|l|l|l|l|l|}
\hline
Species&Tag (Database)&Source&Frequency&E$_{up}$&Quantum No.&A$_{ij}$&V$_{LSR}$&T$_{peak}$&FWHM&$\int$T$_{mb}$.dv\\
&&&(GHz)&(K)&&(s$^{-1}$)&(km.s$^{-1}$)&(K)&(km.s$^{-1}$)&(K.km.s$^{-1}$)\\
\hline
\hline
CH$_3$OH&32003(JPL)&L1544&84.521172&40.4&$5_{-1,0}$ - $4_{0,0}$&	$1.97\times10^{-6}$&7.24$\pm$     0.01&   0.017$\pm$     0.001&   0.441$\pm$     0.024&   0.008$\pm$     0.001\\
(Methanol)
&&&96.744545&20.1&$2_{0,0}$ - $1_{0,0}$&$3.41\times10^{-6}$&7.16$\pm$     0.02&   0.109$\pm$     0.001&   0.358$\pm$     0.004&   0.042$\pm$     0.001\\
&&&96.755501&28.0&$2_{1,0}$ - $1_{1,0}$&$2.62\times10^{-6}$&7.07$\pm$     0.01&   0.013$\pm$     0.001&   0.416$\pm$     0.029&   0.006$\pm$     0.001 \\
&&&97.582798&21.5&$2_{1,-0}$ - $1_{1,-0}$&$2.6\times10^{-6}$&7.18$\pm$     0.01&   0.019$\pm$     0.001&   0.428$\pm$     0.021&   0.009$\pm$     0.001\\
\cline{3-11}
&&B1-b&96.739358&12.5&$2_{-1,0}$ - $1_{-1,0}$&$2.56\times10^{-6}$&6.641$\pm$     0.002&   1.243$\pm$     0.003&   1.355$\pm$     0.005&   1.793$\pm$     0.010\\
&&&	96.741371&6.9&$2_{0,+0}$ - $1_{0,+0}$&$3.41\times10^{-6}$&6.615$\pm$     0.001&   1.694$\pm$     0.003&   1.365$\pm$     0.003&   2.462$\pm$     0.010\\
&&&	96.744545&20.1&	$2_{0,0}$ - $1_{0,0}$&$3.41\times10^{-6}$&6.606$\pm$     0.007&   0.306$\pm$     0.003&   1.365$\pm$     0.018&   0.445$\pm$     0.010\\
&&&	96.755501&28.0&	$2_{1,0}$ - $1_{1,0}$&$2.62\times10^{-6}$&6.569$\pm$     0.033&   0.066$\pm$     0.003&   1.310$\pm$     0.084&   0.092$\pm$     0.010\\
&&&	108.893945&13.1&$0_{0,0}$ - $0_{1,-1}$ &$1.47\times10^{-5}$&6.565$\pm$0.005&   0.377$\pm$     0.003&   1.371$\pm$     0.013&   0.550$\pm$     0.009\\
&&&	145.093754&27.05&$3_{0,0}$ - $2_{0,0}$&$1.23\times10^{-5}$&6.671$\pm$     0.003&   0.479$\pm$     0.003&   0.950 $\pm$    0.008&   0.484$\pm$     0.007\\
&&&	145.103185&13.9&$3_{0,+0}$ - $2_{0,+0}$ &$1.23\times10^{-5}$&6.718$\pm$     0.001&   2.316$\pm$     0.003&   1.005$\pm$     0.002&   2.478$\pm$     0.007\\
&&&	157.270832&15.4&$1_{0,0}$ - $1_{-1,0}$&$2.21\times10^{-5}$&6.513$\pm$0.002&   0.896$\pm$     0.003&   1.012$\pm$  0.004&   0.965$\pm$     0.007\\
&&&	157.276019&20.1&$2_{0,0}$ - $2_{-1,0}$&$2.18\times10^{-5}$&6.533$\pm$0.002&   0.583$\pm$     0.003&   0.978$\pm$     0.006&   0.607$\pm$     0.006\\
&&&	165.050175&23.3&$1_{1,0}$ - $1_{0,0}$&$2.35\times10^{-5}$&6.476$\pm$0.004&   0.334$\pm$     0.003&   0.878$\pm$     0.010&   0.312$\pm$     0.006\\
&&&	165.06113&28.01&$2_{1,0}$ - $2_{0,0}$&$2.34\times10^{-5}$&6.474$\pm$0.005&   0.314$\pm$     0.003&   0.905$\pm$     0.011&   0.302$\pm$     0.006\\
&&&	165.09924&34.9&$3_{1,0}$ - $3_{0,0}$&$2.33\times10^{-5}$&6.524$\pm$0.009&   0.189$\pm$     0.002&   1.201$\pm$     0.023&   0.242$\pm$     0.008\\
&&&	170.060592&36.2&$3_{2,0}$ - $2_{1,0}$&2.55$\times10^{-5}$&6.683$\pm$     0.003&   0.510$\pm$     0.003&   0.742$\pm$     0.006&   0.403$\pm$     0.006\\
&&&	213.427061&23.3&$1_{1,0}$ - $0_{0,0}$&$3.37\times10^{-5}$&6.580$\pm$     0.002&   0.354$\pm$     0.002&   0.903$\pm$     0.005&   0.341$\pm$     0.004\\
&&&	254.015377&20.0&$2_{0,0}$ - $1_{-1,0}$&$1.90\times10^{-5}$&6.621$\pm$     0.002&   0.427$\pm$     0.001&   1.035$\pm$     0.005&   0.471$\pm$     0.004\\
&&&	261.805675&28.0&$2_{1,0}$ - $1_{0,0}$&$5.57\times10^{-5}$&6.623$\pm$     0.002&   0.354$\pm$    0.001&   1.005$\pm$     0.005&   0.379$\pm$     0.003\\
\cline{3-11}
&&IRAS4A&96.755501	&28.0	&	$2_{1,0}$ - $1_{1,0}$&$2.62\times10^{-6}$&7.170$\pm$     0.029&   0.143$\pm$     0.002&   2.600$\pm$     0.118&   0.397$\pm$     0.025\\
&&&108.893945	&13.1	&	$0_{0,0}$ - $1_{-1,0}$&	$1.47\times10^{-6}$&7.402$\pm$     0.011&   0.335$\pm$     0.002&   2.446$\pm$     0.046&   0.871$\pm$     0.022\\
&&&143.865795	&28.3	&	$3_{1,+0}$ - $2_{1,+0}$&$1.07\times10^{-5}$&7.279$\pm$     0.010&   0.313$\pm$     0.002&   2.697$\pm$     0.042&   0.899$\pm$     0.019\\
&&&157.178987	&47.9	&	$5_{0,0}$ - $5_{-1,0}$&	$2.04\times10^{-5}$&7.312$\pm$     0.009&   0.286$\pm$     0.002&   2.352$\pm$     0.034&   0.717$\pm$     0.014\\
&&&165.050175	&23.4	&	$1_{1,0}$ - $1_{0,0}$&	$2.35\times10^{-5}$&7.335$\pm$     0.008&   0.263$\pm$     0.002&   2.259$\pm$     0.028&   0.632$\pm$     0.012\\
&&&165.06113	&28.0	&	$2_{1,0}$ - $2_{0,0}$&	$2.34\times10^{-5}$&7.242$\pm$     0.008&   0.319$\pm$     0.002&   2.547$\pm$     0.027&   0.865$\pm$     0.013\\
&&&165.09924	&35.0	&	$3_{1,0}$ - $3_{0,0}$&	$2.33\times10^{-5}$&7.195$\pm$     0.007&   0.330$\pm$     0.002&   2.553$\pm$     0.026&   0.897$\pm$     0.013\\
&&&213.427061	&23.4	&	$1_{1,0}$ - $0_{0,0}$&	$3.37\times10^{-5}$&7.472$\pm$     0.007&   0.313$\pm$     0.001&   2.254$\pm$     0.023&   0.752$\pm$     0.011\\
&&&230.027047	&39.8	&	$3_{-2,0}$ - $4_{-1,0}$&$1.49\times10^{-5}$&7.345$\pm$     0.019&   0.095$\pm$     0.001&   2.316$\pm$     0.063&   0.233$\pm$     0.010\\
&&&254.015377	&20.1	&	$2_{0,0}$ - $1_{-1,0}$&$1.90\times10^{-5}$&7.073$\pm$     0.005&   0.329$\pm$     0.001&   2.030$\pm$     0.015&   0.711$\pm$     0.008\\
&&&261.805675	&28.0	&	$2_{1,0}$ - $1_{0,0}$&$5.57\times10^{-5}$&7.095$\pm$     0.004&   0.420$\pm$     0.001&   2.275$\pm$     0.014&   1.016$\pm$     0.009	\\
&&&155.320895	&140.6	&	$10_{0,0}$ - $10_{-1,0}$&$1.55\times10^{-5}$&7.289$\pm$     0.077&   0.047$\pm$     0.002&   2.883$\pm$     0.349&   0.145$\pm$     0.022\\
&&&155.997524	&117.5	&	$9_{0,0}$ - $9_{-1,0}$&	$1.67\times10^{-5}$&7.355$\pm$     0.076&   0.060$\pm$     0.002&   4.328$\pm$     0.278&   0.278$\pm$     0.025\\
&&&156.488902	&96.6	&	$8_{8,0}$ - $8_{-1,0}$&	$1.78\times10^{-5}$&7.487$\pm$     0.066&   0.093$\pm$     0.001&   3.158$\pm$     0.253&   0.313$\pm$     0.030\\
&&&156.828517	&78.1	&	$7_{0,0}$ - $7_{-1,0}$&	$1.88\times10^{-5}$&7.305$\pm$     0.031&   0.125$\pm$     0.002&   3.467$\pm$     0.163&   0.460$\pm$     0.027\\
&&&157.048617	&61.8	&	$6_{0,0}$ - $6_{-1,0}$&	$1.96\times10^{-5}$&7.331$\pm$     0.016&   0.192$\pm$     0.002&   2.790$\pm$     0.071&   0.571$\pm$     0.019\\
&&&165.678649	&69.8	&	$6_{1,0}$ - $6_{0,0}$&	$2.30\times10^{-5}$&7.225$\pm$     0.019&   0.148$\pm$     0.002&   2.846$\pm$     0.087&   0.448$\pm$     0.018\\
&&&166.169098	&86.1	&	$7_{1,0}$ - $7_{0,0}$&	$2.28\times10^{-5}$&7.059$\pm$     0.035&   0.119$\pm$     0.001&   3.225$\pm$     0.170&   0.410$\pm$     0.027\\
&&&241.879025	&55.9	&	$5_{1,0}$ - $4_{1,0}$&$5.96\times10^{-5}$&7.350$\pm$     0.006&   0.311$\pm$     0.001&   2.347$\pm$     0.023&   0.777$\pm$     0.011\\
&&&251.738437	&98.5	&	$6_{3,-0}$ - $6_{2,+0}$&$7.46\times10^{-5}$&6.985$\pm$     0.025&   0.113$\pm$     0.001&   3.078$\pm$     0.110&   0.369$\pm$     0.017\\
&&&251.866524	&73.0	&	$4_{3,-0}$ - $4_{2,+0}$&$6.10\times10^{-5}$&6.926$\pm$     0.026&   0.135$\pm$     0.001&   3.119$\pm$     0.090&   0.447$\pm$     0.017\\
&&&251.917065	&63.7	&	$3_{3,+0}$ - $3_{2,-0}$&$4.36\times10^{-5}$&6.867$\pm$     0.032&   0.090$\pm$     0.001&   2.794$\pm$     0.106&   0.267$\pm$     0.014\\
&&&251.984837	&133.4	&	$8_{3,+0}$ - $8_{2,-0}$&$7.99\times10^{-5}$&7.020$\pm$     0.030&   0.073$\pm$     0.001&   2.692$\pm$     0.105&   0.209$\pm$     0.012\\
&&&252.090409	&154.2	&	$9_{3,+0}$ - $9_{2,-0}$&$8.15\times10^{-5}$&6.910$\pm$     0.025&   0.074$\pm$     0.001&   2.341$\pm$     0.080&   0.186$\pm$     0.010\\
\cline{3-11}
&&SVS13A&85.568131&	74.7    &	$6_{-2,0}$ - $7_{-1,0}$&$1.13\times10^{-6}$&8.395$\pm$     0.341&   0.018$\pm$     0.002&   4.570$\pm$     1.399&   0.088$\pm$     0.037\\
&&&	96.755501	&	28.0    &	$2_{1,0}$ - $1_{1,0}$&$2.62\times10^{-6}$&9.336$\pm$     0.460&   0.026$\pm$     0.002&   3.163$\pm$     1.026&   0.088$\pm$     0.036\\
&&&	111.289453	&	102.7	&	$7_{2,+0}$ - $8_{1,+0}$&$2.60\times10^{-6}$&7.288$\pm$     0.078&   0.035$\pm$     0.002&   2.346$\pm$     0.213&   0.088$\pm$     0.013\\
&&&	143.865795	&	28.3	&	$3_{1,+0}$ - $2_{1,+0}$&$1.07\times10^{-5}$&8.528$\pm$     0.022&   0.098$\pm$     0.002&   2.101$\pm$     0.069&   0.219$\pm$     0.011\\
&&&	156.602395	&	21.4	&	$2_{1,+0}$ - $3_{0,+0}$&$8.93\times10^{-5}$&8.250$\pm$     0.032&   0.094$\pm$     0.002&   2.663$\pm$     0.118&   0.267$\pm$     0.016\\
&&&	218.440063	&	45.5	&	$4_{2,0}$ - $3_{1,0}$&$4.69\times10^{-5}$&8.513$\pm$     0.006&   0.246$\pm$     0.001&   1.810$\pm$     0.018&   0.473$\pm$     0.008 \\
&&&	229.758756	&	89.1	&	$8_{-1,0}$ - $7_{0,0}$&$4.19\times10^{-5}$&8.146$\pm$     0.025&   0.155$\pm$     0.001&   3.380$\pm$     0.092&   0.557$\pm$     0.020\\
\hline
\end{tabular}}
\end{table}
\end{landscape}
\begin{landscape}
\begin{table}
\centering
\tiny{
\caption{Observed transitions toward some sources. \label{tab:observation_2}}
\begin{tabular}{|l|l|l|l|l|l|l|l|l|l|l|}
\hline
Species&Tag (Database)&Source&Frequency&E$_{up}$&Quantum No.&A$_{ij}$&V$_{LSR}$&T$_{peak}$&FWHM&$\int$T$_{mb}$.dv\\
&&&(GHz)&(K)&&(s$^{-1}$)&(km.s$^{-1}$)&(K)&(km.s$^{-1}$)&(K.km.s$^{-1}$)\\
\hline
&&&	241.700159	&	47.9	&	$5_{0,0}$ - $4_{0,0}$&$6.04\times10^{-5}$&8.086$\pm$     0.017&   0.184$\pm$     0.001&   2.744$\pm$     0.053&   0.538$\pm$     0.014\\
&&&	241.791352	&	34.8	&	$5_{0,+0}$ - $4_{0,+0}$&$6.05\times10^{-5}$&8.169$\pm$     0.004&   0.430$\pm$     0.001&   1.680$\pm$     0.010&   0.768$\pm$     0.007\\
&&&	241.879025	&	55.9	&	$5_{1,0}$ - $4_{1,0}$&$5.96\times10^{-5}$&8.051$\pm$     0.029&   0.166$\pm$     0.001&   3.094$\pm$     0.085&   0.547$\pm$     0.019\\
&&&	243.915788	&	49.7	&	$5_{1,-0}$ - $4_{1,-0}$&$5.97\times10^{-5}$&8.444$\pm$     0.012&   0.163$\pm$     0.001&   2.659$\pm$     0.043&   0.461$\pm$     0.011\\
&&&	251.738437	&	98.5	&	$6_{3,-0}$ - $6_{0,+0}$&$7.46\times10^{-5}$&8.463$\pm$     0.037&   0.089$\pm$     0.001&   3.380$\pm$     0.197&   0.320$\pm$     0.023\\
&&&	261.805675	&	28.0	&	$2_{1,0}$ - $1_{0,0}$&$5.57\times10^{-5}$&8.915$\pm$     0.009&   0.154$\pm$     0.001&   1.691$\pm$     0.024&   0.277$\pm$     0.006\\
&&&	266.838148	&	57.1	&$5_{2,0}$ - $4_{1,0}$&$7.74\times10^{-5}$&8.357$\pm$     0.009&   0.197$\pm$     0.001&   2.003$\pm$     0.028&   0.420$\pm$     0.009\\
\hline
CH$_3$CHO&44003 (JPL)&L1544&93.5809&15.75&5$_{1,5}$ - 4$_{1,4}$, (A)&$2.53\times10^{-5}$&7.14$\pm$0.009&0.022$\pm$0.001&0.438$\pm$0.02&0.01$\pm$0.001\\
(Acetaldehyde)&&&93.5952&15.82&5$_{1,5}$ - 4$_{1,4}$, (E)&$2.53\times10^{-5}$&7.19$\pm$0.007&0.024$\pm$0.001&0.342$\pm$0.017&0.009$\pm$0.001\\
&&&95.9474&13.93&5$_{0,5}$ - 4$_{0,4}$, (E)&$2.84\times10^{-5}$&7.16$\pm$0.005&0.039$\pm$0.001&0.346$\pm$0.011&0.014$\pm$0.001\\
&&&95.9634&13.84&5$_{0,5}$ - 4$_{0,4}$, (A)&$2.84\times10^{-5}$&7.18$\pm$0.006&0.041$\pm$0.001&0.437$\pm$0.013&0.019$\pm$0.001\\
&&&98.8633&16.59&5$_{1,4}$ - 4$_{1,3}$, (E)&$2.99\times10^{-5}$&7.11$\pm$0.008&0.025$\pm$0.001&0.527$\pm$0.019&0.014$\pm$0.001\\
&&&98.9009&16.51&5$_{1,4}$ - 4$_{1,3}$, (A)&$2.99\times10^{-5}$&7.17$\pm$0.008&0.02$\pm$0.001&0.34$\pm$0.02&0.007$\pm$0.001\\
\cline{3-11}
&&B1-b&93.5809&15.75&5$_{1,5}$ - 4$_{1,4}$, (A)&$2.53\times10^{-5}$&6.57$\pm$0.04&0.051$\pm$0.003&1.428$\pm$0.110&0.078$\pm$0.010\\
&&&93.5952&15.82&5$_{1,5}$ - 4$_{1,4}$, (E)&$2.53\times10^{-5}$&6.58$\pm$0.04&0.050$\pm$0.003&1.568$\pm$0.144&0.084$\pm$0.012\\
&&&95.9634&13.84&5$_{0,5}$ - 4$_{0,4}$, (A)&$2.84\times10^{-5}$&6.66$\pm$0.030&0.073$\pm$0.003&1.461$\pm$0.080&0.114$\pm$0.011\\
&&&96.4256&22.91&5$_{2,4}$ - 4$_{2,3}$, (E)&$2.41\times10^{-5}$&6.63$\pm$0.104&0.023$\pm$0.003&1.661$\pm$0.297&0.041$\pm$0.012\\
&&&96.4755&23.02&5$_{2,3}$ - 4$_{2,2}$, (E)&$2.42\times10^{-5}$&6.52$\pm$0.083&0.022$\pm$0.003&0.973$\pm$0.199&0.022$\pm$0.008\\
&&&98.8633&16.59&5$_{1,4}$ - 4$_{1,3}$, (E)&$2.99\times10^{-5}$&6.51$\pm$0.034&0.054$\pm$0.003&1.040$\pm$0.081&0.059$\pm$0.008\\
&&&98.9009&16.51&5$_{1,4}$ - 4$_{1,3}$, (A)&$2.99\times10^{-5}$&6.55$\pm$0.040&0.049$\pm$0.003&1.249$\pm$0.100&0.066$\pm$0.009\\
&&&138.2849&28.92&7$_{1,6}$ - 6$_{1,5}$, (E)&$5.57\times10^{-5}$&6.82$\pm$0.027&0.054$\pm$0.003&0.870$\pm$0.063&0.050$\pm$0.006\\
&&&138.3196&28.85&7$_{1,6}$ - 6$_{1,5}$, (A)&$8.57\times10^{-5}$&6.65$\pm$0.031&0.048$\pm$0.003&0.775$\pm$0.072&0.039$\pm$0.006\\
&&&152.6352&33.10&8$_{0,8}$ - 7$_{0,7}$, (A)&$1.18\times10^{-4}$&6.28$\pm$0.049&0.034$\pm$0.003&0.750$\pm$0.116&0.027$\pm$0.006\\
&&&155.1796&42.54&8$_{2,6}$ - 7$_{2,5}$, (E)&$1.15\times10^{-4}$&6.28$\pm$0.049&0.034$\pm$0.003&0.750$\pm$0.116&0.027$\pm$0.006\\
\cline{3-11}
&&IRAS4A&74.8917&11.26&4$_{1,4}$ - 3$_{1,3}$, (A)&$1.24\times10^{-5}$&6.84$\pm$0.05&0.051$\pm$0.003&   2.641$\pm$0.216&0.144$\pm$0.021\\
&&&74.9241&11.33&4$_{1,4}$ - 3$_{1,3}$, (E)&$1.24\times10^{-5}$&7.03$\pm$0.07&0.050$\pm$0.003&2.375$\pm$0.225&   0.126$\pm$0.019\\
&&&76.8789&9.23&4$_{0,4}$ - 3$_{0,3}$, (A)&$1.43\times10^{-5}$&7.14$\pm$0.06&0.061$\pm$     0.003&   2.332$\pm$     0.185&   0.152$\pm$     0.019\\
&&&77.0386&18.31&4$_{2,3}$ - 3$_{2,2}$, (A)&$1.08\times10^{-5}$&6.96$\pm$0.19&0.022$\pm$     0.003&   2.653$\pm$     0.507&   0.063$\pm$     0.021\\
&&&77.2183&18.33&4$_{2,2}$ - 3$_{2,1}$, (A)&$1.09\times10^{-5}$&6.51$\pm$0.15&0.029$\pm$     0.003&   2.708$\pm$     0.408&   0.082$\pm$     0.021\\
&&&79.0993&11.84&4$_{1,3}$ - 3$_{1,2}$, (E)&$1.46\times10^{-5}$&7.01$\pm$0.07&0.059$\pm$     0.003&   2.299$\pm$     0.207&   0.144$\pm$     0.020\\
&&&93.5809&15.75&5$_{1,5}$ - 4$_{1,4}$, (A)&$2.53\times10^{-5}$&7.32$\pm$0.05&0.065$\pm$     0.003&   2.298$\pm$     0.143&   0.158$\pm$     0.016\\
&&&93.5952&15.82&5$_{1,5}$ - 4$_{1,4}$, (E)&$2.53\times10^{-5}$&7.44$\pm$0.05&0.062$\pm$     0.003&   2.343$\pm$     0.159&   0.155$\pm$     0.017\\
&&&95.9474&13.93&5$_{0,5}$ - 4$_{0,4}$, (E)&$2.84\times10^{-5}$&7.39$\pm$0.03&0.081$\pm$     0.003&   2.277$\pm$     0.102&   0.197$\pm$     0.016\\
&&&96.2742&22.93&5$_{2,4}$ - 4$_{2,3}$, (A)&$2.41\times10^{-5}$&7.33$\pm$0.07&0.037$\pm$     0.003&   2.148$\pm$     0.222&   0.085$\pm$     0.015\\
&&&96.4256&22.91&5$_{2,4}$ - 4$_{2,3}$, (E)&$2.41\times10^{-5}$&7.54$\pm$0.10&0.031$\pm$     0.003&   2.406$\pm$     0.315&   0.078$\pm$     0.017\\
&&&96.4755&23.03&5$_{2,3}$ - 4$_{2,2}$, (E)&$2.42\times10^{-5}$&7.17$\pm$0.09&0.036$\pm$     0.003&   2.357$\pm$     0.248&   0.092$\pm$     0.016\\
&&&98.8633&16.59&5$_{1,4}$ - 4$_{1,3}$, (E)&$2.99\times10^{-5}$&7.36$\pm$0.04&0.076$\pm$     0.003&   2.242$\pm$     0.115&   0.182$\pm$     0.016\\
&&&112.248&21.13&6$_{1,6}$ - 5$_{1,5}$, (A)&$4.50\times10^{-5}$&7.29$\pm$0.05&0.065$\pm$     0.002&   2.559$\pm$     0.148&   0.178$\pm$     0.017\\
&&&112.2545&21.21&6$_{1,6}$ - 5$_{1,5}$, (E)&$4.50\times10^{-5}$&7.27$\pm$0.06&0.062$\pm$     0.002&   3.061$\pm$     0.199&   0.203$\pm$     0.021\\
&&&133.8305&25.87&7$_{0,7}$ - 6$_{0,6}$, (E)&$7.91\times10^{-5}$&7.33$\pm$0.02&0.084$\pm$     0.002&   2.113$\pm$     0.068&   0.189$\pm$     0.011\\
&&&138.2849&28.92&7$_{1,6}$ - 6$_{1,5}$, (E)&$8.57\times10^{-5}$&7.66$\pm$0.03&0.067$\pm$     0.002&   2.314$\pm$     0.088&   0.165$\pm$     0.011\\
&&&138.3196&28.85&7$_{1,6}$ - 6$_{1,5}$, (A)&$8.57\times10^{-5}$&7.51$\pm$0.03&0.063$\pm$     0.002&   2.541$\pm$     0.102&   0.171$\pm$     0.012\\
&&&152.6352&33.10&8$_{0,8}$ - 7$_{0,7}$, (A)&$1.18\times10^{-4}$&7.48$\pm$0.02&0.077$\pm$     0.002&   2.105$\pm$     0.064&   0.173$\pm$     0.010\\
&&&155.3421&42.50&8$_{2,6}$ - 7$_{2,5}$, (A)&$1.17\times10^{-4}$&7.23$\pm$0.04&0.053$\pm$     0.002&   2.193$\pm$     0.105&   0.123$\pm$     0.010\\
&&&168.0934&42.66&9$_{1,9}$ - 8$_{1,8}$, (A)&$1.57\times10^{-4}$&7.39$\pm$0.03&0.069$\pm$     0.002&   2.622$\pm$     0.095&   0.192$\pm$     0.012\\
&&&208.2285&60.52&11$_{0,11}$ - 10$_{0,10}$, (E)&$3.05\times10^{-4}$&7.36$\pm$0.03&0.054$\pm$     0.002&   2.451$\pm$     0.099&   0.140$\pm$     0.010\\
&&&211.2738&70.00&11$_{2,10}$ - 10$_{2,9}$, (E)&$3.09\times10^{-4}$&7.41$\pm$0.02&0.063$\pm$     0.001&   1.645$\pm$     0.077&   0.110$\pm$     0.008\\
&&&212.2571&81.46&11$_{3,9}$ - 10$_{3,8}$, (A)&$3.00\times10^{-4}$&7.53$\pm$0.02&0.043$\pm$     0.002&   1.306$\pm$     0.059&   0.060$\pm$     0.005\\
&&&214.8450&70.57&11$_{2,9}$ - 10$_{2,8}$, (A)&$3.25\times10^{-4}$&7.47$\pm$0.02&0.055$\pm$     0.001&   1.771$\pm$     0.080&   0.103$\pm$     0.007\\
&&&216.6302&64.81&11$_{1,10}$ - 10$_{1,9}$, (A)&$3.41\times10^{-4}$&7.47$\pm$0.02&0.070$\pm$     0.002&   1.409$\pm$     0.065&   0.105$\pm$     0.007\\
&&&223.6601&72.20&12$_{1,12}$ - 11$_{1,11}$, (A)&$3.77\times10^{-4}$&7.36$\pm$0.02&0.066$\pm$     0.002&   2.027$\pm$     0.061&   0.143$\pm$     0.008\\
&&&242.1060&83.89&13$_{1,13}$ - 12$_{1,12}$, (E)&$4.81\times10^{-4}$&7.13$\pm$0.04&0.051$\pm$     0.001&   2.123$\pm$     0.108&   0.116$\pm$     0.009\\
\hline
\end{tabular}}
\end{table}
\end{landscape}
\begin{landscape}
\begin{table*}
\centering
\tiny{
\caption{Observed transitions toward some sources. \label{tab:observation_3}}
\begin{tabular}{|l|l|l|l|l|l|l|l|l|l|l|}
\hline
Species&Tag (Database)&Source&Frequency&E$_{up}$&Quantum No.&A$_{ij}$&V$_{LSR}$&T$_{peak}$&FWHM&$\int$T$_{mb}$.dv\\
&&&(GHz)&(K)&&(s$^{-1}$)&(km.s$^{-1}$)&(K)&(km.s$^{-1}$)&(K.km.s$^{-1}$)\\
\hline
&&&250.8291&120.33&13$_{4,9}$ - 12$_{4,8}$, (E)&$4.87\times10^{-4}$&7.06$\pm$0.03&0.044$\pm$     0.002&   1.626$\pm$     0.092&   0.077$\pm$     0.007\\
&&&250.9345&104.62&13$_{3,11}$ - 12$_{3,10}$, (A)&$5.10\times10^{-4}$&7.18$\pm$0.03&0.042$\pm$     0.002&   1.497$\pm$     0.093&   0.067$\pm$     0.007\\
&&&251.4893&104.69&13$_{3,10}$ - 12$_{3,9}$, (A)&$5.14\times10^{-4}$&6.85$\pm$0.03&0.055$\pm$     0.001&   1.823$\pm$     0.079&   0.106$\pm$     0.007\\
&&&254.8271&94.09&13$_{2,11}$ - 12$_{2,10}$, (E)&$5.51\times10^{-4}$&7.26$\pm$0.05&0.041$\pm$     0.001&   2.027$\pm$     0.132&   0.089$\pm$     0.009\\
&&&255.3269&88.45&13$_{1,12}$ - 12$_{1,11}$, (E)&$5.64\times10^{-4}$&7.25$\pm$0.03&0.043$\pm$     0.001&   1.884$\pm$     0.115&   0.087$\pm$     0.008\\
&&&262.9601&95.76&14$_{0,14}$ - 13$_{0,13}$, (E)&$6.19\times10^{-4}$&6.73$\pm$0.03&0.056$\pm$     0.001&   1.770$\pm$     0.076&   0.105$\pm$     0.007\\
\cline{3-11}
&&SVS13A&205.1707&61.46&11$_{1,11}$ - 10$_{1,10}$, (A)&$2.90\times10^{-4}$&8.56$\pm$0.03&   0.039$\pm$0.002&   1.550$\pm$     0.094&   0.064$\pm$0.007\\
&&&211.2430&69.99&11$_{2,10}$ - 10$_{2,9}$, (A)&$3.09\times10^{-4}$&8.47$\pm$     0.04&   0.040$\pm$     0.001&   1.972$\pm$     0.141&   0.083$\pm$     0.009\\
&&&211.2738&70.00&11$_{2,10}$ - 10$_{2,9}$, (E)&$3.09\times10^{-4}$&8.79$\pm$     0.05&   0.040$\pm$     0.001&   1.925$\pm$     0.158&   0.081$\pm$     0.010\\
&&&216.5819&64.87&11$_{1,10}$ - 10$_{1,9}$, (E)&$3.41\times10^{-4}$&8.05$\pm$     0.05&   0.044$\pm$     0.001&   1.970$\pm$     0.149&   0.092$\pm$     0.010\\
&&&230.3019&81.04&12$_{2,11}$ - 11$_{2,10}$, (A)&$4.04\times10^{-4}$&8.55$\pm$     0.03&   0.036$\pm$     0.001&   1.630$\pm$     0.104&   0.062$\pm$     0.007\\
&&&242.1060&83.89&13$_{1,13}$ - 12$_{1,12}$, (E)&$4.81\times10^{-4}$&8.08$\pm$     0.03&   0.053$\pm$     0.001&   1.574$\pm$     0.101&   0.090$\pm$     0.008\\
&&&251.4893&104.69&13$_{3,10}$ - 12$_{3,9}$, (A)&$5.14\times10^{-4}$&8.29$\pm$     0.03&   0.030$\pm$     0.002&   1.022$\pm$     0.076&   0.032$\pm$     0.004\\
\hline
CH$_3$OCHO&60003 (JPL)&B1-b&88.851607&17.9&7$_{1,6}$ - $6_{1,5}$ (A)&$9.82\times10^{-6}$&6.56$\pm$0.217&0.013$\pm$0.003&1.915$\pm$0.635&0.026$\pm$0.014\\
(Methyl formate)&&&90.145723&19.68&$7_{2,5}$ - $6_{2,4}$ (E)&$9.74\times10^{-6}$&6.68$\pm$0.186&0.013$\pm$0.003&1.585$\pm$0.490&0.022$\pm$0.011\\
&&&90.156473&19.67&$7_{2,5}$ - $6_{2,4}$ (A)&$9.75\times10^{-6}$&6.57$\pm$0.169&0.014$\pm$0.003&1.514$\pm$0.438&0.023 $\pm$0.011\\
&&&100.294604&27.41&$8_{3,5}$ - $7_{3,4}$ (E)&$1.26\times10^{-5}$&6.51$\pm$0.162&0.012$\pm$0.003&1.172$\pm$0.404&0.016$\pm$0.009\\
&&&100.482241&22.78&$8_{1,7}$ - $7_{1,6}$ (E)&$1.43\times10^{-5}$&6.73$\pm$0.131&0.018$\pm$0.003&1.331$\pm$0.348&0.026$\pm$0.011\\
&&&	103.478663&24.63&$8_{2,6}$ - $7_{2,5}$ (A)	&$1.52\times10^{-5}$&6.61$\pm$0.179&0.013$\pm$0.003&1.636$\pm$0.457&0.022$\pm$0.011\\
\cline{3-11}
&&IRAS4A&129.296357&36.4&$10_{2,8}$ - $9_{2,7}$ (E)&$3.06\times10^{-5}$&7.18$\pm$0.06&0.025$\pm$0.002&1.588$\pm$0.161&0.042$\pm$0.007\\
&&&132.928736&40.4&$11_{1,1}$ - $10_{1,9}$ (A)&$3.36\times10^{-5}$&7.00$\pm$0.14&0.016$\pm$0.001&2.210$\pm$0.429&0.039$\pm$0.011\\
&&&135.921949&55.6&$11_{5,7}$ - $10_{5,6}$ (A)&$2.95\times10^{-5}$&7.34$\pm$0.05&0.027$\pm$0.002&1.200$\pm$0.131&0.034$\pm$0.007\\
&&&141.044354&47.5&$12_{2,11}$ - $11_{2,10}$ (A)&	$4.02\times10^{-5}$&7.23$\pm$0.05&0.028$\pm$0.002&1.342$\pm$0.139&0.040$\pm$0.007\\
&&&158.693722&59.6&$13_{3,11}$ - $12_{3,10}$ (E)&$5.61\times10^{-5}$&7.43$\pm$0.09&0.022$\pm$0.002&2.402$\pm$0.306&0.057$\pm$0.011\\
&&&158.704392&59.6&$13_{3,11}$ - $12_{3,10}$ (A)&$5.61\times10^{-5}$&7.22$\pm$0.09&0.023$\pm$0.002&2.153$\pm$0.313&0.052$\pm$0.011\\
&&&200.956372&97.5&$16_{5,11}$ - $15_{5,10}$ (A)&$1.10\times10^{-4}$&7.03$\pm$0.02&0.045$\pm$0.002&1.110$\pm$0.065&0.054$\pm$0.005\\
&&&206.619476&89.2&$16_{3,13}$ - $15_{3,12}$ (A)&$1.28\times10^{-4}$&7.63$\pm$0.03&0.038$\pm$0.002&1.306$\pm$0.095&0.052$\pm$0.006\\
&&&216.216539&109.3&$19_{1,18}$ - $18_{1,17}$ (A)&$1.49\times10^{-4}$&7.46$\pm$0.03&0.030$\pm$0.002&1.322$\pm$0.094&0.042$\pm$0.005\\
&&&228.628876&118.8&$18_{5,13}$ - $17_{5,12}$ (E)&$1.66\times10^{-4}$&7.19$\pm$0.02&0.049$\pm$0.002&1.172$\pm$0.068&0.061$\pm$0.005\\
&&&240.02114&122.3	&$19_{3,16}$ - $18_{3,15}$ (E)&$2.01\times10^{-4}$& 7.14$\pm$0.032&0.036$\pm$0.002&1.325$\pm$0.078&0.051$\pm$0.005\\
&&&247.044146&139.9&$21_{3,19}$ - $20_{3,18}$ (E)&$2.21\times10^{-4}$&7.38$\pm$0.035&0.033$\pm$0.002&0.964$\pm$0.084&0.034$\pm$0.005\\
&&&249.578117&148.7&$20_{6,15}$ - $19_{6,14}$ (E)&$2.14\times10^{-4}$&7.23$\pm$0.032&0.031$\pm$0.002&1.224$\pm$0.075&0.041$\pm$0.005\\
\cline{3-11}
&&SVS13A&100.490682&22.7&$8_{1,7}$ - $7_{1,6}$ (A)&	$1.43\times10^{-5}$&8.83$\pm$0.23&0.015$\pm$0.002&2.474$\pm$0.779&0.040$\pm$0.018\\
&&&164.205978&64.9&$13_{4,9}$ - $12_{4,8}$ (E)&$5.98\times10^{-5}$&8.96$\pm$0.03&0.040$\pm$0.002&1.453$\pm$0.096&0.061$\pm$0.007\\
&&&210.4632&123.0&17$_{7,10}$ - $16_{7,9}$ (A)&$1.17\times10^{-4}$&8.75$\pm$0.03&0.030$\pm$0.002&1.350$\pm$0.092&0.043$\pm$0.005\\
&&&218.2809&99.7&$17_{3,14}$ - $16_{3,13}$ (E)&$1.51\times10^{-4}$&8.62$\pm$0.04&0.037$\pm$0.001&2.472$\pm$0.168&0.098$\pm$0.010\\
&&&222.421489&143.5&$18_{8,10}$ - $17_{8,9}$ (E)&$1.33\times10^{-4}$&8.63$\pm$0.03&0.037$\pm$0.001&1.996$\pm$0.109&0.079$\pm$0.007\\
&&&269.078001&168.8&$24_{2,23}$ - $23_{2,22}$ (E)&$2.90\times10^{-4}$&8.57$\pm$0.02&0.048$\pm$0.001&1.700$\pm$0.062&0.086$\pm$0.006\\
\hline
$\rm{C_2H_5OH}$ v=0&46524 (CDMS)&IRAS4A&129.6657634&23.8&$5_{3,2}$ - $5_{2,3}$,v$_t$=2 - 2&$1.14\times10^{-5}$&6.89$\pm$0.068&0.015$\pm$0.002& 1.140$\pm$0.172&0.019 $\pm$0.005\\
(Ethanol)&&&133.3234312&23.8&$7_{1,7}$ - $6_{0,6}$,v$_t$=2 - 2&$1.83\times10^{-5}$&7.62$\pm$0.08&0.026$\pm$0.002&2.042$\pm$0.231&0.056$\pm$0.010\\
&&&148.3040357&58.15&$11_{1,10}$ - $10_{2,9}$,v$_t$=2 - 2&$1.21\times10^{-5}$&7.58$\pm$0.06&0.017$\pm$0.002&0.981$\pm$0.165&0.018 $\pm$0.005\\
&&&	159.4140526&101.1&$9_{1,8}$ - $8_{1,7}$,v$_t$=1 - 1&$3.65\times10^{-5}$&7.23$\pm$0.05&0.021$\pm$0.002&0.696$\pm$0.125&0.016$\pm$0.005\\
&&&	205.4584717&64.0&$12_{1,12}$ - $11_{0,11}$,v$_t$=2 - 2&$8.04\times10^{-5}$&7.51$\pm$0.05&0.031$\pm$0.002&1.663$\pm$0.132&0.054$\pm$0.007\\
&&&	209.8652009&132.9&$12_{3,9}$ - $11_{3,8}$,v$_t$=0 - 0&$8.01\times10^{-5}$&7.17$\pm$0.03&0.023$\pm$0.003&0.583$\pm$0.076&0.015$\pm$0.003\\
&&&	227.8919226&140.0&$13_{1,12}$ - $12_{1,11}$,v$_t$=1 - 1&$1.12\times10^{-4}$&7.21$\pm$0.02&0.029$\pm$0.002&0.636$\pm$0.062&0.020$\pm$0.004\\
&&&	230.9913834&85.5&$14_{0,14}$ - $13_{1,13}$,v$_t$=2 - 2&$1.20\times10^{-4}$&7.16$\pm$0.02&0.044 $\pm$0.002&0.957$\pm$0.064&0.045$\pm$0.005\\
\cline{3-11}
&&SVS13A&84.5958861&13.4&$4_{2,3}$ - $4_{1,4}$,v$_t$= 2- 2&$3.25\times10^{--6}$&8.64$\pm$0.17&0.012$\pm$0.002&1.353$\pm$0.475&0.017$\pm$0.009\\
&&&130.2463048&19.7&$4_{3,1}$ - $4_{2,2}$,v$_t$=2 - 2&$9.98\times10^{-6}$&8.68$\pm$0.05&0.016$\pm$0.002&0.719$\pm$0.117&0.012$\pm$0.004\\
&&&153.4842039&99.1&$9_{2,8}$ - $8_{1,8}$,v$_t$=0 - 1&$4.17\times10^{-5}$&8.30$\pm$0.06&0.017$\pm$0.002&0.861$\pm$0.150&0.015$\pm$0.005\\
&&&205.4584717&64.0&$12_{1,12}$ - $11_{0,11}$,v$_t$=2 - 2&$8.04\times10^{-5}$&8.77$\pm$0.02&0.037$\pm$0.002&0.884$\pm$0.066&0.035$\pm$0.005\\
&&&244.633959&151.75&$14_{1,13}$ - $13_{1,12}$,v$_t$=1 - 1&$1.42\times10^{-4}$&8.77$\pm$0.03&0.022$\pm$0.002&0.705$\pm$0.091&0.016$\pm$0.004\\
&&&270.4440852&32.6&$5_{4,2}$ - $4_{3,1}$,v$_t$=2 - 2&$1.59\times10^{-4}$&8.72$\pm$0.01&0.042$\pm$0.002&0.811$\pm$0.044&0.036$\pm$0.004\\
\hline
\hline
\end{tabular}}
\end{table*} 
\end{landscape}
\clearpage
\begin{landscape}
\begin{table*}
\centering
\tiny{
\caption{Observed transitions toward some sources. \label{tab:observation_4}}
\begin{tabular}{|l|l|l|l|l|l|l|l|l|l|l|}
\hline
Species&Tag (Database)&Source&Frequency&E$_{up}$&Quantum No.&A$_{ij}$&V$_{LSR}$&T$_{peak}$&FWHM&$\int$T$_{mb}$.dv\\
&&&(GHz)&(K)&&(s$^{-1}$)&(km.s$^{-1}$)&(K)&(km.s$^{-1}$)&(K.km.s$^{-1}$)\\
\hline
$\rm{HCCCHO}$&54007 (JPL)&L1544&83.7758&20.12&9$_{0,9}$ - 8$_{0,8}$&1.85$\times$10$^{-5}$&7.13 $\pm$0.019&0.011$\pm$0.001&0.393$\pm$0.046&0.004$\pm$0.001\\
(Propynal)&&&99.0391&7.44&4$_{1,4}$ - 3$_{0,3}$&1.13$\times$10$^{-6}$&7.11 $\pm$ 0.02&0.012$\pm$ 0.001& 0.432$\pm$0.045&0.005$\pm$0.001\\
&&&102.2980&29.49&11$_{0,{11}}$ - 10$_{0,{10}}$&3.40$\times$10$^{-5}$&7.08 $\pm$ 0.02&0.013$\pm$0.001&0.486$\pm$0.042&0.007$\pm$0.001\\
\hline
$\rm{CH_3OCH_3}$&46514 (CDMS)&L1544&99.324362&10.21&4$_{1,4}$ - 3$_{0,3}$ (EA)&5.53$\times$ 10$^{-6}$&-&-&-&-\\
(Dimethyl ether)&&&99.324364&10.21&4$_{1,4}$ - 3$_{0,3}$ (AE)&5.53$\times$ 10$^{-6}$&-&-&-&-\\
&&&99.325217&10.21&4$_{1,4}$ - 3$_{0,3}$ (EE)&5.53$\times$ 10$^{-6}$&-&-&-&-\\
&&&99.326072&10.21&4$_{1,4}$ - 3$_{0,3}$ (AA)&5.53$\times$ 10$^{-6}$&-&-&-&-\\
\cline{3-11}
&&B1-b&99.324362&10.21&4$_{1,4}$ - 3$_{0,3}$ (EA)&5.53$\times$ 10$^{-6}$&-&-&-&-\\
&&&99.324364&10.21&4$_{1,4}$ - 3$_{0,3}$ (AE)&5.53$\times$ 10$^{-6}$&-&-&-&-\\
&&&99.325217&10.21&4$_{1,4}$ - 3$_{0,3}$ (EE)&5.53$\times$ 10$^{-6}$&-&-&-&-\\
&&&99.326072&10.21&4$_{1,4}$ - 3$_{0,3}$ (AA)&5.53$\times$ 10$^{-6}$&-&-&-&-\\
\cline{3-11}
&&IRAS4A&162.529582&33.05&8$_{1,8}$ - 7$_{0,7}$, (EE)&2.67$\times$10$^{-5}$&7.18$\pm$0.06&0.041$\pm$0.002&2.668$\pm$0.287&0.117$\pm$0.017\\
&&&209.515644&59.31&11$_{1,11}$ - 10$_{0,10}$, (EE)&6.40$\times$10$^{-5}$&7.05$\pm$0.02&0.053$\pm$0.001&1.640$\pm$0.076&0.093$\pm$0.007\\
&&&225.599126&69.52&12$_{1,12}$ - 11$_{0,11}$, (EE)&8.25$\times$10$^{-5}$&7.42$\pm$0.02&0.068$\pm$0.001&1.360$\pm$0.063&0.099$\pm$0.007\\
&&&241.946542&81.13&13$_{1,13}$ - 12$_{0,12}$, (EE)&1.05$\times$10$^{-4}$&7.37$\pm$0.04&0.047$\pm$0.001&2.025$\pm$0.129&0.103$\pm$0.009\\
\hline
$\rm{CH_3CN}$&41001 (JPL)&L1544&91.985314&20.4&5$_{1}$ - 4$_{1}$&6.08$\times$ 10$^{-5}$&7.280$\pm$0.002&0.087$\pm$0.001&0.542 $\pm$ 0.006&0.050$\pm$0.001\\
 (Methyl cyanide)&&&91.987087&13.2&5$_{0}$ - 4$_{0}$&6.33$\times$ 10$^{-5}$&7.282$\pm$0.002&0.127   $\pm$0.001&0.505$\pm$0.004&0.068$\pm$0.001\\  
\cline{3-11}
&&B1-b&91.985314&20.4&5$_{1}$ - 4$_{1}$&6.08$\times$ 10$^{-5}$&6.719$\pm$    0.143&0.018$\pm$0.003&1.731 $\pm$0.388&0.034 $\pm$0.013\\
 &&&91.987087&13.2&5$_{0}$ - 4$_{0}$&6.33$\times$ 10$^{-5}$&6.654$\pm$     0.090&0.029$\pm$0.003&1.846$\pm$0.253&0.058$\pm$0.013\\ 
 \cline{3-11}
 &&IRAS4A&73.588799&16.0&$4_{1}$ - $3_{1}$&2.97$\times$ 10$^{-5}$&6.726 $\pm$   0.141&0.043  $\pm$  0.003&2.758 $\pm$   0.472&0.127 $\pm$   0.029\\
 &&&73.590218&8.8&$4_{0}$ - 3$_{0}$&3.17$\times$ 10$^{-5}$&7.025 $\pm$    0.121&0.055  $\pm$   0.003&2.858$\pm$    0.480&0.169$\pm$    0.036\\
 &&&91.979994&41.8&$5_{2}$ - $4_{2}$&5.32$\times$ 10$^{-5}$&7.071$\pm$    0.170&0.029$\pm$    0.002&2.661 $\pm$   0.853&0.082 $\pm$   0.033\\
 &&&91.985314&20.4&$5_{1}$ - $4_{1}$&6.08$\times$ 10$^{-5}$&6.943$\pm$    0.089&0.054$\pm$    0.002&3.119 $\pm$   0.399&0.179$\pm$    0.031\\
 &&&91.987087&13.2&$5_{0}$ - $4_{0}$&6.33$\times$ 10$^{-5}$&7.083$\pm$    0.056&0.078$\pm$    0.002&2.658 $\pm$   0.239&0.221 $\pm$   0.027\\
 &&& 110.364353&82.8&$6_{3}$ - $5_{3}$&8.33$\times$ 10$^{-5}$&7.297$\pm$    0.051&0.039 $\pm$   0.003&1.737$\pm$    0.131&0.072$\pm$    0.010\\
 &&&110.3813720&25.7&$6_{1}$ - $5_{1}$&1.08$\times$ 10$^{-4}$&7.406 $\pm$   0.055&0.072$\pm$    0.002&3.018 $\pm$   0.235&0.233 $\pm$   0.025 \\
 &&&110.383499&18.5&$6_{0}$ - $5_{0}$&1.11$\times$ 10$^{-4}$&7.299$\pm$    0.056&0.076$\pm$    0.002&3.180$\pm$    0.253&0.258$\pm$    0.028\\
 &&&128.757030&89.0&$7_{3}$ - $6_{3}$&1.46$\times$ 10$^{-4}$&7.052$\pm$    0.135&0.029$\pm$    0.001&2.902$\pm$    0.464&0.089$\pm$    0.018\\
 &&&128.769436&53.3&$7_{2}$ - $6_{2}$&1.64$\times$ 10$^{-4}$&6.812$\pm$    0.072&0.043$\pm$    0.001&3.134 $\pm$   0.266&0.144$\pm$    0.017\\
 &&&128.776881&31.9&$7_{1}$ - $6_{1}$&1.75$\times$ 10$^{-4}$&6.675$\pm$    0.045&0.078 $\pm$   0.001&2.941$\pm$    0.135&0.244$\pm$    0.015\\
 &&&128.779363&24.7&$7_{0}$ - $6_{0}$&1.78$\times$ 10$^{-4}$&6.990$\pm$    0.050&0.092$\pm$    0.001&2.926 $\pm$   0.209&0.285$\pm$    0.024\\
 &&&147.163244&60.4&$8_{2}$ - $7_{2}$&2.52$\times$ 10$^{-4}$&6.898$\pm$    0.062&0.039$\pm$    0.002&2.349$\pm$    0.211&0.098$\pm$    0.013\\
 &&&147.171751&38.9&$8_{1}$ - $7_{1}$&2.64$\times$ 10$^{-4}$&6.854$\pm$    0.051&0.070$\pm$    0.002&3.131$\pm$    0.186&0.232$\pm$    0.019\\
 &&&147.174588&31.8&$8_{0}$ - $7_{0}$&2.69$\times$ 10$^{-4}$&7.147$\pm$    0.031&0.087$\pm$    0.002&2.787$\pm$    0.134&0.258$\pm$    0.017\\
 &&&165.540377&104.0&$9_{3}$ - $8_{3}$&3.42$\times$ 10$^{-4}$&6.815$\pm$    0.113&0.039 $\pm$   0.001&3.176$\pm$    0.399&0.131$\pm$    0.021\\
 &&&165.556321&68.3&$9_{2}$ - $8_{2}$&3.66$\times$ 10$^{-4}$&7.192$\pm$    0.044&0.042$\pm$    0.002&2.204$\pm$    0.134&0.098$\pm$    0.010\\
 &&&165.565891&46.9&$9_{1}$ - $8_{1}$&3.80$\times$ 10$^{-4}$&7.189$\pm$    0.036&0.070$\pm$    0.002&2.912$\pm$    0.141&0.216$\pm$    0.015\\
 &&&165.569081&39.7&$9_{0}$ - $8_{0}$&3.85$\times$ 10$^{-4}$&7.183$\pm$    0.030&0.078$\pm$    0.002&2.656$\pm$    0.122&0.220$\pm$    0.014\\
 &&&202.320442&122.6&$11_{3}$ - $10_{3}$&6.56$\times$ 10$^{-4}$&7.101$\pm$    0.027&0.059 $\pm$   0.002&2.086$\pm$    0.077&0.131$\pm$    0.008\\
 &&&220.709016&133.2&$12_{3}$ - $11_{3}$&8.66$\times$ 10$^{-4}$&7.324 $\pm$   0.051&0.049 $\pm$    0.001&2.570 $\pm$    0.192&0.135 $\pm$    0.014\\
\hline 

\end{tabular}}
\end{table*}
\begin{table*}
\centering
\tiny{
\caption{Observed transitions toward some sources. \label{tab:observation_4}}
\begin{tabular}{|l|l|l|l|l|l|l|l|l|l|l|}
\hline
Species&Tag (Database)&Source&Frequency&E$_{up}$&Quantum No.&A$_{ij}$&V$_{LSR}$&T$_{peak}$&FWHM&$\int$T$_{mb}$.dv\\
&&&(GHz)&(K)&&(s$^{-1}$)&(km.s$^{-1}$)&(K)&(km.s$^{-1}$)&(K.km.s$^{-1}$)\\
\hline
&&SVS13A&91.979994&41.8&$5_{2}$ - $4_{2}$&3.32$\times$ 10$^{-5}$&7.918$\pm$     0.569&0.020 $\pm$    0.002&3.205$\pm$     1.528&0.067 $\pm$    0.038\\
&&&91.985314&20.4&$5_{1}$ - $4_{1}$&6.08$\times$ 10$^{-5}$&8.428$\pm$     0.133&0.029$\pm$     0.002&2.762$\pm$     0.588&0.086$\pm$     0.024\\
&&&91.987087&13.2&$5_{0}$ - $4_{0}$&36.33$\times$ 10$^{-5}$&8.580$\pm$     0.078 &0.035$\pm$     0.002&2.366$\pm$     0.272&0.088$\pm$     0.015\\
&&&110.364353&82.8&$6_{3}$ - $5_{3}$&8.33$\times$ 10$^{-5}$&8.808$\pm$     0.145&0.025$\pm$     0.002&3.050$\pm$     0.596& 0.080$\pm$     0.022\\
&&&110.374989&47.1&$6_{2}$ - $5_{2}$&9.87$\times$ 10$^{-5}$&8.531$\pm$     0.177&0.030$\pm$     0.002&3.366 $\pm$    1.134&0.109 $\pm$    0.043\\
&&&110.381372&25.7&$6_{1}$ - $5_{1}$&1.08$\times$ 10$^{-4}$&8.284$\pm$     0.083&0.042$\pm$     0.002&2.540$\pm$     0.245&0.114$\pm$     0.016\\
&&&110.383499&18.5&$6_{0}$ - $5_{0}$&1.11$\times$ 10$^{-4}$&8.315$\pm$     0.102&0.033$\pm$     0.002&2.649$\pm$     0.332&0.093$\pm$     0.017\\
&&&128.776881&31.8&$7_{1}$ - $6_{1}$&1.75$\times$ 10$^{-4}$&8.392$\pm$     0.034&0.050$\pm$     0.002&1.719$\pm$     0.094&0.092$\pm$     0.008\\
&&&128.779363&24.7&$7_{0}$ - $6_{0}$&1.78$\times$ 10$^{-4}$&8.280$\pm$     0.036&0.050$\pm$     0.001&1.990$\pm$     0.107&0.105$\pm$     0.009\\
&&&147.171751&38.9&$8_{1}$ - $7_{1}$&2.64$\times$ 10$^{-4}$&8.162$\pm$     0.051&0.062 $\pm$    0.002&2.533 $\pm$    0.164&0.167$\pm$     0.015\\
&&&165.565891&46.9&$9_{1}$ - $8_{1}$&3.80$\times$ 10$^{-4}$&8.343$\pm$     0.030&0.091 $\pm$    0.002&2.307$\pm$     0.099&0.223 $\pm$    0.013\\
&&&220.709016&133.2&$12_{3}$ - $11_{3}$&8.66$\times$ 10$^{-4}$&8.576$\pm$     0.033&0.077 $\pm$    0.001&3.238$\pm$     0.136&0.265 $\pm$    0.016\\
&&&220.730260&97.4&$12_{2}$ - $11_{2}$&8.98$\times$ 10$^{-4}$&8.363$\pm$     0.051&0.071$\pm$     0.001&3.900$\pm$     0.236&0.294$\pm$     0.023\\
&&&220.743010&76.0&$12_{1}$ - $11_{1}$&9.18$\times$ 10$^{-4}$&8.300$\pm$     0.028&0.088$\pm$     0.001&2.951$\pm$     0.110&0.278$\pm$     0.014\\
&&&220.747261&68.9&$12_{0}$ - $11_{0}$&9.24$\times$ 10$^{-4}$&8.382$\pm$     0.040&0.090$\pm$     0.001&3.907$\pm$     0.186&0.373$\pm$     0.023\\
&&&239.119504&108.9&$13_{2}$ - $12_{2}$&1.15$\times$ 10$^{-3}$&8.209$\pm$     0.040&0.091$\pm$     0.001&2.952$\pm$     0.139&0.285$\pm$     0.017\\
&&&257.527383&92.7&$14_{0}$ - $13_{0}$&1.48$\times$ 10$^{-3}$&8.426$\pm$     0.059&0.092$\pm$     0.001&4.037$\pm$     0.321&0.394$\pm$     0.036\\
\hline
\hline
\end{tabular}}
\end{table*}
\end{landscape}
\clearpage
\section{Results \& Discussions} \label{sec:results}
\begin{figure}
\begin{minipage}{0.49\textwidth}
\includegraphics[width=\textwidth]{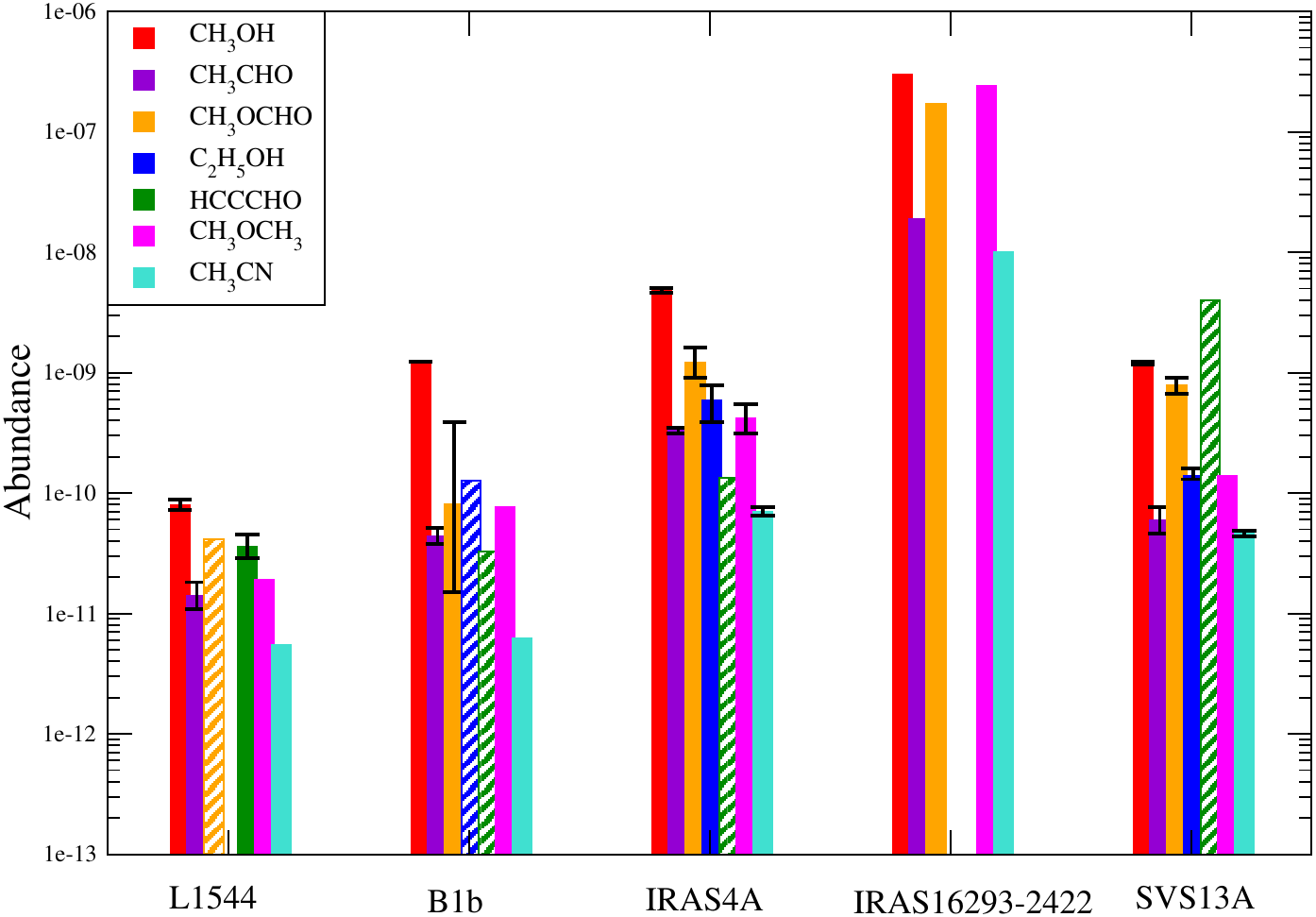}
\end{minipage}
\begin{minipage}{0.49\textwidth}
\includegraphics[width=\textwidth]{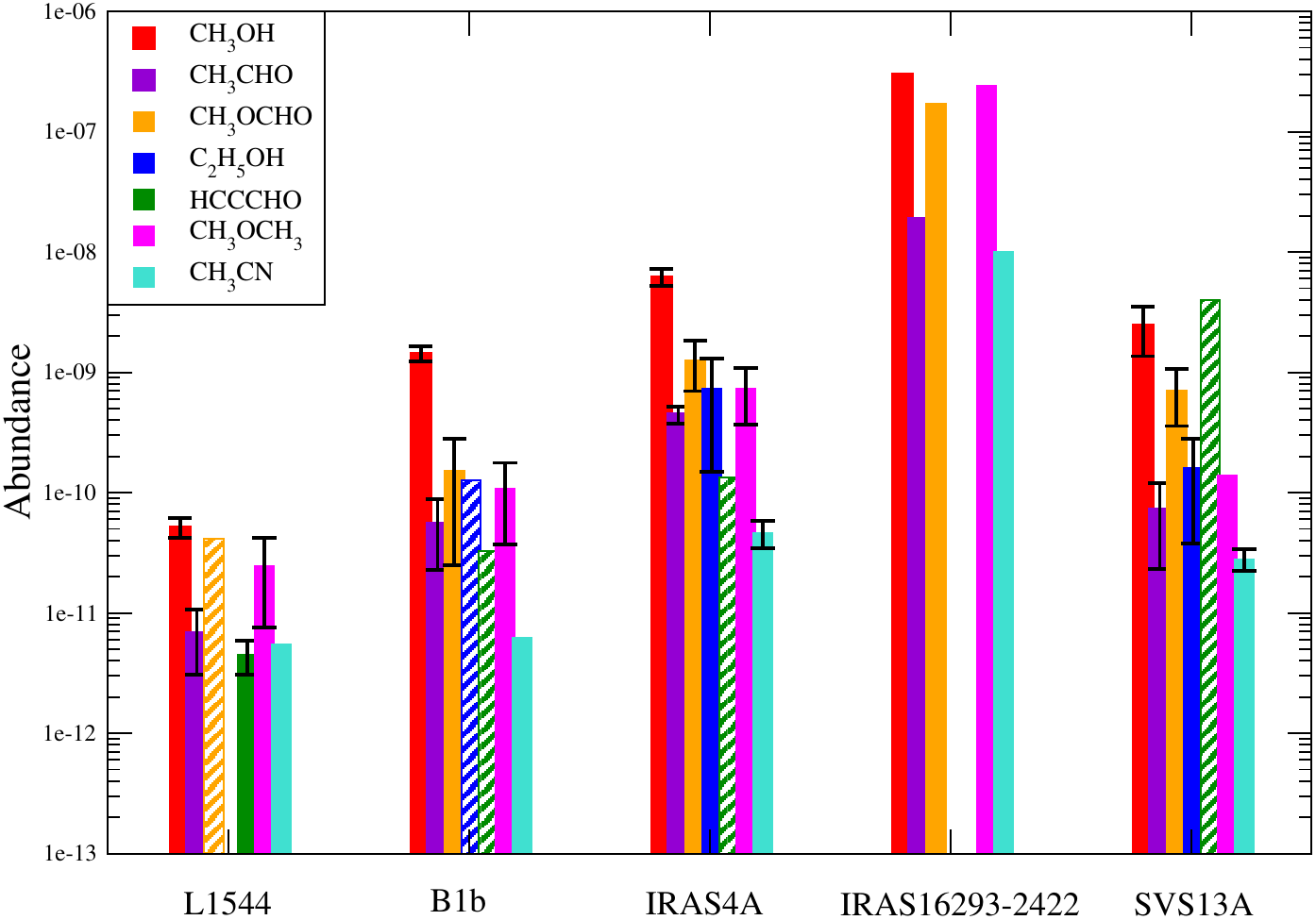}
\end{minipage}
\caption{Evolution of abundances in different stages of low-mass star forming regions. The left panel shows the abundances obtained from the rotational diagram method (dashed panels are the upper limits), whereas the right panel depicts the same that obtained with the MCMC fit. Black vertical lines represent the error bars. For CH$_3$OCH$_3$ in L1544 and B1b, the lines have same upstate energy so rotation diagram cannot be performed and the column density is calculated using simple LTE fitting. For CH$_3$OCH$_3$ in SVS13A the column density is calculated using value from \cite{bian19} scaling it for 30$^{''}$ beam. A new class 0 source, IRAS16293-2422 (22 L$_\odot$), is included where CH$_3$OH, CH$_3$CHO, CH$_3$OCHO, CH$_3$OCH$_3$ and CH$_3$CN is observed \cite{caza03}.}
\label{fig:clmdensity}
\end{figure}

\begin{figure}
\begin{minipage}{0.55\textwidth}
\includegraphics[width=\textwidth]{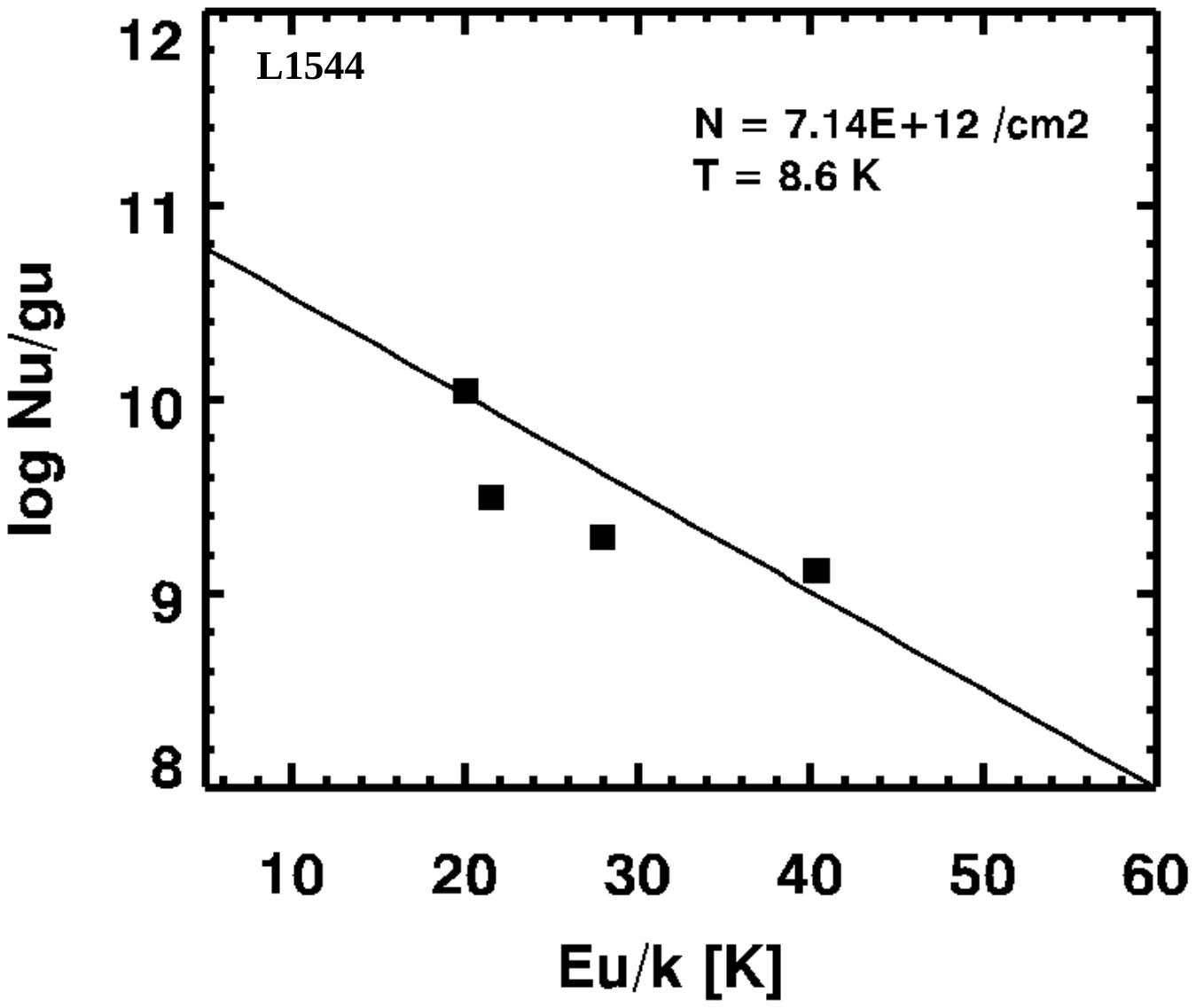}
\end{minipage}
 \begin{minipage}{0.55\textwidth}
 \includegraphics[width=\textwidth]{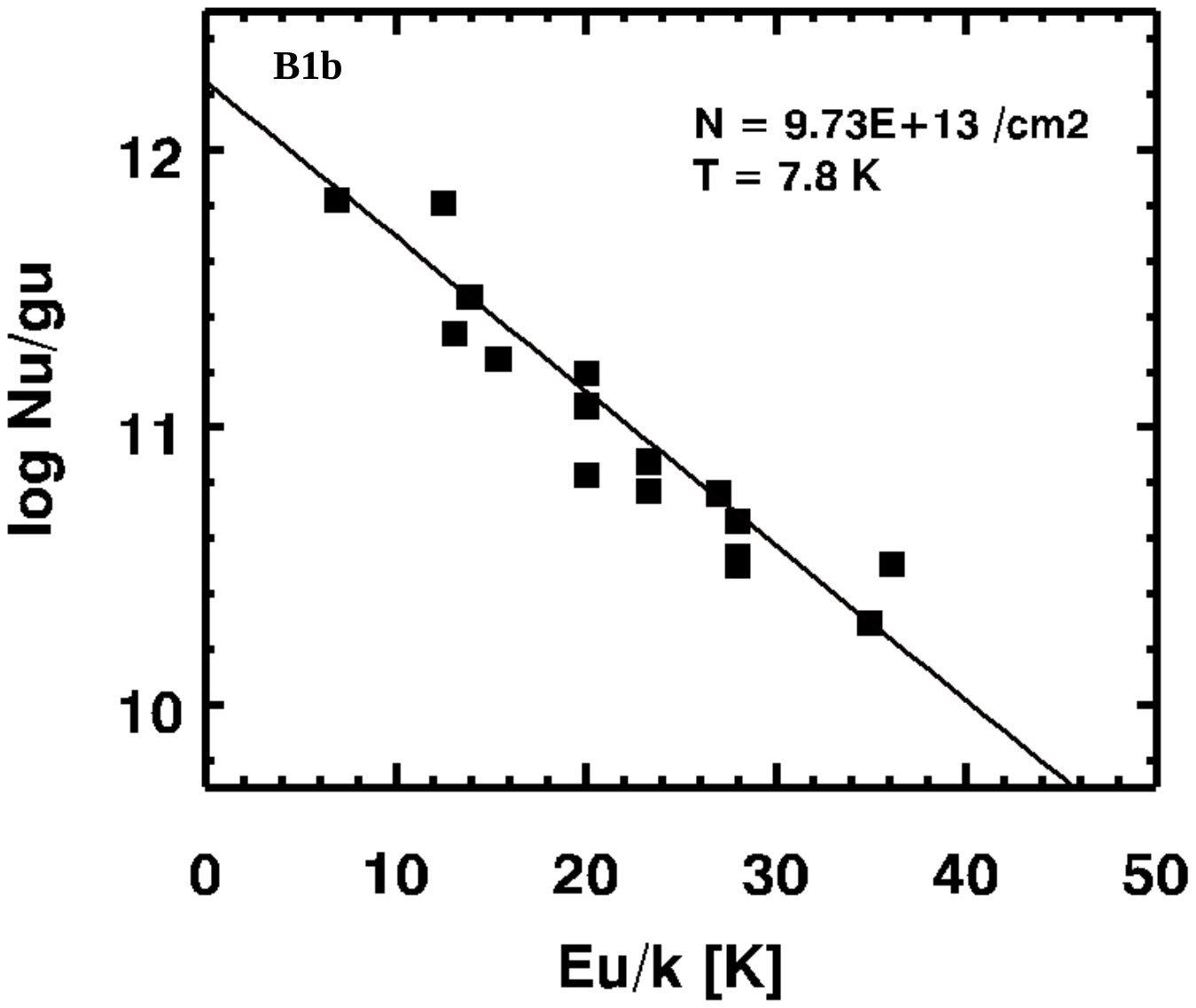}
 \end{minipage}
\begin{minipage}{0.55\textwidth}
 \includegraphics[width=\textwidth]{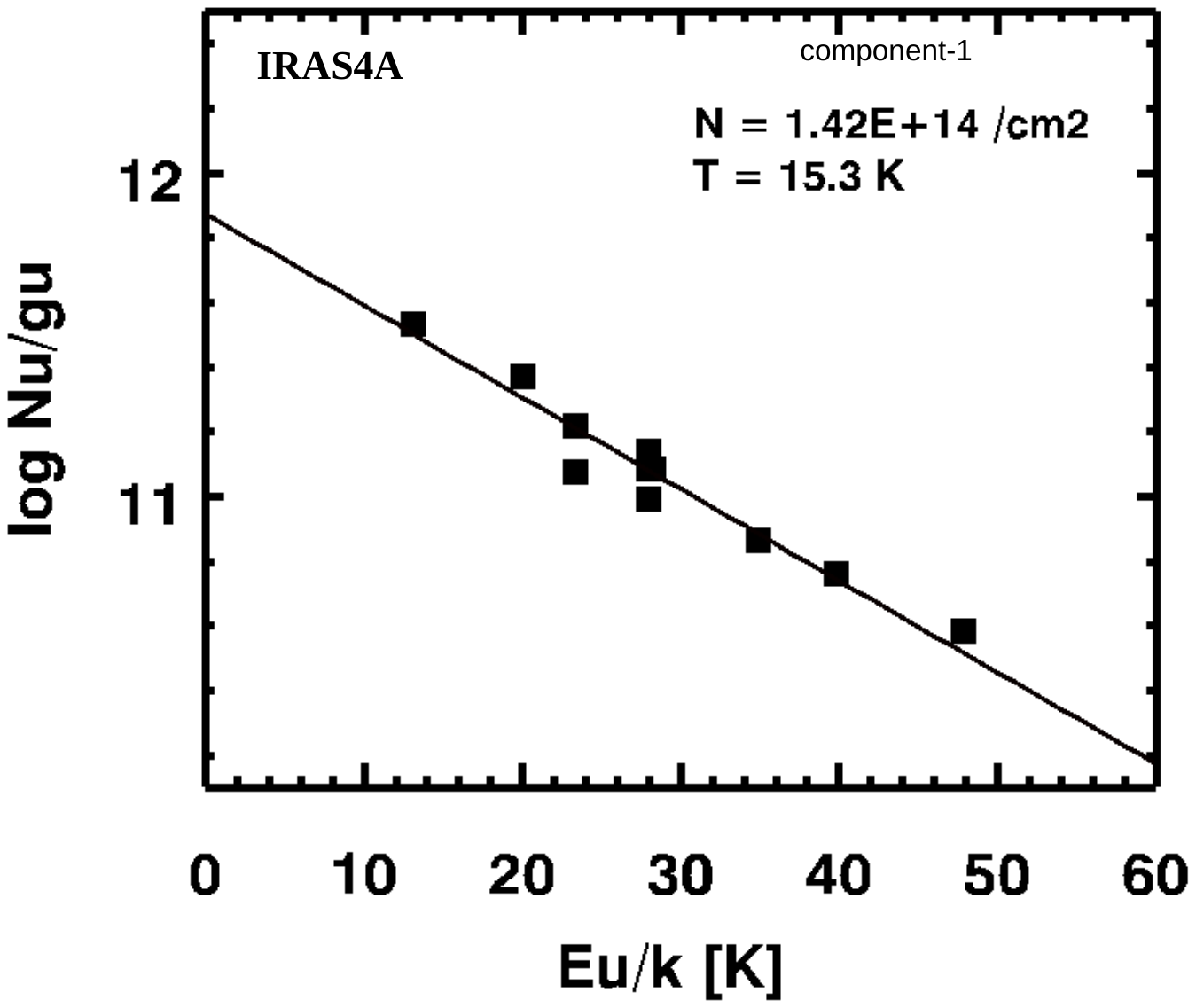}
 \end{minipage}
 \begin{minipage}{0.55\textwidth}
 \includegraphics[width=\textwidth]{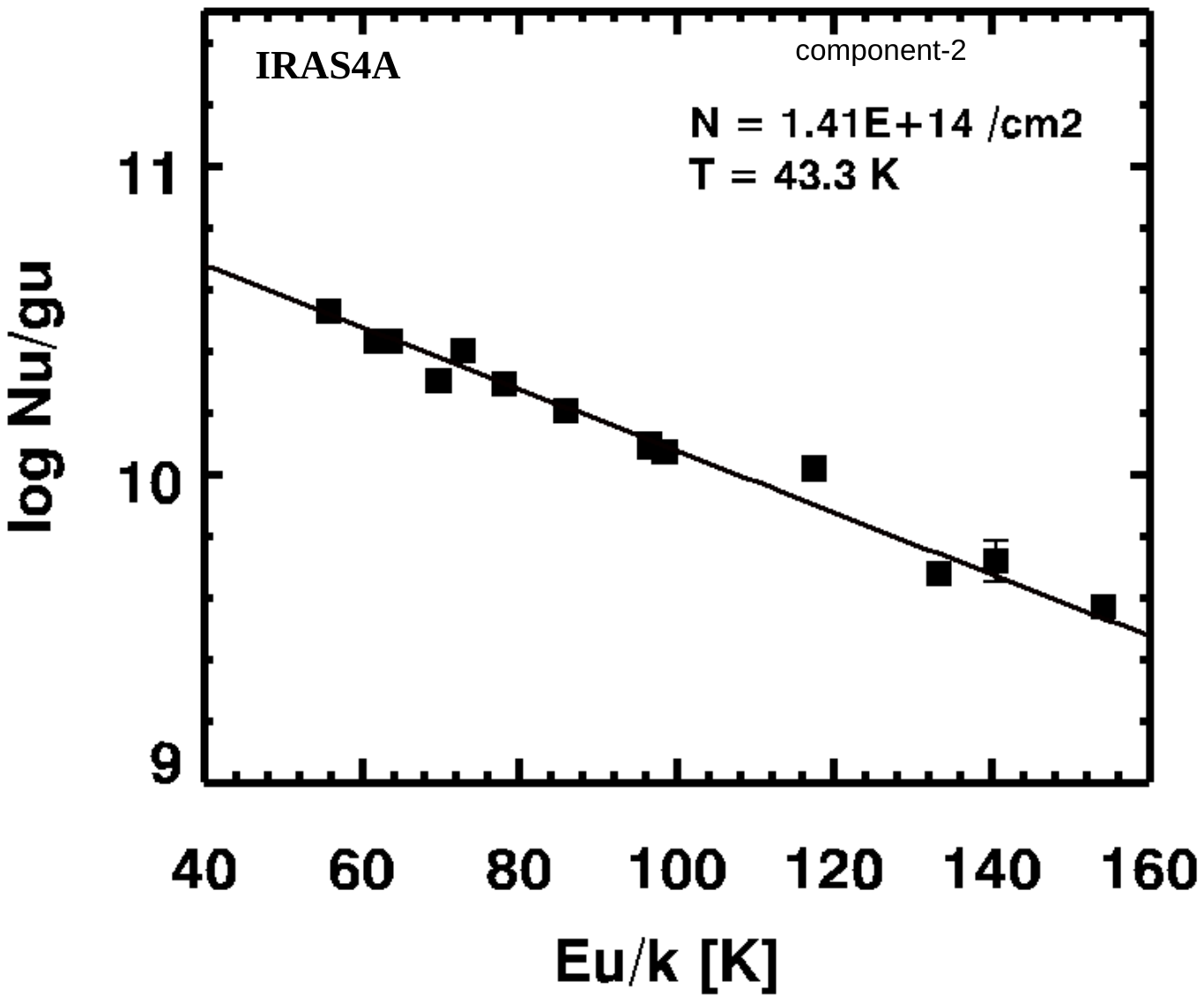}
 \end{minipage}
 \begin{minipage}{0.55\textwidth}
 \includegraphics[width=\textwidth]{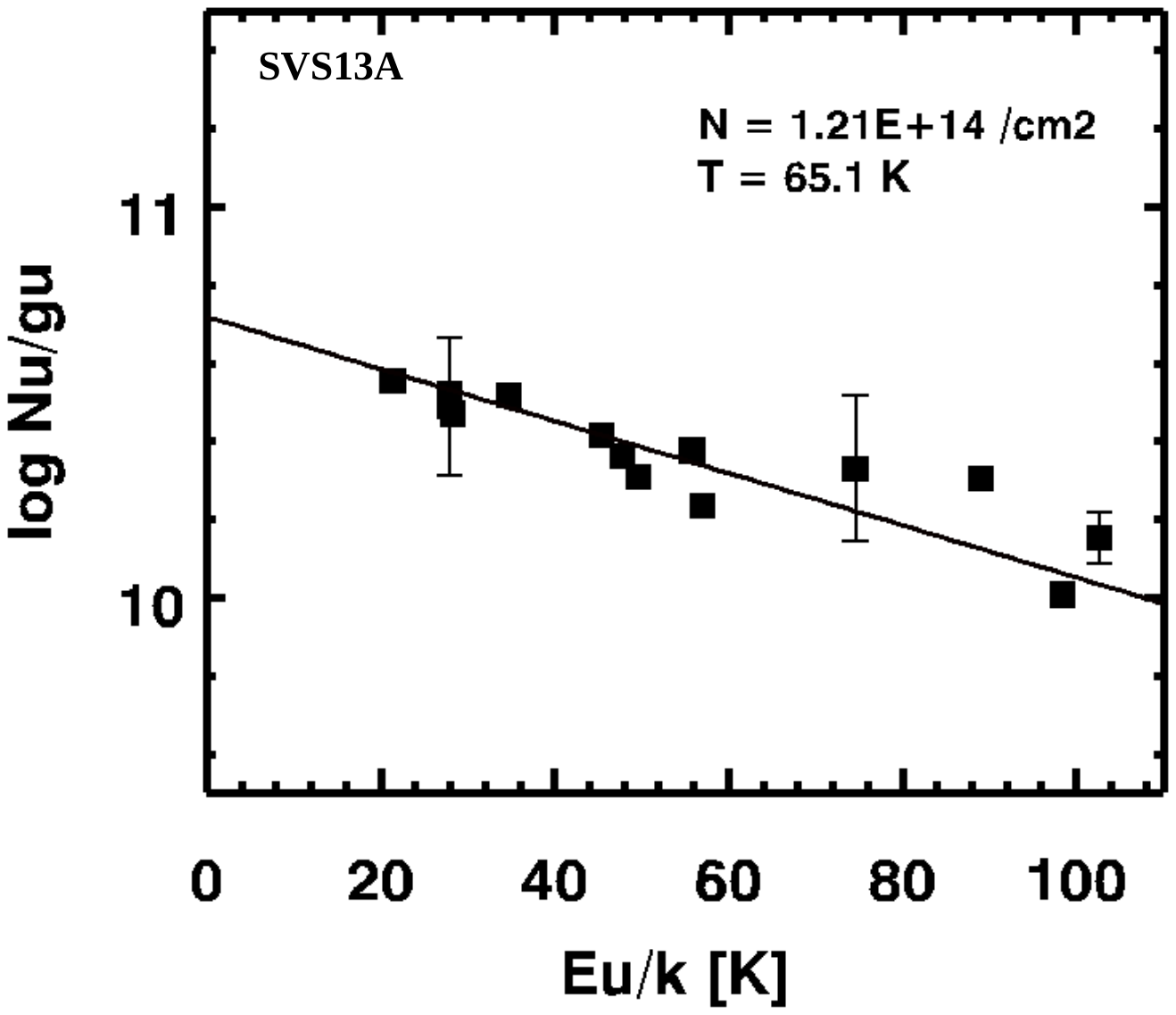}
 \end{minipage}
  \caption{Rotational diagram for CH$_3$OH obtained with various sources. The black points are the position of the data points, the vertical bars are the error bars estimated, and the red lines are the fitted lines to the rotational diagram. The obtained column densities and the excitation temperatures along with the error bars are mentioned in the top right corner of each box. }
\label{fig:rotational_diag_ch3oh_single}
\end{figure}

The major objective of this study is to determine the correlation between the detected COMs in different evolutionary stages of star formation. Of course, it is impossible to observe every instance of a single star-forming region. To get a better understanding of the entire process, it is better to selectively follow specific evolutionary phases of comparable types of ongoing events before the formation of stars. It gives a basic concept of the process of how low-mass stars develop, even though it is not anticipated that all the targeted regions will follow a similar footprint. A cartoon diagram is shown for each of the four evolutionary stages of low-mass star formation, from the prestellar core to the class I stage via the FHSC and class 0 phase in Figure \ref{fig:cartoon}. We were able to identify several transitions of interstellar COMs methanol (CH$_3$OH), acetaldehyde (CH$_3$CHO), methyl formate (CH$_3$OCHO), ethanol (C$_2$H$_5$OH), propynal (HCCCHO), dimethyl ether (CH$_3$OCH$_3$), and methyl cyanide (CH$_3$CN) in different sources. To determine the column density and excitation temperature of a species, we use two robust LTE approaches, rotational diagram bf (RD) and Monte Carlo Markov Chain (MCMC) fitting. The sections \ref{sec:RD} and \ref{sec:mcmc}, respectively, provide a detailed discussion of the RD analysis and the MCMC approach. Table \ref{tab:rotdiag} provides a summary of the column densities and temperatures that were determined by rotational diagram analysis. The rotation diagram plots are shown in \ref{fig:rotational_diag_ch3oh_single}, \ref{fig:rotational_diag_ch3cho}, \ref{fig:rotational_diag_hcooch3}, \ref{fig:rotational_diag_C2H5OH}, \ref{fig:rotational_diag_HCCCHO}, \ref{fig:rotational_diag_CH3OCH3}, \ref{fig:rotational_diag_CH3CN}. The variable parameters and the ranges used for the MCMC method and the best-fitted values are summarized in Table \ref{table:mcmc_lte}. In addition, Table \ref{table:comparison} compares the abundances obtained through MCMC fitting and rotational diagram analysis with earlier findings. It shows that our findings are in agreement with those of past studies. The values obtained by the two approaches deviate from the earlier values by a factor of 0.18 to 7.9. Additionally, the abundances obtained via MCMC fitting and rotational diagram analysis is nearly identical and differ within a narrow range, which supports the accuracy of the data produced by the two techniques. The left panel of Figure \ref{fig:clmdensity} illustrates how abundances produced using the rotating diagram approach vary with different evolutionary phases, whereas the right panel displays findings acquired using the MCMC fitting method that is comparable. Here, certain species are discussed together with the findings from our analysis. The abundances are calculated using the column density values of H$_2$ from various works available in literature. The values are noted in Table \ref{tab:rotdiag} and in section \ref{sec:H2_col}.

\begin{landscape}
\begin{table}
\centering
{\scriptsize
 \caption{Comparison of column densities obtained using two different LTE methods, rotational diagram, and MCMC fitting, along with the values obtained from the literature for COMs observed in different sources. \label{table:comparison}}
\begin{tabular}{|c|c|c|c|c|c|}
  \hline
  \hline
  Species&Column Density&L1544&B1-b&IRAS4A&SVS13A\\
  \hline
  \hline  &Previous&$(2.7\pm0.6)\times10^{13a}$&$(2.5\pm1.3)\times10^{14b}$&$(5.1\pm1.0)\times10^{14c}$&$(1.0\pm0.2)\times10^{17d}$\\
  CH$_3$OH&Rotational Diagram&$(7.1^{+0.7}_{-0.6})\times10^{12}$&$(9.7^{+0.05}_{-0.04})\times10^{13}$&$(1.4^{+0.03}_{-0.02})\times10^{14}$ &$(1.2^{+0.03}_{-0.02})\times10^{14}$\\
  &&&&$(1.4^{+0.04}_{-0.04})\times10^{14}$&\\
&MCMC&$(4.6\pm0.92)\times10^{12}$&$(1.13\pm0.16)\times10^{14}$&$(1.8\pm0.29)\times10^{14}$(C1)&$(2.43\pm1.07)\times10^{14}$\\
  &&&&$(2.0\pm0.28)\times10^{14}$(C2)&\\
 \hline
&Previous&$1.2\times10^{12h}$&$1.5\times10^{12f}$&$2.6\times10^{12g}$-blue lobe&$(1.2\pm0.7)\times10^{16d}$\\
&&&&$7.6\times10^{11g}$-red lobe&\\
  CH$_3$CHO&Rotational Diagram&$(2.35^{+1.0}_{-0.7})\times10^{12}$&$(6.5^{+1.0}_{-0.8})\times10^{12}$&$(1.78^{+0.1}_{-0.09})\times10^{13}$&$(1.1^{+0.3}_{-0.2})\times10^{13}$\\
  &&&&$(1.71^{+0.2}_{-0.1})\times10^{13}$&\\
&MCMC&$(6.1\pm3.55)\times10^{11}$&$(4.4\pm0.56)\times10^{12}$&$(1.3\pm0.21)\times10^{13}$(C1)&$(7.2\pm4.89)\times10^{12}$\\
  &&&&$(1.1\pm0.21)\times10^{13}$(C2)&\\
\hline
&Previous&$(4.4\pm4.0)\times10^{12h}$&$3.0\times10^{12f}$(A+E)&$(5.5\pm2.7)\times10^{16i}$(A)&$(1.3\pm0.1)\times10^{17d}$\\
&&&&$(5.8\pm1.1)\times10^{16i}$(E)&\\
 CH$_3$OCHO&Rotational Diagram&$3.7\times10^{12}x$&$(6.4^{+2.3}_{-5.0})\times10^{12}$&$(3.5^{+0.5}_{-0.4})\times10^{13}$&$(7.8^{+1.3}_{-1.0})\times10^{13}$\\
  &MCMC&-&$(1.2\pm1.57)\times10^{13}$&$(3.65\pm1.63)\times10^{13}$&$(7.1\pm3.52)\times10^{13}$\\
 \hline
&Previous&&$ \leqslant4.8\times10^{12j}$&$(4.4\pm1.4)\times10^{16k}$&$(1.1\pm0.5)\times10^{17d}$\\
  C$_2$H$_5$OH&Rotational Diagram&-&$1.0\times10^{13x}$&$(1.7^{+0.3}_{-0.2})\times10^{13}$&$(1.4^{+1.8}_{-1.6})\times10^{13}$\\
 &MCMC&-&-&$(3.0\pm1.66)\times10^{13}$&$(1.6\pm1.22)\times10^{13}$\\
 \hline
 &Previous&$1.8-6.3\times10^{11h}$&$5.5\times10^{11f}$&-&$\leqslant1.0\times10^{16d}$\\
  HCCCHO&Rotational Diagram&$(3.2^{+0.8}_{-0.7})\times10^{12}$&$2.56\times10^{12x}$&$3.9\times10^{12x}$&$4.0\times10^{14x}$\\
 &MCMC&$(4.0\pm1.33)\times10^{11}$&-&-&-\\
 \hline
&Previous&$ (1.5\pm0.2)\times10^{12h}$&$3.0\times10^{12f}$&$\leqslant4.5\times10^{16i}$&$(1.4\pm0.4)\times10^{17d}$\\
 CH$_3$OCH$_3$&Rotational Diagram&$1.6\times10^{12}$&$6.0\times10^{12}$&$1.95^{+0.4}_{-0.3}\times10^{13}$&$1.4\times10^{13}$\\ &MCMC&$(2.2\pm1.61)\times10^{12}$&$(8.5\pm5.52)\times10^{12}$&$(2.1\pm1.03)\times10^{13}$&\\
\hline
&Previous&$(6.1\pm1.8)\times10^{11l}$&$(3.2^{+1.6}_{-1.4})\times10^{14v}$(B1-b S)&$(6.5\pm2.9)\times10^{15m}$&$2.0\times10^{16n}$\\
 &&&$< 8.5\times10^{13v}$(B1-b N)&&\\
  CH$_3$CN&Rotational Diagram&$4.85\times10^{11y}$&$4.95\times10^{11y}$&$(1.78^{+0.17}_{-0.15})\times10^{12}$&$(4.57^{+0.23}_{-0.22})\times10^{12}$\\
 &MCMC&-&-&$(1.34\pm0.35)\times10^{12}$&$(2.8\pm0.57)\times10^{12}$\\
  \hline
  \hline
 \end{tabular}}\\
 {\scriptsize \noindent $^x$Upper limits, $^a$\cite{bizz14}, $^b$\cite{ober09}, $^c$\cite{mare05}, $^d$\cite{bian19} (Interferometric observation), 
 $^f$\cite{cern12}, $^g$\cite{hold19}, $^h$\cite{jime16}, $^i$\cite{bott04} (Interferometric observation), $^j$\cite{ober10}, $^k$\cite{taqu15} (Interferometric observation), $^l$\cite{nagy19}, $^m$\cite{taqu15}, $^n$\cite{bian22b}, $^v$\cite{yang21}}
 \end{table}
\end{landscape}

\subsection{Upper limit estimation}
We were unable to find suitable transitions for some species to estimate the column density using a rotation diagram and the MCMC approach. We list the guessed upper limits of column density for these species in Table \ref{tab:upper_limit}. Additionally, the input parameters used for this estimation are also noted.

\begin{table}
\centering{
\tiny{
\caption{Estimated upper limit of column density \label{tab:upper_limit}}
\begin{tabular}{|c|c|c|c|c|c|c|c|}
\hline
Molecules&Source&Frequency&Quantum No.&E$_{up}$&A$_{ij}$&N$_{tot}$&T$_{k}$ \\
&&(GHz)&&(K)&(s$^{-1}$)&(cm$^{-2}$)&(K)\\
\hline
CH$_3$OCHO&L1544&90.227659&8$_{0,8}$ - 7$_{0,7}$, (E)&20.1&$1.05\times10^{-5}$&$3.7\times10^{12}$&10.0\\
\hline
C$_2$H$_5$OH&B1-b&131.502781&6$_{3,4}$ - 6$_{2,5}$, v$_t$ = $2-2$ &28.9&$1.27\times10^{-5}$&$1.0\times10^{13}$&20.0\\
\hline
HCCCHO&B1-b&93.0432843&10$_{0,10}$ - 9$_{0,9}$&24.58&$2.55\times10^{-5}$&$2.56\times10^{12}$&20.0\\
&IRAS4A&111.53912&12$_{0,12}$ - 11$_{0,11}$&34.85&$4.43\times10^{-5}$&$3.9\times10^{12}$&40.0\\
&SVS13A&236.6916499&21$_{4,17}$ - 22$_{3,20}$&152.01&$4.96\times10^{-6}$&$4.00\times10^{14}$&100.0\\
\hline
\end{tabular}}\\
}
\end{table}

\subsection{Rotational diagram}\label{sec:RD}

For the optically thin transitions, upper state column density (N$_u^{thin}$) can be expressed as \citep{gold99},
\begin{equation}
\frac{N_u^{thin}}{g_u}=\frac{3k_B\int{T_{mb}dV}}{8\pi^{3}\nu S\mu^{2}},
\label{eqn:clmn}
\end{equation}
\noindent where g$_u$ is the degeneracy of the upper state, k$_B$ is the Boltzmann constant, $\rm{\int T_{mb}dV}$ is the integrated intensity,
$\nu$ is the rest frequency, $\mu$ is the electric dipole moment, and S is the transition line strength. Total column density ($N_{total}$) under LTE condition is, 
\begin{equation}
\frac{N_u^{thin}}{g_u}=\frac{N_{total}}{Q(T_{rot})}\exp(-E_u/k_BT_{rot}),
\end{equation}
where $T_{rot}$ is the rotational temperature, E$_u$ is the upper state energy, $\rm{Q(T_{rot})}$ is the partition function at rotational temperature. This can be rearranged as,
\begin{equation}
ln\Bigg(\frac{N_u^{thin}}{g_u}\Bigg)=-\Bigg(\frac{1}{T_{rot}}\Bigg)\Bigg(\frac{E_u}{k}\Bigg)+ln\Bigg(\frac{N_{total}}{Q(T_{rot})}\Bigg).
\end{equation}
The column density at the upper level and the upper state energy have a linear relationship. The rotational diagram is used to estimate the column density and the rotational temperature.\\
 Only after several transitions ($>$2) with various up-state energies of a molecule are observed then the rotational diagram can be drawn. Table \ref{tab:rotdiag} lists the column densities as well as the predicted rotational temperature (T$_{rot}$). In certain cases, two temperature components can be derived from the rotational diagram (IRAS4A for CH$_3$OH, CH$_3$CHO and CH$_3$CN). The text in the later portion contains information about the components in detail. The error bars (vertical bars) in rotational diagrams are the absolute uncertainty in a log of (N$_u$/g$_u$), which arises from the error of the observed integrated intensity that we measured using a single Gaussian fitting to the observed profile of each transition.

\begin{table}
\centering{
\tiny{
\caption{Results obtained with the rotational diagram analysis.\label{tab:rotdiag}}
\begin{tabular}{|c|c|c|c|c|}
\hline
Molecules&Source&N$_{tot}$&T$_{rot}$&Abundance \\
&&(cm$^{-2}$)&(K)&\\
\hline
CH$_3$OH&L1544&$(7.14^{+0.70}_{-0.60}) \times 10^{12}$&$ (8.60^{+0.40}_{-0.30})$&$(8.02^{+0.80}_{-0.70}) \times 10^{-11}$\\
&B1-b&$(9.73^{+0.05}_{-0.04}) \times 10^{13}$&$(7.80^{+0.20}_{-0.20})$&$(1.23^{+0.01}_{-0.01}) \times 10^{-9}$\\
&IRAS4A&$(1.42^{+0.03}_{-0.02}) \times 10^{14}$&$(15.30^{+0.20}_{-0.20})$&$(4.83^{+0.02}_{-0.02}) \times 10^{-9}$\\
&&$(1.41^{+0.04}_{-0.04}) \times 10^{14}$&$(43.30^{+0.80}_{-0.70})$&$(4.83^{+0.04}_{-0.03}) \times 10^{-9}$\\
&SVS13A&$(1.21^{+0.03}_{-0.03}) \times 10^{14}$&$(65.10^{+2.0}_{-1.9})$&$(1.21^{+0.03}_{-0.03}) \times 10^{-9}$\\
\hline
CH$_3$CHO&L1544&$(1.25^{+0.37}_{-0.28}) \times 10^{12}$  &5.98$^{+0.70}_{-0.57}$&$(1.40^{+0.42}_{-0.32}) \times 10^{-11}$\\
&B1-b&$(3.48^{+0.55}_{-0.47}) \times 10^{12}$&9.09$^{+0.62}_{-0.55}$&$(4.40^{+0.70}_{-0.60}) \times 10^{-11}$\\
&IRAS4A&$(9.47^{+0.56}_{-0.52}) \times 10^{12}$&$(22.09^{+1.03}_{-0.94})$&$(3.30^{+0.19}_{-0.18}) \times 10^{-10}$\\
&&$(9.53^{+1.00}_{-0.95}) \times 10^{12}$&$(64.46^{+5.71}_{-4.85})$&$(3.30^{+0.37}_{-0.33}) \times 10^{-10}$\\
&SVS13A&$(5.89^{+1.60}_{-1.31}) \times 10^{12}$&$ (41.33^{+6.51}_{-4.95})$&$(5.90^{+1.70}_{-1.30}) \times 10^{-11}$\\
\hline
CH$_3$OCHO&L1544&-&-&-\\
&B1-b&$(6.40) \times 10^{12***}$&$(15.50^{***})$&$(8.10) \times 10^{-11***}$\\
&IRAS4A&$(3.49^{+0.50}_{-0.40}) \times 10^{13}$&$(82.30^{+8.40}_{-7.00})$&$(1.20^{+0.4}_{-0.3}) \times 10^{-9}$\\
&SVS13A&$(7.78^{+1.30}_{-1.10}) \times 10^{13}$&$(125.60^{+19.80}_{-15.10})$&$(7.78^{+1.30}_{-1.10}) \times 10^{-10}$\\
\hline
C$_2$H$_5$OH&L1544&-&-&-\\
&B1-b&-&-&-\\
&IRAS4A&$(1.69^{+0.30}_{-0.20}) \times 10^{13}$&$(50.60^{+4.90}_{-4.10})$&$(5.86^{+0.21}_{-0.22}) \times 10^{-10}$\\
&SVS13A&$(1.37^{+0.20}_{-0.10}) \times 10^{13}$&$(59.00^{+7.40}_{-5.90})$&$(1.37^{+0.27}_{-0.17}) \times 10^{-10}$\\
\hline
HCCCHO&L1544&$(3.16^{+0.80}_{-0.70}) \times 10^{12}$&$(5.60^{+0.30}_{-0.30})$&$(3.55^{+0.90}_{-0.70}) \times 10^{-11}$\\
&B1-b&-&-&-\\
&IRAS4A&-&-&-\\
&SVS13A&-&-&-\\
\hline
CH$_3$OCH$_3$&L1544&$1.72\times10^{12}$*&-&$(1.93) \times 10^{-11}$\\
&B1-b&$6.12\times10^{12}$*&-&$(7.74)\times10^{-11}$\\
&IRAS4A&$(1.20^{+0.40}_{-0.30}) \times 10^{13}$&$(61.10^{+20.11}_{-12.10})$&$(4.14^{+1.30}_{-1.00}) \times 10^{-10}$\\
&SVS13A&$1.41\times10^{13}$**&-&$1.41\times10^{-10}$\\
\hline
CH$_3$CN&L1544&$4.85\times10^{11}$*&-&$(5.45) \times 10^{-12}$\\
&B1-b&$4.95\times10^{11}$*&-&$(6.27)\times10^{-12}$\\
&IRAS4A&$(1.78^{+0.17}_{-0.15}) \times 10^{12}$&$(25.78^{+1.9}_{-1.7})$&$(6.10^{+0.59}_{-0.54}) \times 10^{-11}$\\
&&$(2.31^{+0.31}_{-0.28}) \times 10^{12}$&$(61.2^{+5.05}_{-4.34})$&$(8.00^{+1.10}_{-0.96}) \times 10^{-11}$\\
&SVS13A&$(4.57^{+0.23}_{-0.22}) \times 10^{12}$&$(134.69^{+11.8}_{-10.09})$&$(4.6^{+0.23}_{-0.22}) \times 10^{-11}$\\
\hline
\end{tabular}}}\\
{\scriptsize \noindent * Calculated using LTE fitting.\\
** Calculated from \cite{bian19} and scaled for 30$^"$ beam.\\
*** We did not include the errors due to the large uncertainty in data points.\\
The hydrogen column density (N$_{H_2}$) in L1544, B1-b, IRAS4A and SVS13A are $8.9\times10^{22}$ cm$^{-2}$ \citep{hily22}, $7.9\times10^{22}$ cm$^{-2}$ \citep{dani13}, $2.9\times10^{22}$ cm$^{-2}$ \citep{mare02}, 
$1.0\times10^{23}$cm$^{-2}$ \citep{lefl98}, respectively.}
\end{table}
\begin{figure}
\begin{minipage}{0.55\textwidth}
\includegraphics[width=\textwidth]{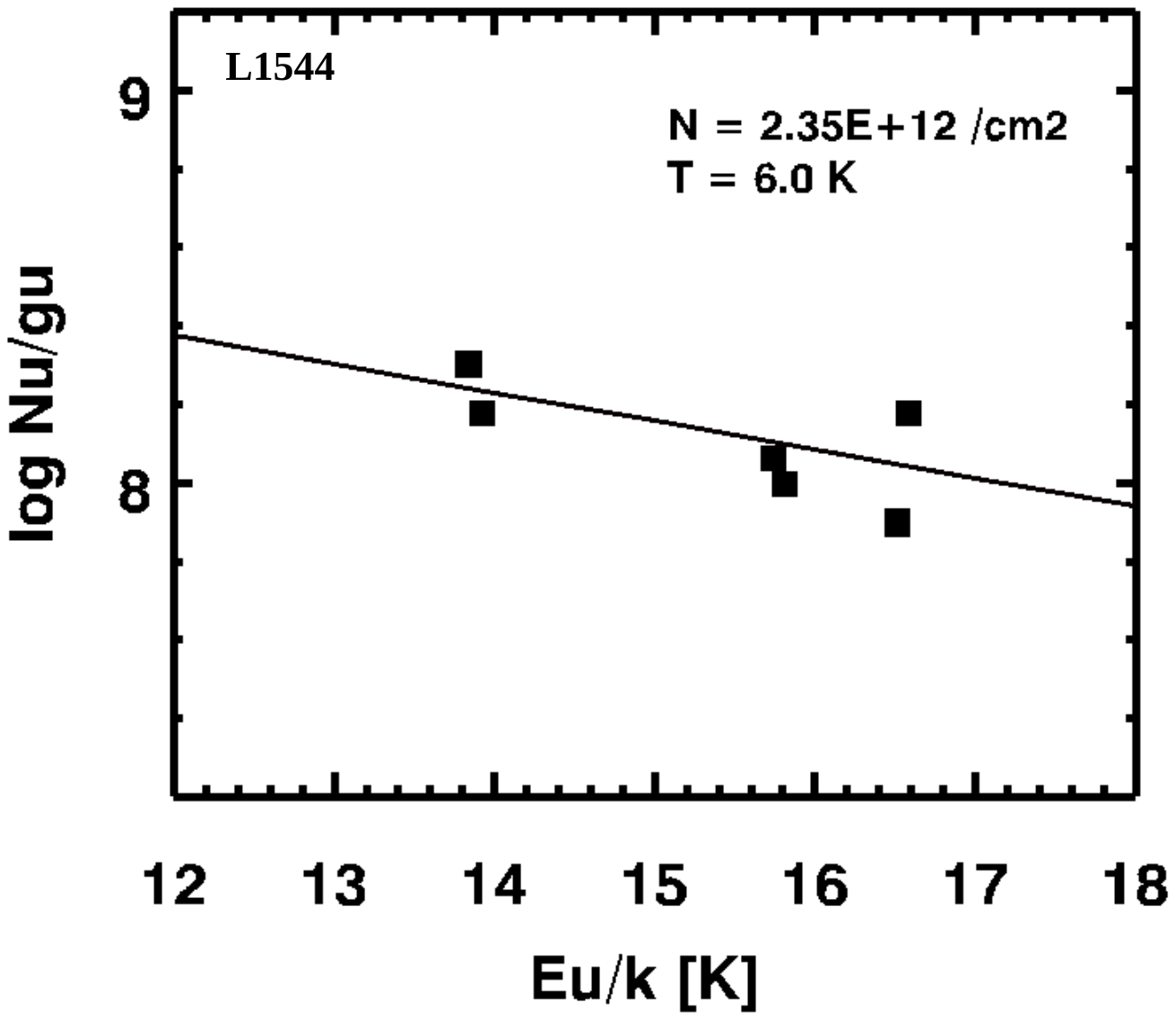}
\end{minipage}
 \begin{minipage}{0.55\textwidth}
 \includegraphics[width=\textwidth]{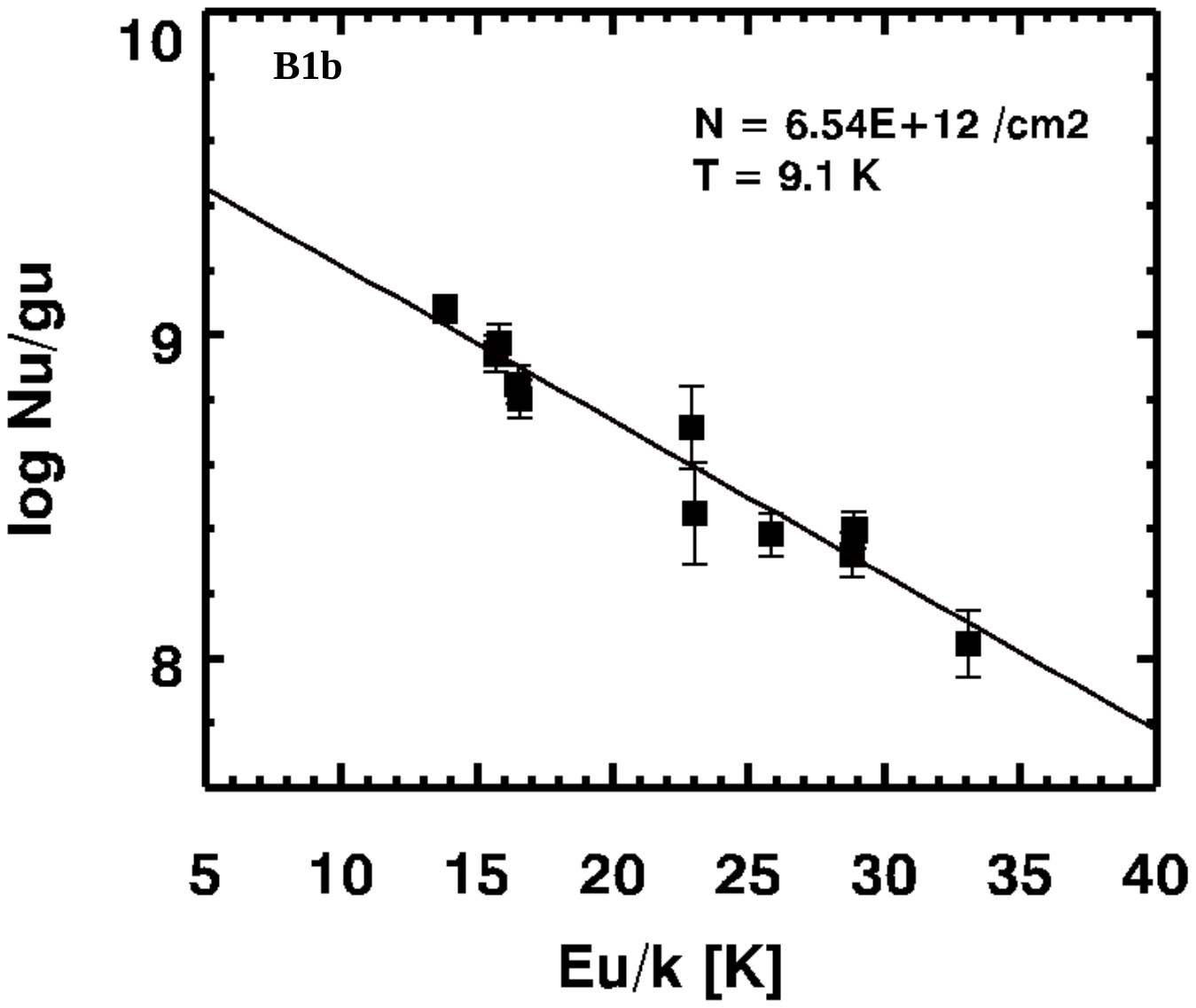}
 \end{minipage}
 \begin{minipage}{0.55\textwidth}
 \includegraphics[width=\textwidth]{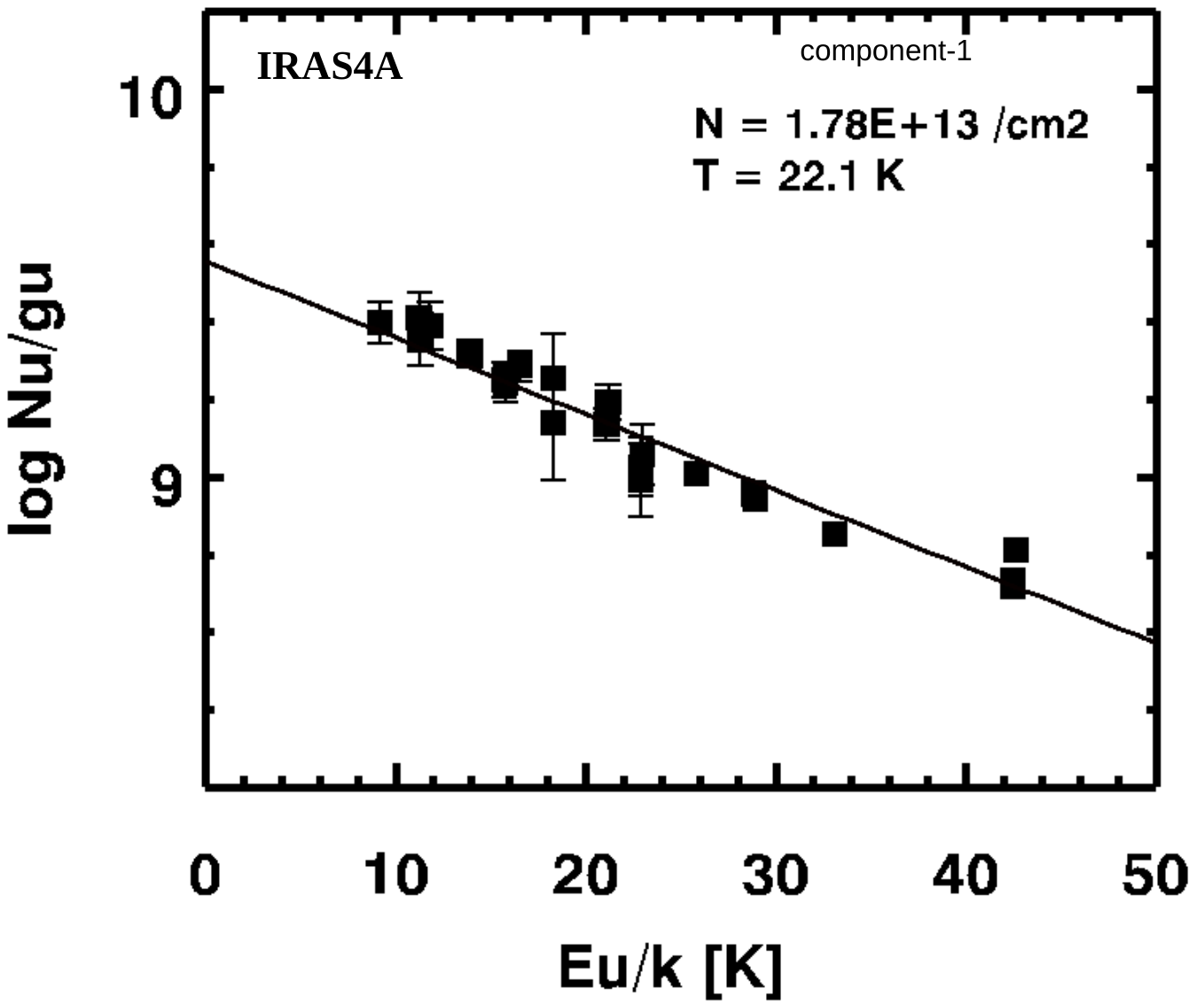}
 \end{minipage}
  \begin{minipage}{0.55\textwidth}
 \includegraphics[width=\textwidth]{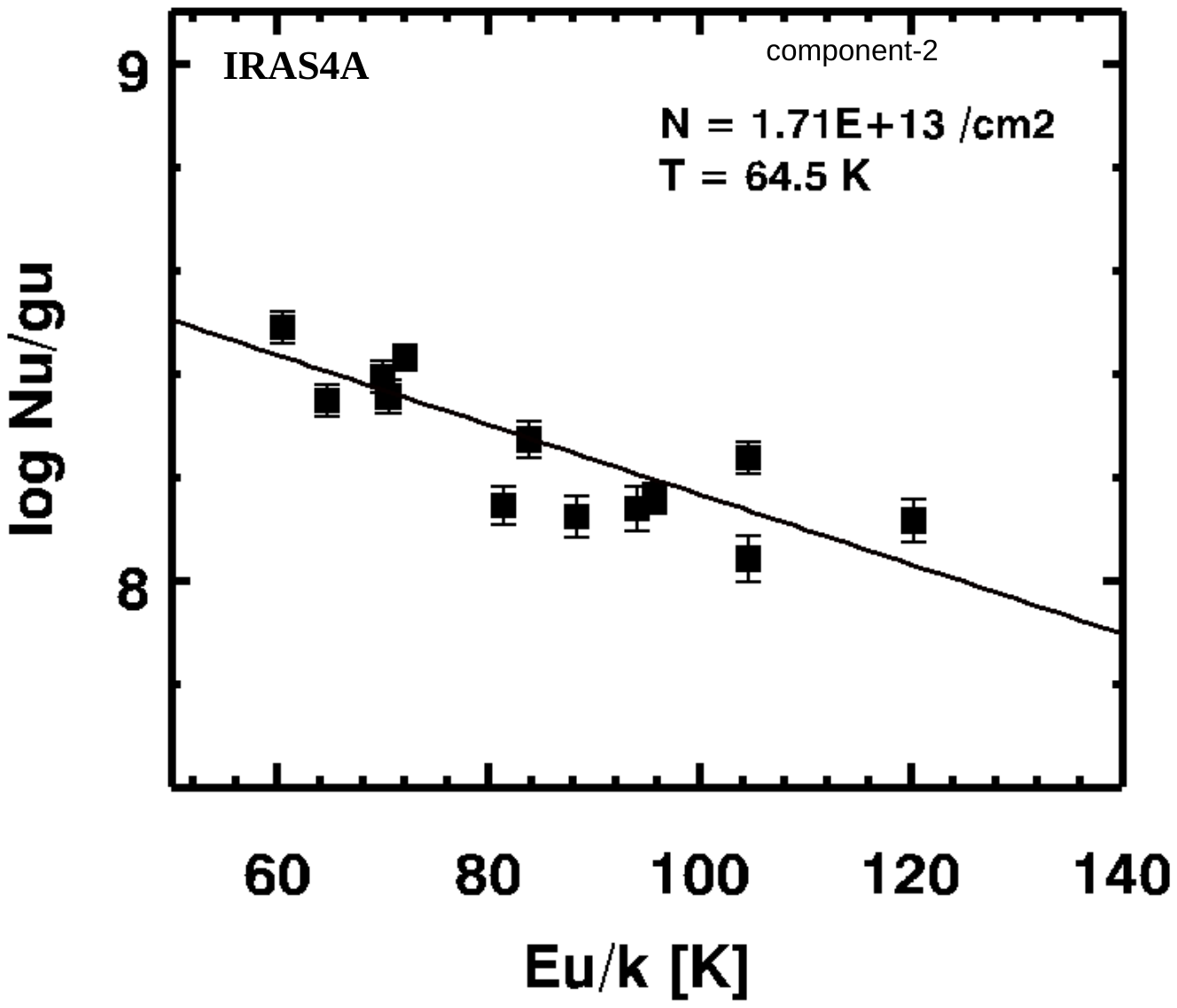}
 \end{minipage}
 \begin{minipage}{0.55\textwidth}
 \includegraphics[width=\textwidth]{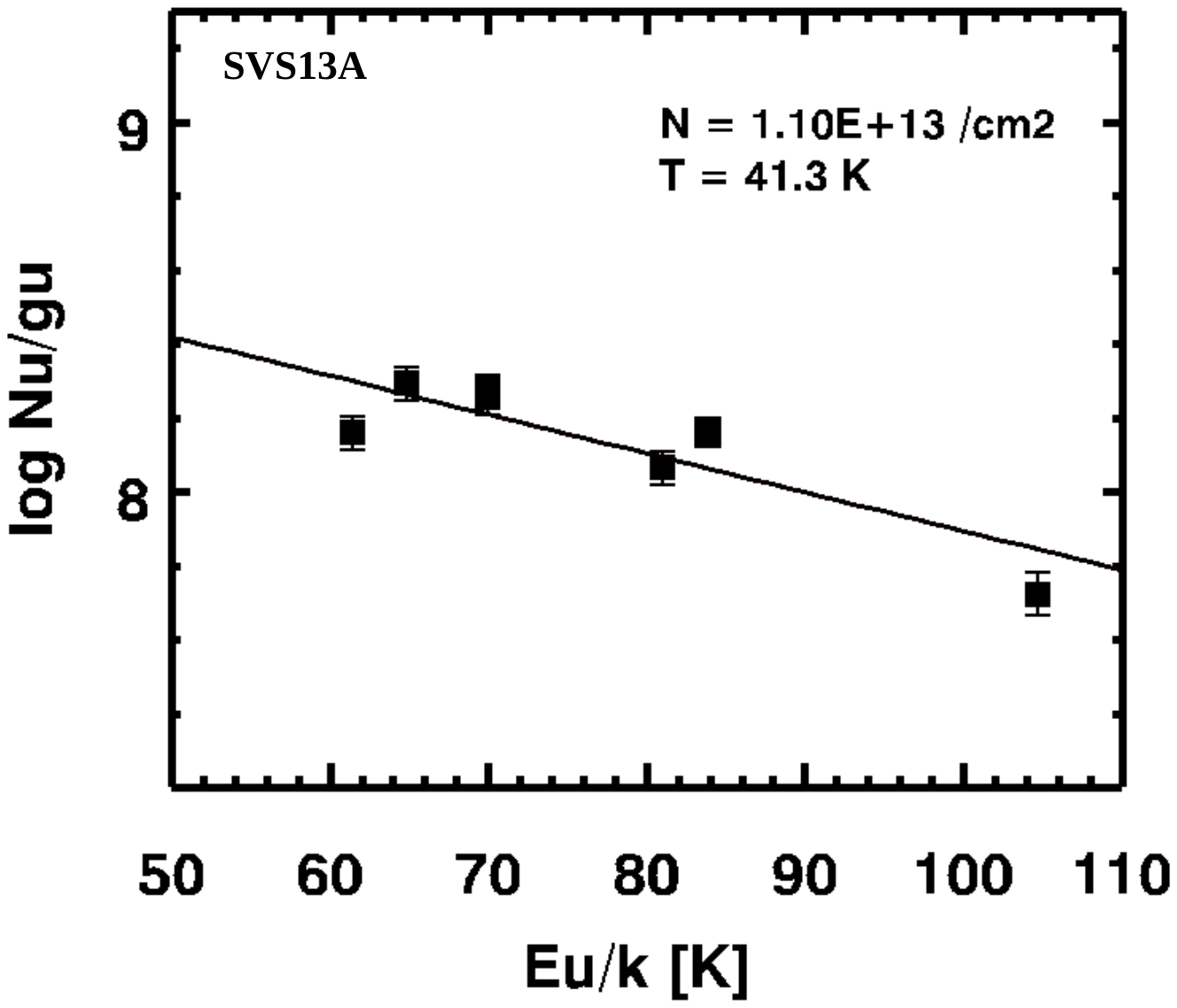}
 \end{minipage}
 \caption{Rotational diagram for CH$_3$CHO obtained for various sources. The symbols represent the same as those depicted in Figure \ref{fig:rotational_diag_ch3oh_single}.}
\label{fig:rotational_diag_ch3cho}
\end{figure}
\begin{figure}
 \begin{minipage}{0.55\textwidth}
 \includegraphics[width=\textwidth]{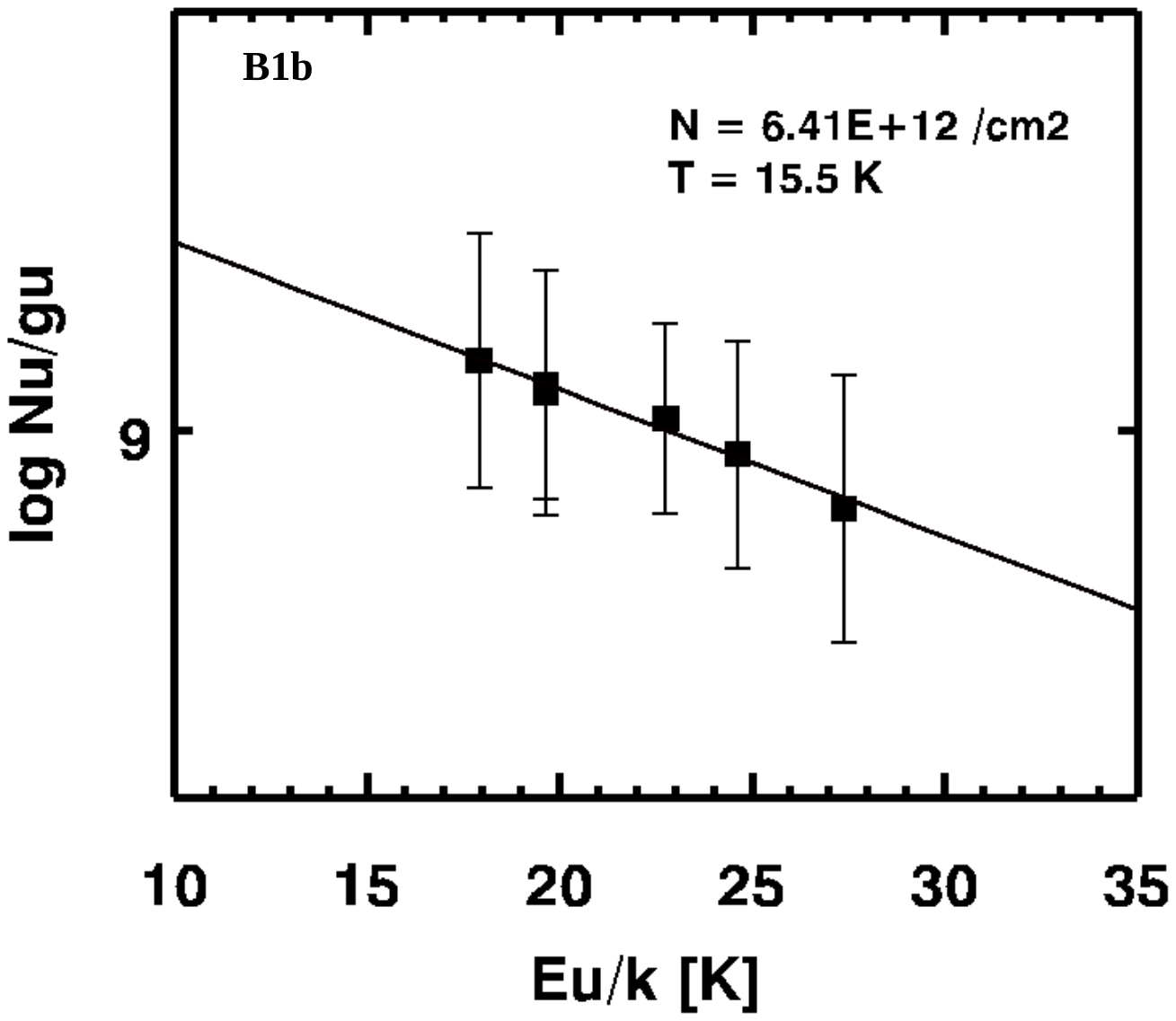}
 \end{minipage}
 \begin{minipage}{0.55\textwidth}
 \includegraphics[width=\textwidth]{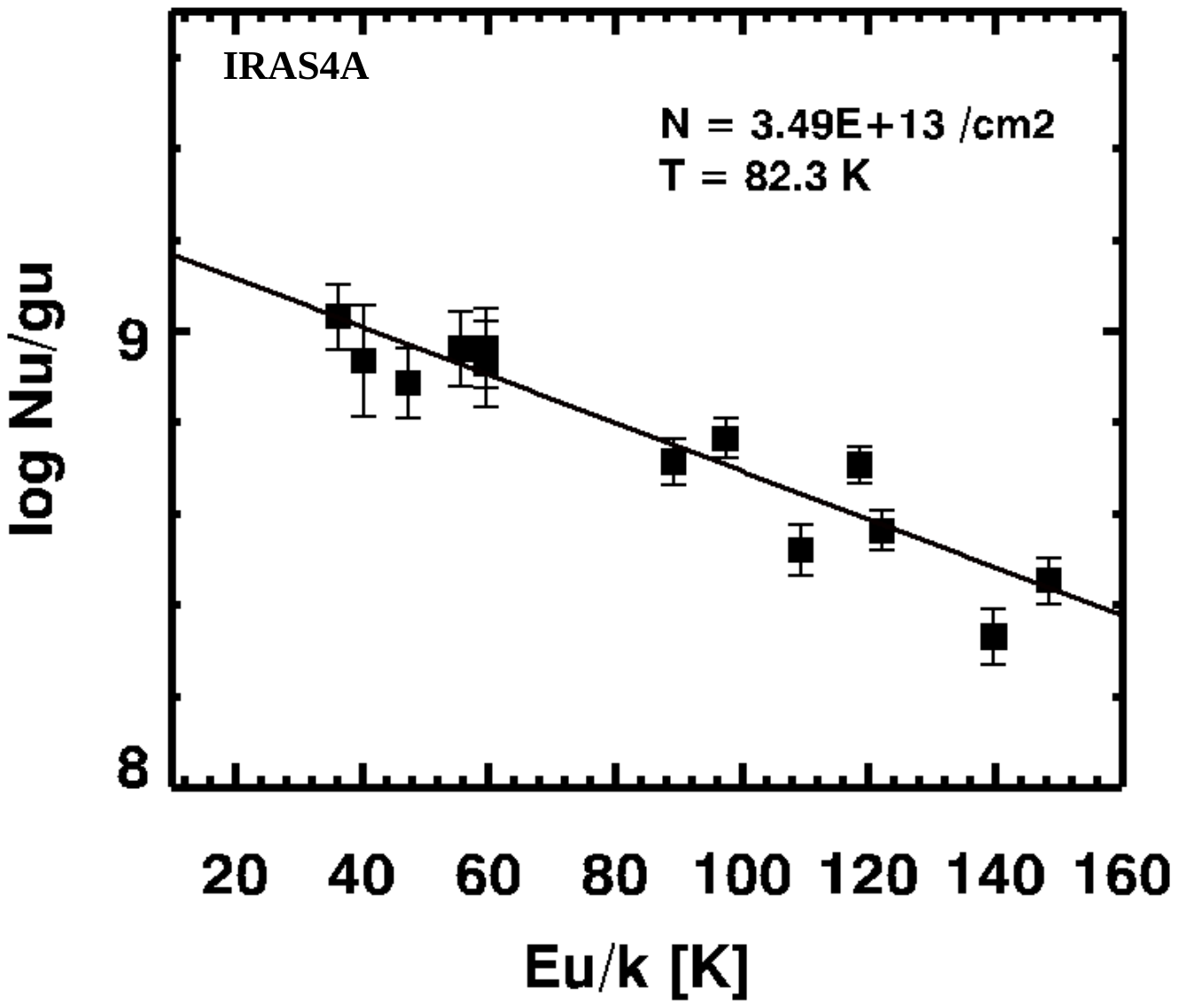}
 \end{minipage}
 \begin{minipage}{0.55\textwidth}
 \includegraphics[width=\textwidth]{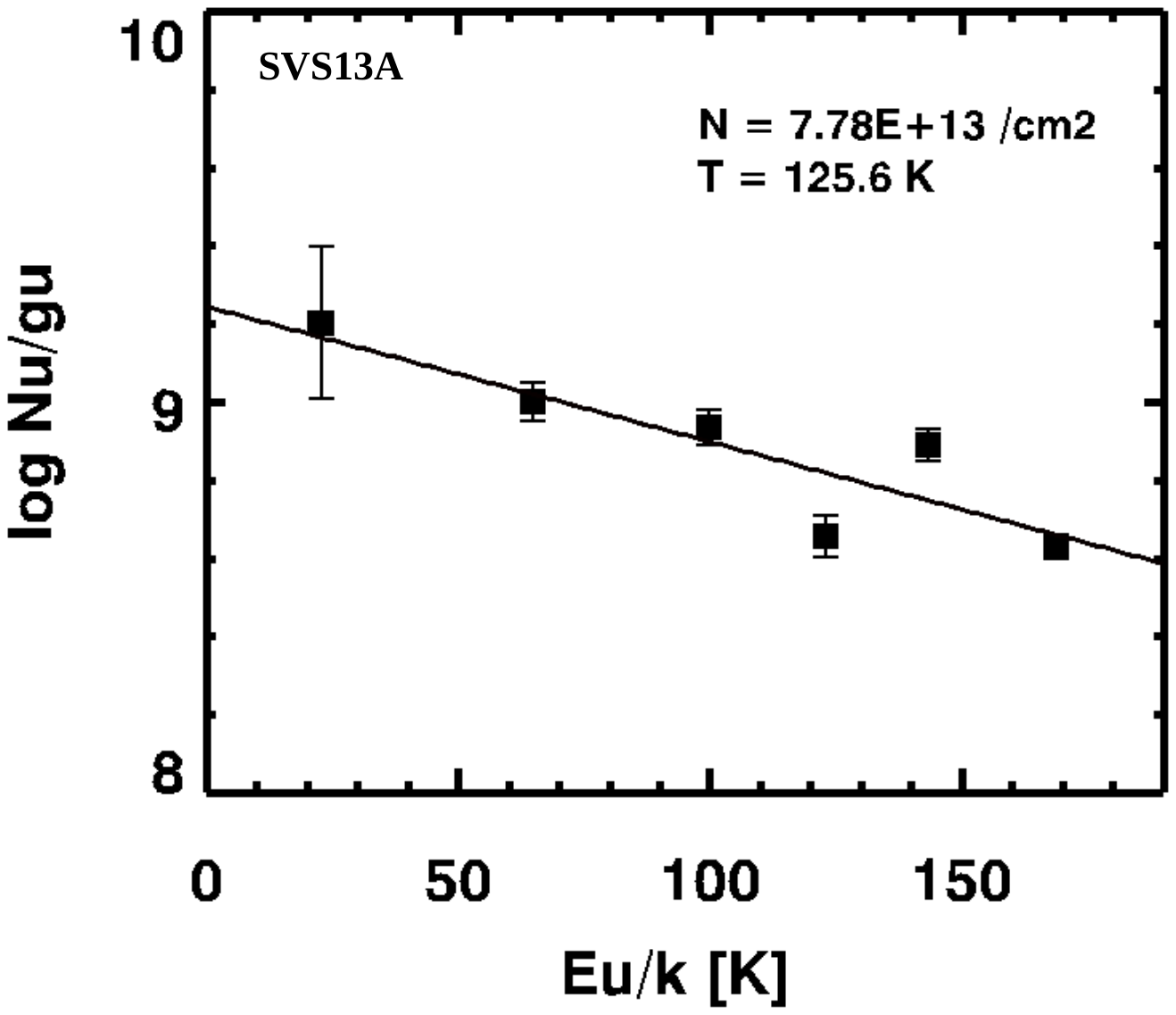}
 \end{minipage}
 \caption{Rotational diagram for CH$_3$OCHO obtained for various sources. The symbols represent the same as those depicted in Figure \ref{fig:rotational_diag_ch3oh_single}.}
\label{fig:rotational_diag_hcooch3}
\end{figure}
\begin{figure}
 \begin{minipage}{0.55\textwidth}
\includegraphics[width=\textwidth]{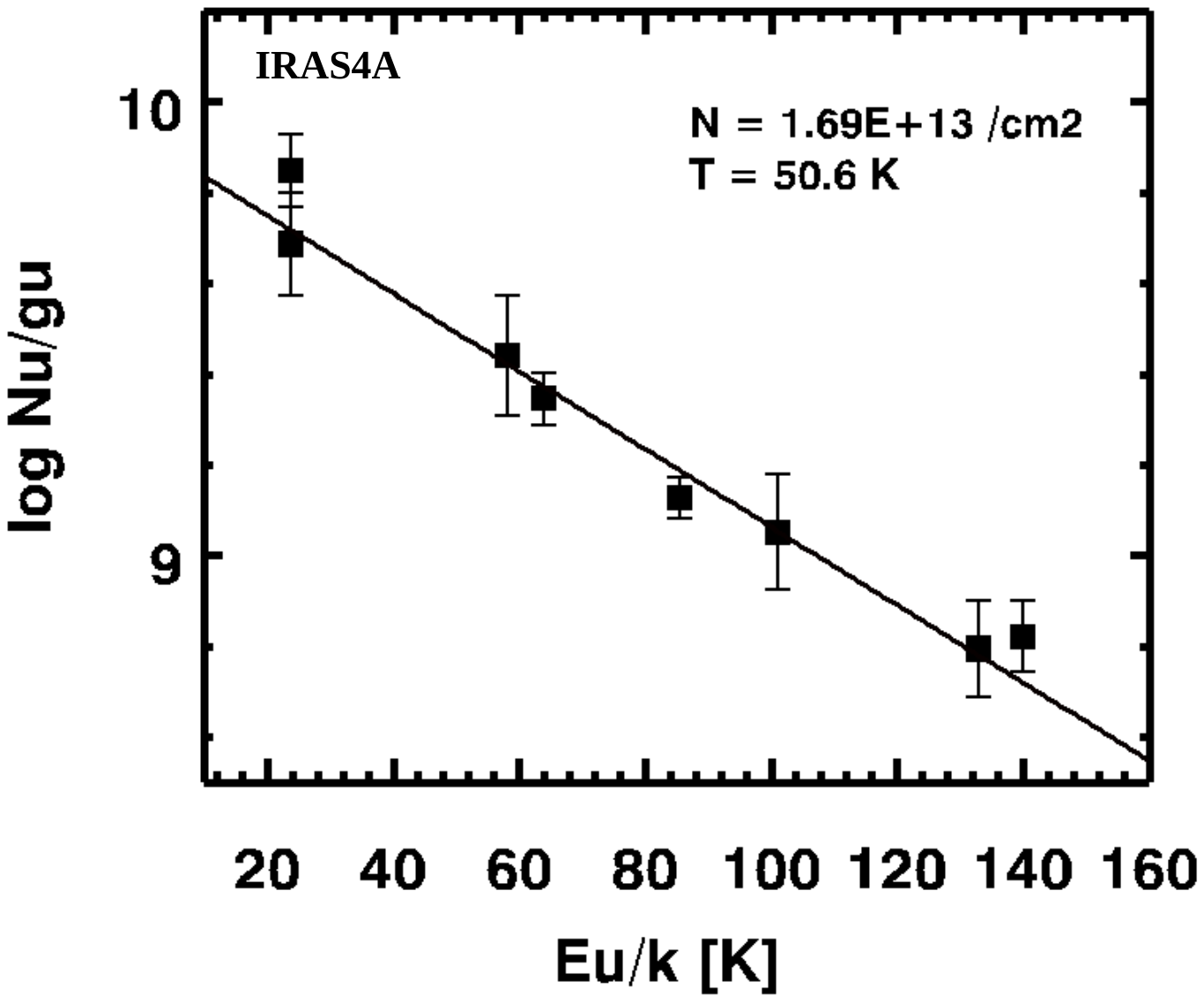}
\end{minipage}
\begin{minipage}{0.55\textwidth}
\includegraphics[width=\textwidth]{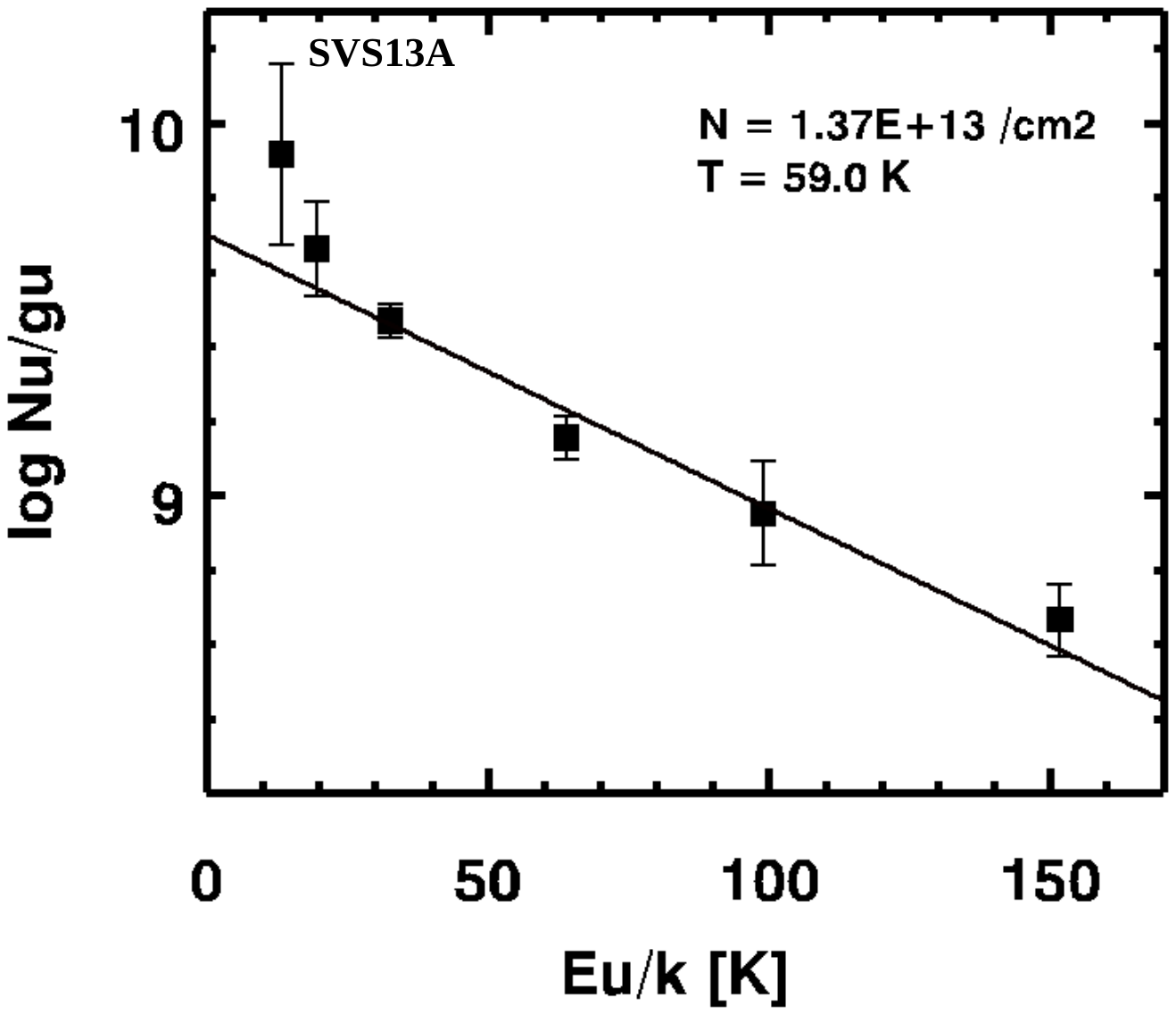}
\end{minipage}
\caption{Rotational diagram for C$_2$H$_5$OH obtained for various sources. Here, we consider the source size and beam size to be the same. The symbols are the same as those used in Figure \ref{fig:rotational_diag_ch3oh_single}.}
\label{fig:rotational_diag_C2H5OH}
\end{figure}
\begin{figure}
\hskip 3cm
\includegraphics[width=8cm]{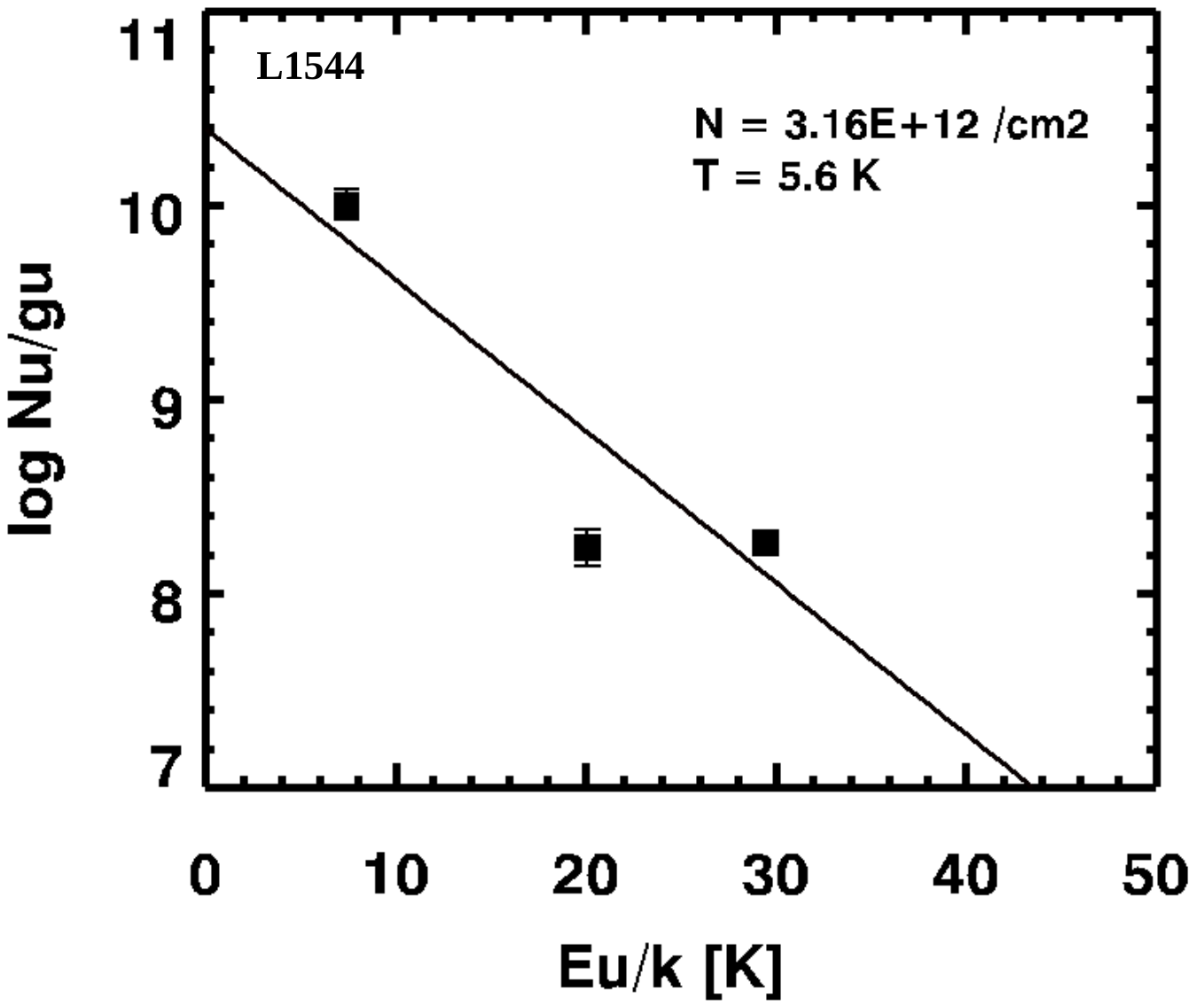}
\caption{Rotational diagram for HCCCHO obtained for one source. Here, we consider the source size and beam size to be the same. The symbols represent the same as those depicted in Figure \ref{fig:rotational_diag_ch3oh_single}.}
\label{fig:rotational_diag_HCCCHO}
\end{figure}

\begin{figure*}
\hskip 3cm
\begin{minipage}{0.55\textwidth}
\includegraphics[width=\textwidth]{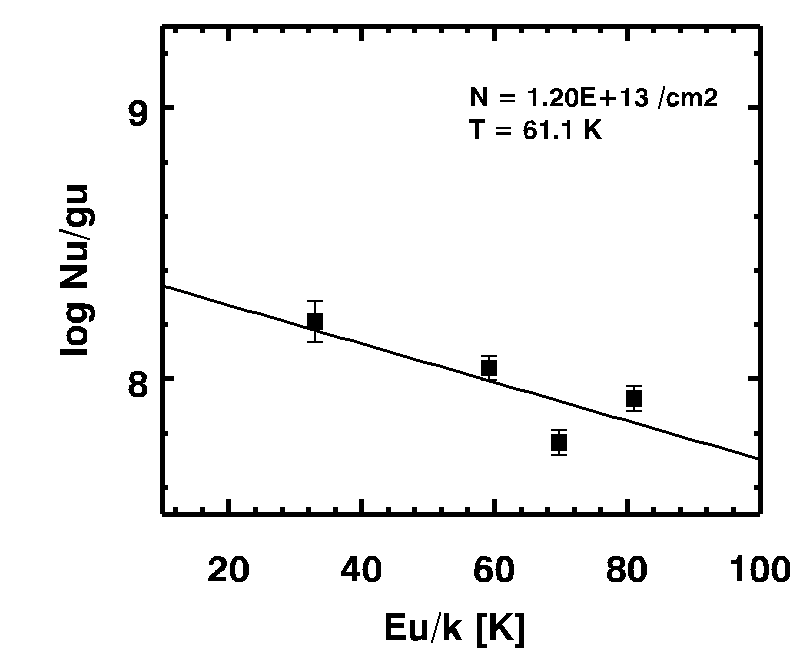}
\end{minipage}
\caption{Rotational diagram for CH$_3$OCH$_3$, v = 0 obtained towards IRAS4A. The symbols represent the same as those depicted in Figure \ref{fig:rotational_diag_ch3oh_single}.}
\label{fig:rotational_diag_CH3OCH3}
\end{figure*}
\begin{figure*}
 \begin{minipage}{0.55\textwidth}
\includegraphics[width=\textwidth]{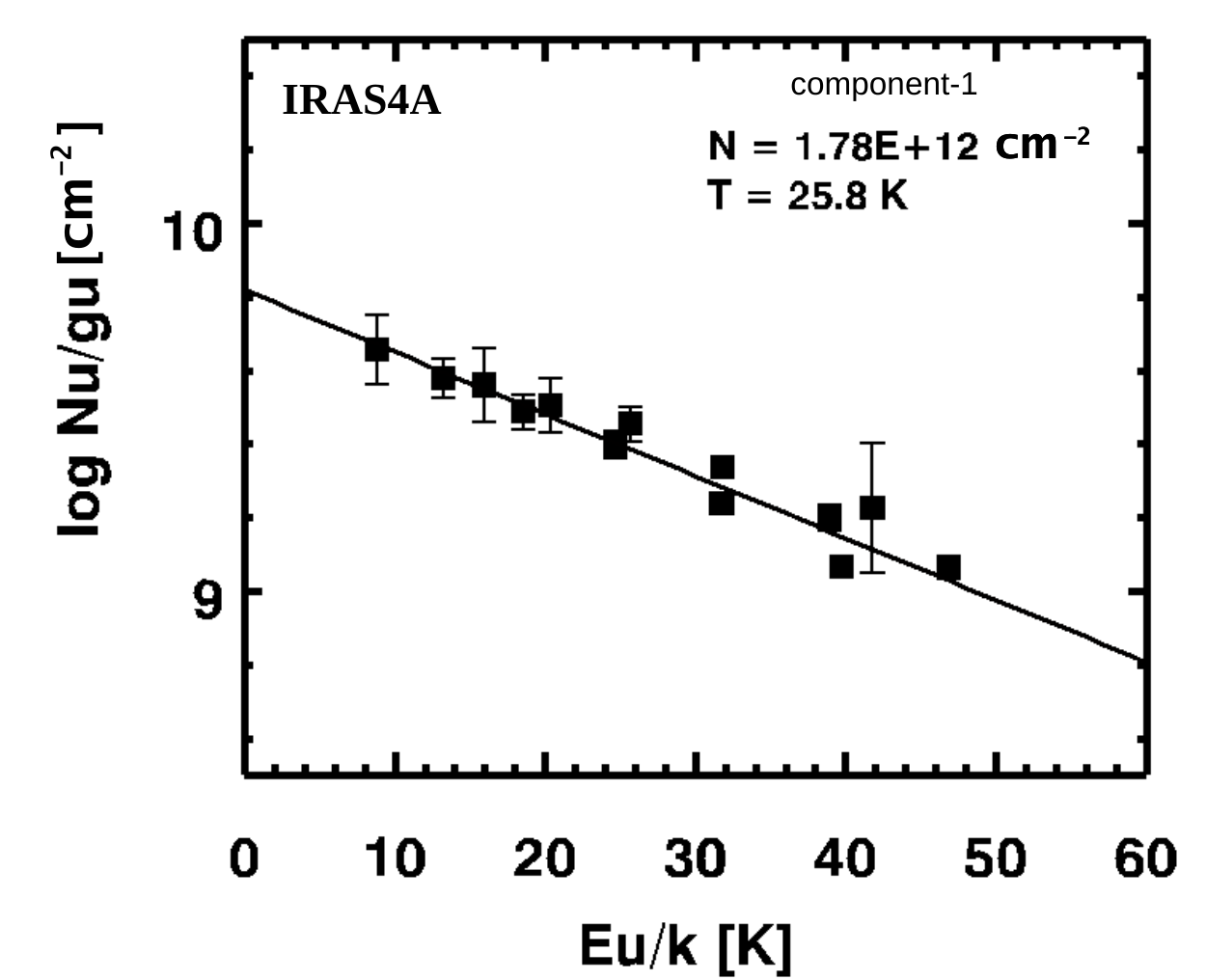}
\end{minipage}
\begin{minipage}{0.55\textwidth}
\includegraphics[width=\textwidth]{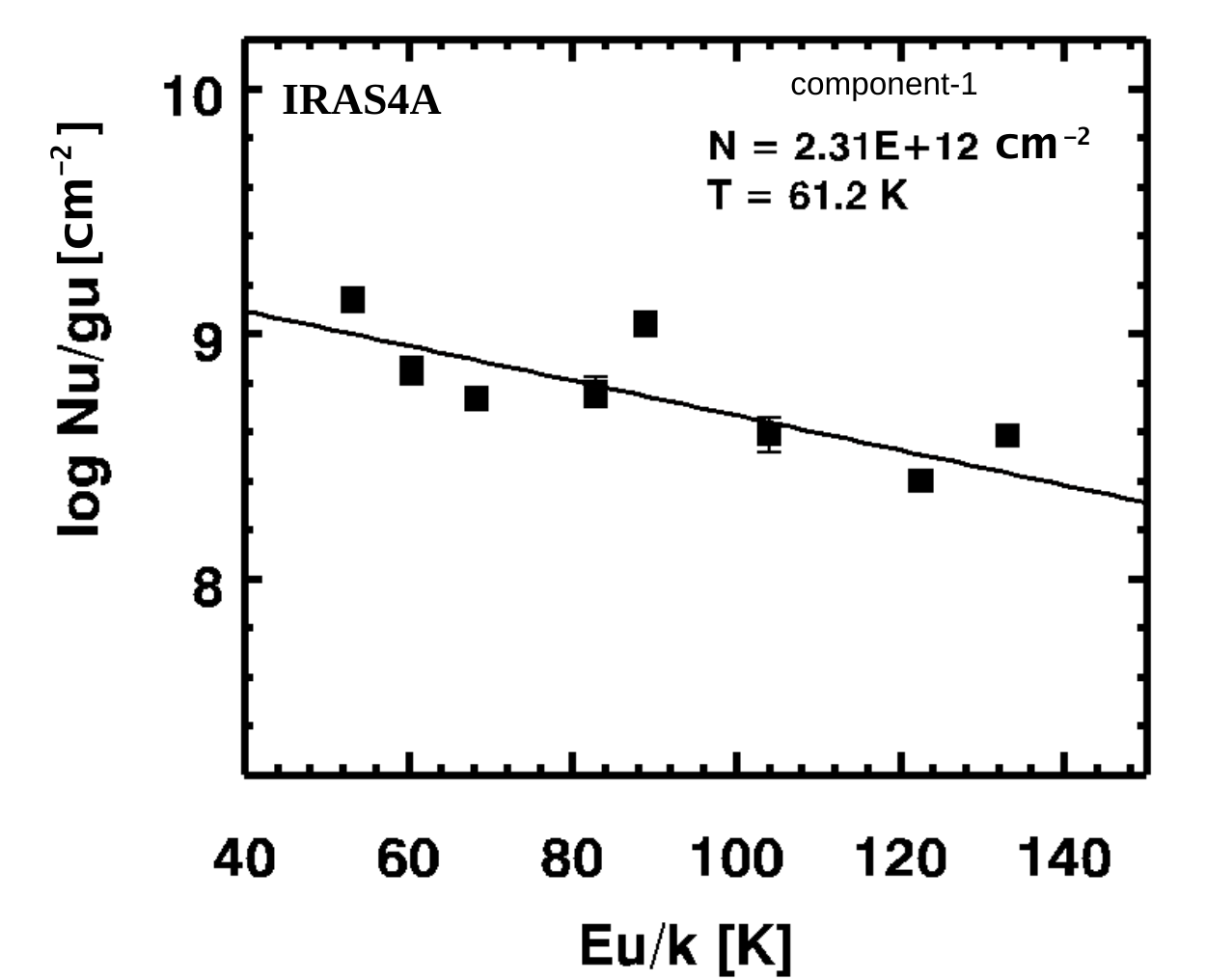}
\end{minipage}
\begin{minipage}{0.55\textwidth}
\includegraphics[width=\textwidth]{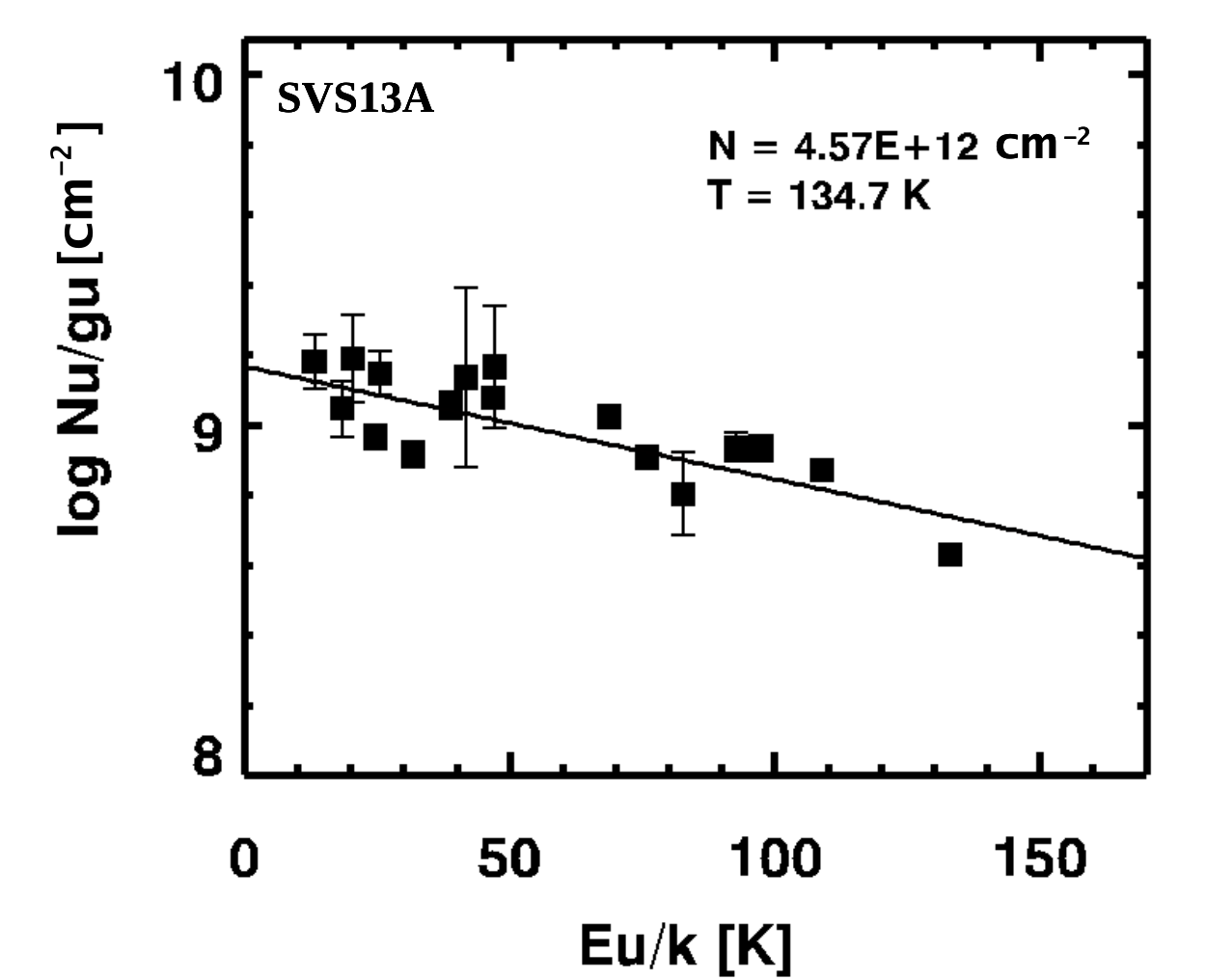}
\end{minipage}
\caption{Rotational diagram for CH$_3$CN obtained for various sources. The symbols represent the same as those depicted in Figure \ref{fig:rotational_diag_ch3oh_single}.}
\label{fig:rotational_diag_CH3CN}
\end{figure*}
\subsection{Monte Carlo Markov chain method}\label{sec:mcmc}
\begin{figure}
\hskip -1.0cm
\includegraphics[width=6cm, height=16cm, angle=270]{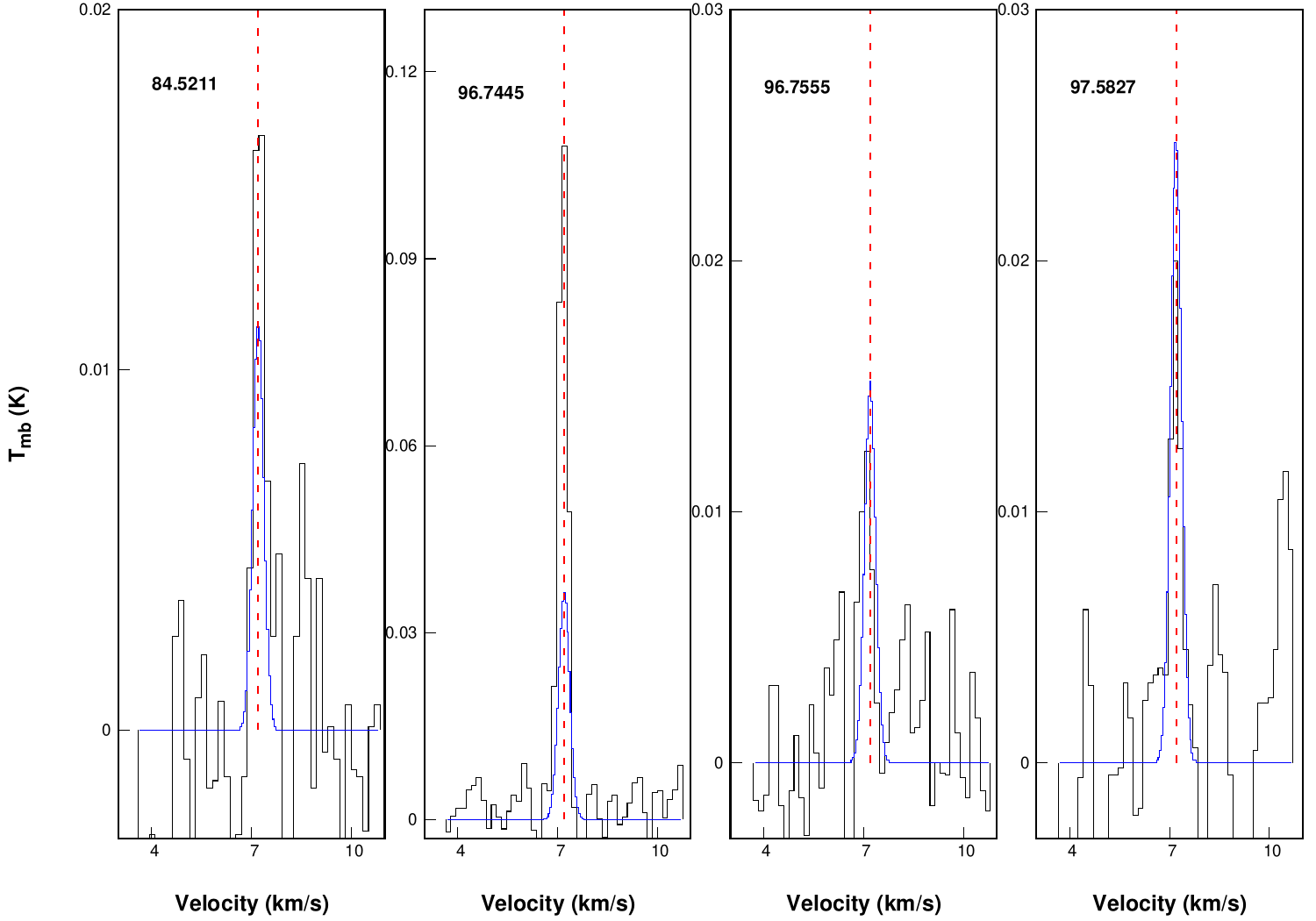}
\caption{MCMC fitting of the observed transitions of $\rm{CH_3OH}$ in L1544. Purple lines represent the modeled spectral profile to the observed spectra (black).}
\label{fig:ch3oh_l1544mcmc}
\end{figure}

\begin{figure*}
\centering
\includegraphics[width=11cm, height=15cm, angle=270]{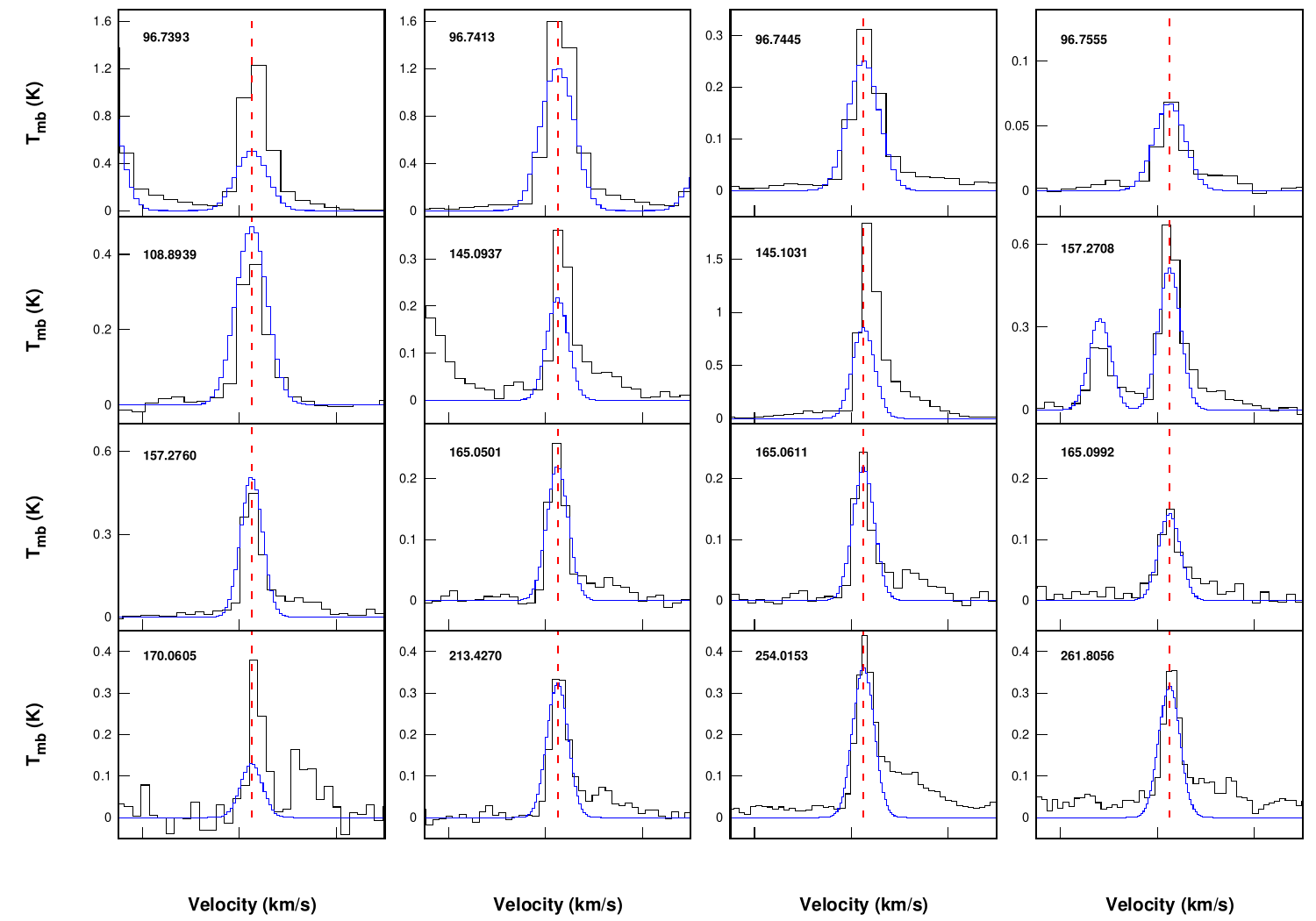}
\caption{MCMC fitting of the observed transitions of $\rm{CH_3OH}$ in B1-b. Purple lines represent the modeled spectral profile to the observed spectra (black).}
\label{fig:ch3oh_barnardmcmc}
\end{figure*}

\begin{figure*}
\centering
\includegraphics[width=11cm, height=15cm, angle=270]{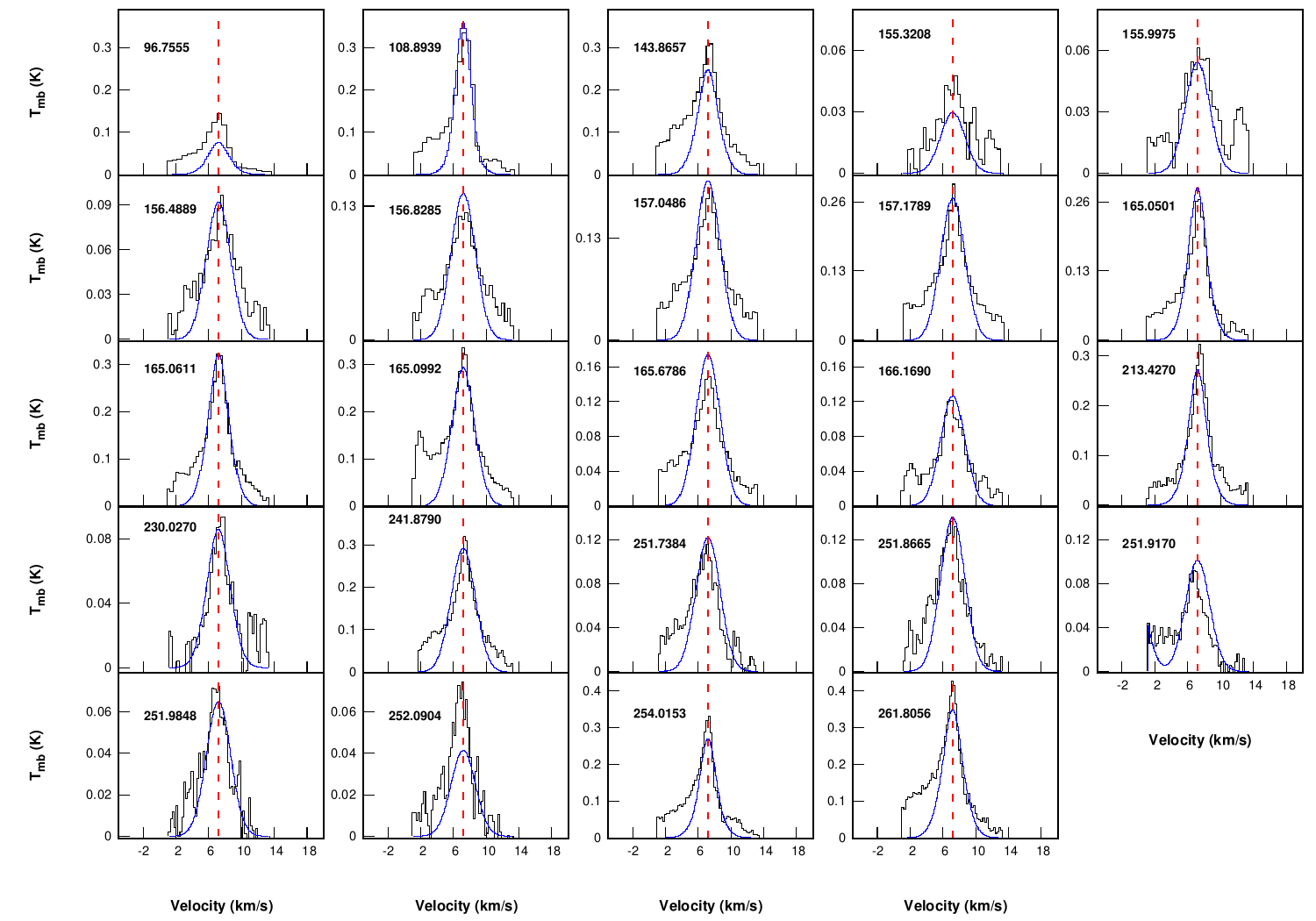}
\caption{MCMC fitting of the observed transitions of $\rm{CH_3OH}$ in IRAS4A. Purple lines represent the modeled spectral profile to the observed spectra (black).}
\label{fig:ch3oh_irasmcmc}
\end{figure*}

\begin{figure*}
\centering
\includegraphics[width=11cm, height=15cm, angle=270]{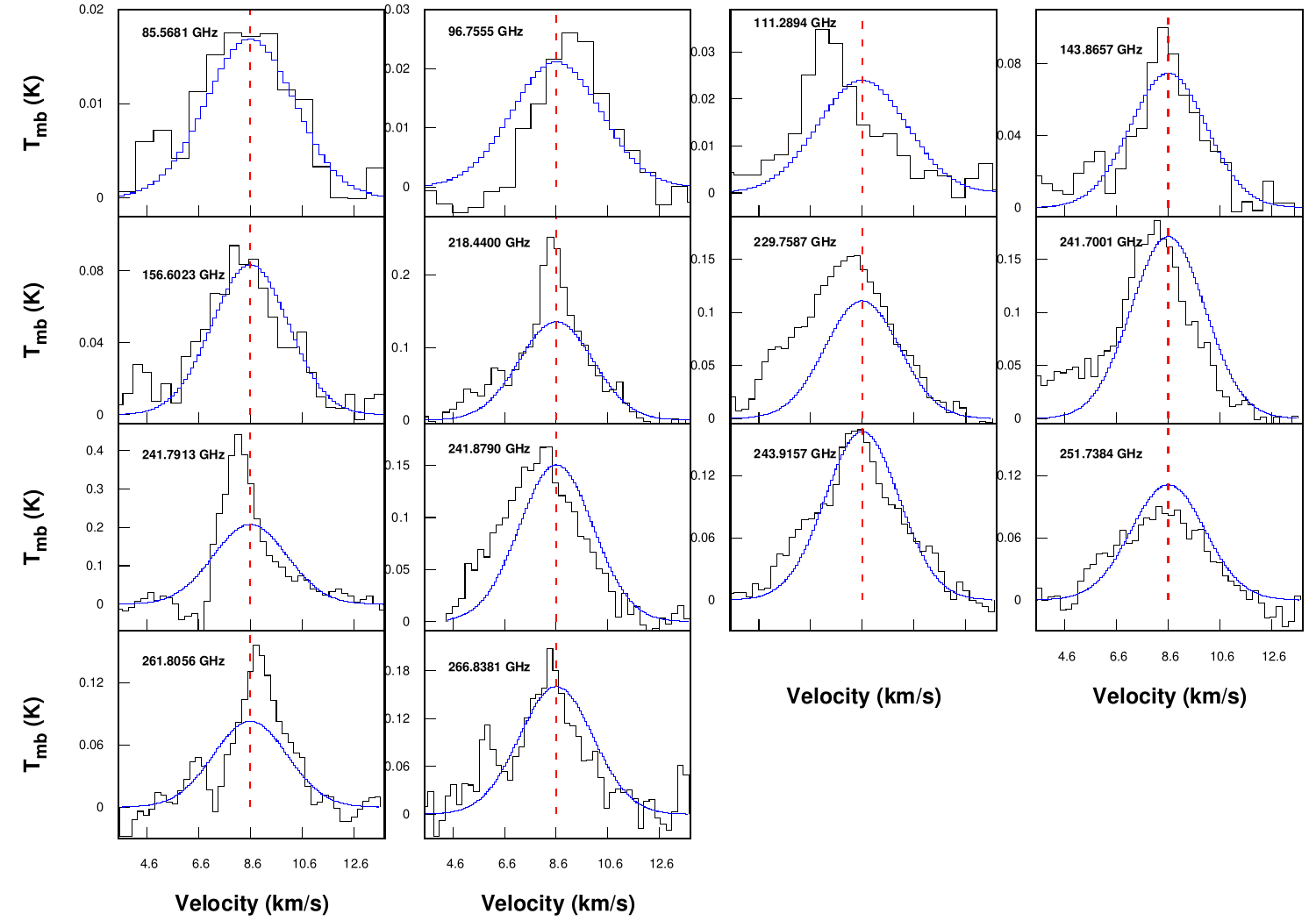}
\caption{MCMC fitting of the observed transitions of $\rm{CH_3OH}$ in SVS13A. Purple lines represent the modeled spectral profile to the observed spectra (black).}
\label{fig:ch3oh_svsmcmc}
\end{figure*}

\begin{figure}
\hskip -1.7cm
\includegraphics[width=17cm, height=19cm, angle=270]{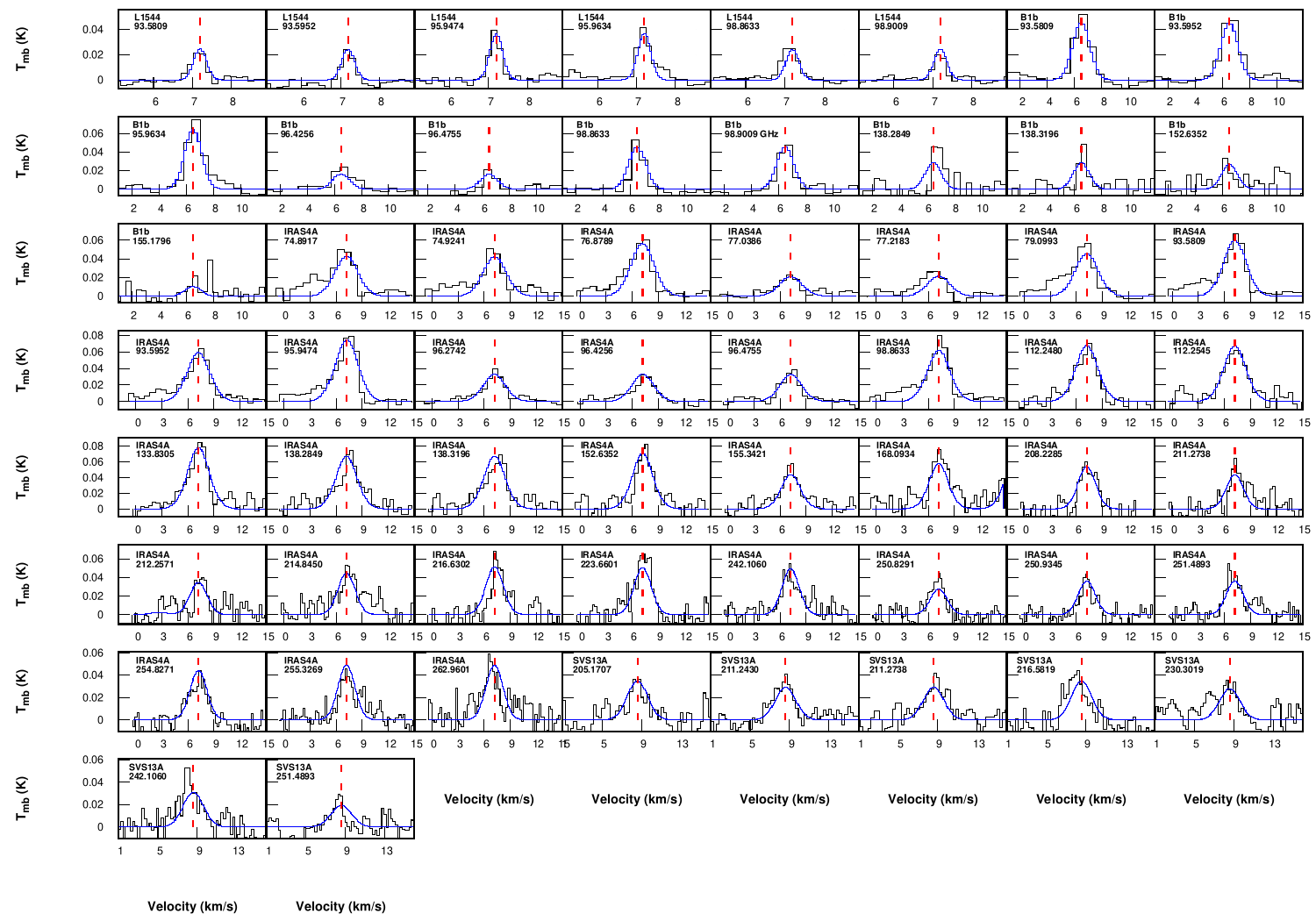}
\caption{MCMC fitting of the observed transitions of $\rm{CH_3CHO}$. Purple lines represent the modeled spectral profile to the observed spectra (black).}
\label{fig:ch3cho_mcmc}
\end{figure}

\begin{figure}
\hskip -1.5 cm
\includegraphics[width=15cm, height=18cm, angle=270]{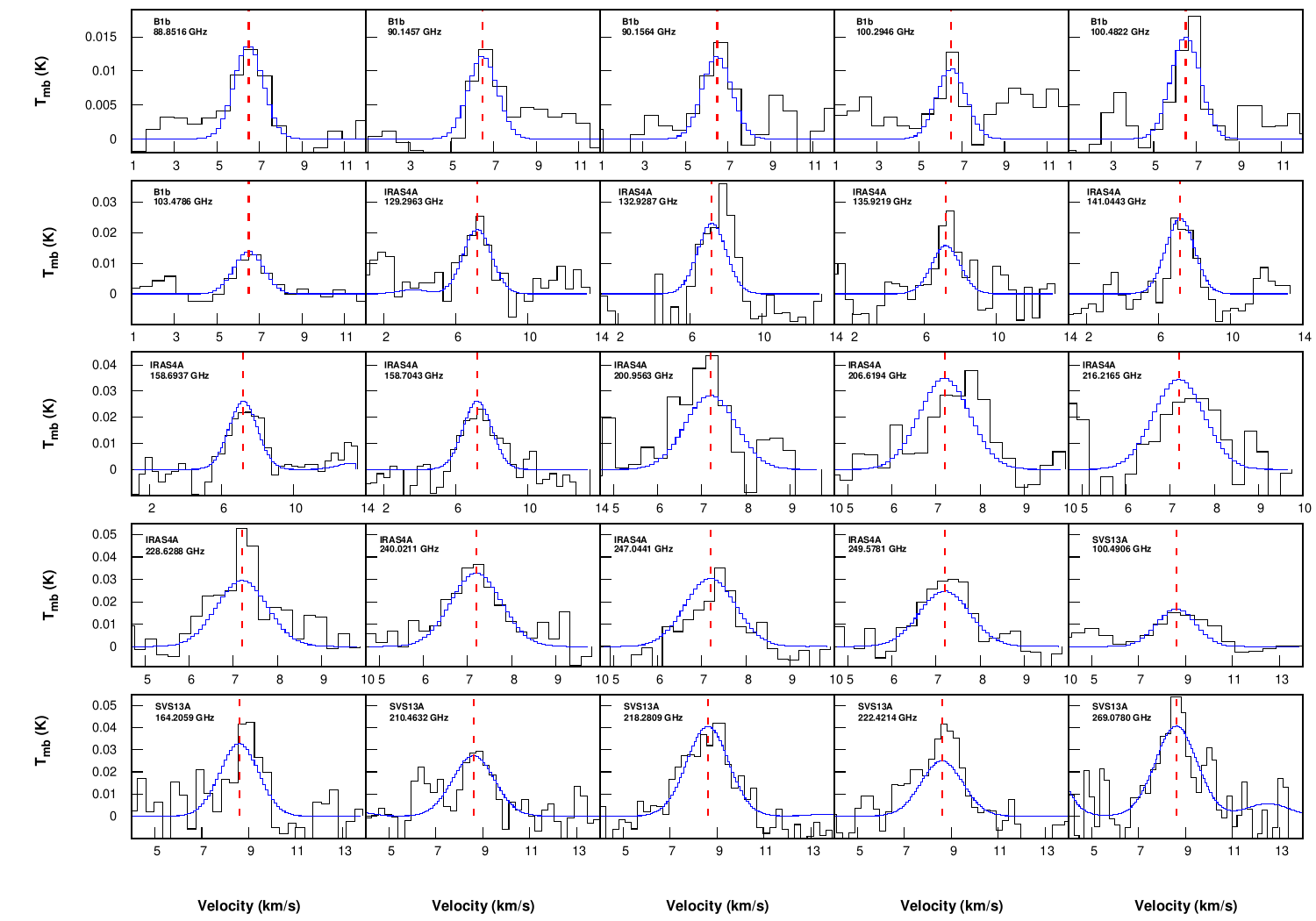}
\caption{MCMC fitting of the observed transitions of $\rm{CH_3OCHO}$. Purple lines represent the modeled spectral profile to the observed spectra (black).}
\label{fig:ch3ocho_mcmc}
\end{figure}

\begin{figure}
\hskip -1.0 cm
\includegraphics[width=11cm, height=15cm, angle=270]{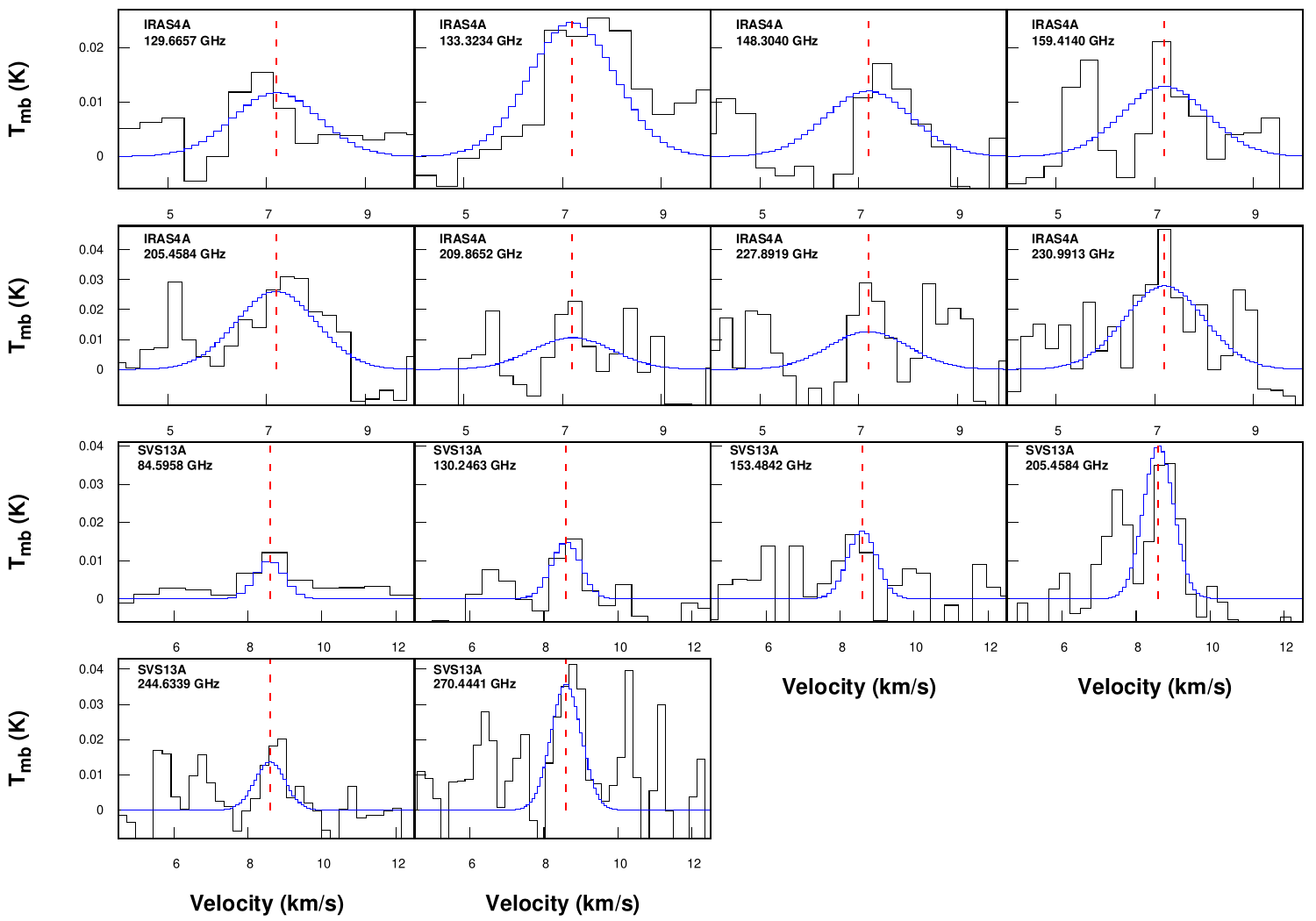}
\caption{MCMC fitting of the observed transitions of $\rm{C_2H_5OH}$. Purple lines represent the modeled spectral profile to the observed spectra (black).}
\label{fig:c2h5oh_mcmc}
\end{figure}

\begin{figure*}
\centering
\includegraphics[width=7cm, height=15cm, angle=270]{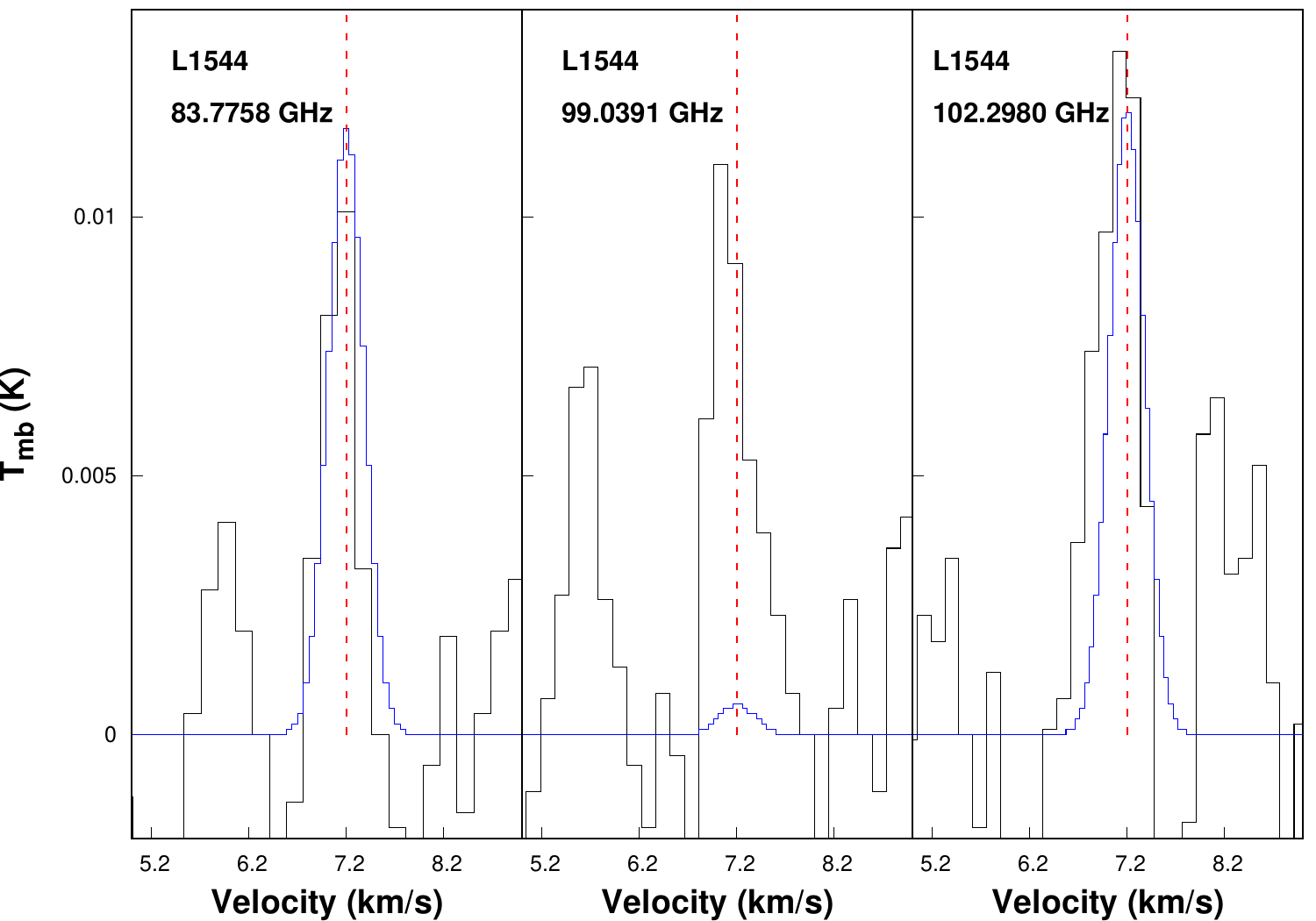}
\caption{MCMC fitting of the observed transitions of $\rm{HCCCHO}$. Purple lines represent the modeled spectral profile to the observed spectra (black).}
\label{fig:hcccho_mcmc}
\end{figure*}

\begin{figure*}
\begin{minipage}{0.35\textwidth}
\includegraphics[width=\textwidth,angle=270]{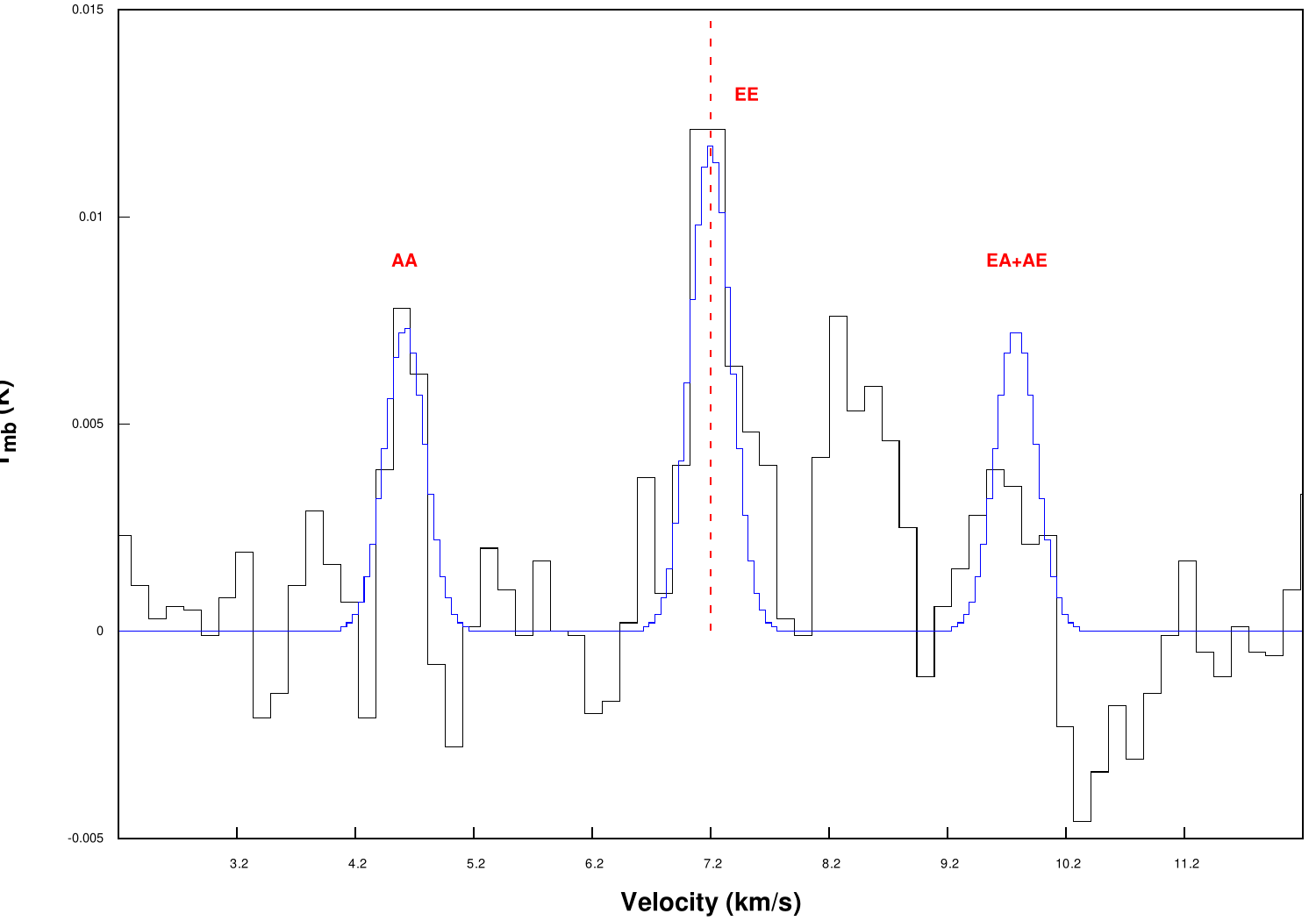}
\end{minipage}
\begin{minipage}{0.35\textwidth}
\includegraphics[width=\textwidth,angle=270]{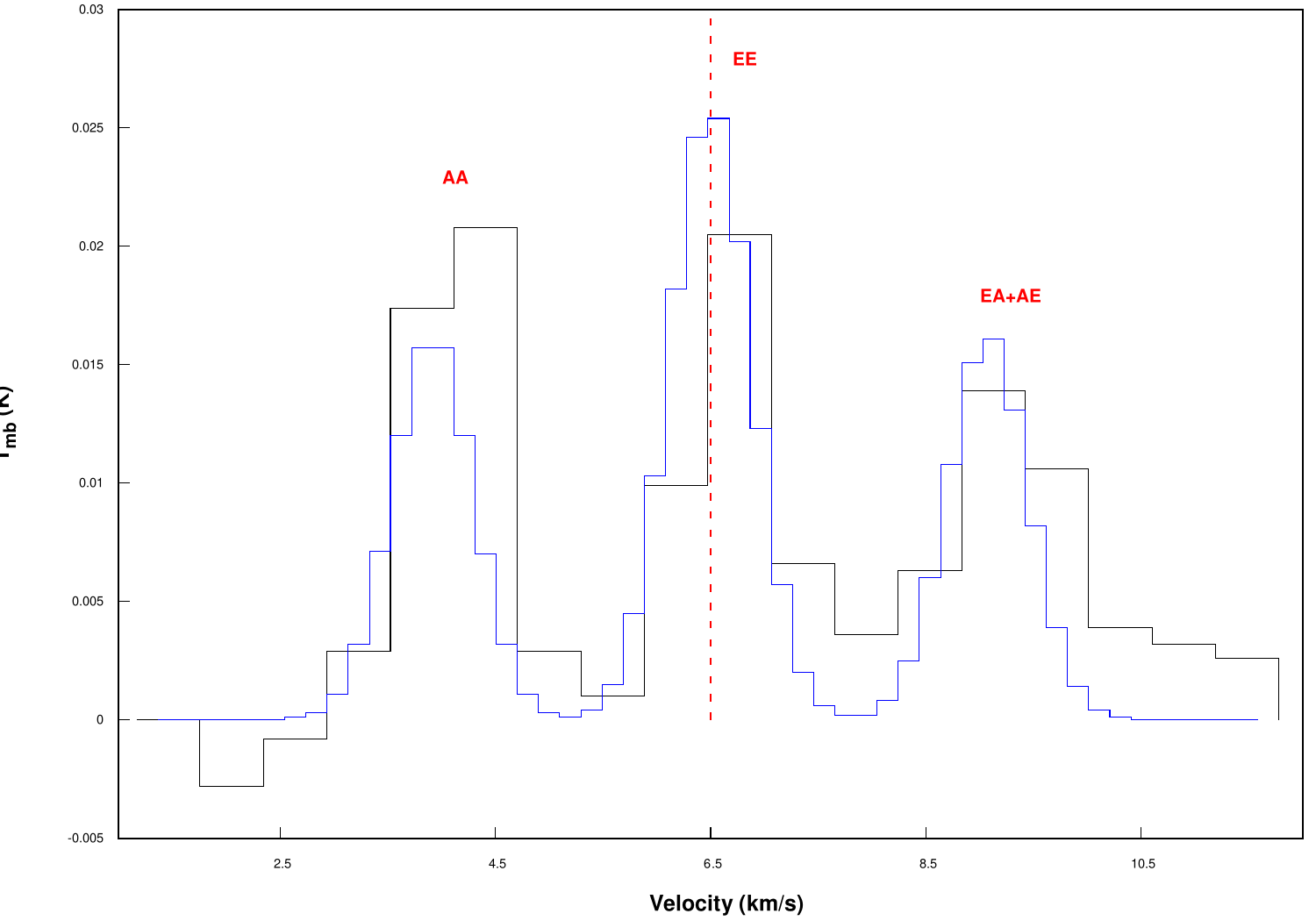}
\end{minipage}
\begin{minipage}{0.35\textwidth}
\includegraphics[width=7cm, height=15cm,angle=270]{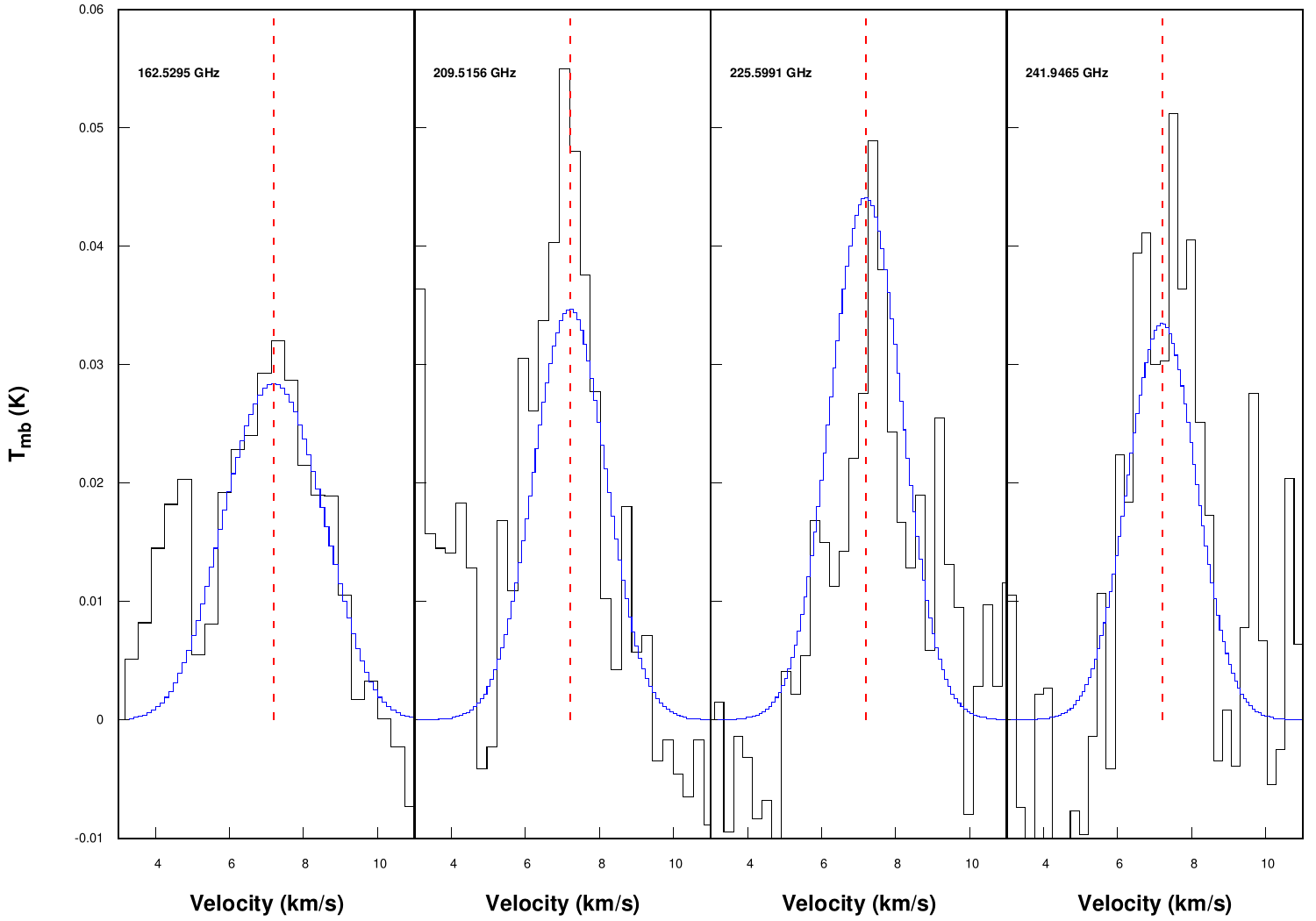}
\end{minipage}
\caption{MCMC fitting of the observed transitions of $\rm{CH_3OCH_3}$ in L1544 (left), B1-b (right) and IRAS4A (bottom). Purple lines represent the modeled spectral profile to the observed spectra (black).}
\label{fig:ch3och3_mcmc}
\end{figure*}

\begin{figure*}
\includegraphics[width=11cm, height=15cm, angle=270]{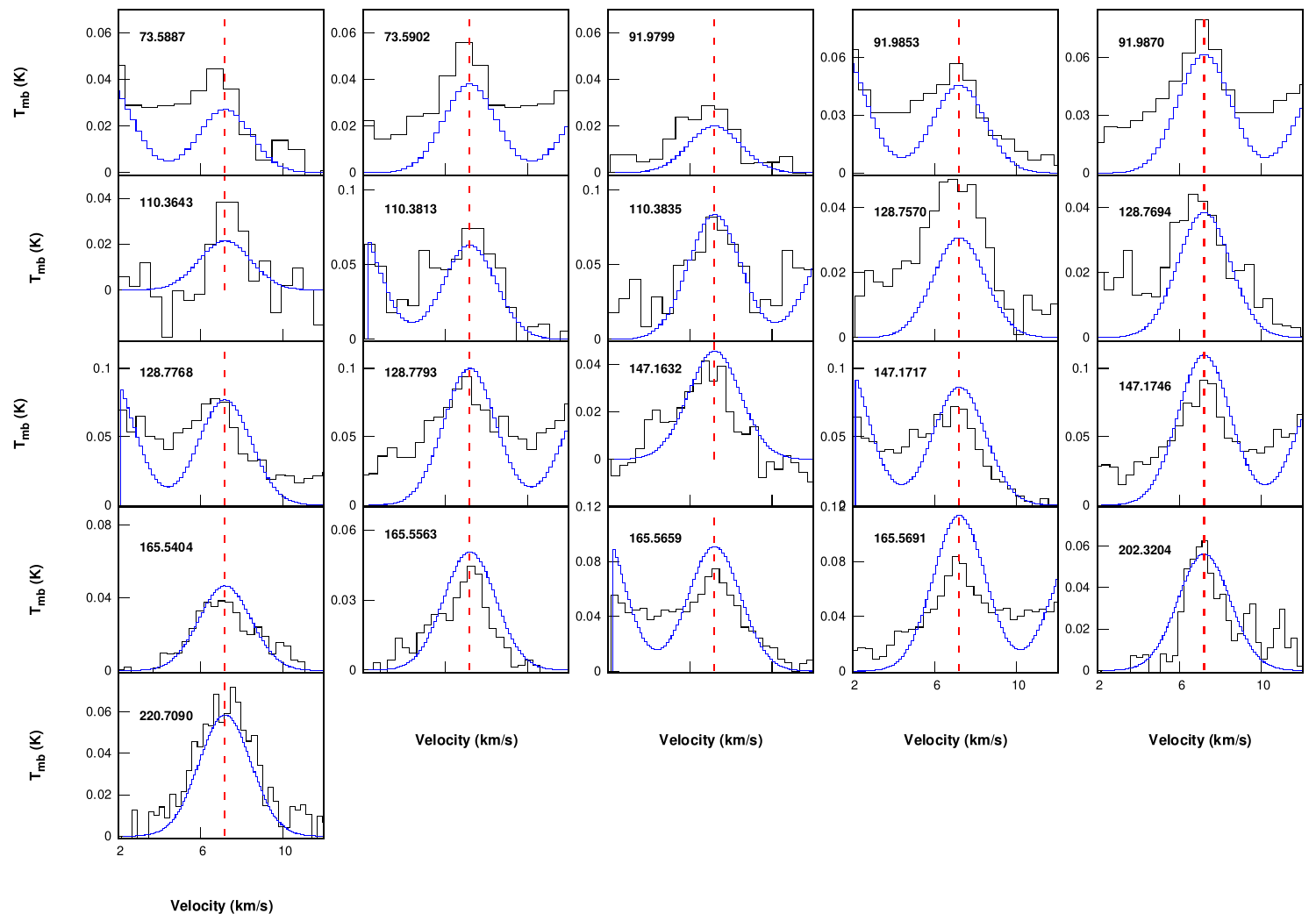}
\caption{Same as Figure \ref{fig:ch3oh_l1544mcmc} for $\rm{CH_3CN}$ in IRAS4A.}
\label{fig:ch3cn_mcmc1}
\end{figure*}

\begin{figure*}
\includegraphics[width=11cm, height=15cm, angle=270]{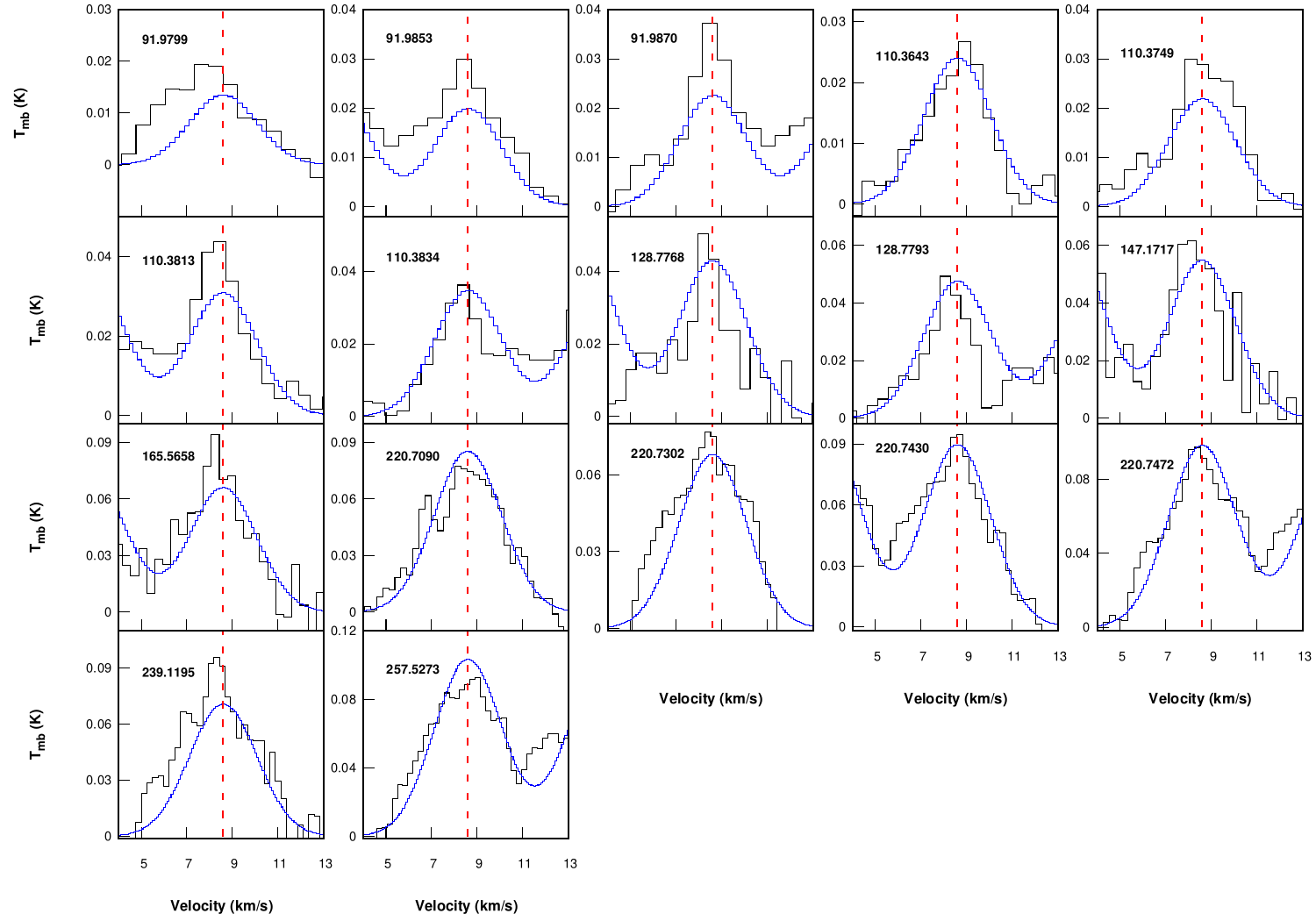}
\caption{Same as Figure \ref{fig:ch3oh_l1544mcmc} for $\rm{CH_3CN}$ in SVS13A.}
\label{fig:ch3cn_mcmc2}
\end{figure*}

Details of this method are described in chapter \ref{chap:mcmc_all}. Here, only those species for whom several transitions are detected have their observed transitions fit by the MCMC approach. Here, we used the MCMC approach from the CASSIS Software interface to match the observed lines of those seven COMs ($\rm{CH_3OH, CH_3CHO, CH_3OCHO,}$ $\rm{C_2H_5OH, HCCCHO, CH_3OCH_3, CH_3CN}$), which we have taken into consideration for various star-forming areas. We neglected the beam filling factor for the MCMC computation because the rotational diagram analysis provided us with a beam average column density. We use a two-component MCMC fitting in certain cases (CH$_3$OH, CH$_3$CHO and CH$_3$CN in IRAS4A) to have a better match with the observation. The components are considered based on their temperature. Section \ref{sec:results} describes the details. For L1544, B1-b, IRAS4A, L1157-mm, and SVS13A, the LSR velocity was maintained at 7.2, 6.5, 7.2, 2.6, and 8.4 km/s, respectively. We changed the FWHM, column density, and excitation temperature (T$_{ex}$) for the fitting and used the MCMC approach to get the best fit parameters from the LTE fitting. In Table \ref{table:mcmc_lte}, all the physical parameters are listed along with the best-fitted values. Spectral fit for all the transitions arose from various molecules are shown in Figure \ref{fig:ch3oh_l1544mcmc}, \ref{fig:ch3oh_barnardmcmc}, \ref{fig:ch3oh_irasmcmc}, \ref{fig:ch3oh_svsmcmc}, \ref{fig:ch3cho_mcmc}, \ref{fig:ch3ocho_mcmc}, \ref{fig:c2h5oh_mcmc}, \ref{fig:hcccho_mcmc}, and \ref{fig:ch3och3_mcmc}, \ref{fig:ch3cn_mcmc1}, \ref{fig:ch3cn_mcmc2}.
\begin{landscape}
\begin{table}
\tiny{
\caption{Summary of the best fitted line parameters obtained by using MCMC method. \label{table:mcmc_lte}}
\begin{tabular}{|c|c|c|c|c|c|c|c|c|c|c|}
\hline
\hline
Species&Source&Frequency&FWHM range&Best fitted&Column density&Best fitted column&T$_{ex}$ range&Best fitted&Best fitted&Optical depth \\
&&(GHz)&used (km.s$^{-1}$)&FWHM (Km.s$^{-1}$)&range used (cm$^{-2}$)&density (cm$^{-2}$) &used (K)& T$_{ex}$ (K)&V$_{LSR}$ (km.s$^{-1}$)& ($\tau$)\\
\hline
\hline
CH$_3$OH&L1544&84.5211&&&&&&&&$1.976\times10^{-3}$\\
&&96.7445&0.3-0.5&0.39$\pm$0.03&$1.0\times10^{11}-1.0\times10^{14}$&(4.6$\pm$0.92)$\times10^{12}$&3.5-17.0&12.9$\pm$3.07&7.2&$5.860\times10^{-3}$\\
&&96.7555&&&&&&&&$2.442\times10^{-3}$\\
&&97.5828&&&&&&&&$3.966\times10^{-3}$\\
\cline{2-11}
&B1-b&96.7393&&&&&&&&$2.084\times10^{-1}$\\
&&96.7413&&&&&&&&$5.978\times10^{-1}$\\
 &&96.7445&1.0-3.0&1.50$\pm$0.10&$1.0\times10^{13}-1.0\times10^{15}$&$(1.6\pm0.12)\times10^{14}$&5.0-25.0&7.27$\pm$0.86&6.5&$9.834\times10^{-2}$\\
&&96.7555&&&&&&&&$2.548\times10^{-2}$\\
&&108.8939&&&&&&&&$1.826\times10^{-1}$\\
\cline{3-11}
&&145.0937&&&&&&&&$5.113\times10^{-2}$\\
&&145.1031&&&&&&&&$2.179\times10^{-1}$\\
&&157.2708&&&&&&&&$1.248\times10^{-1}$\\
&&157.2760&1.0-3.0&1.05$\pm$0.08&$1.0\times10^{13}-1.0\times10^{15}$&$(5.9\pm0.51)\times10^{13}$&5.0-25.0&9.06$\pm$1.26&6.5&$1.232\times10^{-1}$\\
&&165.0501&&&&&&&&$5.152\times10^{-2}$\\
&&165.0611&&&&&&&&$5.128\times10^{-2}$\\
&&165.0992&&&&&&&&$3.312\times10^{-2}$\\
&&170.0605&&&&&&&&$3.038\times10^{-2}$\\
\cline{3-11}
&&213.4270&&&&&&&&$1.466\times10^{-1}$\\
&&254.0153&1.0-3.0&1.0$\pm$0.09&$1.0\times10^{13}-1.0\times10^{15}$&$(1.2\pm0.32)\times10^{14}$&5.0-25.0&6.77$\pm$0.95&6.5&$1.894\times10^{-1}$\\
&&261.8056&&&&&&&&$1.681\times10^{-1}$\\
\cline{2-11}
&IRAS4A&96.7555&&&&&&&&$1.140\times10^{-2}$/$2.615\times10^{-3}$\\
&&108.8939&&&&&&&&$1.517\times10^{-1}$/$3.565\times10^{-3}$\\
&&143.8657&&&&&&&&$3.503\times10^{-2}$/$6.904\times10^{-3}$\\
&&155.3208&&&&&&&&$2.885\times10^{-10}$/$1.066\times10^{-3}$\\
&&155.9975&&&&&&&&$1.726\times10^{-8}$/$1.990\times10^{-3}$\\
&&156.4889&&&&&&&&$6.711\times10^{-7}$/$3.414\times10^{-3}$\\
&&156.8285&&&&&&&&$1.692\times10^{-5}$/$5.364\times10^{-3}$\\
&&157.0486&&&&&&&&$2.754\times10^{-4}$/$7.695\times10^{-3}$\\
&&157.1789&&&&&&&&$2.878\times10^{-3}$/$1.002\times10^{-2}$\\
&&165.0501&&&&$\bf{Component 1}$&&&&$6.794\times10^{-2}$/$5.780\times10^{-3}$\\
&&165.0611&1.0-3.5&3.03$\pm$0.27&$1.0\times10^{13}-1.0\times10^{15}$&$(1.8\pm0.29)\times10^{14}$&5.0-30.0&5.62$\pm$1.68&7.2&$4.939\times10^{-2}$/$8.416\times10^{-3}$\\
&&165.0992&&&&&&&&$1.991\times10^{-2}$/$9.619\times10^{-3}$\\
&&165.6786&&&&$\bf{Component 2}$&&&&$7.341\times10^{-5}$/$6.483\times10^{-3}$\\
&&166.1690&1.0-3.5&3.50$\pm$0.47&$1.0\times10^{13}-1.0\times10^{15}$&$(2.0\pm0.28)\times10^{14}$&30.0-70.0&35.14$\pm$3.86&7.2&$4.651\times10^{-6}$/$4.663\times10^{-3}$\\
&&213.4270&&&&&&&&$7.562\times10^{-2}$/$5.133\times10^{-3}$\\
&&230.0270&&&&&&&&$3.915\times10^{-3}$/$2.880\times10^{-3}$\\
&&241.8790&&&&&&&&$1.371\times10^{-3}$/$1.049\times10^{-2}$\\
&&251.7384&&&&&&&&$9.900\times10^{-7}$/$4.282\times10^{-3}$\\
&&251.8665&&&&&&&&$5.291\times10^{-5}$/$5.011\times10^{-3}$\\
&&251.9170&&&&&&&&$1.536\times10^{-4}$/$3.626\times10^{-3}$\\
&&251.9848&&&&&&&&$4.592\times10^{-9}$/$2.331\times10^{-3}$\\
&&252.0904&&&&&&&&$1.356\times10^{-10}$/$1.485\times10^{-3}$\\
&&254.0153&&&&&&&&$1.130\times10^{-1}$/$3.853\times10^{-3}$\\
&&261.8056&&&&&&&&$7.953\times10^{-2}$/$8.529\times10^{-3}$\\
\hline
\end{tabular}}
\end{table}
\end{landscape}
\begin{landscape}
{\tiny
\centering
\begin{tabular}{|c|c|c|c|c|c|c|c|c|c|c|}
\hline
\hline
Species&Source&Frequency&FWHM range&Best fitted&Column density&Best fitted column&T$_{ex}$ range&Best fitted&Best fitted&Optical depth\\
&&(GHz)&used (km.s$^{-1}$)&FWHM (Km.s$^{-1}$)&range used (cm$^{-2}$)&density (cm$^{-2}$) &used (K)& T$_{ex}$ (K)&V$_{LSR}$ (km.s$^{-1}$)& ($\tau$)\\
\hline
\hline
&SVS13A&85.5681&&&&&&&&$3.080\times10^{-4}$\\
&&96.7555&2.0-4.5&4.04$\pm$0.59&$1.0\times10^{13}-1.0\times10^{15}$&$(3.0\pm2.42)\times10^{14}$&40.0-200.0&96.72$\pm$18.77&8.6&$3.508\times10^{-4}$\\
&&111.2894&&&&&&&&$3.651\times10^{-4}$\\
\cline{3-11}
&&143.8657&&&&&&&&$1.007\times10^{-3}$\\
&&156.6023&2.0-3.5&3.47$\pm$0.39&$1.0\times10^{13}-1.0\times10^{15}$&$(2.8\pm0.62)\times10^{14}$&40.0-100.0&96.70$\pm$14.03&8.6&1.093$\times10^{-3}$\\
\cline{3-11}
&&218.4400&&&&&&&&$2.374\times10^{-3}$\\
&&229.7587&&&&&&&&$1.929\times10^{-3}$\\
&&241.7001&&&&&&&&$2.970\times10^{-3}$\\
&&241.7913&&&&&&&&$3.598\times10^{-3}$\\
&&241.8790&2.0-3.5&3.36$\pm$0.24&$1.0\times10^{13}-1.0\times10^{15}$&$(1.5\pm0.17)\times10^{14}$&40.0-100.0&68.69$\pm$8.41&8.6&$2.609\times10^{-3}$\\
&&243.9157&&&&&&&&$2.813\times10^{-3}$\\
&&251.7384&&&&&&&&$1.920\times10^{-3}$\\
&&261.8056&&&&&&&&$1.430\times10^{-3}$\\
&&266.8381&&&&&&&&$2.758\times10^{-3}$\\
\hline
CH$_3$CHO&L1544&93.5809&&&&&&&&$1.324\times10^{-2}$\\
&&93.5952&&&&&&&&$1.308\times10^{-2}$\\
 &&95.9474&0.3-0.6  &0.39$\pm$0.09&$1.0\times10^{11}-1.0\times10^{13}$&$(6.1\pm3.55)\times10^{11}$&3.5-20.0&6.04$\pm$4.42&7.2&$1.930\times10^{-2}$\\
&&95.9634&&&&&&&&$1.960\times10^{-2}$\\
&&98.8633&&&&&&&&$1.246\times10^{-2}$\\
&&98.9009&&&&&&&&$1.262\times10^{-1}$\\
\cline{2-11}
&B1-b&93.5809&&&&&&&&$1.654\times10^{-2}$\\
&&93.5952&&&&&&&&$1.638\times10^{-2}$\\
&&95.9634&&&&&&&&$2.295\times10^{-2}$\\
&&96.4256&0.5-2.0&1.49$\pm$0.36&$1.0\times10^{11}-5.0\times10^{13}$&$(4.7\pm2.39)\times10^{12}$&6.0-100.0&7.52$\pm$6.61&6.5&$5.777\times10^{-3}$\\
&&96.4755&&&&&&&&$5.695\times10^{-3}$\\
&&98.8633&&&&&&&&$1.591\times10^{-2}$\\
&&98.9009&&&&&&&&$1.608\times10^{-2}$\\
\cline{3-11}
&&138.2849&0.5-1.5&1.38$\pm$0.24&$1.0\times10^{11}-5.0\times10^{13}$&$(4.1\pm2.84)\times10^{12}$&6.0-100.0&10.81$\pm$3.63&6.5&$5.076\times10^{-3}$\\
&&138.3196&&&&&&&&$5.109\times10^{-3}$\\
&&152.6352&&&&&&&&$4.587\times10^{-3}$\\
&&155.1796&&&&&&&&$1.813\times10^{-3}$\\
\cline{2-11}
&IRAS4A&74.8917&&&&&&&&$8.668\times10^{-3}$/$1.886\times10^{-4}$\\
&&74.9241&&&&&&&&$8.599\times10^{-3}$/$1.882\times10^{-4}$\\
&&76.8789&&&&&&&&$1.144\times10^{-2}$/$2.125\times10^{-4}$\\
&&77.0386&&&&&&&&$3.789\times10^{-3}$/$1.407\times10^{-4}$\\
&&77.2183&&&&&&&&$3.795\times10^{-3}$/$1.410\times10^{-4}$\\
&&79.0993&&&&&&&&$8.749\times10^{-3}$/$1.975\times10^{-4}$\\
&&93.5809&&&&&&&&$9.643\times10^{-3}$/$2.850\times10^{-4}$\\
&&93.5952&&&&&&&&$9.580\times10^{-3}$/$2.848\times10^{-4}$\\
&&95.9474&&&&&&&&$1.220\times10^{-2}$/$3.125\times10^{-4}$\\
&&96.2742&&&&&&&&$4.564\times10^{-3}$/$2.322\times10^{-4}$\\
&&96.4256&&&&&&&&$4.556\times10^{-3}$/$2.314\times10^{-4}$\\
&&96.4755&&&&&&&&$4.513\times10^{-3}$/$2.312\times10^{-4}$\\
&&98.8633&&&&&&&&$9.556\times10^{-3}$/$2.981\times10^{-4}$\\
&&112.248&&&&&&&&$9.023\times10^{-3}$/$3.877\times10^{-4}$\\
&&112.2545&&&&&&&&$8.966\times10^{-3}$/$3.874\times10^{-4}$\\
&&133.8305&&&&&&&&$8.846\times10^{-3}$/$5.224\times10^{-4}$\\
\hline
\hline
\end{tabular}}
\end{landscape}
\begin{landscape}
{\tiny
\centering
\begin{tabular}{|c|c|c|c|c|c|c|c|c|c|c|}
\hline
\hline
Species&Source&Frequency&FWHM range&Best fitted&Column density&Best fitted column&T$_{ex}$ range&Best fitted&Best fitted&Optical depth \\
&&(GHz)&used (km.s$^{-1}$)&FWHM (Km.s$^{-1}$)&range used (cm$^{-2}$)&density (cm$^{-2}$) &used (K)& T$_{ex}$ (K)&V$_{LSR}$ (km.s$^{-1}$)& ($\tau$)\\
\hline
\hline
&&138.2849&&&&&&&&$6.880\times10^{-3}$/$5.079\times10^{-4}$\\
&&138.3196&&&&\bf{Component 1}&&&&$6.924\times10^{-3}$/$5.083\times10^{-4}$\\
&&152.6352&2.0-3.5&3.30$\pm$0.22&$1.0\times10^{12}-1.0\times10^{14}$&$(1.3\pm0.21)\times10^{13}$&8.0-30.0&11.07$\pm$3.48&7.2&$6.268\times10^{-3}$/$6.184\times10^{-4}$\\
&&155.3421&&&&&&&&$2.579\times10^{-3}$/$5.186\times10^{-4}$\\
&&168.0934&2.0-3.5&2.3$\pm$0.32&$1.0\times10^{12}-1.0\times10^{14}$&$(1.1\pm0.21)\times10^{13}$&30.0-80.0&71.2$\pm$5.17&7.2&$3.362\times10^{-3}$/$6.662\times10^{-4}$\\
&&208.2285&&&&&&&&$1.131\times10^{-3}$/$8.041\times10^{-4}$\\
&&211.2738&&&&&&&&$4.749\times10^{-4}$/$6.921\times10^{-4}$\\
&&212.2571&&&&&&&&$1.628\times10^{-4}$/$5.677\times10^{-4}$\\
&&214.8450&&&&&&&&$4.634\times10^{-4}$/$7.000\times10^{-4}$\\
&&216.6302&&&&&&&&$8.097\times10^{-4}$/$7.846\times10^{-4}$\\
&&223.6601&&&&&&&&$4.767\times10^{-4}$/$7.997\times10^{-4}$\\
&&242.1060&&&&&&&&$2.039\times10^{-4}$/$8.021\times10^{-4}$\\
&&250.8291&&&&&&&&$7.313\times10^{-6}$/$4.552\times10^{-4}$\\
&&250.9345&&&&&&&&$3.163\times10^{-5}$/$5.938\times10^{-4}$\\
&&251.4893&&&&&&&&$3.155\times10^{-5}$/$5.947\times10^{-4}$\\
&&254.8271&&&&&&&&$8.665\times10^{-5}$/$7.221\times10^{-4}$\\
&&255.3269&&&&&&&&$1.472\times10^{-4}$/$7.966\times10^{-4}$\\
&&262.9601&&&&&&&&$8.633\times10^{-5}$/$8.023\times10^{-4}$\\
\cline{2-11}
&&205.1707&&&&&&&&$9.800\times10^{-4}$\\
&SVS13A&211.2430&&&&&&&&$8.176\times10^{-4}$\\
&&211.2738&&&&&&&&$8.167\times10^{-4}$\\
&&216.5819&0.5-2.5&2.45$\pm$0.33&$1.0\times10^{12}-1.0\times10^{14}$&$(7.2\pm4.89)\times10^{12}$&30.0-200.0&45.16$\pm$11.4&8.6&$9.658\times10^{-4}$\\
&&230.3019&&&&&&&&$7.735\times10^{-4}$\\
&&242.1060&&&&&&&&$8.513\times10^{-4}$\\
&&251.4893&&&&&&&&$5.344\times10^{-4}$\\
\hline
CH$_3$OCHO&B1-b&88.8516&&&&&&&&$1.475\times10^{-3}$\\
&&90.1457&&&&&&&&$1.296\times10^{-3}$\\
&&90.1564&1.0-2.0&1.64$\pm$0.21&$1.0\times10^{11}-1.0\times10^{14}$&$(1.2\pm1.0)\times10^{13}$&15.0-100.0&18.52$\pm$22.81&6.5&$1.298\times10^{-3}$\\
&&100.2946&&&&&&&&$1.026\times10^{-3}$\\
&&100.4822&&&&&&&&$1.492\times10^{-3}$\\
&&103.4786&&&&&&&&$1.354\times10^{-3}$\\
\cline{2-11}
&IRAS4A&129.2963$^{*}$&&&&&&&&$3.921\times10^{-4}$\\
&&132.9287&&&&&&&&$4.241\times10^{-4}$\\
&&135.9219&&&&&&&&$2.900\times10^{-4}$\\
&&141.0443&1.0-3.0&1.86$\pm$0.54&$1.0\times10^{12}-1.0\times10^{14}$&$(3.9\pm1.54)\times10^{13}$&30.0-130.0&73.87$\pm$24.37&7.2&$4.453\times10^{-4}$\\
&&158.6937&&&&&&&&$4.529\times10^{-4}$\\
&&158.7043&&&&&&&&$4.530\times10^{-4}$\\
\cline{3-11}
&&200.9563&&&&&&&&$4.213\times10^{-4}$\\
&&206.6194&&&&&&&&$5.126\times10^{-4}$\\
&&216.2165&&&&&&&&$5.060\times10^{-4}$\\
&&228.6288&1.0-3.0&1.35$\pm$0.50&$1.0\times10^{12}-1.0\times10^{14}$&$(3.4\pm1.72)\times10^{13}$&30.0-130.0&81.32$\pm$18.46&7.2&$4.282\times10^{-4}$\\
&&240.0211&&&&&&&&$4.750\times10^{-4}$\\
&&247.0441&&&&&&&&$4.384\times10^{-4}$\\
&&249.5781&&&&&&&&$3.567\times10^{-4}$\\
\cline{2-11}
&SVS13A&100.4906&1.0-3.0&1.95$\pm$0.53&$1.0\times10^{12}-1.0\times10^{15}$&$(6.4\pm5.17)\times10^{13}$&40.0-200.0&86.94$\pm$33.36&8.6&$3.041\times10^{-4}$\\
&&164.2059&&&&&&&&$4.726\times10^{-4}$\\
\cline{3-11}
&&210.4632&&&&&&&&$2.811\times10^{-4}$\\
&&218.2809&1.0-3.0&2.10$\pm$0.37&$1.0\times10^{12}-1.0\times10^{15}$&$(7.8\pm1.88)\times10^{13}$&40.0-200.0&113.84$\pm$23.41&8.6&$4.153\times10^{-4}$\\
&&222.4214&&&&&&&&$2.545\times10^{-4}$\\
&&269.0780&&&&&&&&$4.058\times10^{-4}$\\
\hline
\hline
\end{tabular}}
\end{landscape}
\clearpage
\begin{landscape}
{\tiny
\begin{tabular}{|c|c|c|c|c|c|c|c|c|c|c|}
\hline
\hline
Species&Source&Frequency&FWHM range&Best fitted&Column density&Best fitted column&T$_{ex}$ range&Best fitted&Best fitted&Optical depth \\
&&(GHz)&used (km.s$^{-1}$)&FWHM (Km.s$^{-1}$)&range used (cm$^{-2}$)&density (cm$^{-2}$) &used (K)& T$_{ex}$ (K)&V$_{LSR}$ (km.s$^{-1}$)& ($\tau$)\\
\hline
\hline
$\rm{C_2H_5OH}$&IRAS4A&129.6657&&&&&&&&$2.734\times10^{-4}$\\
&&133.3234&0.5-2.5&2.05$\pm$0.44&$1.0\times10^{12}-1.0\times10^{14}$&$(3.5\pm1.87)\times10^{13}$&30.0-130.0&59.87$\pm$18.99&7.2&$5.675\times10^{-4}$\\
&&148.3040&&&&&&&&$2.648\times10^{-4}$\\
&&159.4140&&&&&&&&$2.793\times10^{-4}$\\
\cline{3-11}
&&205.4584&&&&&&&&$3.912\times10^{-4}$\\
&&209.8652&0.5 - 2.0&1.91$\pm$0.45&$1.0\times10^{12}-1.0\times10^{14}$&$(2.5\pm1.46)\times10^{13}$&30.0 - 130.0&80.12$\pm$30.81&7.2&$1.582\times10^{-4}$\\
&&227.8919&&&&&&&&$1.864\times10^{-4}$\\
&&230.9913&&&&&&&&$4.115\times10^{-4}$\\
\cline{2-11}
&SVS13A&84.5958&&&&&&&&$4.018\times10^{-4}$\\
&&130.2463&0.5 - 2.0&0.92$\pm$0.41&$1.0\times10^{12}-1.0\times10^{14}$&$(1.7\pm1.59)\times10^{13}$&10.0-100.0&45.63$\pm$17.28&8.6&$4.642\times10^{-4}$\\
&&153.4842&&&&&&&&$5.237\times10^{-4}$\\
\cline{3-11}
&&205.4584&&&&&&&&$8.256\times10^{-4}$\\
&&244.6339&0.5-2.0&0.97$\pm$0.28&$1.0\times10^{12}-1.0\times10^{14}$&$(1.5\pm0.86)\times10^{13}$&10.0-100.0&59.58$\pm$21.85&8.6&$2.775\times10^{-4}$\\
&&270.4441&&&&&&&&$7.212\times10^{-4}$\\
\hline
$\rm{HCCCHO}$&L1544&83.7758&&&&&&&&$1.425\times10^{-3}$\\
&&99.0391&0.3-0.6&0.43$\pm$0.07&$1.0\times10^{11}-1.0\times10^{13}$&$(4.0\pm1.33)\times10^{11}$&3.5-20.0&17.40$\pm$5.24&7.2&$6.276\times10^{-5}$\\
&&102.2980&&&&&&&&$1.275\times10^{-3}$\\
\hline
$\rm{CH_3OCH_3}$&L1544&99.324362&&&&&&&&$3.657\times10^{-4}$\\
&&99.324364&0.1-1.0&$0.41\pm$0.22&$1.0\times10^{11}-1.0\times10^{13}$&$(2.2\pm1.61)\times10^{12}$&3.5-20.0&$15.47\pm2.71$&7.2&$5.486\times10^{-4}$\\
&&99.325217&&&&&&&&$1.463\times10^{-3}$\\
&&99.326072&&&&&&&&$9.145\times10^{-4}$\\
\cline{2-11}
&B1-b&99.324362&&&&&&&&$1.486\times10^{-3}$\\
&&99.324364&0.5-1.0&$0.9\pm0.12$&$6.0\times10^{11}-6.0\times10^{13}$&$(8.5\pm5.52)\times10^{12}$&5.0-25.0&$9.85\pm4.56$&7.2&$2.230\times10^{-3}$\\
&&99.325217&&&&&&&&$5.947\times10^{-3}$\\
&&99.326072&&&&&&&&$3.717\times10^{-3}$\\
\cline{2-11}
\cline{3-11}
&IRAS4A&162.5295&&&&&&&&$5.380\times10^{-4}$\\
&&209.5156&&&&&&&&$6.024\times10^{-4}$\\
&&225.5991&1.0-3.0&2.11$\pm$0.35&$1.0\times10^{12}-1.0\times10^{14}$&$(2.1\pm1.03)\times10^{13}$&30.0-130.0&45.28$\pm$28.29&7.2&$5.828\times10^{-4}$\\
&&241.9465&&&&&&&&$5.454\times10^{-4}$\\
\hline
$\rm{CH_3CN}$&IRAS4A&73.588799&&&&&&&&$1.97\times10^{-3}$\\
 &&73.590218&&&&&&&&$2.96\times10^{-3}$\\
 &&91.979994&&&&&&&&$8.26\times10^{-4}$\\
 &&91.985314&&&&&&&&$2.61\times10^{-3}$\\
 &&91.987087&&&&&&&&$3.83\times10^{-3}$\\
 &&110.364353&&&&&&&&$3.08\times10^{-4}$\\
 &&110.381372&&&&&&&&$3.03\times10^{-3}$\\
 &&110.383499&&&&&&&&$4.38\times10^{-3}$\\
 &&128.757030&&&&&&&&$3.48\times10^{-4}$\\
 &&128.769436&&&&&&&&$1.07\times10^{-3}$\\
 &&128.776881&&&&component1&&&&$3.16\times10^{-3}$\\
&&128.779363&1.5-3.0&$2.9\pm0.19$&$1.0\times10^{11}-1.0\times10^{13}$&$(9.9\pm3.15)\times10^{11}$&20.0-50.0&$21.02\pm3.24$&7.2&$4.54\times10^{-3}$\\
 &&147.163244&&&&component2&&&&$1.04\times10^{-3}$\\
 &&147.171751&1.5-3.0&$2.9\pm0.12$&$1.0\times10^{11}-1.0\times10^{13}$&$(1.7\pm0.38)\times10^{12}$&50.0-80.0&$70.11\pm5.32$&7.2&$3.04\times10^{-3}$\\
 &&147.174588&&&&&&&&$4.34\times10^{-3}$\\
 &&165.540377&&&&&&&&$3.21\times10^{-4}$\\
 &&165.556321&&&&&&&&$9.38\times10^{-4}$\\
 &&165.565891&&&&&&&&$2.70\times10^{-3}$\\
 &&165.569081&&&&&&&&$3.84\times10^{-3}$\\
 &&202.320442&&&&&&&&$2.16\times10^{-4}$\\
 &&220.709016&&&&&&&&$1.61\times10^{-4}$\\
 \hline
\end{tabular}}
\clearpage
{\tiny
\centering
\begin{tabular}{|c|c|c|c|c|c|c|c|c|c|c|}
\hline
\hline
Species&Source&Frequency&FWHM range&Best fitted&Column density&Best fitted column&T$_{ex}$ range&Best fitted&Best fitted&Optical depth \\
&&(GHz)&used (km.s$^{-1}$)&FWHM (Km.s$^{-1}$)&range used (cm$^{-2}$)&density (cm$^{-2}$) &used (K)& T$_{ex}$ (K)&V$_{LSR}$ (km.s$^{-1}$)& ($\tau$)\\
\hline
&SVS13A&91.979994&&&&&&&&$2.705\times10^{-4}$\\
&&91.985314&&&&&&&&$3.996\times10^{-4}$\\
&&91.987087&&&&&&&&$4.535\times10^{-4}$\\
&&110.364353&&&&&&&&$4.281\times10^{-4}$\\
&&110.374989&&&&&&&&$3.890\times10^{-4}$\\
&&110.381372&&&&&&&&$5.499\times10^{-4}$\\
&&110.383499&&&&&&&&$6.162\times10^{-4}$\\
&&128.776881&1.5-3.5&$3.48\pm0.09$&$5.0\times10^{11}-5.0\times10^{13}$&$(2.8\pm0.57)\times10^{12}$&40.0-200.0&$83.55\pm19.34$&8.6&$7.041\times10^{-4}$\\
&&128.779363&&&&&&&&$7.832\times10^{-4}$\\
&&147.171751&&&&&&&&$8.538\times10^{-4}$\\
&&165.565891&&&&&&&&$9.913\times10^{-4}$\\
&&220.709016&&&&&&&&$1.210\times10^{-3}$\\
&&220.730260&&&&&&&&$9.624\times10^{-4}$\\
&&220.743010&&&&&&&&$1.270\times10^{-3}$\\
&&220.747261&&&&&&&&$1.394\times10^{-3}$\\
&&239.119504&&&&&&&&$9.939\times10^{-4}$\\
&&257.527383&&&&&&&&$1.441\times10^{-3}$\\
\hline
\hline
\end{tabular}}
\end{landscape}
\clearpage

\subsection{Observed species}
\subsubsection{CH$_3$OH}
Analysis of methanol has become extremely important since it is a key chemical component of different phases of star formation. From the ASAI data, \cite{vast14} previously detected CH$_3$OH emission in L1544 and calculated a column density of $3 \times 10^{13}$ cm$^{-2}$ using non-LTE approximation. They concluded that the methanol lines most likely arose from $\sim$ 8000 au, where the temperature is $\sim$ 10 K. Using the IRAM 30m telescope, methanol was previously found in B1-b with a column density of $2.5 \times 10^{14}$ cm$^{-2}$. \cite{mare05} obtained a column density of $5.1 \times 10^{14}$ cm$^{-2}$ in IRAS4A.
In the four sample sources, we have found many transitions of CH$_3$OH (see Table \ref{tab:observation_1}). The obtained column densities and excitation temperatures are noted in Table \ref{tab:rotdiag}, and the same is true for MCMC fitting, which is recorded in Table \ref{table:mcmc_lte}. We have used single component MCMC fitting to explore the CH$_3$OH transitions discovered in L1544, B1-b, and SVS13A.
We found that a two-component fit is necessary for the rotational diagram of methanol in the case of the class 0 protostar IRAS4A: (1) having E$_{up}$ $>$ 40 K (hot component), and (2) having E$_{up}$ $<$ 40 K (cool component). In the MCMC method, only when we are taking into account two components we obtain an appropriate match. Red bars in the left (RD) and right (MCMC) panels of Fig. \ref{fig:clmdensity} reflect the chemical evolution of CH$_3$OH with respect to the evolutionary stages of star formation. It reveals that the methanol abundance gradually rises, reaching a maximum in the class 0 phase (IRAS4A), from a minimum value at the prestellar core phase (L1544).
It should be noted that in Fig. \ref{fig:clmdensity}, only the abundance of the cold component of methanol was taken into account in the instance of IRAS4A, despite the fact that the hot component almost has the same value. The gradual rise in temperature may be the reason for the increase in abundance. Since consecutive hydrogen additions to CO can produce methanol at lower temperatures (10–20 K), methanol production during the prestellar core phase may be feasible. 
$$
\rm{CO \rightarrow HCO\rightarrow H_2CO \rightarrow CH_2OH/CH_3O\rightarrow CH_3OH} 
$$
However, the thermal desorption is not effective enough at this point to transfer the methanol content in the ice phase to the gas phase \citep[having a binding energy of 5264 K]{das18}. The observed gas-phase abundance of methanol at this stage is primarily due to the non-thermal desorptions \citep{ober09,garr07}. In addition, the cloud transforms into a protostar phase and, finally, a hydrostatic core. Methanol would also be produced in the protostar phase by the radical-radical surface reaction $\rm{CH_3+OH \rightarrow CH_3OH}$. The temperature rises as a result, gradually improving the probability of production and release by thermal desorption. For the class I object SVS13A, we observe methanol with less abundance. A decrease in the abundance of methanol for the class I object may occur due to the lower methanol formation rate in the class I phase than in the class 0 phase because of the competition between the reaction and thermal desorption of the reactants.

\subsubsection{CH$_3$CHO}
Acetaldehyde is an asymmetric top molecule that is abundant throughout the many evolutionary stages of star formation. In all of the chosen sources, we found several CH$_3$CHO transitions. Table \ref{tab:observation_2} provides a summary of the fitted line parameters for this species. The detected lines' upper state energies range from $9$ to $121$ K. \cite{jime16} found that the column density was $3.2 \times 10^{12}$ cm$^{-2}$ towards a low-density shell at 4000 au from the core center and $1.2 \times 10^{12}$ cm$^{-2}$ towards the dense, highly extinguished continuum peak within the inner 2700 au. However, our study show that the column density for acetaldehyde is $2.35 \times 10^{12}$ cm$^{-2}$. Using the MADEX code by \cite{cern12} in B1-b, a column density of $1.5 \times 10^{12}$ cm$^{-2}$ is obtained while assuming an excitation temperature of 10 K. Our study of the rotation diagram led us to determine the column density of this species to be $6.5 \times 10^{12}$ cm$^{-2}$. Using a single-dish telescope, \cite{hold19} and \cite{bian19} observed CH$_3$CHO in the direction of IRAS4A and SVS13A, respectively, and measured the column densities to be $2.6 \times 10^{12}$ cm$^{-2}$ and $1.2 \times 10^{16}$ cm$^{-2}$. We determine the column densities for IRAS4A and SVS13A to be $1.78 \times 10^{-13}$ cm$^{-2}$ and $1.1 \times 10^{-13}$ cm$^{-2}$, respectively, based on our analysis of the rotation diagram. Similar to the CH$_3$OH, the rotational diagram analysis for CH$_3$CHO shows two components: (1) a hot component with E$_{up}$ $>$ 50 K and (2) a cold component with E$_{up}$ $<$ 50 K (see Figure \ref{fig:rotational_diag_ch3cho}) in IRAS4A. From RD analysis, we determine a temperature of $22.1$ K for the low excitation lines and a temperature of $64.5$ K and the high excitation lines. A similar two-component MCMC fit is carried out, and the results show that components 1 and 2 have respective temperatures of $71.2$ K and $11.1$ K. It is seen that the results of the two procedures are quite comparable. The variation in its abundance is depicted by the violet bar lines in Figure \ref{fig:clmdensity}, and it exhibits the same behavior as $\rm{CH_3OH}$ throughout the stages of star formation.

\subsubsection{$\rm{CH_3OCHO}$}
Methyl formate is a simple asymmetric top complex organic molecule that is found in the majority of star-forming regions. Glycolaldehyde (HCOCH$_2$OH) and acetic acid (CH$_3$COOH), two of its isomers, are comparably less common. Due to the low temperature of the prestellar core, complex molecules are difficult to observe. Using IRAM 30 m, \cite{jime16} found that methyl formate was present in L1544 at two locations: the core's center and the region where methanol abundance is maximum. At these two positions, they obtained column densities of $(4.4 \pm 4.0) \times 10^{12}$ cm$^{-2}$ and $(2.3 \pm 1.4) \times 10^{12}$ cm$^{-2}$, respectively. \cite{cern12} saw this species towards B1-b, calculated a total column density of $3 \times 10^{12}$ cm$^{-2}$. Observing CH$_3$OCHO in the direction of the low-mass protostar IRAS4A, \cite{bott04} calculated column densities of $5.5 \times 10^{16}$ cm$^{-2}$ for A-CH$_3$OCHO and $5.8\times10^{16}$ cm$^{-2}$ for E-CH$_3$OCHO, respectively. \cite{bian19} found CH$_3$OCHO transitions towards SVS13A and measured a column density of $1.3 \times 10^{17}$ cm$^{-2}$. The presence of methyl formate towards prestellar core L1544 is not found in this study. But only one potential transition at 90.22765 GHz has been found in this source. The LTE approach is used to estimate the upper limit of column density, which is $3.7 \times 10^{12}$ cm$^{-2}$ (see Table \ref{tab:upper_limit}). Numerous transitions are found in B1-b, IRAS4A, and SVS13A; the RD analysis and the MCMC approach results are displayed in Tables \ref{tab:rotdiag} and \ref{table:mcmc_lte}, respectively. Figure \ref{fig:rotational_diag_hcooch3} displays the RD plot for CH$_3$OCHO, and the MCMC fitting is shown in Figure \ref{fig:ch3ocho_mcmc}. We have found that $\rm{CH_3OCHO}$ follow a similar pattern as methanol and acetaldehyde in its evolution from the prestellar core phase to the class I stage, as represented by an orange-colored bar line in Figure \ref{fig:clmdensity}.
No ice phase hydrogenation processes are directly involved in the formation of $\rm{CH_3OCHO}$, in contrast to methanol and acetaldehyde. Instead, the ice phase $\rm{CH_3OCHO}$ is formed by the radical-radical reaction between $\rm{CH_3O}$ and HCO:
$$
{\rm CH_3O + HCO \rightarrow CH_3OCHO}.
$$

The activities are constrained at low temperatures because of the high binding energies of these radicals, \citep[4400 K for CH$_3$O]{wake17}, \citep[2206 K for HCO]{das18}. 
It does, however, play an active role in the warmer area. In addition, the binding energy of $\rm{CH_3OCHO}$ (6295 K, \url{https://kida.astrochem-tools.org}) is considerably greater than that of acetaldehyde and methanol (\citep[5264 and 4573, respectively]{das18}). 
The sublimation rate is more favorable than the radical reaction rate and the subsequent gas phase destruction of CH$_3$OCHO resulting from desorption would be responsible for the drop in CH$_3$OCHO abundance at the class I stage.

\subsubsection{$\rm{C_2H_5OH}$}
The first observation of ethanol (C$_2$H$_5$OH) emission was made in 1975 towards Sagittarius B2. It was observed using the 11 m radio telescope of the National Radio Astronomy Observatory (NRAO, \cite{zuck75}), one of the revolutionary radio telescopes of the twentieth century. Both the low-mass and high-mass star-forming regions have been studied using it. Ethanol is composed of two distinct conformers, one of which is anti and the other is gauche, depending on how the OH group is oriented. A gauche$^+$ and gauche $^-$ state forms when the tunnelling between the two equivalent gauche conformers lifts the degeneracy. \cite{bian19} recently carried out ab initio quantum chemical calculations to precisely characterise the geometry and energy of ethanol conformers. Distinct C$_2$H$_5$OH lines are seen in L1544 in the present study. There are several distinct transitions of ethanol found in IRAS4A and SVS13A. There are three possible transitions of C$_2$H$_5$OH that are observed in the direction of B1-b. While the transition at 270.444085 GHz is blended, the transition at 135.989923 GHz is below 3$\sigma$. At 131.502781 GHz, only one unblended transition is found. As a result, we deduced an upper limit of the column density $\sim 1.0 \times 10^{13}$ cm$^{-2}$ and considered the presence of ethanol to be tentatively detected in this source (see Table \ref{tab:upper_limit}). Figure \ref{fig:clmdensity} demonstrates that the abundance is highest in IRAS4A. It would appear on grain surfaces as a result of a radical-radical interaction involving OH and $\rm{C_2H_5}$. According to \cite{das18}, the binding energies of OH and $\rm{C_2H}$ are 2081 K and 3781 K, respectively. The reduced mobility of these two radicals at low temperatures may be the cause of the absence of $\rm{C_2H_5OH}$ in the prestellar core. Additionally, ethanol has a high binding energy \citep[5400 K]{wake17} to release in the gas phase. Due to its lack of production by the radical-radical reaction at such a high temperature and subsequent destruction after its desorption, $\rm{C_2H_5OH}$ would have a relatively lower abundance in the class I stage.

\subsubsection{$\rm{HCCCHO}$}
Using the NRAO's 42.7 m radio telescope at Green Bank \citep{irvi88}, propynal ($\rm{HCCCHO}$) was found in the cold cloud TMC-1. \cite{huds19} presented the first infrared spectra of crystalline and amorphous propynal at various temperatures. This band's intensity and spectral position make it a prime candidate for the search for interstellar ice in astronomy. Here, we only count three HCCCHO transitions in L1544. \cite{jime16} conducted high-sensitivity single-pointing 3 mm observations towards the dust-continuum peak in L1544 using the IRAM 30 m telescope. At 83.775842 GHz, they only found one transition for HCCCHO. We also recognize two more transitions in addition to the same transition (9$_{0,9}$ - 8$_{0,8}$). According to our RD analysis, L1544 has a column density of $3.2 \times 10^{12}$ cm$^{-2}$. No significant transitions are observed in any other sources. For these sources, we estimate an upper limit of HCCCHO (see Table \ref{tab:upper_limit}). \cite{lois16} reported a single line of propynal in B1-b at 83.775832 GHz (9$_{0,9}$ - 8$_{0,8}$), with a column density of 7.9 $\times$ 10$^{11}$ cm$^{-2}$. However, we and \cite{marg20} did not detect this specific transition from the ASAI data. The 102.298 GHz transition used by \cite{marg20} to determine an upper limit of HCCCHO has been observed, but in our observation, we have seen that this transition is shifted and blended with s-Propanal. 
Despite being slightly shifted from the peak, the line at 93.043 GHz is not mixed with any other lines. As a result, we established an upper limit for column density ($\sim$ 2.56 $\times$ 10$^{12}$ cm$^{-2}$) in B1-b based on this transition. 

A green bar in Figure \ref{fig:clmdensity} depicts the variation in HCCCHO abundance during different stages of star formation. In the prestellar phase, we note that the abundance is relatively low. In the first hydrostatic core, B1-b, it practically remains the same. It rises slightly in the case of IRAS4A (class 0 phase).
On the other hand, SVS13A (class I) exhibits a higher abundance. 

A precise conclusion about the evolution of propynal would not be justified since the majority of the abundances presented here are based on the expected upper limit of propynal. However, in the ice phase, the following processes can lead to the production of HCCCHO:
$$
{\rm O + CH_2CCH \rightarrow HCCCHO + H}
$$
$$
{\rm C_2H + H_2CO \rightarrow HCCCHO + H}
$$
The reactants from \cite{das18} have binding energies of 770 K, 3238 K, 3315 K, and 3851 K for O, $\rm{CH_2CCH}$, C$_2$H, and H$_2$CO, respectively. Therefore, at low temperatures, HCCCHO production might be enhanced by the addition of oxygen, whereas in warmer regions, C$_2$H and H$_2$CO could be utilized. Warm chemistry would be effective, according to the fact that maximum column density is found at SVS13A.

\subsubsection{$\rm{CH_3OCH_3}$}

Dimethyl ether (DME) has two CH$_3$ groups and is an asymmetric top molecule that moves with a significant amplitude along CO-bond. Four sub-states, AA, EE, AE, and EA, result from splitting a rotational level by the two internal rotations. \citealt{snyd74} discovered DME in the Orion Nebula. \citealt{blak87} suggested a pathway for DME production in the gas phase. Its presence has been noted in the low-mass binary system \citep{caza03, kuan04}, and high-mass star-forming areas \citep{turn99, sutt95, numm00}. The fate of DME in diverse astrophysical environments was discussed in \cite{peet06} from the experiments, observations, and theoretical viewpoints. In L1544, we have only found one DME transition (3$_{1,3}$ - 2$_{0,2}$) with the four substates AA, AE, EA, and EE. The EE and AA sub-states are clearly resolved in our study, whereas the EA and AE overlap. Due to the identical upper-state energies of those sub-states, RD analysis is not possible. From LTE fitting, we get a column density of $1.6 \times 10^{12}$ cm$^{-2}$. In their investigation, \cite{jime16} also noted the same transition, and a column density of $1.5 \times 10^{12}$ cm$^{-2}$ is obtained. We have identified four sub-states similar to those found in L1544 in B1-b. Using LTE fitting, we estimate the column density in B1-b to be $6.0 \times 10^{12}$ cm$^{-2}$. Four unblended transitions, each of which has four sub-states but which overlap one another, have been identified in IRAS4A. The integrated intensity is calculated using a Gaussian fitting method and then divided according to their $S\mu^2$ values \citep{shim16}. For the rotation diagram analysis to determine the rotation temperature and column density, the transitions between these sub-states with the highest intensities are taken into account. Table \ref{tab:observation_3} only lists these transitions. We obtain a column density of $1.2 \times 10^{13}$ cm$^{-2}$ and a rotational temperature of $61.1$ K for this species in this source. As a result of the asymmetric line profiles of the reported DME transitions in SVS13A (transitions are not addressed here; see \citealt{bian19}), a Gaussian fit cannot be made for these transitions. The column density of DME for an asymmetric line profile has been determined by \cite{bian19} (see the technique presented by \citealt{bian19} in Appendix A1). They calculated the column density of this species to be $1.4 \times 10^{17}$ cm$^{-2}$ using a source size of $0.3^{''}$. We use their value after scaling by beam filling factor\footnote{$ff$ = $\frac{\theta_s^2}{\theta_s^2 + \theta_b^2}$} considering the source size $30^{''}$ and the obtained value is  $\sim 1.4 \times 10^{13}$ cm$^{-2}$. Additionally, we apply LTE fitting and achieve a good fit with a column density several times greater than the scaled value. In this study, we make use of the abovementioned scaled value. A magenta-coloured bar in Figure \ref{fig:clmdensity} depicts the variation in DME abundance during various stages of star formation. As with methanol, acetaldehyde, and methyl formate, we also noticed a growing trend for DME from the prestellar core phase to the class 0 phase. In contrast, this value decreased in class I objects. DME is expected to form either in the gas phase \citep{balu15} or the grain surface \citep{cupp17}. The ice phase production of DME required both the radical-radical and hydrogenation pathways:
$$
{\rm CH_3CHO+H\rightarrow CH_3OCH_2},
$$
$$
{\rm CH_3OCH_2+H\rightarrow CH_3OCH_3},
$$
$$
{\rm CH_3 + CH_3O \rightarrow CH_3OCH_3}.
$$
Dimethyl ether can be produced via hydrogenation or the radical-radical reaction. Its abundance steadily rises during the evolutionary phase up to the class 0 phase and then gradually falls in the class I phase, similar to methanol, acetaldehyde, and methyl formate.

\subsection{$\rm{CH_3CN}$}\label{sec:ch3cn}
Methyl cyanide or CH$_3$CN is a symmetric-top molecule which have a high dipole moment of $\sim$ 3.91 debyes. The K-ladders in the rotational level of CH$_3$CN can be excited only by collisional excitation. Hence CH$_3$CN can be a very good tracer to calculate the kinetic temperature of the molecular clouds. We have detected several transitions of CH$_3$CN in all the selected sources. Similar to the CH$_3$OH and CH$_3$CHO, for CH$_3$CN, a two components fit is required for the rotational diagram of methyl cyanide (see Figure. \ref{fig:rotational_diag_CH3CN}) in IRAS4A, (1) E$_{up}$ $>$ 50 K (hot component) and (2) E$_{up}$ $<$ 50 K (cold component). RD analysis yields a temperature of 61.2 K for the high excitation lines and 25.8 K for the low excitation lines. Similarly, a two-component MCMC fit is performed and yields a temperature of 70.1K and 21.0 K for hot component and cold component, respectively.\\
Taking the ratio between two transitions of the same frequency band can nullify various uncertainties obtained from observation. Here we used the line ratios of different CH$_3$CN transitions observed to calculate the kinetic temperature. Different K$_a$ ladders are connected by collisional excitation. The relative population of two K$_a$ ladders follows the Boltzmann equation at kinetic temperature. Considering the selection rules mentioned in \cite{mang93}, we calculated the line ratio between two transitions. Details of the selection rule and the method are described in \cite{das19,mond23}. Some selected line ratios are calculated ($\frac{J1_{k_{a}}-J2_{k_{a}}}{J3_{k_{a'}}-J4_{k_{a'}}}$). Using LTE approximation, the ratio (R) between two transitions satisfying the conditions mentioned in \cite{mang93} is,
R = S$_{R}$ exp(D/T$_{K}$), where $D = E(J3,k_{a'}) - E(J1, k_{a})$ and $S_{R} =\frac{S_{J1k_{a}}}{S_{J3k_{a'}}}$. 

Under the LTE approximation, as the kinetic temperature is considered equal to the excitation temperature, we calculated the kinetic temperatures for different CH$_3$CN k-ladder transitions for the sources (L1544, Barnard1 b, IRAS4A, SVS13A) using the above-mentioned formula. Calculated values of kinetic temperatures for all the CH$_3$CN transitions are mentioned in Table \ref{table:kladder}. In Figure \ref{fig:temp}, we plotted the average kinetic temperature obtained from this calculations for the transitions having E$_{up}>50$ K with the solid blue line, and the dashed blue line in Figure \ref{fig:temp} is the same for transitions having E$_{up}<50$ K.

\begin{table*}
\centering
{\scriptsize
 \caption{Calculation of kinetic temperature using line ratio of observed CH$_3$CN transitions. \label{table:kladder}}
\begin{tabular}{|c|c|c|c|c|c|c|c|c|c|}
  \hline
  \hline
  Source&Frequency&Quantum No.&E$_{up}$&$\int$T$_{mb}$dv&S$_{ij}$&R&T$_{k}$&Average T$_k$&\\
  &(GHz)&&(K)&(K.km.s$^{-1}$)&&&(K)&(K)&\\
  \hline
  \hline  
L1544&91.985314&$5_1$ - $4_1$&20.4 &0.05&6.61439&$\frac{5_0-4_0}{5_1-4_1}$&5.12&5.12&\\
&91.987087&$5_0$ - $4_0$&13.2 &0.068&2.2051&&&&\\
\hline
B1-b&91.985314&$5_1$ - $4_1$&20.4&0.034&6.61439&$\frac{5_0-4_0}{5_1-4_1}$&4.41&4.41&\\
&91.987087&$5_0$ - $4_0$&13.2&0.058&2.2051&&&&\\
\hline
IRAS4A&128.75703&$7_3$ – $6_3$&89&0.089&15.74921&$\frac{7_2-6_2}{7_3-6_3}$&33.79&&\\
&128.769436&$7_2$ - $6_2$&53.3&0.144&8.85861&&&&\\
\cline{2-8}
&128.769436&$7_2$ - $6_2$&53.3&0.144&8.85861&$\frac{7_1-6_1}{7_2-6_2}$&46.24&&\\
&128.776881&$7_1$ – $6_1$&31.9&0.244&9.44898&&&&\\
\cline{2-8}
&128.75703&$7_3$ – $6_3$&89&0.089&15.74921&$\frac{7_1-6_1}{7_3-6_3}$&37.58&&\\
&128.776881&$7_1$ – $6_1$&31.9&0.244&9.44898&&&&\\
\cline{2-8}
&128.75703&$7_3$ – $6_3$&89&0.089&15.74921&$\frac{7_0-6_0}{7_3-6_3}$&23.19&&\\
&128.779363&$7_0$ - $6_0$&24.7&0.285&3.1501&&&&\\
\cline{2-8}
&128.769436&$7_2$ - $6_2$&53.3&0.144&8.85861&$\frac{7_0-6_0}{7_2-6_2}$&16.66&&\\
&128.779363&$7_0$ - $6_0$&24.7&0.285&3.1501&&&&\\
\cline{2-8}
&147.163244&$8_2$ – $7_2$&60.4&0.098&10.33654&$\frac{8_1-7_1}{8_2-7_2}$&26.44&48.60&E$_{up} >$ 50 K\\
&147.171751&$8_1$ – $7_1$&38.9&0.232&10.85164&&&&\\
\cline{2-8}
&147.163244&$8_2$ – $7_2$&60.4&0.098&10.33654&$\frac{8_0-7_0}{8_2-7_2}$&14.17&&\\
&147.174588&$8_0$ – $7_0$&31.8&0.258&3.61771&&&&\\
\cline{2-8}
&165.540377&$9_3$ – $8_3$&104&0.131&22.05132&$\frac{9_2-8_2}{9_3-8_3}$&106.31&&\\
&165.556321&$9_2$ - $8_2$&68.3&0.098&11.7908&&&&\\
\cline{2-8}
&165.556321&$9_2$ - $8_2$&68.3&0.098&11.7908&$\frac{9_1-8_1}{9_2-8_2}$&28.45&&\\
&165.565891&$9_1$ – $8_1$&46.9&0.216&12.25078&&&&\\
\cline{2-8}
&165.540377&$9_3$ – $8_3$&104&0.131&22.05132&$\frac{9_1-8_1}{9_3-8_3}$&52.49&&\\
&165.565891&$9_1$ – $8_1$&46.9&0.216&12.25078&&&&\\
\cline{2-8}
&165.540377&$9_3$ – $8_3$&104&0.131&22.05132&$\frac{9_0-8_0}{9_3-8_3}$&29.16&&\\
&165.569081&$9_0$ – $8_0$&39.7&0.22	&4.08322&&&&\\
\cline{2-8}
&165.556321&$9_2$ - $8_2$&68.3&0.098&11.7908&$\frac{9_0-8_0}{9_2-8_2}$&15.30&&\\
&165.569081&$9_0$ – $8_0$&39.7&0.22&4.08322&&&&\\
\cline{2-10}
&73.588799&$4_1$ – $3_1$&16&0.127&5.16717&$\frac{4_0-3_0}{4_1-3_1}$&5.20&&\\
&73.590218&$4_0$ – $3_0$&8.8&0.169&1.72263&&&&\\
\cline{2-8}
&91.979994&$5_2$ – $4_2$&41.8&0.082&5.78723&$\frac{5_1-4_1}{5_2-4_2}$&33.07&&\\
&91.985314&$5_1$ – $4_1$&20.4&0.179&6.61439&&&&\\
\cline{2-8}
&91.985314&$5_1$ – $4_1$&20.4&0.179&6.61439&$\frac{5_0-4_0}{5_1-4_1}$&5.50&&\\
&91.987087&$5_0$ – $4_0$&13.2&0.221&2.2051&&&&\\
\cline{2-8}
&91.979994&$5_2$ – $4_2$&41.8&0.082&5.78723&$\frac{5_0-4_0}{5_2-4_2}$&14.62&&\\
&91.987087&$5_0$ – $4_0$&13.2&0.221&2.2051&&&&\\
\cline{2-8}
&110.381372&$6_1$ – $5_1$&25.7&0.233&8.03828&$\frac{6_0-5_0}{6_1-5_1}$&5.99&10.31&E$_{up} <$ 50 K\\
&110.383499&$6_0$ – $5_0$&18.5&0.258&2.6798&&&&\\
\cline{2-8}
&128.776881&$7_1$ – $6_1$&31.9&0.244&9.44898&$\frac{7_0-6_0}{7_1-6_1}$&5.74&&\\
&128.779363&$7_0$ - $6_0$&24.7&0.285&3.1501&&&&\\
\cline{2-8}
&147.171751&$8_1$ – $7_1$&38.9&0.232&10.85164&$\frac{8_0-7_0}{8_1-7_1}$&5.89&&\\
&147.174588&$8_0$ – $7_0$&31.8&0.258&3.61771&&&&\\
\cline{2-8}
&165.565891&$9_1$ – $8_1$&46.9&0.216&12.25078&$\frac{9_0-8_0}{9_1-8_1}$&6.45&&\\
&165.569081&$9_0$ – $8_0$&39.7&0.22&4.08322&&&&\\
\hline
SVS13A&110.364353&$6_3$ - $5_3$&82.8&0.08&12.40333&$\frac{6_2-5_2}{6_3-5_3}$&42.87&&\\
&110.374989&$6_2$ - $5_2$&47.1&0.109&7.34927&&&&\\
\cline{2-8}
&110.364353&$6_3$ - $5_3$&82.8&0.08&12.40333&$\frac{6_1-5_1}{6_3-5_3}$&72.47&&\\
&110.381372&$6_1$ – $5_1$&25.7&0.114&8.03828&&&&\\
\cline{2-8}
&110.364353&$6_3$ - $5_3$&82.8&0.08&12.40333&$\frac{6_0-5_0}{6_3-5_3}$&38.21&&\\
&110.383499&$6_0$ – $5_0$&18.5&0.093&2.6798&&&&\\
\cline{2-8}
&220.709016&$12_3$ – $11_3$&133.2&0.265&31.00599&$\frac{12_2-11_2}{12_3-11_3}$&47.07&&\\
&220.73026&$12_2$ - $11_2$&97.4&0.294&16.07897&&&&\\
\cline{2-8}
&220.74301&$12_1$ - $11_1$&76&0.278&16.4203&$\frac{12_0-11_0}{12_1-11_1}$&24.76&63.90&E$_{up} >$ 50 K\\
&220.747261&$12_0$ - $11_0$&68.9&0.373&16.53946&&&&\\
\cline{2-8}
&220.709016&$12_3$ – $11_3$&133.2&0.265&31.00599&$\frac{12_1-11_1}{12_3-11_3}$&83.68&&\\
&220.74301&$12_1$ - $11_1$&76&0.278&16.4203&&&&\\
\cline{2-8}
&220.709016&$12_3$ – $11_3$&133.2&0.265&31.00599&$\frac{12_0-11_0}{12_3-11_3}$&66.27&&\\
&220.747261&$12_0$ - $11_0$&68.9&0.373&16.53946&&&&\\
\cline{2-8}
&220.73026&$12_2$ - $11_2$&97.4&0.294&16.07897&$\frac{12_0-11_0}{12_2-11_2}$&135.87&&\\
&220.747261&$12_0$ - $11_0$&68.9&0.373&16.53946&&&&\\
\hline
\hline
 \end{tabular}}
 \end{table*}

\begin{table*}
\centering
{\scriptsize
 \caption{Calculation of kinetic temperature using line ratio of observed CH$_3$CN transitions. \label{table:kladder}}
\begin{tabular}{|c|c|c|c|c|c|c|c|c|c|}
  \hline
  \hline
  Source&Frequency&Quantum No.&E$_{up}$&$\int$T$_{mb}$dv&S$_{ij}$&R&T$_{k}$&Average T$_k$&\\
  &(GHz)&&(K)&(K.km.s$^{-1}$)&&&(K)&(K)&\\
  \hline
\hline
&91.979994&$5_2$ – $4_2$&41.8&0.067&5.78723&$\frac{5_1-4_1}{5_2-4_2}$&184.38&&\\
&91.985314&$5_1$ – $4_1$&20.4&0.086&6.61439&&&&\\
\cline{2-8}
&91.985314&$5_1$ – $4_1$&20.4&0.086&6.61439&$\frac{5_0-4_0}{5_1-4_1}$&6.42&&\\
&91.987087&$5_0$ – $4_0$&13.2&0.088&2.2051&&&&\\
\cline{2-8}
&91.979994&$5_2$ – $4_2$&41.8&0.067&5.78723&$\frac{5_0-4_0}{5_2-4_2}$&23.11&&\\
&91.987087&$5_0$ – $4_0$&13.2&0.088&2.2051&&&43.58&E$_{up} <$ 50 K\\
\cline{2-8}
&110.381372&$6_1$ – $5_1$&25.7&0.114&8.03828&$\frac{6_0-5_0}{6_1-5_1}$&8.04&&\\
&110.383499&$6_0$ – $5_0$&18.5&0.093&2.6798&&&&\\
\cline{2-8}
&110.374989&$6_2$ - $5_2$&47.1&0.109&7.34927&$\frac{6_0-5_0}{6_2-5_2}$&33.64&&\\
&110.383499&$6_0$ – $5_0$&18.5&0.093&2.6798&&&&\\
\cline{2-8}
&128.7768817&$7_1$ – $6_1$&31.9&0.092&9.44898&$\frac{7_0-6_0}{7_1-6_1}$&5.85&&\\
&128.779363&$7_0$ - $6_0$&24.7&0.105&3.1501&&&&\\
  \hline
  \hline
 \end{tabular}}
 \end{table*}

\subsection{Abundance variation in different sources}
\begin{figure}
\centering
\begin{minipage}{0.55\textwidth}
\includegraphics[width=\textwidth]{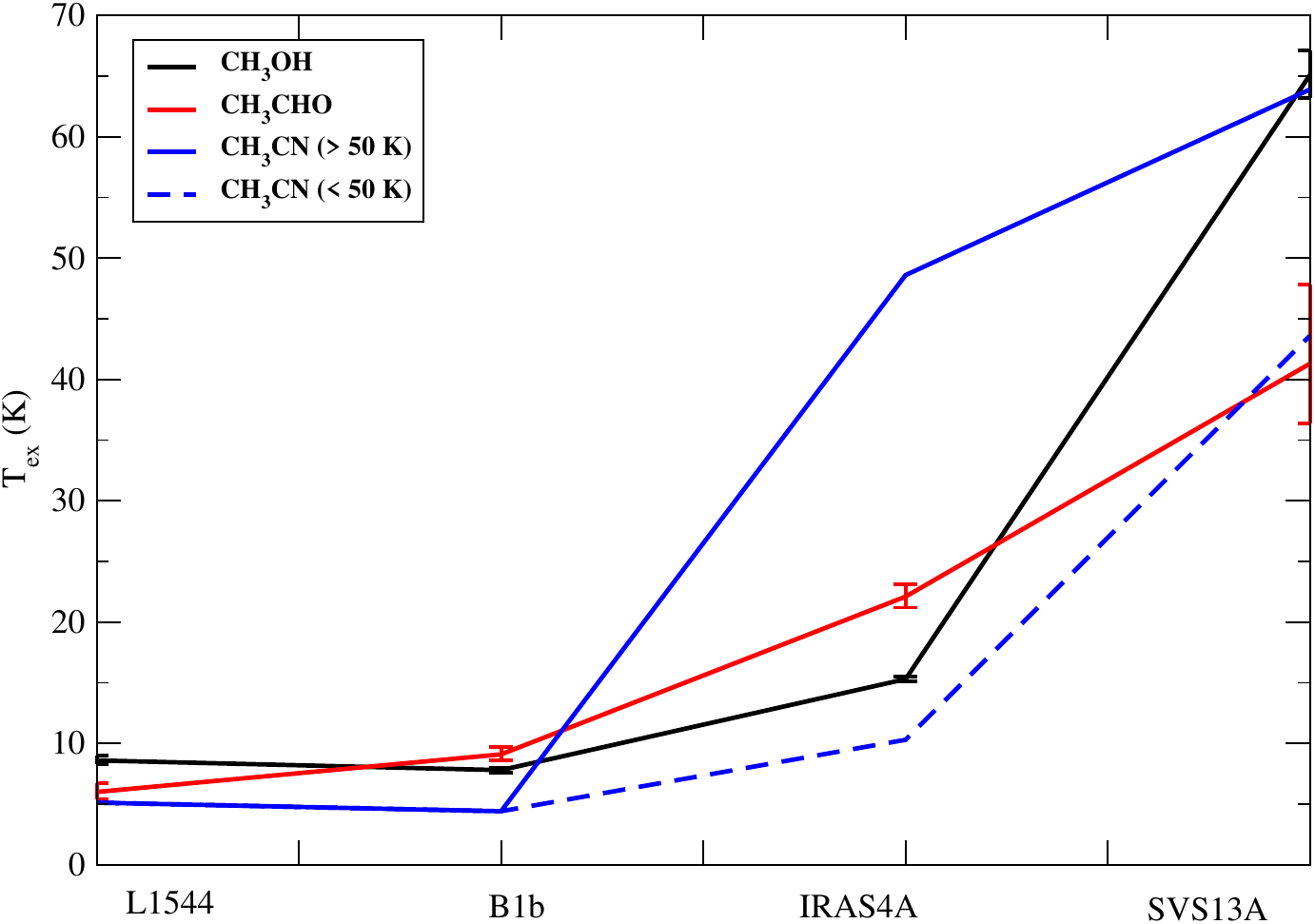}
\end{minipage}
\caption{Excitation Temperature derived from rotation diagram for methanol (in black) and acetaldehyde (in red), respectively. Vertical lines represent the corresponding errors. Kinetic temperature calculated from Table \ref{table:kladder} using CH$_3$CN transitions for high-temperature (solid blue) component and low-temperature (dashed blue) component present in IRAS4A and SVS13A.}
\label{fig:temp}
\end{figure}

\begin{figure}
\centering
\begin{minipage}{0.55\textwidth}
\includegraphics[width=\textwidth]{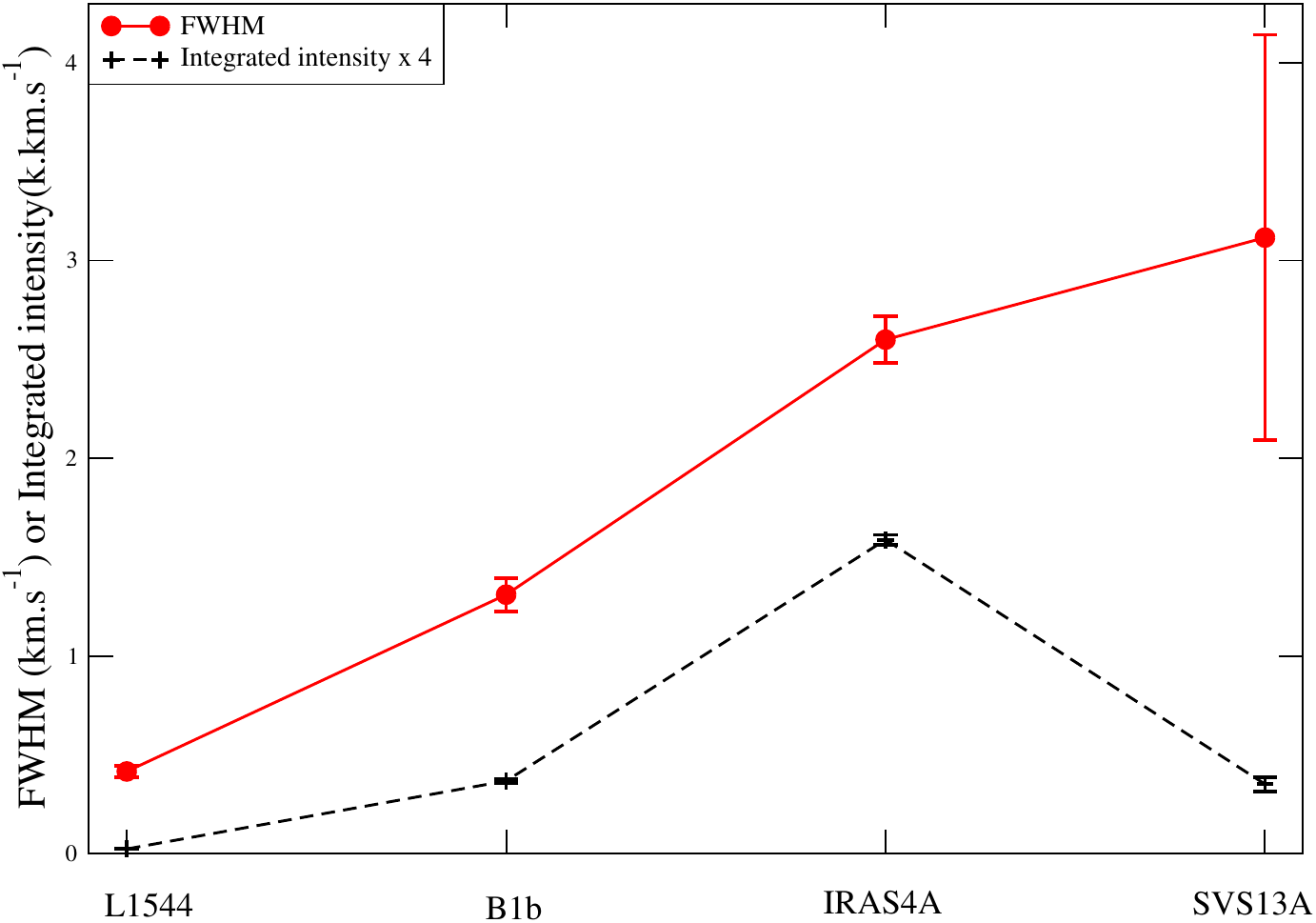}
\end{minipage}
\caption{FWHM (red line) and integrated intensity (black-dashed) for 96.755501 GHz transition of methanol. The vertical lines represent the error bars.}
\label{fig:fwhm}
\end{figure}

For CH$_3$OH, CH$_3$CHO, CH$_3$OCHO, C$_2$H$_5$OH, CH$_3$OCH$_3$, and CH$_3$CN we found an increasing abundance from the prestellar core to the class 0 stage and a decreasing abundance in the class I phase. The abundance of HCCCHO, on the other hand, exhibits an upward tendency up to the class I phase. With the exception for L1544, the trend found for HCCCHO is based on the upper limit derived for the majority of the sources. We were unable to estimate an upper limit in L1544 for C$_2$H$_5$OH.
To calculate the abundances of the species and determine whether they have any relationships with the different evolutionary phases, it is necessary to know the hydrogen column density in each source. We use the hydrogen column density obtained from a beam size comparable to the molecule's source size for the accuracy of the abundance derivation from the obtained column density (see Section \ref{sec:H2_col}). We were unable to distinguish the molecular emission coming from the core because our beam size varied from 30$^{''}$ to 9$^{''}$ in our frequency range. We plot the rotational temperatures of two species found in our sample sources to get an idea. As expected, we observed a rise in temperature from the prestellar to class I phases (see Figure \ref{fig:temp}).
Additionally, we studied a specific transition of a species that is observed in all the sources. Fortunately, all sources have the same CH$_3$OH transition (96.755501 GHz). We plot the integrated intensity of this CH$_3$OH transition, it is shown in Figure \ref{fig:fwhm}. It suggests that the integrated intensity exhibits a similar pattern to the abundances with respect to the evolutionary phases. In addition, we found a transition for acetaldehyde at $211.2738$ GHz for all sources except SVS13A. It demonstrates an upward trend. We found a transition in IRAS4A and SVS13A at a frequency of $216.5819$ GHz and measured a lower integrated intensity in SVS13A than in IRAS4A. Thus, the obtained abundance trend may not just be the result of the adopted N(H$_2$).
Additionally, we examine whether any of the source's physical characteristics have an impact on this transition. So, for all sources, we plot the variation of the FWHM of this transition (96.7555 GHz) (see Figure \ref{fig:fwhm}). It shows an increasing trend from the prestellar core to the class I object, and its value is similar to other transitions that have been observed. We address the relationship between the obtained abundances and source luminosity in section \ref{sec:luminosity} as each of these sources has a different luminosity. 

All of our sample sources include multiple cores except for the L1544. Table \ref{table:source} lists the targeted position for each sample source in the ASAI data. Even at the maximum frequency of our observation, the beam we are observing contains every core that is offered by each source. In order to establish that the cores from each source follow the same pattern, we must first determine whether any chemical differentiation exists among them. Therefore, to confirm the found pattern, high resolution interferometric data is needed. Since it is outside the scope of this work to analyze the interferometric observation, we discuss the interferometric observational results obtained by various authors in these sources in Section \ref{sec:INO}.

\begin{figure}
\begin{minipage}{0.5\textwidth}
\includegraphics[width=\textwidth]{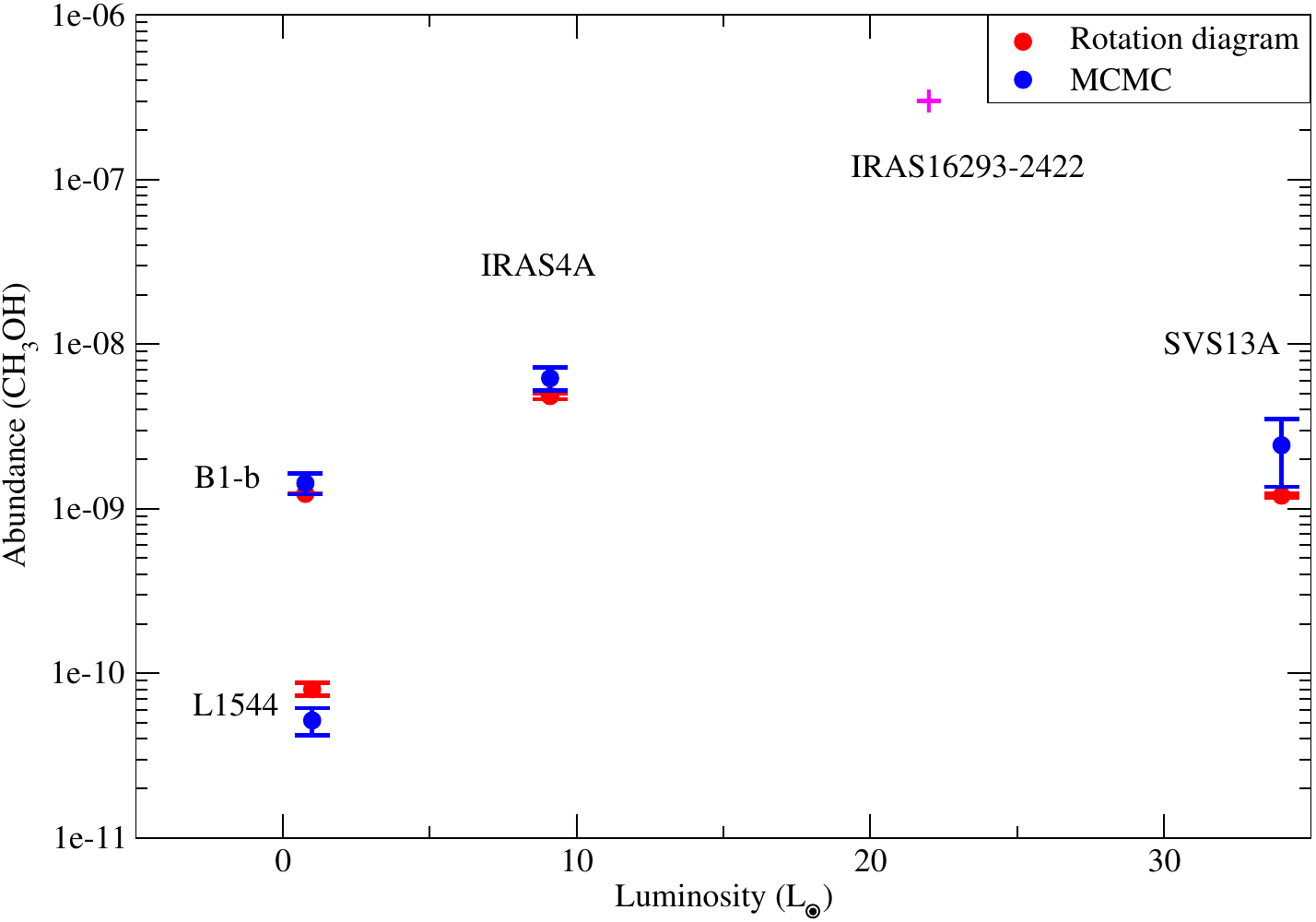}
\end{minipage}
\begin{minipage}{0.5\textwidth}
\includegraphics[width=\textwidth]{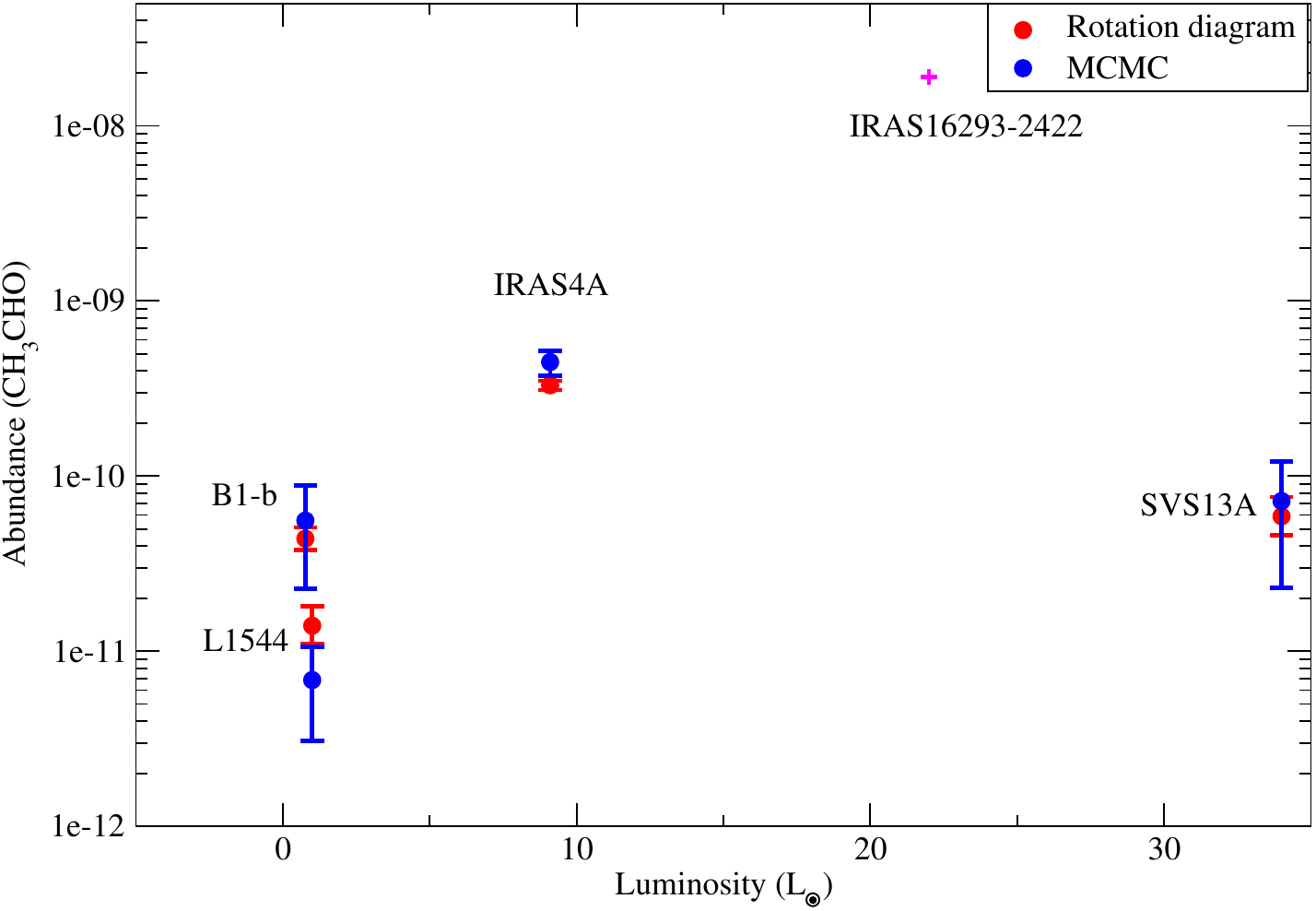}
\end{minipage}
\begin{minipage}{0.5\textwidth}
\includegraphics[width=\textwidth]{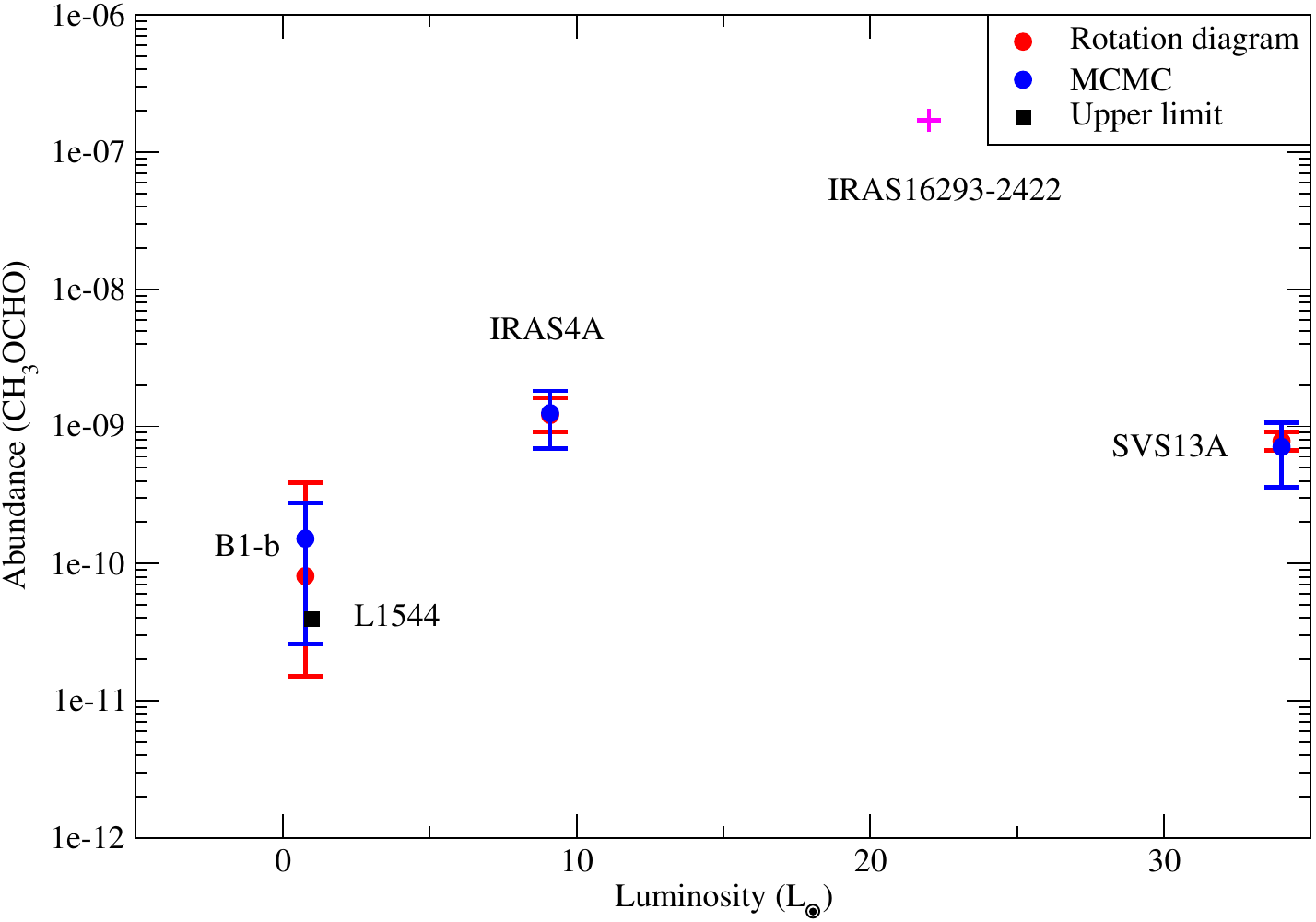}
\end{minipage}
\begin{minipage}{0.5\textwidth}
\includegraphics[width=\textwidth]{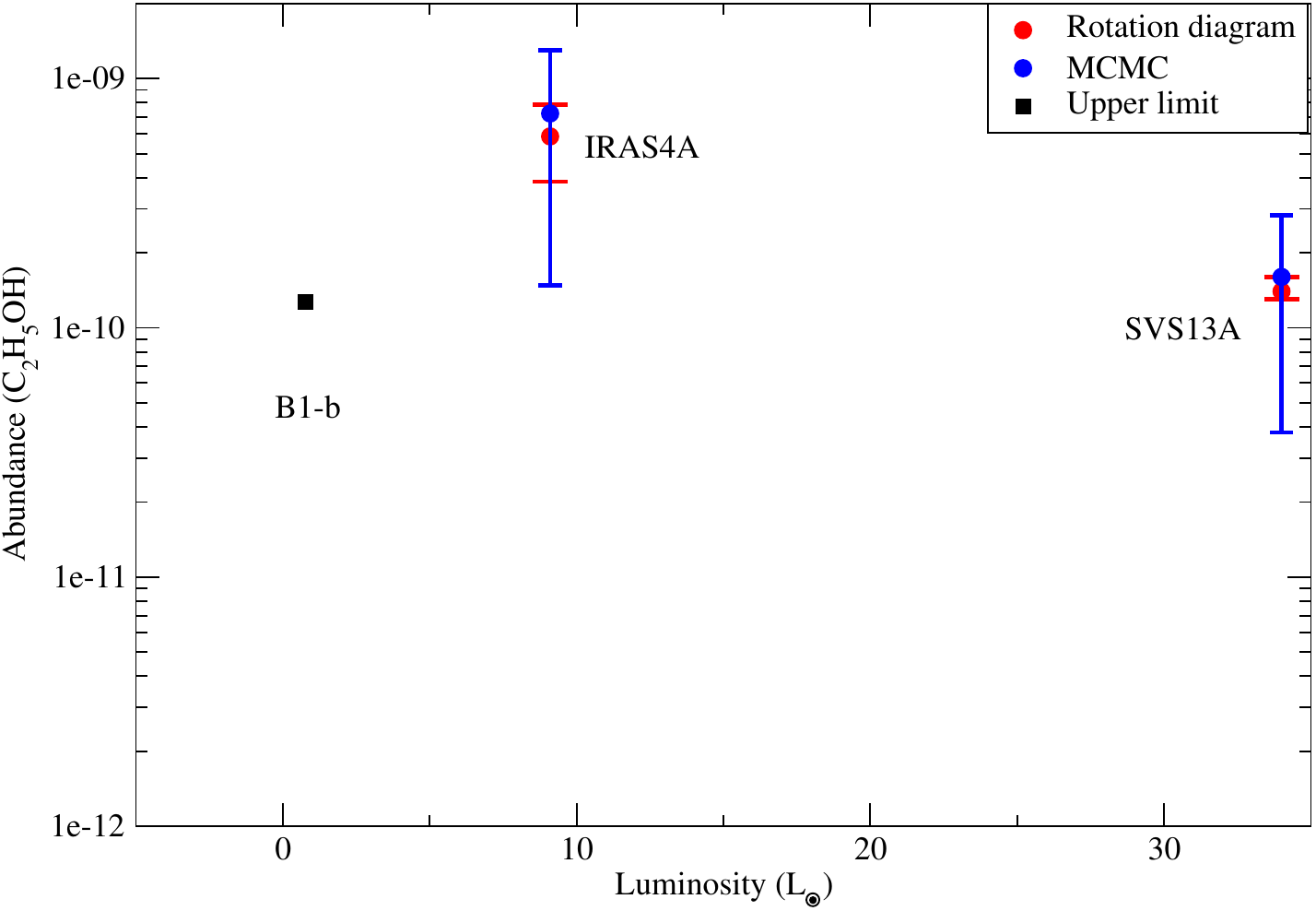}
\end{minipage}
\begin{minipage}{0.5\textwidth}
\includegraphics[width=\textwidth]{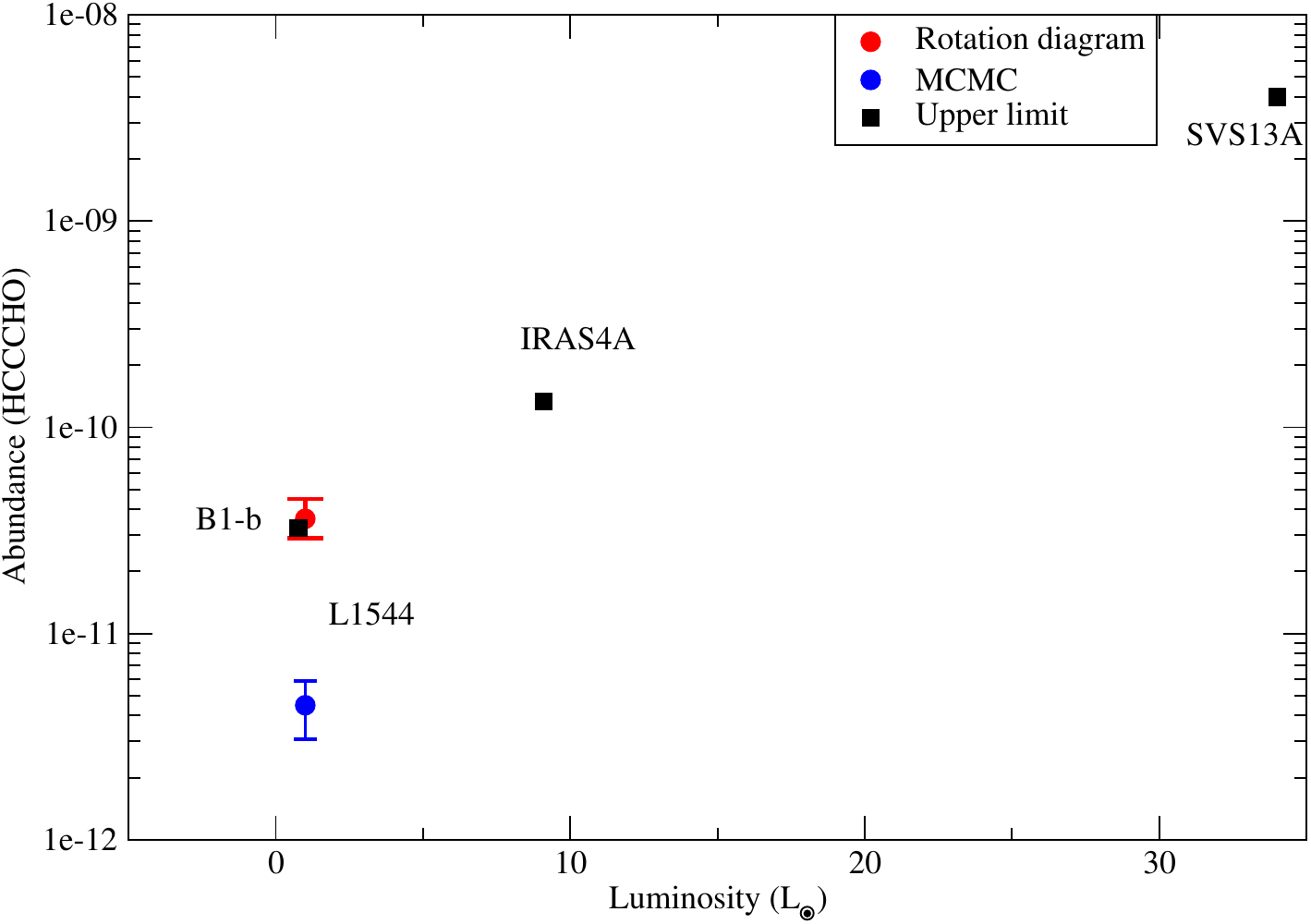}
\end{minipage}
\begin{minipage}{0.5\textwidth}
\includegraphics[width=\textwidth]{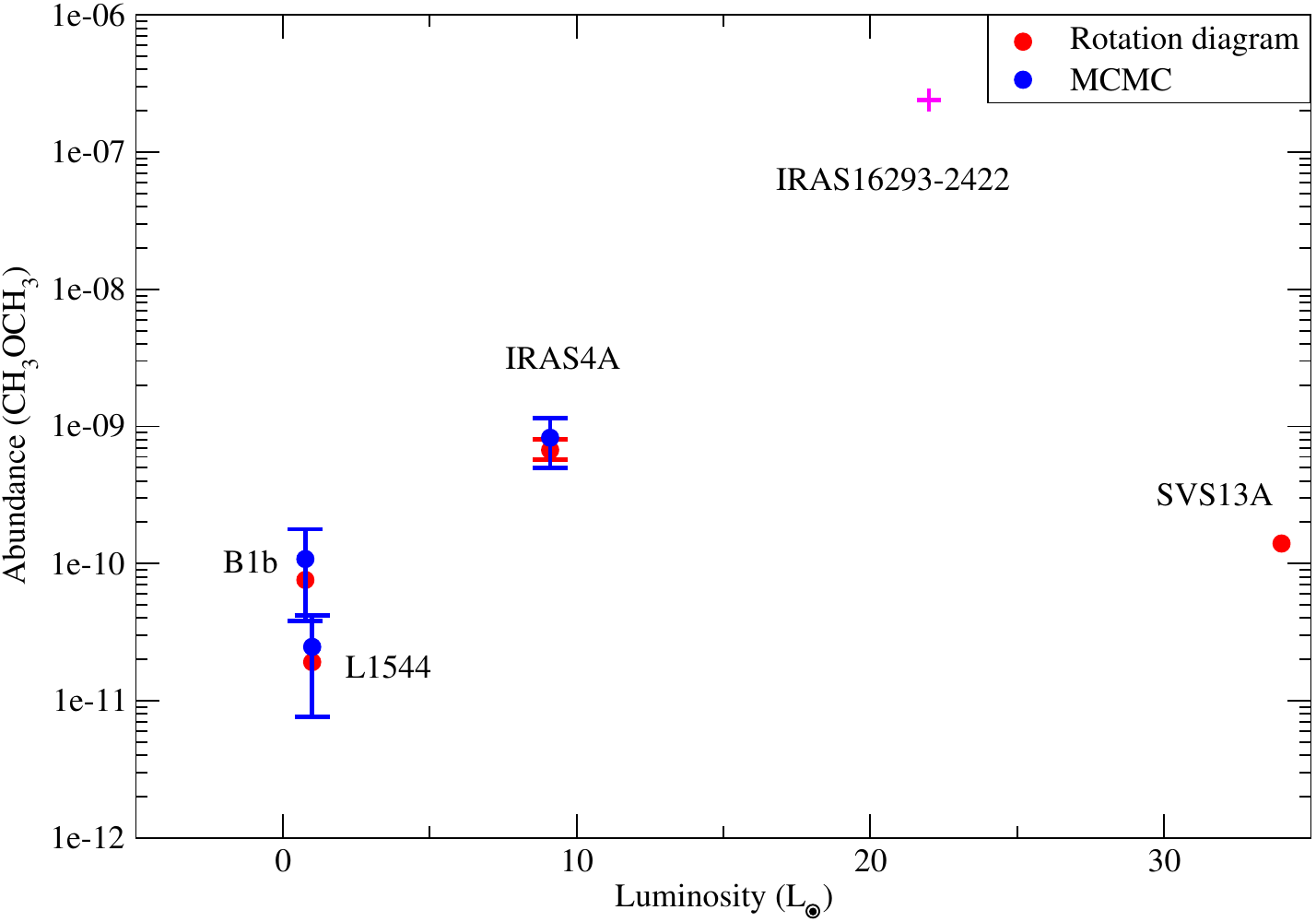}
\end{minipage}
\begin{minipage}{0.5\textwidth}
\includegraphics[width=\textwidth]{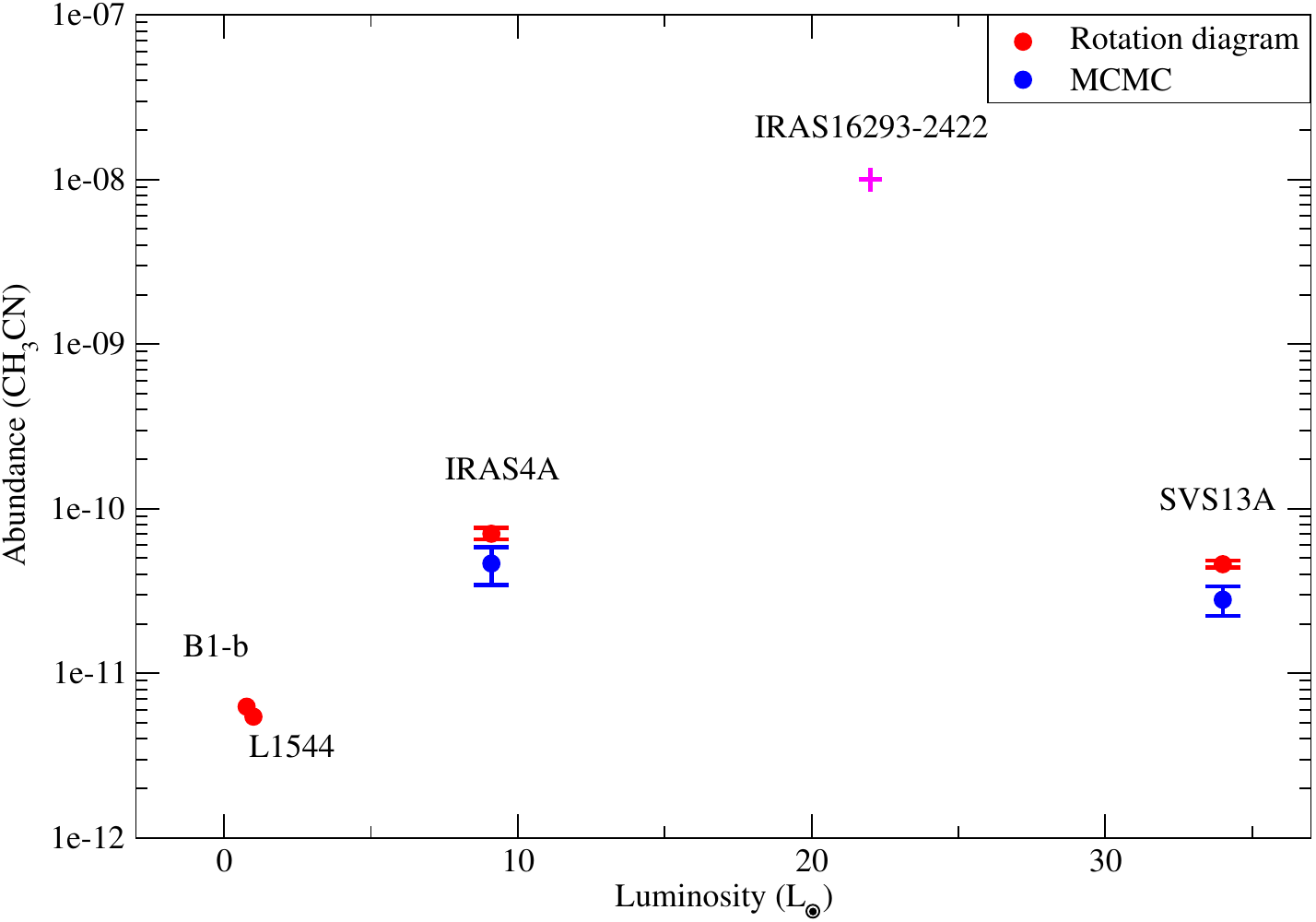}
\end{minipage}
\caption{Abundance variation of CH$_3$OH, CH$_3$CHO, CH$_3$OCHO, C$_2$H$_5$OH, HCCCHO, CH$_3$OCH$_3$, and CH$_3$CN shown with source luminosity. The red circle represents the value obtained from the rotational diagram, and the blue circle represents the same obtained from MCMC. The solid black squares represent the same calculated using upper limits. The plus sign (magenta) represents the abundance for IRAS4A 16293-2422 (22 L$_\odot$) taken from \cite{caza03}. The vertical lines represent the error bars.}
\label{fig:luminosity}
\end{figure}

\begin{figure}
\begin{minipage}{0.5\textwidth}
\includegraphics[width=\textwidth]{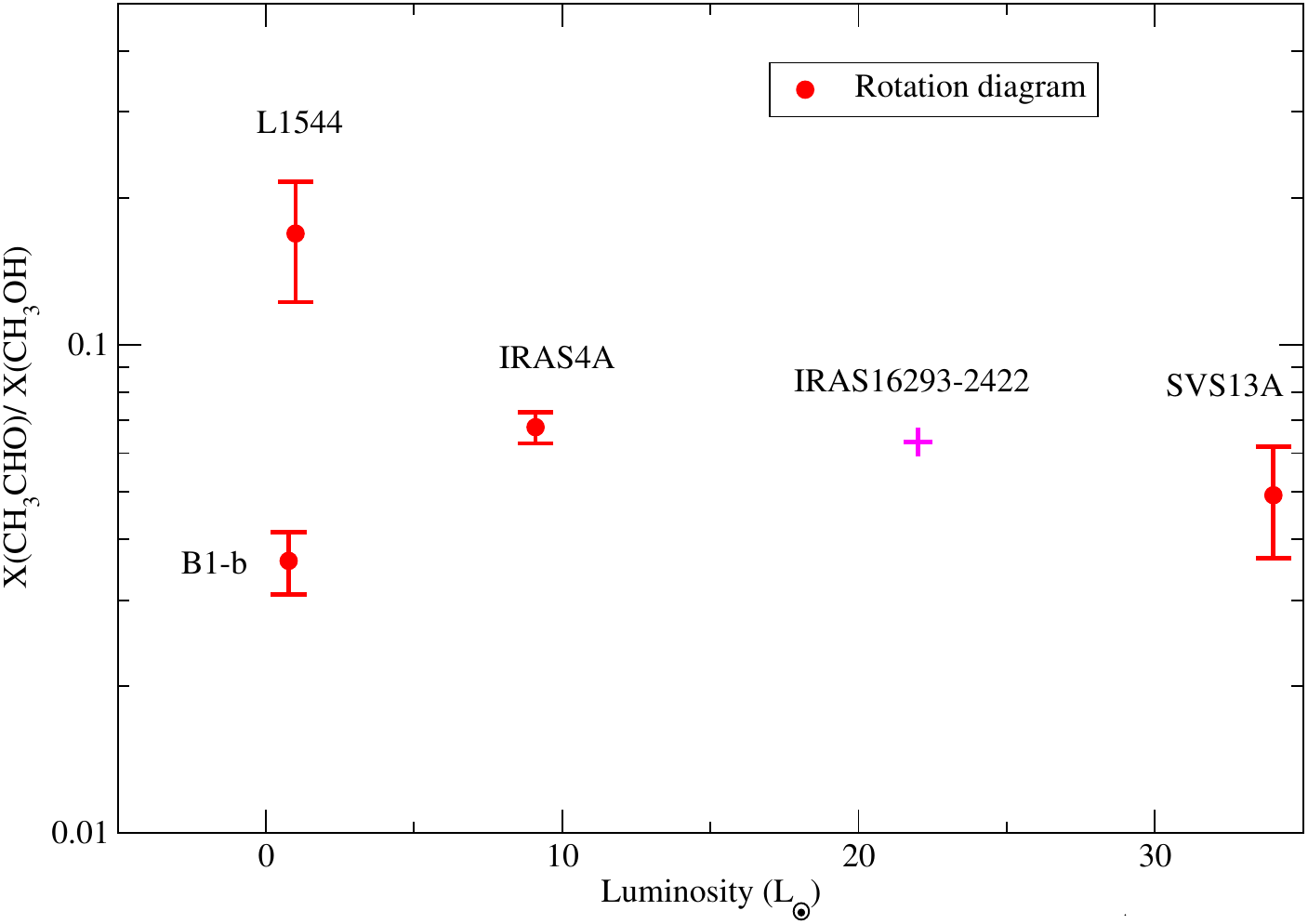}
\end{minipage}
\begin{minipage}{0.5\textwidth}
\includegraphics[width=\textwidth]{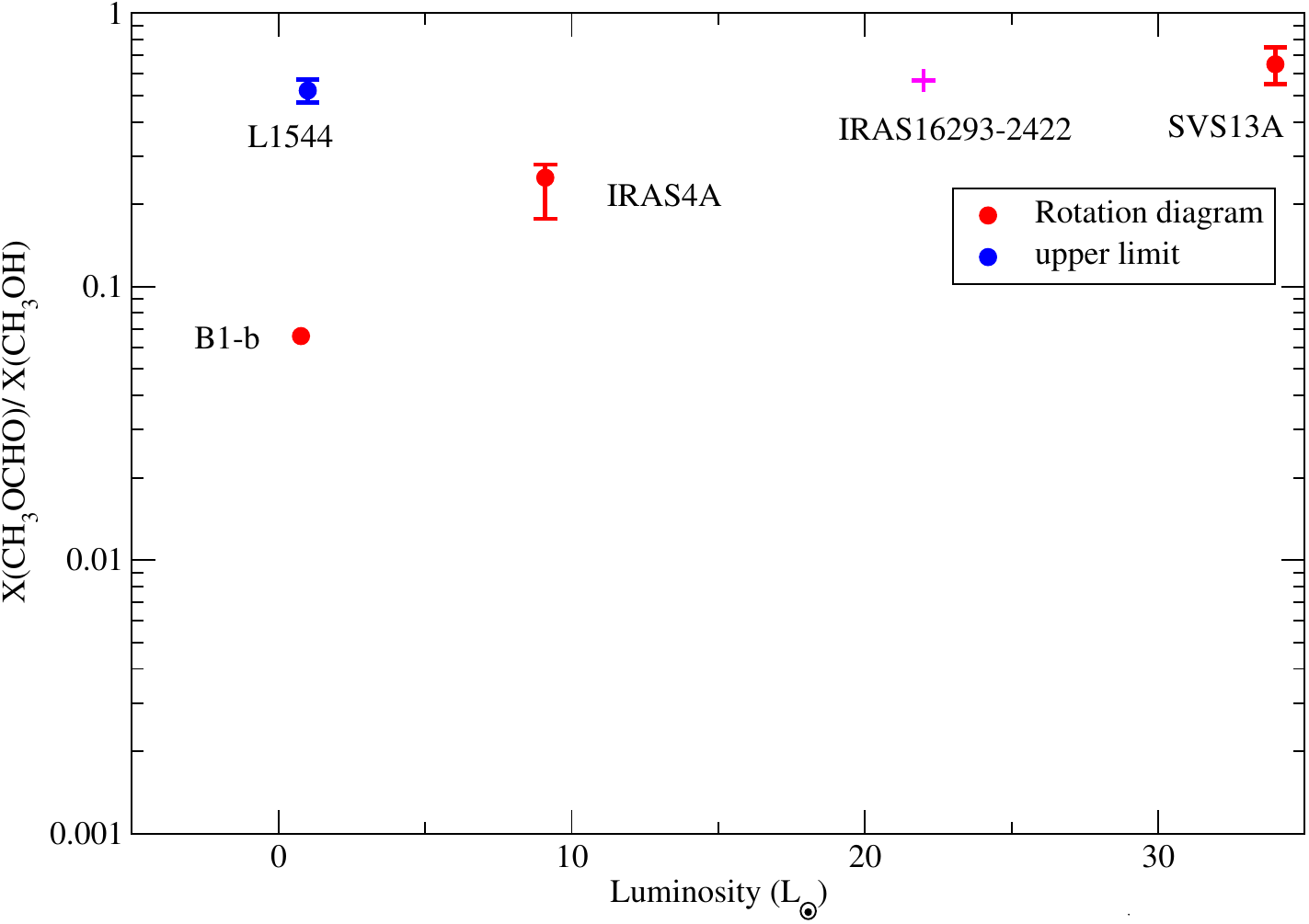}
\end{minipage}
\begin{minipage}{0.5\textwidth}
\includegraphics[width=\textwidth]{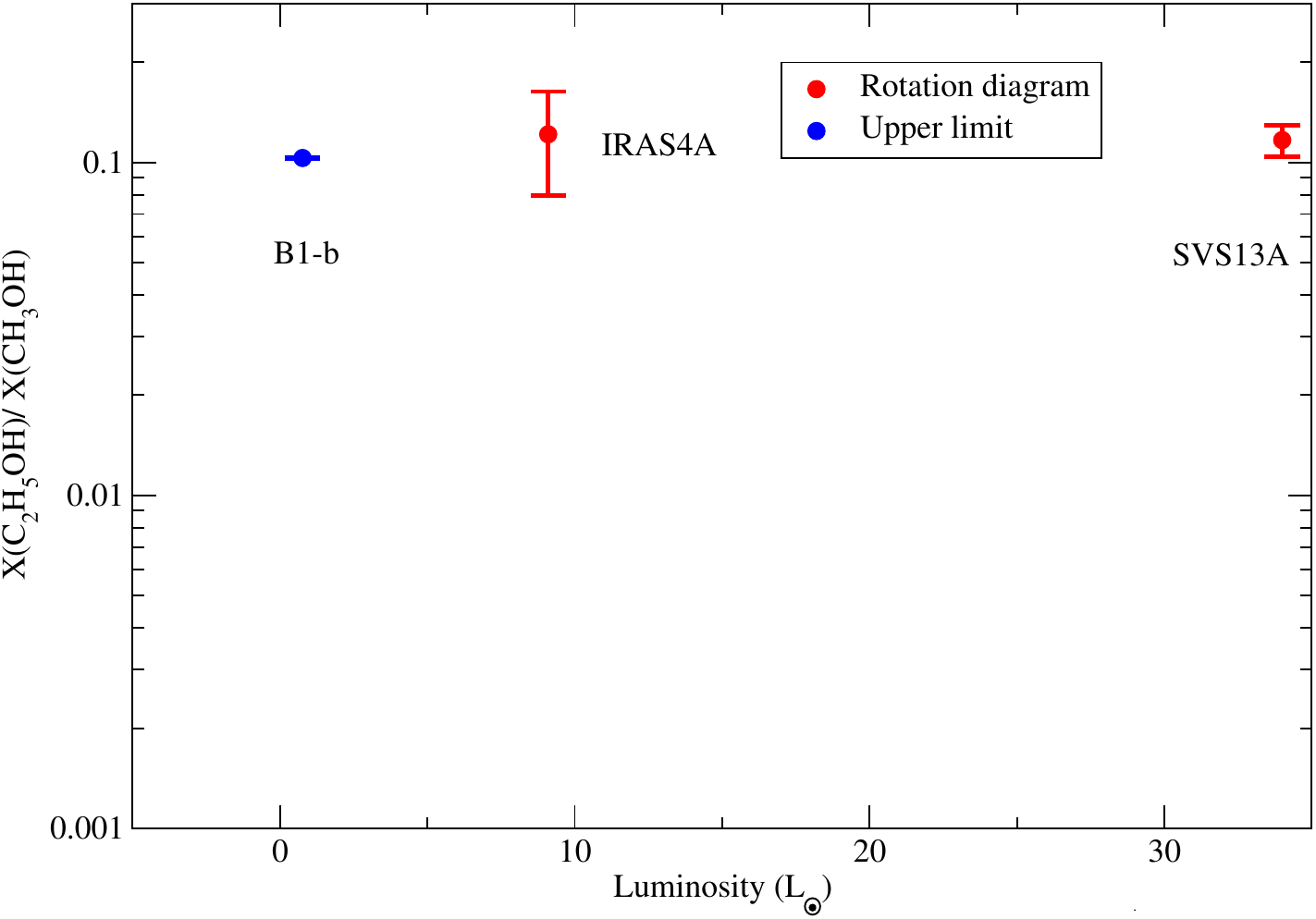}
\end{minipage}
\begin{minipage}{0.5\textwidth}
\includegraphics[width=\textwidth]{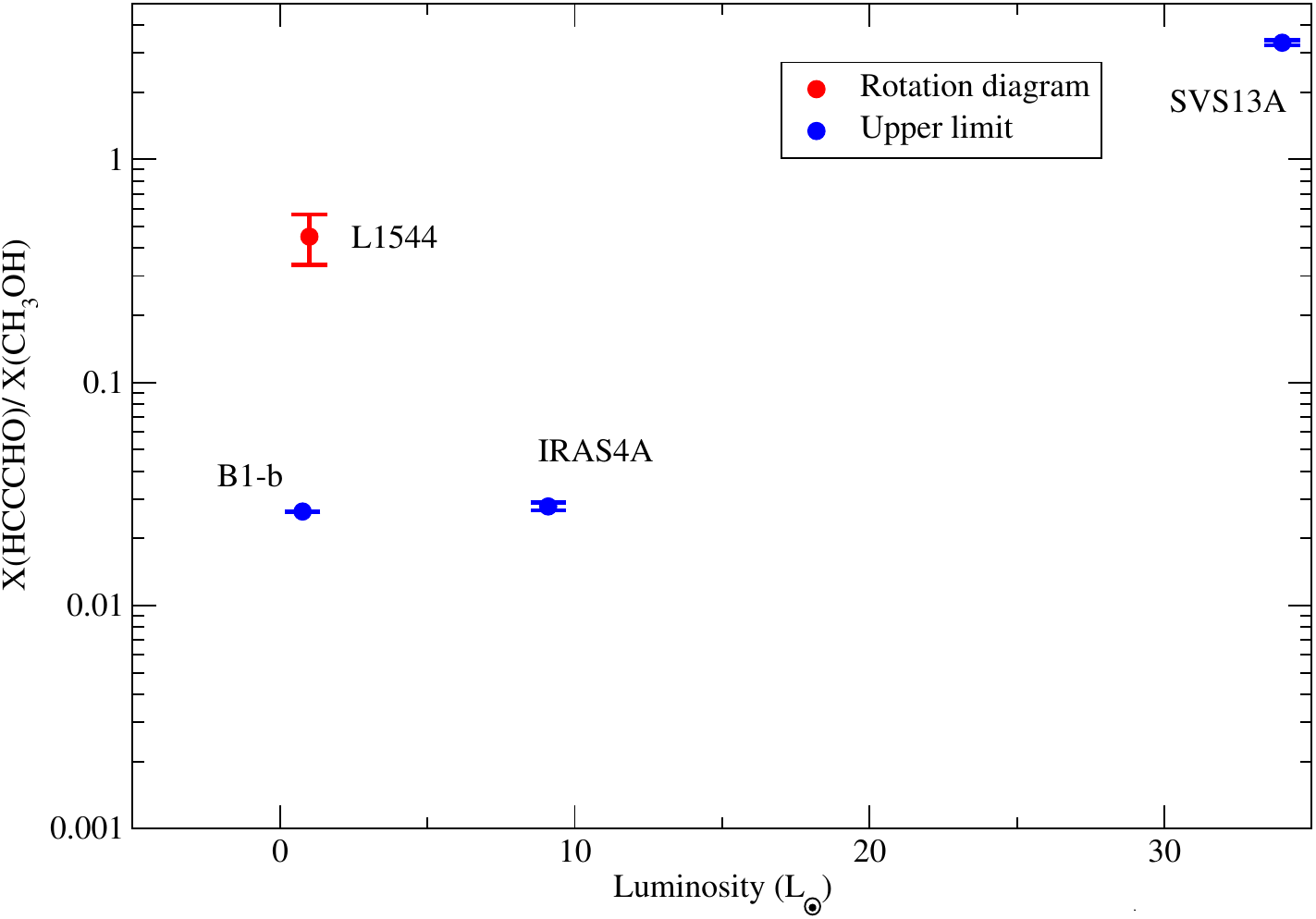}
\end{minipage}
\begin{minipage}{0.5\textwidth}
\includegraphics[width=\textwidth]{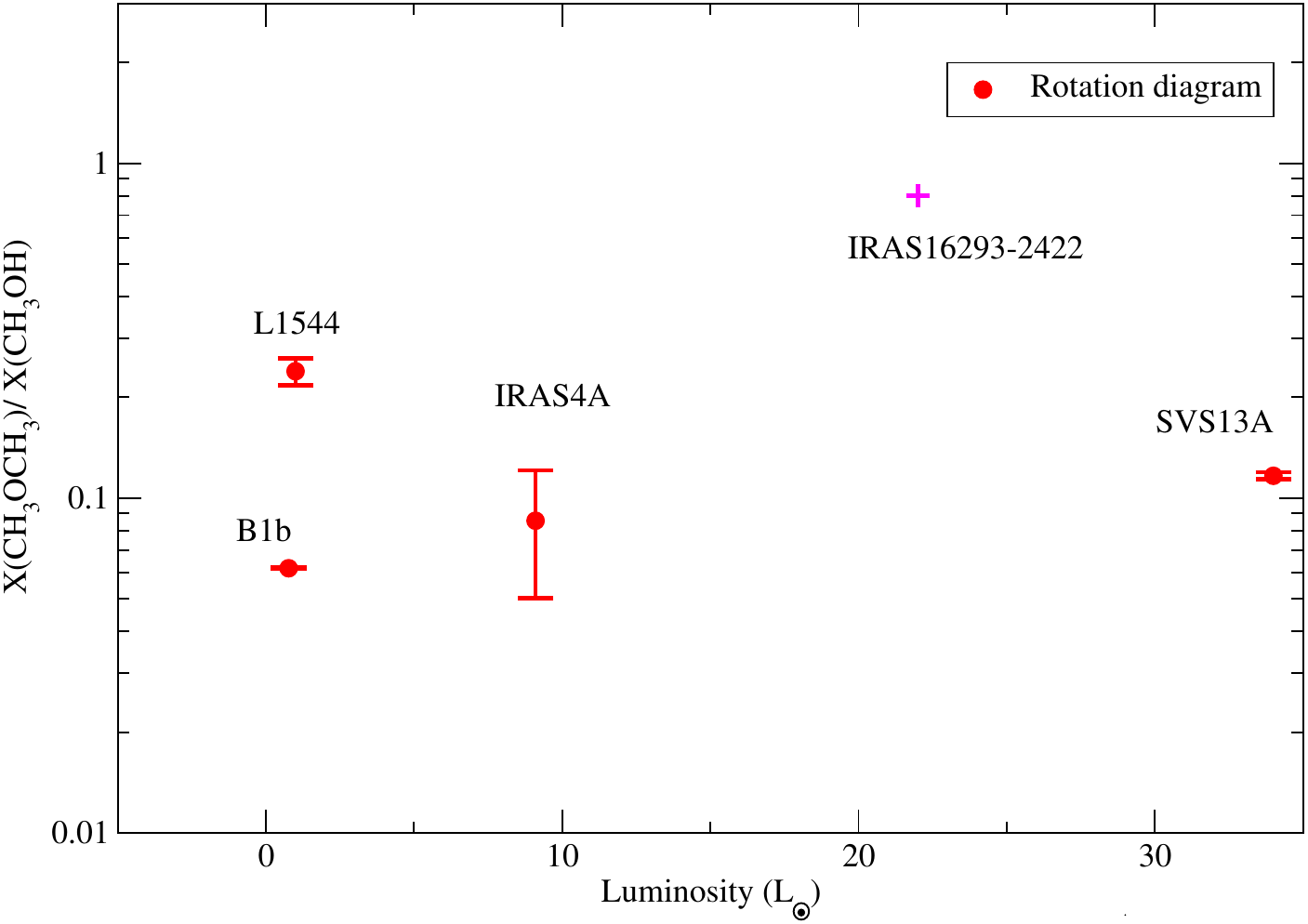}
\end{minipage}
\begin{minipage}{0.5\textwidth}
\includegraphics[width=\textwidth]{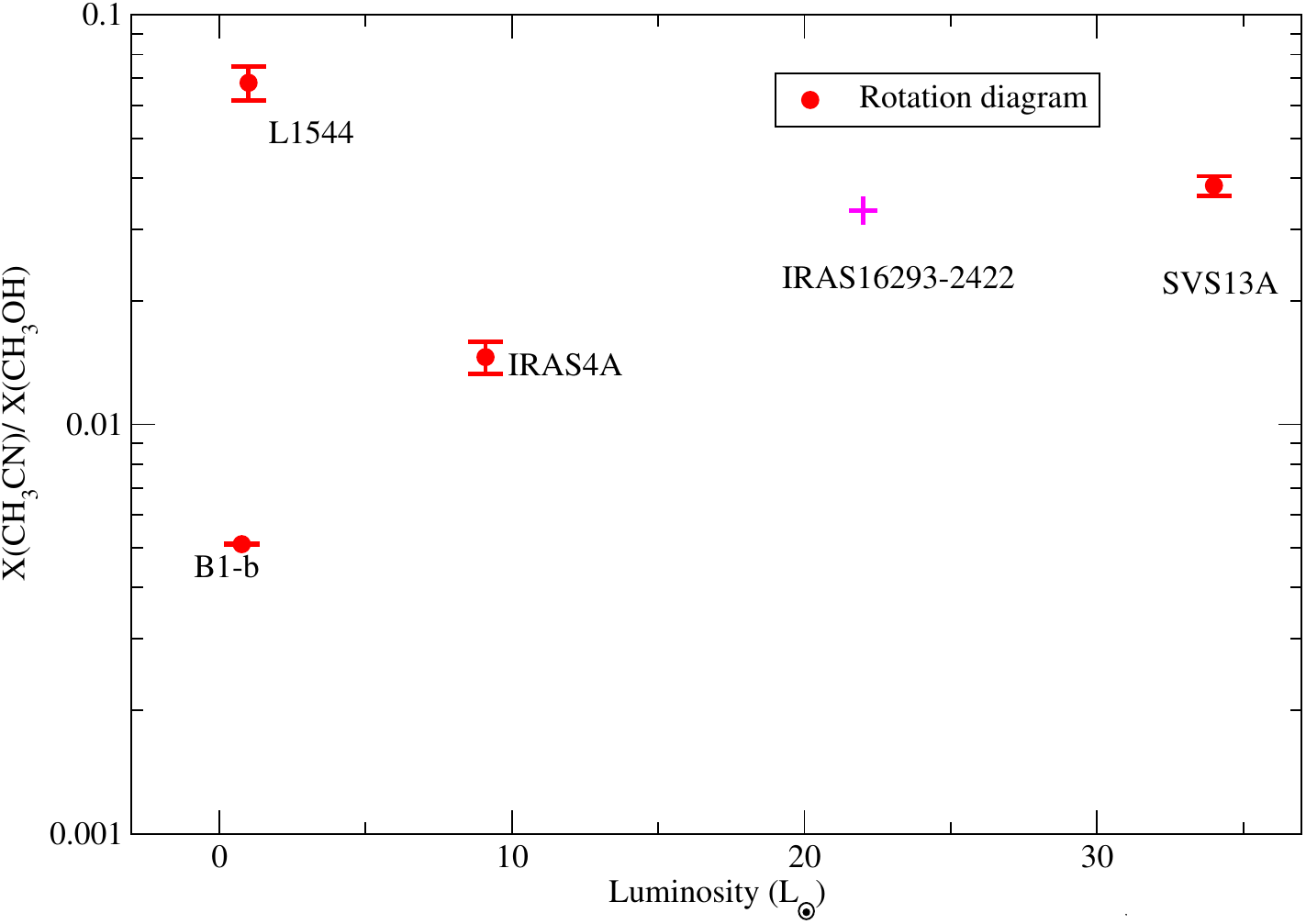}
\end{minipage}
\caption{The abundance ratios of CH$_3$CHO, CH$_3$OCHO, C$_2$H$_5$OH, HCCCHO, CH$_3$OCH$_3$, and CH$_3$CN w.r.t methanol (CH$_3$OH) plotted with source luminosity. The red circles represent the value obtained from the rotational diagram, and the blue circles represent the same obtained from the upper limits. The plus sign (magenta) represents the abundance obtained for IRAS4A 16293-2422 (22 L$_\odot$) taken from \cite{caza03}. Vertical lines represent the error bars. } \label{fig:ratio_methanol}
\end{figure}

\subsection{Luminosity effect \label{sec:luminosity}} 

The various evolutionary stages of star formation can be determined by source luminosity or bolometric luminosities, according to \citep{myer98}. According to \citep{doty05,lefl18}, the luminosities taken into account for L1544, B1-b, IRAS4A, and SVS13A are 1 L$_\odot$, 0.77 L$_\odot$, 9.1 L$_\odot$, and 34 L$_\odot$, respectively.
We take into account another object, IRAS16293-2422, which is located between the IRAS4A and SVS13A in luminosity scale, in order to figure out the trend associated with the luminosity because the luminosity of SVS13A is significantly higher than the other sources studied here. IRAS16293-2422, is a solar-type Class 0 protostar, located at a distance of 120 PC, in the eastern portion of the $\rho$ Ophiuchi star-forming region. It has a 22 L$_\odot$ bolometric luminosity. We take into account the methanol, acetaldehyde, methyl formate, dimethyl ether, and methyl cyanide abundances that were determined for this source from \cite{caza03}. While methanol abundance comes from the JCMT, the abundances of acetaldehyde, methyl formate, and dimethyl ether for this source were obtained from the IRAM 30 m data.
The luminosity effect on the molecules taken into consideration in this study is shown in Figure \ref{fig:luminosity}. In Figure \ref{fig:luminosity}, the abundances derived from rotational diagram analysis are represented by solid red circles, using the MCMC approach by solid blue circles, and the data derived from \cite[IRAS 16293-2422]{caza03} by magenta crosses. The values obtained by applying the upper limit are represented by the solid black squares. It exhibits behavior comparable to that described in Figure \ref{fig:clmdensity}. A higher column density of these species was found in the case of IRAS 16293-2422 compared to another class 0 object, IRAS4A, by \cite{caza03}. It might be because IRAS 16293-2422 has a higher luminosity ($\sim$ 22 L$_\odot$) than IRAS4A ($\sim$ 9.1 L$_\odot$).
We can improve our understanding by examining the molecular ratio of specific pairs of molecules because there are significant uncertainties in determining abundances.
\cite{taqu15} used PdBI to perform multi-line observations of methanol and a number of COMs in the direction of the two low-mass protostars NGC 1333-IRAS2A and NGC 1333-IRAS4A with an angular resolution of 2 $^{''}$. They determined that, for the source size of 0.5 $^{''}$, the ratio of CH$_3$OCH$_3$ to C$_2$H$_5$OH in the low mass protostar IRAS4A is around 0.7. In our case, the rotation diagram analysis data of the IRAS4A yield this ratio with a similar value of 0.7.
The abundance ratio w.r.t methanol is displayed with the luminosity of the source in Figure \ref{fig:ratio_methanol}. By plotting the molecular ratio, we were unable to detect any particular trend. However, we noticed that among the species investigated here, methanol is still the most abundant (ratios are $<$1 with the exception of HCCCHO). The abundance ratio of acetaldehyde and methanol varies with luminosity, as seen in the first panel of Figure \ref{fig:ratio_methanol}. In both the class 0 (IRAS4A, IRAS16293-2422) and class I (SVS13A) phases, the ratio essentially remains unchanged. Figure \ref{fig:ratio_methanol} shows a similar nature in the ratios of methyl formate and methanol in the second panel and ethanol and methanol in the third panel.

\subsubsection{Interferometric observations} \label{sec:INO}
\begin{figure}
    \centering
    \begin{minipage}{0.70\textwidth}
    \includegraphics[width=\textwidth]{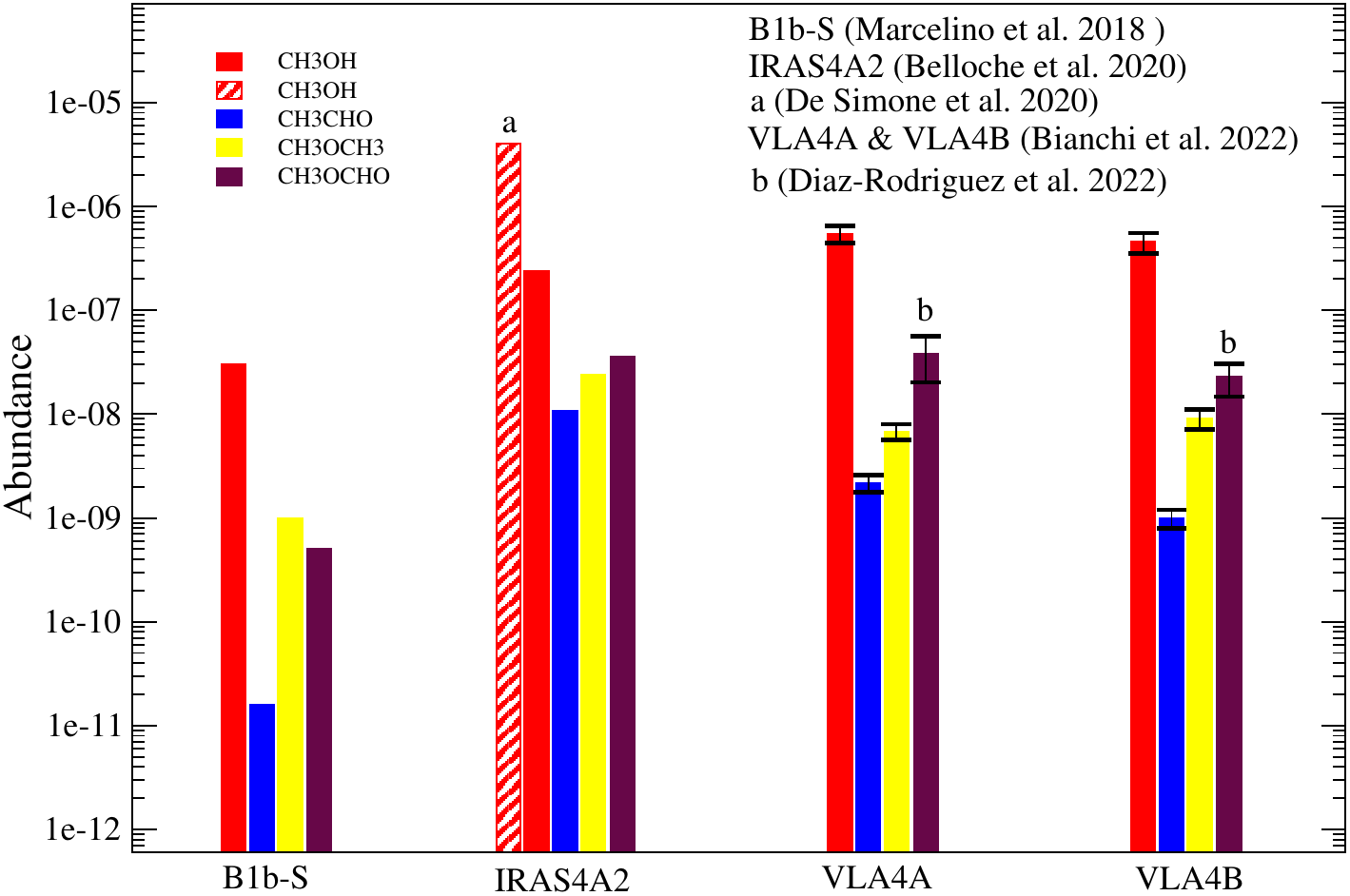}
    \end{minipage}
    \caption{Abundance variation of CH$_3$OH, CH$_3$CHO, CH$_3$OCH$_3$,and CH$_3$OCHO obtained from interferometric observation. Black vertical lines represent the error bar. \label{fig:inter}}
\end{figure}

It is generally accepted that different evolutionary stages of star formation are connected with the evolution of COMs. For this investigation, we make use of the large program ASAI data. To further validate the obtained trend, we also take into account previous interferometric observations. The obtained abundance trend using the interferometric observation is shown in Figure \ref{fig:inter}. We have listed the abundances and H$_2$ column densities obtained from the interferometric data in Table \ref{tab:inter}. There is a great deal of uncertainty when estimating the abundances of these molecules from the obtained column density. The ALMA Band 6 spectral line observations were provided in \cite{marc18b} with an angular resolution of $\sim 0.6$ $^{''}$ towards B1-b. Both the spectra from B1b-S and B1b-N protostars were extracted. However, it has been found that B1b-N emits no COMs, whereas B1b-S is rich in COMs. For sources with diameters of 0.35 $^{''}$, they calculated the column densities of $^{13}$CH$_3$OH. The column density of $^{12}$CH$_3$OH is derived using a value of $^{12}$C/$^{13}$C = 60. For a source size of 0.60 $^{''}$, they also calculated the column densities of CH$_3$CHO, CH$_3$OCH$_3$, CH$_3$OCHO, and N(H$_2$). \cite{simo20} obtained a very high column density ($\sim 10^{19}$ cm$^{-2}$) for methanol with a source size of 0.24 $^{''}$. In the case of IRAS4A2, the column density is derived from the 0$^{''}$.35 region. We use a H$_2$ column density of $2.5 \times 10^{24}$ cm$^{-2}$ derived from the 0.5 $^{''}$ component to determine the abundances of this region.

The column densities for CH$_3$OH, CH$_3$CHO, and CH$_3$OCH$_3$ in SVS13A are obtained from \cite{bian22}. With an angular resolution of 0.106 $^{''}$, they observed SVS13A.
A beam size region of 0$^{''}$.16 $\times$ 0$^{''}$.08 is used to calculate the column density of CH$_3$OCHO towards each component. The H$_2$ column density of $\sim 1.1 \times 10^{25}$ cm$^{-2}$ for the compact component size of 1.0 $^{''}$ is taken from \cite{lope15}.

The general trend (abundances steadily increased up to the class 0 stage and subsequently dropped during the class I phase) discovered by the ASAI survey was also discovered using the interferometric measurements, as shown in Figure \ref{fig:inter}.

\begin{landscape}

\begin{table}
   \caption{Column density of observed molecules obtained from interferometric observations. \label{tab:inter}}
    \begin{tabular}{|c|c|c|c|c|c|}
    \hline
         Sources&\multicolumn{4}{|c|}{Column density [abundance] in the form of \rm{a(b) = a $\times$ 10$^b$}}&N(H$_2$)\\
\cline{2-5}
&CH$_3$OH&CH$_3$CHO&CH$_3$OCH$_3$&CH$_3$OCHO&(cm$^{-2}$)\\
\hline
B1-b-S&3.0(17)$^a$[3.0(-8)]&1.6(14)$^a$[1.6(-11)]&1.0(16)$^a$[1.0(-9)]&5.0(15)$^a$[5.0(-10)]&1.1(25)$^a$\\
\hline
IRAS4A2&6.0(17)$^b$[2.4(-7)]/&2.7(16)$^b$[1.1(-8)]&6.0(16)$^b$[2.4(-8)]&8.9(16)$^b$[3.5(-8)]&2.5(24)$^f$\\
&1.0(19)$^c$[4.0(-6)]&&&&\\
\hline
VLA4A&6.0(18)$^d$[5.5(-7)]&2.4(16)$^d$[2.2(-9)]&7.5(16)$^d$[6.8(-9)]&4.2(17)$^e$[3.8(-8)]&1.1(25)$^f$\\
\hline
VLA4B&5.0(18)$^d$[4.5(-7)]&1.1(16)$^d$[1.0(-9)]&1.0(17)$^d$[9.0(-9)]&2.5(17)$^e$[2.3(-8)]&1.1(25)$^f$\\
\hline
    \end{tabular}
    \\
    {\scriptsize \noindent $^a$\citep{marc18b},$^b$\citep{bell20},$^c$\citep{simo20},$^d$\citep{bian22},$^e$\citep{diaz22},$^f$\citep{lope15}}
\end{table}
\end{landscape}

\section{Beam dilution effect}
\begin{figure}
\centering
\includegraphics[width=14cm, height=8cm]{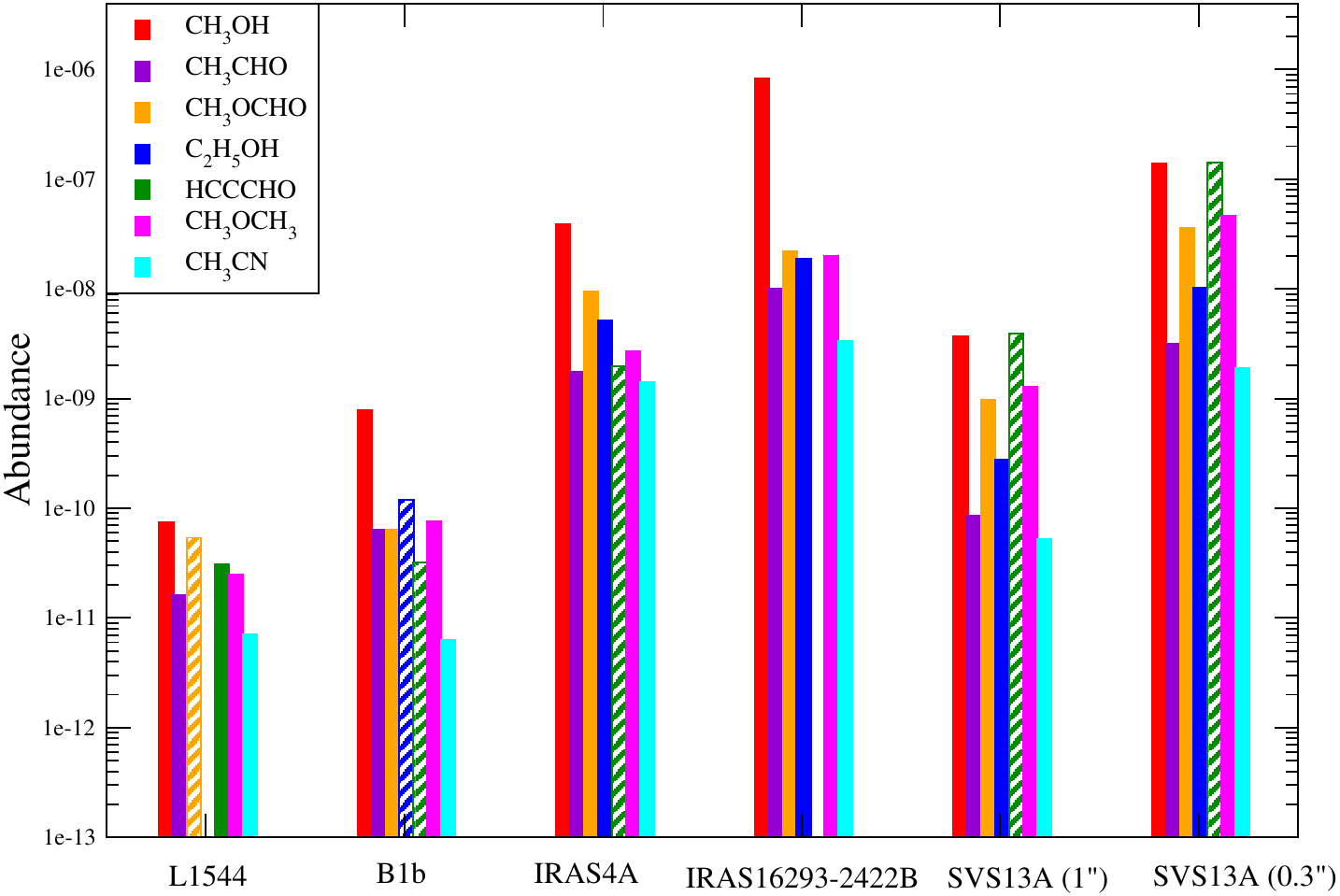}
\caption{Abundances of the COMs at the different evolutionary stages of low-mass star-forming regions. }
\label{fig:beamdil}
\end{figure}

The beam size changes significantly, as the current data have a wide frequency range. In order to observe the effect, we must therefore assume the size of the COMs' emitting region and apply the proper beam dilution factor for each source. By taking into account the beam dilution effect, Figure \ref{fig:beamdil} displays the abundances (obtained from the rotational diagram analysis) at the various evolutionary stages of low-mass star-forming regions. The obtained abundances are also noted in Table \ref{tab:beam-filling-col-density}. By taking the beam dilution factor for each source into account, the obtained intensities in the rotational diagram analysis are scaled. We take into account an average source size for simplicity, and the selection of source size for each source is supported below:\\

{\it L1544:} A compact component of $\sim$ 10$^{''}$ was found in L1544 by \cite{case19, case22}. We assume that COMs are emitting from this region, we take a source size of 10$^{''}$ into account (see Figure \ref{fig:beamdil} and Table \ref{tab:beam-filling-col-density}).\\

{\it B1-b:} The observed line profiles in B1b-S were fitted using a source model with two components, an inner hot and compact component (200 K, 0.35$^{''}$), and an outer and colder component (60 K, 0.6$^{''}$). They also studied using a different component of temperature $\sim$10 K to mimic the envelope. 
In our analysis, we obtain a rotational temperature $\sim$ $10$ K for all molecules in this source, we consider a source size of 10$^"$, which is comparable to the smallest beam size in this data (see Figure \ref{fig:beamdil}).\\ 

{\it IRAS16293-2422:} We consider on the abundances found by the ALMA Protostellar Interferometric Line Survey (PILS) (for further information, see Section \ref{sec:INO}). \\

{\it IRAS4A:} When performing the RD analysis for NH$_2$CHO and HNCO in IRAS4A, \cite{lope15} considered a source size of 0.5$^{''}$. To account for the beam dilution effect, we choose a similar source size of $0.5^{''}$ (see Table \ref{tab:beam-filling-col-density} and Figure \ref{fig:beamdil}). \\

{\it SVS13A:} \cite{lope15} and \cite{bian19} both took source sizes of 1.0$^{''}$ and 0.3$^{''}$ into account when doing RD analyses on the NH$_2$CHO and HNCO in SVS13A, respectively. To account for the beam dilution effect, we use both sizes (see Table \ref{tab:beam-filling-col-density} and Figure \ref{fig:beamdil}). 

Figure \ref{fig:beamdil} illustrates that even with the beam dilution effect, the trend obtained is similar (the abundance is gradually increasing up to class 0 and then decreasing) to that obtained without the beam dilution effect, as shown in Figure \ref{fig:clmdensity}. When we used the abundances from different interferometric observations shown in Figure \ref{fig:inter}, we also found a very similar trend.  

 \begin{table}
     \centering
     {\scriptsize
     \caption{Column density and abundance of the observed species considering the beam dilution factor.}
     \begin{tabular}{|c|c|c|c|c|c|}
          \hline
     Source&Size&N(H$_2$)&Species&Column density&Abundance\\
     \hline
     \hline
          &&&CH$_3$OH&$5.0 \times 10^{13}$&$7.4 \times 10^{-11}$\\
          &&&CH$_3$CHO&$1.1 \times 10^{13}$&$1.6 \times 10^{-11}$\\
          L1544&10$^"$&$6.8 \times 10^{23c}$&CH$_3$OCHO&$3.7 \times 10^{13*}$&$5.4 \times 10^{-11}$\\
         &&&C$_2$H$_5$OH&-&-\\
          &&&HCCCHO&$2.1 \times 10^{13}$&$3.1 \times 10^{-11}$\\
          &&&CH$_3$OCH$_3$&$1.7 \times 10^{13**}$&$2.5 \times 10^{-11}$\\
          &&&CH$_3$CN&$4.85\times 10^{12**}$&$7.1 \times 10^{-12}$\\
          \hline
          &&&CH$_3$OH&$6.2 \times 10^{14}$&$7.8 \times 10^{-10}$\\
          &&&CH$_3$CHO&$5.1 \times 10^{13}$&$6.4 \times 10^{-11}$\\
          B1-b&10$^"$&$7.9 \times 10^{23d}$&CH$_3$OCHO&$5.4 \times 10^{13}$&$6.8 \times 10^{-11}$\\
          &&&C$_2$H$_5$OH&$1.0 \times 10^{14*}$&$1.2 \times 10^{-10}$\\
          &&&HCCCHO&$2.6 \times 10^{13*}$&$3.2 \times 10^{-11}$\\
          &&&CH$_3$OCH$_3$&$6.0 \times 10^{13**}$&$7.6 \times 10^{-11}$\\
          &&&CH$_3$CN&$4.95\times 10^{12**}$&$6.3 \times 10^{-12}$\\
          \hline
           &&&CH$_3$OH (hot)&$9.7 \times 10^{16}$&$3.9 \times 10^{-8}$\\
          &&&CH$_3$CHO&$4.4 \times 10^{15}$&$1.8 \times 10^{-9}$\\
          IRAS4A&0.5$^"$&$2.5 \times 10^{24}$&CH$_3$OCHO&$2.4 \times 10^{16}$&$9.5 \times 10^{-9}$\\
          &\cite{lope15}&\cite{lope15}&C$_2$H$_5$OH&$1.3 \times 10^{16}$&$5.2 \times 10^{-9}$\\
          &&&HCCCHO&$4.9 \times 10^{15*}$&$2.0 \times 10^{-9}$\\
          &&&CH$_3$OCH$_3$&$6.7 \times 10^{15}$&$2.7 \times 10^{-9}$\\
          &&&CH$_3$CN&$3.55\times 10^{15}$&$1.4 \times 10^{-09}$\\
          \hline
           &&&CH$_3$OH&$3.7 \times 10^{16}$&$3.7 \times 10^{-9}$\\
          &&&CH$_3$CHO&$8.6 \times 10^{14}$&$8.6 \times 10^{-11}$\\
          SVS13A&1$^"$&$1.0 \times 10^{25}$&CH$_3$OCHO&$9.7 \times 10^{15}$&$9.7 \times 10^{-10}$\\
          &\cite{lope15}&\cite{lope15}&C$_2$H$_5$OH&$2.8 \times 10^{15}$&$2.8 \times 10^{-10}$\\
          &&&HCCCHO&$3.9 \times 10^{16*}$&$3.9 \times 10^{-9}$\\
          &&&CH$_3$OCH$_3$&$1.3 \times 10^{16a}$&$1.3 \times 10^{-9}$\\
          &&&CH$_3$CN&$5.16\times 10^{14}$&$5.2 \times 10^{-11}$\\
          \hline
           &&&CH$_3$OH&$4.1 \times 10^{17}$&$1.4 \times 10^{-7}$\\
          &&&CH$_3$CHO&$9.6 \times 10^{15}$&$3.2 \times 10^{-9}$\\
          SVS13A&0.3$^"$&$3.0 \times 10^{24}$&CH$_3$OCHO&$1.1 \times 10^{17}$&$3.6 \times 10^{-8}$\\
          &\cite{bian19}&\cite{chen09}&C$_2$H$_5$OH&$3.1 \times 10^{16}$&$1.0 \times 10^{-8}$\\
          &&&HCCCHO&$4.3 \times 10^{17*}$&$1.4 \times 10^{-7}$\\
          &&&CH$_3$OCH$_3$&$1.4 \times 10^{17b}$&$4.7 \times 10^{-8}$\\
          &&&CH$_3$CN&$5.74\times 10^{15}$&$1.9 \times 10^{-09}$\\
          \hline
     \end{tabular}}\\
     { Note: $^*$ indicates upper limit, $^{**}$ indicates LTE derived value, $^{a}$ is scaled value from \cite{bian19}, and $^b$ from \cite{bian19}., $^c$ taken from \cite{hily22} after scaling it for 10$^"$, $^d$ taken from \cite{dani13} after scaling it for 10$^"$.}
     \label{tab:beam-filling-col-density}
 \end{table}

\section{summary} \label{sec:conclusion}
In order to comprehend the chemical and physical evolution of solar-type star-forming regions, we analyse the ASAI large programme data for five sources in this work. In some of these sources, we identify $\rm{CH_3OH}$, $\rm{CH_3CHO}$, $\rm{CH_3OCHO}$, $\rm{C_2H_5OH}$, $\rm{HCCCHO}$, $\rm{CH_3OCH_3}$, and $\rm{CH_3CN}$. We were able to see how evolution has been carried up through the four sources by observing these complex organic molecules. Our early findings in this work are as follows:

$\bullet$ Several species with various transitions have been identified by a thorough analysis of the ASAI large programme data. To measure the excitation temperature and column density of a species, we use a variety of LTE techniques. We tentatively detect certain transitions of $\rm{C_2H_5OH}$ with an upper limit of the column density of $1.0 \times 10^{13}$ cm$^{-2}$ for the first time in B1-b.

$\bullet$ Compared to the prestellar core, we observed a relatively higher column density of these species during the first hydrostatic core phase. Furthermore, no noticeable difference has been seen between class 0 (IRAS4A) and class I phase (SVS13A). There is a significant difference between the two-class 0 objects IRAS4A and IRAS 16293-2422 (perhaps due to the obvious difference in luminosity).
With the exception of HCCCHO (upper limit), COM abundances gradually increase from L1544 to IRAS16293-2422 before decreasing for SVS13A (Figure \ref{fig:clmdensity}). Considering the beam dilution effect (Figure \ref{fig:beamdil}) and interferometric data from the literature (Figure \ref{fig:inter}) both produce a similar pattern.

$\bullet$ With the obtained FWHM for a specific methanol transition, we noticed a pattern. From L1544 (a prestellar core) to SVS13A (a class I object), it seems to rise steadily. Additionally, it shows a consistent rise in FWHM with source luminosity.
Even though the comparison only considers one source from each category, we are unable to clearly identify the trend that has been noticed. To create a trustworthy picture of the chemical evolution during the solar-type star formation process, more high spatial and angular resolution observations of numerous sources at various stages of low-mass star formation are required.

\chapter{Summary, Conclusions and Future Plans} \label{chap:conclusions}
The synthesis of various simple and complex molecules is investigated in this thesis under various astrophysical circumstances. Several species' spectral characteristics are looked at. We used radiative transfer calculations to comprehend the physical parameters of various star-forming regions. Different molecular cloud properties, such as density, temperature, and timescale, are varied to explore the time-dependent and chemical evolution of various exotic, simple, and complex species.\\

The principal findings of this thesis are summarized below:

\section{Summary and Conclusions}
Following are the brief summary and conclusion of this thesis.

\begin{itemize}
\item Monte Carlo Markov Chain (MCMC) method extracts the important physical parameters from the observed line transitions towards different astrophysical environments.

\item We identified three molecules HNCO, NH$_2$CHO, and CH$_3$NCO in G10, which contain peptide-like bonds. Earlier, HNCO and NH$_2$CHO had been identified in G10, but this is the first identification of CH$_3$NCO in this source.

\item Detailed radiative transfer model is implemented towards G31.41+0.31, a hot molecular core. Many COMs are observed toward this hot molecular core. We modeled the observed line transitions of different complex molecules using radiative transfer modeling to generate synthetic spectra and match the observed lines to understand the physical conditions of the source.

\item A detailed time-dependent chemical model is implemented considering different reaction networks available in other databases. The abundances of observed COMs towards G31.41+0.31 are predicted from the chemical modeling.

\item Models simulate the abundances of significant P-bearing species under varied interstellar conditions. For instance, PH$_3$ is minimal in the diffuse cloud region and PDR, whereas PH$_3$ production is noticeable in the hot core region. Furthermore, the destruction of PH$_3$ by H and OH considerably impacts the abundance of PH$_3$. 

\item We analyzed ASAI (an extensive, unbiased, spectral survey) data to understand the evolution of the solar-type star formation process. Five astrophysical sources cover almost all stages of low-mass star formation, from pre-stellar core to FHSC (first hydrostatic core) to class 0 to class I sources. The chemical evolution during this stages are investigated in detail and an overall evolution trend of some COMs are established.

\end{itemize}

\section{Future Research Plans}
I am working on a low-mass star-forming region, a protobinary system. A significant difference is observed between the two binaries of this source. Detailed radiative transfer modeling can be used to investigate the cause of this spectral difference observed.
In future, I plan to work on 3D radiative transfer codes to analyze the observed data in depth. I started using 3D codes like LIME and RADMC3D, which are useful for interpreting the vast interferometric data available.

I also plan to work on modeling the protoplanetary disks in depth to understand the chemical and physical process involved and the chemical evolution of the gas phase and the ice phase molecules with time.

\vfill\eject
 
\appendix
 \chapter{Glossary}

A convenient list of covering units, constants, terms
and acronyms used frequently in this thesis or in cited literature are provided here.

\section{Units and constants}

\begin{quote}
\hskip -2.5 cm
\begin{tabular}{llll}
\hline
\hline
{\bf Symbol} & {\bf Description} & {\bf SI Units} & {\bf CGS Units}\\
\hline
\AA & Angstrom & $\rm{10^{-10}}$ m & $\rm{10^{-8}}$ cm  \\
$\mu$m & Micron & $10^{-6}$ m & $10^{-4}$ cm \\
$G$ &Gravitational constant & $\rm{6.673\times10^{-11} \ N \ m^{2} \ kg^{-2}}$&  $\rm{6.673\times10^{-8} \ dyn \ cm^{2} \ gm^{-2}}$\\
$k_B$ & Boltzmann constant &$\rm{1.3807\times10^{-23}}$ J/K&$\rm{1.3807\times10^{-16}}$ erg/K\\
$h$ & Planck's constant & $\rm{6.6262\times10^{-34}}$ J s&$\rm{6.6262\times10^{-27}}$ erg s\\
$c$ & Speed of light &$\rm{2.997925\times10^{8}\ m \ s^{-1}}$ &$\rm{2.997925\times10^{10} \ cm \ s^{-1}}$\\
eV & Electronvolt & $ 1.602 \times 10^{-19}$ J & $ 1.602 \times 10^{-12}$ erg \\
D & Debye & $3.336\times10^{-30}$ C m & $10^{-18}$ esu cm \\
$m_e$ & Electron mass &$\rm{9.10956\times10^{-31}}$ kg&$\rm{9.10956\times10^{-28}}$ gm\\
$m_p$ & Proton mass &$\rm{1.6726231\times10^{-27}}$ kg&$\rm{1.6726231\times10^{-24}}$ gm\\
$m_H$ & Hydrogen mass &$\rm{1.673534\times10^{-27}}$ kg&$\rm{1.673534\times10^{-24}}$ gm\\
$u$ & Atomic mass unit &$\rm{1.6605402\times10^{-27}}$ kg&$\rm{1.6605402\times10^{-24}}$ gm\\
$E_{\bigoplus}$ & Earth radius & $\rm{6.378\times10^{6}}$ m&$\rm{6.378\times10^{8}}$ cm\\
AU & Astronomical unit & $\rm{1.4959\times10^{11}}$ m&$\rm{1.4959\times10^{13}}$ cm\\
$R_{\odot}$ & Solar radius &$\rm{6.9599\times10^{8}}$ m&$\rm{6.9599\times10^{10}}$ cm\\
ly & Light year &$\rm{9.463\times10^{15}}$ m&$\rm{9.463\times10^{17}}$cm\\
pc & Parsec &$\rm{3.085678\times10^{16}}$ m&$\rm{3.085678\times10^{18}}$ cm\\
$L_{\odot}$ & Solar luminosity & $\rm{3.826\times10^{26} \ J \ s^{-1}}$&$\rm{3.826\times10^{33} \ erg \ s^{-1}}$\\
$M_{\bigoplus}$ & Mass of the Earth & $\rm{5.977\times10^{24}}$ kg&$\rm{5.977\times10^{27}}$ gm\\
$M_{\odot}$ & Solar Mass &$\rm{1.989\times10^{30}}\ kg$&$\rm{1.989\times10^{33}}\ gm$\\
Jy & Jansky & $\rm{1.00\times10^{-26} \ W \ m^{-2} \ Hz^{-1}}$& $\rm{1.00\times10^{-23} \ erg \ sec^{-1}\ cm^{-2} \ Hz^{-1}}$\\
\hline
\hline
\end{tabular} 
\end{quote}

\section{Acronyms}
\begin{quote}
\begin {tabular}{l l}
{\bf ALMA  }&{         Atacama Large Millimeter/submillimeter Array}\\
{\bf ALI  }&{         Accelerated Lambda Iteration}\\
{\bf APEX  }&{         Atacama Pathfinder Experiment}\\ 
{\bf ASAI  }&{         Astrochemical Surveys At IRAM}\\
{\bf BE    }&{         Binding Enegry}\\
{\bf CDMS  }&{         The Cologne Database for Molecular Spectroscopy}\\
{\bf CMB   }&{         Cosmic Microwave Background}\\
{\bf CMMC   }&{         Chemical Model for Molecular Cloud}\\
{\bf COMs  }&{         Complex Organic Molecules}\\
{\bf CR    }&{         Cosmic Rays}\\
{\bf CRESU }&{         $\rm{Cin\acute{e}tique\ de\ R\acute{e}action\ en\ Ecoulement\ Supersonique\ Uniforme}$ (in French)} \\
{\bf CSE   }&{         Circumstellar Envelop}\\
 {\bf DIB   }&{         Diffuse Interstellar Band} \\
 {\bf DNA   }&{         Deoxyribonucleic acid} \\
{\bf EMIR   }&{         Eight MIxer Receiver}\\
{\bf ESA   }&{         European Space Agency}\\
{\bf FHSC   }&{        First Hydrostatic Core}\\
{\bf FWHM  }&{         Full Width at Half Maximum}\\
 {\bf FUV   }&{         Far-Ultraviolet}\\
{\bf GAIA   }&{         Global Astrometric Interferometer for Astrophysics}\\
 {\bf GBT   }&{         Green Bank Telescope}\\
{\bf GREAT   }&{        German REceiver for Astronomy at Terahertz Frequencies}\\
{\bf HIFI  }&{         Heterodyne Instrument for the Far Infrared}\\
{\bf HITRAN}&{         High-resolution Transmission Molecular Absorption Database}\\
{\bf HMC   }&{         Hot Molecular Core}\\
{\bf HMSFRs   }&{         High-mass star-forming regions}\\
{\bf HPBW   }&{         Half Power Beam Width}\\
{\bf HSO   }&{         Herschel Space Observatory}\\
{\bf HST  }&{          Hubble Space Telescope}\\
{\bf IR   }&{         Infrared}\\
{\bf IRAM  }&{         Institute for Radio Astronomy in the Millimeter Range}\\
{\bf ISM   }&{         Interstellar Medium}\\
{\bf ISRF  }&{         Interstellar Radiation Field}\\ 
{\bf JCMT   }&{         James Clerk Maxwell Telescope}\\
{\bf JPL   }&{         Jet Propulsion Laboratory}\\
{\bf JWST  }&{         James Webb Space Telescope}\\
{\bf KIDA  }&{         KInetic Database for Astrochemistry}\\
\end {tabular}
\end{quote}

\begin{table}
\centering
\vskip -1.0cm
\begin{tabular}{l l}

{\bf LAMDA }&{         Leiden Atomic and Molecular Database}\\
{\bf LTE  }&{         Local Thermodynamic Equillibrium}\\
{\bf LVG  }&{         Large Velocity Gradient}\\
{\bf MC    }&{         Monte Carlo}\\
{\bf MCMC    }&{         Monte Carlo Markov Chain}\\
{\bf MP    }&{         M\o ller-Plesset perturbation}\\
{\bf MD    }&{         Molecular Dynamics}\\
{\bf MOPRA    }&{         part of Australia Telescope National Facility}\\
{\bf MRN    }&{         grain modelling was ushered in by
Mathis, Rumpl, \& Nordsieck}\\
{\bf NASA  }&{         National Aeronautics and Space Administration}\\ 
{\bf NOEMA }&{         Northern Extended Millimeter Array}\\
{\bf NGC   }&{         New General Catalog}\\
{\bf NIST  }&{         National Institute of Standards and Technology}\\
{\bf NRAO  }&{         National Radio Astronomy Observatory}\\
{\bf PAHs   }&{         Polycyclic Aromatic Hydrocarbons}\\
{\bf PDR   }&{         Photon-dominated/Photodissociation Region}\\
{\bf PILS   }&{         Protostellar Interferometric Line Survey}\\
{\bf QM/MM   }&{        Quantum Mechanics/Molecular Mechanics}\\
{\bf RD   }&{      Rotational Diagram}\\ 
{\bf RMS   }&{      Route Mean Square}\\
{\bf RNA   }&{      Ribonucleic acid}\\
{\bf ROSINA   }&{      Rosetta Orbiter Spectrometer for Ion and Neutral Analysis}\\ 
{\bf SB   }&{         Surface Brightness}\\
{\bf SED   }&{         Spectral Energy Distribution} \\
{\bf SIFT  }&{         Selected Ion Flow Tube} \\
{\bf SMA   }&{         Submillimeter Array}\\
{\bf SOFIA }&{         The Stratospheric Observatory for Infrared Astronomy}\\
{\bf UC HII   }&{         Ultra-compact HII regions}\\
{\bf UDfA  }&{         UMIST Database for Astrochemistry}\\
{\bf UV    }&{         Ultraviolet}\\
 {\bf VAMDC }&{         Virtual Atomic and Molecular Data Center}\\
{\bf VLA }&{           Very Large Array} \\
{\bf YSO   }&{         Young Stellar Object}\\
\end {tabular}
\end{table}



\vfill\eject
\pagestyle{newheadings}

\backmatter
 \bibliographystyle{aasjournal}
 \scriptsize
 \bibliography{references.bib}

\begin{thebibliography}{}
\expandafter\ifx\csname natexlab\endcsname\relax\def\natexlab#1{#1}\fi
\providecommand{\url}[1]{\href{#1}{#1}}
\providecommand{\dodoi}[1]{doi:~\href{http://doi.org/#1}{\nolinkurl{#1}}}
\providecommand{\doeprint}[1]{\href{http://ascl.net/#1}{\nolinkurl{http://ascl.net/#1}}}
\providecommand{\doarXiv}[1]{\href{https://arxiv.org/abs/#1}{\nolinkurl{https://arxiv.org/abs/#1}}}

\bibitem[{{Ag{\'u}ndez} {et~al.}(2014){Ag{\'u}ndez}, {Cernicharo}, {Decin},
  {Encrenaz}, \& {Teyssier}}]{agun14}
{Ag{\'u}ndez}, M., {Cernicharo}, J., {Decin}, L., {Encrenaz}, P., \&
  {Teyssier}, D. 2014, ApJL, 790, L27, \dodoi{10.1088/2041-8205/790/2/L27}

\bibitem[{{Ag{\'u}ndez} {et~al.}(2007){Ag{\'u}ndez}, {Cernicharo}, \&
  {Gu{\'e}lin}}]{agun07}
{Ag{\'u}ndez}, M., {Cernicharo}, J., \& {Gu{\'e}lin}, M. 2007, ApJL, 662, L91,
  \dodoi{10.1086/519561}

\bibitem[{{Ag{\'u}ndez} {et~al.}(2008){Ag{\'u}ndez}, {Cernicharo}, {Pardo},
  {Gu{\'e}lin}, \& {Phillips}}]{agun08}
{Ag{\'u}ndez}, M., {Cernicharo}, J., {Pardo}, J.~R., {Gu{\'e}lin}, M., \&
  {Phillips}, T.~G. 2008, AAP, 485, L33, \dodoi{10.1051/0004-6361:200810193}

\bibitem[{{Altwegg} {et~al.}(2016){Altwegg}, {Balsiger}, {Bar-Nun},
  {Berthelier}, {Bieler}, {Bochsler}, {Briois}, {Calmonte}, {Combi}, {Cottin},
  {De Keyser}, {Dhooghe}, {Fiethe}, {Fuselier}, {Gasc}, {Gombosi}, {Hansen},
  {Haessig}, {Ja ckel}, {Kopp}, {Korth}, {Le Roy}, {Mall}, {Marty}, {Mousis},
  {Owen}, {Reme}, {Rubin}, {Semon}, {Tzou}, {Waite}, \& {Wurz}}]{altw16}
{Altwegg}, K., {Balsiger}, H., {Bar-Nun}, A., {et~al.} 2016, Science Advances,
  2, e1600285, \dodoi{10.1126/sciadv.1600285}

\bibitem[{{Andre} {et~al.}(1993){Andre}, {Ward-Thompson}, \&
  {Barsony}}]{andr93}
{Andre}, P., {Ward-Thompson}, D., \& {Barsony}, M. 1993, ApJ, 406, 122,
  \dodoi{10.1086/172425}

\bibitem[{{Anglada} {et~al.}(2000){Anglada}, {Rodr{\'\i}guez}, \&
  {Torrelles}}]{angl00}
{Anglada}, G., {Rodr{\'\i}guez}, L.~F., \& {Torrelles}, J.~M. 2000, ApJL, 542,
  L123, \dodoi{10.1086/312933}

\bibitem[{{Aota} \& {Aikawa}(2012)}]{aota12}
{Aota}, T., \& {Aikawa}, Y. 2012, ApJ, 761, 74,
  \dodoi{10.1088/0004-637X/761/1/74}

\bibitem[{{Apr{\`a}} {et~al.}(2020){Apr{\`a}}, {Bylaska}, {de Jong}, {Govind},
  {Kowalski}, {Straatsma}, {Valiev}, {van Dam}, {Alexeev}, {Anchell},
  {Anisimov}, {Aquino}, {Atta-Fynn}, {Autschbach}, {Bauman}, {Becca},
  {Bernholdt}, {Bhaskaran-Nair}, {Bogatko}, {Borowski}, {Boschen}, {Brabec},
  {Bruner}, {Cau{\"e}t}, {Chen}, {Chuev}, {Cramer}, {Daily}, {Deegan},
  {Dunning}, {Dupuis}, {Dyall}, {Fann}, {Fischer}, {Fonari}, {Fr{\"u}chtl},
  {Gagliardi}, {Garza}, {Gawande}, {Ghosh}, {Glaesemann}, {G{\"o}tz},
  {Hammond}, {Helms}, {Hermes}, {Hirao}, {Hirata}, {Jacquelin}, {Jensen},
  {Johnson}, {J{\'o}nsson}, {Kendall}, {Klemm}, {Kobayashi}, {Konkov},
  {Krishnamoorthy}, {Krishnan}, {Lin}, {Lins}, {Littlefield}, {Logsdail},
  {Lopata}, {Ma}, {Marenich}, {Martin del Campo}, {Mejia-Rodriguez}, {Moore},
  {Mullin}, {Nakajima}, {Nascimento}, {Nichols}, {Nichols}, {Nieplocha},
  {Otero-de-la-Roza}, {Palmer}, {Panyala}, {Pirojsirikul}, {Peng}, {Peverati},
  {Pittner}, {Pollack}, {Richard}, {Sadayappan}, {Schatz}, {Shelton},
  {Silverstein}, {Smith}, {Soares}, {Song}, {Swart}, {Taylor}, {Thomas},
  {Tipparaju}, {Truhlar}, {Tsemekhman}, {Van Voorhis},
  {V{\'a}zquez-Mayagoitia}, {Verma}, {Villa}, {Vishnu}, {Vogiatzis}, {Wang},
  {Weare}, {Williamson}, {Windus}, {Woli{\'n}ski}, {Wong}, {Wu}, {Yang}, {Yu},
  {Zacharias}, {Zhang}, {Zhao}, \& {Harrison}}]{apra20}
{Apr{\`a}}, E., {Bylaska}, E.~J., {de Jong}, W.~A., {et~al.} 2020, jcp, 152,
  184102, \dodoi{10.1063/5.0004997}

\bibitem[{{Araya} {et~al.}(2008){Araya}, {Hofner}, {Kurtz}, {Olmi}, \&
  {Linz}}]{aray08}
{Araya}, E., {Hofner}, P., {Kurtz}, S., {Olmi}, L., \& {Linz}, H. 2008, apj,
  675, 420, \dodoi{10.1086/527284}

\bibitem[{{Bachiller} {et~al.}(1998){Bachiller}, {Guilloteau}, {Gueth},
  {Tafalla}, {Dutrey}, {Codella}, \& {Castets}}]{bach98}
{Bachiller}, R., {Guilloteau}, S., {Gueth}, F., {et~al.} 1998, AAP, 339, L49

\bibitem[{{Bains} {et~al.}(2020){Bains}, {Petkowski}, {Seager}, {Ranjan},
  {Sousa-Silva}, {Rimmer}, {Zhan}, {Greaves}, \& {Richards}}]{bain20}
{Bains}, W., {Petkowski}, J.~J., {Seager}, S., {et~al.} 2020, arXiv e-prints,
  arXiv:2009.06499.
\newblock \doarXiv{2009.06499}

\bibitem[{Balucani {et~al.}(2015)Balucani, Ceccarelli, \& Taquet}]{balu15}
Balucani, N., Ceccarelli, C., \& Taquet, V. 2015, Monthly Notices of the Royal
  Astronomical Society: Letters, 449, L16

\bibitem[{Barca {et~al.}(2020)Barca, Bertoni, Carrington, Datta, De~Silva,
  Deustua, Fedorov, Gour, Gunina, Guidez, Harville, Irle, Ivanic, Kowalski,
  Leang, Li, Li, Lutz, Magoulas, Mato, Mironov, Nakata, Pham, Piecuch, Poole,
  Pruitt, Rendell, Roskop, Ruedenberg, Sattasathuchana, Schmidt, Shen,
  Slipchenko, Sosonkina, Sundriyal, Tiwari, Galvez~Vallejo, Westheimer, Wloch,
  Xu, Zahariev, \& Gordon}]{barc20}
Barca, G. M.~J., Bertoni, C., Carrington, L., {et~al.} 2020, The Journal of
  Chemical Physics, 152, 154102, \dodoi{10.1063/5.0005188}

\bibitem[{{Bates} \& {Spitzer}(1951)}]{bate51}
{Bates}, D.~R., \& {Spitzer}, Lyman, J. 1951, ApJ, 113, 441,
  \dodoi{10.1086/145415}

\bibitem[{{Belloche} {et~al.}(2020){Belloche}, {Maury}, {Maret}, {Anderl},
  {Bacmann}, {Andr{\'e}}, {Bontemps}, {Cabrit}, {Codella}, {Gaudel}, {Gueth},
  {Lef{\`e}vre}, {Lefloch}, {Podio}, \& {Testi}}]{bell20}
{Belloche}, A., {Maury}, A.~J., {Maret}, S., {et~al.} 2020, aap, 635, A198,
  \dodoi{10.1051/0004-6361/201937352}

\bibitem[{{Beltr{\'a}n} {et~al.}(2009){Beltr{\'a}n}, {Codella}, {Viti}, {Neri},
  \& {Cesaroni}}]{belt09}
{Beltr{\'a}n}, M.~T., {Codella}, C., {Viti}, S., {Neri}, R., \& {Cesaroni}, R.
  2009, apjl, 690, L93, \dodoi{10.1088/0004-637X/690/2/L93}

\bibitem[{{Beltr{\'a}n} {et~al.}(2018){Beltr{\'a}n}, {Cesaroni}, {Rivilla},
  {S{\'a}nchez-Monge}, {Moscadelli}, {Ahmadi}, {Allen}, {Beuther}, {Etoka},
  {Galli}, {Galv{\'a}n-Madrid}, {Goddi}, {Johnston}, {Klaassen},
  {K{\"o}lligan}, {Kuiper}, {Kumar}, {Maud}, {Mottram}, {Peters}, {Schilke},
  {Testi}, {van der Tak}, \& {Walmsley}}]{belt18}
{Beltr{\'a}n}, M.~T., {Cesaroni}, R., {Rivilla}, V.~M., {et~al.} 2018, aap,
  615, A141, \dodoi{10.1051/0004-6361/201832811}

\bibitem[{{Beltr{\'a}n} {et~al.}(2019){Beltr{\'a}n}, {Padovani}, {Girart},
  {Galli}, {Cesaroni}, {Paladino}, {Anglada}, {Estalella}, {Osorio}, {Rao},
  {S{\'a}nchez-Monge}, \& {Zhang}}]{belt19}
{Beltr{\'a}n}, M.~T., {Padovani}, M., {Girart}, J.~M., {et~al.} 2019, aap, 630,
  A54, \dodoi{10.1051/0004-6361/201935701}

\bibitem[{{Beuther} {et~al.}(2018){Beuther}, {Mottram}, {Ahmadi}, {Bosco},
  {Linz}, {Henning}, {Klaassen}, {Winters}, {Maud}, {Kuiper}, {Semenov},
  {Gieser}, {Peters}, {Urquhart}, {Pudritz}, {Ragan}, {Feng}, {Keto},
  {Leurini}, {Cesaroni}, {Beltran}, {Palau}, {S{\'a}nchez-Monge},
  {Galvan-Madrid}, {Zhang}, {Schilke}, {Wyrowski}, {Johnston}, {Longmore},
  {Lumsden}, {Hoare}, {Menten}, \& {Csengeri}}]{beut18}
{Beuther}, H., {Mottram}, J.~C., {Ahmadi}, A., {et~al.} 2018, aap, 617, A100,
  \dodoi{10.1051/0004-6361/201833021}

\bibitem[{{Bhat} {et~al.}(2022){Bhat}, {Gorai}, {Mondal}, {Chakrabarti}, \&
  {Das}}]{bhat22}
{Bhat}, B., {Gorai}, P., {Mondal}, S.~K., {Chakrabarti}, S.~K., \& {Das}, A.
  2022, Advances in Space Research, 69, 415, \dodoi{10.1016/j.asr.2021.07.011}

\bibitem[{{Bianchi} {et~al.}(2022{\natexlab{a}}){Bianchi},
  {L{\'o}pez-Sepulcre}, {Ceccarelli}, {Codella}, {Podio}, {Bouvier}, \&
  {Enrique-Romero}}]{bian22}
{Bianchi}, E., {L{\'o}pez-Sepulcre}, A., {Ceccarelli}, C., {et~al.}
  2022{\natexlab{a}}, apjl, 928, L3, \dodoi{10.3847/2041-8213/ac5a56}

\bibitem[{{Bianchi} {et~al.}(2019){Bianchi}, {Codella}, {Ceccarelli}, {Vazart},
  {Bachiller}, {Balucani}, {Bouvier}, {De Simone}, {Enrique-Romero}, {Kahane},
  {Lefloch}, {L{\'o}pez-Sepulcre}, {Ospina-Zamudio}, {Podio}, \&
  {Taquet}}]{bian19}
{Bianchi}, E., {Codella}, C., {Ceccarelli}, C., {et~al.} 2019, MNRAS, 483,
  1850, \dodoi{10.1093/mnras/sty2915}

\bibitem[{{Bianchi} {et~al.}(2022{\natexlab{b}}){Bianchi}, {Ceccarelli},
  {Codella}, {L{\'o}pez-Sepulcre}, {Yamamoto}, {Balucani}, {Caselli}, {Podio},
  {Neri}, {Bachiller}, {Favre}, {Fontani}, {Lefloch}, {Sakai}, \&
  {Segura-Cox}}]{bian22b}
{Bianchi}, E., {Ceccarelli}, C., {Codella}, C., {et~al.} 2022{\natexlab{b}},
  aap, 662, A103, \dodoi{10.1051/0004-6361/202141893}

\bibitem[{{Bizzocchi} {et~al.}(2014){Bizzocchi}, {Caselli}, {Spezzano}, \&
  {Leonardo}}]{bizz14}
{Bizzocchi}, L., {Caselli}, P., {Spezzano}, S., \& {Leonardo}, E. 2014, AAP,
  569, A27, \dodoi{10.1051/0004-6361/201423858}

\bibitem[{{Blake} {et~al.}(1995){Blake}, {Sandell}, {van Dishoeck},
  {Groesbeck}, {Mundy}, \& {Aspin}}]{blak95}
{Blake}, G.~A., {Sandell}, G., {van Dishoeck}, E.~F., {et~al.} 1995, ApJ, 441,
  689, \dodoi{10.1086/175392}

\bibitem[{Blake {et~al.}(1987)Blake, Sutton, Masson, \& Phillips}]{blak87}
Blake, G.~A., Sutton, E., Masson, C., \& Phillips, T. 1987, Astrophysical
  Journal, 315, 621

\bibitem[{{Bonnell}(2008)}]{bonn08}
{Bonnell}, I.~A. 2008, in Astronomical Society of the Pacific Conference
  Series, Vol. 390, Pathways Through an Eclectic Universe, ed. J.~H. {Knapen},
  T.~J. {Mahoney}, \& A.~{Vazdekis}, 26

\bibitem[{{Bosco} {et~al.}(2019){Bosco}, {Beuther}, {Ahmadi}, {Mottram},
  {Kuiper}, {Linz}, {Maud}, {Winters}, {Henning}, {Feng}, {Peters}, {Semenov},
  {Klaassen}, {Schilke}, {Urquhart}, {Beltr{\'a}n}, {Lumsden}, {Leurini},
  {Moscadelli}, {Cesaroni}, {S{\'a}nchez-Monge}, {Palau}, {Pudritz},
  {Wyrowski}, \& {Longmore}}]{bosc19}
{Bosco}, F., {Beuther}, H., {Ahmadi}, A., {et~al.} 2019, aap, 629, A10,
  \dodoi{10.1051/0004-6361/201935318}

\bibitem[{{Bottinelli} {et~al.}(2007){Bottinelli}, {Ceccarelli}, {Williams}, \&
  {Lefloch}}]{bott07}
{Bottinelli}, S., {Ceccarelli}, C., {Williams}, J.~P., \& {Lefloch}, B. 2007,
  aap, 463, 601, \dodoi{10.1051/0004-6361:20065139}

\bibitem[{{Bottinelli} {et~al.}(2004){Bottinelli}, {Ceccarelli}, {Lefloch},
  {Williams}, {Castets}, {Caux}, {Cazaux}, {Maret}, {Parise}, \&
  {Tielens}}]{bott04}
{Bottinelli}, S., {Ceccarelli}, C., {Lefloch}, B., {et~al.} 2004, ApJ, 615,
  354, \dodoi{10.1086/423952}

\bibitem[{Bregman {et~al.}(1975)Bregman, Lester, \& Rank}]{breg75}
Bregman, J., Lester, D., \& Rank, D. 1975, The Astrophysical Journal, 202, L55

\bibitem[{{Brinch} \& {Hogerheijde}(2011)}]{brin11}
{Brinch}, C., \& {Hogerheijde}, M.~R. 2011, {LIME: Flexible, Non-LTE Line
  Excitation and Radiation Transfer Method for Millimeter and Far-infrared
  Wavelengths}, Astrophysics Source Code Library, record ascl:1107.012.
\newblock \doeprint{1107.012}

\bibitem[{{Caselli} \& {Ceccarelli}(2012)}]{case12}
{Caselli}, P., \& {Ceccarelli}, C. 2012, AAPR, 20, 56,
  \dodoi{10.1007/s00159-012-0056-x}

\bibitem[{{Caselli} {et~al.}(2002{\natexlab{a}}){Caselli}, {Stantcheva},
  {Shalabiea}, {Shematovich}, \& {Herbst}}]{case02}
{Caselli}, P., {Stantcheva}, T., {Shalabiea}, O., {Shematovich}, V.~I., \&
  {Herbst}, E. 2002{\natexlab{a}}, planss, 50, 1257,
  \dodoi{10.1016/S0032-0633(02)00092-2}

\bibitem[{{Caselli} {et~al.}(1999){Caselli}, {Walmsley}, {Tafalla}, {Dore}, \&
  {Myers}}]{case99}
{Caselli}, P., {Walmsley}, C.~M., {Tafalla}, M., {Dore}, L., \& {Myers}, P.~C.
  1999, ApJL, 523, L165, \dodoi{10.1086/312280}

\bibitem[{{Caselli} {et~al.}(2002{\natexlab{b}}){Caselli}, {Walmsley},
  {Zucconi}, {Tafalla}, {Dore}, \& {Myers}}]{caseb02}
{Caselli}, P., {Walmsley}, C.~M., {Zucconi}, A., {et~al.} 2002{\natexlab{b}},
  ApJ, 565, 344, \dodoi{10.1086/324302}

\bibitem[{{Caselli} {et~al.}(2017){Caselli}, {Bizzocchi}, {Keto}, {Sipil{\"a}},
  {Tafalla}, {Pagani}, {Kristensen}, {van der Tak}, {Walmsley}, {Codella},
  {Nisini}, {Aikawa}, {Faure}, \& {van Dishoeck}}]{case17}
{Caselli}, P., {Bizzocchi}, L., {Keto}, E., {et~al.} 2017, AAP, 603, L1,
  \dodoi{10.1051/0004-6361/201731121}

\bibitem[{{Caselli} {et~al.}(2019){Caselli}, {Pineda}, {Zhao}, {Walmsley},
  {Keto}, {Tafalla}, {Chac{\'o}n-Tanarro}, {Bourke}, {Friesen}, {Galli}, \&
  {Padovani}}]{case19}
{Caselli}, P., {Pineda}, J.~E., {Zhao}, B., {et~al.} 2019, apj, 874, 89,
  \dodoi{10.3847/1538-4357/ab0700}

\bibitem[{{Caselli} {et~al.}(2022){Caselli}, {Pineda}, {Sipil{\"a}}, {Zhao},
  {Redaelli}, {Spezzano}, {Maureira}, {Alves}, {Bizzocchi}, {Bourke},
  {Chac{\'o}n-Tanarro}, {Friesen}, {Galli}, {Harju}, {Jim{\'e}nez-Serra},
  {Keto}, {Li}, {Padovani}, {Schmiedeke}, {Tafalla}, \& {Vastel}}]{case22}
{Caselli}, P., {Pineda}, J.~E., {Sipil{\"a}}, O., {et~al.} 2022, apj, 929, 13,
  \dodoi{10.3847/1538-4357/ac5913}

\bibitem[{{Caux} {et~al.}(2011){Caux}, {Kahane}, {Castets}, {Coutens},
  {Ceccarelli}, {Bacmann}, {Bisschop}, {Bottinelli}, {Comito}, {Helmich},
  {Lefloch}, {Parise}, {Schilke}, {Tielens}, {van Dishoeck}, {Vastel},
  {Wakelam}, \& {Walters}}]{caux11}
{Caux}, E., {Kahane}, C., {Castets}, A., {et~al.} 2011, AAP, 532, A23,
  \dodoi{10.1051/0004-6361/201015399}

\bibitem[{Cazaux {et~al.}(2003)Cazaux, Tielens, Ceccarelli, Castets, Wakelam,
  Caux, Parise, \& Teyssier}]{caza03}
Cazaux, S., Tielens, A., Ceccarelli, C., {et~al.} 2003, The Astrophysical
  Journal, 593, L51

\bibitem[{{Ceccarelli} {et~al.}(2017){Ceccarelli}, {Caselli}, {Fontani},
  {Neri}, {L{\'o}pez-Sepulcre}, {Codella}, {Feng}, {Jim{\'e}nez-Serra},
  {Lefloch}, {Pineda}, {Vastel}, {Alves}, {Bachiller}, {Balucani}, {Bianchi},
  {Bizzocchi}, {Bottinelli}, {Caux}, {Chac{\'o}n-Tanarro}, {Choudhury},
  {Coutens}, {Dulieu}, {Favre}, {Hily-Blant}, {Holdship}, {Kahane}, {Jaber
  Al-Edhari}, {Laas}, {Ospina}, {Oya}, {Podio}, {Pon}, {Punanova}, {Quenard},
  {Rimola}, {Sakai}, {Sims}, {Spezzano}, {Taquet}, {Testi}, {Theul{\'e}},
  {Ugliengo}, {Vasyunin}, {Viti}, {Wiesenfeld}, \& {Yamamoto}}]{cecc17}
{Ceccarelli}, C., {Caselli}, P., {Fontani}, F., {et~al.} 2017, ApJ, 850, 176,
  \dodoi{10.3847/1538-4357/aa961d}

\bibitem[{{Cernicharo} \& {Guelin}(1987)}]{cern87}
{Cernicharo}, J., \& {Guelin}, M. 1987, AAP, 176, 299

\bibitem[{{Cernicharo} {et~al.}(2012){Cernicharo}, {Marcelino}, {Roueff},
  {Gerin}, {Jim{\'e}nez-Escobar}, \& {Mu{\~n}oz Caro}}]{cern12}
{Cernicharo}, J., {Marcelino}, N., {Roueff}, E., {et~al.} 2012, ApJL, 759, L43,
  \dodoi{10.1088/2041-8205/759/2/L43}

\bibitem[{{Cernicharo} {et~al.}(2013){Cernicharo}, {Tercero}, {Fuente},
  {Domenech}, {Cueto}, {Carrasco}, {Herrero}, {Tanarro}, {Marcelino}, {Roueff},
  {Gerin}, \& {Pearson}}]{cern13}
{Cernicharo}, J., {Tercero}, B., {Fuente}, A., {et~al.} 2013, ApJL, 771, L10,
  \dodoi{10.1088/2041-8205/771/1/L10}

\bibitem[{{Cesaroni} {et~al.}(2011){Cesaroni}, {Beltr{\'a}n}, {Zhang},
  {Beuther}, \& {Fallscheer}}]{cesa11}
{Cesaroni}, R., {Beltr{\'a}n}, M.~T., {Zhang}, Q., {Beuther}, H., \&
  {Fallscheer}, C. 2011, aap, 533, A73, \dodoi{10.1051/0004-6361/201117206}

\bibitem[{{Cesaroni} {et~al.}(1994){Cesaroni}, {Churchwell}, {Hofner},
  {Walmsley}, \& {Kurtz}}]{cesa94}
{Cesaroni}, R., {Churchwell}, E., {Hofner}, P., {Walmsley}, C.~M., \& {Kurtz},
  S. 1994, aap, 288, 903

\bibitem[{{Cesaroni} {et~al.}(2010){Cesaroni}, {Hofner}, {Araya}, \&
  {Kurtz}}]{cesa10}
{Cesaroni}, R., {Hofner}, P., {Araya}, E., \& {Kurtz}, S. 2010, aap, 509, A50,
  \dodoi{10.1051/0004-6361/200912877}

\bibitem[{{Cesaroni} {et~al.}(1998){Cesaroni}, {Hofner}, {Walmsley}, \&
  {Churchwell}}]{cesa98}
{Cesaroni}, R., {Hofner}, P., {Walmsley}, C.~M., \& {Churchwell}, E. 1998, aap,
  331, 709

\bibitem[{{Cesaroni} {et~al.}(1992){Cesaroni}, {Walmsley}, \&
  {Churchwell}}]{cesa92}
{Cesaroni}, R., {Walmsley}, C.~M., \& {Churchwell}, E. 1992, aap, 256, 618

\bibitem[{{Chakrabarti} \& {Chakrabarti}(2000a)}]{chak00a}
{Chakrabarti}, S., \& {Chakrabarti}, S.~K. 2000a, aap, 354, L6.
\newblock \doarXiv{astro-ph/0001079}

\bibitem[{{Chakrabarti} \& {Chakrabarti}(2000b)}]{chak00b}
{Chakrabarti}, S.~K., \& {Chakrabarti}, S. 2000b, Indian Journal of Physics
  Section B, 74B, 97.
\newblock \doarXiv{astro-ph/0003271}

\bibitem[{{Chakrabarti} {et~al.}(2006a){Chakrabarti}, {Das}, {Acharyya}, \&
  {Chakrabarti}}]{chak06a}
{Chakrabarti}, S.~K., {Das}, A., {Acharyya}, K., \& {Chakrabarti}, S. 2006a,
  aap, 457, 167, \dodoi{10.1051/0004-6361:20065335}

\bibitem[{{Chakrabarti} {et~al.}(2006b){Chakrabarti}, {Das}, {Acharyya}, \&
  {Chakrabarti}}]{chak06b}
---. 2006b, Bulletin of the Astronomical Society of India, 34, 299.
\newblock \doarXiv{0806.4679}

\bibitem[{{Chakrabarti} {et~al.}(2015){Chakrabarti}, {Majumdar}, {Das}, \&
  {Chakrabarti}}]{chak15}
{Chakrabarti}, S.~K., {Majumdar}, L., {Das}, A., \& {Chakrabarti}, S. 2015,
  APSS, 357, 90, \dodoi{10.1007/s10509-015-2239-1}

\bibitem[{{Chantzos} {et~al.}(2020){Chantzos}, {Rivilla}, {Vasyunin},
  {Redaelli}, {Bizzocchi}, {Fontani}, \& {Caselli}}]{chan20}
{Chantzos}, J., {Rivilla}, V.~M., {Vasyunin}, A., {et~al.} 2020, AAP, 633, A54,
  \dodoi{10.1051/0004-6361/201936531}

\bibitem[{Charnley \& Millar(1994)}]{char94}
Charnley, S.~B., \& Millar, T.~J. 1994, Monthly Notices of the Royal
  Astronomical Society, 270, 570, \dodoi{10.1093/mnras/270.3.570}

\bibitem[{{Charnley} {et~al.}(1992){Charnley}, {Tielens}, \& {Millar}}]{char92}
{Charnley}, S.~B., {Tielens}, A.~G.~G.~M., \& {Millar}, T.~J. 1992, ApJL, 399,
  L71, \dodoi{10.1086/186609}

\bibitem[{{Chastaing} {et~al.}(2001){Chastaing}, {Le Picard}, {Sims}, \&
  {Smith}}]{chas01}
{Chastaing}, D., {Le Picard}, S.~D., {Sims}, I.~R., \& {Smith}, I.~W.~M. 2001,
  aap, 365, 241, \dodoi{10.1051/0004-6361:20000026}

\bibitem[{Chen {et~al.}(2009)Chen, Launhardt, \& Henning}]{chen09}
Chen, X., Launhardt, R., \& Henning, T. 2009, The Astrophysical Journal, 691,
  1729

\bibitem[{{Cheung} {et~al.}(1968){Cheung}, {Rank}, {Townes}, {Thornton}, \&
  {Welch}}]{cheu68}
{Cheung}, A.~C., {Rank}, D.~M., {Townes}, C.~H., {Thornton}, D.~D., \& {Welch},
  W.~J. 1968, PRL, 21, 1701, \dodoi{10.1103/PhysRevLett.21.1701}

\bibitem[{{Choi} {et~al.}(1999){Choi}, {Panis}, \& {Evans}}]{choi99}
{Choi}, M., {Panis}, J.-F., \& {Evans}, Neal~J., I. 1999, ApJS, 122, 519,
  \dodoi{10.1086/313222}

\bibitem[{{Churchwell} {et~al.}(1990){Churchwell}, {Walmsley}, \&
  {Cesaroni}}]{chur90}
{Churchwell}, E., {Walmsley}, C.~M., \& {Cesaroni}, R. 1990, aaps, 83, 119

\bibitem[{{Ciolek} \& {Basu}(2000)}]{ciol00}
{Ciolek}, G.~E., \& {Basu}, S. 2000, ApJ, 529, 925, \dodoi{10.1086/308293}

\bibitem[{Cuppen {et~al.}(2017)Cuppen, Walsh, Lamberts, Semenov, Garrod,
  Penteado, \& Ioppolo}]{cupp17}
Cuppen, H., Walsh, C., Lamberts, T., {et~al.} 2017, Space Science Reviews, 212,
  1

\bibitem[{{Dalgarno}(2008)}]{dalg08}
{Dalgarno}, A. 2008, araa, 46, 1,
  \dodoi{10.1146/annurev.astro.46.060407.145216}

\bibitem[{{Daniel} {et~al.}(2006){Daniel}, {Cernicharo}, \&
  {Dubernet}}]{dani06}
{Daniel}, F., {Cernicharo}, J., \& {Dubernet}, M.~L. 2006, ApJ, 648, 461,
  \dodoi{10.1086/505738}

\bibitem[{Daniel {et~al.}(2013)Daniel, G{\'e}rin, Roueff, Cernicharo,
  Marcelino, Lique, Lis, Teyssier, Biver, \& Bockel{\'e}e-Morvan}]{dani13}
Daniel, F., G{\'e}rin, M., Roueff, E., {et~al.} 2013, Astronomy \&
  Astrophysics, 560, A3

\bibitem[{{Das} {et~al.}(2008){Das}, {Acharyya}, {Chakrabarti}, \&
  {Chakrabarti}}]{das08a}
{Das}, A., {Acharyya}, K., {Chakrabarti}, S., \& {Chakrabarti}, S.~K. 2008,
  AAP, 486, 209, \dodoi{10.1051/0004-6361:20078422}

\bibitem[{{Das} {et~al.}(2010){Das}, {Acharyya}, \& {Chakrabarti}}]{das10}
{Das}, A., {Acharyya}, K., \& {Chakrabarti}, S.~K. 2010, mnras, 409, 789,
  \dodoi{10.1111/j.1365-2966.2010.17343.x}

\bibitem[{{Das} \& {Chakrabarti}(2011)}]{das11}
{Das}, A., \& {Chakrabarti}, S.~K. 2011, mnras, 418, 545,
  \dodoi{10.1111/j.1365-2966.2011.19503.x}

\bibitem[{{Das} {et~al.}(2019){Das}, {Gorai}, \& {Chakrabarti}}]{das19}
{Das}, A., {Gorai}, P., \& {Chakrabarti}, S.~K. 2019, AAP, 628, A73,
  \dodoi{10.1051/0004-6361/201834923}

\bibitem[{{Das} {et~al.}(2013a){Das}, {Majumdar}, {Chakrabarti}, {Saha}, \&
  {Chakrabarti}}]{das13a}
{Das}, A., {Majumdar}, L., {Chakrabarti}, S.~K., {Saha}, R., \& {Chakrabarti},
  S. 2013a, mnras, 433, 3152, \dodoi{10.1093/mnras/stt958}

\bibitem[{{Das} {et~al.}(2015{\natexlab{a}}){Das}, {Majumdar}, {Chakrabarti},
  \& {Sahu}}]{dasa15}
{Das}, A., {Majumdar}, L., {Chakrabarti}, S.~K., \& {Sahu}, D.
  2015{\natexlab{a}}, NA, 35, 53, \dodoi{10.1016/j.newast.2014.07.006}

\bibitem[{{Das} {et~al.}(2015{\natexlab{b}}){Das}, {Majumdar}, {Chakrabarti},
  \& {Sahu}}]{das15a}
---. 2015{\natexlab{b}}, na, 35, 53, \dodoi{10.1016/j.newast.2014.07.006}

\bibitem[{{Das} {et~al.}(2016){Das}, {Sahu}, {Majumdar}, \&
  {Chakrabarti}}]{das16}
{Das}, A., {Sahu}, D., {Majumdar}, L., \& {Chakrabarti}, S.~K. 2016, mnras,
  455, 540, \dodoi{10.1093/mnras/stv2264}

\bibitem[{{Das} {et~al.}(2021){Das}, {Sil}, {Ghosh}, {Gorai}, {Adak},
  {Samanta}, \& {Chakrabarti}}]{das21}
{Das}, A., {Sil}, M., {Ghosh}, R., {et~al.} 2021, Frontiers in Astronomy and
  Space Sciences, 8, 78, \dodoi{10.3389/fspas.2021.671622}

\bibitem[{{Das} {et~al.}(2018){Das}, {Sil}, {Gorai}, {Chakrabarti}, \&
  {Loison}}]{das18}
{Das}, A., {Sil}, M., {Gorai}, P., {Chakrabarti}, S.~K., \& {Loison}, J.~C.
  2018, apjs, 237, 9, \dodoi{10.3847/1538-4365/aac886}

\bibitem[{{De Beck} {et~al.}(2013){De Beck}, {Kami{\'n}ski}, {Patel}, {Young},
  {Gottlieb}, {Menten}, \& {Decin}}]{debe13}
{De Beck}, E., {Kami{\'n}ski}, T., {Patel}, N.~A., {et~al.} 2013, AAP, 558,
  A132, \dodoi{10.1051/0004-6361/201321349}

\bibitem[{{De Simone} {et~al.}(2020{\natexlab{a}}){De Simone}, {Ceccarelli},
  {Codella}, {Svoboda}, {Chandler}, {Bouvier}, {Yamamoto}, {Sakai}, {Caselli},
  {Favre}, {Loinard}, {Lefloch}, {Liu}, {L{\'o}pez-Sepulcre}, {Pineda},
  {Taquet}, \& {Testi}}]{des20a}
{De Simone}, M., {Ceccarelli}, C., {Codella}, C., {et~al.} 2020{\natexlab{a}},
  ApJL, 896, L3, \dodoi{10.3847/2041-8213/ab8d41}

\bibitem[{{De Simone} {et~al.}(2020{\natexlab{b}}){De Simone}, {Ceccarelli},
  {Codella}, {Svoboda}, {Chandler}, {Bouvier}, {Yamamoto}, {Sakai}, {Caselli},
  {Favre}, {Loinard}, {Lefloch}, {Liu}, {L{\'o}pez-Sepulcre}, {Pineda},
  {Taquet}, \& {Testi}}]{simo20}
---. 2020{\natexlab{b}}, apjl, 896, L3, \dodoi{10.3847/2041-8213/ab8d41}

\bibitem[{{Di Francesco} {et~al.}(2001{\natexlab{a}}){Di Francesco}, {Myers},
  {Wilner}, {Ohashi}, \& {Mardones}}]{difr01}
{Di Francesco}, J., {Myers}, P.~C., {Wilner}, D.~J., {Ohashi}, N., \&
  {Mardones}, D. 2001{\natexlab{a}}, apj, 562, 770, \dodoi{10.1086/323854}

\bibitem[{{Di Francesco} {et~al.}(2001{\natexlab{b}}){Di Francesco}, {Myers},
  {Wilner}, {Ohashi}, \& {Mardones}}]{fran01}
---. 2001{\natexlab{b}}, ApJ, 562, 770, \dodoi{10.1086/323854}

\bibitem[{{Diaz-Rodriguez} {et~al.}(2022){Diaz-Rodriguez}, {Anglada},
  {Bl{\'a}zquez-Calero}, {Osorio}, {G{\'o}mez}, {Fuller}, {Estalella},
  {Torrelles}, {Cabrit}, {Rodr{\'\i}guez}, {Lef{\`e}vre}, {Mac{\'\i}as},
  {Carrasco-Gonz{\'a}lez}, {Zapata}, {de Gregorio-Monsalvo}, \& {Ho}}]{diaz22}
{Diaz-Rodriguez}, A.~K., {Anglada}, G., {Bl{\'a}zquez-Calero}, G., {et~al.}
  2022, ApJ, 930, 91, \dodoi{10.3847/1538-4357/ac3b50}

\bibitem[{{Doty} {et~al.}(2005){Doty}, {Everett}, {Shirley}, {Evans}, \&
  {Palotti}}]{doty05}
{Doty}, S.~D., {Everett}, S.~E., {Shirley}, Y.~L., {Evans}, N.~J., \&
  {Palotti}, M.~L. 2005, mnras, 359, 228,
  \dodoi{10.1111/j.1365-2966.2005.08893.x}

\bibitem[{{Dubernet} {et~al.}(2013){Dubernet}, {Alexander}, {Ba},
  {Balakrishnan}, {Balan{\c{c}}a}, {Ceccarelli}, {Cernicharo}, {Daniel},
  {Dayou}, {Doronin}, {Dumouchel}, {Faure}, {Feautrier}, {Flower}, {Grosjean},
  {Halvick}, {K{\l}os}, {Lique}, {McBane}, {Marinakis}, {Moreau}, {Moszynski},
  {Neufeld}, {Roueff}, {Schilke}, {Spielfiedel}, {Stancil}, {Stoecklin},
  {Tennyson}, {Yang}, {Vasserot}, \& {Wiesenfeld}}]{dube13}
{Dubernet}, M.~L., {Alexander}, M.~H., {Ba}, Y.~A., {et~al.} 2013, AAP, 553,
  A50, \dodoi{10.1051/0004-6361/201220630}

\bibitem[{{Endres} {et~al.}(2016){Endres}, {Schlemmer}, {Schilke}, {Stutzki},
  \& {M{\"u}ller}}]{endr16}
{Endres}, C.~P., {Schlemmer}, S., {Schilke}, P., {Stutzki}, J., \&
  {M{\"u}ller}, H. S.~P. 2016, Journal of Molecular Spectroscopy, 327, 95,
  \dodoi{10.1016/j.jms.2016.03.005}

\bibitem[{{Estalella} {et~al.}(2019){Estalella}, {Anglada},
  {D{\'\i}az-Rodr{\'\i}guez}, \& {Mayen-Gijon}}]{esta19}
{Estalella}, R., {Anglada}, G., {D{\'\i}az-Rodr{\'\i}guez}, A.~K., \&
  {Mayen-Gijon}, J.~M. 2019, aap, 626, A84, \dodoi{10.1051/0004-6361/201834998}

\bibitem[{{Ferland} {et~al.}(2017){Ferland}, {Chatzikos}, {Guzm{\'a}n},
  {Lykins}, {van Hoof}, {Williams}, {Abel}, {Badnell}, {Keenan}, {Porter}, \&
  {Stancil}}]{ferl17}
{Ferland}, G.~J., {Chatzikos}, M., {Guzm{\'a}n}, F., {et~al.} 2017, RMxAA, 53,
  385.
\newblock \doarXiv{1705.10877}

\bibitem[{{Fletcher} {et~al.}(2009){Fletcher}, {Orton}, {Teanby}, \&
  {Irwin}}]{flet09}
{Fletcher}, L.~N., {Orton}, G.~S., {Teanby}, N.~A., \& {Irwin}, P.~G.~J. 2009,
  Icarus, 202, 543, \dodoi{10.1016/j.icarus.2009.03.023}

\bibitem[{{Fonfr{\'\i}a Exp{\'o}sito} {et~al.}(2006){Fonfr{\'\i}a
  Exp{\'o}sito}, {Ag{\'u}ndez}, {Tercero}, {Pardo}, \& {Cernicharo}}]{fonf06}
{Fonfr{\'\i}a Exp{\'o}sito}, J.~P., {Ag{\'u}ndez}, M., {Tercero}, B., {Pardo},
  J.~R., \& {Cernicharo}, J. 2006, ApJL, 646, L127, \dodoi{10.1086/507104}

\bibitem[{{Fontani} {et~al.}(2016){Fontani}, {Rivilla}, {Caselli}, {Vasyunin},
  \& {Palau}}]{font16}
{Fontani}, F., {Rivilla}, V.~M., {Caselli}, P., {Vasyunin}, A., \& {Palau}, A.
  2016, ApJL, 822, L30, \dodoi{10.3847/2041-8205/822/2/L30}

\bibitem[{{Fontani} {et~al.}(2019){Fontani}, {Rivilla}, {van der Tak},
  {Mininni}, {Beltr{\'a}n}, \& {Caselli}}]{font19}
{Fontani}, F., {Rivilla}, V.~M., {van der Tak}, F.~F.~S., {et~al.} 2019, MNRAS,
  489, 4530, \dodoi{10.1093/mnras/stz2446}

\bibitem[{Foster \& Chevalier(1993)}]{fost93}
Foster, P.~N., \& Chevalier, R.~A. 1993, The Astrophysical Journal, 416, 303

\bibitem[{Frisch {et~al.}(2013)Frisch, Trucks, Schlegel, Scuseria, Robb,
  Cheeseman, Scalmani, Barone, Mennucci, Petersson, Nakatsuji, Caricato, Li,
  Hratchian, Izmaylov, Bloino, Zheng, Sonnenberg, Hada, Ehara, Toyota, Fukuda,
  Hasegawa, Ishida, Nakajima, Honda, Kitao, Nakai, Vreven, J.~A.~Montgomery,
  Peralta, Ogliaro, Bearpark, Heyd, Brothers, Kudin, Staroverov, Keith,
  Kobayashi, Normand, Raghavachari, Rendell, Burant, Iyengar, Tomasi, Cossi,
  Rega, Millam, Klene, Knox, Cross, Bakken, Adamo, Jaramillo, Gomperts,
  Stratmann, Yazyev, Austin, Cammi, Pomelli, Ochterski, Martin, Morokuma,
  Zakrzewski, Voth, Salvador, Dannenberg, Dapprich, Daniels, Farkas, Foresman,
  Ortiz, Cioslowski, \& Fox}]{fris13}
Frisch, M.~J., Trucks, G.~W., Schlegel, H.~B., {et~al.} 2013, Gaussian 09
  {R}evision {D}.01

\bibitem[{{Fuente} {et~al.}(2016){Fuente}, {Cernicharo}, {Roueff}, {Gerin},
  {Pety}, {Marcelino}, {Bachiller}, {Lefloch}, {Roncero}, \& {Aguado}}]{fuen16}
{Fuente}, A., {Cernicharo}, J., {Roueff}, E., {et~al.} 2016, AAP, 593, A94,
  \dodoi{10.1051/0004-6361/201628285}

\bibitem[{Gagniuc(2017)}]{gagn17}
Gagniuc, P. 2017, Markov Chains: From Theory to Implementation and
  Experimentation (Wiley).
\newblock \url{https://books.google.co.in/books?id=oNYtDwAAQBAJ}

\bibitem[{{Garrod} \& {Herbst}(2006)}]{garr06b}
{Garrod}, R.~T., \& {Herbst}, E. 2006, AAP, 457, 927,
  \dodoi{10.1051/0004-6361:20065560}

\bibitem[{{Garrod} {et~al.}(2009){Garrod}, {Vasyunin}, {Semenov}, {Wiebe}, \&
  {Henning}}]{garr09}
{Garrod}, R.~T., {Vasyunin}, A.~I., {Semenov}, D.~A., {Wiebe}, D.~S., \&
  {Henning}, T. 2009, ApJL, 700, L43, \dodoi{10.1088/0004-637X/700/1/L43}

\bibitem[{{Garrod} {et~al.}(2007){Garrod}, {Wakelam}, \& {Herbst}}]{garr07}
{Garrod}, R.~T., {Wakelam}, V., \& {Herbst}, E. 2007, aap, 467, 1103,
  \dodoi{10.1051/0004-6361:20066704}

\bibitem[{Gerin {et~al.}(2009)Gerin, Marcelino, Biver, Roueff, Coudert,
  Elkeurti, Lis, \& Bockel{\'e}e-Morvan}]{geri09}
Gerin, M., Marcelino, N., Biver, N., {et~al.} 2009, Astronomy \& Astrophysics,
  498, L9

\bibitem[{{Gerin} {et~al.}(2015){Gerin}, {Pety}, {Fuente}, {Cernicharo},
  {Commer{\c{c}}on}, \& {Marcelino}}]{geri15}
{Gerin}, M., {Pety}, J., {Fuente}, A., {et~al.} 2015, AAP, 577, L2,
  \dodoi{10.1051/0004-6361/201525777}

\bibitem[{Girart {et~al.}(2009)Girart, Beltr{\'a}n, Zhang, Rao, \&
  Estalella}]{gira09}
Girart, J.~M., Beltr{\'a}n, M.~T., Zhang, Q., Rao, R., \& Estalella, R. 2009,
  Science, 324, 1408, \dodoi{10.1126/science.1171807}

\bibitem[{{Goldsmith} \& {Langer}(1999)}]{gold99}
{Goldsmith}, P.~F., \& {Langer}, W.~D. 1999, ApJ, 517, 209,
  \dodoi{10.1086/307195}

\bibitem[{{Gorai} {et~al.}(2020){Gorai}, {Bhat}, {Sil}, {Mondal}, {Ghosh},
  {Chakrabarti}, \& {Das}}]{gora20}
{Gorai}, P., {Bhat}, B., {Sil}, M., {et~al.} 2020, ApJ, 895, 86,
  \dodoi{10.3847/1538-4357/ab8871}

\bibitem[{{Gorai} {et~al.}(2017{\natexlab{a}}){Gorai}, {Das}, {Das},
  {Sivaraman}, {Etim}, \& {Chakrabarti}}]{gor17a}
{Gorai}, P., {Das}, A., {Das}, A., {et~al.} 2017{\natexlab{a}}, apj, 836, 70,
  \dodoi{10.3847/1538-4357/836/1/70}

\bibitem[{{Gorai} {et~al.}(2017{\natexlab{b}}){Gorai}, {Das}, {Majumdar},
  {Chakrabarti}, {Sivaraman}, \& {Herbst}}]{gor17b}
{Gorai}, P., {Das}, A., {Majumdar}, L., {et~al.} 2017{\natexlab{b}}, Molecular
  Astrophysics, 6, 36, \dodoi{10.1016/j.molap.2017.01.004}

\bibitem[{{Gorai} {et~al.}(2021){Gorai}, {Das}, {Shimonishi}, {Sahu}, {Mondal},
  {Bhat}, \& {Chakrabarti}}]{gora21}
{Gorai}, P., {Das}, A., {Shimonishi}, T., {et~al.} 2021, AaJ, 907, 108,
  \dodoi{10.3847/1538-4357/abc9c4}

\bibitem[{{Greaves} {et~al.}(2020){Greaves}, {Richards}, {Bains}, {Rimmer},
  {Sagawa}, {Clements}, {Seager}, {Petkowski}, {Sousa-Silva}, {Ranjan},
  {Drabek-Maunder}, {Fraser}, {Cartwright}, {Mueller-Wodarg}, {Zhan},
  {Friberg}, {Coulson}, {Lee}, \& {Hoge}}]{grea20}
{Greaves}, J.~S., {Richards}, A. M.~S., {Bains}, W., {et~al.} 2020, Nature
  Astronomy, \dodoi{10.1038/s41550-020-1174-4}

\bibitem[{{Guan} \& {Krone}(2007)}]{guan07}
{Guan}, Y., \& {Krone}, S.~M. 2007, arXiv Mathematics e-prints, math/0703021.
\newblock \doarXiv{math/0703021}

\bibitem[{{Guelin} {et~al.}(1990){Guelin}, {Cernicharo}, {Paubert}, \&
  {Turner}}]{guel90}
{Guelin}, M., {Cernicharo}, J., {Paubert}, G., \& {Turner}, B.~E. 1990, AAP,
  230, L9

\bibitem[{{G{\"u}sten} {et~al.}(2019){G{\"u}sten}, {Wiesemeyer}, {Neufeld},
  {Menten}, {Graf}, {Jacobs}, {Klein}, {Ricken}, {Risacher}, \&
  {Stutzki}}]{gust19}
{G{\"u}sten}, R., {Wiesemeyer}, H., {Neufeld}, D., {et~al.} 2019, Nature, 568,
  357, \dodoi{10.1038/s41586-019-1090-x}

\bibitem[{{Halfen} {et~al.}(2008){Halfen}, {Clouthier}, \& {Ziurys}}]{half08}
{Halfen}, D.~T., {Clouthier}, D.~J., \& {Ziurys}, L.~M. 2008, ApJL, 677, L101,
  \dodoi{10.1086/588024}

\bibitem[{{Han} {et~al.}(2014){Han}, {Wang}, {Wright}, {Feng}, {Zhao},
  {Fakhouri}, {Brown}, \& {Hancock}}]{han14}
{Han}, E., {Wang}, S.~X., {Wright}, J.~T., {et~al.} 2014, pasp, 126, 827,
  \dodoi{10.1086/678447}

\bibitem[{{Hasegawa} {et~al.}(1992){Hasegawa}, {Herbst}, \& {Leung}}]{hase92}
{Hasegawa}, T.~I., {Herbst}, E., \& {Leung}, C.~M. 1992, ApJS, 82, 167,
  \dodoi{10.1086/191713}

\bibitem[{{Hasegawa} \& {Pudritz}(2014)}]{hase14}
{Hasegawa}, Y., \& {Pudritz}, R.~E. 2014, apj, 794, 25,
  \dodoi{10.1088/0004-637X/794/1/25}

\bibitem[{{Hastings}(1970)}]{hast70}
{Hastings}, W.~K. 1970, Biometrika, 57, 97, \dodoi{10.1093/biomet/57.1.97}

\bibitem[{{Herbst} \& {van Dishoeck}(2009)}]{herb09}
{Herbst}, E., \& {van Dishoeck}, E.~F. 2009, ARAA, 47, 427,
  \dodoi{10.1146/annurev-astro-082708-101654}

\bibitem[{{Herv{\'\i}as-Caimapo} {et~al.}(2019){Herv{\'\i}as-Caimapo},
  {Merello}, {Bronfman}, {{\r{A}}ke-Nyman}, {Garay}, {Lo}, {Evans},
  {L{\'o}pez-Calder{\'o}n}, \& {Mendoza}}]{herv19}
{Herv{\'\i}as-Caimapo}, C., {Merello}, M., {Bronfman}, L., {et~al.} 2019, apj,
  872, 200, \dodoi{10.3847/1538-4357/aaf9ac}

\bibitem[{Hily-Blant {et~al.}(2022)Hily-Blant, Des~For{\^e}ts, Faure, \&
  Lique}]{hily22}
Hily-Blant, P., Des~For{\^e}ts, G.~P., Faure, A., \& Lique, F. 2022, Astronomy
  and Astrophysics-A\&A, 658, A168

\bibitem[{{Hogerheijde} \& {van der Tak}(2000)}]{hoge00}
{Hogerheijde}, M.~R., \& {van der Tak}, F.~F.~S. 2000, AAP, 362, 697.
\newblock \doarXiv{astro-ph/0008169}

\bibitem[{{Holdship} {et~al.}(2019){Holdship}, {Viti}, {Codella}, {Rawlings},
  {Jimenez-Serra}, {Ayalew}, {Curtis}, {Habib}, {Lawrence}, {Warsame}, \&
  {Horn}}]{hold19}
{Holdship}, J., {Viti}, S., {Codella}, C., {et~al.} 2019, ApJ, 880, 138,
  \dodoi{10.3847/1538-4357/ab1f8f}

\bibitem[{{Huang} \& {Hirano}(2013)}]{huan13}
{Huang}, Y.-H., \& {Hirano}, N. 2013, ApJ, 766, 131,
  \dodoi{10.1088/0004-637X/766/2/131}

\bibitem[{Hudson \& Gerakines(2019)}]{huds19}
Hudson, R.~L., \& Gerakines, P.~A. 2019, Monthly Notices of the Royal
  Astronomical Society, 482, 4009

\bibitem[{{Hung} {et~al.}(2019){Hung}, {Liu}, {Su}, {He}, {Lee}, {Takahashi},
  \& {Chen}}]{hung19}
{Hung}, T., {Liu}, S.-Y., {Su}, Y.-N., {et~al.} 2019, apj, 872, 61,
  \dodoi{10.3847/1538-4357/aafc23}

\bibitem[{{Hwang} {et~al.}(2008){Hwang}, {Wei}, {Chang}, {Wang}, {Shi}, \&
  {Chen}}]{hwan08}
{Hwang}, Y.-J., {Wei}, T., {Chang}, S.-W., {et~al.} 2008, in 2008 Global
  Symposium on Millimeter Waves, 205--208

\bibitem[{Irvine {et~al.}(1988)Irvine, Brown, Cragg, Friberg, Godfrey, Kaifu,
  Matthews, Ohishi, Suzuki, \& Takeo}]{irvi88}
Irvine, W.~M., Brown, R., Cragg, D., {et~al.} 1988, The Astrophysical Journal,
  335, L89

\bibitem[{{Jaber} {et~al.}(2014){Jaber}, {Ceccarelli}, {Kahane}, \&
  {Caux}}]{jabe14}
{Jaber}, A.~A., {Ceccarelli}, C., {Kahane}, C., \& {Caux}, E. 2014, ApJ, 791,
  29, \dodoi{10.1088/0004-637X/791/1/29}

\bibitem[{{Jacobsen} {et~al.}(2019){Jacobsen}, {J{\o}rgensen}, {Di Francesco},
  {Evans}, {Choi}, \& {Lee}}]{jaco19}
{Jacobsen}, S.~K., {J{\o}rgensen}, J.~K., {Di Francesco}, J., {et~al.} 2019,
  AAP, 629, A29, \dodoi{10.1051/0004-6361/201833214}

\bibitem[{{Jarosewich}(1990)}]{jaro90}
{Jarosewich}, E. 1990, Meteoritics, 25, 323,
  \dodoi{10.1111/j.1945-5100.1990.tb00717.x}

\bibitem[{Jim{\'{e}}nez-Serra {et~al.}(2018)Jim{\'{e}}nez-Serra, Viti,
  Qu{\'{e}}nard, \& Holdship}]{jime18}
Jim{\'{e}}nez-Serra, I., Viti, S., Qu{\'{e}}nard, D., \& Holdship, J. 2018, The
  Astrophysical Journal, 862, 128, \dodoi{10.3847/1538-4357/aacdf2}

\bibitem[{{Jim{\'e}nez-Serra} {et~al.}(2016){Jim{\'e}nez-Serra}, {Vasyunin},
  {Caselli}, {Marcelino}, {Billot}, {Viti}, {Testi}, {Vastel}, {Lefloch}, \&
  {Bachiller}}]{jime16}
{Jim{\'e}nez-Serra}, I., {Vasyunin}, A.~I., {Caselli}, P., {et~al.} 2016, ApJL,
  830, L6, \dodoi{10.3847/2041-8205/830/1/L6}

\bibitem[{{Johnstone} {et~al.}(2010){Johnstone}, {Rosolowsky}, {Tafalla}, \&
  {Kirk}}]{john10}
{Johnstone}, D., {Rosolowsky}, E., {Tafalla}, M., \& {Kirk}, H. 2010, apj, 711,
  655, \dodoi{10.1088/0004-637X/711/2/655}

\bibitem[{{J{\o}rgensen} {et~al.}(2004{\natexlab{a}}){J{\o}rgensen},
  {Hogerheijde}, {van Dishoeck}, {Blake}, \& {Sch{\"o}ier}}]{jorg04a}
{J{\o}rgensen}, J.~K., {Hogerheijde}, M.~R., {van Dishoeck}, E.~F., {Blake},
  G.~A., \& {Sch{\"o}ier}, F.~L. 2004{\natexlab{a}}, AAP, 413, 993,
  \dodoi{10.1051/0004-6361:20031550}

\bibitem[{{J{\o}rgensen} {et~al.}(2004{\natexlab{b}}){J{\o}rgensen},
  {Sch{\"o}ier}, \& {van Dishoeck}}]{Jorg04}
{J{\o}rgensen}, J.~K., {Sch{\"o}ier}, F.~L., \& {van Dishoeck}, E.~F.
  2004{\natexlab{b}}, AAP, 416, 603, \dodoi{10.1051/0004-6361:20034440}

\bibitem[{{Keto} \& {Caselli}(2010)}]{keto10b}
{Keto}, E., \& {Caselli}, P. 2010, MNRAS, 402, 1625,
  \dodoi{10.1111/j.1365-2966.2009.16033.x}

\bibitem[{Keto \& Rybicki(2010)}]{keto10}
Keto, E., \& Rybicki, G. 2010, The Astrophysical Journal, 716, 1315,
  \dodoi{10.1088/0004-637x/716/2/1315}

\bibitem[{{Klaassen} \& {Wilson}(2007)}]{klaa07}
{Klaassen}, P.~D., \& {Wilson}, C.~D. 2007, apj, 663, 1092,
  \dodoi{10.1086/518760}

\bibitem[{{Kramers} \& {Ter Haar}(1946)}]{kram46}
{Kramers}, H.~A., \& {Ter Haar}, D. 1946, BAIN, 10, 137

\bibitem[{Kuan {et~al.}(2004)Kuan, Huang, Charnley, Hirano, Takakuwa, Wilner,
  Liu, Ohashi, Bourke, Qi, {et~al.}}]{kuan04}
Kuan, Y.-J., Huang, H.-C., Charnley, S.~B., {et~al.} 2004, The Astrophysical
  Journal, 616, L27

\bibitem[{{Kurtz} {et~al.}(2000){Kurtz}, {Cesaroni}, {Churchwell}, {Hofner}, \&
  {Walmsley}}]{kurt00}
{Kurtz}, S., {Cesaroni}, R., {Churchwell}, E., {Hofner}, P., \& {Walmsley},
  C.~M. 2000, in Protostars and Planets IV, ed. V.~{Mannings}, A.~P. {Boss}, \&
  S.~S. {Russell}, 299--326

\bibitem[{{Lattanzi} {et~al.}(2020){Lattanzi}, {Bizzocchi}, {Vasyunin},
  {Harju}, {Giuliano}, {Vastel}, \& {Caselli}}]{latt20}
{Lattanzi}, V., {Bizzocchi}, L., {Vasyunin}, A.~I., {et~al.} 2020, AAP, 633,
  A118, \dodoi{10.1051/0004-6361/201936884}

\bibitem[{{Lefloch} {et~al.}(1998){Lefloch}, {Castets}, {Cernicharo}, {Langer},
  \& {Zylka}}]{lefl98}
{Lefloch}, B., {Castets}, A., {Cernicharo}, J., {Langer}, W.~D., \& {Zylka}, R.
  1998, AAP, 334, 269

\bibitem[{{Lefloch} {et~al.}(2016){Lefloch}, {Vastel}, {Viti}, {Jimenez-Serra},
  {Codella}, {Podio}, {Ceccarelli}, {Mendoza}, {Lepine}, \&
  {Bachiller}}]{lefl16}
{Lefloch}, B., {Vastel}, C., {Viti}, S., {et~al.} 2016, MNRAS, 462, 3937,
  \dodoi{10.1093/mnras/stw1918}

\bibitem[{{Lefloch} {et~al.}(2018){Lefloch}, {Bachiller}, {Ceccarelli},
  {Cernicharo}, {Codella}, {Fuente}, {Kahane}, {L{\'o}pez-Sepulcre}, {Tafalla},
  {Vastel}, {Caux}, {Gonz{\'a}lez-Garc{\'\i}a}, {Bianchi}, {G{\'o}mez-Ruiz},
  {Holdship}, {Mendoza}, {Ospina-Zamudio}, {Podio}, {Qu{\'e}nard}, {Roueff},
  {Sakai}, {Viti}, {Yamamoto}, {Yoshida}, {Favre}, {Monfredini},
  {Quiti{\'a}n-Lara}, {Marcelino}, {Boechat-Roberty}, \& {Cabrit}}]{lefl18}
{Lefloch}, B., {Bachiller}, R., {Ceccarelli}, C., {et~al.} 2018, MNRAS, 477,
  4792, \dodoi{10.1093/mnras/sty937}

\bibitem[{{Lodders}(2003)}]{lodd03}
{Lodders}, K. 2003, ApJ, 591, 1220, \dodoi{10.1086/375492}

\bibitem[{Loison {et~al.}(2016)Loison, Ag{\'u}ndez, Marcelino, Wakelam,
  Hickson, Cernicharo, Gerin, Roueff, \& Gu{\'e}lin}]{lois16}
Loison, J.-C., Ag{\'u}ndez, M., Marcelino, N., {et~al.} 2016, Monthly Notices
  of the Royal Astronomical Society, 456, 4101

\bibitem[{{Looney} {et~al.}(2007){Looney}, {Tobin}, \& {Kwon}}]{loon07}
{Looney}, L.~W., {Tobin}, J.~J., \& {Kwon}, W. 2007, ApJL, 670, L131,
  \dodoi{10.1086/524361}

\bibitem[{L{\'o}pez-Sepulcre {et~al.}(2015)L{\'o}pez-Sepulcre, Jaber, Mendoza,
  Lefloch, Ceccarelli, Vastel, Bachiller, Cernicharo, Codella, Kahane, Kama, \&
  Tafalla}]{lope15}
L{\'o}pez-Sepulcre, A., Jaber, A.~A., Mendoza, E., {et~al.} 2015, Monthly
  Notices of the Royal Astronomical Society, 449, 2438,
  \dodoi{10.1093/mnras/stv377}

\bibitem[{{L{\'o}pez-Sepulcre} {et~al.}(2017){L{\'o}pez-Sepulcre}, {Sakai},
  {Neri}, {Imai}, {Oya}, {Ceccarelli}, {Higuchi}, {Aikawa}, {Bottinelli},
  {Caux}, {Hirota}, {Kahane}, {Lefloch}, {Vastel}, {Watanabe}, \&
  {Yamamoto}}]{lope17}
{L{\'o}pez-Sepulcre}, A., {Sakai}, N., {Neri}, R., {et~al.} 2017, AAP, 606,
  A121, \dodoi{10.1051/0004-6361/201630334}

\bibitem[{Lucas \& Gu{\'e}lin(1990)}]{luca90}
Lucas, R., \& Gu{\'e}lin, M. 1990, in Submillimetre Astronomy, ed. G.~D. Watt
  \& A.~S. Webster (Dordrecht: Springer Netherlands), 97--102

\bibitem[{Maci{\'a}(2005)}]{maci05}
Maci{\'a}, E. 2005, Chemical Society Reviews, 34, 691

\bibitem[{Maci{\'a} {et~al.}(1997)Maci{\'a}, Hern{\'a}ndez, \&
  Or{\'o}}]{maci97}
Maci{\'a}, E., Hern{\'a}ndez, M., \& Or{\'o}, J. 1997, Origins of Life and
  Evolution of the Biosphere, 27, 459

\bibitem[{{Majumdar} {et~al.}(2014a){Majumdar}, {Das}, \&
  {Chakrabarti}}]{maju14a}
{Majumdar}, L., {Das}, A., \& {Chakrabarti}, S.~K. 2014a, aap, 562, A56,
  \dodoi{10.1051/0004-6361/201322473}

\bibitem[{{Majumdar} {et~al.}(2015){Majumdar}, {Gorai}, {Das}, \&
  {Chakrabarti}}]{maju15}
{Majumdar}, L., {Gorai}, P., {Das}, A., \& {Chakrabarti}, S.~K. 2015, APSS,
  360, 18, \dodoi{10.1007/s10509-015-2567-1}

\bibitem[{{Mangum} \& {Wootten}(1993)}]{mang93}
{Mangum}, J.~G., \& {Wootten}, A. 1993, apjs, 89, 123, \dodoi{10.1086/191841}

\bibitem[{{Marcelino} {et~al.}(2009){Marcelino}, {Cernicharo}, {Tercero}, \&
  {Roueff}}]{marc09}
{Marcelino}, N., {Cernicharo}, J., {Tercero}, B., \& {Roueff}, E. 2009, ApJL,
  690, L27, \dodoi{10.1088/0004-637X/690/1/L27}

\bibitem[{{Marcelino} {et~al.}(2018){Marcelino}, {Gerin}, {Cernicharo},
  {Fuente}, {Wootten}, {Chapillon}, {Pety}, {Lis}, {Roueff}, {Commer{\c{c}}on},
  \& {Ciardi}}]{marc18b}
{Marcelino}, N., {Gerin}, M., {Cernicharo}, J., {et~al.} 2018, AAP, 620, A80,
  \dodoi{10.1051/0004-6361/201731955}

\bibitem[{{Maret} {et~al.}(2002){Maret}, {Ceccarelli}, {Caux}, {Tielens}, \&
  {Castets}}]{mare02}
{Maret}, S., {Ceccarelli}, C., {Caux}, E., {Tielens}, A.~G.~G.~M., \&
  {Castets}, A. 2002, aap, 395, 573, \dodoi{10.1051/0004-6361:20021334}

\bibitem[{{Maret} {et~al.}(2003){Maret}, {Ceccarelli}, {Caux}, {Tielens}, \&
  {Castets}}]{mare03}
---. 2003, arXiv e-prints, astro.
\newblock \doarXiv{astro-ph/0302305}

\bibitem[{{Maret} {et~al.}(2005){Maret}, {Ceccarelli}, {Tielens}, {Caux},
  {Lefloch}, {Faure}, {Castets}, \& {Flower}}]{mare05}
{Maret}, S., {Ceccarelli}, C., {Tielens}, A.~G.~G.~M., {et~al.} 2005, AAP, 442,
  527, \dodoi{10.1051/0004-6361:20052899}

\bibitem[{Margul{\`e}s {et~al.}(2020)Margul{\`e}s, McGuire, Evans, Motiyenko,
  Remijan, Guillemin, Wong, \& McNaughton}]{marg20}
Margul{\`e}s, L., McGuire, B., Evans, C., {et~al.} 2020, Astronomy \&
  Astrophysics, 642, A206

\bibitem[{{Marseille} {et~al.}(2010){Marseille}, {van der Tak}, {Herpin},
  {Wyrowski}, {Chavarr{\'\i}a}, {Pietropaoli}, {Baudry}, {Bontemps},
  {Cernicharo}, {Jacq}, {Frieswijk}, {Shipman}, {van Dishoeck}, {Bachiller},
  {Benedettini}, {Benz}, {Bergin}, {Bjerkeli}, {Blake}, {Braine}, {Bruderer},
  {Caselli}, {Caux}, {Codella}, {Daniel}, {Dieleman}, {di Giorgio}, {Dominik},
  {Doty}, {Encrenaz}, {Fich}, {Fuente}, {Gaier}, {Giannini}, {Goicoechea}, {de
  Graauw}, {Helmich}, {Herczeg}, {Hogerheijde}, {Jackson}, {Javadi}, {Jellema},
  {Johnstone}, {J{\o}rgensen}, {Kester}, {Kristensen}, {Larsson}, {Laauwen},
  {Lis}, {Liseau}, {Luinge}, {McCoey}, {Megej}, {Melnick}, {Neufeld}, {Nisini},
  {Olberg}, {Parise}, {Pearson}, {Plume}, {Risacher}, {Roelfsema},
  {Santiago-Garc{\'\i}a}, {Saraceno}, {Siegel}, {Stutzki}, {Tafalla}, {van
  Kempen}, {Visser}, {Wampfler}, \& {Y{\i}ld{\i}z}}]{mars10}
{Marseille}, M.~G., {van der Tak}, F.~F.~S., {Herpin}, F., {et~al.} 2010, AAP,
  521, L32, \dodoi{10.1051/0004-6361/201015103}

\bibitem[{{Martinez} {et~al.}(2008){Martinez}, {Betts}, {Villano}, {Eyet},
  {Snow}, \& {Bierbaum}}]{mart08}
{Martinez}, Oscar, J., {Betts}, N.~B., {Villano}, S.~M., {et~al.} 2008, apj,
  686, 1486, \dodoi{10.1086/591548}

\bibitem[{{Marvel} {et~al.}(2008){Marvel}, {Wilking}, {Claussen}, \&
  {Wootten}}]{marv08}
{Marvel}, K.~B., {Wilking}, B.~A., {Claussen}, M.~J., \& {Wootten}, A. 2008,
  ApJ, 685, 285, \dodoi{10.1086/590465}

\bibitem[{{Maxia} {et~al.}(2001){Maxia}, {Testi}, {Cesaroni}, \&
  {Walmsley}}]{maxi01}
{Maxia}, C., {Testi}, L., {Cesaroni}, R., \& {Walmsley}, C.~M. 2001, aap, 371,
  287, \dodoi{10.1051/0004-6361:20010338}

\bibitem[{{Mayen-Gijon} {et~al.}(2014){Mayen-Gijon}, {Anglada}, {Osorio},
  {Rodr{\'\i}guez}, {Lizano}, {G{\'o}mez}, \& {Carrasco-Gonz{\'a}lez}}]{maye14}
{Mayen-Gijon}, J.~M., {Anglada}, G., {Osorio}, M., {et~al.} 2014, mnras, 437,
  3766, \dodoi{10.1093/mnras/stt2172}

\bibitem[{{McElroy} {et~al.}(2013){McElroy}, {Walsh}, {Markwick}, {Cordiner},
  {Smith}, \& {Millar}}]{mcel13}
{McElroy}, D., {Walsh}, C., {Markwick}, A.~J., {et~al.} 2013, aap, 550, A36,
  \dodoi{10.1051/0004-6361/201220465}

\bibitem[{{McMullin} {et~al.}(2007){McMullin}, {Waters}, {Schiebel}, {Young},
  \& {Golap}}]{mcmu07}
{McMullin}, J.~P., {Waters}, B., {Schiebel}, D., {Young}, W., \& {Golap}, K.
  2007, in Astronomical Society of the Pacific Conference Series, Vol. 376,
  Astronomical Data Analysis Software and Systems XVI, ed. R.~A. {Shaw},
  F.~{Hill}, \& D.~J. {Bell}, 127

\bibitem[{{Milam} {et~al.}(2008){Milam}, {Halfen}, {Tenenbaum}, {Apponi},
  {Woolf}, \& {Ziurys}}]{mila08}
{Milam}, S.~N., {Halfen}, D.~T., {Tenenbaum}, E.~D., {et~al.} 2008, ApJ, 684,
  618, \dodoi{10.1086/589135}

\bibitem[{{Millar}(1991)}]{mill91}
{Millar}, T.~J. 1991, AAP, 242, 241

\bibitem[{{Miller} \& {Urey}(1959)}]{mill59}
{Miller}, S.~L., \& {Urey}, H.~C. 1959, Science, 130, 245,
  \dodoi{10.1126/science.130.3370.245}

\bibitem[{{Mininni} {et~al.}(2018){Mininni}, {Fontani}, {Rivilla},
  {Beltr{\'a}n}, {Caselli}, \& {Vasyunin}}]{mini18}
{Mininni}, C., {Fontani}, F., {Rivilla}, V.~M., {et~al.} 2018, MNRAS, 476, L39,
  \dodoi{10.1093/mnrasl/sly026}

\bibitem[{{Mondal} {et~al.}(2023){Mondal}, {Iqbal}, {Gorai}, {Bhat}, {Wakelam},
  \& {Das}}]{mond23}
{Mondal}, S.~K., {Iqbal}, W., {Gorai}, P., {et~al.} 2023, aap, 669, A71,
  \dodoi{10.1051/0004-6361/202243802}

\bibitem[{{Morales} {et~al.}(2011){Morales}, {Bennett}, {Le Picard}, {Canosa},
  {Sims}, {Sun}, {Chen}, {Chang}, {Kislov}, {Mebel}, {Gu}, {Zhang},
  {Maksyutenko}, \& {Kaiser}}]{mora11}
{Morales}, S.~B., {Bennett}, C.~J., {Le Picard}, S.~D., {et~al.} 2011, apj,
  742, 26, \dodoi{10.1088/0004-637X/742/1/26}

\bibitem[{{Mordasini} {et~al.}(2012){Mordasini}, {Alibert}, {Georgy},
  {Dittkrist}, {Klahr}, \& {Henning}}]{mord12}
{Mordasini}, C., {Alibert}, Y., {Georgy}, C., {et~al.} 2012, aap, 547, A112,
  \dodoi{10.1051/0004-6361/201118464}

\bibitem[{{Mottram} {et~al.}(2013){Mottram}, {van Dishoeck}, {Schmalzl},
  {Kristensen}, {Visser}, {Hogerheijde}, \& {Bruderer}}]{mott13}
{Mottram}, J.~C., {van Dishoeck}, E.~F., {Schmalzl}, M., {et~al.} 2013, aap,
  558, A126, \dodoi{10.1051/0004-6361/201321828}

\bibitem[{{Mueller} {et~al.}(2002){Mueller}, {Shirley}, {Evans}, \&
  {Jacobson}}]{muel02}
{Mueller}, K.~E., {Shirley}, Y.~L., {Evans}, Neal~J., I., \& {Jacobson}, H.~R.
  2002, apjs, 143, 469, \dodoi{10.1086/342881}

\bibitem[{{M{\"u}ller} {et~al.}(2005){M{\"u}ller}, {Schl{\"o}der}, {Stutzki},
  \& {Winnewisser}}]{mull05}
{M{\"u}ller}, H. S.~P., {Schl{\"o}der}, F., {Stutzki}, J., \& {Winnewisser}, G.
  2005, Journal of Molecular Structure, 742, 215,
  \dodoi{10.1016/j.molstruc.2005.01.027}

\bibitem[{{M{\"u}ller} {et~al.}(2001){M{\"u}ller}, {Thorwirth}, {Roth}, \&
  {Winnewisser}}]{mull01}
{M{\"u}ller}, H.~S.~P., {Thorwirth}, S., {Roth}, D.~A., \& {Winnewisser}, G.
  2001, AAP, 370, L49, \dodoi{10.1051/0004-6361:20010367}

\bibitem[{{Myers} {et~al.}(1998){Myers}, {Adams}, {Chen}, \& {Schaff}}]{myer98}
{Myers}, P.~C., {Adams}, F.~C., {Chen}, H., \& {Schaff}, E. 1998, apj, 492,
  703, \dodoi{10.1086/305048}

\bibitem[{{Myers} {et~al.}(1996){Myers}, {Mardones}, {Tafalla}, {Williams}, \&
  {Wilner}}]{myer96}
{Myers}, P.~C., {Mardones}, D., {Tafalla}, M., {Williams}, J.~P., \& {Wilner},
  D.~J. 1996, apjl, 465, L133, \dodoi{10.1086/310146}

\bibitem[{{Nagy} {et~al.}(2019){Nagy}, {Spezzano}, {Caselli}, {Vasyunin},
  {Tafalla}, {Bizzocchi}, {Prudenzano}, \& {Redaelli}}]{nagy19}
{Nagy}, Z., {Spezzano}, S., {Caselli}, P., {et~al.} 2019, aap, 630, A136,
  \dodoi{10.1051/0004-6361/201935568}

\bibitem[{{Nguyen} {et~al.}(2020){Nguyen}, {Oba}, {Shimonishi}, {Kouchi}, \&
  {Watanabe}}]{nguy20}
{Nguyen}, T., {Oba}, Y., {Shimonishi}, T., {Kouchi}, A., \& {Watanabe}, N.
  2020, ApJL, 898, L52, \dodoi{10.3847/2041-8213/aba695}

\bibitem[{{Noll} \& {Marley}(1997)}]{noll97}
{Noll}, K.~S., \& {Marley}, M.~S. 1997, in Astronomical Society of the Pacific
  Conference Series, Vol. 119, Planets Beyond the Solar System and the Next
  Generation of Space Missions, ed. D.~{Soderblom}, 115

\bibitem[{Nummelin {et~al.}(2000)Nummelin, Bergman, Hjalmarson, Friberg,
  Irvine, Millar, Ohishi, \& Saito}]{numm00}
Nummelin, A., Bergman, P., Hjalmarson, {\AA}., {et~al.} 2000, The Astrophysical
  Journal Supplement Series, 128, 213

\bibitem[{{{\"O}berg} {et~al.}(2010){{\"O}berg}, {Bottinelli}, {J{\o}rgensen},
  \& {van Dishoeck}}]{ober10}
{{\"O}berg}, K.~I., {Bottinelli}, S., {J{\o}rgensen}, J.~K., \& {van Dishoeck},
  E.~F. 2010, ApJ, 716, 825, \dodoi{10.1088/0004-637X/716/1/825}

\bibitem[{{{\"O}berg} {et~al.}(2009){{\"O}berg}, {Bottinelli}, \& {van
  Dishoeck}}]{ober09}
{{\"O}berg}, K.~I., {Bottinelli}, S., \& {van Dishoeck}, E.~F. 2009, AAP, 494,
  L13, \dodoi{10.1051/0004-6361:200811228}

\bibitem[{{Olmi} {et~al.}(1996){Olmi}, {Cesaroni}, \& {Walmsley}}]{olmi96}
{Olmi}, L., {Cesaroni}, R., \& {Walmsley}, C.~M. 1996, aap, 307, 599

\bibitem[{{Ortiz-Le{\'o}n} {et~al.}(2018){Ortiz-Le{\'o}n}, {Loinard}, {Dzib},
  {Galli}, {Kounkel}, {Mioduszewski}, {Rodr{\'\i}guez}, {Torres}, {Hartmann},
  {Boden}, {Evans}, {Brice{\~n}o}, \& {Tobin}}]{orti18}
{Ortiz-Le{\'o}n}, G.~N., {Loinard}, L., {Dzib}, S.~A., {et~al.} 2018, ApJ, 865,
  73, \dodoi{10.3847/1538-4357/aada49}

\bibitem[{{Osorio} {et~al.}(2009){Osorio}, {Anglada}, {Lizano}, \&
  {D'Alessio}}]{osor09}
{Osorio}, M., {Anglada}, G., {Lizano}, S., \& {D'Alessio}, P. 2009, apj, 694,
  29, \dodoi{10.1088/0004-637X/694/1/29}

\bibitem[{{Ossenkopf} \& {Henning}(1994)}]{osse94}
{Ossenkopf}, V., \& {Henning}, T. 1994, AAP, 291, 943

\bibitem[{Palau {et~al.}(2014)Palau, Estalella, Girart, Fuente, Fontani,
  Commer{\c{c}}on, Busquet, Bontemps, S{\'{a}}nchez-Monge, Zapata, Zhang,
  Hennebelle, \& di~Francesco}]{pala14}
Palau, A., Estalella, R., Girart, J.~M., {et~al.} 2014, The Astrophysical
  Journal, 785, 42, \dodoi{10.1088/0004-637x/785/1/42}

\bibitem[{{Pasek}(2019)}]{pase19}
{Pasek}, M.~A. 2019, Icarus, 317, 59, \dodoi{10.1016/j.icarus.2018.07.011}

\bibitem[{Peeters {et~al.}(2006)Peeters, Rodgers, Charnley, Schriver-Mazzuoli,
  Schriver, Keane, \& Ehrenfreund}]{peet06}
Peeters, Z., Rodgers, S., Charnley, S., {et~al.} 2006, Astronomy \&
  Astrophysics, 445, 197

\bibitem[{{Pezzuto} {et~al.}(2012){Pezzuto}, {Elia}, {Schisano}, {Strafella},
  {Di Francesco}, {Sadavoy}, {Andr{\'e}}, {Benedettini}, {Bernard}, {di
  Giorgio}, {Facchini}, {Hennemann}, {Hill}, {K{\"o}nyves}, {Molinari},
  {Motte}, {Nguyen-Luong}, {Peretto}, {Pestalozzi}, {Polychroni}, {Rygl},
  {Saraceno}, {Schneider}, {Spinoglio}, {Testi}, {Ward-Thompson}, \&
  {White}}]{pezz12}
{Pezzuto}, S., {Elia}, D., {Schisano}, E., {et~al.} 2012, AAP, 547, A54,
  \dodoi{10.1051/0004-6361/201219501}

\bibitem[{{Pickett} {et~al.}(1998){Pickett}, {Poynter}, {Cohen}, {Delitsky},
  {Pearson}, \& {M{\"u}ller}}]{pick98}
{Pickett}, H.~M., {Poynter}, R.~L., {Cohen}, E.~A., {et~al.} 1998, JQSRT, 60,
  883, \dodoi{10.1016/S0022-4073(98)00091-0}

\bibitem[{{Redaelli} {et~al.}(2019){Redaelli}, {Bizzocchi}, {Caselli},
  {Sipil{\"a}}, {Lattanzi}, {Giuliano}, \& {Spezzano}}]{rade19}
{Redaelli}, E., {Bizzocchi}, L., {Caselli}, P., {et~al.} 2019, AAP, 629, A15,
  \dodoi{10.1051/0004-6361/201935314}

\bibitem[{{Reid} {et~al.}(2019){Reid}, {Menten}, {Brunthaler}, {Zheng}, {Dame},
  {Xu}, {Li}, {Sakai}, {Wu}, {Immer}, {Zhang}, {Sanna}, {Moscadelli}, {Rygl},
  {Bartkiewicz}, {Hu}, {Quiroga-Nu{\~n}ez}, \& {van Langevelde}}]{reid19}
{Reid}, M.~J., {Menten}, K.~M., {Brunthaler}, A., {et~al.} 2019, apj, 885, 131,
  \dodoi{10.3847/1538-4357/ab4a11}

\bibitem[{{Reipurth} {et~al.}(1993){Reipurth}, {Chini}, {Krugel}, {Kreysa}, \&
  {Sievers}}]{reip93}
{Reipurth}, B., {Chini}, R., {Krugel}, E., {Kreysa}, E., \& {Sievers}, A. 1993,
  AAP, 273, 221

\bibitem[{{Rivilla} {et~al.}(2017){Rivilla}, {Beltr{\'a}n}, {Cesaroni},
  {Fontani}, {Codella}, \& {Zhang}}]{rivi17}
{Rivilla}, V.~M., {Beltr{\'a}n}, M.~T., {Cesaroni}, R., {et~al.} 2017, aap,
  598, A59, \dodoi{10.1051/0004-6361/201628373}

\bibitem[{{Rivilla} {et~al.}(2016){Rivilla}, {Fontani}, {Beltr{\'a}n},
  {Vasyunin}, {Caselli}, {Mart{\'\i}n-Pintado}, \& {Cesaroni}}]{rivi16}
{Rivilla}, V.~M., {Fontani}, F., {Beltr{\'a}n}, M.~T., {et~al.} 2016, ApJ, 826,
  161, \dodoi{10.3847/0004-637X/826/2/161}

\bibitem[{{Rivilla} {et~al.}(2020){Rivilla}, {Drozdovskaya}, {Altwegg},
  {Caselli}, {Beltr{\'a}n}, {Fontani}, {van der Tak}, {Cesaroni}, {Vasyunin},
  {Rubin}, {Lique}, {Marinakis}, {Testi}, {Rosina Team}, {Balsiger},
  {Berthelier}, {de Keyser}, {Fiethe}, {Fuselier}, {Gasc}, {Gombosi},
  {S{\'e}mon}, \& {Tzou}}]{rivi20}
{Rivilla}, V.~M., {Drozdovskaya}, M.~N., {Altwegg}, K., {et~al.} 2020, MNRAS,
  492, 1180, \dodoi{10.1093/mnras/stz3336}

\bibitem[{{Rolffs} {et~al.}(2010){Rolffs}, {Schilke}, {Comito}, {Bergin}, {van
  der Tak}, {Lis}, {Qin}, {Menten}, {G{\"u}sten}, {Bell}, {Blake}, {Caux},
  {Ceccarelli}, {Cernicharo}, {Crockett}, {Daniel}, {Dubernet},
  {Emprechtinger}, {Encrenaz}, {Gerin}, {Giesen}, {Goicoechea}, {Goldsmith},
  {Gupta}, {Herbst}, {Joblin}, {Johnstone}, {Langer}, {Latter}, {Lord},
  {Maret}, {Martin}, {Melnick}, {Morris}, {M{\"u}ller}, {Murphy}, {Ossenkopf},
  {Pearson}, {P{\'e}rault}, {Phillips}, {Plume}, {Schlemmer}, {Stutzki},
  {Trappe}, {Vastel}, {Wang}, {Yorke}, {Yu}, {Zmuidzinas}, {Diez-Gonzalez},
  {Bachiller}, {Martin-Pintado}, {Baechtold}, {Olberg}, {Nordh}, {Gill}, \&
  {Chattopadhyay}}]{rolf10}
{Rolffs}, R., {Schilke}, P., {Comito}, C., {et~al.} 2010, AAP, 521, L46,
  \dodoi{10.1051/0004-6361/201015106}

\bibitem[{{Ruaud} {et~al.}(2016){Ruaud}, {Wakelam}, \& {Hersant}}]{ruau16}
{Ruaud}, M., {Wakelam}, V., \& {Hersant}, F. 2016, mnras, 459, 3756,
  \dodoi{10.1093/mnras/stw887}

\bibitem[{{Rybicki} \& {Hummer}(1991)}]{rybi91}
{Rybicki}, G.~B., \& {Hummer}, D.~G. 1991, AAP, 245, 171

\bibitem[{{Sahu} {et~al.}(2019){Sahu}, {Liu}, {Su}, {Li}, {Lee}, {Hirano}, \&
  {Takakuwa}}]{sahu19}
{Sahu}, D., {Liu}, S.-Y., {Su}, Y.-N., {et~al.} 2019, ApJ, 872, 196,
  \dodoi{10.3847/1538-4357/aaffda}

\bibitem[{{Sahu} {et~al.}(2018){Sahu}, {Minh}, {Lee}, {Liu}, {Das},
  {Chakrabarti}, \& {Sivaraman}}]{sahu18}
{Sahu}, D., {Minh}, Y.~C., {Lee}, C.-F., {et~al.} 2018, MNRAS, 475, 5322,
  \dodoi{10.1093/mnras/sty190}

\bibitem[{{Sandford} \& {Allamandola}(1993)}]{sand93}
{Sandford}, S.~A., \& {Allamandola}, L.~J. 1993, ApJ, 417, 815,
  \dodoi{10.1086/173362}

\bibitem[{{Santangelo} {et~al.}(2015){Santangelo}, {Codella}, {Cabrit},
  {Maury}, {Gueth}, {Maret}, {Lefloch}, {Belloche}, {Andr{\'e}}, {Hennebelle},
  {Anderl}, {Podio}, \& {Testi}}]{sant15}
{Santangelo}, G., {Codella}, C., {Cabrit}, S., {et~al.} 2015, AAP, 584, A126,
  \dodoi{10.1051/0004-6361/201526323}

\bibitem[{{Schlafly} {et~al.}(2014){Schlafly}, {Green}, {Finkbeiner}, {Rix},
  {Bell}, {Burgett}, {Chambers}, {Draper}, {Hodapp}, {Kaiser}, {Magnier},
  {Martin}, {Metcalfe}, {Price}, \& {Tonry}}]{schl14}
{Schlafly}, E.~F., {Green}, G., {Finkbeiner}, D.~P., {et~al.} 2014, ApJ, 786,
  29, \dodoi{10.1088/0004-637X/786/1/29}

\bibitem[{{Sch{\"o}ier} {et~al.}(2005){Sch{\"o}ier}, {van der Tak}, {van
  Dishoeck}, \& {Black}}]{scho05}
{Sch{\"o}ier}, F.~L., {van der Tak}, F.~F.~S., {van Dishoeck}, E.~F., \&
  {Black}, J.~H. 2005, AAP, 432, 369, \dodoi{10.1051/0004-6361:20041729}

\bibitem[{{Shimonishi} {et~al.}(2016){Shimonishi}, {Onaka}, {Kawamura}, \&
  {Aikawa}}]{shim16}
{Shimonishi}, T., {Onaka}, T., {Kawamura}, A., \& {Aikawa}, Y. 2016, apj, 827,
  72, \dodoi{10.3847/0004-637X/827/1/72}

\bibitem[{Shirley(2015)}]{shir15}
Shirley, Y.~L. 2015, Publications of the Astronomical Society of the Pacific,
  127, 299–310, \dodoi{10.1086/680342}

\bibitem[{{Shu} {et~al.}(1994){Shu}, {Najita}, {Ostriker}, {Wilkin}, {Ruden},
  \& {Lizano}}]{shu94}
{Shu}, F., {Najita}, J., {Ostriker}, E., {et~al.} 1994, apj, 429, 781,
  \dodoi{10.1086/174363}

\bibitem[{{Shu}(1977)}]{shuf77}
{Shu}, F.~H. 1977, ApJ, 214, 488, \dodoi{10.1086/155274}

\bibitem[{{Shu} {et~al.}(1987){Shu}, {Adams}, \& {Lizano}}]{shu87}
{Shu}, F.~H., {Adams}, F.~C., \& {Lizano}, S. 1987, araa, 25, 23,
  \dodoi{10.1146/annurev.aa.25.090187.000323}

\bibitem[{{Shu} \& {Li}(1997)}]{shu97}
{Shu}, F.~H., \& {Li}, Z.-Y. 1997, apj, 475, 251, \dodoi{10.1086/303521}

\bibitem[{{Shu} \& {Lizano}(1988)}]{shu88a}
{Shu}, F.~H., \& {Lizano}, S. 1988, Astrophysical Letters and Communications,
  26, 217

\bibitem[{{Shu} {et~al.}(1988){Shu}, {Lizano}, {Ruden}, \& {Najita}}]{shu88b}
{Shu}, F.~H., {Lizano}, S., {Ruden}, S.~P., \& {Najita}, J. 1988, apjl, 328,
  L19, \dodoi{10.1086/185152}

\bibitem[{{Sil} {et~al.}(2018){Sil}, {Gorai}, {Das}, {Bhat}, {Etim}, \&
  {Chakrabarti}}]{sil18}
{Sil}, M., {Gorai}, P., {Das}, A., {et~al.} 2018, apj, 853, 139,
  \dodoi{10.3847/1538-4357/aa984d}

\bibitem[{{Sil} {et~al.}(2017){Sil}, {Gorai}, {Das}, {Sahu}, \&
  {Chakrabarti}}]{sil17}
{Sil}, M., {Gorai}, P., {Das}, A., {Sahu}, D., \& {Chakrabarti}, S.~K. 2017,
  European Physical Journal D, 71, 45, \dodoi{10.1140/epjd/e2017-70610-4}

\bibitem[{{Sil} {et~al.}(2021){Sil}, {Srivastav}, {Bhat}, {Mondal}, {Gorai},
  {Ghosh}, {Shimonishi}, {Chakrabarti}, {Sivaraman}, {Pathak}, {Nakatani},
  {Furuya}, \& {Das}}]{sil21}
{Sil}, M., {Srivastav}, S., {Bhat}, B., {et~al.} 2021, AJ, 162, 119,
  \dodoi{10.3847/1538-3881/ac09f9}

\bibitem[{{Sims} {et~al.}(1993){Sims}, {Queffelec}, {Travers}, {Rowe},
  {Herbert}, {Karth{\"a}user}, \& {Smith}}]{sims93}
{Sims}, I.~R., {Queffelec}, J.-L., {Travers}, D., {et~al.} 1993, Chemical
  Physics Letters, 211, 461, \dodoi{10.1016/0009-2614(93)87091-G}

\bibitem[{Snellen {et~al.}(2020)Snellen, Guzman-Ramirez, Hogerheijde, Hygate,
  \& van~der Tak}]{snell20}
Snellen, I. A.~G., Guzman-Ramirez, L., Hogerheijde, M.~R., Hygate, A. P.~S., \&
  van~der Tak, F. F.~S. 2020, Re-analysis of the 267-GHz ALMA observations of
  Venus: No statistically significant detection of phosphine.
\newblock \doarXiv{2010.09761}

\bibitem[{Snyder {et~al.}(1974)Snyder, Buhl, Schwartz, Clark, Johnson, Lovas,
  \& Giguere}]{snyd74}
Snyder, L., Buhl, D., Schwartz, P., {et~al.} 1974, The Astrophysical Journal,
  191, L79

\bibitem[{{Sobolev}(1960)}]{sobo60}
{Sobolev}, V.~V. 1960, {Moving envelopes of stars}

\bibitem[{{Sousa-Silva} {et~al.}(2020){Sousa-Silva}, {Seager}, {Ranjan},
  {Petkowski}, {Zhan}, {Hu}, \& {Bains}}]{sous20}
{Sousa-Silva}, C., {Seager}, S., {Ranjan}, S., {et~al.} 2020, Astrobiology, 20,
  235, \dodoi{10.1089/ast.2018.1954}

\bibitem[{{Stahler} \& {Palla}(2004)}]{stah04}
{Stahler}, S.~W., \& {Palla}, F. 2004, {The Formation of Stars}

\bibitem[{Sutton {et~al.}(1995)Sutton, Peng, Danchi, Jaminet, Sandell, \&
  Russell}]{sutt95}
Sutton, E., Peng, R., Danchi, W., {et~al.} 1995, The Astrophysical Journal
  Supplement Series, 97, 455

\bibitem[{{Tafalla} {et~al.}(1998){Tafalla}, {Mardones}, {Myers}, {Caselli},
  {Bachiller}, \& {Benson}}]{tafa98}
{Tafalla}, M., {Mardones}, D., {Myers}, P.~C., {et~al.} 1998, ApJ, 504, 900,
  \dodoi{10.1086/306115}

\bibitem[{{Taquet} {et~al.}(2015){Taquet}, {L{\'o}pez-Sepulcre}, {Ceccarelli},
  {Neri}, {Kahane}, \& {Charnley}}]{taqu15}
{Taquet}, V., {L{\'o}pez-Sepulcre}, A., {Ceccarelli}, C., {et~al.} 2015, ApJ,
  804, 81, \dodoi{10.1088/0004-637X/804/2/81}

\bibitem[{{Tarrago} {et~al.}(1992){Tarrago}, {Lacome}, {L{\'e}vy},
  {Guelachvili}, {B{\'e}zard}, \& {Drossart}}]{tarr92}
{Tarrago}, G., {Lacome}, N., {L{\'e}vy}, A., {et~al.} 1992, Journal of
  Molecular Spectroscopy, 154, 30, \dodoi{10.1016/0022-2852(92)90026-K}

\bibitem[{{Tenenbaum} {et~al.}(2007){Tenenbaum}, {Woolf}, \& {Ziurys}}]{tene07}
{Tenenbaum}, E.~D., {Woolf}, N.~J., \& {Ziurys}, L.~M. 2007, ApJL, 666, L29,
  \dodoi{10.1086/521361}

\bibitem[{Tenenbaum \& Ziurys(2008)}]{tene08}
Tenenbaum, E.~D., \& Ziurys, L.~M. 2008, The Astrophysical Journal, 680, L121,
  \dodoi{10.1086/589973}

\bibitem[{{Turner}(1991)}]{turn91}
{Turner}, B.~E. 1991, ApJS, 76, 617, \dodoi{10.1086/191577}

\bibitem[{{Turner} \& {Bally}(1987)}]{turn87}
{Turner}, B.~E., \& {Bally}, J. 1987, ApJL, 321, L75, \dodoi{10.1086/185009}

\bibitem[{{Turner} {et~al.}(1999){Turner}, {Terzieva}, \& {Herbst}}]{turn99}
{Turner}, B.~E., {Terzieva}, R., \& {Herbst}, E. 1999, ApJ, 518, 699,
  \dodoi{10.1086/307300}

\bibitem[{{Turner} {et~al.}(1990){Turner}, {Tsuji}, {Bally}, {Guelin}, \&
  {Cernicharo}}]{turn90}
{Turner}, B.~E., {Tsuji}, T., {Bally}, J., {Guelin}, M., \& {Cernicharo}, J.
  1990, ApJ, 365, 569, \dodoi{10.1086/169511}

\bibitem[{{Valenti} \& {Fischer}(2008)}]{vale08}
{Valenti}, J.~A., \& {Fischer}, D.~A. 2008, Physica Scripta Volume T, 130,
  014003, \dodoi{10.1088/0031-8949/2008/T130/014003}

\bibitem[{{van der Tak} {et~al.}(2007){van der Tak}, {Black}, {Sch{\"o}ier},
  {Jansen}, \& {van Dishoeck}}]{vand07}
{van der Tak}, F.~F.~S., {Black}, J.~H., {Sch{\"o}ier}, F.~L., {Jansen}, D.~J.,
  \& {van Dishoeck}, E.~F. 2007, AAP, 468, 627,
  \dodoi{10.1051/0004-6361:20066820}

\bibitem[{{van der Tak} {et~al.}(2013){van der Tak}, {Chavarr{\'\i}a},
  {Herpin}, {Wyrowski}, {Walmsley}, {van Dishoeck}, {Benz}, {Bergin},
  {Caselli}, {Hogerheijde}, {Johnstone}, {Kristensen}, {Liseau}, {Nisini}, \&
  {Tafalla}}]{vand13}
{van der Tak}, F.~F.~S., {Chavarr{\'\i}a}, L., {Herpin}, F., {et~al.} 2013,
  aap, 554, A83, \dodoi{10.1051/0004-6361/201220976}

\bibitem[{{van Dishoeck}(2006)}]{vand06}
{van Dishoeck}, E.~F. 2006, Proceedings of the National Academy of Science,
  103, 12249, \dodoi{10.1073/pnas.0602207103}

\bibitem[{{van Dishoeck} \& {Blake}(1998)}]{vand98}
{van Dishoeck}, E.~F., \& {Blake}, G.~A. 1998, ARAA, 36, 317,
  \dodoi{10.1146/annurev.astro.36.1.317}

\bibitem[{{Vastel} {et~al.}(2015){Vastel}, {Bottinelli}, {Caux}, {Glorian}, \&
  {Boiziot}}]{vast15}
{Vastel}, C., {Bottinelli}, S., {Caux}, E., {Glorian}, J.~M., \& {Boiziot}, M.
  2015, in SF2A-2015: Proceedings of the Annual meeting of the French Society
  of Astronomy and Astrophysics, 313--316

\bibitem[{{Vastel} {et~al.}(2014){Vastel}, {Ceccarelli}, {Lefloch}, \&
  {Bachiller}}]{vast14}
{Vastel}, C., {Ceccarelli}, C., {Lefloch}, B., \& {Bachiller}, R. 2014, ApJL,
  795, L2, \dodoi{10.1088/2041-8205/795/1/L2}

\bibitem[{{Vastel} {et~al.}(2018){Vastel}, {Qu{\'e}nard}, {Le Gal}, {Wakelam},
  {Andrianasolo}, {Caselli}, {Vidal}, {Ceccarelli}, {Lefloch}, \&
  {Bachiller}}]{vast18}
{Vastel}, C., {Qu{\'e}nard}, D., {Le Gal}, R., {et~al.} 2018, MNRAS, 478, 5514,
  \dodoi{10.1093/mnras/sty1336}

\bibitem[{Villanueva {et~al.}(2020)Villanueva, Cordiner, Irwin, de~Pater,
  Butler, Gurwell, Milam, Nixon, Luszcz-Cook, Wilson, Kofman, Liuzzi, Faggi,
  Fauchez, Lippi, Cosentino, Thelen, Moullet, Hartogh, Molter, Charnley, Arney,
  Mandell, Biver, Vandaele, de~Kleer, \& Kopparapu}]{vill20}
Villanueva, G., Cordiner, M., Irwin, P., {et~al.} 2020, No phosphine in the
  atmosphere of Venus.
\newblock \doarXiv{2010.14305}

\bibitem[{{Visscher} {et~al.}(2006){Visscher}, {Lodders}, \& {Fegley}}]{viss06}
{Visscher}, C., {Lodders}, K., \& {Fegley}, Bruce, J. 2006, ApJ, 648, 1181,
  \dodoi{10.1086/506245}

\bibitem[{{Wakelam} {et~al.}(2017){Wakelam}, {Loison}, {Mereau}, \&
  {Ruaud}}]{wake17}
{Wakelam}, V., {Loison}, J.~C., {Mereau}, R., \& {Ruaud}, M. 2017, Molecular
  Astrophysics, 6, 22, \dodoi{10.1016/j.molap.2017.01.002}

\bibitem[{{Wang} \& {Fischer}(2015)}]{wang15}
{Wang}, J., \& {Fischer}, D.~A. 2015, aj, 149, 14,
  \dodoi{10.1088/0004-6256/149/1/14}

\bibitem[{{Weinreb} {et~al.}(1963){Weinreb}, {Barrett}, {Meeks}, \&
  {Henry}}]{wein63}
{Weinreb}, S., {Barrett}, A.~H., {Meeks}, M.~L., \& {Henry}, J.~C. 1963,
  Nature, 200, 829, \dodoi{10.1038/200829a0}

\bibitem[{{Wilson} {et~al.}(1970){Wilson}, {Jefferts}, \& {Penzias}}]{wils70}
{Wilson}, R.~W., {Jefferts}, K.~B., \& {Penzias}, A.~A. 1970, ApJL, 161, L43,
  \dodoi{10.1086/180567}

\bibitem[{{Wyrowski} {et~al.}(2016){Wyrowski}, {G{\"u}sten}, {Menten},
  {Wiesemeyer}, {Csengeri}, {Heyminck}, {Klein}, {K{\"o}nig}, \&
  {Urquhart}}]{wyro16}
{Wyrowski}, F., {G{\"u}sten}, R., {Menten}, K.~M., {et~al.} 2016, aap, 585,
  A149, \dodoi{10.1051/0004-6361/201526361}

\bibitem[{{Yamaguchi} {et~al.}(2011){Yamaguchi}, {Takano}, {Sakai}, {Sakai},
  {Liu}, {Su}, {Hirano}, {Takakuwa}, {Aikawa}, {Nomura}, \&
  {Yamamoto}}]{yama11}
{Yamaguchi}, T., {Takano}, S., {Sakai}, N., {et~al.} 2011, PASJ, 63, L37,
  \dodoi{10.1093/pasj/63.5.L37}

\bibitem[{{Yamamoto}(2017)}]{yama17}
{Yamamoto}, S. 2017, {Introduction to Astrochemistry: Chemical Evolution from
  Interstellar Clouds to Star and Planet Formation},
  \dodoi{10.1007/978-4-431-54171-4}

\bibitem[{{Yang} {et~al.}(2021){Yang}, {Sakai}, {Zhang}, {Murillo}, {Zhang},
  {Higuchi}, {Zeng}, {L{\'o}pez-Sepulcre}, {Yamamoto}, {Lefloch}, {Bouvier},
  {Ceccarelli}, {Hirota}, {Imai}, {Oya}, {Sakai}, \& {Watanabe}}]{yang21}
{Yang}, Y.-L., {Sakai}, N., {Zhang}, Y., {et~al.} 2021, apj, 910, 20,
  \dodoi{10.3847/1538-4357/abdfd6}

\bibitem[{{Zhu} {et~al.}(2011){Zhu}, {Zhao}, \& {Wright}}]{zhul11}
{Zhu}, L., {Zhao}, J.-H., \& {Wright}, M.~C.~H. 2011, apj, 740, 114,
  \dodoi{10.1088/0004-637X/740/2/114}

\bibitem[{{Zinnecker} \& {Yorke}(2007)}]{zinn07}
{Zinnecker}, H., \& {Yorke}, H.~W. 2007, araa, 45, 481,
  \dodoi{10.1146/annurev.astro.44.051905.092549}

\bibitem[{{Ziurys}(1987)}]{ziur87}
{Ziurys}, L.~M. 1987, ApJL, 321, L81, \dodoi{10.1086/185010}

\bibitem[{{Ziurys} {et~al.}(2018){Ziurys}, {Schmidt}, \& {Bernal}}]{ziur18}
{Ziurys}, L.~M., {Schmidt}, D.~R., \& {Bernal}, J.~J. 2018, ApJ, 856, 169,
  \dodoi{10.3847/1538-4357/aaafc6}

\bibitem[{{Zucker} {et~al.}(2018){Zucker}, {Schlafly}, {Speagle}, {Green},
  {Portillo}, {Finkbeiner}, \& {Goodman}}]{zuck18}
{Zucker}, C., {Schlafly}, E.~F., {Speagle}, J.~S., {et~al.} 2018, ApJ, 869, 83,
  \dodoi{10.3847/1538-4357/aae97c}

\bibitem[{Zuckerman {et~al.}(1975)Zuckerman, Turner, Johnson, Clark, Lovas,
  Fourikis, Palmer, Morris, Lilley, Ball, {et~al.}}]{zuck75}
Zuckerman, B., Turner, B., Johnson, D., {et~al.} 1975, The Astrophysical
  Journal, 196, L99

\end{thebibliography}

\end{document}